\documentclass[11pt]{article}
\usepackage[mathscr]{eucal}
\usepackage{epsfig,amsfonts}
\usepackage{amsmath}
\usepackage{amsthm,amssymb}
\usepackage{wasysym}
\usepackage{bm}
\usepackage{graphicx}
\usepackage{relsize}
\usepackage{titlesec}
\usepackage{hhline}
\usepackage{comment}
\usepackage{cite}
\usepackage{psfrag}
\usepackage{mathrsfs} 
\usepackage{hyperref}
\usepackage{bm}
\usepackage{array}
\usepackage{tikz}
\usetikzlibrary{patterns}
\usepackage{enumerate}
\usepackage{textcomp}
\usepackage{hyperref}
\usepackage{mathrsfs} 
\usepackage{pifont}
\DeclareRobustCommand{\bmT}[1]{\bm{#1}}
\usepackage{lettrine}

\DeclareMathAlphabet{\mathpzc}{OT1}{pzc}{m}{it}
\DeclareMathAlphabet\mathbfcal{OMS}{cmsy}{b}{n} 
\makeatletter
\@addtoreset{equation}{section}
\makeatother

\def\bea{\begin{eqnarray}}
\def\eea{\end{eqnarray}}
\def\be{\begin{equation}}
\def\ee{\end{equation}}

\def\be{\begin{equation}}
\def\ee{\end{equation}}
\def\bdm{\begin{displaymath}}
\def\edm{\end{displaymath}}
\def\bea{\begin{eqnarray}}
\def\eea{\end{eqnarray}}

\def\sgn{{\rm sgn}}

\def\ri{{\rm i}}
\def\half{\textstyle\frac{1}{2}}

\def\XXint#1#2#3{{\setbox0=\hbox{$#1{#2#3}{\int}$}
    \vcenter{\hbox{$#2#3$}}\kern-.5\wd0}}

\newcommand{\rd}{\mbox{d}}
\newcommand{\re}{\mbox{e}}

\newcommand{\Ecal}{{\mathcal E}}

\newcommand{\Hcal}{{\mathcal H}}

\newcommand{\leng}{{N}}
\DeclareMathAlphabet{\mathpzc}{OT1}{pzc}{m}{it}

\begin{document}

\begin{titlepage}
\begin{flushright}
$\phantom{{\it tresrtfdsgqw }}$\\
\end{flushright}
\begin{flushright}
DESY 20-166\\
\end{flushright}

\vspace{0.8cm}

\begin{center}
\begin{LARGE}

{\bf Scaling limit of the ${\cal Z}_2$ invariant
inhomogeneous six-vertex model }

\end{LARGE}

\vspace{1.3cm}
\begin{large}

{\bf Vladimir V. Bazhanov$^{1}$,  Gleb A.  Kotousov$^{2}$  \\
\bigskip
Sergii M. Koval$^{1}$ and Sergei  L. Lukyanov$^{3,4}$}

\end{large}

\vspace{1.cm}
$^1$Department of Theoretical Physics\\
         Research School of Physics and Engineering\\
    Australian National University, Canberra, ACT 2601, Australia\\\ \\
$^2$DESY, Theory Group, Notkestrasse 85, Hamburg 22607, Germany\\
\vspace{.4cm}

${}^{3}$NHETC, Department of Physics and Astronomy\\
     Rutgers University\\
     Piscataway, NJ 08855-0849, USA\\
\vspace{.2cm}
and\\
\vspace{.2cm}
${}^{4}$Kharkevich Institute for Information Transmission Problems\\
Moscow, 127994, Russia
\vspace{1.0cm}

\end{center}

\begin{center}
\centerline{\bf Abstract} \vspace{.8cm}

\parbox{13cm}{%
The work contains a detailed study of the scaling limit of a certain
critical, integrable  inhomogeneous six-vertex model
 subject to twisted boundary conditions.
It is based on a numerical analysis of the Bethe ansatz equations
as well as the powerful analytic technique of the ODE/IQFT correspondence.
The results indicate that the critical behaviour of the lattice system is described by
the gauged ${\rm SL}(2)$ WZW model with certain boundary  and reality
conditions imposed on the fields. Our proposal revises and extends the
conjectured relation  between the
lattice system and the Euclidean black hole non-linear sigma model
that was made in the 2011 paper of 
 Ikhlef, Jacobsen and Saleur.
}
\end{center}

\vfill

\end{titlepage}
\setcounter{page}{2}

\tableofcontents
\newpage

\section{Introduction}
The seminal work of Polyakov on the
$O(n)$ models \cite{Polyakov:1975rr} opened an era in
 the study of quantum Non-Linear
 Sigma Models (NLSM) in $1+1$ dimensions. 
Among their most prominent physical applications
is the description of the universality class of phase transitions
in disordered electronic systems \cite{Efetov:1983xg,Pruisken:1984ni,Read:2001pz}.
Taking inspiration from the AdS/CFT correspondence \cite{Berkovits:1999im,Bershadsky:1999hk}, 
an interesting proposal
was made in ref.\cite{Zirnbauer} for the NLSM 
that would describe
the transition between the
plateaus in the quantum Hall effect in a $2D$ disordered electron gas.
One of the basic principles for identifying the target space background,
in the author's own words, was the following
\medskip

``\emph{In trying to solve the statistical
physics problem at hand, we have to be very discriminating about which functional
integral to accept as well-defined and which to not. In concrete terms, we
are looking for a field theory defined over $\bm{ Euclidean}$ two-space, and with a target
space of $\bm{ Euclidean}$ signature. This constraint eliminates candidate theories
with an action functional that is bounded neither from below nor from above.
Among these are the above supergroups, the natural supergeometry of which is
non-Riemann, or of indefinite signature. (The natural geometry is forced on us
by symmetry considerations.)}''
\medskip

\noindent
The requirement of Euclidean signature for the target space,
which is closely related to the unitarity of the model,
 is well motivated from the technical point of view. 
However, the original heuristic treatment of the problem 
relied  on a fermionic version of the replica trick, leading to the 
Pruisken model -- a $G/H$ NLSM where $G=U(2n)$ is gauged by $H=U(n)\times U(n)$
with $n=0$ \cite{Pruisken:1984ni}. In light of this the above requirement
may seem as too severe.
\bigskip

As was explained  in  \cite{Zirnbauer} the Pruisken model
 shares the same infra-red behaviour as a certain 
one dimensional spin chain, whose degrees of freedom
take values in an alternating sequence of
 modules ${ V}$ and ${ V}^*$ for the super Lie algebra  $\mathfrak{gl}(2,2)$.
This super spin chain turns out not to be integrable  in the Yang-Baxter sense,
and there has been little progress towards its solution.
Nevertheless, interest was prompted  into studying
integrable  critical ``alternating'' spin chains \cite{Essler:2005ag,Jacobsen:2005xz,Ikhlef:2008zz,
Ikhlef:2011ay, Frahm:2012eb,Frahm:2013cma,Candu:2013fva,Bazhanov:2019xvy}.
Perhaps the most remarkable output
of this study was the conjecture formulated in ref.\cite{Ikhlef:2011ay}.
\bigskip

The proposal of Ikhlef, Jacobsen and Saleur concerns a critical spin chain,
belonging to the integrability class of 
a ${\cal Z}_2$ invariant inhomogeneous six-vertex model, which is
 a special case of the lattice system introduced by Baxter in 1971 \cite{Baxter:1971}.
They present highly non-trivial arguments, including numerical evidence, 
that the  infra-red behaviour of the spin chain is governed by the
so-called Euclidean black hole NLSM\cite{Elitzur:1991cb,Mandal:1991tz,Witten:1991yr,Dijkgraaf:1991ba,ZAM,Maldacena:2000hw,
Maldacena:2000kv,Hanany:2002ev,Ribault:2003ss,Schomerus:2005aq}. 
However their proposal raises an immediate question. For a spin chain of
finite length,
 the energy spectrum is complex so that 
there does not exist any positive definite inner product w.r.t. which
the spin chain Hamiltonian is Hermitian. On the other hand the 
Euclidean black hole NLSM is a unitary CFT \cite{Dixon:1989cg,Witten:1991yr}.
Of course one could argue that unitarity is restored
in the scaling limit of the non-unitary lattice model.
The same argument can be employed to
explain why the infra-red fixed point of the non-unitary Pruisken model
is controlled by an NLSM with a Riemannian target space manifold.
However, if one were not to simply brush aside this issue,
it could be taken as a signal that the conjecture from ref.\cite{Ikhlef:2011ay}
is not quite correct. An interesting alternative would be
 that the scaling behaviour of the spin chain
is still described by a NLSM, but with a non-Riemannian target space.
This would open a way of assigning a meaning to a  quantum NLSM whose target space
metric has a Lorentzian type signature. 
Apart from Condensed Matter Physics applications, that would be of  interest
for understanding the physics of black holes \cite{Witten:1991yr,Dijkgraaf:1991ba}. This work was motivated by such an 
exciting possibility.
\bigskip

Our study 
essentially employs the Yang-Baxter integrable structures of
the lattice system.
Due to the heavy amount of technical details involved, we moved
the part of the work that considers the
formal algebraic aspects of the general inhomogeneous Baxter
model to a separate publication \cite{Bazhanov:2020new}. Some formulae from that paper, which are
directly relevant to the ${\cal Z}_2$ invariant inhomogeneous
six-vertex model are collected, for the reader's convenience, in
the Preliminaries section of this work. 
\bigskip

The key tool in our
 analysis of the scaling limit is the ODE/IQFT correspondence.
The first part of the paper serves to illustrate
the technique for the critical homogeneous six-vertex model. No original results are contained therein.
It gives us an opportunity  to explain the ODE/IQFT approach 
\cite{Bazhanov:1994ft,Bazhanov:1996dr,Bazhanov:1998dq,Voros:1999,Dorey:1998pt,Bazhanov:1998wj,
Bazhanov:2003ni}
and  to set-up the notation.
Moreover, following the recent paper \cite{Kotousov:2019ygw},
we discuss the Hermitian structures consistent
with the integrable one for the homogeneous six-vertex model.
Then the link is explained between these Hermitian structures
and those that they induce in the scaling limit. 
Our elaboration of this example would be important 
for the conceptual understanding of the non-unitarity issue for the ${\cal Z}_2$ invariant
inhomogeneous six-vertex model.
\bigskip

Part II contains the main results of this paper. Using the ODE/IQFT correspondence
we identify the algebra of extended conformal  symmetry and 
describe the linear and Hermitian structures of the space of states
occurring in the scaling limit  of the ${\cal Z}_2$ invariant
inhomogeneous six-vertex model.
The final Part III is devoted to a discussion of the CFT underlying
the critical behaviour of the lattice model. In particular,
we put forward a modified version of the conjecture of 
Ikhlef, Jacobsen and Saleur. 
A list of the central results of this work is given in the Summary section.

\section{Preliminaries\label{sec1}}
In this work we follow the conventions and use the results of  \cite{Bazhanov:2020new},
which  discusses some general aspects of the inhomogeneous six vertex model. 
Here, for the convenience of the reader, we collect some  basic formulae from 
that paper.
\bigskip

Let $\sigma^A_m$ $(A=\pm,z)$ be the standard Pauli matrices acting 
on the $m$-th factor of the tensor product
 \be\label{vec1}
{\mathscr V}_N=\mathbb{C}^2_N\otimes
 \mathbb{C}^2_{N-1}\otimes\cdots\otimes\mathbb{C}^2_1\ .
\ee
Introduce the monodromy matrix
\be\label{M-in}
{\bm{M}}(\zeta)=
q^{-\frac{N}{2}}\,\bm{R}_{N}\big(q\zeta/\eta_N\big)\,
\bm{R}_{N-1}\big(q\zeta/\eta_{N-1}\big)\cdots
\bm{R}_{1}\big(q\zeta/\eta_{1}\big)\,,
\ee
where the $N$ complex numbers
$\{\eta_J\}_{J=1}^N$ parameterize the inhomogeneities, while $\bm{R}_{m}$
stands for the $2\times2$ matrix
\be\label{rmat2}
\bm{R}_m(q\zeta)=
\left(\begin{array}{cc} q^{\frac{1}{2}(1+\sigma_m^z)}+q^{\frac{1}{2}(1-\sigma_m^z)}\,\zeta & 
-(q-q^{-1})\,q\, \zeta\,\sigma^-_m \\[0.2cm]
(q-q^{-1})\,\sigma^+_m & q^{\frac{1}{2}(1-\sigma_m^z)}+q^{\frac{1}{2}(1+\sigma_m^z)}\,\zeta
\end{array}\right)\,,
\ee
whose entries act in the $\mathbb{C}^2_m$ factor in the tensor product \eqref{vec1}.
We'll be considering twisted boundary conditions parameterized by $\omega$.
Then the transfer matrix for the inhomogeneous six vertex model on the 
square lattice with $N$ columns is given by the trace
\be\label{Tmat1}
\mathbb{T}(\zeta)={\rm Tr}\big[\,\omega^{\sigma^z}\,\bm{M}(\zeta)\,\big]\ .
\ee
The transfer matrix satisfies a number of operator valued relations.
The latter involve the matrices $\mathbb{A}_\pm(\zeta)$ which  
together with $\mathbb{T}(\zeta)$ form a commuting family
\be\label{commrel1a}
[\mathbb{A}_\pm(\zeta),\mathbb{A}_\mp(\zeta')]=[\mathbb{A}_\pm(\zeta),\mathbb{A}_\pm(\zeta')]=
[\mathbb{A}_\pm(\zeta),\mathbb{T}(\zeta')]=0\ .
\ee
The construction of $\mathbb{A}_\pm(\zeta)$ along with their properties
may be found in  sec.\,3 of \cite{Bazhanov:2020new}. 
Here we just mention that
\be\label{TArel1a}
\mathbb{T}(\zeta)\,\mathbb{A}_\pm(\zeta)=\omega^{\pm1}\,q^{\pm\mathbb{S}^z}\,f(q^{-1}\zeta)\,
\mathbb{A}_\pm(q^2\zeta) +
\omega^{\mp1}\,q^{\mp\mathbb{S}^z}\,f(q^{+1}\zeta)\,\mathbb{A}_\pm(q^{-2}\zeta)\,,
\ee
where $f(\zeta)$ is given by
\be\label{ffunc1a}
f(\zeta)=\prod_{J=1}^N(1+\zeta/\eta_J)
\ee
and $\mathbb{S}^z$ stands for the $z$ projection of the total spin operator,
\be
\mathbb{S}^z=\tfrac{1}{2}\sum_m \sigma^z_m\ :\ \ \  \ [\mathbb{S}^z,\,\mathbb{A}_\pm(\zeta)]=
 [\mathbb{S}^z,\,\mathbb{T}(\zeta)]= 0\ .
\ee
\bigskip

It follows from the definition \eqref{M-in}\,-\,\eqref{Tmat1} 
that the matrix elements of $\mathbb{T}(\zeta)$ are polynomials of order $N$
 in the variable $\zeta$. Due to the mutual commutativity, $[\mathbb{T}(\zeta),\,\mathbb{T}(\zeta')]=0$,
the eigenvectors of the transfer matrix do not depend on this variable and hence its eigenvalues
are also $N$-th order polynomials in $\zeta$.
It turns out  that the eigenvalues of $\mathbb{A}_\pm(\zeta)$, which will be denoted as
$A_\pm(\zeta)$ below, are polynomials of order $N/2\mp S^z$, respectively,
where $S^z$ denotes the eigenvalue of $\mathbb{S}^z$.
Let $\{\zeta_m\}_{m=1}^M$  with $M=N/2-S^z$ be the set of roots of $A_+(\zeta)$.
For generic values of the parameters, none of the $\zeta_m$ are equal to zero, and it will be convenient to choose the normalization convention for
$\mathbb{A}_+$ such that
\be\label{Aeig1bb1}
A_+(\zeta)=\prod_{m=1}^M\,(1-\zeta/\zeta_m)\,,\qquad M= \tfrac{1}{2}\,N-S^z\ .
\ee
Applying both sides of the relation  \eqref{TArel1a}
to a common eigenvector
and setting $\zeta=\zeta_m$, 
yields the system of
 algebraic equations \cite{Lieb:1967,Baxter:1971}
\be\label{bae}
\prod_{J=1}^{N}
\frac{\eta_J+\,q\,\zeta_m}
{q\,\eta_J+\zeta_m }
=-\,\omega^2\,
\prod_{j=1}^M\,
\frac{q^{-1}\,\zeta_j-q^{+1}\,\zeta_m }
{q^{+1}\,\zeta_j-q^{-1}\,\zeta_m }
\,\qquad (m=1,2,\ldots,M)
\ee
for the set of zeroes  of $A_+(\zeta)$.
Having a solution of the above equations, the eigenvalue
of the
transfer matrix is given by
\bea\label{teigen}
T^{(N)}(\zeta)&=&\omega^{+1}\,q^{+S^z}\ 
\bigg(\prod_{J=1}^N \big(1+q^{-1}\,\zeta/\eta_J\big)\bigg)
\prod_{j=1}^M\frac{\zeta_j-q^{+2}\,\zeta}{\zeta_j-\zeta}\nonumber\\[0.2cm]
&+&\omega^{-1}\,q^{-S^z}\
\bigg(\prod_{J=1}^N \big(1+q^{+1}\,\zeta/\eta_J\big)\bigg)
\prod_{j=1}^M\frac{\zeta_j-q^{-2}\,\zeta}{\zeta_j-\zeta}\ .
\eea
\bigskip

Of course, there are similar formulae  involving the roots of  ${A}_-(\zeta)$.
However, we will mainly focus on $\mathbb{A}_+(\zeta)$ for the following reason.
It will be assumed that the inhomogeneities satisfy the constraints
\be\label{ogranich1}
\eta_{N+1-J}=\eta_{J}^{-1}\ \qquad\ \  (J=1,2,\ldots,N)\ .
\ee
In this case the model possesses the so-called global ${{\cal C}}{\cal P}$ invariance
(the explicit formula for the generators $\hat{\cal C}$ and $\hat{\cal P}$
are quoted in eqs.\,\eqref{oaspodp9012} and \eqref{oaisodi899832}, respectively).
The ${\cal C P}$ transformation intertwines the sectors with $S^z$ and
$-S^z$.
Moreover, it relates the operators $\mathbb{A}_+(\zeta)$ and $\mathbb{A}_-(\zeta)$ as
\bea\label{CP}
{\mathbb A}_-(\zeta)
=\zeta^{\frac{N}{2}-\,\mathbb{S}^z}\ { \hat {\cal C}}{\hat  {\cal P}}\,{\mathbb A}_+
\big(\zeta^{-1}\big)\,{ \hat {\cal C}}{\hat  {\cal P}} \ \mathbb{A}^{(\infty)}_+\ ,
\ 
\eea
where
\be\label{8d8d91029a}
\mathbb{A}^{(\infty)}_+=\ \lim_{\zeta\to\infty}\ \zeta^{-\frac{1}{2}N+\mathbb{S}^z}\ \mathbb{A}_+(\zeta)\ .
\ee
Therefore, for the diagonalization problem of the commuting family \eqref{commrel1a},  it is sufficient to 
consider  $\mathbb{A}_+$ and focus on the sector $S^z\ge 0$.
Thus, in the Bethe ansatz equations \eqref{bae} we will always assume
\be
M\le\tfrac{1}{2}\,N \, .
\ee
Note that combining \eqref{CP} with the operator relation \eqref{TArel1a} one finds
\be\label{CP2}
 { \hat {\cal C}}{\hat  {\cal P}}
\,{{\mathbb T}}(\zeta)\,{ \hat {\cal C}}{\hat  {\cal P}}=
\zeta^{N}\ {{\mathbb T}}\big(\zeta^{-1}\big)\ .
\ee

\bigskip

The eigenvectors  
can be constructed within the framework
of the algebraic Bethe ansatz \cite{Faddeev:1979gh}. To this end, introduce the following notation for the 
entries of the monodromy matrix
\be\label{Mmat}
\bm{M}(\zeta)=
\left(\begin{array}{cc} { \hat{ {\mathsf A}}}(\zeta) & 
a(\zeta)\, { \hat{ {\mathsf B}}} (\zeta)\\[0.2cm] d(\zeta)\,{ \hat{ {\mathsf C}}} (\zeta) & 
{ \hat{ {\mathsf D}}}(\zeta)\end{array}\right)\,,
\ee
where $ { \hat{ {\mathsf A}}},\,{ \hat{ {\mathsf B}}},\, { \hat{ {\mathsf C}}},\,{ \hat{ {\mathsf D}}}$
are operators acting in \eqref{vec1},
while $a(\zeta)$, $d(\zeta)$ stand for the polynomials
\be\label{addef1}
a(\zeta)=-\ri\,\omega^{+1}\, q^{+\frac{N+1}{2}}\,\prod_{J=1}^N\big(1+q^{-1}\zeta/\eta_J\big)\,,\qquad
d(\zeta)=+\ri\,\omega^{-1}\, q^{-\frac{N+1}{2}}\,\prod_{J=1}^N\big(1+q\,\zeta/\eta_J\big)\ .
\ee
Let $ \bm{\Psi}^{(0)}\in\mathbb{C}^2_N\otimes
 \mathbb{C}^2_{N-1}\otimes\cdots\otimes\mathbb{C}^2_1$ be the pseudovacuum
\bea\label{iaususa}
\bm{\Psi}^{(0)}=
\underbrace{|\uparrow\rangle\otimes |\uparrow\rangle\otimes \ldots\otimes |\uparrow\rangle}_{N}\ .
\eea
Then the state
\be\label{Bstate1}
\boldsymbol{\Psi}\big(\{\zeta_j\}\big)={ \hat{ {\mathsf B}}}(\zeta_M)\,\cdots\,
{ \hat{ {\mathsf B}}}(\zeta_2)\, { \hat{ {\mathsf B}}}(\zeta_1)\,\bm{\Psi}^{(0)}
\qquad\qquad\qquad (M= \tfrac{1}{2}\,N-S^z)
\ee
is a common eigenstate for the commuting family of operators provided that
 the set $\{\zeta_j\}_{j=1}^M$ satisfies the Bethe ansatz equations \eqref{bae}.
\bigskip

In this work we consider the case where $q$ and $\omega$ are unimodular:
\be\label{qwuni1}
q^*=q^{-1}\,,\qquad \omega^*=\omega^{-1} \ .
\ee
If the inhomogeneities satisfying \eqref{ogranich1}
 are also taken to be unimodular,
\be\label{ogranich2}
\eta_{J}^*=\eta_J^{-1}=\eta_{N+1-J}
\ee
then the system possesses ${\cal T}$-invariance.
The time reversal transformation
 is realized as an anti-unitary operator acting on
an arbitrary state   $\bm{\Xi}\in{\mathscr V}_N$ as
\bea\label{Tdef}
\hat{\cal T}\,\bm{\Xi}=\hat{\mathsf  U}\,\bm{\Xi}^*\ \qquad\qquad {\rm with}
\qquad\qquad 
\hat{\mathsf  U}=
\prod_{m=1}^N\,\sigma^x_m\ .
\eea
Similar to the ${\cal CP}$ conjugation, it flips the sign of $S^z$
so that 
${\cal {C}{P}{T}}$  acts invariantly in the sector with given $S^z$.
Moreover,
for the state $\bm{\Psi}$ \eqref{Bstate1} corresponding to the set $\{\zeta_j\}_{j=1}^M$ solving \eqref{bae},
one has
\be\label{CPTBethe1}
{\cal \hat{C}\hat{P}\hat{T}}\,\boldsymbol{\Psi}\big(\{\zeta_j\}\big)=
\,\boldsymbol{\Psi}\big(\{\zeta_j^*\}\big)\ .
\ee
The Bethe state in the r.h.s. of the above formula is built using the
Bethe roots for the complex conjugated set $\{\zeta_j^*\}_{j=1}^M$, which is also
a solution of the Bethe ansatz equations.

\bigskip

When further restrictions are placed on the inhomogeneities,
additional global symmetries appear in the model.
In particular, suppose that $N$ is divisible by the integer $r$,
\be\label{rdef1}
N=rL\,,
\ee
and the  $\eta_J$ are taken to satisfy the periodicity condition
\be\label{period1}
\eta_{J+r}=\eta_J\qquad (J=1,2,\ldots,N)\,,
\ee
where $\eta_{J+N}\equiv\eta_J$.
Then one can introduce the lattice translation operator
\be\label{Kformula0}
{\mathbb K}\  :\ \   \ \ \big[\mathbb{K},\,\mathbb{T}(\zeta)\big]=\big[\mathbb{K},\,\mathbb{A}_\pm(\zeta)\big]=0\ , 
\qquad 
{\mathbb K}^L=\re^{2\pi\ri{\tt k}\,\mathbb{S}^z}\ .
\ee
Its matrix elements read explicitly as
\be\label{Kformula1}
\big({{\mathbb K}}\big)_{a_{N} a_{N-1}\ldots
  a_1}^{b_{N}b_{N-1}\ldots b_1}=\re^{\ri\pi{\tt k}\,(a_1+a_2+\ldots +a_r)}
\,\delta_{a_N}^{b_{N-r}}\,\delta_{a_{N-1}}^{b_{N-r-1}}\,\ldots\,
\delta_{a_1}^{b_{N-r+1}}
\ee
and its eigenvalue $K$ corresponding to the Bethe state \eqref{Bstate1} is
expressed in terms of $A_+(\zeta)$ \eqref{Aeig1bb1}  as
\be\label{Keigeq1a}
K=\prod_{\ell =1}^r\ \omega\, q^{-\frac{N}{2}+\mathbb{S}^z}\ \frac{A_+(-q^{+1}\eta_\ell)}{A_+(-q^{-1}\eta_{\ell})}\ .
\ee
\bigskip

The transfer matrix and $\mathbb{A}_\pm(\zeta)$ are not Hermitian
w.r.t. the standard matrix conjugation,
 $\hat{{\mathsf O}}^\dag=(\hat{{\mathsf O}}^*)^T$ with
$\hat{{\mathsf O}}\in{\rm End}(\mathscr{V}_N)$.
Nevertheless it is possible to introduce the  Hermitian structure in the
$2^N$ dimensional linear space $\mathscr{V}_N$, which is consistent with
the integrable structure of the inhomogeneous six\,-\,vertex model.
Such Hermitian structures were discussed in the work \cite{Bazhanov:2020new}.
A special r$\hat{{\rm o}}$le belongs to the one associated with
the  conjugation
\be\label{odoappo}
\hat{{\mathsf O}}^\star=
\hat{{\mathsf X}}_\star^{-1}\ \hat{{\mathsf O}}^\dag\ \hat{{\mathsf X}}_\star^{}\ .
\ee
Here  $\hat{{\mathsf X}}_\star^{}=\hat{{\mathsf X}}_\star^{\dag}$ stands for the matrix
\be
\hat{{\mathsf X}}_\star^{}=\hat{{\mathsf X}}\ 
\re^{\ri\pi ({\mathbb S}^z-\frac{N}{2})}\ \mathbb{A}^{({\infty})}_+
\ee
with  $\mathbb{A}^{({\infty})}_+$  given in \eqref{8d8d91029a}, while
$\hat{{\mathsf X}}$ 
is defined through the ordered product
\be\label{Xcase1}
\hat{\mathsf  X}=
\ \bigg(\prod_{J=1}^N (\eta_J)^{\frac{1}{2}\sigma_J^z}\bigg)\ 
\overset{{\displaystyle \curvearrowright}}{\prod_{m=2}^{\leng}} \,
\Bigg[\,\overset{{\displaystyle \curvearrowright}}{\prod_{n=N-m+1}^{\leng-1}}\,
\check{\bm{R}}_{n+1,n}\big(\eta_{n+m-N}/\eta_{m}\big)\Bigg]\ \ .
\ee
In the above formula we use the notation
\be\label{Rcheckdef1}
\check{\bm{R}}_{n+1,n}(\zeta)=\frac{1}{q-q^{-1}\zeta}\ {\bm{R}}_{n+1,n}(-\zeta)\,{\bm P}_{n+1,n}\,,
\ee
where ${\bm{R}}_{n+1,n}(-\zeta)$ is the matrix \eqref{rmat2} acting on the $n+1$-st and $n$-th
components of the tensor product \eqref{vec1}, while ${\bm P}_{n+1,n}$ is the permutation matrix that interchanges
the two components. 
Assuming the conditions \eqref{qwuni1},\,\eqref{ogranich2}
it is possible to show that under the $\star$\,-\,conjugation \eqref{odoappo}
the transfer matrix
as well as $\mathbb{A}_\pm(\zeta)$ satisfy
\be\label{8d88f00a91}
\big[\, \mathbb{T}(\zeta)\,\big]^\star=\mathbb{T}(\zeta^*)\ ,\ \ \ \ \ \ \ 
\big[\, \mathbb{A}_\pm(\zeta)\,\big]^\star=\mathbb{A}_\pm(\zeta^*)
\ee
(for details see sec.\,5 in ref.\cite{Bazhanov:2020new}).
\bigskip

For the conjugation \eqref{odoappo}
there exists a unique sesquilinear form, which is defined
through the relations
\be\label{is8s8s8dia}
\big(\bm{\Xi}_2,\hat{{\mathsf O}}\,\bm{\Xi}_1\big)_{\star}=
\big(\hat{{\mathsf O}}^\star\,\bm{\Xi}_2,\bm{\Xi}_1\big)_{\star}\ \ \ \ \ (\forall\ \bm{\Xi}_1,\bm{\Xi}_2\in\mathscr{V}_N)
\ee
together with   the overall  normalization
\bea\label{kassausa}
(\bm{\Psi}^{(0)},\bm{\Psi}^{(0)})_{\star}=1\ ,
\eea
where $\bm{\Psi}^{(0)}$\,\eqref{iaususa} is the pseudovacuum.
 Then it follows from 
\eqref{8d88f00a91} as well as   the relations
\be\label{CPT1asdasda}
{\cal \hat{C}\hat{P}\hat{T}}\,\mathbb{T}(\zeta)\,{\cal \hat{C}\hat{P}\hat{T}}=\mathbb{T}(\zeta^*)\,,\qquad
{\cal \hat{C}\hat{P}\hat{T}}\,\mathbb{A}_\pm(\zeta)\,{\cal \hat{C}\hat{P}\hat{T}}=\mathbb{A}_\pm(\zeta^*)
\ee
that w.r.t.
the sesquilinear form the Bethe states satisfy the orthogonality condition
\be\label{ortho1}
\big({\boldsymbol \Psi}^{(2)},{\boldsymbol \Psi}^{(1)} \big)_{\star}=0\ \ \ \ \ {\rm unless} \ \ \ \ \ 
\ {\boldsymbol \Psi}^{(2)}={ \hat {\cal C}}{\hat  {\cal P}}{ \hat {\cal T}}\,{\boldsymbol \Psi}^{(1)}\ .
\ee
The ``norm'' of the Bethe state \eqref{Bstate1},
 in terms of the  corresponding set $\{\zeta_m\}$,
is given by \cite{Gaudin:1972,Gaudin:1981cyg,Korepin:1982ej}
\bea\label{FinalNorm}
&&\big({\cal\hat{ C}\hat{P}\hat{T}}
\bm{\Psi},\bm{\Psi}\big)_{\star}=\
\big(q-q^{-1}\big)^{2M}\ 
\prod_{m\ne j}^{M}
\frac{q\zeta_j-q^{-1}\zeta_m}{\zeta_m-\zeta_j}
 \\[0.2cm]
&&\times\, {\rm det}\Bigg[\delta_{j,m}\bigg(\kappa(\zeta_j)+\sum_{l=1}^{M}
\frac{(q+q^{-1})\,\zeta_j\zeta_{l}}
{(q\zeta_l-q^{-1}\,\zeta_j)(q\zeta_j-q^{-1}\,\zeta_l)}\bigg)
-\frac{(q+q^{-1})\,\zeta_j\zeta_{m}}
{(q\zeta_m-q^{-1}\,\zeta_j)(q\zeta_j-q^{-1}\,\zeta_m)}\Bigg]\nonumber
\eea
with
\bea
{\kappa}(\zeta)=-\sum_{J=1}^N
\frac{\zeta}{
\eta_J(1+q^{-1}\zeta/\eta_J)(1+q^{+1}\,\zeta/\eta_J)}\ .\nonumber
\eea

\newpage

\part{Homogeneous six-vertex model}
\section{The Hamiltonian}

The purpose of this work is the study of the scaling limit of the 
alternating six-vertex model, where the 
$\eta_J$ are fixed to be 
$\eta_J=\ri\,(-1)^{J-1}$.
This is a special case of \eqref{ogranich2} and \eqref{period1}  (with $r=2$).
However,
since many of our considerations are parallel  to those for the
homogeneous model, where all the $\eta_J=1$, we'll begin our discussion
with this more familiar example.
In this case the transfer matrix $\mathbb{T}(\zeta)$ \eqref{Tmat1} and
the translation operator $\mathbb{K}$ \eqref{Kformula0},\,\eqref{Kformula1} ($r=1$, $L=N$) commute with the 
 spin $\frac{1}{2}$ $XXZ$ Hamiltonian
\be\label{asiisaias}
\mathbb{ H}_{XXZ}=-\frac{1}{2\sin(\pi\beta^2)}\,
\sum_{i=1}^N\Big( \sigma_i^x\sigma_{i+1}^x+\sigma_i^y\sigma_{i+1}^y+\cos(\pi\beta^2)
\, \big(\sigma_i^z\sigma_{i+1}^z-\hat{\bm{1}}\big)\Big)
\ee
with
\be
\sigma^x_{N+1}\pm\ri\sigma^y_{N+1}=\re^{2\pi\ri{\tt k}}\,
\big(\sigma^x_{1}\pm\ri\sigma^y_{1}\,\big)\ ,
\qquad \qquad
\sigma_{N+1}^z=\sigma^z_1 \ .
\ee
Here we have parameterized the unimodular numbers $q$ and $\omega$ as 
\be\label{qomegaho1}
q=\re^{\ri\pi\beta^2}\,,\qquad\qquad\omega=\re^{\ri\pi{\tt k}}\,,
\ee
where $\beta$ and ${\tt k}$ lie in the domains
\be
0<\beta<1\,,\qquad -\tfrac{1}{2}<{\tt k}\le\tfrac{1}{2}\ .
\ee
The eigenvalue of $\mathbb{H}_{XXZ}$ on the state $\bm{\Psi}$ \eqref{Bstate1}
is given in terms of the Bethe roots by
\be\label{altEnerg1}
{\cal E}=-\sum_{m=1}^M\frac{4\sin(\pi\beta^2)}{
\zeta_m+\zeta_m^{-1}+
2\cos(\pi\beta^2)}\ ,
\ee
while for the eigenvalue of $\mathbb{K}$, see eq.\,\eqref{Keigeq1a} with $r=1$ and $\eta_\ell=1$.

\section{RG flow for the Bethe states\label{sec2.1}}

The  scaling limit is a certain large $N$ limit for a particular class of ``low energy'' states.
The latter are defined w.r.t. a reference state -- the vacuum.
In the case of the homogeneous six-vertex model  the reference state is  
the lowest energy state of the Hamiltonian \eqref{asiisaias}. In turn,
the class of states we'll be considering are those whose energy counted from the vacuum energy
is sufficiently low. 
 It is well known that, with the parameter $\beta^2$ \eqref{qomegaho1} lying in the interval
$0<\beta^2<1$,
the system is critical and as $N\to\infty$ the low energy part of the spectrum
organizes into the conformal towers \cite{Cardy:1986ie}.
In a given tower,
 the eigenvalues of the Hamiltonian \eqref{asiisaias} 
and the lattice translation operator \eqref{Kformula1} are described
by the formulae:
  \bea\label{tower1}
  {\cal E}&=&e_\infty N+\frac{2\pi v_{\tt F}}{N }\ \big(P^2+{\bar P}^2-\tfrac{1}{12}+
 {\tt L}+\bar{\tt L}\,\big)+o\big(N^{-1}\big)\\[0.2cm]
K&=& \sigma\exp\bigg(\frac{2\pi\ri}{N}\,\big(P^2-{\bar P}^2+ {\tt L}-\bar{\tt L}\big)\bigg)\ .\nonumber
  \eea
Here  $e_\infty$ is the specific bulk energy, while $v_{{\rm F}}$ is usually referred to as the 
Fermi velocity and in our conventions
for the Hamiltonian \eqref{asiisaias}  they read
explicitly as
\bea
e_{\infty}&=& -\frac{2v_{{\rm F}}}{\pi}\,\int_0^\infty{\rm d}t\ \frac{\sinh\big(\frac{\beta^2 t}{1-\beta^2}\big)}
{\sinh\big(\frac{ t}{1-\beta^2}\big)\,\cosh(t)}\\[0.2cm]
v_{{\rm F}}&=&\frac{1}{1-\beta^2}\ .\nonumber
\eea
Contrary to $e_{\infty}$ and $v_{{\rm F}}$, which are the same for all the 
low energy states, the factor $\sigma=\pm 1$ 
so that the low energy states are splitted into two sectors corresponding to different values of
the sign.
The pair $(\bar{P},{P})$ labels the different conformal towers, ${\cal V}_{\bar{P},{P}}$,
and its admissible values are described by eq.\,\eqref{PPbareq1}
below.
Each level subspace of the tower, ${\cal V}_{\bar{P},{P}}^{(\bar{\tt L},{{\tt L}})}$,
 is specified by the non-negative integers
${\tt L},\bar{\tt L}=0,1,2,\ldots$ and has dimensions
\be
 {\rm dim}\ {\cal V}_{\bar{P},{P}}^{(\bar{\tt L},{{\tt L}})}={\tt par}_1(\bar{{\tt L}})\,{\tt par}_1({\tt L})\,,
\ee
where 
${\tt par}_1({\tt L})$ and ${\tt par}_1(\bar{{\tt L}})$ are 
 the number of integer partitions of ${\tt L}$ and $\bar{{\tt L}}$ respectively.
Furthermore, it turns out that
\be\label{Vfac1}
{\cal V}_{\bar{P},{P}}^{(\bar{{\tt L}},{{\tt L}})}=
\bar{{\cal F}}^{({\bar{\tt L}})}_{\bar{ P}}\otimes
 {\cal F}^{(\tt L)}_P
\ee
with  $ {\cal F}^{(\tt L)}_P$ standing for the level subspace of the
Fock space ${\cal F}_P$:
\be
{\cal F}^{(\tt L)}_P = {\rm span}\big\{a_{-n_1}\ldots a_{-n_j}\,|P\rangle \ : \ 
n_1+\ldots +n_j={\tt L}\,, \ \ \ \forall n_j>0\ \big\}\,, \qquad
{\rm dim}\,{\cal F}^{(\tt L)}_P={\tt par}_1({\tt L})\ .
\ee
We'll use the conventions that the Heisenberg algebra generators $\{a_n\}$ obey the commutation relations
\bea\label{acomm1}
[a_m,a_n]=\tfrac{m}{2}\ \delta_{m+n,0}\ ,
\eea
while the highest weight vector is defined through the conditions
\bea
a_n\, |P\rangle=0\ \ \ (\forall n>0)\ ,\ \ \ \  a_0\,|P\rangle=P\, |P\rangle\ .
\eea
The  factor $\bar{{\cal F}}^{({\bar{\tt L}})}_{\bar{ P}}$ in the tensor product in the r.h.s. of eq.\,\eqref{Vfac1}
denotes the level subspace of the highest weight representation
of the Heisenberg algebra, generated by 
the operators $\bar{a}_m$ that commute with
 $\{a_m\}$ and
 satisfy the same commutation relations as in
 \eqref{acomm1},
with highest weight vector $|{\bar P}\rangle$.

\bigskip

The zero-mode momenta $P,\bar{P}$ labeling the  conformal tower
are not arbitrary, but take a certain discrete set of values. Namely, in the sector
characterized by the eigenvalue $S^z$ of the $z$\,-\,component of the total
spin operator, they are given by
\bea\label{PPbareq1}
 P=\tfrac{1}{2}\, \big(\beta S^z+\beta^{-1}({\tt k}+{\tt w})\,\big)\ ,\  \ \ \ \  
 {\bar  P}=\tfrac{1}{2}\, \big(\beta S^z-\beta^{-1}({\tt k}+{\tt w})\,\big)\ .
\eea
 The integer
  ${\tt w}=0,\pm1,\pm2,\ldots$ appearing in eq.\,\eqref{PPbareq1} will be referred to below  as the  winding number.
 Together with the non-negative integers ${\tt L}$ and $\bar{\tt L}$, the winding number is
 an important characteristic of the low energy stationary states.
In particular, the sign factor $\sigma$ in the second line of eq.\,\eqref{tower1}
coincides with the parity of ${\tt w}$, i.e.,
\be\label{sigmadef1aaaa}
\sigma=(-1)^{{\tt w}} \ .
\ee
Note that  formulae \eqref{tower1},\,\eqref{PPbareq1} imply that
the eigenvalues of the lattice translation operator $\mathbb{K}$
satisfy $K^N=(-1)^{{\tt w}(N-2S^z)}\,\re^{2\pi\ri{\tt k}S^z}$.  
Since $N-2S^z=2M$ is always even  it follows that
$\mathbb{K}^N=\re^{2\pi\ri {\tt k}\,\mathbb{S}^z}$ 
(see the last equality in \eqref{Kformula0} with $L=N$).
\bigskip

The natural question arises, 
for a \emph{given} Bethe state $\bm{\Psi}$ what would be its scaling limit?
In other words, what particular state in 
${\cal V}_{\bar{P},{P}}^{(\bar{\tt L},{{\tt L}})}$ \eqref{Vfac1},\,\eqref{PPbareq1}
would appear in the large $N$ limit of $\bm{\Psi}$ \eqref{Bstate1}.
In fact, to formulate this question  meaningfully, one should first organize the Bethe states for
different $N$ into a one parameter family, i.e., define
an individual Renormalization Group (RG) trajectory $\bm{\Psi}_N$.
This procedure could only make sense for the low energy part of the spectrum as the
Hilbert spaces $\mathscr{V}_N$ \eqref{vec1} are not isomorphic for different lattice sizes.

\bigskip

For a  general lattice system it is not clear  how to assign the size dependence
for individual low energy stationary states. However,
in the case under consideration,  one can exploit the integrability of the model for
the construction of the RG trajectory $\bm{\Psi}_N$.
For this end, first re-write the Bethe ansatz equations \eqref{bae} with all the $\eta_J=1$
in the logarithmic form:
\bea\label{klkwqnmsd}
L p(\zeta_m)=2\pi{\tt k}-2\pi\, I_m-\sum_{j=1}^M \Theta(\zeta_m, \zeta_j)\ ,
\eea
where
\bea\label{asisaisa}
p(\zeta)=-\ri\, \log\bigg(\!\frac{1+q\,\zeta}{q+\zeta}\!\bigg)\ ,\ \ \ \ 
\Theta(\zeta,\zeta')=-\ri\,  \log\bigg(\!\frac{q\,\zeta'-q^{-1}\zeta}{q\,\zeta-q^{-1}\zeta'}\!\bigg)
\eea
and
$I_m$ are the so-called Bethe numbers which are integers or half-integers for $M$ odd or even respectively. 
An unambiguous definition of $I_m$  requires fixing the
branches of the logarithms in \eqref{asisaisa}.
Although this  is an important step in  any practical calculation, we will not touch on it here
and only mention  that 
\bea\label{kssusu}
I^{\rm (vac)}_m=-\tfrac{1}{2}\ (M+1)+m\ \ \ \ \ \ \ \ \ \ \ \big(\, m=1,\ldots, M=\tfrac{1}{2}\,N-S^z\,\big)
\eea
for the vacuum state in the sector with fixed value of $S^z$.
For sufficiently large $N$ the Bethe numbers corresponding to the low  energy states are given
by $I^{\rm (vac)}_m+\delta I_m$, where the variation $\delta I_m$ from the ``vacuum'' distribution \eqref{kssusu}
are nonzero only in the vicinity  of the  edges, i.e., for $m\ll M$  or $M-m\ll M$. 
The set  $\{\delta I_m\}$ can  be used to define
the individual  RG flow trajectories   $\boldsymbol{\Psi}_N$ in the following way.
\bigskip

Starting with a spin chain for relatively small $N$ one performs the
numerical diagonalization of the Hamiltonian.
Together with the energies \eqref{altEnerg1}, one should also compute the 
eigenvalues  of $\mathbb{A}_+(\zeta)$.
The explicit construction of the $2^N\times 2^N$
 matrix $\mathbb{A}_+(\zeta)$ can be found in sec.\,3 of \cite{Bazhanov:2020new}.
Its eigenvalues are
polynomials whose zeroes coincide with the corresponding Bethe roots (see eq.\,\eqref{Aeig1bb1}). 
This allows one to extract the set $\{\zeta_m\}_{m=1}^M$ for  a particular Bethe state $\boldsymbol{\Psi}_N$
and, using \eqref{klkwqnmsd}, also
 the set of $\{\delta I_m\}$.
For the state $\boldsymbol{\Psi}_{N+2}$,
the Bethe ansatz equations  are taken to have the same
 $\{\delta I_m\}$ in the vicinity of the edges.
Moreover, for their iterative solution the initial approximation
can be constructed using the Bethe roots for $\boldsymbol{\Psi}_N$.
This procedure provides a way for defining
the RG flow of an individual low energy Bethe state. 
Having at hand the RG trajectory
$\boldsymbol{\Psi}_N$ and taking its large $N$ limit,
 our previous discussion means that
\bea\label{Psi1a}
\boldsymbol{\Psi}_N\asymp \Omega_N
\ \bar{\boldsymbol \psi}_{\bar P}(\bar{\boldsymbol v})\otimes {\boldsymbol \psi}_P({\boldsymbol v}) \ \ \ \ \ \ \ {\rm as}\ \ \ \ \ \ \  N\to\infty\ .
\eea
Here the limiting state $\bar{\boldsymbol \psi}_{\bar P}(\bar{\boldsymbol v})\otimes {\boldsymbol \psi}_P({\boldsymbol v})$
does not depend on $N$ and belongs to  the subspace ${\cal V}_{\bar{P},{P}}^{(\bar{\tt L},{{\tt L}})}$
\eqref{Vfac1}, while
  the constant $\Omega_N$ (in fact a functional,  $\Omega_N=\Omega[\boldsymbol{\Psi}_N]$, whose value depends on the 
 Bethe state)   diverges in the large $N$ limit and
will be discussed in details later.
In the next subsection we'll describe the state
$ \bar{\boldsymbol \psi}_{\bar P}(\bar{\boldsymbol v})\otimes {\boldsymbol \psi}_P({\boldsymbol v})$
 in \eqref{Psi1a}. It is sufficient to focus on the ``right'' 
vector ${\boldsymbol \psi}_P({\boldsymbol v})$ since
there is only a notational difference between the left and right component 
in the tensor product. 
\bigskip

\section{Identification of the  RG trajectory with a state in the conformal tower}
\subsection{The sum rules for the scaled Bethe roots}
For finite $N$ the Bethe states are unambiguously characterized by the eigenvalues
of $\mathbb{A}_+(\zeta)$. 
Therefore the chiral state ${\boldsymbol \psi}_P({\boldsymbol v})$  may be determined 
through the study of the large $N$ behaviour of
the eigenvalue \eqref{Aeig1bb1} corresponding to
${\boldsymbol \Psi}_N$. 
The scaling limit for $A_+(\zeta)$
was discussed
 in the series of papers 
\cite{Bazhanov:1994ft,Bazhanov:1996dr,Bazhanov:1998dq,Bazhanov:1998wj,
Bazhanov:2003ni,Kotousov:2019ygw}.
 Below we present the results relevant to this work.
\bigskip

Let $\boldsymbol{\Psi}_N$
be  the  RG trajectory formed by  the 
 low energy Bethe states, whose energy and momentum are
described  by  the  asymptotic formula \eqref{tower1}, and consider the
eigenvalue $A_+(\zeta)$ \eqref{Aeig1bb1} computed on this family.
Its logarithm can
be expanded in the infinite series,
\be\label{logAdef1a}
\log A_+(\zeta)=-\sum_{j=1}^\infty\,h_j^{(N)}\,\zeta^j\,,
\ee
where the coefficients are given by the finite sums
\be\label{Apmdef1}
h_{j}^{(N)}=j^{-1}\,\sum_{m=1}^{\frac{N}{2}-S^z} (\zeta_m)^{-j}\ .
\ee
Keeping $S^z\ge 0$ fixed, consider the large $N$ limit of $h_{j}^{(N)}$ with given $j=1,2,3,\ldots\ $.
Despite that the r.h.s of eq.\,\eqref{Apmdef1} is a symmetric function of the Bethe roots,
for analysing this limit it is useful to impose an ordering 
for the set $\{\zeta_m\}$. For the case of the vacuum states 
 all $\zeta_m$ are real 
and, with the Bethe numbers given by eq.\,\eqref{kssusu},
they are ordered as $\zeta_1<\zeta_2<\ldots<\zeta_M$. 
In general, the Bethe roots are complex numbers and we can order them w.r.t. to their real part
 \bea
 \Re e(\zeta_1)\leq \Re e(\zeta_2)\leq\ldots  \leq \Re e(\zeta_M)\nonumber
 \eea
(the ordering prescription for  the Bethe roots with coinciding real parts is not essential for our purposes).
As was discussed in the work \cite{Kotousov:2019ygw}
for fixed $S^z$ and $m=1,2,\ldots$ the following limits exist:
\be\label{Brootlim1}
s_m=\lim\limits_{N\to\infty}\big(\tfrac{N}{\pi}\big)^{2(1-\beta^2)}\,\zeta_m\ .
\ee
 Furthermore the numbers $s_m$ grow according to
\be\label{sasymp1a}
(s_m)^{\frac{1}{2(1-\beta^2)}}=m+O(1)\qquad {\rm as} \qquad m\to +\infty\ .
\ee
For $0<\beta^2<\tfrac{1}{2}$ the above formulae  imply the existence of the limit
\be\label{gmlimit1}
h_{j}^{(\infty)}=\lim_{N\to\infty} N^{-2j(1-\beta^2)}\,h_{j}^{(N)}=\pi^{-2j(1-\beta^2)}\ 
j^{-1}\  \sum_{m=1}^\infty
 (s_m)^{-j}
\ee
with fixed $j=1,2,\ldots$\ .
Combining \eqref{logAdef1a} and \eqref{gmlimit1} one arrives at
\be\label{Aeigscale1}
\lim_{N\to\infty}\log\,A_+\big(N^{-2(1-\beta^2)}\,{\tilde \zeta}\,\big)=-\sum_{j= 1}^\infty\,h_{j}^{(\infty)}\,\tilde{\zeta}^j\,,
\ee
where the r.h.s. is understood as a formal power series expansion in $\tilde{\zeta}$ without any reference to its convergence.
It turns out there exists a mutually commuting set of operators
${\bf H}_j^{(+)}$ that act in the Fock space $ {\cal F}_P$ and whose eigenvalues for
a certain common eigenvector ${\boldsymbol \psi}_P({\boldsymbol v})\in {\cal F}_P^{({\tt L})}$ 
coincide with $h_{j}^{(\infty)}$
up to an overall multiplicative factor.
The construction of these operators was discussed in the works 
\cite{Bazhanov:1996dr,Bazhanov:1998dq}
 and goes
along the following line.

\bigskip

\bigskip 
Consider the chiral Bose field
\bea\label{bosefiled1}
\varphi(u)=\varphi_0+a_0\, u+\ri \sum_{m\not=0}\frac{a_{m}}{m}\ \re^{-\ri m u}\ ,
\eea
where $a_m$ are the Heisenberg generators satisfying \eqref{acomm1} and
the additional operator $\varphi_0$ obeys the commutation relations
\bea
[\varphi_0,a_m]=\tfrac{\ri}{2}\ \delta_{m,0}\ .
\eea
Introduce the path-ordered exponent
\bea\label{Lop2a}
{\boldsymbol L}_\pm(\lambda)=\re^{\pm\ri\pi \beta a_0{\cal H}}\ \overset{\leftarrow}{{\cal P}}\exp\bigg(\int_0^{2\pi}\rd u\,
\Big(V_-(u)\ q^{\pm\frac{\cal H}{2}}\,{\cal E}_\pm+\lambda^2\, V_+(u)\ q^{\mp\frac{\cal H}{2}}\,{\cal E}_\mp\Big)\bigg)
\eea
involving the vertex operators
\bea\label{vert1a}
V_{\pm}(u)=\re^{\pm 2\ri \beta\varphi}(u)
\eea
as well as   the generators of the
$q$-oscillator algebra ${\cal E}_\pm$ and ${\cal H}$:
\be\label{qosc}
[\Hcal,\Ecal_\pm]=\pm2\,\Ecal_\pm\,,\qquad q\, \Ecal_+\Ecal_--
q^{-1}\,\Ecal_-\Ecal_+=\frac{1}{q-q^{-1}} \ .
\ee
Let $\rho_\pm$  be representations of this algebra such that the traces 
\be\label{tr1}
{{\rm Tr}}_{\rho_\pm}\big[
\re^{\pm 2\ri\pi\beta P{\cal H}}\big]\ne 0 \qquad {\rm with} \qquad {\Im} m(P)<0
\ee
exist and are non-vanishing.
Then
one  may introduce the operators ${\mathlarger{\mathlarger{\mathlarger{\mathlarger {\bf \it a}}}}}_\pm(\lambda)$
as 
\bea\label{soso1a}
{\mathlarger{\mathlarger{\mathlarger {\bf \it a}}}}_\pm(\lambda)=
\frac{{\rm Tr}_{\rho_\pm}\big[\,\re^{\pm\ri\pi\beta  a_0{\cal H}}
{{\boldsymbol L}}_\pm(\lambda)\,\big]}{{\rm Tr}_{\rho_\pm}\big[\,\re^{\pm2\ri\pi \beta a_0{\cal H}}\,\big]}\ .
\eea
Formula \eqref{soso1a}
defines  a  power series in $\lambda^2$. Since
 ${\mathlarger {\mathlarger{\mathlarger{\mathlarger {\bf \it a}}}}}_\pm(0)=\bm{1}$,
its logarithm obeys the formal power series expansion
\be\label{logaseries1}
\log{\mathlarger{\mathlarger{\mathlarger {\bf \it a}}}}_\pm(\lambda)=-\sum_{j=1}^\infty {\bf H}_j^{(\pm)}\,\lambda^{2j}\,.
\ee  
Each of the coefficients ${\bf H}_j^{(\pm)}$  is expressed in terms of ordered integrals over the vertex
operators \eqref{vert1a}. A simple analysis gives that
for $0<\beta^2<\frac{1}{2}$ all these integrals converge and each term in the power
series expansion is well-defined. 
It is possible to show \cite{Bazhanov:1998dq} that ${\bf H}_j^{(\pm)}$
act invariantly  in the level subspaces $ {\cal F}^{({\tt L})}_P$ and mutually commute,
\be
{\bf H}_j^{(\pm)}\ : 
\ {\cal F}^{({\tt L})}_P\mapsto {\cal F}^{({\tt L})}_P\, , \qquad \big[\,{\bf H}_j^{(\pm)},\,{\bf H}_{j'}^{(\pm)}\,\big]=
\big[\,{\bf H}_j^{(\pm)},\,{\bf H}_{j'}^{(\mp)}\,\big]=0\ . 
\ee
Note that,
although  to take the trace in eq.\,\eqref{soso1a}  it is required that $\Im m(P)<0$,
the matrix elements of ${\bf H}_j^{(\pm)}$
 restricted to ${\cal F}^{({\tt L})}_P$
 may be analytically continued to
any complex $P$, except for
\be\label{popso1}
P=\mp\,\tfrac{1}{2}\,\big(m\beta^{-1}+j\beta\,\big)\,,\qquad {\rm where} \qquad m=0,1,2,\ldots\ .
\ee
\bigskip

The simultaneous diagonalization of the mutually commuting
operators ${\bf H}_j^{(\pm)}$, being restricted to ${\cal F}^{({\tt L})}_P$,
becomes a diagonalization problem of finite 
${\tt par}_1({\tt L})\times {\tt par}_1({\tt L})$ 
dimensional matrices.
Then for an RG trajectory $\bm{\Psi}_N$ with given $P$, ${\tt L}$ and characterized by 
the set $\big\{h_j^{(N)}\big\}_{j=1}^\infty$, 
there exists a common eigenvector
${\boldsymbol \psi}_P({\boldsymbol v})\in{\cal F}^{({\tt L})}_P$ such that the eigenvalues
of the operators ${\bf H}_j^{(+)}$,
\be\label{Heigdef1a}
{\bf H}_j^{(+)}\, {\boldsymbol \psi}_P({\boldsymbol v})=H_j(\bm{v})\,{\boldsymbol \psi}_P({\boldsymbol v})\ ,
\ee
are related to $h_j^{(\infty)}$ \eqref{gmlimit1} as
\be\label{hHeq1a}
h_j^{(\infty)}=
\Bigg[\, \Gamma(1-\beta^2)\
\Bigg(\frac{\sqrt{\pi}\,\Gamma\big(\frac{\beta^2}{2-2\beta^2}\big)}
{\Gamma\big(\frac{1}{2-2\beta^2}\big)}
\, \Bigg)^{1-\beta^2}\ \Bigg]^{-2j}\ \ H_j(\bm{v})\,.
\ee
In writing the above it is assumed that the exponential operators \eqref{vert1a} are normalized 
in the following way 
\be
q\,\re^{\pm 2\ri \beta\varphi}(u_1)\,\re^{\mp 2\ri \beta\varphi}(u_2)\big|_{(u_1-u_2)\to 0^+}\to (u_1-u_2)^{-2\beta^2}>0\ ,
\ee
i.e., we set the coefficient of the most singular term in the operator product expansion to be one.
It is clear that the  eigenvector ${\boldsymbol \psi}_P({\boldsymbol v})$
should be identified with the chiral state, which appears in the r.h.s. of \eqref{Psi1a}.
A similar analysis can be repeated to specify
the barred state  ${\bar {\boldsymbol \psi}}_{\bar P}({\bar {\boldsymbol v}})
\in\bar{{\cal F}}^{(\bar{{\tt L}})}_{\bar{P}}$ in that relation.
\bigskip

The l.h.s. of \eqref{hHeq1a} is an infinite sum over inverse powers of the scaled Bethe roots \eqref{Brootlim1}.
Since the eigenvalues  $H_j(\bm{v})$ may be calculated independently using the definition \eqref{soso1a},\,\eqref{logaseries1}, 
the relation \eqref{hHeq1a} provides a set of
sum rules for $s_m$.
It is instructive to consider the explicit formulae for $H_j(\bm{v})$
corresponding to the Fock vacuum $|P\rangle$.
As explained in \cite{Bazhanov:1996dr} they are expressed
 in terms of the $2m$\,-\,fold integrals 
\bea\label{QQeig1a}
Q_m(h,g)&=& \int_0^{2\pi}\rd u_1\int_0^{u_1}\rd v_1\int_0^{v_1}\rd u_2\int_0^{u_2}\rd v_2\ \ldots \int_0^{v_{m-1}}\rd u_m
\int_0^{u_m}\rd v_m  \nonumber \\[0.2cm]
&\times& \prod_{j>i}^m\bigg[\Big(4\sin\big(\tfrac{u_i-u_j}{2}\big)\,\sin\big(\tfrac{v_i-v_j}{2}\big)\Big)^{2g}\bigg] \ \ 
\prod_{j\ge i}^m\Big(2\sin\big(\tfrac{u_i-v_j}{2}\big)\Big)^{-2g}\\[0.2cm]
&\times&
\prod_{j> i}^m\Big(2\sin\big(\tfrac{v_i-u_j}{2}\big)\Big)^{-2g}\ 
 2\cos\Big(2h\,\big(\pi+\sum_{i=1}^m(v_i-u_i)\big)\Big)
\nonumber
\eea
 with $m\le j$.
For instance
\bea\label{eq1a2}
{ H}_1^{({\rm vac})}\!\!&=&\!\!\frac{Q_1(\beta P,\beta^2)}{4\sin(\pi\beta^2)\,\sin\!\big(\pi\beta(\beta+2 P)\big)}\\[0.2cm]
{ H}_2^{({\rm vac})}\!\!&=&\!\!\frac{Q_2(\beta P,\beta^2)}{4\sin(2\pi\beta^2)\sin\!\big(2\pi\beta(\beta+  P)\big)}+
\frac{\cos\!\big(2\pi\beta(\beta + P)\big)\,\big(Q_1(\beta P,\beta^2)\big)^2}{16\sin(2\pi\beta^2)\,\sin^2\!\big(\pi\beta(\beta+ 2P)\big)\,
\sin\!\big(2\pi\beta(\beta+  P)\big)}\nonumber
\eea
Taking a brief look at eq.\,\eqref{QQeig1a}, one finds  that the
integrals converge as $0<g<\frac{1}{2}$ and, in this parametric domain,
$Q_m(h,g)=Q_m(-h,g)$ are entire functions of $h^2$. 
However
the multifold integrals  \eqref{QQeig1a} are not well suited for numerical purposes.
In the Appendix of  ref.\,\cite{Bazhanov:1998za} a technique is developed which brings 
 ${ H}_j^{({\rm vac})}$ to a form that is convenient for computation.
Following that work, introduce the functions
\bea\label{apsodpoa10a}
f_1(h,g)&=&                                          
                                                  \frac{\pi\,\Gamma(1-2g)\,\Gamma(g+2 h)}{\sin(\pi g)\,
                                                  \Gamma(1-g+2 h)}
                                                          \\[0.2cm]\label{mmwenmn12}
f_2(h, g)&=&2^{1-4g}\,\frac{\Gamma^2(1-g)}{\Gamma^2(\frac{1}{2}+g)}\
\frac{\Gamma(2g+2 h)}{\Gamma(1-2g+2 h)}\,\int_{ -\infty}^\infty\frac{\rd x}{2\pi}\,\frac{S_1(x)}{x+\ri h}\ \ \ \ \ \  
\big(0<g<\tfrac{1}{2},\ \Re e(h)>0\big)\ ,\nonumber
\eea
where
\be\label{Sdef1a}
S_1(x)=\sinh(2\pi x)\Gamma(1-2g+2\ri x)\Gamma(1-2g-2\ri x)\,\big(\Gamma(g+2\ri x)\Gamma(g-2\ri x)\big)^2\ .
\ee
Then for the first two eigenvalues, one has
\bea\label{oasoidoi1a}
{ H}_j^{({\rm vac})}=f_j\big(\beta P,\beta^2\big)\ .
\eea
Notice that, although the ordered integral $Q_1(h,g)$ converges only for
$0<g<\frac{1}{2}$, the expression \eqref{apsodpoa10a},\,\eqref{oasoidoi1a}
gives
 an analytic continuation of ${ H}_1^{({\rm vac})}$
to the domain $\frac{1}{2}<\beta^2<1$, which possesses  a simple pole at $\beta^2=\frac{1}{2}$. 
The 
analytic continuation of  $f_2(h,g)$
yields \cite{Bazhanov:1998za}
\bea\label{jassusau}
f_2(h,g)&=&2^{1-4 g}\,\frac{\Gamma^2(1-g)}{\Gamma^2(\frac{1}{2}+g)}\
\frac{\Gamma(2g+2 h)}{\Gamma(1-2g+2h)}\,\Bigg(\int_{-\infty}^\infty\frac{\rd x}{2\pi}\,
\frac{S_1(x)}{x+\ri h}  \\[0.2cm]
&-&
\frac{\sin(2\pi g)\Gamma(3-4 g)\Gamma^2(1-g)\Gamma^2(3g-1)}{(2h+1-2g)
(2h-1+2 g)}\Bigg)
 \qquad \big(\,\tfrac{1}{2}<g<1, \ \Re e(h)>0\, \big)\ .\nonumber
\eea
The above expression shows that ${ H}_2^{({\rm vac})}$ possesses a simple pole at $\beta^2=\frac{3}{4}$.
Similarly, by means of 
analytical continuation in $\beta^2$, the functions ${ H}_j^{({\rm vac})}$ with given $j=1,2,3,\ldots$
may be defined for any $0<\beta^2<1$  except 
 $\beta^2=1-\frac{1}{2j}$. 
\bigskip

 A natural question arises, is it possible 
to extend the definition, not only of the vacuum eigenvalues ${ H}_j^{({\rm vac})}$, but of the operators  
${\bf H}_j^{(\pm)}$  themselves to the domain $0<\beta^2<1$?
Let us reiterate  that for given $j=1,2,\ldots$   formulae \eqref{Lop2a},\,\eqref{soso1a}
and \eqref{logaseries1} define  ${\bf H}_j^{(\pm)}$  in terms of the ordered  integrals over the vertex operators, which
can be taken literally for  $0<\beta^2<\frac{1}{2}$ only. 
As discussed in the work \cite{Bazhanov:1998dq} (see also \cite{Bazhanov:2018xzh}),
by re-expressing the  ordered integrals in terms of contour integrals,
it is possible to extend the applicability of these formulae to any 
 $0<\beta^2<1$ except the points
 $\beta^2=1-\frac{1}{2k}$ with $k=1,2,\ldots\ $. 
At  $\beta^2=1-\frac{1}{2k}$
all the operators ${\bf H}_j^{(\pm)}$ with $j\ne k$ remain non-singular. However
${\bf H}_k^{(\pm)}$ 
possesses a simple pole,
 whose residue is proportional to the identity operator:
\be\label{Hpole1}
{\bf H}_k^{(\pm)}= -
\frac{ \Gamma(\frac{1}{2}+k)\,\Gamma^{2k}(\frac{1}{2k})}{\sqrt{\pi}\,(2 k-1) k\,\Gamma(1+k)}\ \ 
 \frac{\bm{1}}{\beta^2-1+\tfrac{1}{2k}}+O(1)\ \qquad (k=1,2,3,\ldots)\ .
\ee 
By subtracting the singular term in \eqref{Hpole1}
from ${\bf H}_k^{(\pm)}$ one may introduce the
regularized operator ${\bf H}_k^{(\pm,{\rm reg})}$.
The latter is defined up to an overall
additive constant, which should be fixed by imposing some normalization condition.

\begin{figure}
\centering
\scalebox{0.95}{
\begin{tikzpicture}
\node at (0,0) {\includegraphics[width=7cm]{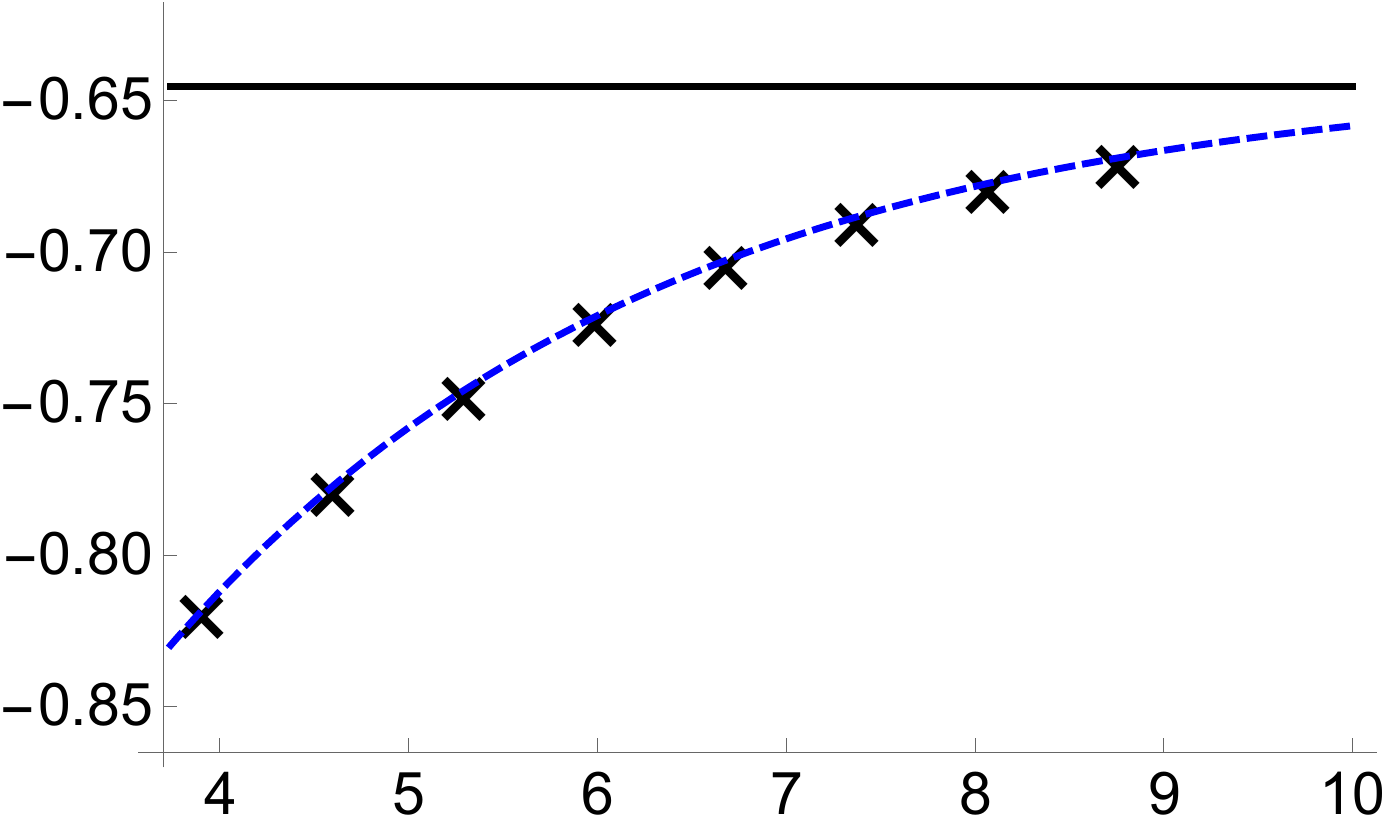}};
\node at (9,0.0) {\includegraphics[width=7.2cm]{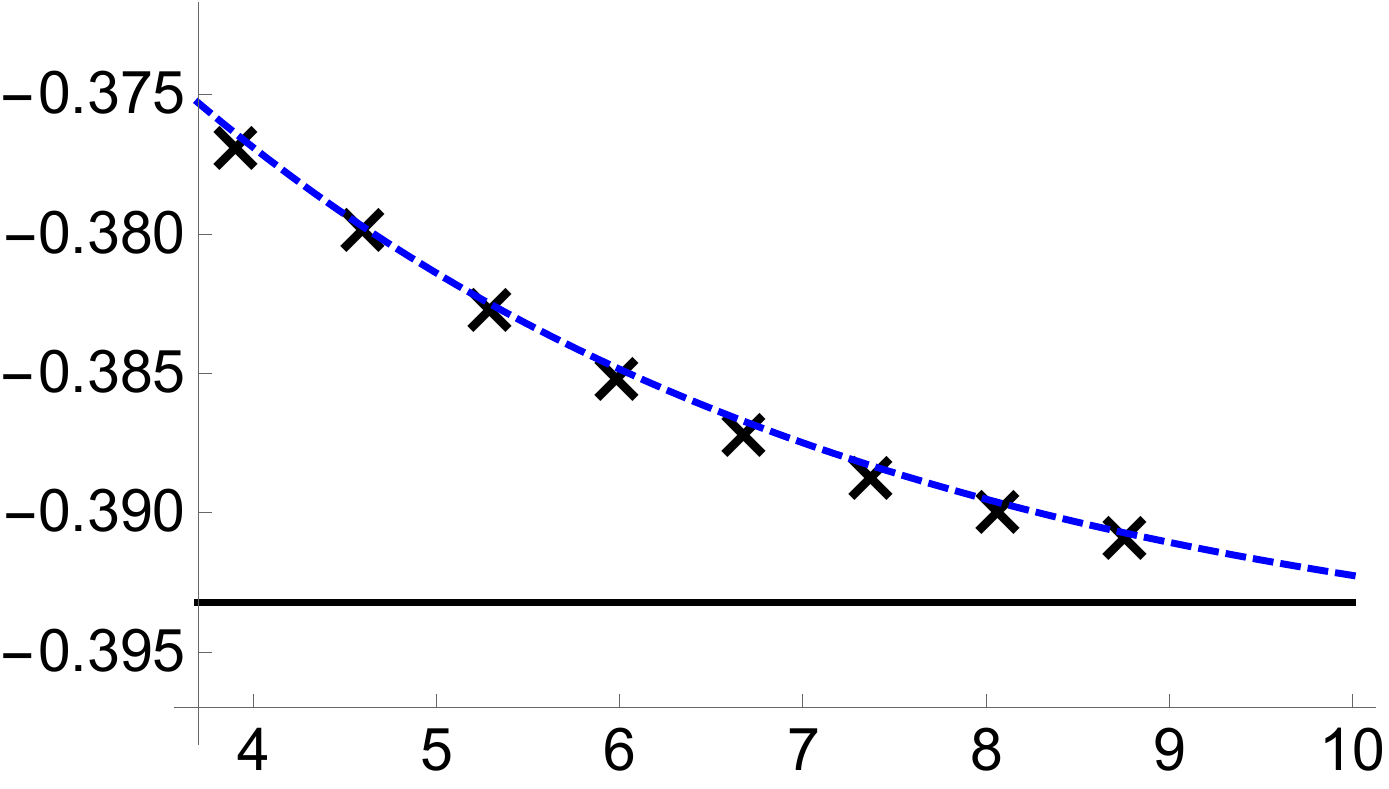}};
\node at (4.3,-1.7) {$\log(N)$};
\node at (13.3,-1.7) {$\log(N)$};
\node at (-1.8,2.5) {$N^{-2(1-\beta^2)}\,h^{(N,{\rm reg})}_1$};
\node at (7.3,2.5) {$N^{-4(1-\beta^2)}\,h^{(N,{\rm reg})}_2$};
\end{tikzpicture}
}
\caption{\small
The crosses come from numerical data that was obtained from
the solution of the Bethe ansatz equations with
$N=50,100,200,\ldots$
corresponding to
 an RG trajectory  $\bm{\Psi}_N$. The latter is  characterized by
${\tt L}=\bar{\tt L}=0$ and $P$, $\bar{P}$ given by eq.\,\eqref{PPbareq1} with
$S^z=0$, ${\tt w}=1$.
 The parameters  are taken to be 
 $\beta^2=\frac{9}{10}$ and ${\tt k}=-\frac{1}{20}$.
 Notice that, since $\beta^2$ is close to one, it is necessary to
perform the subtraction $h_j^{(N,{\rm reg})}\equiv
h_j^{(N)}+\frac{(-1)^{j+1}\,N}{2j\cos(\pi j\beta^2)}$   as in  eq.\,\eqref{poaspo1a333}, so that
$N^{-2j(1-\beta^2)}h_j^{(N,{\rm reg})}$ with $j=1,2$ tends to a finite number in the large $N$ limit.
The solid  line represents
the limiting value given by the r.h.s. of eq.\,\eqref{hHeq1a}, where $H_1^{({\rm vac})}$, $H_2^{({\rm vac})}$ were computed
using \eqref{oasoidoi1a} with $f_1$ as in \eqref{apsodpoa10a}  and $f_2$ from \eqref{jassusau}. 
The blue dashed line comes from fitting the numerical data and was included for visualization.
\label{fig1}}
\end{figure}
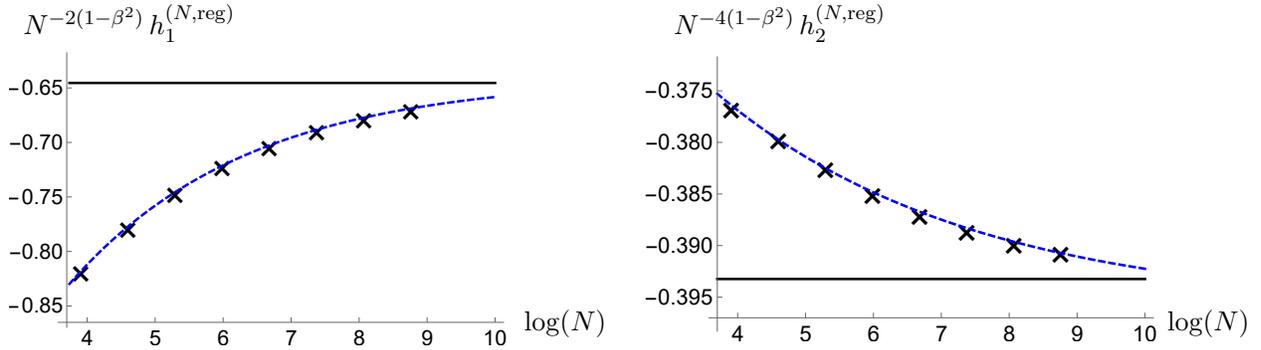

\bigskip

In the parametric domain $\frac{1}{2}\le\beta^2<1$ 
both \eqref{Brootlim1} and \eqref{sasymp1a} continue to hold.
However the large $N$ limit
 \eqref{gmlimit1} involving the coefficients
$h_j^{(N)}$  \eqref{Apmdef1}  no  longer exists when  $1\leq j\leq\frac{1}{2(1-\beta^2)}$.
To properly define $h_j^{(\infty)}$, 
a certain subtraction needs to be made from
$N^{-2j(1-\beta^2)}\,h_j^{(N)}$ 
so that its  large $N$ limit
 can be taken. Namely,
\be\label{poaspo1a333}
h_{j}^{(\infty)}=\lim_{N\to\infty} N^{-2j(1-\beta^2)}\,
\Bigg[h^{(N)}_j+\frac{(-1)^{j+1}\, N}{2j\cos(\pi j\beta^2)}\Bigg]\, , \qquad\qquad  j=1,2,\ldots  
<\tfrac{1}{2(1-\beta^2)}\, .
\ee
Without going into details, we just mention that the existence of the above limit follows from the Bethe ansatz equations 
\eqref{bae} with $\eta_J=1$.
When $\beta^2=1-\frac{1}{2k}$ with $k=1,2,\ldots\,$, not only  $h_j^{(N)}$ with $j=1,\ldots, k-1$ but also
  $h_k^{(N)}$ requires regularization:
\bea\label{hk1aa}
h_{k}^{(\infty)}=  \lim_{N\to\infty}\,
\bigg[N^{-1}\,h_k^{(N)} -\frac{1 }{\pi k}\, \log(N B_k) \bigg]\ \ \ \ \ 
\ \ \  \big(\beta^2=1-\tfrac{1}{2k}\big)\ ,
\eea
where $B_k$ is an arbitrary ($k$\,-\,dependent) constant.
\bigskip

The validity of the relation
\eqref{hHeq1a} may be extended to the domain $0<\beta^2<1$
provided that for $\beta^2>\frac{1}{2}$ the coefficients
$h^{(\infty)}_j$  are defined as in \eqref{poaspo1a333}, while the eigenvalues $H_j(\bm{v})$
are understood via analytic continuation.
Note that in the case when $\beta^2=1-\frac{1}{2k}$, eq.\,\eqref{hHeq1a} with $j=k$
becomes a relation between $h_{k}^{(\infty)}$ \eqref{hk1aa} and the 
eigenvalues of the regularized operator 
${\bf H}_k^{(+,{\rm reg})}$. The  arbitrary constant $B_k$ in \eqref{hk1aa} is related to 
the ambiguity in the definition of ${\bf H}_k^{(+,{\rm reg})}$ discussed above.
In what follows we define the  regularized operator as 
\bea\label{isisaiasi4443}
{\bf H}_k^{(\pm,{\rm reg})}=\lim_{\beta^2\to 1-\frac{1}{2k}} \bigg[\, {\bf H}_k^{(\pm)}+
\frac{\Gamma(\frac{1}{2}+k)\,\Gamma^{2k}(\frac{1}{2k})}{\sqrt{\pi}\, (2 k-1) k\,\Gamma(1+k)}\
\  \frac{1}{\beta^2-1+\tfrac{1}{2k}}\,\bigg]\ .
\eea
Then it is not difficult to show that
\bea\label{iaosi89891}
B_1=\frac{\re^{\gamma_{\rm E}}}{\pi}\ ,
\eea
where $\gamma_{\rm E}$ stands for the Euler constant.
The analytical expression for  $B_k$ with $k\ge2$ is not known.
Numerical calculations yield
\bea\label{iaosi89892}
\log(B_2/B_1)=
3.57079634 \ .
\eea
Also, in fig.\,\ref{fig1} some numerical data illustrating  \eqref{hHeq1a},\,\eqref{poaspo1a333} is presented.

\bigskip

\subsection{Scaling limit of $A_+(\zeta)$}

The set of ``scaled'' Bethe roots $\{s_m\}_{m=1}^\infty$ \eqref{Brootlim1}
admits a remarkable interpretation. 
Namely, following the works \cite{Bazhanov:2003ni,Kotousov:2019ygw}, consider
 the Schr$\ddot{\rm o}$dinger equation
\bea\label{jasys}
\bigg(- {\frac{{\rd}^2}{{\rd} x^{2}}} +V(x)-E\bigg)\, \Phi=0
\eea
with the so-called Monster potentials of the form
\bea\label{MonstP1}
V(x)=\frac{16(\alpha+1)P^2-1}{4x^2}+x^{2\alpha}-2\, \frac{{\rd}^2}{{\rd} x^2}\sum_{b=1}^{\tt L}\log\big(x^{2\alpha+2}
-\tfrac{\alpha+1}{\alpha}\, v_b\big)\ .
\eea
Here the set of complex numbers $\{v_a\}_{a=1}^{\tt L}$ obeys  the
 system of ${\tt L}$ algebraic equations:
\bea\label{jsaysssysa}
\sum_{b\not=a}\frac{v_a\, (\, v_a^2+(3+\alpha)(1+2\alpha) v_a v_b + \alpha(1+2\alpha)\, v_b^2\,)}{(v_a-v_b)^3}
-\frac{ v_a}{4}+P^2-\frac{\alpha^2}{4 (\alpha+1)}
=0\ .
\eea 
With these constraints imposed on the positions of the singularities 
 any solution of the Schr$\ddot{\rm o}$dinger equation 
 is monodromy free  everywhere  except for $x=0$ and $x=\infty$ for any value of $E$.
In other words the solutions remain single-valued 
in the vicinity of each singularity specified by $v_a$.
For this reason the complex numbers $\{v_a\}$ are referred to as  apparent  singularities.
 Assuming that $\alpha>0$, one can consider the standard spectral problem for
the ODE defined on the  ray $x>0$.  This leads to a discrete spectral set 
$\{E_m\}_{m=1}^\infty$. Then for an RG trajectory with given $P$ and ${\tt L}$
and characterized by the set of scaled Bethe numbers $\{s_m\}_{m=1}^\infty$ \eqref{Brootlim1},
there exists a Monster potential of the form \eqref{MonstP1},\,\eqref{jsaysssysa} 
such that
\be\label{Energj1}
E_m =\big(N_0/\pi\big)^{-2(1-\beta^2)}\,s_m\ .
\ee
Here
\bea\label{N0def1}
N_0=\frac{\sqrt{\pi}\,\Gamma\big(1+\frac{\beta^2}{2-2\beta^2}\big)}
{2 \Gamma\big(\frac{3}{2}+\frac{\beta^2}{2-2\beta^2}\big)}
\eea
and the parameter $\alpha$ is related to $\beta$ as
\be\label{alphbet1a}
\alpha=\beta^{-2}-1>0\ .
\ee
It was conjectured in the work  \cite{Bazhanov:2003ni} and  proven
 by Conti and  Masoero \cite{Masoero}
that for generic (complex) values of $P$ and $\alpha$
the number of distinct, up to the action of the symmetric group $S_N$,
 solutions of  \eqref{jsaysssysa} coincides with 
${\tt par}_1({\tt L})$. 
In other words, for given ${\tt L}$, the number of Monster potentials 
is equal to the number of states in the level subspace 
${\cal F}_{P}^{({\tt L})}$.
This allows one to label the chiral state 
${\boldsymbol\psi}_P({\bm{v}})$  entering in the scaling limit \eqref{Psi1a}
by the unordered set of solutions
of the system \eqref{jsaysssysa},
\be\label{aspos1a}
{\boldsymbol\psi}_P({\bm{v}})\ \  :\ \  \ \ \bm{v}=\{v_a\}_{a=1}^{{\tt L}}\ .
\ee
\bigskip

Introduce the  spectral determinant
\be\label{saoisoi1a}
D_+(E\,|\,{\boldsymbol v})=\prod_{m=1}^\infty\bigg(1-\frac{E}{E_m}\bigg)\,.
\ee
Due to \eqref{Energj1},\,\eqref{sasymp1a}
$
E_m\sim m^{2(1-\beta^2)}$ as $m\to\infty$,
so that the infinite product in the r.h.s. converges when $0<\beta^2<\frac{1}{2}$
for any value of $E$.
Hence in this domain \eqref{saoisoi1a} defines an entire function. 
Expanding $\log\,D_+(E)$ in an infinite series, it follows from eqs.\,\eqref{gmlimit1} and
\eqref{Energj1} that the resulting series expansion coincides term by term with the r.h.s. of \eqref{Aeigscale1}
where
$\tilde{\zeta}=N_0^{2(1-\beta^2)}\,E$.
This immediately yields that the series in \eqref{Aeigscale1} has a non-vanishing  radius of convergence and
\bea\label{iisisi}
 \lim_{N\to\infty} A_+\big(\, (N/N_0)^{2(\beta^2-1)} E \,\big)=D_+({ E}\,|\,{\boldsymbol v})
\qquad\qquad  (0<\beta^2<\tfrac{1}{2})\ .
\eea

\bigskip
To introduce $D_+(E\,|\,{\boldsymbol v})$ 
for $\frac{1}{2}\le \beta^2<1$, Weierstrass factors must be included in the
infinite product \eqref{saoisoi1a} in order to ensure its convergence.
Alternatively one can define the spectral determinant for $0<\beta^2<1$ 
through the set of conditions
\begin{enumerate}\label{spec1a}
\item[$({\rm i})$] $D_+(E\,|\,{\boldsymbol v})$ is an entire function whose zeroes coincide with $\{E_m\}_{m=1}^\infty$\,.
\item[$({\rm ii})$] $D_+(E\,|\,{\boldsymbol v})$ satisfies the normalization condition:
\be
D_+(0\,|\,{\boldsymbol v})=1\, .
\ee
\item[$({\rm iii})$] It possesses the asymptotic 
behaviour as $E\to\infty$,
$ |\arg(-E)|<\pi$: 
\bea\label{hassaast}
\log D_+(E\,|\,{\boldsymbol v})\asymp
\begin{cases}
\dfrac{N_0}{\cos(\frac{\pi \beta^2}{2-2\beta^2})}
\ (- E)^{\frac{1}{2-2\beta^2}}+o\big(E\big)\ \ \  &{\rm for}\ \ \ \ \beta^2\not= 1-\frac{1}{2k}\\[0.8cm]
 \dfrac{\Gamma(\frac{1}{2}+k)}{2\sqrt{\pi}\Gamma(1+k)} \, E^k\ \big(\log(-E)+c_k\big)+o\big(E\big)
 \ \ \  &{\rm for}\ \ \ \ \beta^2= 1-\frac{1}{2k}
 \end{cases}
\eea
with $k=1,2,3,\ldots$ and
\bea\label{ckdef1a}
c_k=\psi\Big(\frac{1}{2}+k\Big)-\psi(1+k)-\frac{1}{k}+\frac{1}{k}\, \bigg[\, \log\Big(\frac{2k-1}{4k}\Big)-\frac{1}{2k-1}-
\psi\Big(\frac{1}{2k}\Big)\,\bigg]\ .
\eea
\end{enumerate}
The function $D_+(E\,|\,{\boldsymbol v})$, thus defined, coincides with the eigenvalue of the operator
${\mathlarger{\mathlarger{\mathlarger{\mathlarger {\bf \it a}}}}}_+(\lambda)$ for the vector ${\boldsymbol\psi}_P({\bm{v}})$
\eqref{aspos1a}
\be\label{oiasodio1ia}
{\mathlarger{\mathlarger{\mathlarger {\bf \it a}}}}_+(\lambda)\,{\boldsymbol\psi}_P({\bm{v}})
=D_+(E\,|\,{\boldsymbol v})\,{\boldsymbol\psi}_P({\bm{v}})
\ .
\ee
As it follows from eqs.\,\eqref{Aeigscale1},\,\eqref{logaseries1},\,\eqref{hHeq1a},\,\eqref{iisisi}
$E$ and $\lambda$ are related as
\bea\label{kassaisa}
\lambda^2= \frac{(\beta^2/2)^{2-2\beta^2}}{\Gamma^2(1-\beta^2)}\ E\ .
\eea
The above conditions $({\rm i})-({\rm iii})$ 
fully specify $D_+(E\,|\,{\boldsymbol v})$, i.e., all the 
eigenvalues of ${\mathlarger{\mathlarger{\mathlarger{\mathlarger {\bf \it a}}}}}_+(\lambda)$.
Notice that with
the choice of the constant $c_k$ as in \eqref{ckdef1a},
for $\beta^2=1-\frac{1}{2k}$  with $k=1,2,3,\ldots$
the operator ${\mathlarger{\mathlarger{\mathlarger{\mathlarger {\bf \it a}}}}}_+(\lambda)$
 is defined as
\bea
{\mathlarger{\mathlarger{\mathlarger {\bf \it a}}}}_+\big|_{\beta^2=1-\frac{1}{2k}}&=&\lim_{\beta^2\to 1-\frac{1}{2k}\atop
E-{\rm fixed}} 
{\mathlarger{\mathlarger{\mathlarger {\bf \it a}}}}_+\big(\lambda(E)\big)\ 
\exp\bigg[\, -
\frac{\Gamma(\frac{1}{2}+k)\,\Gamma^{2k}(\frac{1}{2k})}{\sqrt{\pi}\, (2 k-1) k\,\Gamma(1+k)}\
\  \frac{\big(\lambda(E)\big)^{2k}}{\beta^2-1+\tfrac{1}{2k}}\,\bigg] \nonumber \\[0.4cm]
&=&\exp\bigg(-{\bf H}_k^{(+,{\rm reg})}\,\lambda^{2k}\,-\sum_{j\ne k}\,{\bf H}_j^{(+)}\,\lambda^{2j}\bigg)\,,
\eea
where  ${\bf H}_k^{(+,{\rm reg})}$ is
given by eq.\eqref{isisaiasi4443}.
\bigskip

Now we can describe the scaling limit of the eigenvalue $A_+(\zeta)$ corresponding to the RG trajectory $\bm{\Psi}_N$
for any $0<\beta^2<1$. Namely,
\bea\label{iisis1a}
 \lim_{N\to\infty} G^{(N)}\big(E\,|\,\beta^2\big)\ A_+\big(\, (N/N_0)^{2(\beta^2-1)} E \big)=D_+({ E}\,|\,{\boldsymbol v})
\qquad\qquad  \big(0<\beta^2<1\big)
\eea
with
\bea\label{saysysa}
G^{(N)}(E\,|\,g)=\begin{cases} \exp\Bigg(\
{\displaystyle \sum_{m=1}^{\big[\frac{1}{2(1-g)}\big]}}\
\dfrac{(-1)^{m}\,N}{2m\cos(\pi m g)}\  (N/N_0)^{2m(g-1)}\  E^m\Bigg) &  {\rm for} \ \ \ \  g\ne 1-\tfrac{1}{2k}\\[1cm]
\exp\Bigg(
\dfrac{N_0 E^k}{\pi k}\,\log(N B_k)+
{\displaystyle\sum_{m=1}^{k-1}}\
\dfrac{N}{2m\cos(\frac{\pi m}{2k})}\  (N/N_0)^{-\frac{m}{k}}\  E^m\Bigg) &  {\rm for}\ \ \ \   g= 1-\tfrac{1}{2k}
\end{cases}
\eea
Here $k=1,2,\ldots$ and $[...]$ stands for the integer part.
\bigskip

The operator 
 ${\mathlarger{\mathlarger{\mathlarger{\mathlarger {\bf \it a}}}}}_-(\lambda)$ \eqref{soso1a}
appears in  
the scaling limit of  $\mathbb{A}_-$, which was briefly mentioned
in the Preliminaries (for a further discussion,
 including its definition, see sec.\,3 of \cite{Bazhanov:2020new}).
Similar to  \eqref{oiasodio1ia}, its eigenvalues are related
to the spectral determinant $D_-(E)$ corresponding to 
another spectral problem for the  same Schr$\ddot{\rm o}$dinger equation.
\bigskip

There is an efficient way of computing $D_\pm(E)$.
To describe it, first
introduce 
two solutions of \eqref{jasys},\,\eqref{MonstP1} satisfying the asymptotic condition
\be\label{psidef1aa}
\Phi_{ \pm P}(x)\to
\frac{1}{\sqrt{\pi}}\
(\beta^2/2)^{\frac{1}{2}\pm 2\beta P}\ \Gamma(\mp 2\beta P)\ 
 x^{\frac{1}{2}\pm\frac{2}{\beta}\,P} \ \ \ \ \quad {\rm as}\quad \ \ \ x\to 0  \ \ \ \ \ \ \ \ \
\big(\,0<\Re e( 2P)<\beta\big)
\ee
where $\beta=(\alpha+1)^{-\frac{1}{2}}>0$.
This unambiguously defines the solutions in the  strip in the complex $P$ plane.
It turns out that through analytic continuation in $P$  it is possible to 
introduce the solutions $\Phi_{\pm P}$ 
for any complex values of $P$, except for the set $P=\tfrac{1}{2}\,(m\beta^{-1}+n\beta)$ with 
$m,n$ integers. 
 Let $\Xi$ be another solution
that decays at large positive $x$ 
according to the asymptotic formula
\be\label{chilargezeq0}
\Xi(x)\asymp\,  x^{-\frac{1}{2\beta^2}(1-\beta^2)}\,\exp\Big(-\beta^2\,x^{\frac{1}{\beta^2}}+o(1)\Big)
\qquad\qquad {\rm as}\qquad\qquad x\to+\infty\ .
\ee
This condition specifies $\Xi(x)$ for $0<\beta^2<\frac{1}{2}$.
In the parametric domain $\frac{1}{2}< \beta^2<1$  a more accurate 
description of the large\,-\,$x$ asymptotic is required. Namely, the argument in the exponent in
\eqref{chilargezeq0} should be replaced by $-\beta^2\,x^{1/\beta^2}+
x^{1/\beta^2}\,\sum_{m\ge1} d_m\,x^{2m(1-1/\beta^{2})}$, where the 
 coefficients $d_m$ are easily obtained using the standard WKB technique.
This way the solution $\Xi(x)$ may be introduced for any $0<\beta^2<1$ 
 except the points $\beta^2\ne1-\frac{1}{2k}$ with $k=1,2,3,\ldots$ by means of the 
 asymptotic condition
\be\label{chilargezeq1}
\Xi(x)\asymp  x^{-\frac{1}{2\beta^2}(1-\beta^2)}\,\exp\Big[-\beta^2 x^{\beta^{-2}}\,
{}_2F_1\big( -\tfrac{1}{2},  -\tfrac{1}{2(1-\beta^2)},  
1 -\tfrac{1}{2(1-\beta^2)}\,\big|\, E\,x^{2(1-\beta^{-2})}  \,\big)+o(1)\Big]\,,
\ee
 where ${}_2F_1$ is the Gauss hypergeometric function. 
The case $\beta^2=1-\frac{1}{2k}$  requires
further attention and will not  be discussed here.
It turns out that the spectral determinant $D_+(E\,|\,{\boldsymbol v})$ defined by
$({\rm i})-({\rm iii})$ above as well as $D_-(E\,|\,{\boldsymbol v})$, which can be introduced by
a similar set of conditions, are given by
\be\label{Dconneq1a}
D_\pm(E\,|\,{\boldsymbol v})=\mp\,\sin(2\pi\beta P)
\ W[\,\Phi_{\pm P},\Xi\,]
\ee
with $W[\,\Phi_{\pm P},\Xi\,]=\Xi\,\partial_x\Phi_{\pm P}-\Phi_{\pm P}\,\partial_x\Xi$ being the Wronskian.
Indeed, using basic facts from the analytic theory of differential equations, it is easy to show that \eqref{Dconneq1a}
defines entire functions of $E$. 
When $D_+(E\,|\,{\boldsymbol v})$ vanishes, the solutions $\psi_+$ and $\chi$ are linearly dependent,
so that $\{E_m\}_{m=1}^\infty: \, D_+(E_m\,|\,{\boldsymbol v})=0$, is the spectral set for the corresponding spectral problem.
Similarly, the zeroes of $D_-(E\,|\,{\boldsymbol v})$ form the spectral set for the problem, where $\Xi$ becomes proportional to $\Phi_-$.
The normalization of the solutions $\Phi_\pm$ \eqref{psidef1aa} and  the overall factor in \eqref{Dconneq1a} ensure that
$D_\pm(0\,|\,{\boldsymbol v})=1$. Using the WKB technique one can check that the functions  \eqref{Dconneq1a}
satisfy the large\,-\,$E$ asymptotic \eqref{hassaast}.
Finally we note 
\be
\Xi(x)=D_+(E\,|\,{\boldsymbol v})\ \Phi_{-P}(x)
+ D_-(E\,|\,{\boldsymbol v})\ \Phi_{+P}(x)
\ee
so   that $D_\pm(E\,|\,{\boldsymbol v})$ are the connection coefficients in the expansion of $\Xi$
in terms of the fundamental set of solutions $\{\Phi_{\pm P}\}$.

\subsection{Scaling limit of the transfer matrix\label{sec24}}

Let $U_q({\mathfrak {sl}}_2)$ be  the quantum universal enveloping algebra whose generators satisfy
the commutation relations 
 \bea\label{iosdif090as}
[\,{\tt h}, {\tt e}_\pm\,]=\pm 2\, {\tt e}_{\pm}\ ,\ \ \ \ \ \ \ [\,{\tt e}_+,{\tt e}_-\,]=\frac{q^{\tt h}-q^{-{\tt h}}}{q-q^{-1}}\ .
\eea
Following ref.\cite{Bazhanov:1994ft} consider the formal path ordered exponent 
built out  of the vertex operators \eqref{vert1a},
\bea\label{9sd090as}
\bm{{ L}}(\lambda)=\lambda^{+\frac{1}{2}{\tt h}}\, 
\re^{\ri\pi \beta a_0{\tt h}}\ \overset{\leftarrow}{{\cal P}}\exp\bigg(\int_0^{2\pi}\rd u\,
\Big(V_-(u)\ q^{+\frac{{\tt h}}{2}}\,{\tt e}_++
\lambda^2\, V_+(u)\ q^{-\frac{{\tt h}}{2}}\,{\tt e}_-\Big)\bigg)\, \lambda^{-\frac{1}{2}{\tt h}}\ .
\eea
In the fundamental representation of $U_q(\mathfrak{sl}_2)$,  such that  
$\pi_{\text{\textonehalf}}({\tt e}_\pm)=\sigma^\pm$ and $\pi_{\text{\textonehalf}}({\tt h})=\sigma^z$,
$\bm{{ L}}(\lambda)$ becomes an operator valued $2\times 2$ matrix
\be
\bm{L}_{\text{\textonehalf}}(\lambda)=\pi_{\text{\textonehalf}}\big(\bm{{ L}}(\lambda)\big)\ .
\ee
As was shown in the work \cite{Bazhanov:1998dq}, the trace
\bea\label{asusuya231}
{\boldsymbol{\tau}}(\lambda)={\rm Tr}
\Big[\re^{\ri\pi \beta a_0\sigma^3}\,{\boldsymbol L}_{\text{\textonehalf}}(\lambda)\Big]\
\eea
commutes with  the operator \eqref{soso1a},
$\big[\,\bm{\tau}(\lambda),\,{\mathlarger{\mathlarger{\mathlarger{\mathlarger {\bf \it a}}}}}_+(\lambda')\,\big]=0$,
and furthermore satisfies the relation
  \bea\label{iaisaisa11a}
  \boldsymbol{\tau}(\lambda){\mathlarger{\mathlarger{\mathlarger {\bf \it a}}}}_+(\lambda)=
  \re^{+2\ri\pi \beta a_0}\,{\mathlarger{\mathlarger{\mathlarger {\bf \it a}}}}_+(q\lambda)  +
  \re^{- 2\ri\pi \beta a_0}\,{\mathlarger{\mathlarger{\mathlarger {\bf \it a}}}}_+(q^{-1}\lambda)\ .
  \eea
\smallskip

It should be
clear that \eqref{iaisaisa11a} is the scaling counterpart of \eqref{TArel1a},
where
 $\boldsymbol{\tau}(\lambda)$ appears  in the scaling limit of the transfer matrix
${\mathbb T}(\zeta)$. 
One can obtain a formula that describes  the scaling limit of the eigenvalues of  ${\mathbb T}(\zeta)$ 
 via a comparison of these two relations. Let $T^{(N)}(\zeta)$ be the eigenvalue
of the transfer matrix for an RG trajectory ${\boldsymbol\Psi}_N$ and consider
 eq.\,\eqref{TArel1a}   specialized to that common eigenvector.
Substituting the parameter  $\zeta$  by $(N/N_0)^{2(\beta^2-1)}\,E$ and then
using formulae  \eqref{iisis1a} and   \eqref{oiasodio1ia} 
one finds
\bea\label{sysaysaaa}
 \lim_{N\to\infty} G^{(N)}\big(q^2 E \,|\,\beta^2\big)\, G^{(N)}\big( q^{-2} E\,|\,\beta^2\big)\
 T^{(N)}\big(\, (N/N_0)^{2(\beta^2-1)} E \big)=(-1)^{\tt w}\ \tau(\lambda)\ .
\eea
Here we take into account that in the large $N$ limit,
\be
f\big((N/N_0)^{2(\beta^2-1)}\,E\,\big)=G^{(N)}\big(\,qE\,|\,\beta^2\,\big)\,
G^{(N)}\big(\,q^{-1}E\,|\,\beta^2\,\big)\,\big(\,1+o(1)\,\big)\ ,
\ee
where $f(\zeta)$ is the function \eqref{ffunc1a}  with all the inhomogeneities set to one.
Also recall that $q=\re^{\ri\pi\beta^2}$ and $E\propto \lambda^2$  as in \eqref{kassaisa}. 
\bigskip

The sign factor
in the r.h.s. of \eqref{sysaysaaa} appears for the following reason.
According to \eqref{PPbareq1}, the eigenvalues of the operators $\re^{+\ri\pi \beta a_0}$
entering into eq.\,\eqref{iaisaisa11a}  are given by 
$\re^{\ri\pi({\tt k}+{\tt w})+\pi\beta^2S^z}$.
This differs from the eigenvalues of
the  corresponding factors $\omega\ q^{+{\mathbb S}^z}$ in \eqref{TArel1a}
 by $(-1)^{\tt w}$. Notice that the
same sign factor enters 
 into the 
asymptotic formula \eqref{tower1}
for the eigenvalues of the lattice translation operator ${\mathbb K}$, where 
it is denoted by $\sigma=(-1)^{{\tt w}}$.
Thus
eq.\,\eqref{sysaysaaa} can be rewritten in the operator form as:
\bea
 \lim_{N\to\infty} G^{(N)}\big(q^2 E \,|\,\beta^2\big)G^{(N)}\big( q^{-2} E\,|\,\beta^2\big)
 \mathbb{T}\big(\, (N/N_0)^{2(\beta^2-1)} E \big)\,{\mathbb K}={\boldsymbol  \tau}(\lambda)\ .
\eea

 \bigskip
Since both $\boldsymbol{\tau}$ and  ${\mathlarger{\mathlarger{\mathlarger{\mathlarger {\bf \it a}}}}}_\pm$ admit
a regular power series expansion in $\lambda^2$, eq.\,\eqref{iaisaisa11a}
 allows one to express the operators ${\bf H}_j^{(\pm)}$
\eqref{logaseries1} in terms of 
\be
{\bf Q}_j\ :\  \ \ \ \  \ \ \ \ \boldsymbol{\tau}(\lambda)=\sum_{j=0}^\infty {\bf Q}_j\, \lambda^{2j}\ .
\ee
This leads to relations between the corresponding eigenvalues $H_j^{(\pm)}({\boldsymbol v})$ and 
$Q_j({\boldsymbol v})$. 
For the case of the Fock vacuum, 
formula \eqref{eq1a2} gives
the first few eigenvalues $H_j^{({\rm vac})}$ 
 in terms of the $2m$ fold integrals  $Q_m(h,g)$ \eqref{QQeig1a}
with $h=\beta P$ and $g=\beta^2$,
which coincide with the vacuum eigenvalues of ${\bf Q}_m$.

\subsection{Chiral states $\bmT{\psi}_{P}(\bmT{v})$\label{54}}
We have yet to discuss an important practical problem:  
having at hand the Bethe roots corresponding to the
family $\bm{\Psi}_N$ for a few values of $N$, how to identify
the states ${\boldsymbol \psi}_P({\boldsymbol v})$ and $\bar{\boldsymbol \psi}_{\bar P}(\bar{\boldsymbol v})$ appearing in
 the r.h.s. of eq.\,\eqref{Psi1a}.

\bigskip

To obtain the set $\bm{v}=\{v_a\}_{a=1}^{{\tt L}}$ labeling the state
${\boldsymbol \psi}_P({\boldsymbol v})$
one can in principle compute
the  connection coefficient $D_+(E\,|\,\bm{v})$ from the Bethe roots for $\bm{\Psi}_N$ 
using eq.\,\eqref{iisis1a}.
In practice, however, 
instead of using  the  full spectral determinant it is sufficient to focus on
its   large\,-\,$E$ asymptotic expansion. Eq.\,\eqref{hassaast}  describes just  the  leading  large\,-\,$E$  behaviour.
A more detailed description involves 
the asymptotic coefficient $R_{ P}$, which depends on the set of apparent singularities:
\bea\label{iasisai}
D_+(E|{\boldsymbol v})= R_{ P}({\boldsymbol v})\ (-E)^{-P/\beta}\
\exp\Bigg(\, \frac{N_0}{\cos(\frac{\pi \beta^2}{2-2\beta^2})}\ (- E)^{\frac{1}{2-2\beta^2}}+o(1)\Bigg) 
\eea
(here $\beta^2\not= 1-\tfrac{1}{2k}$).
In the recent work \cite{Kotousov:2019nvt} a  closed   expression  for $R_{ P}({\boldsymbol v})$ was 
obtained for  the Schr$\ddot{\rm o}$dinger equation  with Monster potentials involving
an arbitrary number of apparent
singularities ${\tt L}=0,1,2,\ldots\ $.
It takes the form
\be
R_{ P}({\boldsymbol v})=R_{ P}^{(0)}\,\check{R}_{ P}({\boldsymbol v})
\ee
with
 \bea\label{Reig1a}
 R^{(0)}_P= \beta^{1+4P\beta}\ 2^{2P(\beta^{-1}-\beta) }\ 
 \frac{\Gamma(1+\frac{2P}{\beta})}{\Gamma(1+2P\beta)}
 \eea
and $\check{R}_{ P}({\boldsymbol v})$ is given in eq.\,(5.19) in \cite{Kotousov:2019nvt}.
On the other hand, formula  \eqref{iisis1a} implies an important relation 
which allows one to extract the asymptotic
coefficient $R_{ P}({\boldsymbol v})$ numerically from the Bethe roots for
sufficiently large $N$ \cite{Kotousov:2019ygw}:
\be\label{prodeqA}
\prod_{m=1}^M\big(\zeta_m^{-1}+q\big)\big(\zeta_m^{-1}+q^{-1}\big)\asymp\big (R_{ P}({\boldsymbol v})\big)^2\  \ 
\big(N/N_0\big)^{-4(\beta^{-1}-\beta) {P}}\ 
\big(4(1-\beta^2)\big)^{N}\,\big(1+o(1)\big)\, .
\ee
Together with the similar relation
\be\label{prodeqB}
\prod_{m=1}^M\big(\zeta_m+q\big)\big(\zeta_m+q^{-1}\big)\asymp\big (R_{\bar {P}}({\bar {\boldsymbol v}})\big)^2\ \  
\big(N/N_0\big)^{-4(\beta^{-1}-\beta){\bar { P}}}\ 
\big(4(1-\beta^2)\big)^{N}\,\big(1+o(1)\big)\ ,
\ee
this provides a way of identifying the sets
${\boldsymbol v}=\{v_a\}_{a=1}^{\tt L}$ and $\bar{\boldsymbol v}=\{\bar{v}_a\}_{a=1}^{\bar {\tt L}}$,
which label
the state
$\bar{\boldsymbol \psi}_{\bar P}(\bar{\boldsymbol v})\otimes {\boldsymbol \psi}_P({\boldsymbol v})\in \bar{{\cal F}}^{({\bar{\tt L}})}_{\bar{ P}}\otimes
 {\cal F}^{(\tt L)}_P$
that occurs  in the scaling limit of ${\boldsymbol \Psi}_N$ \eqref{Psi1a}.
In practice we found this to be an effective procedure for small ${\tt L}$ and $\bar{\tt L}$ $(\le 5)$.

\bigskip 
The state ${\boldsymbol \psi}_P({\boldsymbol v})\in{\cal F}_P^{({\tt L})}$ can be constructed,
in principle, through the diagonalization problem of 
the operator ${\mathlarger{\mathlarger{\mathlarger{\mathlarger {\bf \it a}}}}}_+$. 
However the computation of its matrix elements
using eqs.\,\eqref{Lop2a},\,\eqref{soso1a} is an unduly complicated task.
It turns out that 
 in practice the most effective way of determining
the states ${\boldsymbol \psi}_P({\boldsymbol v})\in{\cal F}^{({\tt L})}_P$ 
for small values of ${\tt L}$
is based on the diagonalization of the so-called reflection
operator. The latter commutes with ${\mathlarger{\mathlarger{\mathlarger{\mathlarger {\bf \it a}}}}}_\pm(\lambda)$ 
and its eigenvalues  coincide
with the subleading coefficients $ R_{ P}({\boldsymbol v})$ entering into  the asymptotic formula
\eqref{iasisai}.
There is a simple algebraic procedure for
 constructing the reflection operator restricted  to  a  level subspace $ {\cal F}^{({\tt L})}_P$
 with given ${\tt L}$. 
For ${\tt L}=1,2,3$ some  explicit formulae can be found in the Appendix of ref.\cite{Kotousov:2019nvt}.
\bigskip

As an illustration  here we quote the explicit expression for the states ${\boldsymbol \psi}_P({\boldsymbol v})
\in {\cal F}^{({\tt L})}_P$ for the first
two levels.
For ${\tt L}=1$, when the Monster potential contains only one
apparent singularity,
the system \eqref{jsaysssysa} dramatically simplifies. Its solution is
\bea
v_1=(2P-\rho)(2P+\rho)\ \qquad {\rm with} \qquad \rho=\beta^{-1}-\beta
\eea
and $\beta$ is related to $\alpha$ as in \eqref{alphbet1a}. 
Since ${\dim {\cal F}}_{P}^{(1)}=1$ one has
\bea\label{sjssusu}
{\boldsymbol \psi}_P\big({\boldsymbol v}^{(1)}\big)=\frac{1}{2P+\rho}\ \ a_{-1}\, |\,{P}\rangle\ .
\eea
For ${\tt L}=2$ there are two solutions of \eqref{jsaysssysa}, which we denote as
${\boldsymbol v}^{(2,+)}=(v^{+}_1,v^{+}_2)$  and  ${\boldsymbol v}^{(2,-)}=(v^{-}_1,v^{-}_2)$.
They read explicitly as
\bea
v_1^{\pm}&=&2\, \omega_\pm \big(\omega_\pm+\beta^{-1}\big)\big(\omega^2_\pm
+\beta^2-\beta^{-2}\big)\\[0.2cm]
v_2^{\pm}&=&
2\, \omega_\pm \big(\omega_\pm-\beta^{-1}\big)\big(\omega^2_\pm
+\beta^2-\beta^{-2}\big)
\ ,\nonumber
\eea
where
\bea
\omega_\pm
&=&\frac{1}{2}\ \sqrt{(1+2\beta^2)(2\beta^{-2}-1)\pm B}\,,\qquad\qquad
B=\sqrt{(2\rho^2-1)^2+32 P^2}\ >0\ .
\eea
The corresponding  basis states $|{\boldsymbol v}^{(2,\pm)}\rangle\in {\cal F}_{p}^{(2)}$ are given by
\bea\label{aisaiisiass}
{\boldsymbol \psi}_P\big({\boldsymbol v}^{(2,\pm)}\big)=
\big((4P+2\rho)^2-2\rho^2-1\pm B\big)\ \Big(\ \frac{1}{4}\ a_{-1}^2-\frac{P}{1-2\rho^2\mp B}\ a_{-2}\,\Big)\, |{ P}\rangle\ .
\eea
The normalization of the states \eqref{sjssusu} and \eqref{aisaiisiass} will be explained  in the next section.

\section{Scaling limit of the Bethe state norms \label{sec26}}

The chiral states ${\boldsymbol \psi}_P({\boldsymbol v})$
appearing in the scaling limit of $\bm{\Psi}_N$ \eqref{Psi1a}
have been identified as 
 eigenstates of the
operators ${\mathlarger{\mathlarger{\mathlarger{\mathlarger {\bf \it a}}}}}_\pm(\zeta)$ that act in
the Fock space ${\cal F}_{ P}$.
Of course $\bar{\boldsymbol \psi}_{\bar P}(\bar{\boldsymbol v})\in\bar{{\cal F}}_{\bar P}$ may be specified
similarly. On the other hand, for a given $N$, $\bm{\Psi}_N$  is a state
in the finite dimensional space $\mathscr{V}_N$ \eqref{vec1}.  
In order to assign a precise meaning to the asymptotic formula \eqref{Psi1a}  
we should equip $\mathscr{V}_N$ 
and the Fock spaces with suitable Hermitian structures. 
For $\mathscr{V}_N$ we take 
the Hermitian structure to be one, which is
consistent with the integrable structure in the model.
As was already mentioned in sec.\,\ref{sec1},
this means that the sesquilinear form is such that the condition \eqref{ortho1}
is obeyed.
An important feature of the homogeneous six-vertex model
is that any set of solutions to the Bethe ansatz equations
coincides with the complex conjugated set so that eq.\,\eqref{CPTBethe1} becomes
\be\label{CPTBethehom1}
{\cal \hat{C}\hat{P}\hat{T}}\,\bm{\Psi}=\bm{\Psi}\ .
\ee
Hence, for the
homogeneous case, consistency between
the integrable and Hermitian structures implies that different
Bethe states are orthogonal to each other. The corresponding
Hermitian structure in 
the chiral Fock space should be chosen
so that
$\big({\boldsymbol \psi}_P({\boldsymbol v}'), {\boldsymbol \psi}_P({\boldsymbol v})\big)=0$ for $\bm{v}'\ne\bm{v}$.
Then specifying the norms of the Bethe
states as well as the norms of 
${\boldsymbol \psi}_P({\boldsymbol v})$ and $\bar{\boldsymbol \psi}_{\bar P}(\bar{\boldsymbol v})$,
one may obtain
 the constant  $\Omega_N$ as the ratio of the norms 
of the states appearing on both sides of eq.\,\eqref{Psi1a}.

\bigskip

The Fock space  ${\cal F}_P$ admits  an inner product that is consistent with
the natural conjugation condition for the Heisenberg generators:
\bea\label{iisisa}
a_{m}^\dagger=a_{-m}\ \ \ \qquad (\forall m)\ .
\eea
Using the definitions \eqref{soso1a},\,\eqref{asusuya231}
one can show that  for real $\lambda^2$
the operators ${\mathlarger{\mathlarger{\mathlarger{\mathlarger {\bf \it a}}}}}_\pm(\lambda)$ and $\bm{\tau}(\lambda)$
are Hermitian:
\bea\label{Hermapm}
\big[{\mathlarger{\mathlarger{\mathlarger {\bf \it a}}}}_\pm(\lambda)\big]^\dagger=
{\mathlarger{\mathlarger{\mathlarger {\bf \it a}}}}_\pm(\lambda^*)\ ,\ \ \ \  \ \ \ 
\big[\bm{\tau}(\lambda)\big]^\dagger=\bm{\tau}(\lambda^*)\ .
\eea
Assuming that the spectrum of ${\mathlarger{\mathlarger{\mathlarger{\mathlarger {\bf \it a}}}}}_\pm(\lambda)$
is non-degenerate, one concludes that different  states ${\boldsymbol \psi}_P({\boldsymbol v})$
and ${\boldsymbol \psi}_P({\boldsymbol v}')$
 are orthogonal
w.r.t. the inner product  associated  with this conjugation, i.e.,
\bea\label{innerprod1}
\big({\boldsymbol \psi}_P({\boldsymbol v}'), {\boldsymbol \psi}_P({\boldsymbol v})\big)_{\dag}={F}_P({\boldsymbol v})\ 
\delta_{{\boldsymbol v}',{\boldsymbol v} }\ .
\eea
Here the ``$\dag$''  subscript is used to emphasize that the inner product corresponds to
the  conjugation  \eqref{iisisa} that is consistent with the Heisenberg algebra commutation relations.
\bigskip

It is possible to introduce another natural inner product in
${\cal F}_P$ such that the orthogonality condition similar to \eqref{innerprod1}
is satisfied.
To describe it, we'll
use the fact that the Fock space  admits the structure
of the highest weight representation
of the Virasoro algebra.
Consider the composite field $T(u)$ built from  $\partial\varphi$ 
\eqref{bosefiled1} 
\bea\label{Tasdasd}
T(u)=(\partial\varphi)^2-\ri\rho\,\partial^2\varphi\ ,
\eea
where $\rho$ is a real parameter and $\partial\equiv\frac{\partial}{\partial u}$.
The Fourier coefficients
 \bea\label{Texpansion1}
 T(u)=-\frac{c}{24}+\sum_{m=-\infty}^\infty L_m\ \re^{-\ri mu}\ ,
 \eea 
are generators of the Virasoro algebra
\be
[L_n,L_m]=(n-m)\, L_{n+m}+\tfrac{c}{12}\ n(n^2-1)\, \delta_{n+m,0}
\ee
with central charge
\be
c=1-6\rho^2\ .
\ee
The above relations define the structure of the
Verma module  for the Virasoro algebra ${\cal V}_\Delta$ on ${\cal F}_P$ with highest weight
\be
\Delta=P^2-\tfrac{1}{4}\,\rho^2\ .
\ee
One can introduce the 
 inner product  $(\cdot,\cdot)_{\star}$
in ${\cal F}_P\cong{\cal V}_\Delta$
that is consistent with the natural conjugation condition
for the Virasoro algebra generators:
\be\label{starconj1}
L_m^\star=L_{-m}\ .
\ee
Although the $\star$\,-\,conjugation,
as it follows from eqs.\,\eqref{bosefiled1},\,\eqref{Tasdasd} and \eqref{Texpansion1},
acts non-trivially on the Heisenberg generators $\{a_m\}$,
it turns out that
${\mathlarger{\mathlarger{\mathlarger{\mathlarger {\bf \it a}}}}}_\pm(\lambda)$ and $\tau(\lambda)$
satisfy the Hermiticity conditions
\bea\label{hausidhu888s}
\big[{\mathlarger{\mathlarger{\mathlarger {\bf \it a}}}}_\pm(\lambda)\big]^\star=
{\mathlarger{\mathlarger{\mathlarger {\bf \it a}}}}_\pm(\lambda^*)\ ,\ \ \ \ \ \ \ \ 
\big[\bm{\tau}(\lambda)\big]^\star=\bm{\tau}(\lambda^*)
\eea
provided  that
\be
\rho=\beta^{-1}-\beta\ .
\ee
Thus
one has 
\bea\label{Vnorm1}
\big({\boldsymbol \psi}_P({\boldsymbol v}'), {\boldsymbol \psi}_P({\boldsymbol v})\big)
_{\star}= {V}_P({\boldsymbol v})\ 
\delta_{{\boldsymbol v}',{\boldsymbol v}}\ .
\eea
Here the ``Virasoro norm'' ${V}_P(\bm{v})$ is of course different from
the Heisenberg one ${F}_P(\bm{v})$ from eq.\eqref{innerprod1}.

\bigskip

The states ${\boldsymbol \psi}_P({\boldsymbol v})$ have been defined as
  eigenvectors of 
${\mathlarger{\mathlarger{\mathlarger{\mathlarger {\bf \it a}}}}}_+(\lambda)$
which specifies them up to an overall factor.
It will be convenient for us to fix this last ambiguity
by imposing 
\bea\label{isisissis}
{\boldsymbol \psi}_P({\boldsymbol v})
=\big(\, (L_{-1})^{\tt L}+\ldots \,\big)\, |P\rangle\ , \ \ \ \ \ \ \ \ \ \ \ \qquad
\bm{v}=\{v_a\}_{a=1}^{\tt L}\ \,,
\eea
where the dots denote the terms
involving $L_{-m}$ with $2\le m\leq  {\tt L}$.
Let us emphasize that, though
the condition  \eqref{isisissis} is written in terms of the Virasoro algebra generators,
 the vector ${\boldsymbol \psi}_P({\boldsymbol v})$  is considered as
 a state in the Fock space ${\cal F}_P$
with the operators $\{L_{-m}\}$  expressed in terms of the Heisenberg generators
via eqs.\,\eqref{bosefiled1},\,\eqref{Tasdasd} and \eqref{Texpansion1}.
For instance,  the  formulae \eqref{sjssusu} and \eqref{aisaiisiass}
 give the states ${\boldsymbol \psi}_P({\boldsymbol v})\in{\cal F}_P^{({\tt L})}$
with ${\tt L}=1$ and ${\tt L}=2$, respectively,
that are normalized according to \eqref{isisissis}.
Having imposed a normalization condition for ${\boldsymbol \psi}_P({\boldsymbol v})$,
each of the norms ${F}_P(\bm{v})$ \eqref{innerprod1} and ${V}_P(\bm{v})$ \eqref{Vnorm1}
is determined up to an overall multiplicative factor that does not depend on the state
in ${\cal F}_P$.
The latter may be fixed by specifying the value of the norms
of the Fock vacuum ${\boldsymbol \psi}^{\rm(vac)}_P\equiv |P\rangle$, i.e., 
$({\boldsymbol \psi}^{\rm(vac)}_P, {\boldsymbol \psi}^{\rm(vac)}_P)_{\dag}$
and $({\boldsymbol \psi}^{\rm(vac)}_P, {\boldsymbol \psi}^{\rm(vac)}_P)_{\star}$, respectively.
For the states of the other chirality 
$\bar{\boldsymbol \psi}_{\bar P}(\bar{\boldsymbol v})\in
\bar{{\cal F}}_{{\bar P}}^{(\bar{\tt{L}})}$ such that
$\bar{\boldsymbol \psi}_{\bar P}(\bar{\boldsymbol v})
=\big((\bar{L}_{-1})^{\bar{\tt{L}}}+\ldots\big)|\bar{P}\rangle$, the Heisenberg and Virasoro norms 
can be introduced similarly and will be uniquely defined
 up to the  choice of the factors
 $(\bar{\boldsymbol \psi}^{\rm(vac)}_{\bar P}, \bar{\boldsymbol \psi}^{\rm(vac)}_{\bar P})_{\dag}$
 and
$(\bar{\boldsymbol \psi}^{\rm(vac)}_{\bar P}, \bar{\boldsymbol \psi}^{\rm(vac)}_{\bar P})_{\star}$.

\bigskip

The large $N$ limit  of the norms of the
low energy Bethe states for the homogeneous six-vertex model
was studied in the work \cite{Kotousov:2019ygw}.
The results imply that in the scaling limit 
the Hermitian form
\bea\label{Hform1a}
\big(\bm{\Psi}^{(2)},\bm{\Psi}^{(1)}\big)_{\star}=
\big(\bm{\Psi}^{(1)},\bm{\Psi}^{(1)}\big)_{\star}\ 
 \delta_{\bm{\Psi}^{(2)},\bm{\Psi}^{(1)}}\,,
\eea
where $(\bm{\Psi}^{(1)},\bm{\Psi}^{(1)})_{\star}=
({\cal \hat{C}\hat{P}\hat{T}}\bm{\Psi}^{(1)},\bm{\Psi}^{(1)})_{\star}$ is given by \eqref{FinalNorm}  with $\eta_J=1$,
induces  the Hermitian form  in the space  $\bar{\cal F}_{\bar P}\otimes{\cal F}_{P} $
defined by the conditions \eqref{Vnorm1}, \eqref{isisissis} and the similar relations for
the barred counterpart. 
Furthermore, the natural choice for the norms of the Fock vacua
turns out to be
\be\label{asd1afa}
\big({\boldsymbol \psi}^{\rm(vac)}_P, {\boldsymbol \psi}^{\rm(vac)}_P\big)_{\star}
=Z_+(P\,|\,\beta)\,,\qquad
\big(\bar{\boldsymbol \psi}^{\rm(vac)}_{\bar P}, \bar{\boldsymbol \psi}^{\rm(vac)}_{\bar P}\big)_{\star}
=Z_+(\bar{P}\,|\,\beta)\ ,
\ee
where 
\bea\label{asusausss}
Z_+(P\,|\,\beta)&=&(A_{\rm G})^{-2 \beta^2}\ 
(2\pi)^{\frac{1}{2}-2P\beta}\ \beta^{h(P)+4P\beta+1}\ 
\frac{
\re^{-(\frac{2P}{\beta}+h(P)+\frac{1}{2}-\frac{1}{6}\, \beta^2)\gamma_{\rm E}}
}{\Gamma(1+\frac{2P}{\beta})\, \Gamma(1+2P\beta)}\nonumber\\[0.2cm]
&\times&
 \prod_{m=1}^\infty
\frac{2\pi\, (m\beta^2)^{2m\beta^2+4P\beta+1}\, \re^{-2m\beta^2+\frac{1}{m}(\frac{2P}{\beta}
+h({ P)}+\frac{1}{2}-\frac{1}{ 6}\beta^2)}}{\Gamma^2(1+2P\beta+m \beta^2)}\ .
\eea
Here we use the notation 
\be
h(P)=4 P^2+\tfrac{1}{6}\, (\beta^2+\beta^{-2}-3)
\ee
and  $A_{\rm G}$, $\gamma_{{\rm E}}$ stand for the Glaisher  and Euler constants, respectively.
Now that the norms of the states on both sides of the asymptotic formula \eqref{Psi1a} are unambiguously 
specified, the constant $\Omega_N$ can be obtained through the study
of the large $N$ behaviour of the ratio $(\bm{\Psi},\bm{\Psi})_{\star}
/\big({V}_{\bar P}(\bar{\bm{v}})\,V_P(\bm{v})\big)$.
Numerical work  from ref.\cite{Kotousov:2019ygw} suggests 
\be\label{Omegaeq1}
|\,\Omega_N\,|^2=C_0^2(\beta)\  
N^{\frac{1}{6}}\,
\big(N/C(\beta)\big)^{-h(P)-h(\bar{P})-4{\tt L}-4\bar{{\tt L}}}\  \re^{{\cal A}_2 N^2}\, ,
\ee
where ${\cal A}_2$, $C_0(\beta)$ and $C(\beta)$ are the same for all the low energy states
and only depend on $\beta$. 
The constant ${\cal A}_2$ is given by the integral
\bea\label{aisiasjjs}
{\cal A}_2=\int_{0}^\infty\frac{\rd t}{t}
\ \frac{\sinh( \frac{\beta^2 t}{1-\beta^2} )\ \sinh(t)}{2\sinh(\frac{t}{1-\beta^2})\, \cosh^2(t)}\ .
\eea
For $C_0(\beta)$ and $C(\beta)$ the explicit analytical form is currently unknown.
Numerical data for these constants is presented in  Appendix \ref{app1}.
\bigskip

Formula \eqref{Omegaeq1}  specifies $\Omega_N$  up to an overall phase factor.
As usual, this ambiguity can be fixed by using  global ${\cal {C}{P}{T}}$\,-\,symmetry.
The generators  $\hat{\cal {C}}$,  $\hat{\cal{P}}$ and $\hat{\cal{T}}$ acting in the tensor product
$\bar{{\cal F}}_{\bar{P}}\otimes{\cal F}_P$ may be introduced in such a way that the combination 
$\hat{\cal {C}}\hat{\cal{P}}\hat{\cal{T}}$ commutes with all of the Heisenberg modes,
\be\label{CPTeq1}
\big[{\cal {\hat{C}}{\hat{P}}{\hat{T}}},\,a_m\big]=\big[{\cal {\hat{C}}{\hat{P}}{\hat{T}}},
\,\bar{a}_m\big]=0\qquad \ \ \ \ \  (\forall m)\,,
\ee
and acts identically on the vacuum
\be\label{CPTeq2}
{\cal \hat{C}\hat{P}\hat{T}}\,
|\bar{P}\rangle\otimes |P\rangle=
|\bar{P}\rangle\otimes |P\rangle\ .
\ee
Then the ${\cal CPT}$ conjugation acts as the identity operator on any state
$\bar{\boldsymbol \psi}_{\bar P}(\bar{\boldsymbol v})\otimes{\boldsymbol \psi}_P({\boldsymbol v})$, where
${\boldsymbol \psi}_P({\boldsymbol v})$ is normalized by the condition
 \eqref{isisissis}  and similarly for $\bar{\boldsymbol \psi}_{\bar P}(\bar{\boldsymbol v})$.
In other words all the coefficients of the states ${\boldsymbol \psi}_P({\boldsymbol v})$  expanded in
the basis $\{a_{-i_1}\,\ldots a_{-i_m}|P\rangle\, : \, 1\le i_1\le i_2\le \ldots\le i_m, \, P {\rm\,-\,real}\}$ are real numbers 
(for an illustration see eqs.\,\eqref{sjssusu}-\eqref{aisaiisiass}).
Since the  ${\cal CPT}$ conjugation  also acts trivially on the Bethe states \eqref{CPTBethehom1}, it follows that
the constant $\Omega_N$ must be real, and without loss of generality we can take it to be positive:
\be\label{iasisaias}
\Omega_N=\sqrt{{\cal K}^{({\tt L})} _N(P) \, {\cal K}^{(\bar{\tt L})} _N(\bar{P})}\ \ \ \ \ 
{\rm with}\ \ \ \  {\cal  K}^{({\tt L})}_N(P)=C_0(\beta)\, 
N^{\frac{1}{12}}\
\big(N/C(\beta)\big)^{-h(P)-4{\tt L}}\  \re^{\frac{1}{2}\,{\cal A}_2 N^2}
\ .
\ee
\bigskip

The following comment is in order here. Together with the ${\cal CPT}$\,-\,invariance 
the system possesses global ${\cal CP}$ and ${\cal T}$ symmetry separately.
The action of the ${\cal CP}$ transformation intertwines the spaces
 $\bar{{\cal F}}_{\bar{P}}\otimes{\cal F}_P\mapsto \bar{{\cal F}}_{-P}\otimes{\cal F}_{-\bar{P}}$ and is
defined by the following relations
\be
{\cal \hat{C}\hat{P}}\,a_m=\bar{a}_m\,{\cal \hat{C}\hat{P}} \qquad {\rm and} \qquad
{\cal \hat{C}\hat{P}}\,|\bar{P}\rangle\otimes |P\rangle=|-{P}\rangle\otimes |-\bar{P}\rangle\ .
\ee
It was already mentioned that for the lattice model,
the  ${\cal CP}$ and ${\cal T}$ transformations acting in $\mathscr{V}_N$ intertwine the sectors
with $+S^z$ and $-S^z$, while we only focus on the Bethe states \eqref{Bstate1} 
in the sector $S^z\ge 0$ (see sec.\,4 in \cite{Bazhanov:2020new} for the explicit formulae
for the action of the ${\cal C}$, ${\cal P}$ and ${\cal T}$ conjugations in $\mathscr{V}_N$).
The state ${\cal \hat{C}\hat{P}}\,\bm{\Psi}\in \mathscr{V}_N$
can be written in the form similar to \eqref{Bstate1}, but with   the 
set $\{\zeta_m\}$ being the zeroes of the corresponding eigenvalue $A_-(\zeta)$ of the operator
$\mathbb{A}_-(\zeta)$. Recall that the  solutions sets
of the Bethe ansatz equations \eqref{bae} are roots of
${A}_+(\zeta)$.

\bigskip
The Hermitian form \eqref{Hform1a} in the finite dimensional space 
$\mathscr{V}_N$ is not positive definite.
At the same time this space can be equipped with a  positive definite inner product, which
is consistent with the integrable structure. This is a special property of the homogeneous case.
The positive definite Hermitian form in 
$\mathscr{V}_N=\mathbb{C}_N^2\otimes\mathbb{C}_{N-1}^2\otimes\ldots\otimes\mathbb{C}_1^2$
 is induced by that of each two-dimensional component in the tensor product. The latter is defined as 
$\langle\sigma|\sigma'\rangle=\delta_{\sigma,\sigma'}$, where 
$|\pm \rangle\in\mathbb{C}^2$ stand for the two basis vectors such that $\sigma^z|\pm\rangle=\pm|\pm\rangle$. 
To make connection with the works \cite{Gaudin:1981cyg,Korepin:1982ej}
 let's  change the overall normalization of the Bethe
state \eqref{Bstate1} and introduce
\be\label{asdasd1a}
\bm{\Psi}'=
\alpha(\zeta_1,\ldots,\zeta_M)\,
\bm{\Psi}\big(\{\zeta_j\}\big)\ ,
\ee
where  $\alpha(\zeta_1,\ldots,\zeta_M)$ is given by
\be\label{Gdef1a}
\alpha(\zeta_1,\ldots,\zeta_M)=
\big(-\ri q^{-\frac{1}{2}}\,\re^{-\ri\pi{\tt k}}\ (q-q^{-1})\big)^{-M}\ A_+(-q^{-1})
\ee
with $A_+(-q^{-1})=\prod_{m=1}^M\big(1+1/(q\zeta_m)\big)$.
Then the wavefunction
\bea\label{cba}
{\Psi}'(x_M,\ldots,x_1)\ :\ \ \ \qquad \boldsymbol{\Psi}'=
\sum_{1\leq  x_1<x_2<\ldots<x_M\leq N}\Psi'(x_M,\ldots,x_1)\, \sigma_{x_M}^-\cdots
\sigma_{x_1}^- \,|\,0\,\rangle
\eea
can be written in the form
\be
\Psi'(x_M,\ldots,x_1)=\sum_{\hat P}A_{\hat P}\  \re^{\ri \sum_{m=1}^M p_{{\hat P}m}x_m}\ .
\ee
Here the summation is taken over all $M!$ permutations ${\hat  P}$ of the
integers $(1,2,\ldots,M)$, and we use the notation
\be\label{ap}
A_{\hat  P}=\prod_{1\le j < m\le M}
\frac{q\zeta_{\hat Pj}-q^{-1}\zeta_{\hat Pm}}
{\zeta_{\hat Pj}-\zeta_{\hat Pm}}\,,
\ee
while $p_{m}=p(\zeta_{m})$ which was defined in eq.\,\eqref{asisaisa}. 
The norm of the Bethe state $\bm{\Psi}'$ \eqref{cba} w.r.t. 
the positive definite inner product reads as
\be\label{aosido22893}
\parallel\!\bm{\Psi}'\!\parallel^2=\sum_{1\leq  x_1<x_2<\ldots<x_M\leq N}
\big|\Psi'(x_1,\ldots,x_M)\big|^2\ .
\ee
There exists a remarkable formula for this norm, which
was originally conjectured by Gaudin, McCoy  and Wu in ref.\cite{Gaudin:1981cyg} 
and proven by Korepin  in \cite{Korepin:1982ej}.
In our notation it reads as
\be\label{norm1aaa}
\parallel\!\bm{\Psi}'\!\parallel^2=
\big|\alpha(\zeta_1,\ldots,\zeta_M)\big|^{2}
\ (\bm{\Psi},\bm{\Psi})_{\star}\ \prod_{m=1}^M \zeta_m\ ,
\ee
where $(\bm{\Psi},\bm{\Psi})_{\star}=({\cal \hat{C}\hat{P}\hat{T}}\bm{\Psi},\bm{\Psi})_{\star}$
is given by  \eqref{FinalNorm}  with $\eta_J=1$ and 
 $\alpha(\zeta_1,\ldots,\zeta_M)$ is as in \eqref{Gdef1a}.

\bigskip

The scaling limit of the norm  \eqref{norm1aaa}
for the RG trajectory $\bm{\Psi}_N$ was studied
in ref.\cite{Kotousov:2019ygw}.
It was found that the positive definite inner product in the space
$\mathscr{V}_N$ becomes the positive definite Hermitian form in
 $\bar{\cal F}_{\bar P}\otimes{\cal F}_{P}$ consistent with the 
conjugation conditions $a^\dag_m=a_{-m}$  and $\bar{a}^\dag_m=\bar{a}_{-m}$.
 In this case it is convenient to fix the norms of the highest states
in the Fock spaces as
\be\label{Heis1a}
\begin{array}{c}
\big({\boldsymbol \psi}^{\rm(vac)}_P, {\boldsymbol \psi}^{\rm(vac)}_P\big)_{\dag}=
(2/\beta)^{2P(\beta^{-1}-\beta)}\ 
Z^2(P\,|\,\beta) \\[0.4cm]
\big(\bar{\boldsymbol \psi}^{\rm(vac)}_{\bar P}, \bar{\boldsymbol \psi}^{\rm(vac)}_{\bar P}\big)_{\dag}=(2/\beta)^{2{\bar P}(\beta^{-1}-\beta)}\ 
Z^2(\bar{P}\,|\,\beta)
\end{array}\ ,
\ee
where  we use   the special function $Z(P\,|\,\beta)$  from ref.\cite{Kotousov:2019ygw}. 
The latter can be represented through the convergent product
similar to \eqref{asusausss}:
\bea\label{asss}
Z(P\,|\,\beta)&=&(A_{\rm G})^{- \beta^2}\ 
(2\pi)^{\frac{1}{4}-P\beta}\ \beta^{\frac{1}{2} h(P)+1+P(\beta^{-1}+3\beta)}\ 
\frac{
\re^{-\frac{1}{2} (\frac{2P}{\beta}+h(P)+\frac{1}{2}-\frac{1}{6}\, \beta^2)\gamma_{\rm E}}
}{\Gamma(1+2P\beta)}\nonumber\\[0.2cm]
&\times&
 \prod_{m=1}^\infty
\frac{\sqrt{2\pi}\, (m\beta^2)^{m\beta^2+2P\beta+\frac{1}{2}}\, \re^{-m\beta^2+\frac{1}{2m}(\frac{2P}{\beta}
+h( P)+\frac{1}{2}-\frac{1}{ 6}\beta^2)}}{\Gamma(1+2P\beta+m \beta^2)}\ .
\eea
The norms $F_{\bar P}({\bar{\bm v}})F_{P}({\bm v})$ of the 
 states
 $\bar{\boldsymbol \psi}_{\bar P}(\bar{\boldsymbol v})
\otimes{\boldsymbol \psi}_P({\boldsymbol v})\in\bar{\cal F}_{\bar P}\otimes{\cal F}_{P}$
are fully determined by eqs.\,\eqref{innerprod1} and \eqref{isisissis}, their barred counterparts
and formula \eqref{Heis1a}. 
As was pointed out in \cite{Kotousov:2019ygw} the  Heisenberg $F_{P}({\bm v})$ \eqref{innerprod1}
and Virasoro $V_{P}({\bm v})$ \eqref{Vnorm1}  norms are   related 
to each other through the eigenvalues of the
reflection operator \eqref{iasisai}-\eqref{Reig1a}. With the norms of the Fock vacua 
$\big({\boldsymbol \psi}^{\rm(vac)}_P, {\boldsymbol \psi}^{\rm(vac)}_P\big)_{\star}$ and 
$\big({\boldsymbol \psi}^{\rm(vac)}_P, {\boldsymbol \psi}^{\rm(vac)}_P\big)_{\dag}$ fixed  as in
\eqref{asd1afa} and \eqref{Heis1a}, respectively, one has the relation
\bea
\frac{F_{P}({\bm v})}{V_{P}({\bm v})}=
R_P({\bm v})\ .
\eea
A numerical study  
leads to the following asymptotic formula describing the large $N$ behaviour of the
positive definite norm
\eqref{aosido22893}:
\bea
\parallel\!\bm{\Psi}_N'\!\parallel^2\ \asymp\,F_{\bar P}({\bar{\bm v}})F_{P}({\bm v})\ 
\big(2\sin(\pi\beta^2)\big)^{\frac{2}{\beta} (P+{\bar P})}\  (N/N_0)^{-2(\beta^{-1}-\beta)(P+\bar{P})}\ 
{\cal K}^{({\tt L})}_N(P)\,{\cal K}^{(\bar{{\tt L}})}_N(\bar{P})\ \re^{{\cal A}_1 N}\ .
\qquad
\eea
Here 
 ${\cal K}^{({\tt L})}_N(P)$  is defined in eq.\eqref{iasisaias},
the constant $N_0$ is given by  eq.\,\eqref{N0def1}, while
\bea\label{asdasd1122}
 {\cal A}_1=\log\bigg(\!\frac{2(1-\beta^2)}{\sin(\pi \beta^2)}\!\bigg)\ .
 \eea

\newpage

\part{Inhomogeneous six-vertex model with 
global ${\cal Z}_2$ symmetry\label{sec3}}

\section{Introduction}
As was mentioned in the Preliminaries,
additional global symmetries in the inhomogeneous six-vertex model arise
when  certain constraints are  imposed on the inhomogeneities.
For example, the restrictions $\eta_{N+1-J}=\eta_{J}^{-1}$ and $(\eta_J)^*=\eta_J^{-1}$ lead to
${\cal C}{\cal P}$ and ${\cal T}$ invariance, 
while imposing the condition $\eta_{J+r}=\eta_J$ with $N=rL$ gives rise to the lattice translation symmetry.
Proceeding further and completely fixing the 
 inhomogeneities as
\be\label{zsym1}
\eta_{J}=(-1)^r\re^{\frac{\ri\pi}{r} (2J-1)}\ \ \ \ \ \ (J=1,\ldots, rL)
\ee
one arrives at a model
possessing global ${\cal Z}_r$ invariance (for further details
see sec.\,7 of ref.\cite{Bazhanov:2020new}).
It turns out that, like in the homogeneous case, which formally corresponds
to $r=1$, the model is critical when $q$ and $\omega$ are unimodular \eqref{qwuni1}.
However for $r\ge 2$, different types of  critical behaviour occur
depending on the value of $\arg(q^2)$.
For instance, in the case of the ${\cal Z}_2$ invariant model 
there are two such domains with $\arg(q^2)\in(0,\pi)$ and 
$\arg(q^2)\in(\pi,2\pi)$.

\bigskip

This work is devoted to the study of the 
 critical behaviour of  the   ${\cal Z}_2$ invariant six-vertex model
with $\arg(q^2)\in(0,\pi)$.
The  
generator of the extra symmetry $\hat{{\cal D}}\in{\rm End}(\mathscr{V}_N)$
is built out of the matrices \eqref{Rcheckdef1} as
\be\label{Dz2case}
\hat{\cal D}= \prod_{m=1}^{N/2}\, \check{\bm
  R}_{2m,2m-1}(-1)\ :\qquad\qquad\qquad \hat{\cal D}^2=1\ .
\ee
Its adjoint  action  on the local spin operators is given by
\begin{subequations}\label{Dadjoint1}
\bea\label{Dadjoint1a}
\hat{\cal D}\,\sigma^\pm_m\,\hat{\cal D}&=&\frac{1}{q+q^{-1}}\ \Big( 2\,\sigma^{\pm}_{m+1}-\,(q-q^{-1})\,
\sigma^z_{m+1}\sigma^{\pm}_m\,\Big)\\[0.2cm]
\hat{\cal D}\,\sigma^z_m\,\hat{\cal D}&=&
\frac{1}{(q+q^{-1})^2}\,\Big(
4\,\sigma^z_{m+1}+(q-q^{-1})^2\,\sigma^z_m+4\,(q-q^{-1})\,
\big(\sigma^+_{m+1}\,\sigma^-_m+\sigma^-_{m+1}\,\sigma^+_m\big)\,\Big)\nonumber
\eea
for odd $m$ and
\bea\label{Dadjoint2a}
\hat{\cal D}\,\sigma^\pm_m\,\hat{\cal D}&=&\frac{1}{q+q^{-1}}\ \Big(2\,\sigma^{\pm}_{m-1}+\,(q-q^{-1})\,
\sigma^{\pm}_m\sigma^{z}_{m-1}\,\Big)\\[0.2cm]
\hat{\cal D}\,\sigma^z_m\,\hat{\cal D}&=&
\frac{1}{(q+q^{-1})^2}\,\Big(
4\,\sigma^z_{m-1}+(q-q^{-1})^2\,\sigma^z_m-4\,(q-q^{-1})\,
\big(\sigma^-_{m}\,\sigma^+_{m-1}+\sigma^+_{m}\,\sigma^-_{m-1}\big)\,\Big) \nonumber
\eea
\end{subequations}
for even $m$.
On the transfer matrix and the
operators $\mathbb{A}_\pm(\zeta)$, the adjoint action of  $\hat{{\cal D}}$  reads as
\be\label{DTApmcomm1}
\hat{{\cal D}}\,\mathbb{T}(\zeta)\,\hat{{\cal D}}=\mathbb{T}(-\zeta)\,,\qquad 
\hat{{\cal D}}\,\mathbb{A}_\pm(\zeta)\,\hat{{\cal D}}=\mathbb{A}_\pm(-\zeta)\ .
\ee
Note that the above equation implies that for the Bethe state \eqref{Bstate1} corresponding  to the solution set $\{\zeta_j\}$
of the Bethe ansatz equations,
\be\label{8s887f87d87fd}
\hat{{\cal D}}\,\boldsymbol{\Psi}\big(\{\zeta_j\}\big)=\boldsymbol{\Psi}\big(\{-\zeta_j\}\big)\ .
\ee
Since the system \eqref{bae} with $\eta_J=\ri\,(-1)^{J-1}$ is invariant under the substitution
$\zeta_j\mapsto -\zeta_j$, the set $\{-\zeta_j\}$ also solves the Bethe ansatz equations.
\bigskip

Despite that $\hat{{\cal D}}$ does not commute with the transfer matrix, it is a symmetry of the model in the following sense.
The transfer matrix commutes with the Hamiltonian%
\footnote{%
This form for the Hamiltonian, up to 
an overall multiplicative factor and an additive constant, 
appeared in ref.\cite{IJS2}.
The one defined by eq.\,(2) in the work \cite{Bazhanov:2019xvy} coincides with
$
\hat{\mathsf{V}}\,\mathbb{H}\,\hat{\mathsf{V}}^{-1}
$,
where $\mathbb{H}$ is as in \eqref{aioiisa}, while 
$\hat{\mathsf{V}}=\prod_{m=1}^{N/2}\,\exp\big(\frac{\ri\pi}{4}\,\sigma^z_{2m-1}\big)$.
}
\bea\label{aioiisa}
{\mathbb H}&=&-\frac{\ri}{q^2-q^{-2}}\
\sum_{m=1}^{N}\,\Big((q-q^{-1})^2\ \sigma^z_m\,\sigma^z_{m+1}+
\,2\,\big(\sigma^x_m\,\sigma^x_{m+2}+\sigma^y_m\,\sigma^y_{m+2}+
\sigma^z_m\,\sigma^z_{m+2}\big)\nonumber
\\[0.2cm]
&+&\,(q-q^{-1}) \big(\sigma_m^x\sigma_{m+1}^x+
\sigma_m^y\sigma_{m+1}^y\big)
\big(\sigma^z_{m-1}-\sigma^z_{m+2}\big)
\,\Big)
+\ri\,N\,\frac{q^2+q^{-2}}{q^2-q^{-2}}
\   \hat{{\bf 1}}\,,
\eea
where
\be\label{BC1a}
\sigma^x_{N+\ell}\pm\ri\sigma^y_{N+\ell}=\re^{\pm2\pi\ri{\tt k}}\,\big(\sigma^x_{\ell}\pm\ri\sigma^y_{\ell}\big)\,,
\qquad \qquad \sigma_{N+\ell}^z=\sigma^z_\ell \qquad \qquad (\ell=1,2)
\ee
and this Hamiltonian commutes with the generator of the ${\cal Z}_2$\,-\,symmetry
\be
\big[\,\hat{{\cal D}},\,\mathbb{H}\,\big]=0\ .
\ee
Recall that the parameter ${\tt k}$ entering into the boundary conditions \eqref{BC1a} is related to $\omega$ from \eqref{Tmat1}
as $\omega^2=\re^{2\pi\ri{\tt k}}$.
The eigenvalue of the Hamiltonian \eqref{aioiisa} for the state $\bm{\Psi}$
is given in terms of the Bethe roots by
\be\label{Eeq1}
{\cal E}=\sum_{m=1}^M\ \frac{4\ri\,(q^2-q^{-2})}{\zeta_m^2+\zeta^{-2}_m+q^2+q^{-2}}\ .
\ee
In this work we'll use the parameterization 
\be
q=\re^{\frac{\ri\pi}{n+2}}\qquad\qquad {\rm with}\qquad\qquad n>0\ .
\ee

\section{The low energy Bethe states \label{sec8}}
The class of states for which we'll be considering the large $N$ limit
is taken to be the low energy states for the Hamiltonian \eqref{aioiisa}. 
Similar to what was discussed in the 
homogeneous case, the Bethe ansatz
equations allow one to organize the low energy Bethe states 
for different $N$ into
the RG trajectories $\bm{\Psi}_N$. 
For technical details of the construction of these trajectories
and some specific examples
see the work \cite{Bazhanov:2019xvy}.
In the large $N$ limit the low energy Bethe states 
form the conformal towers similar to those in the $XXZ$ spin chain.
In particular, each such tower  is characterized by a set of quantum numbers
which includes the value of $S^z$ and the winding number ${\tt w}=0,\pm1,\pm2,\ldots$\ .
Following ref.\cite{Bazhanov:2019xvy} we will employ the notation
\bea\label{oaisoi1093}
p=\half\ \big(S^z+({\tt k}+{\tt w})(n+2)\big)\, ,\ \ \ \ \ \ \ \ \bar{p}=\half\ \big(S^z-({\tt k}+{\tt w})(n+2)\big)\ .
\eea
Any state in the conformal tower can be assigned
a pair of non-negative integers -- the chiral  levels $({\bar{\tt L}},{\tt L})$.
The extensive numerical work performed in refs.\cite{Jacobsen:2005xz,Ikhlef:2008zz, Ikhlef:2011ay,
Frahm:2012eb,Frahm:2013cma,Candu:2013fva,Bazhanov:2019xvy}
suggests that
 the large $N$ behaviour of the eigenvalues of $\mathbb{H}$ \eqref{aioiisa} and the lattice
 translation operator
$\mathbb{K}$ 
\eqref{Kformula0},\eqref{Kformula1} with $r=2$,
for the RG trajectory with given $p$, $\bar{p}$, ${\tt L}$ and $\bar{\tt L}$, are described by the formulae
\begin{subequations}\label{tower31}
  \bea\label{tower1a}
  {\cal E}&=&e_\infty\,N  +\frac{4\pi v_{\tt F}}{N }\ \bigg(\frac{{p}^2+{\bar p}^2}{n+2}+\frac{2 b^2}{n}-\frac{1}{6}+
 {\tt L}+\bar{\tt L}\bigg)+o\big(N^{-1-\epsilon}\big)\\[0.2cm]
\label{tower1b}
K&=& \exp\bigg(\frac{4\pi\ri}{N}\,\bigg( \frac{{p}^2-{\bar p}^2}{n+2}+ {\tt L}-\bar{\tt L}\bigg)\bigg)\ .
  \eea
\end{subequations}
Here
 \bea\label{uassaysa}
e_{\infty}= -\frac{2 v_{{\rm F}}}{\pi}\ \int_0^\infty{\rm d}t\ \frac{\sinh\big(\frac{2 t}{n}\big)}
{\sinh\big(\frac{(n+2)t}{n}\big)\,\cosh(t)}\ ,\qquad \qquad
v_{{\rm F}}=\frac{2(n+2)}{n}\ ,
\eea
while the correction
 term $o\big(N^{-1-\epsilon}\big)$ contains an infinitesimally small positive $\epsilon>0$
 (for a  more detailed  description of the correction term see ref.\cite{Bazhanov:2019xvy}).
An important difference of eq.\,\eqref{tower1a} compared with the homogeneous case \eqref{tower1}
is the presence of the additional $N$\,-\,dependent term $\propto b^2$ with $b=b(N)$.  It turns out that $b(N)$ is
 related to the
eigenvalue of the so-called quasi-shift operator, that was introduced in ref.\cite{Ikhlef:2011ay}. 
The latter  is  expressed in terms of 
 the transfer matrix \eqref{Tmat1} as 
\begin{equation}\label{qshift}
{\mathbb B}={\mathbb T}(-\ri q^{-1})
\big[{\mathbb T}(+\ri q^{-1})\big]^{-1}
\end{equation}
and its eigenvalues are given by
\bea\label{Beq1}
B=\frac{A_+(-\ri q)A_+(+\ri q^{-1})}{A_+(+\ri q)A_+(-\ri q^{-1})}
\  \,.
\eea
Then $b$ entering into eq.\,\eqref{tower1a}
and $B$ are related as
\be\label{poapso1a}
b(N)=\frac{n}{4\pi}\log(B)\ ,
\ee
where $B=B(N)$ denotes the 
eigenvalue of the quasi-shift operator corresponding to $\bm{\Psi}_N$.

\medskip

Since $B$ is in general a complex number,
the definition \eqref{poapso1a} requires the specification of the branch of the logarithm. 
It turns out that fixing the branch such that $b(N)$ is real whenever $B>0$
ensures that \eqref{poapso1a} is consistent with formula  \eqref{tower1a} that describes
the low energy spectrum. 
Thus we define $b(N)$
 for all the low energy Bethe states  with $|{ \arg}(B)|<\pi$
by supplementing \eqref{poapso1a} with the condition
\be\label{aisodio1231}
-\frac{n}{4}<\Im m\big(b(N)\big)<\frac{n}{4}\ .
\ee
Special attention is needed for
the Bethe states with $|{ \arg}(B)|=\pi$.
The explicit diagonalization of the commuting families of operators
for small $N$ reveals the existence of states with $B=-1$, see fig.\,\ref{fig3}. 
Although for such states 
$\delta{\cal E}\equiv\frac{N}{4\pi v_{{\rm F}}}\,({\cal E}-N\,e_\infty)$
is of order one,
we found that 
computing
$\delta{\cal E}$ for increasing $N$ 
through the solution of the Bethe ansatz equations,
$|\delta{\cal E}|$
grows logarithmically with $N$
and hence the states are not counted as low energy ones.
In addition, there are states for which 
$B$ is a complex number that tends to $-1$
in the large $N$ limit.
For most of these, $|\delta{\cal E}|$
goes to infinity similar as with the states where $B=-1$. However,
there do exist the RG trajectories for which 
the energy follows eq.\,\eqref{tower1a},
while $\lim_{N\to\infty}b(N)=\pm\frac{\ri n}{4}$ 
(see fig.\,\ref{fig2A}).
\bigskip

It is worth mentioning how the value of $b(N)$ transforms under the action of the ${\cal CPT}$ and
${\cal D}$ conjugations on the Bethe state $\bm{\Psi}_N$.
The quasi-shift operator satisfies the following relations with
their generators
\be\label{ioasido1a}
{\cal \hat{C}\hat{P}\hat{T}}\ \mathbb{B}\ {\cal \hat{C}\hat{P}\hat{T}}=\mathbb{B}\,,
\qquad\qquad 
{\cal \hat{D}}\,\mathbb{B}\,{\cal \hat{D}}=\mathbb{B}^{-1}\ .
\ee
The first equation implies that the eigenvalue of $\mathbb{B}$ for the Bethe state and the
${\cal CPT}$\,-\,transformed state \eqref{CPTBethe1} are complex conjugate of each other.
In turn, 
\be\label{CPTtrans1a}
{\cal CPT}\ : \ \ \ b(N)\mapsto b^*(N)\ .
\ee
 The last equation in \eqref{ioasido1a} combined with \eqref{poapso1a} yields
 that under the ${\cal Z}_2$ symmetry
transformation 
\be\label{Dtrans1a}
{\cal D}\ :\ \ \  b(N)\mapsto-b(N)\ .
\ee
Note that both the ${\cal CPT}$ and ${\cal D}$ conjugations
preserve the strip \eqref{aisodio1231}.

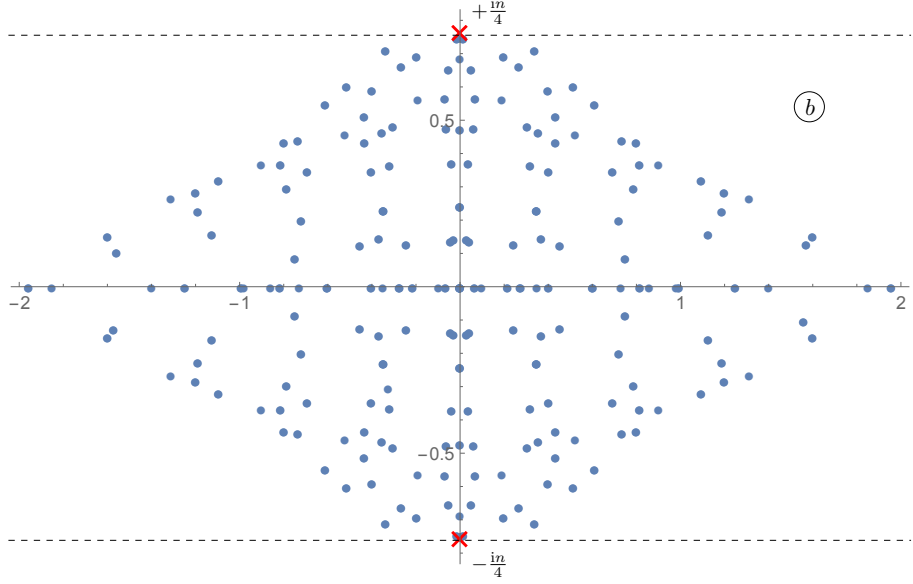
\begin{figure}
\centering
\scalebox{0.8}{
\begin{tikzpicture}
\node at (0.5,4.6) {\small $+\frac{\ri n}{4}$};
\node at (0.5,-4.6) {\small $-\frac{\ri n}{4}$};
\node (0,0) {\includegraphics[width=15cm]{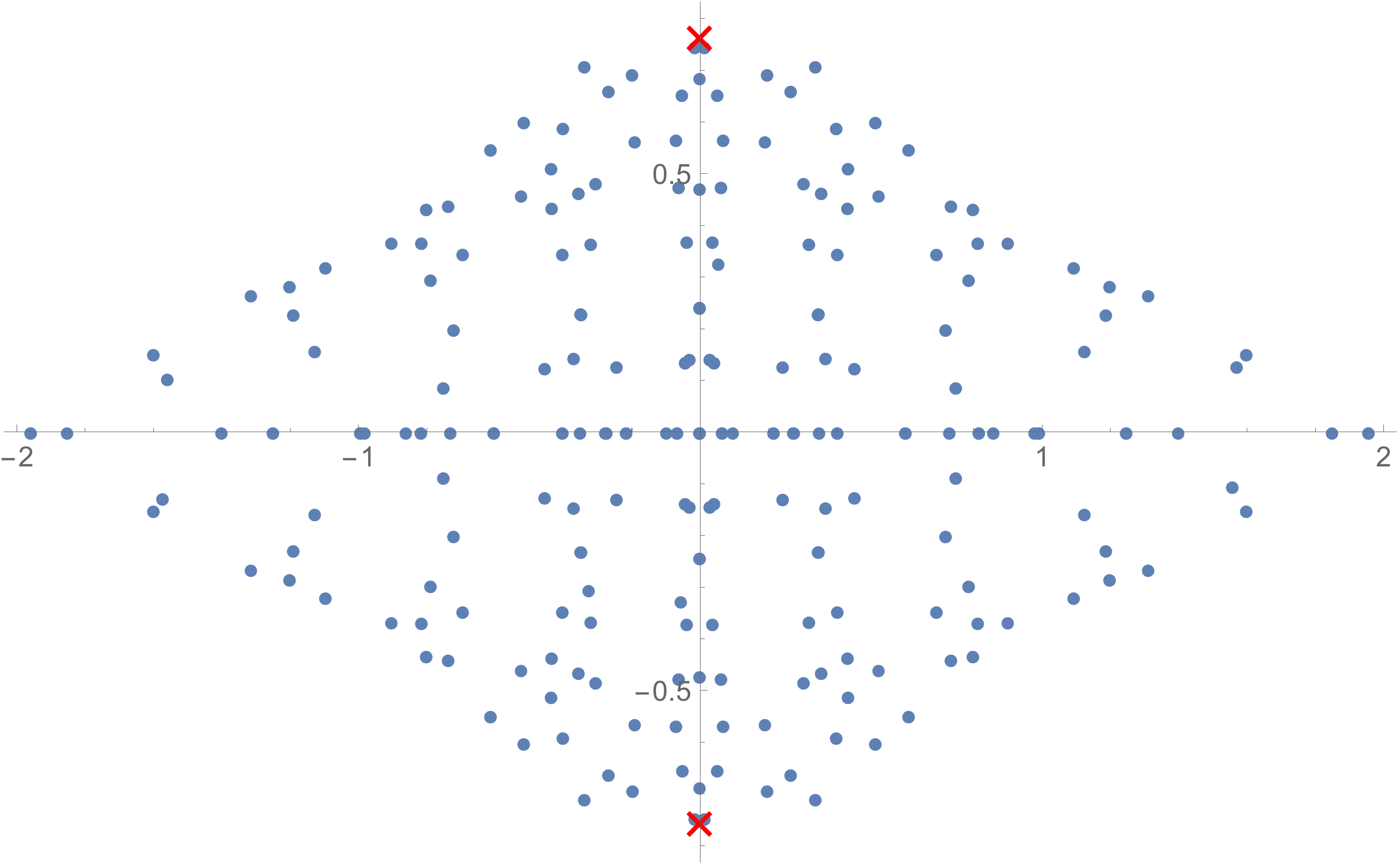}};
\draw[dashed,line width = 0.05mm] (-7.5,4.2) -- (7.5,4.2);
\draw[dashed,line width = 0.05mm] (-7.5,-4.2) -- (7.5,-4.2);
\node at (5.82,3) {$b$};
\draw  (5.81,3.01) circle [radius=0.25];
\draw[white,fill = white] (0.2,1.8) circle (0.1cm);
\draw[white,fill = white] (-0.2,-1.8) circle (0.1cm);
\end{tikzpicture}
}
\caption{\small
The blue dots mark the values of $b=\frac{n}{4\pi}\log(B)$ in the complex plane
for the
first 400 lowest energy states of the Hamiltonian \eqref{aioiisa},
\,\eqref{BC1a} with $N = 24$  in the sector $S^z=0$.
The branch of the  logarithm of $B$ is chosen such that $|\Im m\big(\!\log B\big)|\le \pi$. 
The parameters were taken to be $n=3$, $\pi{\tt k}=18/100$. There are four states with 
$B=-1$, which are represented by the red crosses in the figure.
Two of them have the same complex energy and they are related to each other
through the ${\cal Z}_2$ transformation \eqref{8s887f87d87fd}. The other two 
have the complex conjugated energy, they may be obtained from the 
${\cal Z}_2$ doublet by means of the ${\cal CPT}$ conjugation \eqref{CPTBethe1}.
The typical pattern of
Bethe roots for one of these states  is depicted in the left panel of fig.\,\ref{fig4}.
The right panel of that figure plots the absolute value of
$\delta{\cal E}\equiv\frac{N}{4\pi v_{{\rm F}}}\,({\cal E}-N\,e_\infty)$
as a function of $N$.
Clearly the energy is not described by \eqref{tower1a}.
Moreover, the eigenvalue of the lattice translation operator $\mathbb{K}$ 
remains fixed at $K=-1$ for any $N$, which does not follow   \eqref{tower1b}.
Due to this we do not count these states as low energy ones.
\label{fig3}}
\end{figure}
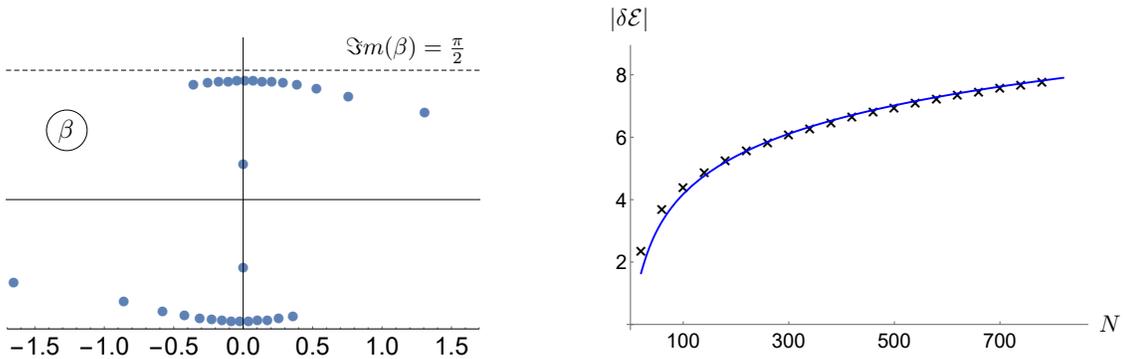
\begin{figure}
\centering
\scalebox{0.9}{
\begin{tikzpicture}
\node at (0,0) {\includegraphics[width=7cm]{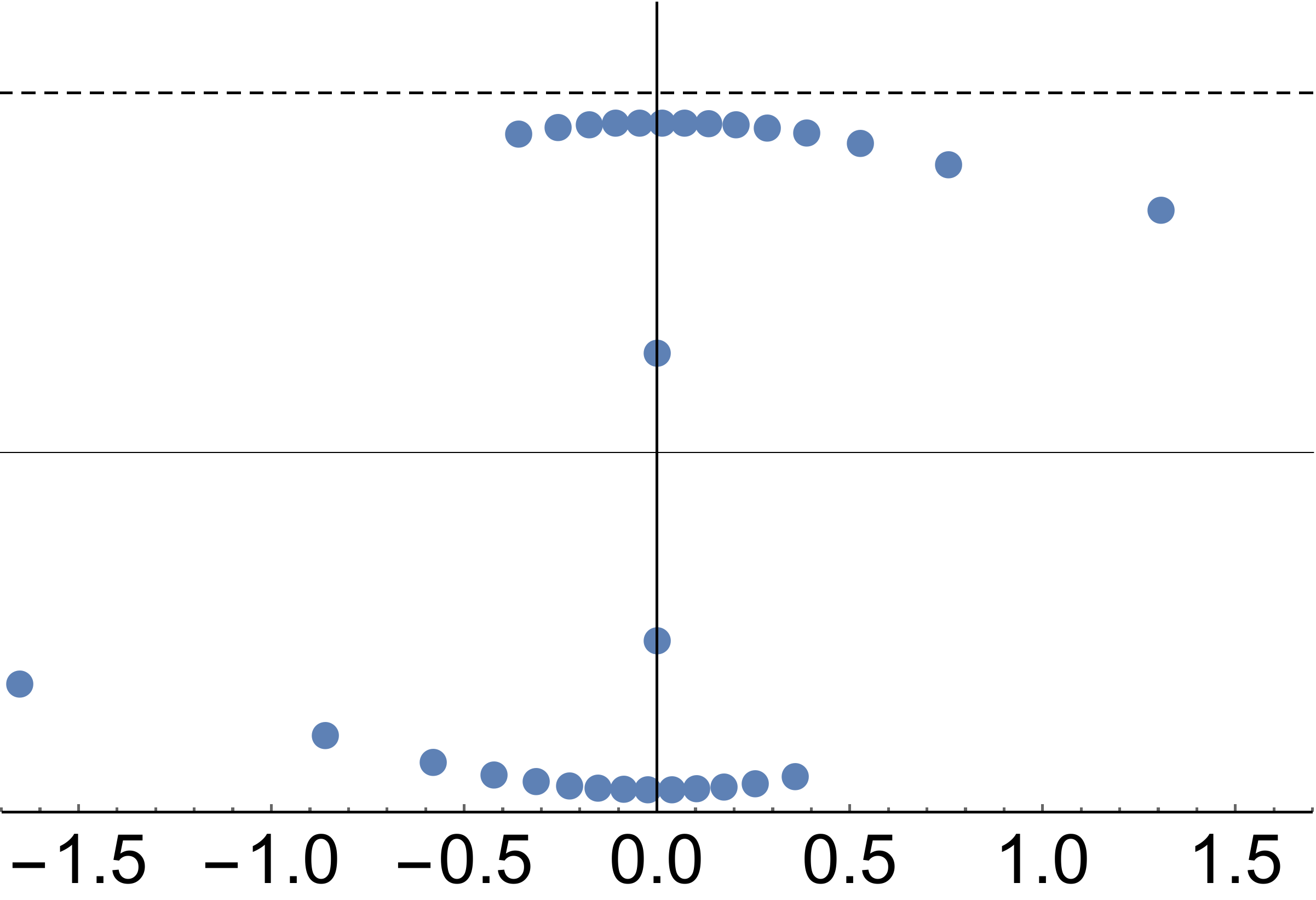}};
\node at (2.4,2.2) {\small $\Im m(\beta)=\frac{\pi}{2}$};
\node at (-2.62,1) {$\beta$};
\draw  (-2.61,1.01) circle [radius=0.3];
\node at (9,0) {\includegraphics[width=7cm]{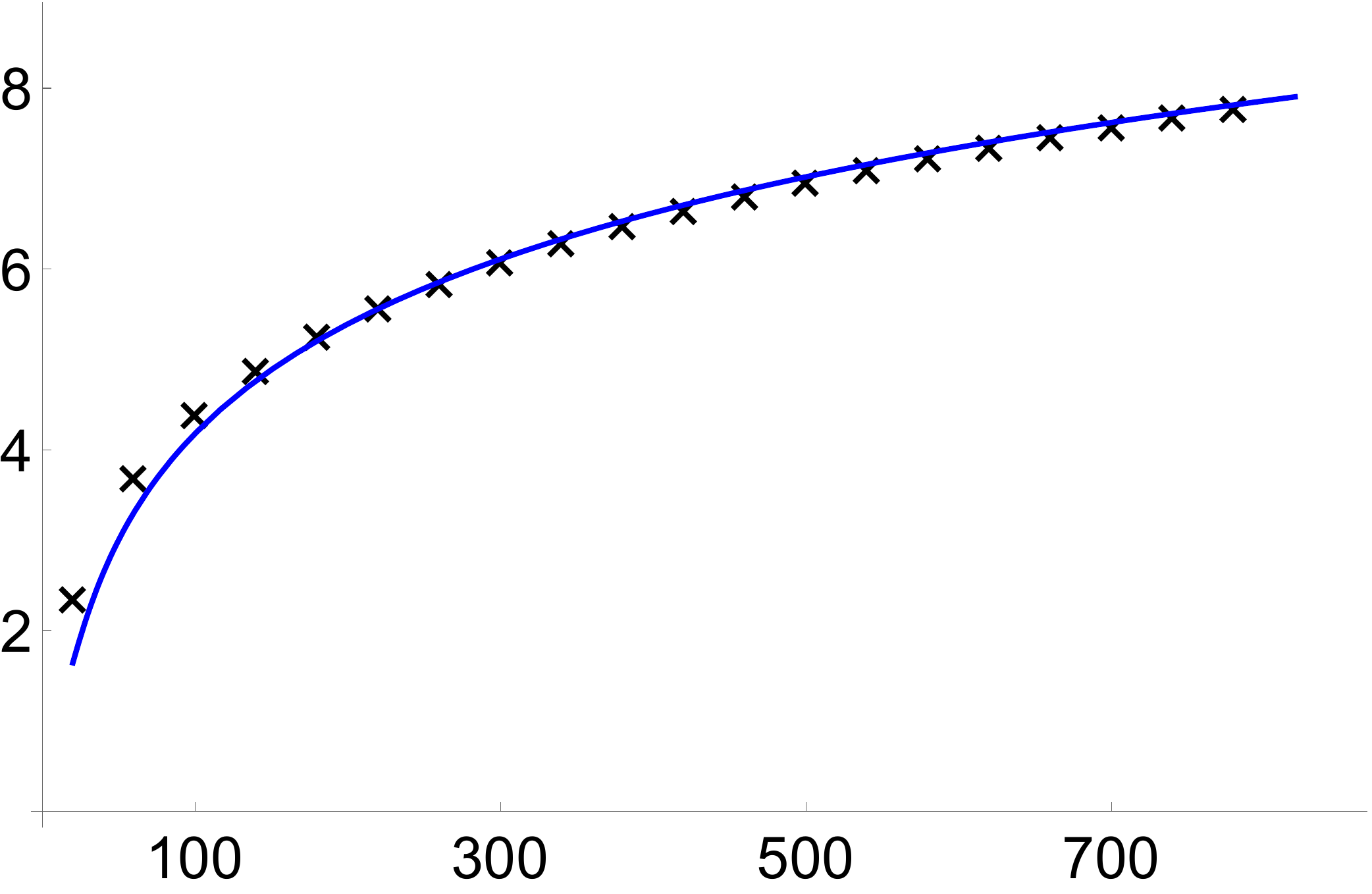}};
\node at (5.7,2.65) {\small $|\delta{\cal E}|$};
\node at (12.8,-1.85) {\small $N$};
\end{tikzpicture}
}
\caption{\small
The typical pattern of Bethe roots $\beta_j=-\frac{1}{2}\log\zeta_j$ for one of the four states having $B=K=-1$,
with $n=3$, $\pi{\tt k}=18/100$ and $N=60$ (left panel). 
The scaled energy $\delta{\cal E}=\frac{N}{4\pi v_{{\rm F}}}\,({\cal E}-N\,e_\infty)$ 
for these states grows logarithmically for large $N$. The crosses  in the right figure depict
 the numerical values of $|\delta{\cal E}|$ found via the solution of the Bethe ansatz equations. 
The solid line comes from the fit 
$\delta{\cal E}\approx-4.5702 - 0.2272\,\ri + (1.7724 - 0.4110\,\ri ) \log(N)$.
\label{fig4}}
\end{figure}

\begin{figure}
\centering
\scalebox{0.99}{
\begin{tikzpicture}
\node at (-5.5,0.7) {\small$\beta$};
\draw  (-5.5,0.73) circle [radius=0.3];
\node at (-5.5,1.75) {\small $\Im m(\beta)=\frac{\pi}{2}$};
\node at (-5.5,-1.3) {\small$\Im m(\beta)=0$};
\node at (-8,0) {\includegraphics[width=7cm]{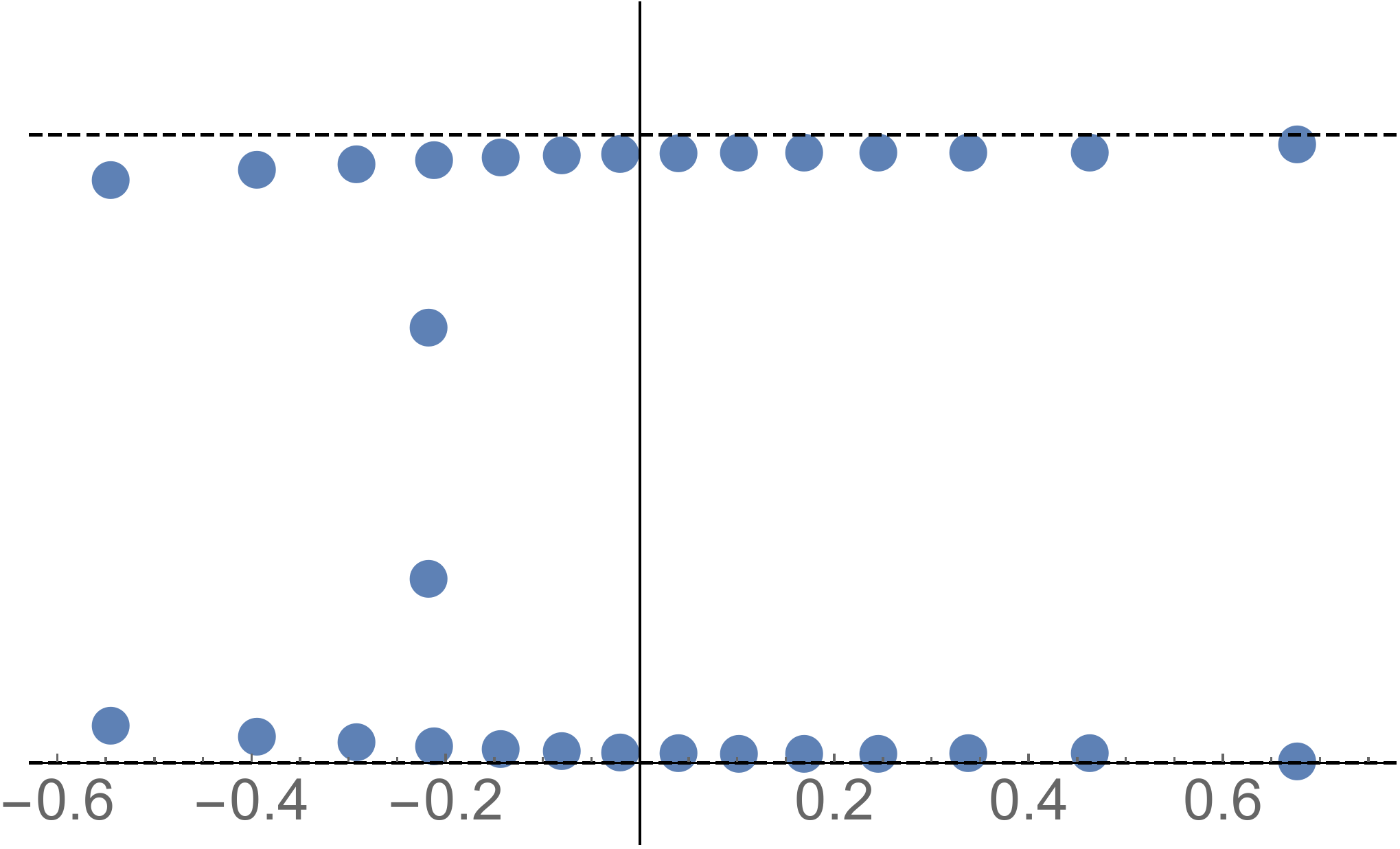}};
\node at (0.3,0) {\includegraphics[width=7cm]{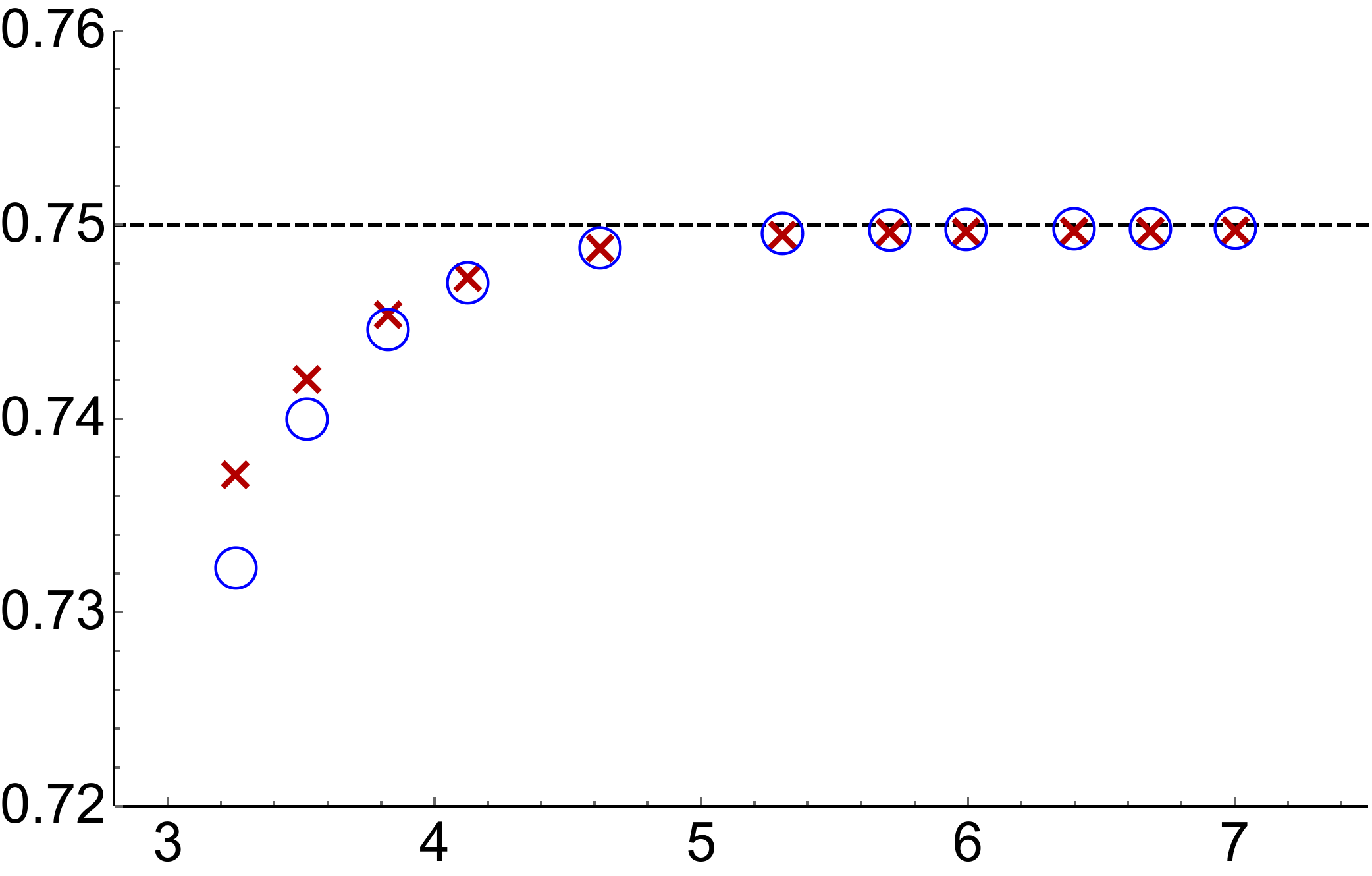}};
\node at (4.6,-1.9) {\small$\log(N)$};
\node at (-2.7,2.6) {\small$-\ri b(N)$};
\end{tikzpicture}
}
\caption{\label{fig2A}%
\small The left panel depicts the pattern of Bethe roots
$\beta_j=-\frac{1}{2}\log(\zeta_j)$ 
for a low energy state $\bm{\Psi}_N$ for which 
$\lim_{N\to\infty}b(N)=\frac{\ri n}{4}$, i.e., belongs to the boundary of the strip \eqref{aisodio1231}. 
The parameters are $n=3$, ${\tt k}=\frac{1}{10}$,
while the state belongs to the sector $S^z={\tt w}=0$
and ${\tt L}=\bar{\tt L}=1$. 
On the right panel $-\ri b(N)$ (which turns out to be real)
is plotted as a function of $\log(N)$
for this RG trajectory.
 The red crosses correspond
to $b(N)$ calculated from the eigenvalue of the quasi-shift operator
using eq.\,\eqref{poapso1a}, while the dashed line 
shows the limiting value $\lim_{N\to\infty}b(N)=0.75\,\ri$. To illustrate that the energy for 
this state obeys eq.\,\eqref{tower1a}, we depict via the
open circles the values of $b(N)$
computed from the numerical data for the energy by inverting \eqref{tower1a}
for $b(N)$ with the correction terms ignored.
}
\end{figure}
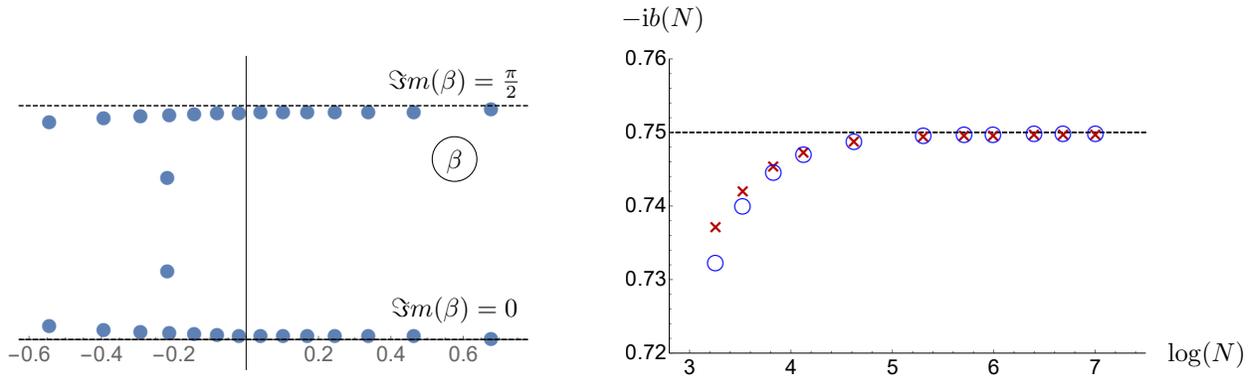

\bigskip

To summarize this section, let us emphasize that 
the definition of the low energy states for the
${\cal Z}_2$ invariant inhomogeneous six-vertex model
is far from evident.
In what follows we'll use the ``working'' definition that
a low energy Bethe state $\bm{\Psi}_N$ is one, whose energy and momentum
is described by eqs.\eqref{oaisoi1093}-\eqref{uassaysa},
with some ${\tt w}=0,\pm1,\pm2$ and 
non-negative integers ${\tt L}$, $\bar{\tt L}$, while
$b(N)$ is defined through eq.\,\eqref{poapso1a}
along with the condition \eqref{aisodio1231}.
Also it is important to note that the case of periodic boundary conditions 
${\tt k}=0$ requires special attention. In our  analysis,
unless explicitly stated,
 it will always be assumed that
$(n+2)\,{\tt k}\notin\mathbb{Z}$.  
The results for periodic boundary conditions
may be obtained through taking  the limit ${\tt k}\to 0$.
If necessary, we'll include comments regarding this limit separately.

\section{The RG invariant $s$ \label{sec31}}

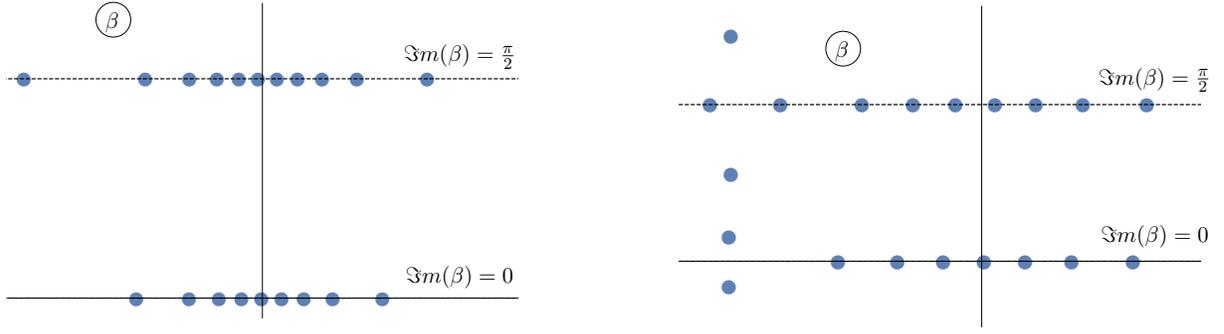
\begin{figure}
\centering
\scalebox{0.77}{
\begin{tikzpicture}
\node at (0.3,0.1) {\includegraphics[width=0.52\textwidth]{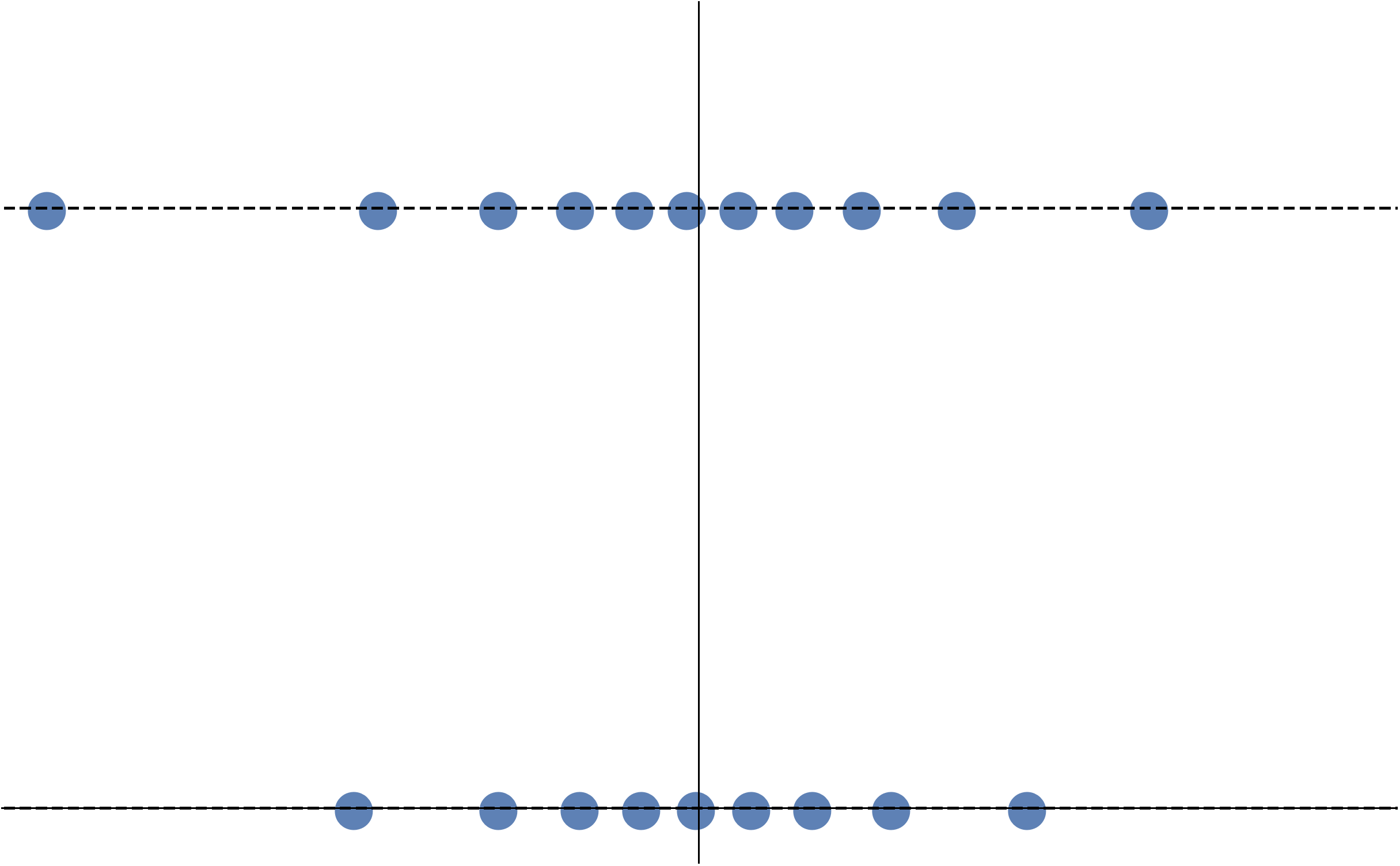}};
\node at (-2.3,2.5) {$\beta$};
\draw  (-2.28,2.54) circle [radius=0.3];
\node at (3.7,1.9) {$\Im m(\beta)=\frac{\pi}{2}$};
\node at (3.7,-1.9) {$\Im m(\beta)=0$};
\node at (12,0) {\includegraphics[width=0.53\textwidth]{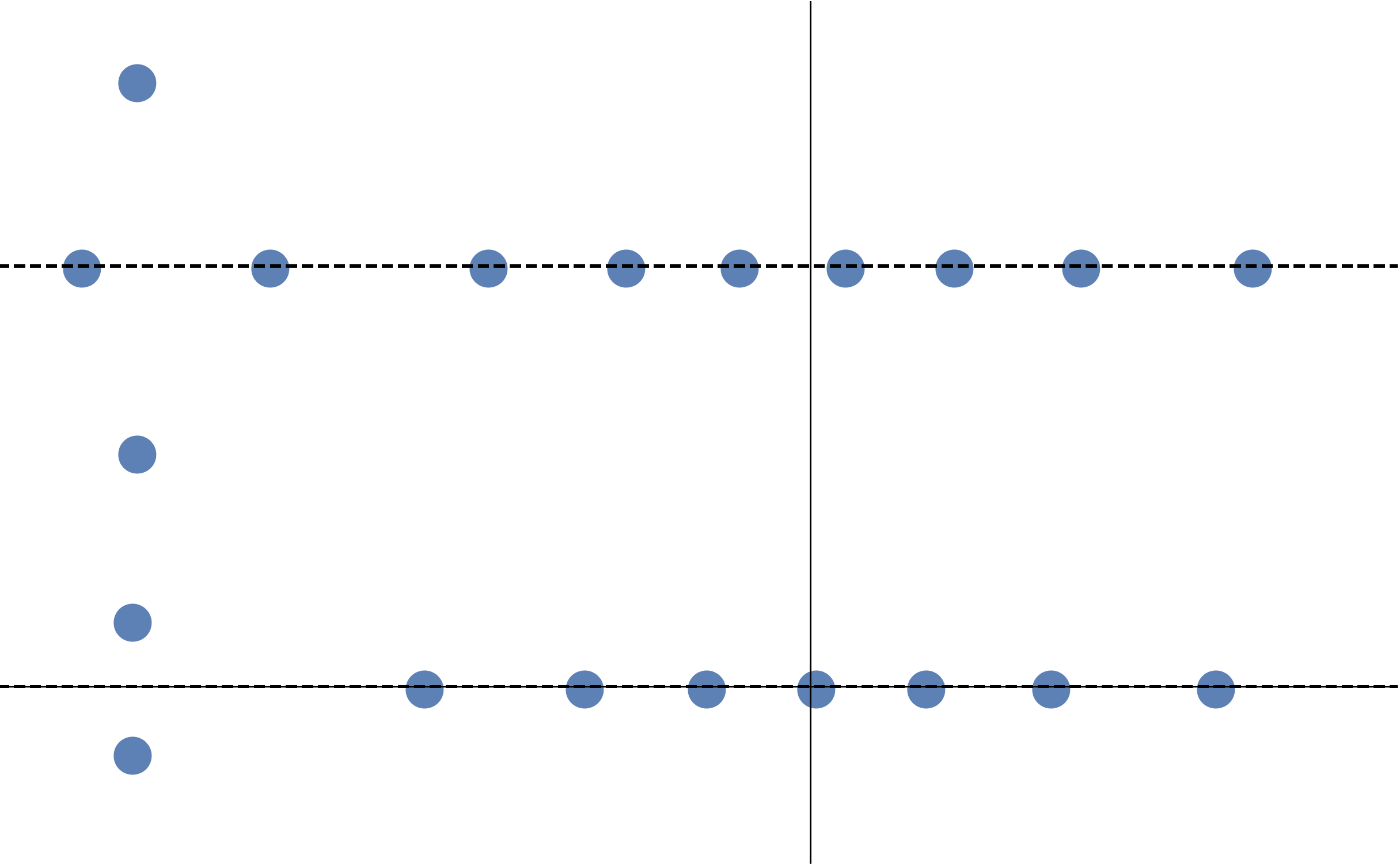}};
\node at (10.3,2) {$\beta$};
\draw  (10.32,2.04) circle [radius=0.3];
\node at (15.7,1.5) {$\Im m(\beta)=\frac{\pi}{2}$};
\node at (15.7,-1.2) {$\Im m(\beta)=0$};
\end{tikzpicture}
}
\caption{\small
The left panel shows the pattern of Bethe roots in the complex $\beta$ plane with $\beta=-\frac{1}{2}\log(\zeta)$
for the primary Bethe state with $N=40$, $n=3$, ${\tt k}=\frac{1}{10}$ and $S^z={\tt w}=0$.
The value of $b$ obtained from \eqref{Beq1},\,\eqref{poapso1a}
is consistent with the asymptotic relation \eqref{poaospo1qa1a} with ${\tt m}=2$.
The right panel depicts the pattern of Bethe roots for the eigenstate of the
 Hamiltonian \eqref{aioiisa},\,\eqref{BC1a}
 with $n=3$, ${\tt k}=\frac{1}{25}$ and $N=40$ characterized by ${\tt w}=1$, $S^z=0$,
${\tt L}=\bar{\tt L}=0$. For the RG trajectory continued from this state,
$\lim_{N\to\infty}b(N)=-\frac{\ri}{10}$, for further details
see fig.\,8 in ref.\cite{Bazhanov:2019xvy}.
\label{BAplot1}
}
\end{figure}

The specification of the RG trajectory for the ${\cal Z}_2$ invariant six-vertex model
has some essential differences to the homogeneous case.
In particular, for  the ``primary'' Bethe states where ${\tt L}=\bar{\tt L}=0$,
there exist many RG trajectories $\bm{\Psi}_N$, which correspond to the same values of
the RG invariants $p$ and $\bar{p}$ \eqref{oaisoi1093} and are distinguished 
by the eigenvalue of the quasi-shift operator \eqref{Beq1}.
Following ref.\cite{Ikhlef:2008zz}, let's illustrate this 
on a class of  Bethe states which occur  when $|{\tt k}|<\frac{2}{n+2}$.
Fixing $N$ and $S^z$, the corresponding 
Bethe roots $\{\zeta_m\}_{m=1}^{M}$ are real, while the states are distinguished by the integers
$M_--M_+$, where
$M_-$ stands for the number of negative roots,
$\zeta_m<0$, while $M_+$ is the number of positive ones,
$\zeta_m>0$.
An example of such a pattern is depicted in the left panel of fig.\,\ref{BAplot1}
in the complex $\beta$ plane with $\beta=-\frac{1}{2}\log(\zeta)$.
Although in principle  one can construct a state with $M_--M_+$ being
any integer from $-\tfrac{N}{2}+S^z$  to
$\tfrac{N}{2}-S^z$, it should be emphasized that
the states will only be low energy ones provided that $|M_--M_+|\ll N$.
With this restriction they turn out to be primary Bethe states all having the same
$p$, $\bar{p}$
given by eq.\,\eqref{oaisoi1093} with ${\tt w}=0$.
The value of  $b(N)$ \eqref{poapso1a} is always real 
and possesses the following leading large $N$  behaviour
\be\label{iaosido121} 
b(N)\asymp\frac{\pi{\tt m}}{4\log(N)}\,,\qquad N\to\infty\quad 
{\rm with} \quad {\tt m}-{\rm fixed}\qquad ({\tt L}=\bar{\tt L}=0)\ ,
\ee
where  ${\tt m}=M_- - M_+$.
\bigskip

Formula \eqref{iaosido121} resembles the quantization condition of a quantum
mechanical particle in a potential well of length $\propto\log(N)$. 
A more accurate quantization condition is achieved by taking into account the
phase shift that the particle picks up in the vicinity of the turning points.\footnote{%
The analogy with the potential well may seem rather artificial here. However in the third part of this work,
in sec.\,\ref{sec213},
it will arise naturally in the discussion of the CFT underlying the scaling behaviour of the
${\cal Z}_2$ invariant six-vertex model.}
In ref.\cite{Ikhlef:2011ay}, based on a numerical analysis, the following remarkable formula
was proposed for describing the large $N$ behaviour of $b(N)$ for an RG trajectory
with ${\tt L}=\bar{\tt L}=0$:
\bea\label{poaospo1qa1a}
8\,b(N)\,\log\bigg(\frac{N}{2N_0}\bigg)
+\delta\,\big|_{s=b(N)}=2\pi\,{\tt m}+O\big((\log N)^{-\infty}\big)\ .
\eea
Here the phase shift $\delta$ is a
continuous function of $s\in(-\infty,+\infty)$
such that $\delta|_{s=0}=0$ and 
\be\label{oisaoid132}
\re^{\frac{\ri}{2}\delta}=2^{\frac{4\ri s(n+2)}{n}}
\ \frac{\Gamma(\frac{1}{2}+p-{\ri s})\,\Gamma(\frac{1}{2}+{\bar p}-{\ri s})}
{\Gamma(\frac{1}{2}+p+{\ri s})\,\Gamma(\frac{1}{2}+{\bar p}+{\ri s})}\qquad
\qquad
\qquad ({\tt L}=\bar{\tt L}=0)\ .
\ee
The integer ${\tt m}$ takes even values if $\frac{N}{2}-S^z$ is even
and odd values otherwise, so that
\be
(-1)^{\tt m}=(-1)^{\frac{N}{2}-S^z}\,,
\ee
while the symbol $O\big((\log N)^{-\infty}\big)$ indicates that
\eqref{poaospo1qa1a} holds true  up to power law corrections in $N$.
The explicit formula for the $n$ dependent constant $N_0$ was found in the later work \cite{Bazhanov:2019xvy}  and reads as
\bea\label{N0altdef1}
N_0=\frac{\sqrt{\pi}\,\Gamma\big(1+\frac{1}{n}\big)}
{2\Gamma\big(\frac{3}{2}+\frac{1}{n}\big)}\ .
\eea
Since the above expression coincides with $N_0$ from \eqref{N0def1} upon the substitution
$\beta^2\mapsto \frac{2}{n+2}$, with some abuse of notation we
use the same symbol for these two constants.

\bigskip

\begin{figure}
\centering
\scalebox{0.96}{
\begin{tikzpicture}
\node at (0,0) {\includegraphics[width=7.5cm]{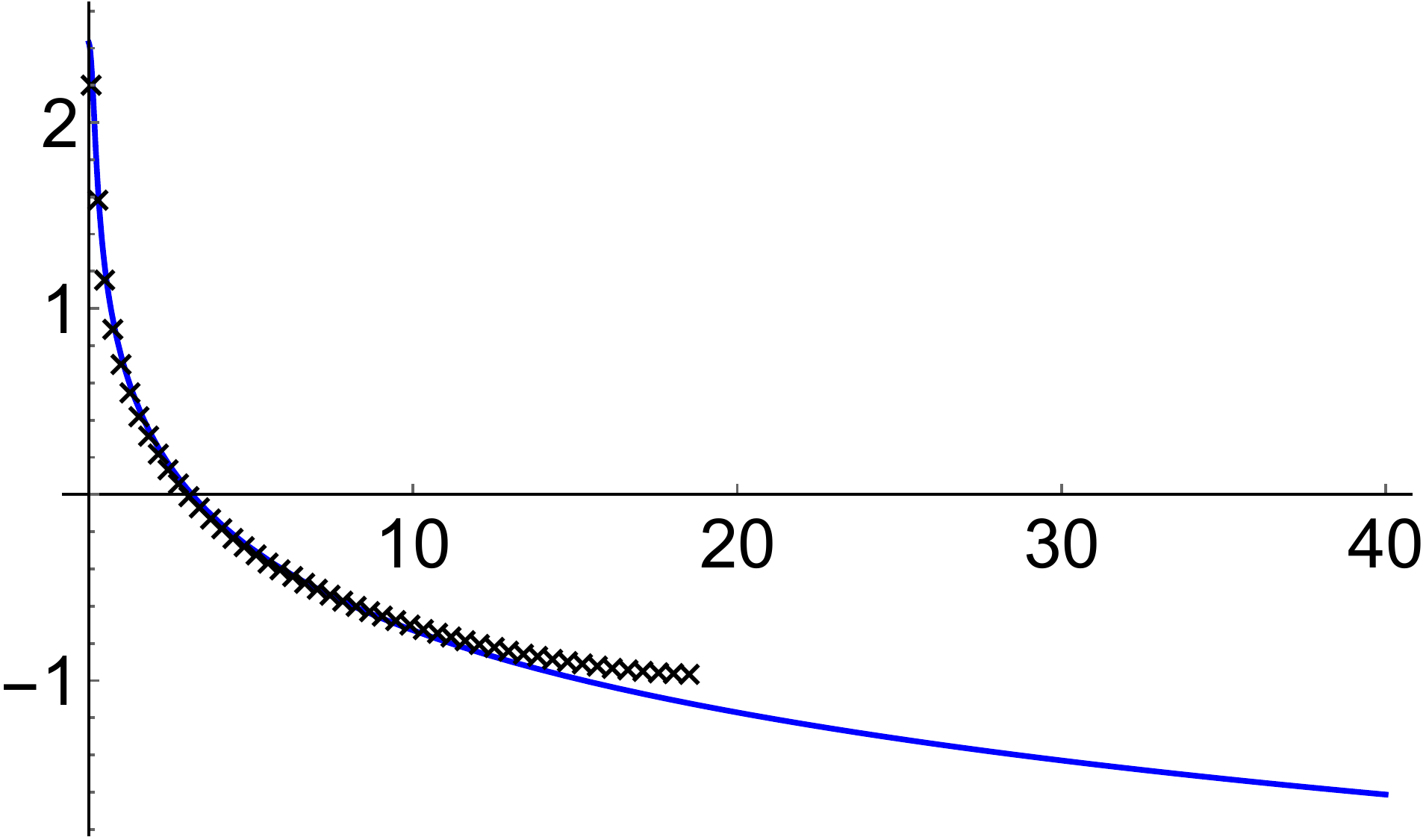}};
\node at (9,0) {\includegraphics[width=7.5cm]{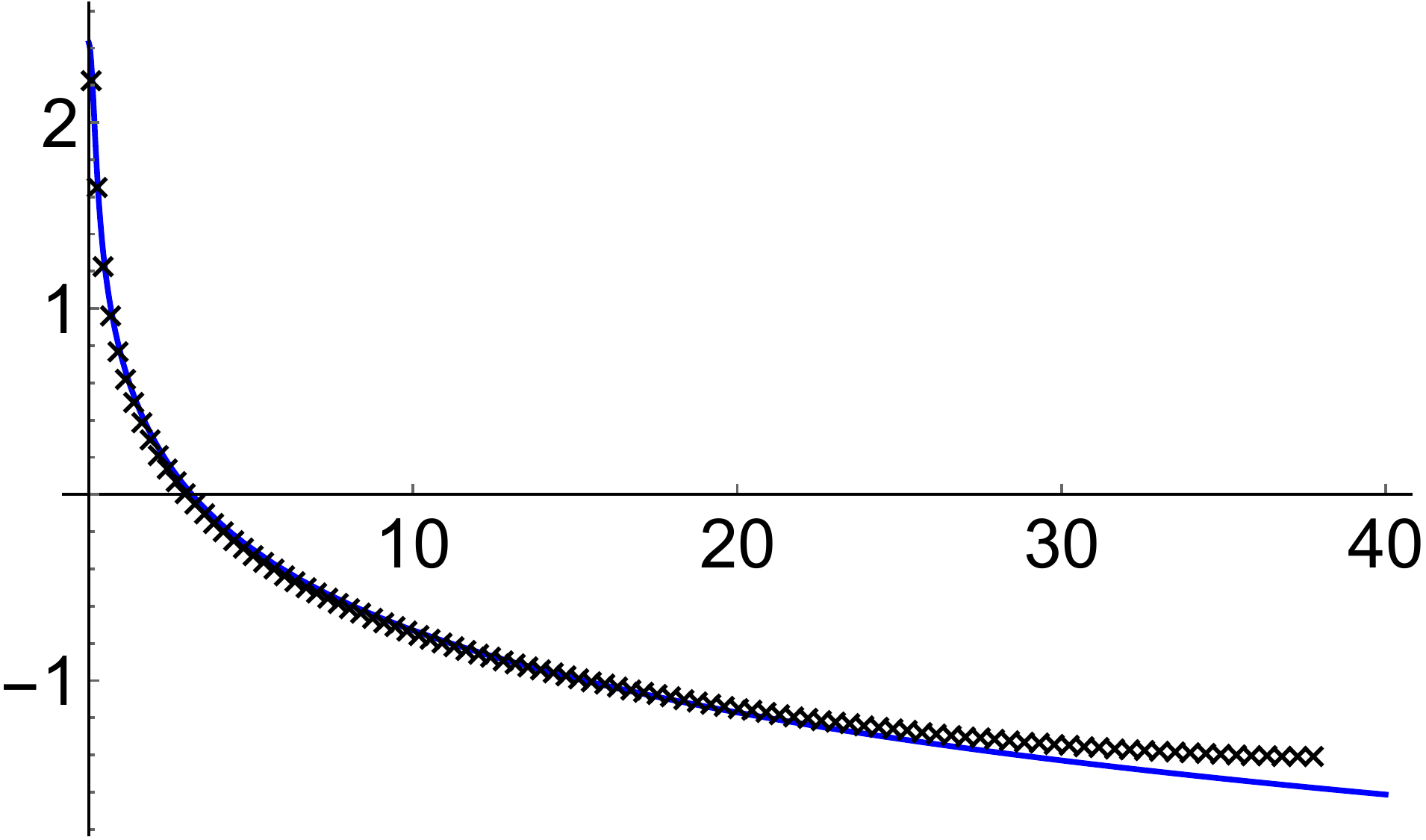}};
\node at (1,3.3) {$N=200$};
\draw (0.2,3) -- (1.8,3);
\node at (10,3.3) {$N=400$};
\draw (9.2,3) -- (10.8,3);
\node at (4,-0.4) {$s$};
\node at (13,-0.4) {$s$};
\node at (-3,2.6) {$\rho^{({\rm reg})}_{\bar{p},p}(s)$};
\node at (6,2.6) {$\rho^{({\rm reg})}_{\bar{p},p}(s)$};
\end{tikzpicture}
}
\caption{\small
For the two cases $N=200,400$ with the parameters taken to be
 $S^z=0$, ${\tt k}=\frac{1}{10}$, $n=3$,
the Bethe states were constructed for which the Bethe roots are real
with $M_+$ of them being positive and
$M_-$ being negative, i.e., $\{\zeta_m\}_{m=1}^M=\{\zeta_m^{(+)}>0\}_{m=1}^{M_+}\cup
 \{\zeta_m^{(-)}<0\}_{m=1}^{M_-}$. 
Note that when $|M_--M_+|\ll N$ these states are part of the low energy spectrum
with ${\tt L}=\bar{\tt L}={\tt w}=0$.
The corresponding values of $b=b_{\tt m}$
were computed for ${\tt m}\equiv M_--M_+=0,2,4,\ldots\frac{N}{2}$
and the crosses on the plots represent 
$(b_{{\tt m}+1}-b_{{\tt m}})^{-1}-\frac{2}{\pi}\log\big(\frac{N}{2N_0}\big)$
as a function of $s=(b_{{\tt m}+1}+b_{{\tt m}})/2$.
The blue line depicts
$\rho^{({\rm reg})}_{\bar{p},p}=\rho_{\bar{p},p}^{(0,0)}(s)-\frac{2}{\pi}\log\big(\frac{N}{2N_0}\big)$,
where $\rho_{\bar{p},p}^{(0,0)}(s)$ is the density of primary Bethe states given in eq.\,\eqref{rho01a}.
\label{rhofig1}}
\end{figure}
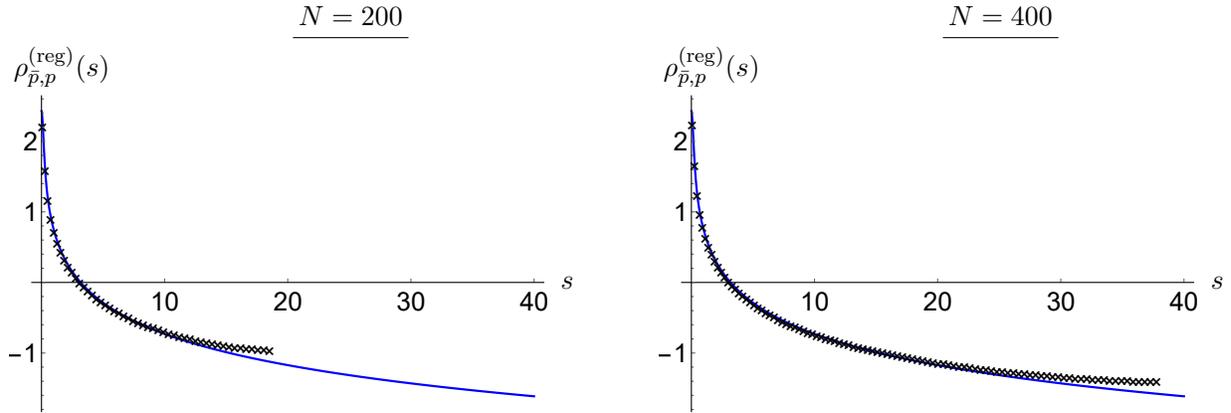

For a primary Bethe state with $|{\tt k}|>\frac{2}{n+2}$ or non-zero
 ${\tt w}$ some of the Bethe roots $\zeta_j$ become
complex. Nevertheless,
numerical work shows that eq.\,\eqref{poaospo1qa1a} holds true for
the primary Bethe states 
for any generic value of the twist parameter $-\frac{1}{2}<{\tt k}<\frac{1}{2}$, 
the positive integer $S^z=0,1,2,\ldots$ as well as the winding number
${\tt w}=0,\pm1,\pm2\ldots\ $.
However 
there is a possibility that there could be multiple primary Bethe states having
distinct $b(N)$, which satisfy eq.\,\eqref{poaospo1qa1a} with the same integer ${\tt m}$.
We observed that for sufficiently large $N$ for one of these states
$b(N)$ is always real, while for the rest it is pure imaginary.
This is tied to the fact that for  $N\gg 1$ the l.h.s. of \eqref{poaospo1qa1a}
becomes a monotonic continuous function of real $b$.
Thus for given $N\gg 1$, $p$ and $\bar{p}$,  one can
 distinguish  the primary Bethe states $\bm{\Psi}_N$  having \emph{real} $b(N)$
via the integer ${\tt m}$ from eq.\,\eqref{poaospo1qa1a}.
Moreover   $b=b_{\tt m}(N)$ obeys the ordering 
\be
b_{\tt m}(N)<b_{{\tt m}'}(N)\qquad
{\rm whenever} \qquad {\tt m}<{\tt m}'\ .
\ee
For ${\tt m}$ in \eqref{poaospo1qa1a}  to correspond to a low energy state,
it should be bounded as $|{\tt m}|\le{\tt m}_{\rm max}$ with some positive
 integer ${\tt m}_{\rm max}={\tt m}_{\rm max}(N)\ll N$.
This is similar to the case with ${\tt w}=0$ and $|{\tt k}|<\frac{2}{n+2}$ discussed above.
Again numerical work suggests that it is possible to construct a  Bethe state for any
${\tt m}=-{\tt m}_{\rm max},-{\tt m}_{\rm max}+2,\ldots,{\tt m}_{\rm max}-2,\,{\tt m}_{\rm max}$.
Eq.\,\eqref{iaosido121} implies that $b_{\tt m+1}(N)-b_{{\tt m}}(N)\propto 1/\log(N)$ and
hence for $N\gg 1$ 
the $b_{\tt m}(N)$ become densely distributed 
within the segment $(-b_{\rm max}(N),+b_{\rm max}(N))$
where $b_{\rm max}(N)=b_{\tt m}(N)$ with ${\tt m}={\tt m}_{\rm max}(N)$.
Though it is difficult to give an accurate estimate of 
$b_{{\rm max}}(N)$, one may expect that
\be
\lim_{N\to\infty}\ \frac{\log(N)}{{\tt m}_{{\rm max}}(N)}=0
\ee
and hence $\lim_{N\to\infty}b_{\rm max}(N)=\infty$.
This way we conclude that in the scaling limit the spectrum   develops
a continuous component, which we will label  by the parameter $-\infty<s<+\infty$.
As it follows from \eqref{poaospo1qa1a}, for $N\gg 1$ the number of primary Bethe states
with real $b(N)$ lying in the segment $(s,s+\Delta s)$ is approximated by
$\rho_{\bar{p},p}^{(0,0)}(s)\,\Delta s$ with
\be\label{rho01a}
\rho_{\bar{p},p}^{(0,0)}(s)=
\frac{2}{\pi}\ \log\bigg(\frac{N}{2N_0}\bigg)+\frac{1}{2\pi\ri}\ 
\partial_s\log\bigg[2^{\frac{4\ri s(n+2)}{n}}
\ \frac{\Gamma(\frac{1}{2}+p-{\ri s})\,\Gamma(\frac{1}{2}+{\bar p}-{\ri s})}
{\Gamma(\frac{1}{2}+p+{\ri s})\,\Gamma(\frac{1}{2}+{\bar p}+{\ri s})}\bigg]\ .
\ee
For an illustration see fig.\,\ref{rhofig1}.
The parameter $s$ can be understood as an RG invariant 
along with $p$ and $\bar{p}$.
Then the scaling limit of a family of primary 
Bethe states $\bm{\Psi}_N$ corresponding
to a given value of $s$ can be achieved by 
assigning an $N$ dependence to the integer ${\tt m}$
via the formula 
\bea\label{oaspo190099}
8\,s\,\log\bigg(\frac{N}{2N_0}\bigg)
+\delta=2\pi\,{\tt m}(N)+O\big((\log N)^{-\infty}\big)\ .
\eea
With this understanding of the scaling limit it follows 
from \eqref{oaspo190099},\,\eqref{poaospo1qa1a}
 that ${\rm s}\!\lim_{N\to\infty}b(N)= s$ for a RG trajectory labeled by $s$ and hence
\be\label{oaisodi192019299}
{\rm s}\!\!\!\lim_{N\to\infty}\ \, 
\frac{N}{4\pi v_{\rm F}}\ \Big({\cal E}-e_\infty\,N\Big)=
\frac{p^2+\bar{p}^2}{n+2}+\frac{2s^2}{n}-\frac{1}{6}\qquad \qquad ({\tt L}=\bar{\tt L}=0)\ .
\ee
\bigskip

Numerical studies show that there exist 
the RG trajectories $\bm{\Psi}_N$
with $b(N)$ tending to a pure imaginary number in the large $N$ limit.
The pattern of Bethe roots for one  such
trajectory with ${\tt L}=\bar{\tt L}=0$
is shown in the right panel of fig.\,\ref{BAplot1}.
In this case the quantum number ${\tt m}$ is not well defined
(besides its parity). However, $b(N)$ still satisfies the exponential form of 
 eq.\,\eqref{poaospo1qa1a},
\bea\label{quantC1}
\bigg(\frac{N}{2N_0}\bigg)^{4\ri b(N)}\ \re^{\frac{\ri}{2}\delta}\,\big|_{s=b(N)}=\sigma+
O\big((\log N)^{-\infty}\big)
\eea
with $\re^{\frac{\ri}{2}\delta}$ as in \eqref{oisaoid132}
and $\sigma$ is a sign factor, which coincides with the parity of 
$\frac{N}{2}-S^z$:
\be\label{sigmadef1a}
\sigma = (-1)^{\frac{N}{2}-S^z}\ .
\ee
In constructing a RG trajectory the value of $\sigma$ should be kept fixed.
\bigskip

Let's consider a trajectory with $\lim_{N\to\infty} b(N)=s$ such that
 $\Im m(s)\ne 0$.
The factor $ N^{4\ri b(N)}$ in the l.h.s. of
 eq.\,\eqref{quantC1}  goes to zero  as $N\to\infty$  if $\Im m(s)>0$ or 
tends to infinity when $\Im m(s)<0$. Hence in order for \eqref{quantC1} to be obeyed,
one must have that
\bea\label{deltacond1a}
\re^{-\frac{\ri}{2}\delta}&=&0\qquad {\rm for} \qquad \Im m(s)>0 \\[0.2cm]
\re^{+\frac{\ri}{2}\delta}&=&0\qquad {\rm for} \qquad \Im m(s)<0\ .\nonumber
\eea
This condition,  combined with
the explicit formula for $\re^{\frac{\ri}{2}\delta}$ \eqref{oisaoid132}, yields
that the limiting values $s=\lim_{N\to\infty}b(N)$
with $\Im m(s)\ne 0$ 
must be of the form 
\be\label{oiasiodio98089}
s=\pm\ri\,\big(-p_{\rm min}-\tfrac{1}{2}-a\big)\,,
\ee
where $p_{\rm min}=\min(p,\bar{p})$ and
$a$ may be any non-negative integer provided that
\be\label{aossdido1a}
a\ : \ a\ge 0 \ \ {\rm and} \ \ \ -p_{{\rm min}}-\tfrac{n+2}{4}\le a<-p_{\rm min}-\tfrac{1}{2}\ .
\ee
In writing the above inequalities on $a$ we've taken into account the restriction \eqref{aisodio1231}.
Note that  $-p_{\rm min}-\tfrac{1}{2}-a$ must be a positive number.
Hence such values \eqref{oiasiodio98089} are only possible if either $p<-\frac{1}{2}$, in
which case $\bar{p}>\frac{1}{2}$ or the other way around:
$\bar{p}<-\frac{1}{2}$ and $p>\frac{1}{2}$.
\bigskip

\begin{figure}
\centering
\scalebox{0.855}{
\begin{tikzpicture}
\node at (0,0) {\includegraphics[width=10cm]{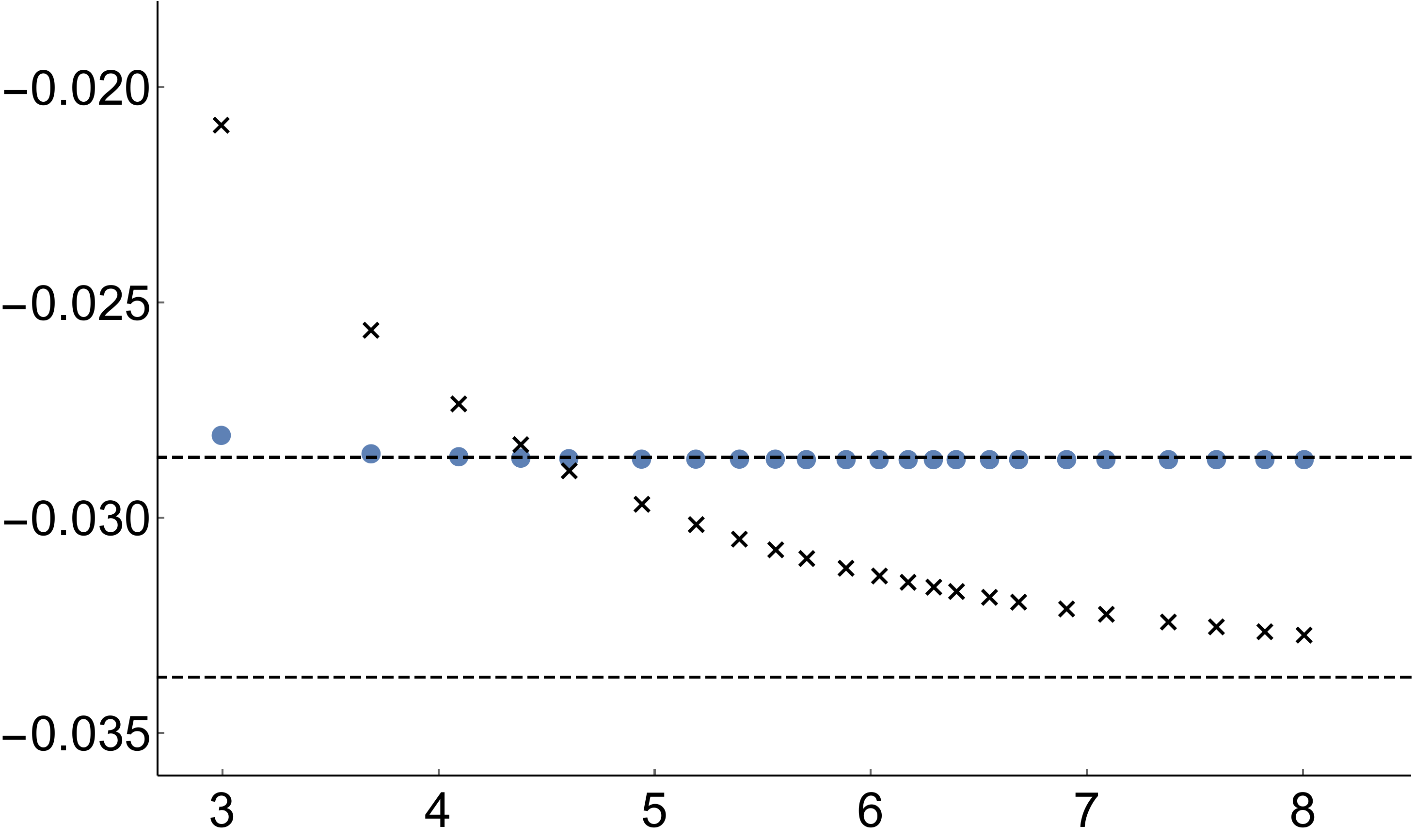}};
\node at (5.8,-2.5) { $\log(N)$};
\node at (-3.85,3.3) {$\delta{\cal E}$};
\end{tikzpicture}
}
\caption{\label{sec9fig}\small
The value of $\delta{\cal E}=\frac{N}{4\pi { v}_{\rm F}}\,({\cal E}-e_\infty\,N)$
 is plotted as a function of $\log(N)$ for two RG trajectories
in the case $n=3$ and
${\tt k}=0.235>\frac{1}{n+2}$. The blue dots depict $\delta{\cal E}$ for the trajectory
having ${\tt L}=\bar{\tt L}=S^z={\tt w}=0$ and with $s=0$. 
If $\log(N)\lessapprox 4.5$ the corresponding Bethe state 
is  the ground state
of the Hamiltonian $\mathbb{H}$. 
For the second RG trajectory (crosses)
the RG invariants ${\tt L}$, $\bar{{\tt L}}$, $S^z$, ${\tt w}$ are also zero
but  $s=\frac{\ri}{2}\big((n+2)\,{\tt k}-1\big)=\frac{7\ri}{80}$ is a pure imaginary number.
The dashed lines denote the limiting values 
$\delta{\cal E}=-0.0286\ldots$ for the RG trajectory with zero $s$ and
$\delta{\cal E}=-0.0337\ldots$ for the other one.
}
\end{figure}

The RG trajectory  for which $b(N)$ tends to a pure
imaginary number  can
be labeled by the  limiting value of $b(N)$, i.e., $s$ from \eqref{oiasiodio98089},\,\eqref{aossdido1a}. 
The latter
should be treated as an RG invariant along with
$p$ and $\bar{p}$. The scaling limit of the energy
for such states
is still described by eq.\,\eqref{oaisodi192019299}.
It should be mentioned that when $\frac{1}{n+2}<|{\tt k}|<\frac{1}{2}$,
the ${\cal Z}_2$ doublet of the primary Bethe states with
${\tt w}=S^z=0$ and  
$s=\pm\frac{\ri}{2}\big((n+2)\,|{\tt k}|-1\big)$ turn out to be the lowest energy states
of the lattice  Hamiltonian $\mathbb{H}$ for $N\gg 1$. Their energy is lower than that of the
primary Bethe state with ${\tt w}=S^z=s=0$, which is the ground state 
in the case $|{\tt k}|\le\frac{1}{n+2}$. An example is provided in fig.\,\ref{sec9fig}.

\bigskip

\section{Summary of numerical work: basic conjectures \label{sec10}}
From the study of the primary Bethe states
we've found that, with a proper understanding of the
scaling limit, the RG trajectories with ${\tt L}=\bar{\tt L}=0$ are labeled by $p$, $\bar{p}$ and $s$.
The last RG invariant may take any real values $s\in(-\infty,+\infty)$ as
well as a finite discrete set of pure imaginary numbers given by
 eqs.\,\eqref{oiasiodio98089} and \eqref{aossdido1a}.
For the low energy Bethe states with ${\tt L}+\bar{\tt L}>0$ 
the same qualitative picture is expected to hold true
as well. In particular, these trajectories  may
be assigned the number $s={\rm s}\!\lim_{N\to\infty}b(N)$, subject to the constraint
\be\label{scon3iuewi1}
-\frac{n}{4}\le \Im m(s)\le \frac{n}{4}
\ee
(to be compared with \eqref{aisodio1231}).
This way the conformal towers appearing in the scaling 
limit are labeled by $\bar{p}$, $p$, 
as well as $s$,  whose set of admissible values  contains both
 a  continuous and a discrete component. 
An explicit description of both these components is given in this section.
Since our analysis involves many assumptions, which are
 mostly justified through the numerical work,
we formulate our findings regarding the conformal towers  as a series of conjectures.
\bigskip

Clearly, when ${\tt L}+\bar{\tt L}>0$, the RG invariants $p$, $\bar{p}$ and $s$ 
are insufficient
for the unambiguous specialization of a RG trajectory or, equivalently, a
state in
the level subspace of the conformal tower.
The latter is achieved by means of 
the two non-ordered sets
\bea\label{jssy}
\bm{w}=\{w_a\}_{a=1}^{\tt L}\ \,,\ \ \ \ \ \ \  \qquad \qquad\bar{\bm{w}}=\{ \bar{w}_a\}_{a=1}^{\bar {\tt L}}\ \,,
\eea
which play a r$\hat{\rm{o}}$le similar to
 $\bm{v}=\{v_a\}_{a=1}^{\tt L}$ and $\bar{{\bm v}}=\{\bar{v}_a\}_{a=1}^{\bar{\tt L}}$ 
in the homogeneous case.
For the ${\cal Z}_2$ invariant six-vertex model, the algebraic systems satisfied  by
$\bm{w}$ and $\bar{\bm{w}}$ read, respectively, as
\begin{subequations}\label{sksksk10}
\bea\label{sksksk1}
4 n\, w_a^2\!\!&+&\!\!8\ri s\, (n+1)\, w_a-(n+2)\ \big((n+1)^2-4p^2\big)\\[0.2cm]
&+&\!\!
4\ \sum_{b\not=a}^{{\tt L}}\frac{w_a\, (\, (n+2)^2\, w_a^2- n(2n+5)\, w_a w_b + n(n+1)\, w_b^2\,)}{(w_a-w_b)^3}=0\  \ \ \ \ \ \ \ \ (a=1,\ldots,
{\tt L})\,,\nonumber
\eea
\bea\label{sksksk1bar}
4 n\, \bar{w}_a^2\!\!&+&\!\!8\ri s\, (n+1)\, \bar{w}_a-(n+2)\ \big((n+1)^2-4\bar{p}^2\big)\\[0.2cm]
&+&\!\!
4\ \sum_{b\not=a}^{\bar{{\tt L}}}
\frac{\bar{w}_a\, (\, (n+2)^2\, \bar{w}_a^2- n(2n+5)\, \bar{w}_a 
\bar{w}_b + n(n+1)\, \bar{w}_b^2\,)}{(\bar{w}_a-\bar{w}_b)^3}=0\  \ \ \ \ \ \ \ \ 
(a=1,\ldots,
\bar{{\tt L}})\,.\nonumber
\eea
\end{subequations}
It was conjectured in the work \cite{Bazhanov:2019xvy} that for fixed ${\tt L}$ and for generic values of  $p$, $n$
the number of solutions of \eqref{sksksk1}, up to the action of the permutation group, is 
equal to ${\tt par}_2({\tt L})$ --  the number of bipartitions of ${\tt L}$,
\be\label{aisuasau}
\sum_{{\tt L}=0}^\infty {\tt par}_2({\tt L})\,{\tt q}^{\tt L}
=\frac{1}{({\tt q},{\tt q})_\infty^{2}}=1+2\,{\tt q}
+5\,{\tt q}^2+10\,{\tt q}^3+20\,{\tt q}^4+36\,{\tt q}^5+\ldots \ .
\ee
Here and below we use the notation
\bea\label{aiosiod1929812}
(z,{\tt q})_\infty=\prod_{m=0}^\infty(1-z\, {\tt q}^m)\ .
\eea
In turn,
the number of solutions of \eqref{sksksk1bar} is expected to be ${\rm par}_2(\bar{{\tt L}})$.
\bigskip

In the description of the scaling limit for the primary Bethe states
a key r${\rm{\hat o}}$le was played by eq.\,\eqref{quantC1}.
This relation was extended to the
RG trajectories with any values of the non-negative integers ${\tt L}$ and $\bar{\tt L}$ in the work \cite{Bazhanov:2019xvy}.
All that is required is a modification of the phase shift $\delta$, which is now
a  function of the sets
$\bm{w}$ and $\bar{\bm{w}}$ solving eqs.\,\eqref{sksksk10}, as well as $\bar{p}$, $p$ and $s$  so that
$\delta=\delta(\bar{\bm{w}},\bm{w}\,|\,\bar{p},p,s)$.
In that same work a 
 formula was proposed, which expresses $\re^{\frac{\ri}{2}\delta}$
in terms of the  connection coefficients
of a certain ODE (see also the next section and, in particular, eq.\,\eqref{ioasd2989823} below).
It reads as
\be\label{phasedef1a}
\re^{\frac{\ri}{2}\delta(\bar{\bm{w}},\bm{w}|\bar{p},p,s)}=
{ D}_{{\bar{p}},s}(\bar{{\boldsymbol  w}})\,{ D}_{p,s}({\boldsymbol  w})\ ,\qquad
\ee
where
\be\label{oiaodi1a1a}
{ D}_{p,s}({\boldsymbol  w})=
2^{\frac{2\ri(n+2)s}{n}}\ \frac{\Gamma(\frac{1}{2}+p-\ri s)}{\Gamma(\frac{1}{2}+p+\ri s)}\  \ 
\check{ D}_{p,s}({\boldsymbol  w})
\ee
and $\check{ D}_{p,s}({\boldsymbol  w})$ are  normalized to be one for ${\tt L}=0$.
For general ${\tt L}$, the explicit expression for $\check{ D}_{p,s}({\boldsymbol  w})$
as a function of $p$, $s$ and the set $\bm{w}$ 
was derived in ref.\cite{Kotousov:2019nvt}.
It's quoted in formula \eqref{Dformula1} in 
 Appendix \ref{app2}.
\bigskip

\subsection{Continuous spectrum\label{sec101}}
In general  \eqref{quantC1}, regarded as
an equation determining
 the large $N$ dependence of $b(N)$, has complex solutions. Nevertheless there
exists a class of them such that $\lim_{N\to\infty}\Im m\big(b(N)\big)=0$.
For their description 
it is useful to take the logarithm of both sides of \eqref{quantC1} 
and bring it to the form \eqref{poaospo1qa1a}.
The phase shift $\delta$ entering into that equation depends on $s$ both
explicitly and implicitly through the solution sets
 $\bm{w}$, $\bar{\bm{w}}$ of \eqref{sksksk10}.
One should choose $\bm{w}$, $\bar{\bm{w}}$ 
in such a way so that they are continuous functions of $s$. 
It will be argued later that for real $p$, $\bar{p}$ the product 
${ D}_{{\bar{p}},s}(\bar{{\boldsymbol  w}})\,{ D}_{p,s}({\boldsymbol  w})$
in \eqref{phasedef1a} is never zero or infinity for any  $s\in(-\infty,+\infty)$.
Due to this 
$\delta$  can be made to be a uniformly bounded continuous function of  real $s$.
\bigskip

Suppose that the term $\propto\log(N)$ in  the l.h.s. of  eq.\,\eqref{poaospo1qa1a}
dominates. Then an iterative solution yields
\be\label{asymp1a}
b_{\tt m}(N)\asymp\frac{\pi{\tt m}-\frac{1}{2}\,\delta_0}{4\log\big(N\re^{\frac{1}{8}\delta'_0}/(2N_0)\big)}
+O\big((\log N)^{-3}\big)
\,,\qquad N\to\infty\quad {\rm with}\quad {\tt m}-{\rm fixed}\,,
\ee
where
\be
\delta_0=\delta\big|_{s=0}\,,\qquad\qquad \delta'_0=\partial_s\,\delta\big|_{s=0}\ .
\ee
These last two numbers are typically complex 
so that 
$\Im m\big(b_{\tt m}(N)\big)=O\big(1/\log(N)\big)$.
The solutions of this class can be labeled by the integer ${\tt m}$ and, in addition, obey
the ordering
\be
\Re e\big(b_{\tt m}(N)\big)<\Re e\big(b_{{\tt m}'}(N)\big)\,,\quad {\rm for}
\quad {\tt m}<{\tt m}'\qquad (N\gg 1)\ .
\ee
\bigskip

\begin{figure}
\centering
\scalebox{1}{
\begin{tikzpicture}
\node at (4,4.5) {$b$};
\draw  (4,4.5) circle [radius=0.3];
\node at (0,0) {\includegraphics[width=17cm]{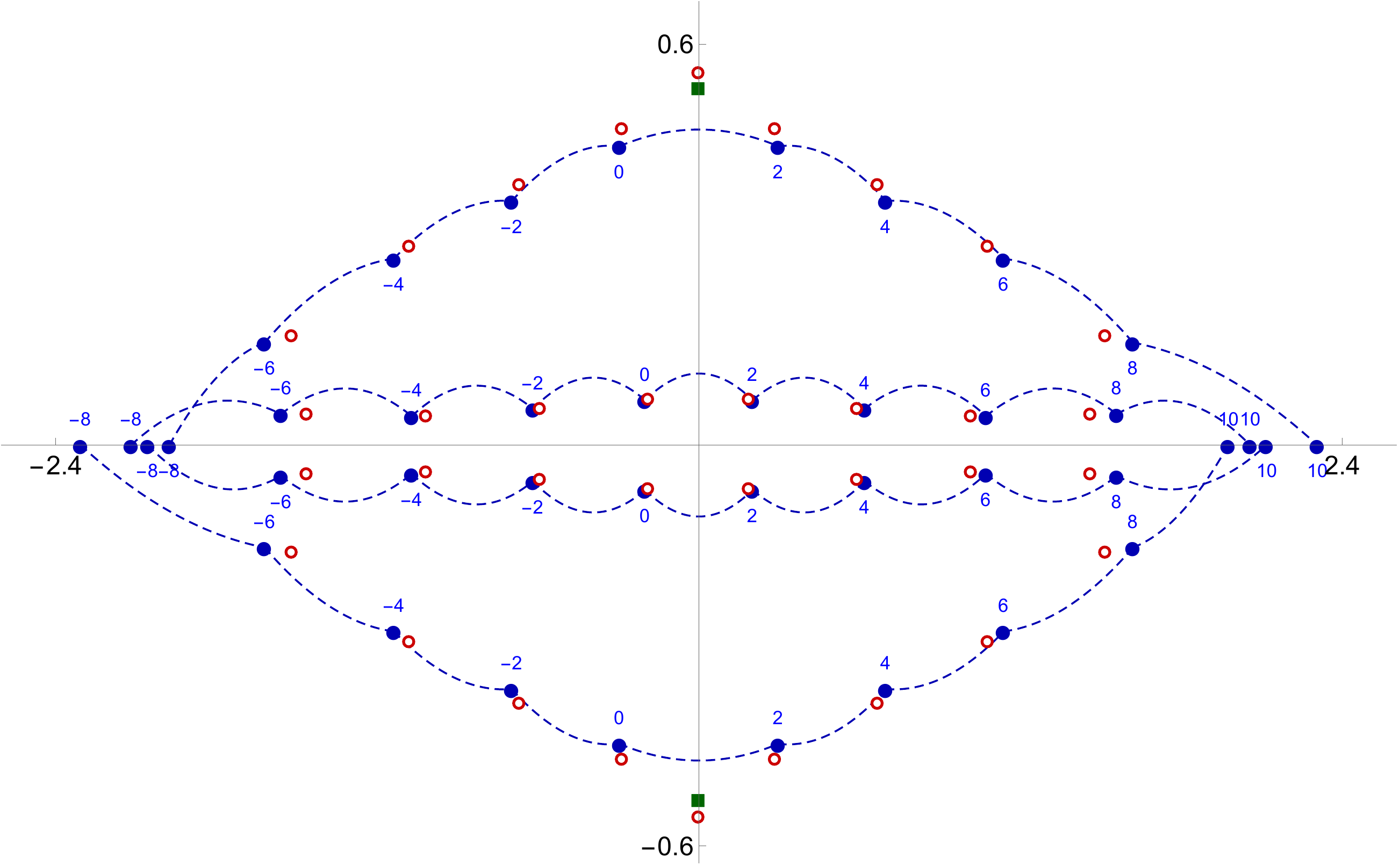}};
\end{tikzpicture}
}
\caption{\small
The figure depicts all the solutions of \eqref{quantC1}, regarded as an equation for
$b(N)$ with the correction term ignored, in the rectangle $-0.6<\Im m\big(b(N)\big)<0.6$
and $-2.4<\Re e\big(b(N)\big)<2.4$.
The case being considered is $N=22$ and ${\tt L}=\bar{\tt L}=1$, 
while the parameters are $n=3$,  ${\tt k}=-0.18$, $S^z=1$ and ${\tt w}=0$
so that $p=\frac{1}{20}$, $\bar{p}=\frac{19}{20}$.
The algebraic system \eqref{sksksk1} for ${\tt L}=1$ 
becomes a quadratic equation on $w\equiv w_1$, whose two solutions are given by
$w_\pm=-\frac{n+1}{2n}\,\big(2\ri\,s\pm \sqrt{C}\,\big)$, where
 $C=n(n+2)\,\big(1-\frac{4p^2}{(n+1)^2}-\frac{4s^2}{n(n+2)}\big)$.
Here the branch of the square root is taken so that $\sqrt{C}$ is positive when $C>0$.
In the case when $C<0$, 
we set $w_\pm=-\ri\, \frac{n+1}{2n}\, \big(2s\mp \sqrt{-C}\,\big)$.
Similar formulae, with $p$ substituted by $\bar{p}$, are used to define
$\bar{w}_\pm$ which solve \eqref{sksksk1bar} with $\bar{{\tt L}}=1$.
The phase shift entering into
\eqref{quantC1} may be any one of the
four functions $\delta(\bar{w}_{\sigma},{w}_{\sigma'})\equiv \delta(\bar{w}_{\sigma},w_{\sigma'}\,|\,\bar{p},p,s)$
with $\sigma,\,\sigma'=\pm 1$. 
The filled circles corresponding to the solutions of \eqref{quantC1}  with the same function $\delta$ are grouped
together by the dashed line for visualization.
Those from the top set  represent solutions of
 \eqref{poaospo1qa1a} with $\delta=\delta(\bar{w}_+,w_+)$. The integer ${\tt m}$ entering into that equation is indicated by the label beside each circle. The lower set of connected filled circles corresponds to 
$\delta=\delta(\bar{w}_-,w_+)$,
the next lowest (just below the real axis)
 to   $\delta=\delta(\bar{w}_+,w_-)$, while for the bottom most set of circles
$\delta=\delta(\bar{w}_-,w_-)$.
The green boxes also depict solutions of \eqref{quantC1} with $\delta=\delta(\bar{w}_+,w_+)$ for the 
top box and $\delta=\delta(\bar{w}_-,w_-)$ for the bottom one. However, whereas all the filled circles correspond to
the RG trajectories with
$\lim_{N\to\infty}\Im m\big(b(N)\big)=0$, for the green boxes $b(N)$ tends to a non-vanishing pure imaginary value
$\lim_{N\to\infty}\Im m \big(b(N)\big)=\pm\ri\,(p+\frac{1}{2})=\pm0.55\,\ri $. For these solutions
the definition of the integer ${\tt m}$ from \eqref{poaospo1qa1a}
is ambiguous, since  $\delta$ turns out to have a logarithmic branch point at $s=\pm\ri\,(p+\frac{1}{2})$.
Finally, the empty circles represent the value of $b(N)$ \eqref{poapso1a} that was obtained by means of direct diagonalization
of the quasi-shift operator \eqref{qshift} within the sector ${\tt L}=\bar{\tt L}=1$, $S^z=1$ and ${\tt w}=0$.
Note that the states corresponding to ${\tt m}=-8$ and ${\tt m}=10$ were not observed among the first 700
lowest energy states. 
\label{sdist1}}
\end{figure}

Let ${\cal H}_{N|S^z}^{({\rm cont})}$ be the set of low energy Bethe states in the sector
with given $S^z$ such that
$\Im m \big(b(N)\big)\to0$ as $N\to\infty$. 
Appealing to  numerical work, 
we expect that for fixed $N\gg1$, and
given ${\tt L}$, $\bar{\tt L}$,
$\bm{w}$ and $\bar{\bm{w}}$, 
the states from ${\cal H}_{N|S^z}^{({\rm cont})}$  can be labeled by the integer ${\tt m}$, which is defined through
eq.\,\eqref{poaospo1qa1a}.
This integer takes the values
${\tt m}=-{\tt m}_{\rm max},\,-{\tt m}_{\rm max}+2,\,\ldots,
{\tt m}_{\rm max}-2,\,{\tt m}_{\rm max}$ with some ${\tt m}_{\rm max}={\tt m}_{\rm max}(N)\ll N$.
For an illustration see fig.\,\ref{sdist1}. 
The asymptotic condition \eqref{asymp1a} implies that $b_{{\tt m}+1}(N)-b_{\tt m}(N)\propto 1/\log(N)$
so that the set  $\{b_{\tt m}(N)\}$ becomes densely distributed within the segment
$\big(-b_{\rm max}(N),+b_{\rm max}(N)\big)$. 
As $N$ tends to infinity we suppose that $b_{\rm max}(N)\to +\infty$.
All the above properties are analogous to those of the primary Bethe states with real $b(N)$ discussed before,
except that now $\Im m \big(b(N)\big)$ vanishes only in the limit $N\to\infty$.
This way we come to the conjecture:
\begin{enumerate}[(I)]
\item For fixed $N\gg 1$ let
$\Delta{\cal N}^{(\bar{\tt L},{\tt L})}_{\bar{p},p,s}$ be
the  number of Bethe states from the set ${\cal H}_{N|S^z}^{({\rm cont})}$ 
with given ${\tt L}$, $\bar{\tt L}$, $p$, $\bar{p}$
such that
$\Re e\big(b(N)\big)$
 lies in the interval
$(s,s+\Delta s)\subset\big(-b_{\rm max}(N),+b_{\rm max}(N)\big)$.
Then
\be
\Delta{\cal N}^{(\bar{\tt L},{\tt L})}_{\bar{p},p,s}\approx\rho_{\bar{p},p}^{(\bar{\tt L},{\tt L})}(s)\,\Delta s
\qquad \qquad(\Delta s\ll 1)
\ee
with the density
\bea\label{aisodio12311}
&&\rho_{\bar{p},p}^{(\bar{\tt L},{\tt L})}(s)={\rm par}_2({\tt L})\,{\rm par}_2(\bar{\tt L})\,
\rho_{\bar{p},p}^{(0,0)}(s) +\\[0.2cm]
&&\frac{1}{2\pi\ri}\ \partial_s\Bigg(
{\rm par}_2({\tt L})\log \bigg(\prod_{\bar{{\bm w}}\atop\bar{\tt L}-{\rm fixed}}\check{ D}_{\bar{p},s}(\bar{{\boldsymbol  w}})\bigg)
+{\rm par}_2(\bar{{\tt L}})\log\bigg( \prod_{{\bm w}\atop{\tt L}-{\rm fixed}}\check{ D}_{p,s}({\boldsymbol  w})\bigg)\Bigg)\ .\nonumber
\eea
The density of primary Bethe states $\rho_{\bar{p},p}^{(0,0)}(s)$
is quoted 
 in eq.\,\eqref{rho01a}. Also
the product in $\prod_{{\bm w}}\check{ D}_{p,s}(\bm{w})$ goes  over all the ${\rm par}_2({\tt L})$ 
solutions $\bm{w}$ of eq.\,\eqref{sksksk1} with fixed ${\tt L}$ and similarly for 
$\prod_{\bar{{\bm w}}}\check{ D}_{\bar{p},s}(\bar{{\boldsymbol  w}})$.
\end{enumerate}
In ref.\cite{Kotousov:2019nvt} the following explicit formula for the product 
$\prod_{{\bm w}}\check{ D}_{p,s}(\bm{w})$ 
was obtained:
\be\label{iaoio898aaf}
\prod_{\bm{w}\atop{\tt L}-{\rm fixed}}\check{D}_{p,s}(\bm{w})
=\prod_{m=1}^{\tt L}\!
 \prod_{1\leq j,k\atop jk\leq m}\!
\bigg[\frac{(2p-2\ri s+2k-j)\,(2p+2\ri s-2k+j) }
{(2p+2\ri s+2k-j)\,(2p-2\ri s-2k+j)}\bigg]^{{\tt par}_1(m-kj)\,{\tt par}_1({\tt L}-m)}
\ee
\bigskip

\begin{figure}
\centering
\scalebox{0.9}{
\begin{tikzpicture}
\node at (-6.1,4.4) {$\Re e\big(b(N)\big)$};
\node at (-6.1,-5.6) {$\Im m\big(b(N)\big)$};
\node at (0,0) {\includegraphics[width=13cm]{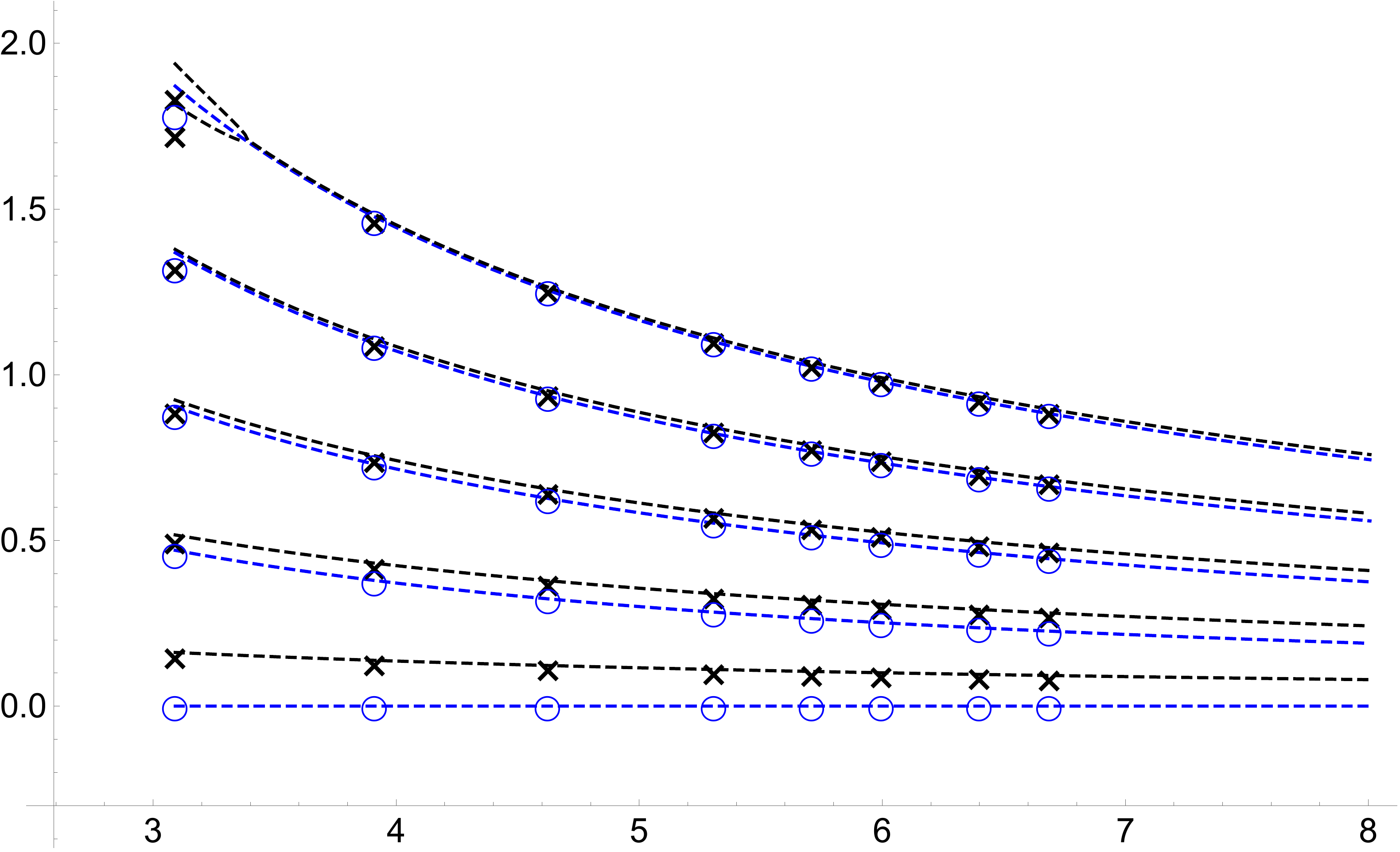}};
\node at (0,-10) {\includegraphics[width=13cm]{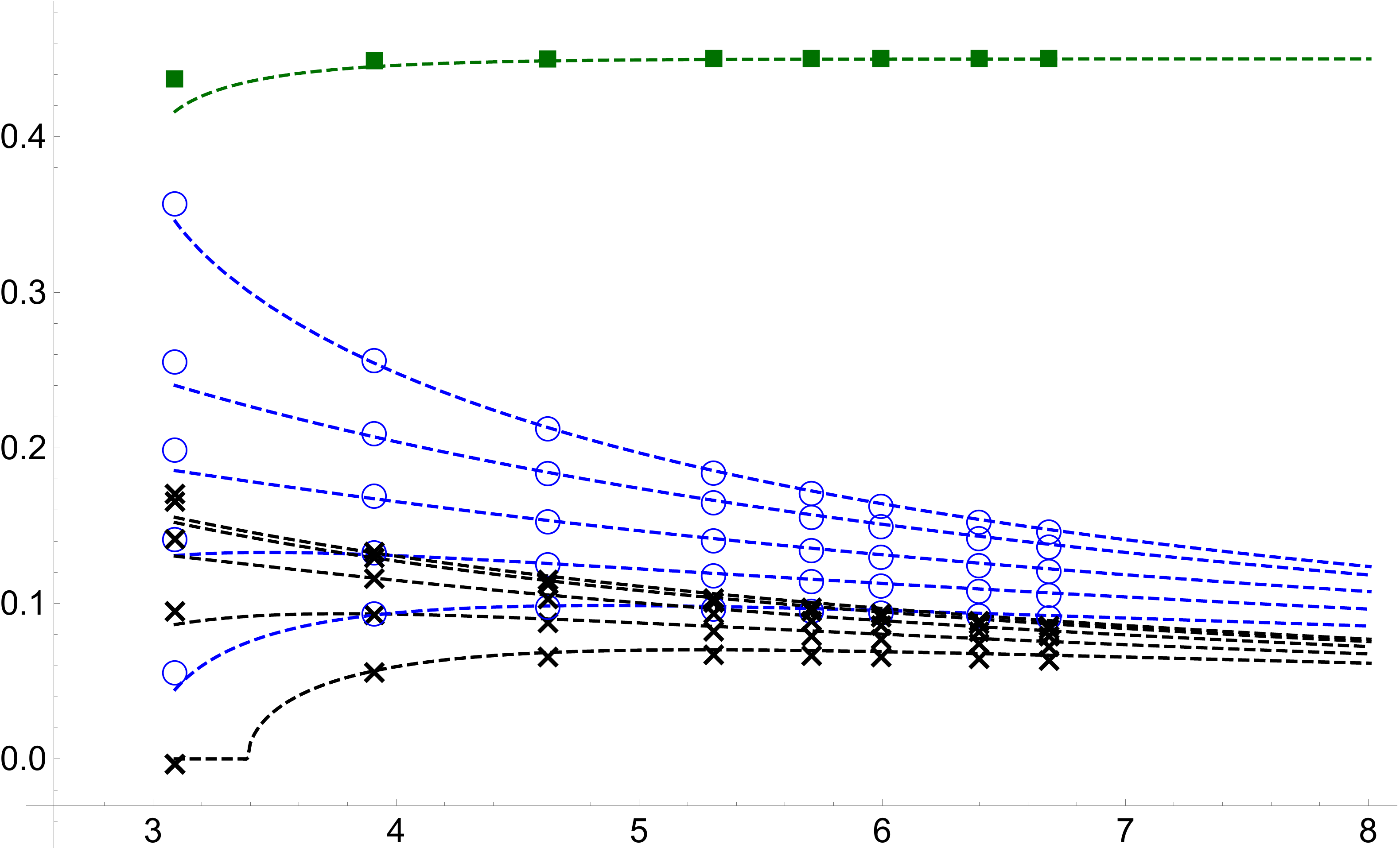}};
\node at (7.3,-3.5) {$\log(N)$};
\node at (7.3,-13.5) {$\log(N)$};
\node at (-5.3,3.5) {$11$};
\node at (-5.3,3.05) {$10$};
\node at (-5.3,2.6) {$9$};
\node at (-5.3,1.7) {$8$};
\node at (-5.3,1.3) {$7$};
\node at (-5.3,0.3) {$6$};
\node at (-5.3,-0.1) {$5$};
\node at (-5.3,-0.9) {$4$};
\node at (-5.3,-1.3) {$3$};
\node at (-5.3,-2.1) {$2$};
\node at (-5.3,-2.7) {$1$};
\node at (-5.3,-12.95) {$11$};
\node at (-5.3,-13.35) {$9$};
\node at (-5.3,-12.5) {$10$};
\node at (-5.3,-12.0) {$8$};
\node at (-5.7,-11.6) {$6$};
\node at (-5.7,-11.2) {$7$};
\node at (-5.7,-10.8) {$4$};
\node at (-5.7,-10.45) {$2$};
\node at (-5.3,-10.0) {$5$};
\node at (-5.3,-9.1) {$3$};
\node at (-5.3,-7.7) {$1$};
\node at (-5.3,-6.5) {$12$};
\draw (-5.4,-11.6) -- (-5.1,-11.3);
\draw (-5.4,-11.2) -- (-5.1,-11.1);
\draw (-5.4,-10.8) -- (-5.1,-10.7);
\draw (-5.4,-10.45) -- (-5.1,-10.55);
\end{tikzpicture}
}
\caption{\small%
The plots present the real and imaginary parts of $b(N)$, as functions of $\log (N)$,
for twelve RG trajectories. The initial points of the trajectories  
correspond to all the low energy states in the sector 
 $S^z=1$, ${\tt w}=0$, ${\tt L}+\bar{\tt L}=1$ and 
with $\Re e\big(b(N)\big)\ge 0$, $\Im m\big(b(N)\big)\ge 0$ 
(see also fig.\,\ref{fig20AA}). These
 were found from the 
numerical diagonalization of the Hamiltonian and quasi-shift operator
for the spin chain of length $N=22$ and with $n=3$, ${\tt k}=-0.18$.
The points for $N>22$
were obtained via the solution of the Bethe ansatz equations. 
The dashed lines represent $b(N)$, which solves \eqref{quantC1}
regarded as an equation for $b(N)$ with the correction terms ignored.
For the first eleven trajectories $\lim_{N\to\infty}\Im m\big(b(N)\big)=0$
(the different symbols crosses/circles and   colours black/blue 
are used only to improve the readability of the plot).
Note that for
trajectories $9$ and $11$, $b(N)$  is real for $N\le 26 $
and forms a complex conjugated pair as $N\ge 30$ (the trajectory with $\Im m\big(b(N)\big)<0$
is not depicted for $N\ge 30$).
For the $12$-th trajectory (green squares and  dashed line)
$b(N)$ is pure imaginary for any $N\ge 22$ and $\lim_{N\to\infty}b(N)= 0.45\,\ri$.
\label{fig999}}
\end{figure}
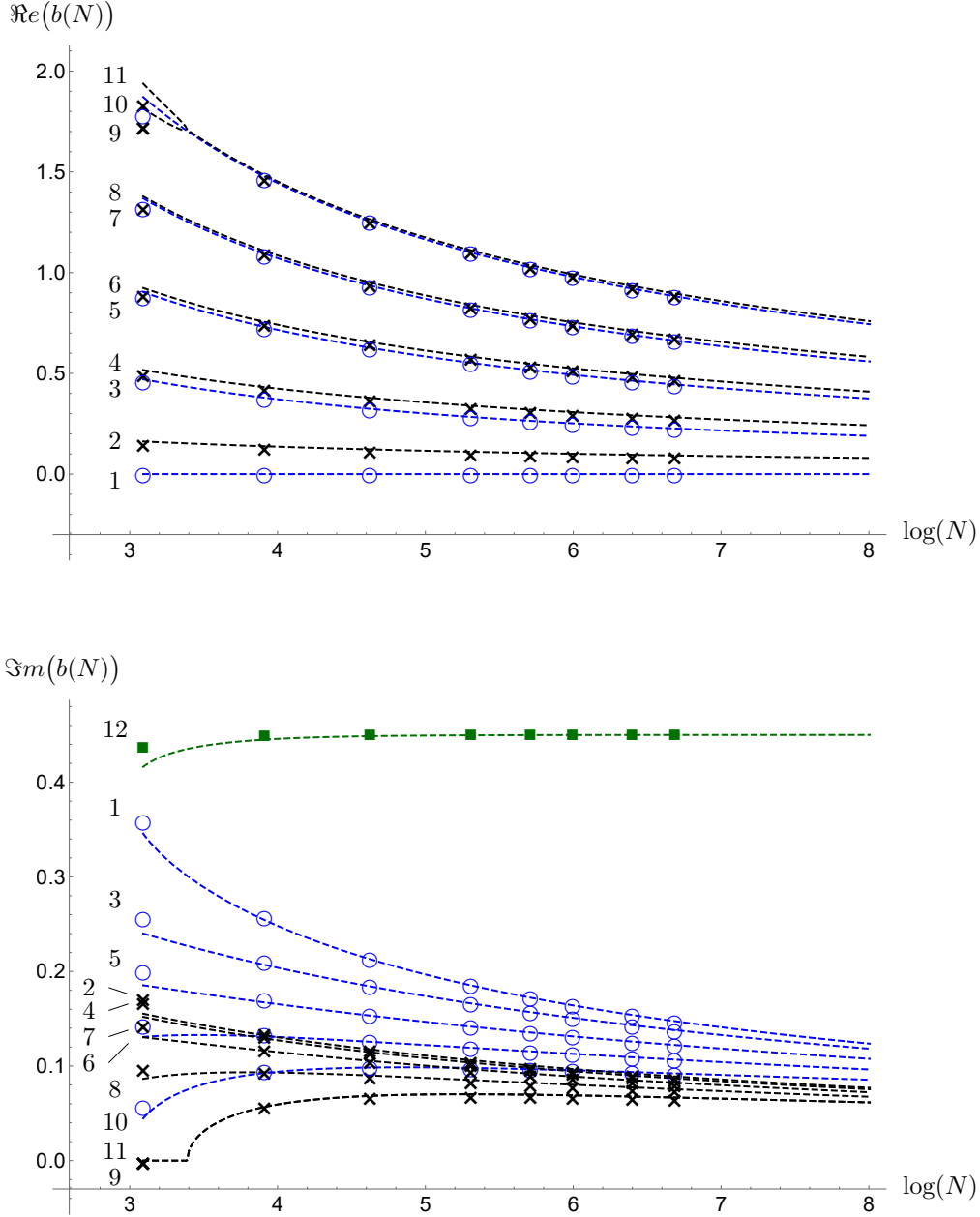

Since $\rho_{\bar{p},p}^{(0,0)}(s)=O\big(\log(N)\big)$, see eq.\,\eqref{rho01a},
Conjecture (I)
 implies that
\be
\frac{\Delta{\cal N}^{(\bar{\tt L},{\tt L})}_{\bar{p},p,s}}{\Delta{\cal N}^{(0,0)}_{\bar{p},p,s}}=
 {\rm par}_2({\tt L})\, {\rm par}_2(\bar{{\tt L}})+O\big(1/\log(N)\big)\ .
\ee
Like for the primary Bethe states,
the scaling limit of any state from ${\cal H}_{N|S^z}^{({\rm cont})}$
can be defined by assigning the
 integer ${\tt m}$ an $N$ dependence via eq.\,\eqref{oaspo190099} so that
 ${\rm s}\!\lim_{N\to\infty}b(N)= s$ with some real $s$.
Although ${\cal H}_{N|S^z}^{({\rm cont})}$ has been so far regarded merely as the formal 
set of 
all the low energy Bethe states with $\lim_{N\to\infty}\Im m\big(b(N)\big)=0$,
it is natural to introduce the structure of the linear space on ${\cal H}_{N|S^z}^{({\rm cont})}$.
Then, with the above understanding of the scaling limit,
  it follows that for given $S^z=0,1,2,\ldots\ $,
the linear space
\be
{\cal H}^{({\rm cont})}_{S^z}=
{\rm s}\!\!\!\lim_{N\to\infty}{\cal H}_{N|S^z}^{({\rm cont})}
\ee
 admits the decomposition:
\be\label{iaosidoi192009aaaaa}
{\cal H}^{({\rm cont})}_{S^z}=
\bigoplus_{{\tt w}\in\mathbb{Z}}
\int^{\oplus}_{\mathbb{R}}\!\rd s\
\Bigg[\,\bigoplus_{{\tt L},\bar{\tt L}=0}^\infty\ 
{\cal H}_{\bar{p},p,s}^{(\bar{{\tt L}},{\tt L})}\,\Bigg]\,,
\qquad  
{\rm where}
\qquad 
\begin{array}{l} p=\frac{1}{2}\,S^z+\frac{1}{2}\,(n+2)\,({\tt k}+{\tt w})\\[0.2cm]
                                       \bar{p}=\frac{1}{2}\,S^z-\frac{1}{2}\,(n+2)\,({\tt k}+{\tt w})
\end{array}
\ee
and each level subspace has dimensions
\be\label{893289121}
{\rm dim}\big({\cal H}_{\bar{p},p,s}^{(\bar{{\tt L}},{\tt L})}\big)={\rm par}_2({\tt L})\, {\rm par}_2(\bar{{\tt L}})\ .
\ee
In  the scaling limit, the low energy Bethe states in ${\cal H}_{N|S^z}^{({\rm cont})}$
with given $\bar{p}$, $p$, $\bar{{\tt L}}$, ${\tt L}$ and $s$
form a basis in ${\cal H}_{\bar{p},p,s}^{(\bar{{\tt L}},{\tt L})}$.
For all of them the scaled energy is
\be\label{oaisodi192}
E={\rm s}\!\!\!\lim_{N\to\infty}\ \, 
\frac{N}{4\pi v_{\rm F}}\ \Big({\cal E}-e_\infty\,N\Big)=
\frac{p^2+\bar{p}^2}{n+2}+\frac{2s^2}{n}-\frac{1}{6}+{\tt L}+\bar{{\tt L}}\ .
\ee
Thus, one can introduce the CFT Hamiltonian $\hat{H}\in{\rm End}\big({\cal H}_{S^z}^{({\rm cont})}\big)$
such that any state from ${\cal H}_{\bar{p},p,s}^{(\bar{{\tt L}},{\tt L})}$ is
an eigenstate of this operator with energy given by the r.h.s. of \eqref{oaisodi192}.
Symbolically, 
\be\label{aiosid892392}
\hat{H} ={\rm s}\!\!\!\lim_{N\to\infty}\ \, 
\frac{N}{4\pi v_{\rm F}}\ \Big(\mathbb{H}-e_\infty\,N\Big)\,.
\ee

\bigskip

\subsection{Discrete spectrum\label{sec102}}

The states from  ${\cal H}^{({\rm cont})}_{N|S^z}$ do not cover
all the low energy states in the lattice model.
As we saw previously,
there exist RG trajectories such that
$\lim_{N\to\infty}b(N)=s$ with
$\Im m(s)\ne 0$.
The possible limiting values  of $b(N)$
are still determined by the condition \eqref{deltacond1a},
which already appeared in our discussion of the primary Bethe states.
In view of eq.\,\eqref{phasedef1a}, for a RG trajectory $\bm{\Psi}_N$ labeled by 
${\tt L}$, $\bar{\tt L}$, $p$, $\bar{p}$ as well as the solutions sets 
$\bm{w}$, $\bar{\bm{w}}$ of the
algebraic
system  \eqref{sksksk10}, this condition 
can be rewritten as
\bea\label{pdpd000d}
\big({ D}_{{\bar{p}},s}(\bar{{\boldsymbol  w}})\,{ D}_{p,s}({\boldsymbol  w})\big)^{-1}
&=&0\qquad {\rm for}\qquad  \Im m(s)>0  \\[0.2cm]
\big({ D}_{{\bar{p}},s}(\bar{{\boldsymbol  w}})\,{ D}_{p,s}({\boldsymbol  w})\big)^{+1}
&=&0 \qquad {\rm for}\qquad  \Im m(s)<0\ .\nonumber
\eea
The formula for  ${ D}_{p,s}({\boldsymbol  w})$  
follows from eq.\,\eqref{oiaodi1a1a} 
as well as \eqref{Dformula1} from Appendix \ref{app2}.
Notice that 
 the algebraic system \eqref{sksksk10}  is invariant w.r.t.
the substitution $s\mapsto-s$, $\bm{w}\mapsto-\bm{w}$
and $\bar{\bm{w}}\mapsto-\bar{\bm{w}}$.
Furthermore it turns out that
\be
{ D}_{{{p}},s}({{\boldsymbol  w}})=
\big({ D}_{{{p}},-s}(-{{\boldsymbol  w}})\big)^{-1}\ ,
\qquad
{ D}_{{\bar{p}},s}({\bar{{\boldsymbol  w}}})=
\big({ D}_{{\bar{p}},-s}(-{\bar{{\boldsymbol  w}}})\big)^{-1}\ .
\ee
Thus, without loss of generality, one can always focus on the domain
$\Im m(s)>0$ and consider only the first line of \eqref{pdpd000d}.
\bigskip

The classification of 
the possible values of non-real $s$ appearing in the scaling limit
reduces to a study of the condition \eqref{pdpd000d}, where
$\bm{w}$ and $\bar{\bm{w}}$ are solution sets of
 the joint algebraic system \eqref{sksksk10} with given 
non-negative integers ${\tt L}$ and $\bar{\tt L}$.
Due to eqs.\,\eqref{oaisoi1093},\eqref{scon3iuewi1}
the parameters  $p$, $\bar{p}$ and $s$
will be restricted to the domain
\be\label{io89sa89}
\Im m(p)=\Im m(\bar{p})=0\  , \qquad
p+\bar{p}=S^z=0,1,2,\ldots\ ; \qquad 0<\Im m(s)\le\frac{n}{4}\,,
\ee
while $n$ is a generic positive number.
The position of the singularities of
${D}_{p,s}(\bm{w})$ as a function of $s$
can in principle be found
using the explicit formulae \eqref{oiaodi1a1a} and \eqref{Dformula1}.
However
since $s$ enters into the algebraic system \eqref{sksksk1} that is solved by $\bm{w}$,
such an analysis of ${D}_{p,s}(\bm{w})$ 
is rather difficult except for the first few levels. 
Nevertheless, one can make use of \eqref{iaoio898aaf}, which provides 
a simple expression for
the product of ${D}_{p,s}(\bm{w})$
 over all the ${\rm par}_2({\tt L})$ solutions of 
\eqref{sksksk1} with fixed ${\tt L}$.
It is possible to re-write it 
 in the form 
\be\label{osidoisoi90}
\prod_{\bm{w}\atop{\tt L}-{\rm fixed}}{D}_{p,s}(\bm{w})=
\bigg(2^{\frac{2\ri(n+2)s}{n}}\ \frac{\Gamma(\frac{1}{2}+p-\ri s)}{\Gamma(\frac{1}{2}+p+\ri s)}
\bigg)^{{\tt par}_2({\tt L})}\  \prod_{\bm{w}\atop{\tt L}-{\rm fixed}}\check{D}_{p,s}(\bm{w}) \ ,
\ee
where
\be\label{iaosidoi43232}
\prod_{\bm{w}\atop{\tt L}-{\rm fixed}}\check{D}_{p,s}(\bm{w}) = 
\prod_{a=0}^{{\tt L}-1}
\Bigg[\frac{\big(\tfrac{1}{2}+a+p-\ri s\big)\,\big(\tfrac{1}{2}+a-p-\ri s\big)}
{\big(\tfrac{1}{2}+a+p+\ri s\big)\,\big(\tfrac{1}{2}+a-p+\ri s\big)}\Bigg]^{{\rm par}_2({\tt L})-d_{a}({\tt L})} \ .
\ee
Here the  generating function for the integers
 $0\le d_{a}({\tt L})\le{\rm par}_2({\tt L})$ reads as
\be\label{Zdef1b}
\chi_{a}({\tt q})\equiv ({\tt q},{\tt q})^{-2}_\infty\  
\sum_{m=0}^\infty (-1)^{m}\ {\tt q}^{m a+\frac{m(m+1)}{2}}=
\sum_{{\tt L}=0}^\infty {d}_{a}({\tt L})\,{\tt q}^{\tt L}
\ee
(for details see Appendix \ref{app2}).
Assuming 
that  all the singularities of ${D}_{p,s}(\bm{w})$  are poles
and that there is no mutual cancellation of poles and zeroes
in the product $\prod_{\bm{w}}\check{D}_{p,s}(\bm{w}) $
 in the r.h.s. of \eqref{osidoisoi90}
one concludes that the poles of  ${D}_{p,s}(\bm{w})$ 
may only be at
$s=\pm\ri\,(p+\frac{1}{2}+a)$ with $a$ an integer.
Provided some further assumptions are made (again see Appendix \ref{app2}), 
an analysis of eq.\,\eqref{osidoisoi90} together
with the analogous formula for $\prod_{\bar{\bm{w}}}{D}_{\bar{p},s}(\bar{\bm{w}})$
leads one to the following conjectures:
\begin{enumerate}[(A)]
\item 
Let  the parameters be such that
$0<\Im m(s)\le\frac{n}{4}$, $\Im m(p)=\Im m(\bar{p})=0$ with
$p+\bar{p}=S^z=0,1,2,\ldots$ and 
 $n$ a generic positive number.
For any sets $\bm{w}$ and $\bar{\bm{w}}$,
the values of $s$ at which
\eqref{pdpd000d} is satisfied must be of the form
$s=\ri\mathfrak{q}_a,\,\ri\bar{\mathfrak{q}}_a$ with
\be\label{isoid8989}
\mathfrak{q}_a=-p-\tfrac{1}{2}-a\,,\qquad
\bar{\mathfrak{q}}_a=-\bar{p}-\tfrac{1}{2}-a\ .
\ee
Here $a$ is an integer such that
\bea\label{oiasio8998}
&&
 -p-\tfrac{n+2}{4}\le a<-\tfrac{1}{2}-p\qquad {\rm for} \qquad s=\ri\mathfrak{q}_a\\[0.2cm]
 &&
 -\bar{p}-\tfrac{n+2}{4}\le a<-\tfrac{1}{2}-\bar{p}\qquad {\rm for} \qquad s=\ri\bar{\mathfrak{q}}_a\ .\nonumber
\eea 
With the last restriction $0<\mathfrak{q}_a,\bar{\mathfrak{q}}_a\le\frac{n}{4}$.

\item 
There are ${\rm par}_2({\tt L})\times{\rm par}_2(\bar{\tt L})$
solutions of the joint system \eqref{sksksk10}.
Let ${\cal N}_a^{(\bar{{\tt L}},{\tt L})}$ and $\bar{\cal N}_a^{(\bar{{\tt L}},{\tt L})}$ denote the number of
them for which  \eqref{pdpd000d}  is obeyed at 
 $s=\ri\mathfrak{q}_a$ and $s=\ri\bar{\mathfrak{q}}_a$, respectively.
Then
\bea\label{99898dsjksj}
{\cal N}_a^{(\bar{{\tt L}},{\tt L})}=
{d}_{S^z+a}(\bar{\tt L})\ {d}_{a}({\tt L})\,,\qquad
\bar{{\cal N}}_a^{(\bar{{\tt L}},{\tt L})}=
{d}_{a}(\bar{{\tt L}})\ {d}_{S^z+a}({\tt L})\,,
\eea 
where the integers $d_a({\tt L})$ 
 are 
defined via \eqref{Zdef1b}.\footnote{\label{ft4}%
The generating function $\chi_a({\tt q})$  obeys the identity
\be
\chi_a({\tt q})+\chi_{-1-a}({\tt q})=({\tt q},{\tt q})_\infty^{-2}
\ ,\nonumber
\ee
which in turn implies that
\be
d_a({\tt L})+d_{-1-a}({\tt L})={\rm par}_2({\tt L})\ .\nonumber
\ee
This makes the definition of  $0\le d_a({\tt L})\le{\rm par}_2({\tt L})$  \eqref{Zdef1b}
applicable for the case of negative $a$. Also there exists  the following
integral representation for $\chi_a({\tt q})$:
$$\chi_a({\tt q})= \oint_{|z|<1}\frac{\rd z}{2\pi\ri}\ \frac{z^{-a-1}}{
(z,{\tt q})_\infty (z^{-1}\,{\tt q},{\tt q})_\infty}\ .
$$
}
\end{enumerate}
Since the condition \eqref{pdpd000d} as well as the algebraic equations
satisfied by $\bm{w}$ and $\bar{\bm{w}}$ are 
invariant w.r.t. the substitutions
 $s\mapsto -s$, $\bm{w}\mapsto-\bm{w}$ and $\bar{\bm{w}}\mapsto-\bar{\bm{w}}$,
all the above follows through essentially
unchanged for $-\frac{n}{4}\le\Im m(s)<0$. Thus,  to take into account  the full domain
 $|\Im m(s)|\le \frac{n}{4}$,
one just needs to replace $s=\ri\mathfrak{q}_a,\ri\bar{\mathfrak{q}}_a$ appearing
in the above two conjectures with $s=\pm\ri\mathfrak{q}_a,\pm\ri\bar{\mathfrak{q}}_a$.
\bigskip

Let ${\cal H}_{N|S^z}^{({\rm disc})}$ be the set of low energy states   such that 
in the $N\to\infty$ limit $b(N)$ tends to a pure imaginary number
 $\pm\ri\mathfrak{q}_a$ or $\pm\ri\bar{\mathfrak{q}}_a$  defined through eqs.\,\eqref{isoid8989} and \eqref{oiasio8998}.
In our investigations of the low energy states of the lattice model we found only the RG trajectories belonging
to ${\cal H}_{N|S^z}^{({\rm disc})}$, or those with 
$\lim_{N\to\infty}\Im m\big(b(N)\big)=0$ which are members of ${\cal H}_{N|S^z}^{({\rm cont})}$.
For instance,  we performed a numerical check 
by  explicitly constructing the RG trajectories
for all the low energy states in the lattice model with $N=22$ sites 
in the sector $S^z=1$, ${\tt w}=0$ and ${\tt L}+\bar{\tt L}=1$.
For those states having both $\Im m\big(b(N)\big)\ge 0$ and $\Re e\big(b(N)\big)\ge 0$,
a plot of $b$ as a function of $N$ is provided in fig.\,\ref{fig999}.
A systematic exposition of our numerical work   is given in sec.\ref{sec18} below.
This way we come to expect
\begin{enumerate}[(I)]
\setcounter{enumi}{1}
\item For any  low energy Bethe state $\bm{\Psi}_N$ belonging to the sector with $S^z=0,1,2,3\ldots$
either $\lim_{N\to\infty}\Im m\big(b(N)\big)=0$ and $\bm{\Psi}_N\in {\cal H}_{N|S^z}^{({\rm cont})}$,
or else $\lim_{N\to\infty}\Im m\big(b(N)\big)=s$ with $s=\pm\ri\mathfrak{q}_a,\,\pm\ri\bar{\mathfrak{q}}_a$
\eqref{isoid8989},\,\eqref{oiasio8998}.
In the latter case $\bm{\Psi}_N\in{\cal H}_{N|S^z}^{({\rm disc})}$.
\end{enumerate}

\bigskip

It should be kept in mind that eq.\,\eqref{pdpd000d} is 
a necessary  condition for the
existence of an RG trajectory  belonging to the
set ${\cal H}_{N|S^z}^{({\rm disc})}$.
Establishing that such a trajectory actually 
exists can not be done based on the formal analysis of 
this equation alone.
Moreover, one may imagine that there could be multiple RG trajectories,
labeled by the identical sets $\bm{w}$ and $\bar{\bm{w}}$,
whose $b(N)$ tends to the same pure imaginary value of $s$.
Nevertheless in our numerical work we have always 
observed that  for every  $\bm{w}$, $\bar{\bm{w}}$
and $s=\pm\ri\mathfrak{q}_a,\pm\ri\bar{\mathfrak{q}}_a$ at which \eqref{pdpd000d} is obeyed,
there exists one and only one RG trajectory  with $\lim_{N\to\infty}b(N)=s$.\footnote{%
Recall that in constructing an RG trajectory $\bm{\Psi}_N$ with pure imaginary $s=\lim_{N\to\infty}b(N)$ 
the parity of $\frac{N}{2}-S^z$ must be kept fixed. 
There exist the RG trajectories with even $\frac{N}{2}-S^z$ and
odd $\frac{N}{2}-S^z$ which have the same value of $s$. 
However since for fixed parity of  $\frac{N}{2}-S^z$ only one of these trajectories
is present, we do not count this as a degeneracy.
}
For instance, the right panel of fig.\,\ref{BAplot1}
depicts the typical pattern of Bethe roots 
for $\bm{\Psi}_N$ with $\lim_{N\to\infty}b(N)=-\ri\bar{\mathfrak{q}}_2$.
Among others, fig.\,\ref{fig999} presents numerical data for
an RG trajectory with ${\tt L}=0$, $\bar{\tt L}=1$ and for which
$b(N)\to\ri\mathfrak{q}_{-1}$. 
This leads us to the conjecture:
\begin{enumerate}[(I)]
\setcounter{enumi}{2}
\item 
For sufficiently large $N$ and fixed  ${\tt L}$,
$\bar{\tt L}$, $p$, $\bar{p}$, the number of
low energy Bethe states $\bm{\Psi}_N$  
such that $\lim_{N\to\infty}b(N)=\pm\ri\mathfrak{q}_a$
is given by ${\cal N}_a^{(\bar{{\tt L}},{\tt L})}={d}_{S^z+a}(\bar{\tt L})\ {d}_{a}({\tt L})$,
where ${d}_{a}({\tt L})$ are defined through eq.\,\eqref{Zdef1b}.
Similarly, there are  $\bar{\cal N}_a^{(\bar{{\tt L}},{\tt L})}={d}_{a}(\bar{{\tt L}})\ {d}_{S^z+a}({\tt L})$
 trajectories with $\lim_{N\to\infty}b(N)=\pm\ri\bar{\mathfrak{q}}_a$.
\end{enumerate}
\bigskip

It should be pointed out that the integers ${\cal N}_a^{({\bar{\tt L}},{\tt L})}$ satisfy the condition
\be
{\cal N}_a^{(0,0)}=\begin{cases} 1 & {\rm for}\quad a\ge 0 \\[0.2cm]
0 & {\rm for} \quad a<0
\end{cases}
\ee
and in
describing the conformal towers 
corresponding to pure imaginary  $s$,
it is necessary to distinguish the cases
 $a\ge 0$ and $a<0$. For this purpose we denote by
${\cal H}_{N|S^z}^{({\rm disc},+)}$ 
the set of RG trajectories with $b(N)\to\pm\ri\mathfrak{q}_a,\pm\ri\bar{\mathfrak{q}}_a$
and $a\ge 0$, while  ${\cal H}_{N|S^z}^{({\rm disc},-)}$ is the set of trajectories 
labeled by $s=\pm\ri\mathfrak{q}_a,\pm\ri\bar{\mathfrak{q}}_a$ with  $a<0$.
As before  the sets 
${\cal H}_{N|S^z}^{({\rm disc},\pm)}$ may be equipped with the structure of a linear space. 
Their scaling limits,
\be
{\cal H}^{({\rm disc},\pm)}_{S^z}={\rm s}\!\!\!\lim_{N\to\infty}{\cal H}_{N|S^z}^{({\rm disc},\pm)}\,,
\ee
are decomposed into
a direct sum over finite dimensional subspaces, similar to eq.\,\eqref{iaosidoi192009aaaaa}, but with the 
direct integral replaced by a sum over the admissible values of the RG invariant 
$s=\pm\ri\mathfrak{q}_a,\,\pm\ri\bar{\mathfrak{q}}_a$.
We'll postpone 
a detailed account of this decomposition to sec.\,\ref{sec12}
as it requires a discussion of the scaling limit 
of the eigenvalues of the lattice operators $\mathbb{A}_\pm(\zeta)$.

\section{Scaling limit of ${A}_\pm(\zeta)$\label{sec11}}

In ref.\cite{Bazhanov:2019xvy} it was proposed that the scaling limit of the eigenvalues  of 
$\mathbb{A}_\pm(\zeta)$ for the low energy Bethe states is given in terms of the  
connection coefficients of a certain ODE.
Let $A_\pm(\zeta)$ be the eigenvalue corresponding to $\bm{\Psi}_N$, labeled by
the full set of RG invariants
$p$, $\bar{p}$,  $\bm{w}$, $\bar{\bm{w}}$ and $s$.
Then 
\bea\label{as56d1a}
{\rm s}\!\!\!\!\lim_{N\to \infty\atop b(N)\to s}
G^{(N/2)}\big(-\mu^2\,|\,\tfrac{2}{n+2}\big)\, 
 A_\pm\big(\, \ri\, \big(N/(2N_0)\big)^{-\frac{n}{n+2}}\ \mu \big)={D}_\pm(\mu\,|\,\bm{w},p,s)\ ,
\eea
where  $G^{(N)}(E\,|\,g)$  and $N_0$ are  given in eqs.\,\eqref{saysysa} and  \eqref{N0altdef1}, respectively.
The functions  $D_\pm(\mu\,|\,\bm{w},p,s)$ coincide with
 the connection coefficients for the linear differential equation:
\bea\label{aisausau}
\Bigg[\,-\frac{\rd ^2}{{\rd z}^2}+\frac{p^2-\frac{1}{4}}{z^2}+\frac{2\ri  s}{z}+1+
\sum_{a=1}^{\tt L}\bigg(\frac{2}{(z-w_a)^2}+\frac{n}{z(z-w_a)}\bigg)+\mu^{-2-n}\ z^n\,
\Bigg]\, \Phi=0\ .
\eea
The fact that the set $\bm{w}\equiv\{w_a\}_{a=1}^{\tt L}$  satisfies the algebraic system \eqref{sksksk1}
 ensures that any solution of this ODE
 is monodromy free in the vicinity of each apparent singularity at $z=w_a$. 
To specify the connection coefficients introduce the two basis solutions,
$\Phi_{ \pm p}(z)$, of  \eqref{aisausau} 
such that
\be\label{psidef1}
\Phi_{ \pm p}(z)\to 
\frac{1}{\sqrt{\pi}}\
(n+2)^{\mp\frac{2p}{n+2}-\frac{1}{2}}\ \mu^{\mp p-\frac{1}{2}}\,\Gamma(\mp\tfrac{2p}{n+2})\ \
z^{\frac{1}{2}\pm p} \ \ \ \ {\rm as} \ \ z\to 0  \qquad \big(0<\Re e(p)<1\big)\ .
\ee
For large $z$ the term $\mu^{-2-n}\,z^n$ in \eqref{aisausau} 
becomes dominant and one can define another solution through the $z\to+\infty$  asymptotic
(to be compared with \eqref{chilargezeq1})
\be
\Xi(z)\asymp  \Big(\frac{z}{\mu}\Big)^{-\frac{n}{4}}\,\exp\bigg[-\frac{2}{n+2}\  \Big(\frac{z}{\mu}\Big)^{\frac{n}{2}+1}\, 
{}_2F_1\big( -\tfrac{1}{2}, -\tfrac{n+2}{2n},  
\tfrac{n-2}{2n}\,\big|-\mu^{n+2}\, z^{-n}  \big)\ +\ o(1)\,\bigg]\,.
\ee
Here we make the technical assumption that $\mu>0$ and $n\ne\frac{2}{2k-1}$
with $k=1,2,\ldots\ $. 
The connection coefficients $D_\pm(\mu\,|\,\bm{w},p,s) $ are given by
\be\label{Ddef1a}
D_\pm(\mu\,|\,\bm{w},p,s)=\mp\mu\,\sin(\tfrac{2\pi p}{n+2}) \, W[\,\Phi_{\pm p},\Xi\,]
\ee
with $W[\,\Phi_{\pm p},\Xi\,]=\Xi\partial_z\Phi_{\pm p}-\Phi_{\pm p}\partial_z\Xi$ being the Wronskian.
The overall factor in \eqref{Ddef1a} has been chosen so that,
for generic values of $p$,
\be\label{iaosid91209}
D_\pm(0\,|\,\bm{w},p,s)= 1\ .
\ee
It can be shown that when  $n>0$ and  $p$ is a generic complex number,
$D_\pm(\mu\,|\,\bm{w},p,s)$
are entire functions of $\mu$.
\bigskip

Unfortunately the formula  \eqref{as56d1a}, where the r.h.s. coincides with the 
connection coefficient \eqref{Ddef1a},   
at the current moment remains a conjecture.
Below we'll discuss some possible ways of checking this relation.
However, before doing so
let us explain at the formal level the link  between the ODE
\eqref{aisausau},\,\eqref{sksksk1}, and the one 
appearing in the homogeneous case \eqref{jasys}-\eqref{jsaysssysa}.
When ${\tt L}$ is even and $s=0$, the algebraic system \eqref{sksksk1} admits solutions 
such that
$
w_{{\tt L}+1-a}=-w_{a}
$.
Then the substitution
\be
 s\mapsto 0\,,\qquad 
w_a=-w_{{\tt L}+1-a}\mapsto\,\ri\,\sqrt{\tfrac{v_a}{\alpha(\alpha+1)}}\,,\qquad 
n\mapsto-\tfrac{2\alpha}{\alpha+1}\,,\qquad
p^2\mapsto \tfrac{4}{\alpha+1}\,P^2
\ee
 brings \eqref{sksksk1} to the form \eqref{jsaysssysa}.
With these specializations  and
upon the change of variables
$
\Phi(z)\mapsto x^{\frac{\alpha}{2}}\,{\Phi}(x)$, 
$z\mapsto\frac{x^{\alpha+1}}{\alpha+1}$,
the ODE \eqref{aisausau}
 is transformed to the one given by \eqref{jasys},\,\eqref{MonstP1}
with 
$E= (1+\alpha)^{\frac{2\alpha}{1+\alpha}}\,\mu^{-\frac{2}{1+\alpha}}$.
\bigskip

Expanding both sides of the relation \eqref{as56d1a} in a Taylor series 
in $\mu$ leads to an infinite set of sum rules for the
Bethe roots. In particular, the series expansion of
$A_+(\zeta)$ involves the finite sums
\be\label{hjN1a}
h_{j}^{(N)}=j^{-1}\sum_{m=1}^{\frac{N}{2}-S^z} (\zeta_m)^{-j}
\ee
computed on the corresponding RG trajectory $\bm{\Psi}_N$.
On the other hand, in view of eq.\,\eqref{iaosid91209}
and that the connection coefficients are entire functions of $\mu$,
one has
\be\label{logaseries1abbc}
\log{D}_\pm(\mu\,|\,\bm{w},p,s)=-\sum_{j=1}^\infty 
J_j^{(\pm)}(\bm{w},p,s)\,\lambda^{j}\,.
\ee  
Here, for future convenience, we swap $\mu$ for the parameter  $\lambda$
defined via the relation
\bea\label{mulambda1a}
\mu=-\ri\ (n+2)^{-\frac{2}{n+2}}\ \Gamma^2\big(-\tfrac{1}{n+2}\big)\ \lambda\ .
\eea
Through the perturbation theory for the differential equation \eqref{aisausau}, one can in principle
derive an explicit expression for the expansion coefficients $J_j^{(\pm)}(\bm{w},p,s)$.
The computations turn out to be quite cumbersome, however, 
for the case ${\tt L}=0$ when there are no apparent singularities,
the first two $J^{({\rm vac})}_{j}(p,s)\equiv J_j^{(+)}(\bm{w},p,s)\big|_{{\tt L}=0}$
are given by
\bea\label{Jvaceig1a}
J^{({\rm vac})}_{1}(p,s)&=&-2 s\ f_1\big(\tfrac{p}{n+2},\tfrac{1}{n+2}\big)\\[0.2cm]
J^{({\rm vac})}_{2}(p,s)&=&2^{\frac{4}{n+2}}\ \frac{\pi \Gamma^2(-\frac{1}{n+2})}
{\Gamma^2(\frac{1}{2}-\frac{1}{n+2})}
\ f_1\big(\tfrac{p}{n+2},\tfrac{2}{n+2}\big)+4 s^2\, f_2\big(\tfrac{p}{n+2},\tfrac{1}{n+2}\big)\ ,\nonumber
\eea
where the functions $f_j$ are defined in eqs.\,\eqref{apsodpoa10a},\,\eqref{jassusau}.
The analogous expression for $J_{1,2}^{(-)}(\bm{w},p,s)\big|_{{\tt L}=0}$
may be obtained from \eqref{Jvaceig1a} through the substitution $p\to-p$.
Note that for general $j=1,2,\ldots$ the coefficients
 $J^{({\rm vac})}_{j}(p,s)$ turn out to be polynomials in $s$ of order $j$.
A quick inspection of \eqref{Jvaceig1a} as well as the formula \eqref{apsodpoa10a} for $f_1$
shows that $J_2^{({\rm vac})}(p,s)$ contains a simple pole when
 $\frac{2}{n+2}$ is equal to $\frac{1}{2}$.
In fact, similar to the homogeneous case, the coefficients $J_{2k}^{(\pm)}(\bm{w},p,s)$
with $k=1,2,3,\ldots$
possess a simple pole if $\frac{2}{n+2}=1-\frac{1}{2k}$. In this case
one can define the regularized coefficient
 through the subtraction
\be
{ J}_{2k}^{(\pm,{\rm reg})}=\lim_{\frac{2}{n+2}\to 1-\frac{1}{2k}} \bigg[\, { J}_{2 k}^{(\pm)}+
\ \bigg(\frac{\Gamma^2(-\frac{1}{n+2})\,\Gamma(\frac{1}{2k})}
{(n+2)\,\Gamma(1-\frac{2}{n+2})}\bigg)^{2k}\ 
\frac{\Gamma(-\frac{1}{2}+k)\,}{2\sqrt{\pi}\, k\,\Gamma(1+k)}\
\  \frac{1}{\frac{2}{n+2}-1+\tfrac{1}{2k}}\,\bigg]\ .
\ee

\bigskip

Formulae \eqref{as56d1a} and \eqref{logaseries1abbc}  imply the infinite  set of relations
for $h_{j}^{(N)}$ \eqref{hjN1a}:
\bea\label{poaspo1ddda}
{\rm s}\!\!\!\!\lim_{N\to \infty\atop b(N)\to s}
N^{-\frac{j n}{n+2}}\ h_{j}^{(N)}
=
 \Big[(2N_0)^{\frac{n}{n+2}}\,(n+2)^{-\frac{2}{n+2}}\ \Gamma^2\big(-\tfrac{1}{n+2}\big)\Big]^{-j}
 \ J_{j}(\bm{w},p,s)\ ,
 \eea
 which hold true for  any odd $j=1,3,\ldots$ and for all even $j>1+\frac{2}{n}$.
 When $j$ is even and $j<1+\frac{2}{n}$ \eqref{poaspo1ddda} should be replaced by
 \be\label{poaspo1da2}
{\rm s}\!\!\!\!\lim_{N\to \infty\atop b(N)\to s}
N^{-\frac{j n}{n+2}}\ 
\bigg[\,h_{j}^{(N)}+ \frac{(-1)^{\frac{j}{2}+1}\, N}{2j\cos(\frac{\pi j}{n+2})}
\ \bigg]=
 \Big[(2N_0)^{\frac{n}{n+2}}(n+2)^{-\frac{2}{n+2}}\ \Gamma^2\big(-\tfrac{1}{n+2}\big)\Big]^{-j}
 \ J_{j}(\bm{w},p,s)\ .
 \ee
 Finally in the case $n=\frac{2}{2k-1}$
 \be\label{poaspo1da}
{\rm s}\!\!\!\!\!\lim_{N\to \infty\atop b(N)\to s}
\bigg[N^{-1}\,h_{2k}^{(N)}-\frac{1 }{2\pi k}\, \log(N B_k/2)
 \bigg]=
(2N_0)^{-1} \Big[(n+2)^{-\frac{2}{n+2}}\ \Gamma^2\big(-\tfrac{1}{n+2}\big)\Big]^{-2k}
\  J_{2k}(\bm{w},p,s)\,.
 \ee
Here $B_k$
are the  constants that enter in to the function $G^{(N)}(E\,|\,g)$ \eqref{saysysa},
which appears in the relation \eqref{as56d1a}. 
The first two of them are given by eqs.\,\eqref{iaosi89891} and \eqref{iaosi89892}.
\bigskip

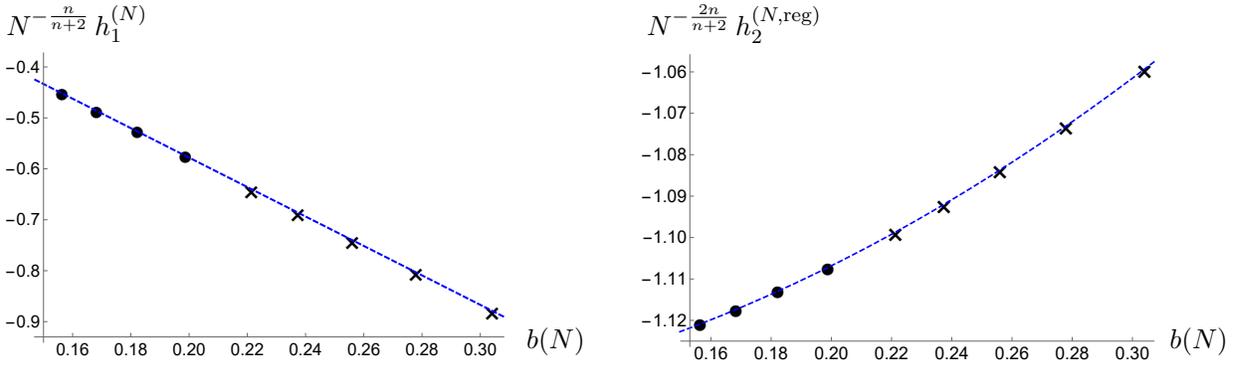
\begin{figure}
\centering
\scalebox{0.95}{
\begin{tikzpicture}
\node at (0,0.05) {\includegraphics[width=7cm]{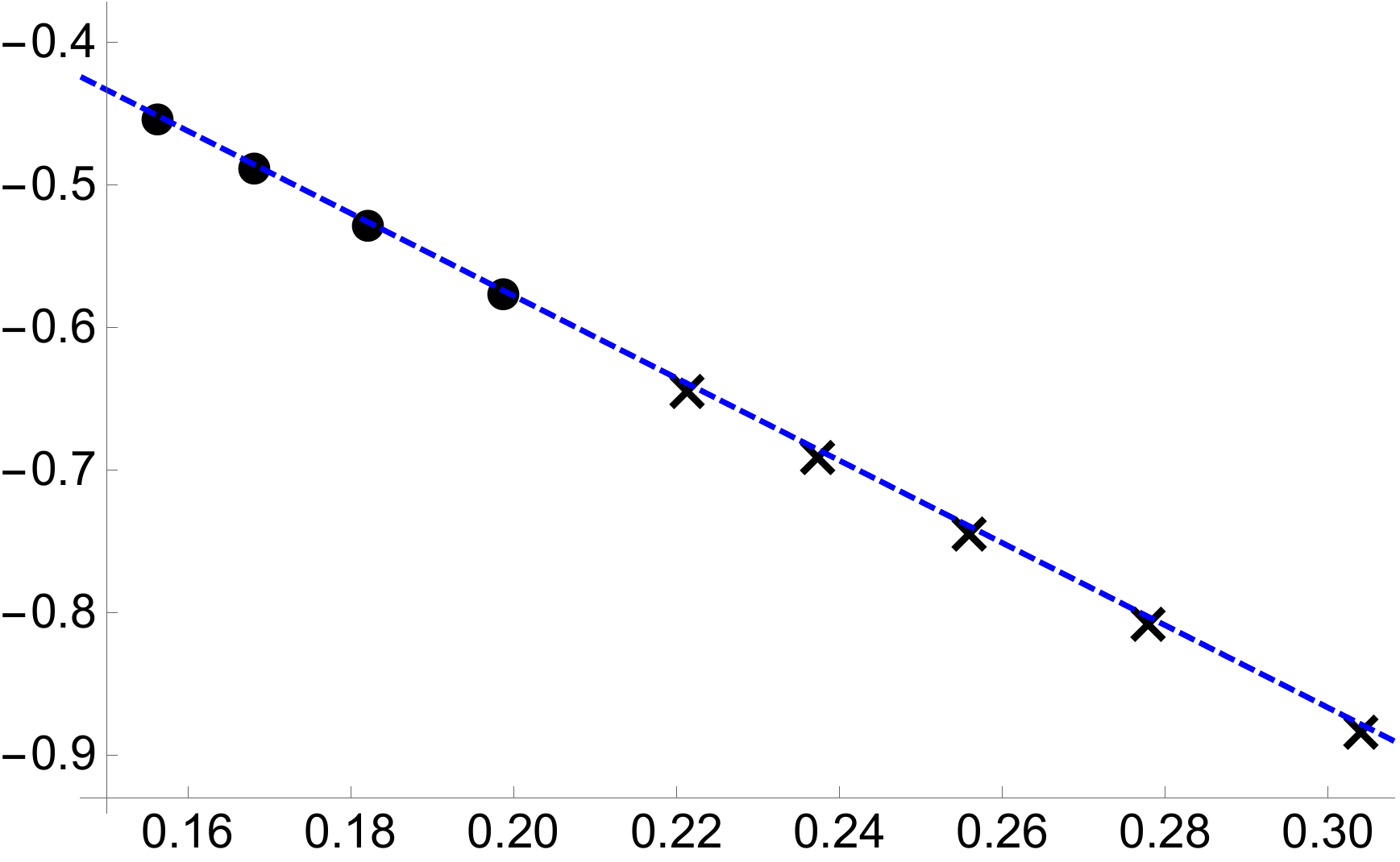}};
\node at (9,0) {\includegraphics[width=7.2cm]{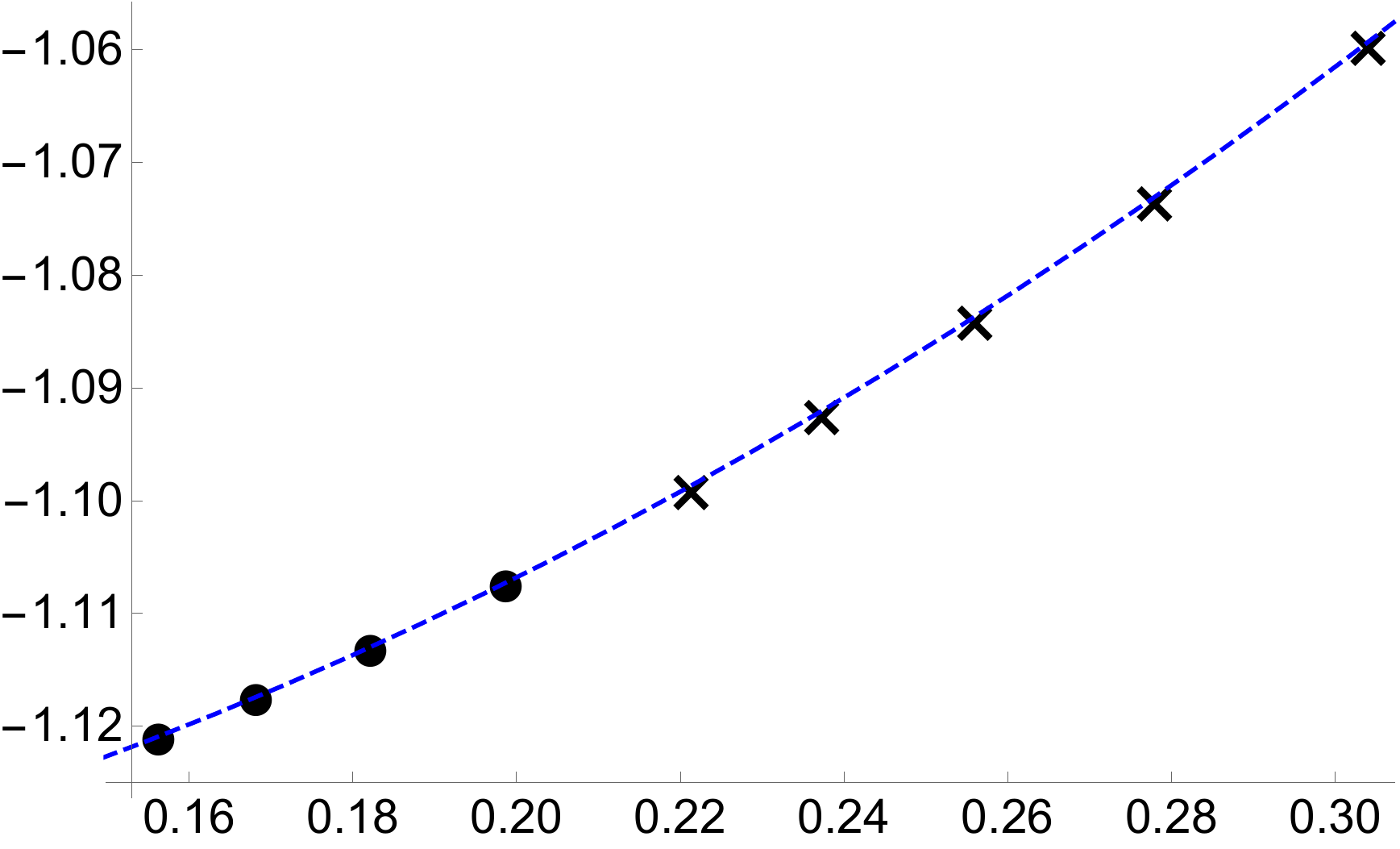}};
\node at (13.2,-1.85) {$b(N)$};
\node at (4.2,-1.85) {$b(N)$};
\node at (-2.5,2.6) {$N^{-\frac{n}{n+2}}\,h_1^{(N)}$};
\node at (6.7,2.6) {$N^{-\frac{2n}{n+2}}\,h_2^{(N,{\rm reg})}$};
\end{tikzpicture}
}
\caption{\small The sums $h_1^{(N)}$ and $h_2^{(N,{\rm reg})}\equiv
h_{2}^{(N)}+ \frac{N}{4\cos(\frac{2\pi}{n+2})}$ were computed by solving the
Bethe ansatz equations with the value of the parameters $n=\frac{3}{2}$, ${\tt k}=\frac{1}{10}$
and $S^z=0$.
The solution sets were taken to be the 
ones corresponding to the primary  Bethe states, which were
discussed at the beginning of sec.\,\ref{sec31}.
For the black dots,
the integer ${\tt m}=M_--M_+$  was set to
${\tt m}=2$, while  $N=100,200,400,800$. For the crosses
${\tt m}=3$ and $N=102,202,402,802,1602$.
The numerical data is plotted versus $b(N)$. The latter
 was computed from the Bethe roots  via eqs.\,\eqref{Beq1} and \eqref{poapso1a}.
The dashed blue lines come from the predictions \eqref{poaspo1ddda},\,\eqref{poaspo1da2}
with the vacuum eigenvalues $J^{({\rm vac})}_{1,2}(p,s)$ calculated through eq.\,\eqref{Jvaceig1a}.
Note that the relative error between the numerical data and the analytical formula is of the order
of $10^{-4}-10^{-6}$.
\label{fig2}}
\end{figure}

We performed extensive numerical checks of \eqref{poaspo1ddda}-\eqref{poaspo1da}   
considering both the Bethe
 states belonging to the space ${\cal H}_{N|S^z}^{({\rm cont})}$
and ${\cal H}_{N|S^z}^{({\rm disc},\pm)}$,
where the RG invariant $s$ is real and pure imaginary, respectively.
Some results concerning the primary Bethe states for which ${\tt L}=\bar{\tt L}=0$
are shown in fig.\,\ref{fig2}.
\bigskip

 It is worth mentioning that  formula
\eqref{as56d1a}  recovers the universal properties of 
the eigenvalues $A_\pm(\zeta)$ in the vicinity of $\zeta=0$.
By organizing the scaling limit differently one can describe
the universal behaviour of $A_\pm(\zeta)$ near the point $\zeta=\infty$.
The latter involves the connection coefficients for an ODE
similar to \eqref{aisausau}, but with $p$ and $\{w_a\}_{a=1}^{{\tt L}}$ replaced by their
barred counterparts $\bar{p}$ and $\{\bar{w}_a\}_{a=1}^{\bar{{\tt L}}}$, respectively.
 All the above could, of course, be repeated for this case as well.
In the scaling limit the low energy Bethe states take the form
 $\bar{\bm{\psi}}_{\bar{p},s}(\bar{\bm{w}})\otimes\bm{\psi}_{{p},s}({\bm{w}})$,
where the chiral states $\bar{\bm{\psi}}_{\bar{p},s}(\bar{\bm{w}})$ and
$\bm{\psi}_{{p},s}({\bm{w}})$
are specified by the connection coefficients 
${D}_\pm(\mu\,|\,\bar{\bm{w}},\bar{p},s)$ and ${D}_\pm(\mu\,|\,\bm{w},p,s)$,
respectively.

\bigskip

Apart from the sum rules there is another way of
checking the key relation \eqref{as56d1a} (and its barred counterpart), which turns out to be
more convenient for the case of the non-primary Bethe states.
It involves the coefficients, which occur in the 
 large $\mu$ asymptotic expansion of $D_\pm(\mu\,|\,\bm{w},p,s)$.
For
 $D_+\equiv D_+(\mu\,|\,\bm{w},p,s)$, the latter takes the form
\be\label{Apeq1}
D_+\,\asymp\,  { \mathfrak C}_{p,s}^{(\pm)}(\bm{w})\, \big(\pm \mu\big)^{\pm\frac{\ri (n+2) s}{n}-p}\,\exp\bigg(\, 
\frac{N_0}{\cos(\frac{\pi}{n})}\ 
\big(\pm \mu\big)^{\frac{n+2}{n}}+o(1)\,\bigg) \ \ \ \, {\rm for}\ \ \ \  \Re e(\pm \mu)>0\ .
 \ee
At the special values $n=\frac{2}{2k-1}$ with $k=1,2,\ldots\ $:
\be\label{Apeq1b}
D_+\,\asymp\,{ \mathfrak C}_{p,s}^{(\pm)}(\bm{w})\, \big(\pm \mu\big)^{\pm 2\ri k s-p}\, \exp\bigg(\, 
\frac{(-1)^k\, \Gamma(\frac{1}{2}+k)}{\sqrt{\pi}\,\Gamma(1+k)}\ \mu^{2k}\, \big(\log(\pm \mu)+\tfrac{1}{2}\,c_k\big)+o(1)\,\bigg)
\ee
for $\Re e(\pm\mu)>0$ and $c_k$ are the same as in eq.\,\eqref{ckdef1a}.
Note that the  large $\mu$ behaviour of $D_-(\mu\,|\,\bm{w},p,s)$ can be obtained from
that of $D_+$ by means of the substitution $p\mapsto -p$.
It is possible to compute the  coefficients ${ \mathfrak C}_{p,s}^{(\pm)}(\bm{w})$ 
explicitly through an  analysis of the ODE \eqref{aisausau}. In particular, when there are no apparent singularities,
\be\label{iaosdi012891}
\mathfrak{C}_{p,s}^{(0,\pm)}= 
\sqrt{\frac{2\pi}{n+2}}\ \ \ 2^{-p\pm\frac{\ri(n+2)s}{n}}\ 
(n+2)^{-\frac{2p}{n+2}}\ 
\frac{\Gamma(1+2p)}{\Gamma(1+\frac{2p}{n+2})\,\Gamma(\frac{1}{2}+p\pm\ri s)}\ \,.
\ee
For a general set $\bm{w}$ satisfying the algebraic system \eqref{sksksk10}, 
the expression for ${ \mathfrak C}_{p,s}^{(\pm)}(\bm{w})$  is provided by eq.\,(3.7)
in ref.\cite{Kotousov:2019nvt}.
\bigskip

The coefficients ${ \mathfrak C}_{p,s}^{(\pm)}(\bm{w})$,  
and the similarly defined ${ \mathfrak C}_{\bar{p},s}^{(\pm)}(\bar{\bm{w}})$,  
occur in the  large $N$ asymptotic
formulae for the products over the Bethe roots which resemble the relation
\eqref{prodeqA} for the homogeneous case:
\bea\label{aiosdi1209112}
\prod_{m=1}^M q\,\big(\zeta_m\mp\ri q^{-1}\big)\,\big(\zeta_m^{-1}\mp\ri q^{-1}\big)&\asymp&
\re^{\pm\frac{2\pi s}{n}}\ 
\mathfrak{C}_{\bar{p},s}^{(\pm)}(\bar{\bm{w}})\,
\mathfrak{C}_{p,s}^{(\pm)}(\bm{w}) \  \bigg(\frac{N}{2N_0}\bigg)^{-\frac{n(\bar{p}+p)}{n+2}\pm2\ri s}\,\bigg|_{s=b(N)}
\nonumber \\[0.2cm]
&\times&\, \Big(\frac{4n}{n+2}\Big)^{N/2}\ \big(1+o(1)\big)
\eea
as well as
\bea\label{oasodi1121}
\prod_{m=1}^M\,\zeta_m^2\,&\asymp&\,
\frac{\mathfrak{C}_{\bar{p},s}^{(+)}(\bar{\bm{w}})\,\mathfrak{C}_{\bar{p},s}^{(-)}(\bar{\bm{w}})}
{\mathfrak{C}_{{p},s}^{(+)}({\bm{w}})\,\mathfrak{C}_{{p},s}^{(-)}({\bm{w}})}\,\bigg|_{s=b(N)}\ 
\ \bigg(\frac{N}{2N_0}\bigg)^{\frac{2n(p-\bar{p})}{n+2} }\ \big(1+o(1)\big)\,.
\eea
Note that 
in these formulae we substitute
the RG invariant $s$  by the ``running coupling'' $b(N)=\frac{n}{4\pi}\,\log(B)$
with $B$ being the eigenvalue of the quasi shift operator \eqref{Beq1}.
This significantly improves their accuracy.

\bigskip

There are many  consequences of \eqref{aiosdi1209112} and  \eqref{oasodi1121}.
An important one follows from taking the ratio of
the asymptotic relation \eqref{aiosdi1209112}
corresponding to  ``$+$'' with that corresponding to ``$-$''. 
Keeping in mind  that the eigenvalues of the quasi-shift operator are given by \eqref{Beq1}, 
it is easy to see that the  l.h.s. of the ratio
coincides with $(-1)^{\frac{N}{2}-S^z}\,B$.
Then since  $B=\re^{\frac{4\pi}{n}b(N)}$ one finds
\bea
\bigg(\frac{N}{2N_0}\bigg)^{4\ri s}\ 
\frac{\mathfrak{C}_{\bar{p},s}^{(+)}(\bar{\bm{w}})\,
\mathfrak{C}_{p,s}^{(+)}(\bm{w})}
{\mathfrak{C}_{\bar{p},s}^{(-)}(\bar{\bm{w}})\,
\mathfrak{C}_{p,s}^{(-)}(\bm{w})}\,\bigg|_{s=b(N)}&\asymp& (-1)^{\frac{N}{2}-S^z}\big(1+o(1)\big)\ .
\eea
Upon the identification
\be\label{ioasd2989823}
{ D}_{p,s}({\boldsymbol  w})=\frac{{\mathfrak C}_{p,s}^{(+)}(\bm{w})}{{\mathfrak C}_{p,s}^{(-)}(\bm{w})}\ ,
\qquad\qquad
{ D}_{\bar{p},s}(\bar{{\boldsymbol  w}})=\frac{{\mathfrak C}_{\bar{p},s}^{(+)}(\bar{\bm{w}})}
{{\mathfrak C}_{\bar{p},s}^{(-)}(\bar{\bm{w}})}
\ee
this is nothing but the asymptotic formula \eqref{quantC1},
where $\re^{\frac{\ri}{2}\delta}={ D}_{p,s}({\boldsymbol  w})\,{ D}_{\bar{p},s}(\bar{{\boldsymbol  w}})$.
\bigskip

Another  important outcome
 of   the relations \eqref{aiosdi1209112} and \eqref{oasodi1121}
 involves the products
\be\label{iaosido9898}
{ R}_{p,s}({\boldsymbol  w})=
{\mathfrak C}_{p,s}^{(+)}(\bm{w})\,{\mathfrak C}_{p,s}^{(-)}(\bm{w})\ ,\qquad\qquad
{ R}_{\bar{p},s}(\bar{{\boldsymbol  w}})=
{\mathfrak C}_{\bar{p},s}^{(+)}(\bar{\bm{w}})\,{\mathfrak C}_{\bar{p},s}^{(-)}(\bar{\bm{w}})\ .
\ee
Namely, 
\bea\label{eq3331}
\!\!\!\!\!\!\!\!\!\!\prod_{m=1}^M \big(\zeta^{-2}_m+q^2\big) \big(\zeta^{-2}_m+q^{-2}\big)
\asymp
\big({ R}_{p,s}({\boldsymbol  w})\big)^2\,\Big|_{s=b(N)}
\ \bigg(\frac{N}{2N_0}\bigg)^{-\frac{4 n p}{n+2} }\  \bigg(\frac{4n}{n+2}\bigg)^{N}\,\big(1+o(1)\big)\ \,
\eea
and
\bea\label{eq3332}
\!\!\!\!\!\!\!\!\!\!\prod_{m=1}^M \big(\zeta_m^{+2}+q^2\big) \big(\zeta_m^{+2}+q^{-2}\big)
\asymp
\big({  R}_{{\bar p},s}({\bar {\boldsymbol  w}})\big)^2\,\Big|_{s=b(N)}
\ \bigg(\frac{N}{2N_0}\bigg)^{-\frac{4 n {\bar p}}{n+2} }\  \bigg(\frac{4n}{n+2}\bigg)^{N}\,\big(1+o(1)\big)\ .
\eea
The advantage of \eqref{eq3331} compared with the sum rules 
\eqref{poaspo1ddda}-\eqref{poaspo1da}  is that, contrary to 
the expansion coefficients $J_j^{(\pm)}(\bm{w},p,s)$ from the Taylor series 
\eqref{logaseries1abbc}, there exists a closed expression for
${ R}_{p,s}({\boldsymbol  w})$ in terms of the set $\bm{w}$.
In the case with no apparent singularities,
eq.\,\eqref{iaosdi012891} implies that
\be\label{opsod898s}
{ R}_{p,s}^{(0)}= 2^{1+2p}\,(n+2)^{-1-\frac{4p}{n+2}}\ 
\bigg[\frac{\Gamma(1+p)}{\Gamma(1+\frac{2p}{n+2})}\bigg]^2\ 
\frac{\Gamma^2(\frac{1}{2}+p)}{\Gamma(\frac{1}{2}+p+\ri s)\,\Gamma(\frac{1}{2}+p-\ri s)}\ .
\ee
For  ${\tt L}\ge 0$,
\be\label{oasid89129812}
{ R}_{p,s}({\boldsymbol  w})={ R}_{p,s}^{(0)}\,\check{R}_{p,s}(\bm{w})
\ee
and eq.\,\eqref{aosid981212} in  Appendix  \ref{app2} gives 
$\check{R}_{p,s}(\bm{w})$ in terms of $p$, $s$ and $\bm{w}$
(the latter, up to notation, coincides with (3.11) from ref.\cite{Kotousov:2019nvt}).

\bigskip

\begin{figure}
\centering
\scalebox{0.9}{
\begin{tikzpicture}
\node at (0,0) {\includegraphics[width=10cm]{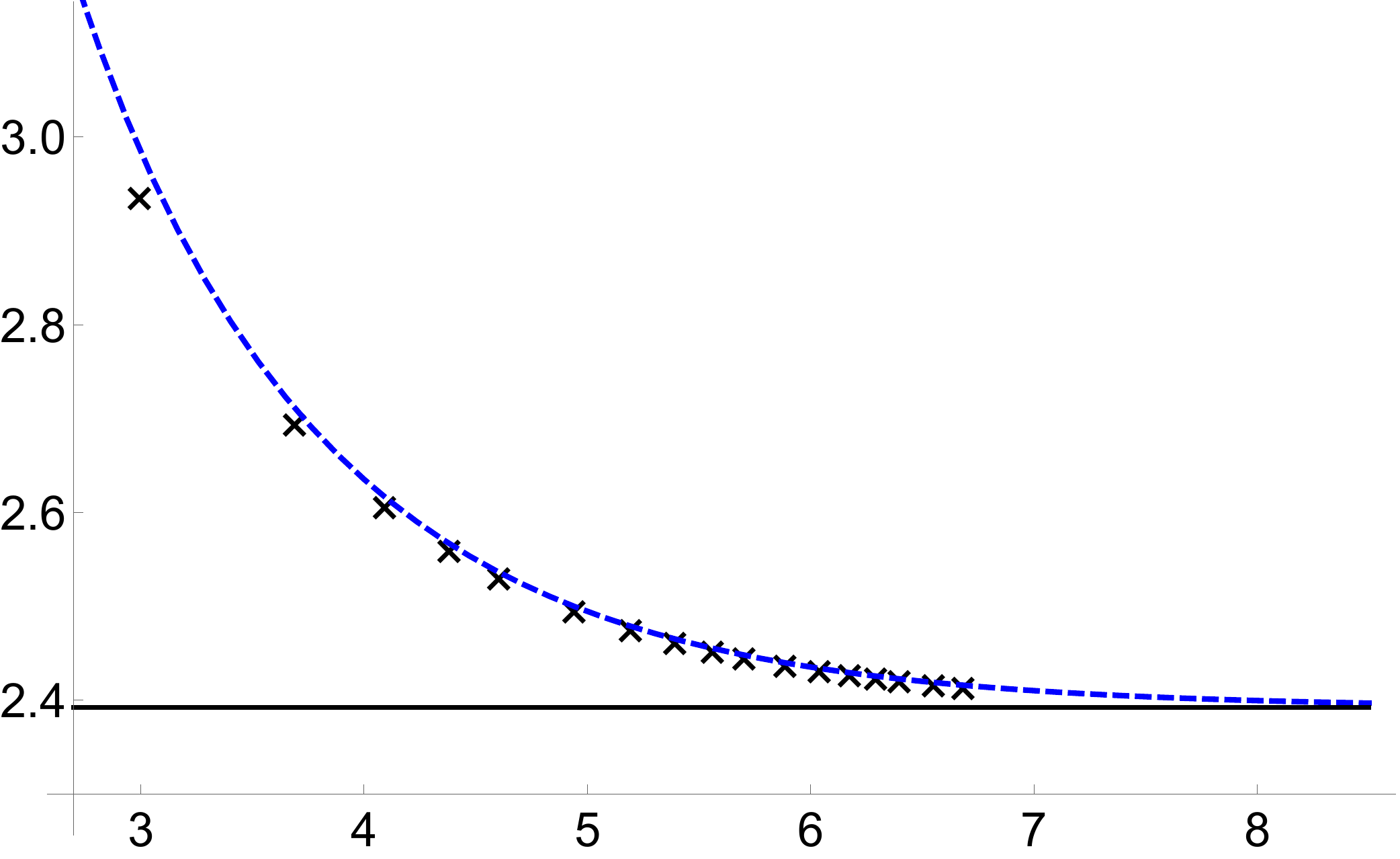}};
\node at (5.8,-2.65) {$\log(N)$};
\end{tikzpicture}
}
\caption{\label{sec11fig1}\small
Numerical data for an RG trajectory $\bm{\Psi}_N$ with ${\tt L}=\bar{\tt L}=S^z={\tt w}=0$
and labeled by pure imaginary $s=\ri\,(-p-\frac{1}{2})=\frac{\ri}{4}$   is used to illustrate
the asymptotic formula \eqref{aoisid981921}. The  parameters have been set to be
$n=3$, ${\tt k}=-\frac{3}{10}$.
Depicted by the crosses is the r.h.s. of  \eqref{aoisid981921}, calculated from the  solution of the
Bethe ansatz equations corresponding to  $\bm{\Psi}_N$ and then
divided by the leading and subleading large $N$ asymptotic that is given in the second line of that
relation, i.e.,
 $\big(\frac{N}{2N_0}\big)^{-\frac{4np}{n+2}-8|s|}
\,\big(\frac{4n}{n+2}\big)^{-N}\prod(\zeta_m^{-2}+q^2)(\zeta_m^{-2}+q^{-2})$.
The predicted limiting value of the last quantity,
$2^{-\frac{1}{3}}\,5^{-\frac{4}{5}}\,\pi^3/\,\Gamma^4(\frac{7}{10})=2.392\ldots\,$,
is represented by the solid line. The dashed line corresponds to $\big(\frac{N}{2N_0}\big)^{-8|b|}\,
\big({ R}_{p,b}^{(0)}\big)^2$, where $b=b(N)$ was obtained by 
solving eq.\,\eqref{quantC1} with the correction terms ignored, 
$\sigma=+1$ and $\re^{\frac{\ri}{2}\delta}$
is given by \eqref{oisaoid132}. 
 }
\end{figure}

It should be pointed out that for the RG trajectories with pure imaginary $s$,
the large $N$ asymptotic formulae 
\eqref{aiosdi1209112},\,\eqref{oasodi1121} as well as their 
derivatives
 require some special attention.
As an illustration, let's consider  
\eqref{eq3331} applied
to such a  trajectory with ${\tt L}=\bar{\tt L}=0$. 
In this case the admissible values of $s$ 
 are described by eqs.\,\eqref{oiasiodio98089},\,\eqref{aossdido1a}.
When $p<-\frac{1}{2}$ this gives $s=\pm\ri\,(-p-\frac{1}{2}-a)$ and $a$ is a non-negative integer
such that
$-p-\tfrac{n+2}{4}\le a<-p-\tfrac{1}{2}$.
At these values of $s$
the vacuum eigenvalue $R^{(0)}_{p,s}$  vanishes.
Nevertheless \eqref{eq3331} continues to hold 
if one follows the prescription of replacing
 $s$
\ by $b(N)$ computed from the Bethe roots for $\bm{\Psi}_N$.
Relation \eqref{quantC1} with 
$\re^{\frac{\ri}{2}\delta}$ as in \eqref{oisaoid132}
implies that
\be\label{oiaos901012}
b(N)=\pm\,\ri\bigg[\mathfrak{q}_a+ \frac{\sigma\, (-1)^a}{a!\,(S^z+a)!}\ \ 
2^{-\frac{4}{n}(n+2)\mathfrak{q}_a}\  \ \frac{\Gamma(\bar{p}-p-a)}{\Gamma(1+2p+a)}
\ \bigg(\frac{N}{2N_0}\bigg)^{-4\mathfrak{q}_a}+o\big(N^{-4\mathfrak{q}_a}\big)\bigg]\ ,
\ee
where $\sigma=(-1)^{\frac{N}{2}-S^z}$.
Then one finds
\bea\label{aoisid981921}
&&\prod_{m=1}^M \big(\zeta^{-2}_m+q^2\big) \big(\zeta^{-2}_m+q^{-2}\big)
\asymp
\bigg[ \frac{2^{-\frac{4}{n}\,(n+2)\,\mathfrak{q}_a}\ \Gamma(\bar{p}-p-a)}{ 
(n+2)^{1+\frac{4p}{n+2}}\ (S^z+a)!}\bigg]^2
\ \bigg[\frac{2^{\frac{1}{2}+p}\,
\Gamma(1+p)\,\Gamma(\frac{1}{2}+p)}{\Gamma(1+\frac{2p}{n+2})\,\Gamma(1+a+2p)}\bigg]^4 \nonumber
\\[0.3cm]
&&\times\  \bigg(\frac{N}{2N_0}\bigg)^{-\frac{4np}{n+2}-8\mathfrak{q}_a}\Big(\frac{4n}{n+2}\Big)^{N}
\big(1+o(1)\big)\ \  \ \ \qquad\qquad \big({\tt L}=\bar{\tt L}=0,\, s=\pm\ri\mathfrak{q}_a\big)\ .
\eea
In fig.\,\ref{sec11fig1} the prediction coming from the last relation is compared with the numerical data.
\smallskip

A second comment  regarding  \eqref{aiosdi1209112},\,\eqref{oasodi1121}  concerns the remainder term 
denoted by $o(1)$.
It is expected to be a double series of the form $\sum_{i,j}C_{i,j}\,N^{-i-jn}$.
The expansion coefficients $C_{i,j}$ may become singular
 when $s$ belongs to the discrete set of admissible values, $s=\pm\ri\mathfrak{q}_a,\pm\ri\bar{\mathfrak{q}}_a$, which
results  in
a change of the leading large $N$ behaviour of the products over the Bethe roots.

\section{Conformal towers for pure imaginary $s$ \label{sec12}}
The low energy Bethe states characterized by  pure imaginary 
values of the RG invariant
$s=\sigma\ri\mathfrak{q}_a,\sigma\ri\bar{\mathfrak{q}}_a$ ($\sigma$ is a sign factor) were split into the
two sectors ${\cal H}_{N|S^z}^{({\rm disc},+)}$ and 
${\cal H}_{N|S^z}^{({\rm disc},-)}$ according to whether 
$a$ was a non-negative or a negative integer.
The reason for doing so was
motivated by the following observation.
Together with the integer $a$ the states  are labeled by the 
pair of non-negative integers $(\bar{\tt L},{\tt L})$.
For the case of ${\cal H}_{S^z}^{({\rm disc},+)}=
{\rm slim}_{N\to\infty}{\cal H}_{N|S^z}^{({\rm disc},+)}$
there exists at least one state
for any value of ${\tt L}$ and $\bar{\tt L}$.
Contrary to this,
there are no states belonging to ${\cal H}_{S^z}^{({\rm disc},-)}$
at the levels ${\tt L}=0,1,2,\ldots, |a|-1$ for $s=\sigma\ri\mathfrak{q}_a$ and
$\bar{{\tt L}}=0,1,2,\ldots, |a|-1$  when $s=\sigma\ri\bar{\mathfrak{q}}_a$.
To be more precise, introduce the notation
${\cal H}_{\bar{p},p,\sigma\ri\mathfrak{q}_a}^{(\bar{\tt L},{\tt L},\pm)}$ 
and
${\cal H}_{\bar{p},p,\sigma\ri\bar{\mathfrak{q}}_a}^{(\bar{\tt L},{\tt L},\pm)}$,
for the 
level subspaces of the corresponding
conformal towers
belonging to ${\cal H}_{S^z}^{({\rm disc},\pm)}$.
According to conjecture (III) from sec.\,\ref{sec102} the dimensions of these level
subspaces are given by 
${\cal N}_a^{(\bar{{\tt L}},{\tt L})}={d}_{S^z+a}(\bar{\tt L})\ {d}_{a}({\tt L})$
and 
 $\bar{\cal N}_a^{(\bar{{\tt L}},{\tt L})}={d}_{a}(\bar{{\tt L}})\ {d}_{S^z+a}({\tt L})$,
respectively.
It follows from the definition \eqref{Zdef1b} of the integers
${d}_{a}({\tt L})$
that 
${\rm dim}\big({\cal H}_{\bar{p},p,\sigma\ri\mathfrak{q}_a}^{(\bar{\tt L},{\tt L},+)}\big)$ 
is always greater or equal to one, while 
\be\label{ospo32093}
{\rm dim}\big({\cal H}_{\bar{p},p,\sigma\ri\mathfrak{q}_a}^{(\bar{\tt L},{\tt L},-)}\big)
=\begin{cases} 0 & {\rm for}\qquad {\tt L}<|a|\ \ {\rm or} \ \ \ \ \bar{\tt L}<|a|-S^z\quad \\[0.2cm]
1 &{\rm for} \qquad{\tt L}=|a|\ \ {\rm and}\ \ \bar{\tt L}={\rm max}(0,|a|-S^z)  \\[0.2cm]
\ge 1 &{\rm otherwise}
\end{cases}\qquad\quad (a=-1,-2,\ldots) \ .
\ee
For ${\rm dim}\big({\cal H}_{\bar{p},p,\sigma\ri\bar{\mathfrak{q}}_a}^{(\bar{\tt L},{\tt L},-)}\big)$  similar conditions
 hold true with
${\tt L}$  and $\bar{\tt L}$ interchanged.
\bigskip

To avoid excessive technical details and cumbersome notation,
let's first focus on the conformal tower with $s=+\ri\mathfrak{q}_a$ and $|a|\le S^z$.
Then the lowest energy state would occur at ${\tt L}=|a|$, $\bar{\tt L}=0$ and 
be characterized by the set $\bm{w}$ solving \eqref{sksksk1} and subject  to the extra 
constraint 
$\big({ D}_{p,\ri\mathfrak{q}_a}({\boldsymbol  w})\big)^{-1}=0$
 (see eq.\,\eqref{pdpd000d}). Since the space ${\cal H}_{\bar{p},p,\ri\mathfrak{q}_a}^{(0,|a|,-)}$
is one dimensional, these  conditions must uniquely determine $\bm{w}=\{w_j\}_{j=1}^{|a|}$.
 It turns out to be possible
to give an explicit description of this set by showing that
the $|a|$ numbers $2w_j$  are roots of the generalized Laguerre polynomial\footnote{%
Recall the definition of the generalized Laguerre polynomials:
\be
L_{m}^{(\alpha)}(x)=\frac{x^{-\alpha}\,\re^x}{m!}\ \frac{\rd^m}{\rd x^m}\ \big(\re^{-x}\,x^{m+\alpha}\big)\ . 
\nonumber
\ee
}
\be\label{nullw1}
 L_{|a|}^{(-2p-n-2)}(2w_j)=0\ .
\ee
As will be explained shortly, the connection coefficients
of the ODE \eqref{aisausau} with the apparent singularities  as in \eqref{nullw1} 
coincide with the connection coefficients 
of a similar ODE having no apparent singularities:
\be\label{ioasid91221}
D_\pm(\mu\,|\,\bm{w},p,\ri\mathfrak{q}_a)=D_\pm^{({\rm vac})}(\mu\,|\,p',\ri\mathfrak{q}_a')\,,
\ee
where
\be\label{ioa899099}
p'=p+\tfrac{1}{2}\,(n+2)\,, \qquad\qquad  \mathfrak{q}_a'=\mathfrak{q}_a-\tfrac{n}{2}\,.
\ee
This suggests that the
lowest energy states in the conformal towers
from ${\cal H}_{S^z}^{({\rm disc},-)}$
can be described by the differential equations
without apparent singularities,
similar to the primary states of the conformal towers 
from ${\cal H}_{S^z}^{({\rm disc},+)}$.
\bigskip

Relation \eqref{ioasid91221} between the connection coefficients
and its generalization
arises from a simple relation for the corresponding ODEs.
Let's consider the  case where
the level ${\tt L}$, or equivalently the number of apparent singularities,
is greater or equal to $|a|$, i.e., can be written in the form
${\tt L}=|a|+{\tt L}'$ with
${\tt L}'$ a non-negative integer.  
If $\Phi$ is a solution of the ODE \eqref{aisausau} with $s=\ri\mathfrak{q}_a$
and $a<0$,
 one can show via a straightforward computation that the function
\be\label{iaosido189812}
\Phi'=z^{-\frac{n}{2}}\,\bigg(\frac{\rd }{\rd z}-1-\frac{p+\frac{1}{2}}{z}+
\sum_{j=1}^{{\tt L}}\frac{1}{z-w_j}-\sum_{j=1}^{{\tt L}'}\frac{1}{z-w_j'}\,\bigg)\Phi
\ee
satisfies the differential equation
\be\label{ODE1b}
\bigg[-\frac{\rd^2}{\rd z^2}+\frac{(p')^2-\frac{1}{4}}{z^2}+\frac{2\ri s'}{z}+1
+\sum_{j=1}^{{\tt L}'}\bigg(\frac{2}{(z-w_j')^2}+\frac{n}{z\,(z-w_j')}\bigg)
+\mu^{-2-n}\ z^n\bigg]\Phi'=0\ .
\ee
Here  $s'=\ri\mathfrak{q}_a'$ \eqref{ioa899099},
and the sets $\bm{w}=\{w_j\}_{j=1}^{\tt L}$, 
$\bm{w}'=\{w_j'\}_{j=1}^{{\tt L}'}$ must obey the  coupled algebraic system 
\bea\label{eqsym1a}
\frac{1+n+2p}{2w_l}&\!\!\!+\!\!\!&
1-\sum_{j\ne l}^{{\tt L}}\frac{1}{w_l-w_j}+\sum_{j= 1}^{{\tt L}'}\frac{1}{w_l-w_j'}=0\,,\qquad
l=1,2,\ldots {\tt L} \nonumber\\[-0.3cm]
\\[-0.3cm]
\frac{1+2p}{2w_l'}&\!\!\!+\!\!\!&
1+\sum_{j\ne l}^{{\tt L}'}\frac{1}{w_l'-w_j'}-\sum_{j=1}^{{\tt L}}\frac{1}{w_l'-w_j}=0\,,\qquad
l=1,2,\ldots {\tt L}'\ . \nonumber
\eea
\bigskip

In the case ${\tt L}'=0$ the above equations simplify to 
\be
\frac{1+n+2p}{2w_l}+
1-\sum_{j\ne l}^{{\tt L}}\frac{1}{w_l-w_j}=0\,,\qquad
l=1,2,\ldots {\tt L} \ .
\ee
It turns out that their solution is unique (up to permutation of the $w_j$)
and that $2w_j$ coincide with the roots of the generalized Laguerre polynomial as 
prescribed by \eqref{nullw1}.
For ${\tt L}'>0$ the  rigorous analysis of the solutions 
of the algebraic system \eqref{eqsym1a}
 is an interesting mathematical problem. 
However, it would take us well beyond the original aim
of describing the states in the level subspaces
${\cal H}_{\bar{p},p,\ri\mathfrak{q}_a}^{(\bar{\tt L},{\tt L},-)}$.
Our intuition, supported by a numerical study, leads us to the following picture.
If the set $\bm{w}$ obeys \eqref{sksksk1} as well as the extra condition 
 $\big({ D}_{p,\ri\mathfrak{q}_a}({\boldsymbol  w})\big)^{-1}=0$, then
\eqref{eqsym1a} reduces to just ${\tt L}'$ independent equations that uniquely determine
the set $\bm{w}'$, which satisfies
\bea
4 n\, (w_a')^2\!\!&+&\!\!8\ri s'\, (n+1)\, w_a'-(n+2)\ \big((n+1)^2-4(p')^2\big)\\[0.2cm]
&+&\!\!
4\ \sum_{b\not=a}^{{\tt L}'}\frac{w_a'\, (\, (n+2)^2\, (w_a')^2- n(2n+5)\, w_a' w_b' +
 n(n+1)\, (w_b')^2\,)}{(w_a'-w_b')^3}=0 \nonumber
\eea
along with $\big({ D}_{p',\ri\mathfrak{q}_a'}({\boldsymbol  w}')\big)^{-1}=0$.
It is clear from \eqref{iaosido189812} that the functions   $\psi$ and $z^{\frac{n}{2}}\psi'$
possess the same monodromy properties and hence
\be
D_\pm(\mu\,|\,\bm{w},p,\ri\mathfrak{q}_a)=D_\pm(\mu\,|\,\bm{w}',p',\ri\mathfrak{q}_a')
\ee
for any two sets $\bm{w},\,\bm{w}'$ satisfying \eqref{eqsym1a}.
\bigskip

We are now in a position to describe the conformal
towers from ${\cal H}_{S^z}^{({\rm disc},+)}$ and ${\cal H}_{S^z}^{({\rm disc},-)}$ in 
a uniform way. For the case of ${\cal H}_{S^z}^{({\rm disc},+)}$ each conformal tower
is decomposed into the level subspaces, which themselves can be expressed as
a tensor product of the form
\be\label{iasod28398}
{\cal H}_{\bar{p},p,\sigma\ri\mathfrak{q}_a}^{(\bar{\tt L},{\tt L},+)}=
\bar{{\cal V}}_{\bar{p},\sigma\ri\mathfrak{q}_a}^{(\bar{{\tt L}})}\otimes {\cal V}_{{p},\sigma\ri\mathfrak{q}_a}^{({{\tt L}})}\,,
\qquad\qquad
{\cal H}_{\bar{p},p,\sigma\ri\bar{\mathfrak{q}}_a}^{(\bar{\tt L},{\tt L},+)}=
\bar{{\cal V}}_{\bar{p},\sigma\ri\bar{\mathfrak{q}}_a}^{(\bar{{\tt L}})}\otimes
 {\cal V}_{{p},\sigma\ri\bar{\mathfrak{q}}_a}^{({{\tt L}})}\ .
\ee
Here $\sigma=\pm$ while $\mathfrak{q}_a=-p-\frac{1}{2}-a$,
$\bar{\mathfrak{q}}_a=-\bar{p}-\frac{1}{2}-a$ and $a$ is a non-negative integer
subject to the restrictions \eqref{oiasio8998} which ensure that 
\be\label{iasoid12909324}
0<\mathfrak{q}_a,\bar{\mathfrak{q}}_a\le\tfrac{n}{4}\ .
\ee
The chiral components in \eqref{iasod28398} are finite dimensional linear spaces
whose dimensions are given by
\be\label{dim1777a}
\arraycolsep=1cm
\begin{array}{ll}
\dim\big(\bar{{\cal V}}_{\bar{p},\sigma\ri\mathfrak{q}_a}^{(\bar{{\tt L}})}\big)=d_{S^z+a}(\bar{{\tt L}})\,, &
\dim\big({{\cal V}}_{{p},\sigma\ri\mathfrak{q}_a}^{({{\tt L}})}\big)=d_{a}({{\tt L}}) \\[0.3cm]
\dim\big(\bar{{\cal V}}_{\bar{p},\sigma\ri\bar{\mathfrak{q}}_a}^{(\bar{{\tt L}})}\big)=d_{a}(\bar{{\tt L}})\,, &
\dim\big({{\cal V}}_{{p},\sigma\ri\bar{\mathfrak{q}}_a}^{({{\tt L}})}\big)=d_{S^z+a}({{\tt L}}) 
\end{array} \!\!\!\!\!\!\!\!.
\ee
\bigskip

To describe the level subspaces of the conformal towers from  ${\cal H}_{S^z}^{({\rm disc},-)}$
one should distinguish the two cases $s=\sigma\ri\mathfrak{q}_a$ and 
$s=\sigma\ri\bar{\mathfrak{q}}_a$. When $s=\sigma\ri\mathfrak{q}_a$
it is useful to introduce the notation $p_+$, $\bar{p}_+$, ${\tt L}_+$, $\bar{{\tt L}}_+$ and $\mathfrak{q}_a'$ 
as
\be\label{iaosido2109}
\arraycolsep=0.5cm
\begin{array}{ll}
p_+=p+\tfrac{1}{2}\,(n+2)\,, & {\tt L}_+={\tt L}-|a|
\\[0.2cm]
\bar{p}_+=\bar{p}-\tfrac{1}{2}\,(n+2)\,, & \bar{\tt L}_+=\bar{\tt L}-|a|+S^z
\end{array}\qquad{\rm and}\qquad \qquad{\mathfrak{q}}_a'={\mathfrak{q}}_a-\tfrac{n}{2}\ ,
\ee
where $a$ is negative and obeys the same constraints \eqref{oiasio8998}.
Then the analogue of \eqref{iasod28398} is given by
\be\label{aiso102}
{\cal H}_{\bar{p},p,\sigma\ri\mathfrak{q}_a}^{(\bar{\tt L},{\tt L},-)}=
\begin{cases}
\bar{{\cal V}}_{\bar{p},\sigma\ri\mathfrak{q}_a}^{(\bar{{\tt L}})}\otimes
{\cal V}_{p_+,\sigma\ri\mathfrak{q}_a'}^{({\tt L}_+)}&{\rm for} \qquad  |a|\le S^z 
\\[0.2cm]
\bar{{\cal V}}_{\bar{p}_+,\sigma\ri\mathfrak{q}_a'}^{(\bar{{\tt L}}_+)}\otimes
{\cal V}_{p_+,\sigma\ri\mathfrak{q}_a'}^{({\tt L}_+)}&{\rm for} \qquad  |a|> S^z
\end{cases}\,.
\ee
Here the dimension of each chiral component reads as
\bea\label{dim1a}
&&\dim\big(\bar{{\cal V}}_{\bar{p},\sigma\ri\mathfrak{q}_a}^{(\bar{{\tt L}})}\big)=d_{|S^z+a|}(\bar{{\tt L}})\,, \qquad\qquad
\dim\big(\bar{{\cal V}}_{\bar{p}_+,\sigma\ri\mathfrak{q}_a'}^{(\bar{{\tt L}}_+)}\big)=d_{|S^z+a|}(\bar{{\tt L}}_+) \nonumber\\[0.2cm]
&&\dim\big({\cal V}_{p_+,\sigma\ri\bar{\mathfrak{q}}_a'}^{({\tt L}_+)}\big)=d_{|a|}({{\tt L}_+}) \ .
\eea
For the case $s=\sigma\ri\bar{\mathfrak{q}}_a$ 
we define
$p_-$, $\bar{p}_-$, ${\tt L}_-$, $\bar{{\tt L}}_-$ and $\bar{\mathfrak{q}}_a'$,
through the formulae
\be\label{iaosido2109b}
\arraycolsep=0.5cm
\begin{array}{ll}
{p}_-={p}-\tfrac{1}{2}\,(n+2)\,, & {\tt L}_-={\tt L}-|a|+S^z 
\\[0.2cm]
\bar{p}_-=\bar{p}+\tfrac{1}{2}\,(n+2)\,, & \bar{{\tt L}}_-=\bar{{\tt L}}-|a|
\end{array}\qquad{\rm and}\qquad\qquad \bar{\mathfrak{q}}_a'=\bar{\mathfrak{q}}_a-\tfrac{n}{2}\ .
\ee
With this notation, the decomposition of the level subspace looks similar to that for
$s=\sigma\ri\mathfrak{q}_a$. Namely,
\be\label{aiso103}
{\cal H}_{\bar{p},p,\sigma\ri\bar{\mathfrak{q}}_a}^{(\bar{\tt L},{\tt L},-)}=
\begin{cases}
\bar{{\cal V}}_{\bar{p}_-,\sigma\ri\bar{\mathfrak{q}}_a'}^{(\bar{{\tt L}}_-)}\otimes
{\cal V}_{p,\sigma\ri\bar{\mathfrak{q}}_a}^{({\tt L})}&{\rm for} \qquad |a|\le S^z 
\\[0.2cm]
\bar{{\cal V}}_{\bar{p}_-,\sigma\ri\bar{\mathfrak{q}}_a'}^{(\bar{{\tt L}}_-)}\otimes
{\cal V}_{p_-,\sigma\ri\bar{\mathfrak{q}}_a'}^{({\tt L}_-)}& {\rm for} \qquad |a|> S^z
\end{cases}
\ee
and the dimensions of the chiral subspaces are given by
\bea\label{dim1b}
&&\dim\big({\cal V}_{p,\sigma\ri\bar{\mathfrak{q}}_a}^{({\tt L})}\big)=d_{|S^z+a|}({{\tt L}})\,, \qquad
\dim\big({\cal V}_{p_-,\sigma\ri\bar{\mathfrak{q}}_a'}^{({\tt L}_-)}\big)=d_{|S^z+a|}({{\tt L}_-}) \nonumber \\[0.2cm]
&&\dim\big(\bar{{\cal V}}_{\bar{p}_-,\sigma\ri\bar{\mathfrak{q}}_a'}^{(\bar{{\tt L}}_-)}\big)=d_{|a|}(\bar{{\tt L}}_-) \ .
\eea
\smallskip

The following comment is in order here.  It is simple to check the identities
\be\label{id3009a}
\frac{p^2}{n+2}-\frac{\mathfrak{q}_a^2}{n}+{\tt L}=
\frac{(p_+)^2}{n+2}-\frac{(\mathfrak{q}_a')^2}{n}+{\tt L}_+\,,\qquad
\frac{\bar{p}^2}{n+2}-\frac{{\mathfrak{q}}_a^2}{n}+\bar{{\tt L}}=
\frac{(\bar{p}_+)^2}{n+2}-\frac{({\mathfrak{q}}_a')^2}{n}+\bar{{\tt L}}_+\,,
\ee
where $p_+$, $\bar{p}_+$, ${\tt L}_+$, $\bar{\tt L}_+$ and $\mathfrak{q}_a'$ are given in \eqref{iaosido2109}. 
This makes it possible to re-write the 
 scaled energy, defined by eq.\,\eqref{oaisodi192}, for the level subspaces 
${\cal H}_{\bar{p},p,\sigma\ri\mathfrak{q}_a}^{(\bar{\tt L},{\tt L},-)}$
 in terms of the numbers labeling the chiral components
 $\bar{\cal V}$ and ${\cal V}$ in the r.h.s. of eq.\,\eqref{aiso102}.
For example, for the case $s=\sigma\ri\mathfrak{q}_a$ and $|a|>S^z$ one has
\be
E=\frac{(p_+)^2+(\bar{p}_+)^2}{n+2}-\frac{2(\mathfrak{q}_a')^2}{n}-\frac{1}{6}+{\tt L}_++\bar{\tt L}_+\ .
\ee
The same can be done for ${\cal H}_{\bar{p},p,\sigma\ri\bar{\mathfrak{q}}_a}^{(\bar{\tt L},{\tt L},-)}$ using
the similar relations to \eqref{id3009a} involving $p_-$, $\bar{p}_-$, ${\tt L}_-$, $\bar{\tt L}_-$ and $\bar{\mathfrak{q}}_a'$.
It should be emphasized that $\mathfrak{q}_a'$ and $\bar{\mathfrak{q}}_a'$ do not lie in the strip
from \eqref{iasoid12909324}, but rather
\be
-\tfrac{n}{2}<\mathfrak{q}_a',\bar{\mathfrak{q}}_a'\le -\tfrac{n}{4}\ .
\ee
As a result $s'=\pm\ri\mathfrak{q}_a',\pm\ri\bar{\mathfrak{q}}_a'$ does not obey the constraint
 \eqref{scon3iuewi1}.

\section{Scaling limit of the lattice operators $\mathbb{A}_\pm(\zeta)$ \label{CTsec}}
The key r$\hat{\rm{o}}$le in the description of the scaling limit of the ${\cal Z}_2$ invariant inhomogeneous six-vertex model 
 is played by the relation \eqref{as56d1a}. Much of our numerical work, which was outlined in sec.\,\ref{sec11}, 
was devoted to 
its verification.
Accepting   that \eqref{as56d1a} holds true, it can be interpreted
as an operator relation
\bea\label{scalingrel1a}
{\rm s}\!\!\!\!\lim_{N\to \infty\atop b(N)\to s}
G^{(N/2)}\big(-\mu^2\,|\,\tfrac{2}{n+2}\big)\, 
 \mathbb{A}_\pm\big(\, \ri\, \big(N/(2N_0)\big)^{-\frac{n}{n+2}}\ \mu \big)=
{\mathlarger{\mathlarger{\mathlarger {\bf \it a}}}}_\pm(\lambda)
\eea
with
 ${\mathlarger {\mathlarger{\mathlarger{\mathlarger {\bf \it a}}}}}_\pm(\lambda)$
acting invariantly in the right chiral component of the level subspaces of the conformal towers.
In view of \eqref{logaseries1abbc}, the operators 
$\log{\mathlarger{\mathlarger{\mathlarger{\mathlarger {\bf \it a}}}}}_\pm(\lambda)$
 possess the series expansion
\be\label{logaseries1a}
\log{\mathlarger{\mathlarger{\mathlarger {\bf \it a}}}}_\pm(\lambda)=-\sum_{j=1}^\infty 
{\bf J}_j^{(\pm)}\,\lambda^{j}
\ee  
and the eigenvalues of ${\bf J}_j^{(\pm)}$ 
 coincide with $J^{(\pm)}_{j}(\bm{w},p,s)$.
Recall that the parameters $\lambda$ and $\mu$ are proportional to each other as in eq.\,\eqref{mulambda1a}.
The dimensions of the level subspaces of the conformal towers 
have already been described in sections \ref{sec101} and \ref{sec12}. In particular
\eqref{893289121} suggests that
for real $s$,
\be\label{aijdskfdkjj222}
{\cal H}_{\bar{p},p,s}^{(\bar{\tt L},{\tt L})}=\bar{{\cal V}}_{\bar{p},s}^{(\bar{{\tt L}})}\otimes{\cal V}_{p,s}^{({\tt L})}\qquad{\rm with} \qquad
\dim\big({\cal V}_{p,s}^{({\tt L})}\big)={\rm par}_2({\tt L})\,,\quad 
\dim\big(\bar{{\cal V}}_{\bar{p},s}^{(\bar{{\tt L}})}\big)={\rm par}_2(\bar{{\tt L}})\,.
\ee
As this  is simpler than
for the case of pure imaginary  $s$, where the corresponding dimensions are given  by
 eqs.\,\eqref{dim1777a},\,\eqref{dim1a} and \eqref{dim1b}, we start by describing 
the operators ${\mathlarger{\mathlarger{\mathlarger{\mathlarger {\bf \it a}}}}}_\pm(\lambda):\, 
{{\cal V}}_{{p},s}^{({{\tt L}})}\mapsto {{\cal V}}_{{p},s}^{({{\tt L}})}$ 
 with $s$ being a real number. 

\subsection{The case of real $s$}
For real $s$ the dimensions of
${\cal V}_{p,s}^{({\tt L})}$ 
coincide with those of
 the level subspace
of the Fock space generated by two independent copies of the
Heisenberg algebra. Hence one can identify them as linear spaces:
\be\label{FOck39933}
{\cal V}_{p,s}^{({\tt L})}={\cal F}_{\bf P}^{({\tt L})} \ \qquad \qquad \qquad (s\in\mathbb{R})\ .
\ee
We take the commutation relations for the
Heisenberg algebra generators to be
\bea\label{hasaast}
[a_m,a_j]=\tfrac{m}{2}\ \delta_{m+j,0}\ ,\ \ \ [b_m,b_j]=\tfrac{m}{2}\ \delta_{m+j,0}\ ,\ \ \  [a_m,b_j]=0\,,
\eea
while ${\bf P}$ stands for 
the highest weight, i.e., the values of $a_0$ and $b_0$ in ${\cal F}_{\bf P}$.
The dimensions of ${\cal F}_{\bf P}^{({\tt L})}$ of course do not depend on the value of 
the highest weight. We'll set
\be\label{oasidoioais311}
{\bf P}=\big(\tfrac{p}{\sqrt{n+2}},\tfrac{s}{\sqrt{n}}\big)\ .
\ee
\medskip

The construction of the operators 
${\mathlarger{\mathlarger{\mathlarger {\mathlarger{\bf \it a}}}}}_\pm(\lambda)$  
parallels that for the homogeneous case. 
Formulae \eqref{Lop2a},\,\eqref{soso1a} remain essentially unchanged, but the vertex operators are now given by
\bea\label{vert1ad}
V_{+}(u)=\re^{+ \frac{2\ri\varphi}{\sqrt{n+2}}}(u)\ ,\ \ \ \ \ \ \ \qquad
V_{-}(u)=-2\sqrt{n}\ \partial \vartheta\, \re^{- \frac{2\ri\varphi}{\sqrt{n+2}}}(u)\,.
\eea
Here $\varphi(u)$ is the same as in eq.\,\eqref{bosefiled1} and the additional chiral field
\bea\label{teteiqiqi1a}
\partial\vartheta(u)= \sum_{m=-\infty}^\infty b_{m}\ \re^{-\ri m u}
\eea
 involves the Heisenberg generators $\{b_m\}$ \eqref{hasaast}.
Then it turns out that
\bea\label{soso1aa}
{\mathlarger{\mathlarger{\mathlarger {\bf \it a}}}}_\pm(\lambda)=
\frac{{\rm Tr}_{\rho_\pm}\big[\,\re^{\pm \frac{\ri\pi}{\sqrt{n+2}} \, a_0{\cal H}}\
{{\boldsymbol L}}_\pm(\lambda)\,\big]}
{{\rm Tr}_{\rho_\pm}\big[\,\re^{\pm \frac{2\ri\pi}{\sqrt{n+2}}\, a_0{\cal H}}\,\big]}
\eea
with
\bea\label{Lop2ad}
{\boldsymbol L}_\pm(\lambda)=\re^{\pm\frac{\ri\pi}{\sqrt{n+2}}\,  
a_0\,{\cal H}}\ \overset{\leftarrow}{{\cal P}}\exp\bigg(\int_0^{2\pi}\rd u\,
\Big(\,V_-(u)\ q^{\pm\frac{\cal H}{2}}\,{\cal E}_\pm+
\lambda\, V_+(u)\ q^{\mp\frac{\cal H}{2}}\,{\cal E}_\mp\,\Big)\bigg)\ .
\eea
As before ${\cal E}_\pm$ and ${\cal H}$ stand for the generators of the
$q$-oscillator algebra
\eqref{qosc} and $\rho_\pm$   are 
representations of this algebra -- the same as in \eqref{soso1a}. 
Since  ${\mathlarger{\mathlarger{\mathlarger{\mathlarger {\bf \it a}}}}}_\pm(0)=\bm{1}$
the formal power series \eqref{soso1aa} can be rewritten as
the Taylor series \eqref{logaseries1a} for
$\log{\mathlarger{\mathlarger{\mathlarger{\mathlarger {\bf \it a}}}}}_\pm(\lambda)$.
The  expansion coefficients ${\bf J}_j^{(\pm)}$ 
involve the ordered multifold integrals. Like in  the homogeneous case, expressing these in terms
of the contour integrals  makes the operators    ${\bf J}_j^{(\pm)}$
 well defined for any $n>0$ except $n=\frac{2}{2k-1}$
with $k=1,2,\ldots\ $. In the latter case  ${\bf J}_{2k}^{(\pm)}$ requires regularization and
we define ${\bf J}_{2 k}^{(\pm,{\rm reg})}$ through a subtraction of the counterterm
of the unit operator, similar to eq.\,\eqref{isisaiasi4443}. 
\bigskip

Following the lines of ref.\cite{Bazhanov:1998dq} one can prove that 
${\mathlarger{\mathlarger{\mathlarger{\mathlarger {\bf \it a}}}}}_\pm(\lambda)$, defined as above,
act invariantly in ${\cal F}_{{\bf P}}^{({\tt L})}$ and form a commuting family
\be
\big[{\mathlarger{\mathlarger{\mathlarger {\bf \it a}}}}_\pm(\lambda),\,
{\mathlarger{\mathlarger{\mathlarger {\bf \it a}}}}_\pm(\lambda')\big]=0\ .
\ee
In addition, it is possible to derive  a set of operator relations for 
${\mathlarger{\mathlarger{\mathlarger{\mathlarger {\bf \it a}}}}}_\pm(\lambda)$,
which in turn become functional relations for their eigenvalues.
The latter are identical to those
satisfied by the connection coefficients, which 
follow from the basic properties of the ODE \eqref{aisausau}. Among these is the 
so-called quantum Wronskian 
relation 
\be
q^{2p}\,D_+(q^{+1}\mu)\,D_-(q^{-1}\mu)-
q^{-2p}\,D_-(q^{+1}\mu)\,D_+(q^{-1}\mu)=2\ri\sin\big(\tfrac{2\pi p}{n+2}\big)\ ,
\ee
where $D_\pm(\mu)\equiv {D}_\pm(\mu\,|\,\bm{w},p,s)$.
The fact that  the functional relations coincide is not sufficient
to prove that each of the ${\rm par}_2({\tt L})$ eigenvalues of
${\mathlarger{\mathlarger{\mathlarger{\mathlarger {\bf \it a}}}}}_\pm(\lambda)$ in
${\cal F}^{({\tt L})}_{\bf P}$ is given by a connection coefficient
${D}_\pm(\mu)$ for one of the ${\rm par}_2({\tt L})$ solution
sets $\bm{w}$ of the algebraic system \eqref{sksksk1}. Nevertheless 
we confirmed, for instance, that the vacuum eigenvalues of 
${\bf J}_1^{(\pm)}$ and
${\bf J}_2^{(\pm)}$, computed  from the definition \eqref{soso1aa},\,\eqref{Lop2ad},
coincide with  $J_1^{({\rm vac})}(p,s)$ 
and $J_2^{({\rm vac})}(p,s)$ from \eqref{Jvaceig1a}, which were obtained via the 
perturbation theory of the ODE \eqref{aisausau}.
This strongly suggests that
\be\label{iaosid9898}
{\mathlarger{\mathlarger{\mathlarger {\bf \it a}}}}_\pm(\lambda)\,\bm{\psi}_{p,s}(\bm{w})=
{D}_\pm(\mu\,|\,\bm{w},p,s)\,\bm{\psi}_{p,s}(\bm{w})\,,
\ee
where $\bm{\psi}_{p,s}(\bm{w})\in {\cal F}^{({\tt L})}_{\bf P}$ stands for the corresponding
eigenvector and with the $\lambda$-$\mu$ relation as in \eqref{mulambda1a}.

\subsection{The case of pure imaginary $s$\label{iaso132218s}}
It should be pointed out that formulae \eqref{soso1aa} and \eqref{Lop2ad}
define the operators
${\mathlarger{\mathlarger{\mathlarger{\mathlarger {\bf \it a}}}}}_\pm(\lambda)$, acting 
invariantly in the level subspace of the Fock space,
\be\label{iaosi988988}
{\mathlarger{\mathlarger{\mathlarger {\bf \it a}}}}_\pm(\lambda): \ \ \ {\cal F}_{{\bf P}}^{({\tt L})}\mapsto
{\cal F}_{{\bf P}}^{({\tt L})}\,,
\ee
for any value of the highest weight  ${\bf P}\equiv (P_1,P_2 )$
 except when  $q^{2\sqrt{n+2}\,P_1}=\pm q^m$ and $m=0,\pm 1,\pm 2\ldots\ $.
In the latter case ${\mathlarger{\mathlarger{\mathlarger{\mathlarger {\bf \it a}}}}}_\pm(\lambda)$
may still be introduced, though some special treatment is required.
The same holds true for the connection coefficients and
it is expected that \eqref{iaosid9898} is valid for any
 complex $P_1$ and $P_2$.
On the other hand, the operators which appear in the scaling limit  \eqref{scalingrel1a} 
act in the chiral components of the conformal towers. For the 
discrete spectrum, where $P_1$ and $P_2$ are related as
\be\label{ioasid8998}
\sqrt{n+2}\ P_1+\tfrac{1}{2}\pm\ri \sqrt{n}\ P_2\in\mathbb{Z}\,,
\ee
these have dimensions that are typically less than ${\rm dim}\,\big({\cal F}_{\bf P}^{({\tt L})}\big)$.
In this case the operators from \eqref{scalingrel1a} 
should   be understood   to be
the ones in \eqref{iaosi988988} restricted to a  certain subspace of 
${\cal F}_{{\bf P}}^{({\tt L})}$.
Since a rigorous treatment of these restrictions
involves many technical details,
here we just give a sketch of the underlying ideas.
\bigskip

Let's take $P_1$ and $P_2$ , satisfying \eqref{ioasid8998}, in the form
\be\label{ioasid899283}
P_1=\tfrac{1}{\sqrt{n+2}}\, \big(\,p+\half (n+2)\,\ell\,\big)\,,
\qquad P_2=-\frac{\ri\sigma}{\sqrt{n}}\, (\,p+\half+a +\frac{n}{2}\,\ell\,) \ ,
\ee
where $a$, $\ell$ are  integers, $\sigma=\pm 1$, while
 $p$ can be arbitrary. 
 The Fock space ${\cal F}_{\bf P}$ corresponding to this value of the highest weight ${\bf P}=(P_1,P_2)$ will be denoted as
${\cal F}\big[{}_{p,a}^{\sigma,\ell}\big]$.
 Consider the so-called  ``screening charge'' built from the chiral fields
 $\varphi$ and $\vartheta$ \cite{Fateev:1987vh} (see also \cite{Jayaraman:1989tu,Griffin:1990fg})
  \bea\label{ioasd90023}
\hat{{\tt Q}}_\sigma=\int_0^{2\pi}\rd u \  \re^{\ri\sqrt{n+2}\,\varphi+\sigma \sqrt{n}\,\vartheta}(u)\ .
\eea
The field  $\vartheta$ is defined via a formula similar to eq.\,\eqref{bosefiled1}
and involves the mode  $\vartheta_0$ conjugated to $b_0$ such that
$[\vartheta_0,b_m]=\frac{\ri}{2}\delta_{m,0}$.
It turns out that for a fixed choice of
the sign factor  $\sigma=\pm1$  the following  holds true:
\begin{enumerate}
\item[(a)]
 The screening charge 
is a well defined operator in the direct sum of the Fock spaces
\bea\label{iaosid09120912}
\hat{{\tt Q}}_\sigma\in {\rm End}\Big(\bigoplus_{{\ell=-\infty}}^\infty{\cal F}\big[{}_{p,a}^{\sigma,\ell}\big]\Big)
\eea
and acts as the intertwiner
 \bea
\hat{{\tt Q}}_\sigma\ :\ \ \ \ {\cal F}\big[{}_{p,a}^{\sigma,\ell}\big]\mapsto {\cal F}[{}_{p,\ \,a}^{\sigma,\ell+1}\big]\ .
\eea
 \item[(b)] The operator  $\hat{{\tt Q}}_\sigma$ is nilpotent
\bea
\hat{{\tt Q}}^2_\sigma=0\ .
\eea
\item[(c)]
The action of   $\hat{{\tt Q}}_\sigma$  commutes  with  the action  of 
${\mathlarger{\mathlarger{\mathlarger{\mathlarger {\bf \it a}}}}}_\pm(\lambda)$:
\bea
\hat{{\tt Q}}_\sigma{\mathlarger{\mathlarger{\mathlarger {\bf \it a}}}}_\pm(\lambda)=
{\mathlarger{\mathlarger{\mathlarger {\bf \it a}}}}_\pm(\lambda)\,\hat{{\tt Q}}_\sigma\ .
\eea
\end{enumerate}

\bigskip

\begin{figure}
\centering
\scalebox{0.9}{
\begin{tikzpicture}
\draw[->, line width = 0.5mm] (-3,-4.5) -> (-0.5,-4.5) ;
\node at (-3.5,-4.5) {$\cdots$};
\draw (0,-2) -- (-2,-5.5);
\draw (0,-2) -- (2,-5.5);
\node at (0.05,-1.5) {${\cal F}\big[{}_{p,\ \,a}^{\sigma,\ell'-1}\big]$};
\draw (0,-4.5) -- (-0.5,-5.5);
\draw (0,-4.5) -- (0.5,-5.5);
\fill[pattern=north west lines, pattern color=gray!50!white] (0,-4.5) -- (-0.5,-5.5) -- (0.5,-5.5) -- cycle;
\draw[black,fill=black] (0,-4.5) circle (0.05cm);
\draw (-2.5,-2) -- (-2,-2);
\draw (-2.5,-4.5) -- (-2,-4.5);
\node at (-2.25,-3.25) {\small $a-\ell'+2$};
\draw[->,thick] (-2.25,-3.6) -> (-2.25,-4.4);
\draw[->,thick] (-2.25,-2.9) -> (-2.25,-2.1);
\draw[->, line width = 0.5mm] (0.5,-2) -> (4.5,-2) ;
\node at (2,-1.6) {$\hat{{\tt Q}}_\sigma$};
\draw (5,0) -- (3,-4);
\draw (5,0) -- (7,-4);
\node at (5.05,0.5) {${\cal F}\big[{}_{p,a}^{\sigma,\ell'}\big]$};
\draw (5,-2) -- (4,-4);
\draw (5,-2) -- (6,-4);
\fill[pattern=north west lines, pattern color=gray!50!white] (5,-2) -- (4,-4) -- (6,-4) -- cycle;
\draw[black,fill=black] (5,-2) circle (0.05cm);
\draw (3.2,0) -- (3.7,0);
\node at (3.45,-1) {\small $a-\ell'+1$};
\draw[->,thick] (3.45,-0.65) -> (3.45,-0.1);
\draw[->,thick] (3.45,-1.35) -> (3.45,-1.9);
\draw[->, line width = 0.5mm] (5.5,0) -> (9.7,0) ;
\node at (7,0.4) {$\hat{{\tt Q}}_\sigma$};
\draw (10,1.5) -- (8,-2.5);
\draw (10,1.5) -- (12,-2.5);
\node at (10.05,2) {${\cal F}\big[{}_{p,\ \,a}^{\sigma,\ell'+1}\big]$};
\draw (10,0) -- (9,-2.5);
\draw (10,0) -- (11,-2.5);
\fill[pattern=north west lines, pattern color=gray!50!white] (10,0) -- (9,-2.5) -- (11,-2.5) -- cycle;
\draw[black,fill=black] (10,0) circle (0.05cm);
\draw (8.2,1.5) -- (8.7,1.5);
\node at (8.45,0.75) {\small $a-\ell'$};
\draw[->,thick] (8.45,1.0) -> (8.45,1.4);
\draw[->,thick] (8.45,0.5) -> (8.45,0.1);
\draw[->,line width = 0.5mm] (10.5,1.5) -> (12.5,1.5) ;
\node at (13,1.5) {$\cdots$};
\node at (12,1.9) {$\hat{{\tt Q}}_\sigma$};
\end{tikzpicture}
}
\caption{\small
A depiction of a fragment of the half-infinite chain complex
for the action of the screening charges \eqref{iaosid09120912}
with $\ell'\le a$.
The shaded regions represent the  proper 
subspaces
${\cal K}\big[{}_{p,a}^{\sigma,\ell}\big]={\cal I}\big[{}_{p,a}^{\sigma,\ell}\big]
\subset{\cal F}\big[{}_{p,a}^{\sigma,\ell}\big]$ with $\ell=\ell'-1,\ell',\ell'+1$ and the bullet 
at each vertex corresponds to the state \eqref{iaosid9823}.
The chain is infinitely extended to the left.
If $\ell'=a+1$, the chain terminates
since, according to eq.\eqref{iodi889981},
the whole Fock space  ${\cal F}\big[{}_{p,\ \,a}^{\sigma,a+1}\big]$ lies in the kernel of $\hat{{\mathtt Q}}_\sigma$.
\label{fig10}}
\end{figure}
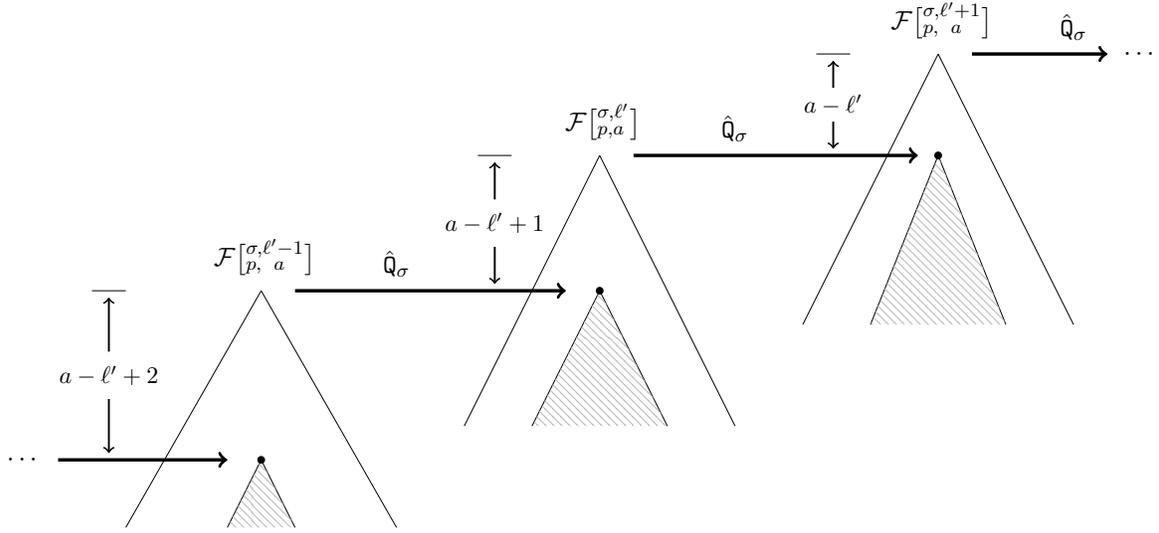

Introduce the notation
\bea\label{iaosid899382}
{\cal K}\big[{}_{p,a}^{\sigma,\ell}\big]={\rm Ker}(\hat{{\tt Q}}_\sigma)\cap{\cal F}\big[{}_{p,a}^{\sigma,\ell}\big]\ ,\ \ \ \ \ 
{\cal I}\big[{}_{p,a}^{\sigma,\ell}\big]={\rm Im}(\hat{{\tt Q}}_\sigma)\cap{\cal F}\big[{}_{p,a}^{\sigma,\ell}\big]\ .
\eea 
Property (b) implies ${\cal I}\big[{}_{p,a}^{\sigma,\ell}\big]\subseteq{\cal K}\big[{}_{p,a}^{\sigma,\ell}\big] $,
while (c) gives that \eqref{iaosid899382}
are  invariant  subspaces  of ${\cal F}\big[{}_{p,a}^{\sigma,\ell}\big]$  w.r.t. the action of  
${\mathlarger{\mathlarger{\mathlarger{\mathlarger {\bf \it a}}}}}_\pm(\lambda)$.
Moreover one expects that 
\bea\label{iodi889981}
&&{\cal K}\big[{}_{p,a}^{\sigma,\ell}\big]={\cal F}\big[{}_{p,a}^{\sigma,\ell}\big]\,,\qquad
\quad{\cal I}\big[{}_{p,a}^{\sigma,\ell}\big]=\emptyset
\qquad \qquad (\ell >a+1) \\[0.2cm]
&&{\cal K}\big[{}_{p,\ \,a}^{\sigma,a+1}\big]={\cal I}\big[{}_{p,\ \,a}^{\sigma,a+1}\big]=
{\cal F}\big[{}_{p,\ \,a}^{\sigma,a+1}\big]\ .
\nonumber
\eea
In the case $\ell \le a$ the subspaces \eqref{iaosid899382} 
 coincide,
\be\label{ioasiu112}
{\cal K}\big[{}_{p,a}^{\sigma,\ell}\big]={\cal I}\big[{}_{p,a}^{\sigma,\ell}\big] \qquad \qquad (\ell\le a)
\ee
and  are proper subspaces in the sense that 
they are neither the empty set nor
equal to the  Fock space itself.
In particular, it is easy to
check that the highest state in ${\cal F}\big[{}_{p,a}^{\sigma,\ell}\big]$ does not
belong to the kernel of $\hat{{\tt Q}}_\sigma$,
while the $\hat{{\tt Q}}_\sigma$-image of the highest state from
${\cal F}\big[{}_{p,\ \,a}^{\sigma,\ell-1}\big]$
takes the form
\bea\label{iaosid9823}
\Bigg[\bigg(\frac{\partial}{\partial z}\bigg)^{a+1-\ell}
 \
\exp\bigg(\sum_{m=1}^\infty\big(\sqrt{n+2}\ a_{-m}-\sigma\ri\,\sqrt{n}\,b_{-m}\big)\ \frac{z^m}{m}
\bigg)\Bigg]_{z=0}\big| {\bf P}\big\rangle\in{\cal I}\big[{}_{p,a}^{\sigma,\ell}\big]\ .
\eea
Fig.\,\ref{fig10} provides a visualization of the action of the screening charges on the Fock spaces.
With the above properties one can show (see, e.g.,\cite{Jayaraman:1989tu,Griffin:1990fg})
that the dimensions of the level subspaces of the factor space
${\cal V}\big[{}_{p,a}^{\sigma,\ell}\big]\equiv
{\cal F}\big[{}_{p,a}^{\sigma,\ell}\big]/{\cal I}\big[{}_{p,a}^{\sigma,\ell}\big]$ are given by 
\bea
\dim \Big({\cal V}^{(L)}\big[{}_{p,a}^{\sigma,\ell}\big]\Big)=
d_{a-\ell}({\tt L})\ \ \ \ \ \ \ (\ell\leq a)
\eea
with  $d_a({\tt L})$ as in   \eqref{Zdef1b}.

\smallskip

From here on out, without loss of generality, we set the parameter 
$\ell$ in \eqref{ioasid899283} to be zero. In fact, $\ell$ is a fake parameter that was 
introduced only for convenience.
Then using our previous notation, $\mathfrak{q}_a=-p-\frac{1}{2}-a$,
one has $P_1=\frac{p}{\sqrt{n+2}}$, $P_2=\frac{\sigma\ri\mathfrak{q}_a}{\sqrt{n}}$ so that
\be\label{iao9023900921}
\sqrt{n+2}\ P_1+\tfrac{1}{2}-\sigma\,\ri \sqrt{n}\ P_2=-a\ .
\ee
In view of \eqref{iodi889981} and \eqref{ioasiu112}
 one should  distinguish the cases $a<0$ and $a\ge0$.
 Let's first take $a$ to be a non-negative integer.
The eigenstates  $\bm{\psi}_{p,\sigma\ri\mathfrak{q}_a}(\bm{w})$ \eqref{iaosid9898} form a basis in the level subspace of
${\cal F}\big[{}_{p,a}^{\sigma,0}\big]$. Those which are annihilated by the screening charge,
$
\hat{{\tt Q}}_\sigma \bm{\psi}_{p,\sigma\ri\mathfrak{q}_a}(\bm{w})=0\,,
$
are a basis in the level subspace of ${\cal K}\big[{}_{p,a}^{\sigma,0}\big]$.
Among them is the state \eqref{iaosid9823} occurring at the level ${\tt L}=a+1$.
The eigenstates $\bm{\psi}_{p,\sigma\ri\mathfrak{q}_a}(\bm{w})$ which do not belong to the kernel
of $\hat{{\tt Q}}_\sigma $ provide a basis for the level subspace 
${\cal V}_{{p},\sigma\ri{\mathfrak{q}}_a}^{({\tt L})}$ of the space
\be\label{ioasido1212}
 {\cal V}_{{p},\sigma\ri{\mathfrak{q}}_a}=
\big(\hat{\bf{1}}-\hat{\Pi}_{{\cal K}_a}\big)\,\big(
{\cal F}\big[{}_{p,a}^{\sigma,0}\big]\big)\ ,
\qquad \qquad
 {\cal V}_{{p},\sigma\ri{\mathfrak{q}}_a}=\bigoplus_{{\tt L}\ge 0} {\cal V}_{{p},\sigma\ri{\mathfrak{q}}_a}^{({{\tt L}})}\ 
\qquad \qquad (a\ge 0)\ .
\ee
Here $\hat{\Pi}_{{\cal K}_a}$ stands for the projector onto ${\cal K}\big[{}_{p,a}^{\sigma,0}\big]$.
The dimensions of ${\cal V}_{{p},\sigma\ri{\mathfrak{q}}_a}^{({{\tt L}})}$
are given by $d_a({\tt L})$  and, moreover, they can be identified
with the chiral components appearing in the decomposition of the conformal tower
 \eqref{iasod28398}.
\bigskip

When $a$ from \eqref{iao9023900921} is a negative integer, the above analysis does not
follow through  literally since the full Fock space ${\cal F}\big[{}_{p,a}^{\sigma,0}\big]$
belongs to the kernel of the screening charge. Nevertheless, one can use the fact
that  the algebraic equations \eqref{sksksk1}, obeyed by the sets $\bm{w}$ labeling the eigenbasis
of ${\mathlarger{\mathlarger{\mathlarger{\mathlarger {\bf \it a}}}}}_\pm(\lambda)$, do not depend on the sign of $p$.
This allows one to introduce the operator $\hat{{\tt C}}_{\rm R}$ via the formula
\be\label{ioasi19020129}
\hat{{\tt C}}_{\rm R}\,\bm{\psi}_{p,s}(\bm{w})=\bm{\psi}_{-p,s}(\bm{w})\ .
\ee
The precise specification of 
$\hat{{\tt C}}_{\rm R}$ requires fixing 
the normalization of the states
$\bm{\psi}_{\pm p,s}(\bm{w})$.
However, this is not important for our purposes as
all that is needed is that $\hat{{\tt C}}_{\rm R}$
 intertwines the Fock spaces,
\be
\hat{{\tt C}}_{\rm R}\big( {\cal F}_{(\pm P_1,P_2)}\big)
= {\cal F}_{(\mp P_1,P_2)}
\ee
and obeys the following commutation relations
\be\label{iaos98a9s8d}
\hat{{\tt C}}_{\rm R}{\mathlarger{\mathlarger{\mathlarger {\bf \it a}}}}_\pm(\lambda)=
{\mathlarger{\mathlarger{\mathlarger {\bf \it a}}}}_\mp(\lambda)\,\hat{{\tt C}}_{\rm R}\ .
\ee
It is easy to see that for $P_2$ as in eq.\,\eqref{ioasid899283} with $\ell =0$,  a change in the sign of $p$
is equivalent to the substitutions $\sigma\mapsto-\sigma$ and $a\mapsto -a-1$. Hence
\be
\hat{{\tt C}}_{\rm R}\,\Big(\,{\cal F}\big[{}_{\pm p,\,a}^{\ \,\sigma,\,0}\big]\Big)=
{\cal F}\big[{}_{\mp p\,,-a-1}^{-\sigma,\ \  0}\big]\ .
\ee
The transformed space contains the proper subspace
 ${\cal K}\big[{}_{\mp p\,,-a-1}^{-\sigma,\ \  0}\big]$,
which is  invariant w.r.t. the action of
${\mathlarger{\mathlarger{\mathlarger{\mathlarger {\bf \it a}}}}}_\pm(\lambda)$.
Then instead of \eqref{ioasido1212} one should introduce
${\cal V}_{{p},\sigma\ri{\mathfrak{q}}_a}$ for negative $a$ as
\be\label{ioasido1212a}
 {\cal V}_{{p},\sigma\ri{\mathfrak{q}}_a}=
\hat{{\tt C}}_{\rm R}\,
\big(\hat{\bf{1}}-\hat{\Pi}_{{\cal K}_{-a-1}}\big)\,\big(\,{\cal F}\big[{}_{-p\,,-a-1}^{-\sigma,\ \  0}\big]\big)
\qquad 
 {\cal V}_{{p},\sigma\ri{\mathfrak{q}}_a}=\bigoplus_{{\tt L}\ge 0} {\cal V}_{{p},\sigma\ri{\mathfrak{q}}_a}^{({{\tt L}})}\ 
\qquad \quad (a< 0)\,,
\ee
where $\hat{\Pi}_{{\cal K}_{-a-1}}$ stands for the projector onto the subspace ${\cal K}\big[{}_{-p\,,-a-1}^{-\sigma,\ \  0}\big]$.
The dimensions of the level subspaces
are given by $\dim\big({\cal V}_{{p},\sigma\ri{\mathfrak{q}}_a}^{({{\tt L}})}\big) = d_{|a+1|}({\tt L})$.

\bigskip

The right chiral level subspaces of the conformal towers from ${\cal H}_{S^z}^{({\rm disc},\pm)}$,
appearing in \eqref{iasod28398},\,\eqref{aiso102} and \eqref{aiso103},
may be organized into the graded linear spaces
\be\label{iaoisd8182}
\arraycolsep=0.6cm
\begin{array}{ll}
{\cal V}_{{p},\sigma\ri{\mathfrak{q}}_a} = \bigoplus\limits_{{\tt L}\ge 0}
{\cal V}_{{p},\sigma\ri{\mathfrak{q}}_a}^{({{\tt L}})}\,, &
{\cal V}_{p_+,\sigma\ri\mathfrak{q}_a'}=\bigoplus\limits_{{\tt L}_+\ge 0}{\cal V}_{p_+,\sigma\ri\mathfrak{q}_a'}^{({\tt L}_+)}
\\[0.6cm]
 {\cal V}_{{p},\sigma\ri\bar{\mathfrak{q}}_a}= \bigoplus\limits_{{\tt L}\ge 0}
{\cal V}_{{p},\sigma\ri\bar{\mathfrak{q}}_a}^{({{\tt L}})}\,, & 
{\cal V}_{p_-,\sigma\ri\bar{\mathfrak{q}}_a'}=\bigoplus\limits_{{\tt L}_-\ge 0} 
{\cal V}_{p_-,\sigma\ri\bar{\mathfrak{q}}_a'}^{({\tt L}_-)}
\end{array}.
\ee
In all four cases the corresponding combination
$\sqrt{n+2}\ P_1+\tfrac{1}{2}\pm\,\ri \sqrt{n}\ P_2\in \mathbb{Z}$, for 
some choice of the sign $\pm$,
so that formulae \eqref{ioasido1212} and \eqref{ioasido1212a} provide a description of these spaces
 in terms of the Fock spaces.
In turn, this defines  the action of ${\mathlarger{\mathlarger{\mathlarger{\mathlarger {\bf \it a}}}}}_\pm(\lambda)$
in the chiral components of the conformal towers for pure imaginary $s$.
\bigskip

Finally we 
note that the description of ${\cal V}_{{p},\sigma\ri{\mathfrak{q}}_a}$  with $a<0$ requires, in addition
to the screening charges, the intertwiner $\hat{{\tt C}}_{\rm R}$.
The latter was defined rather formally through  the eigenbasis of 
${\mathlarger{\mathlarger{\mathlarger{\mathlarger {\bf \it a}}}}}_\pm(\lambda)$.
As will be explained in sec.\,\ref{sec16}, the operator $\hat{{\tt C}}_{\rm R}$ may be introduced
in an invariant way that does not require the choice of a specific basis.

\section{Scaling limit of the transfer matrix\label{sec33}}
In our previous discussion  the scaling limit was taken in such a way  that resulted in the
 operators acting in the chiral Fock space ${\cal F}_{\bf P}$.
Of course it is possible to organize the scaling limit of 
$\mathbb{A}_\pm(\zeta)$ and the
transfer matrix  $\mathbb{T}(\zeta)$ 
that yields the operators acting in the
barred chiral space $\bar{{\cal F}}_{\bar{{\bf P}}}$.
Since the corresponding formulae are very similar
they were omitted up to this point.
However here, for future references, we will need to describe
both  $\bm{\tau}(\lambda)\in{\rm End}({\cal F}_{{\bf P}})$ and
$\bar{\bm{\tau}}(\bar{\lambda})\in{\rm End}(\bar{{\cal F}}_{\bar{{\bf P}}})$
that appear in the scaling limit of the transfer matrix.
For this purpose let us introduce the
barred counterpart of the vertex operators from \eqref{vert1ad}:
\bea\label{vert1ada}
\bar{V}_{+}({\bar u})=\re^{+ \frac{2\ri{\bar \varphi}}{\sqrt{n+2}}}(\bar{u})\ ,\ \ \ \ \ \ \ 
\bar{V}_{-}(\bar{u})=-2\sqrt{n}\ \bar{\partial}\bar{\vartheta}\, \re^{- \frac{2\ri{\bar\varphi}}{\sqrt{n+2}}}(\bar{u})\ ,
\eea
where
\bea\label{bosefiledss1}
{\bar \varphi}(\bar{u})&=&\bar{\varphi}_0+\bar{a}_0\, {\bar u}+\ri \sum_{m\not=0}\frac{\bar{a}_{m}}{m}\ \re^{-\ri m\bar{ u}}
\ \ \ \ \ \ \ \ 
\big(\,[\bar{\varphi}_0,\bar{a}_m]=\tfrac{\ri}{2}\ \delta_{m,0}\,\big)\\
\bar{\vartheta}(\bar{u})&=&\bar{\vartheta}_0+{\bar b}_0\, \bar{u}+\ri \sum_{m\not=0}\frac{\bar{b}_{m}}{m}\ \re^{-\ri m \bar{u}}
\ \ \ \ \ \ \ \ 
\big(\,[\bar{\vartheta}_0,\bar{b}_m]=\tfrac{\ri}{2}\ \delta_{m,0}\,\big)\ .\nonumber
\eea
Then consider  the two formal path ordered exponents, which are
defined similarly as in the homogeneous case
(see sec.\,\ref{sec24}):
\bea\label{Laltcase1a}
\!\!\!\!\!\!\!\!\!\!{\boldsymbol L}(\lambda)&=&\lambda^{+\frac{{\tt h}}{4}}\, 
\re^{\frac{\ri\pi}{\sqrt{n+2}}\,a_0{\tt h}}\
 \overset{\leftarrow}{{\cal P}}\exp\bigg(\int_0^{2\pi}\rd u\,
\Big(V_-(u)\, q^{+\frac{{\tt h}}{2}}\,{\tt e}_++\lambda\, V_+(u)\,
q^{-\frac{{\tt h}}{2}}\,{{\tt e}_-}\, \Big)\bigg)\, \lambda^{-\frac{{\tt h}}{4}}  \\[0.2cm]
\!\!\!\!\!\!\!\!\!\!\bar{\boldsymbol L}(\bar{\lambda})&=&\bar{\lambda}^{+\frac{{\tt h}}{4}}\
 \overset{\rightarrow}{{\cal P}}\exp\bigg(\int_0^{2\pi}\rd \bar{u}\,
\Big(\bar{V}_-(\bar{u})\, q^{+\frac{{\tt h}}{2}}\,{\tt e}_++\bar{\lambda}\, \bar{V}_+(\bar{u})\,
q^{-\frac{{\tt h}}{2}}\,{{\tt e}_-}\, \Big)\bigg)\, \re^{-\frac{\ri\pi}{\sqrt{n+2}}\,{\bar a}_0{\tt h}}\,
\bar{\lambda}^{-\frac{{\tt h}}{4}} \ .\nonumber
\eea
The universal enveloping algebra  $U_q(\mathfrak{sl}_2)$ admits a $2j+1$ dimensional representation
$(j=\frac{1}{2},1,\frac{3}{2},\ldots)$, so that
\be
\bm{L}_j(\lambda)=\pi_j\big({\boldsymbol L}(\lambda)\big)\,,\qquad\qquad
\bar{\bm{L}}_j(\bar{\lambda})=\pi_j\big(\bar{{\boldsymbol L}}(\bar{\lambda})\big)
\ee
are $(2j+1)\times (2j+1)$ operator valued matrices.
Following the  same line of arguments as in ref.\cite{Bazhanov:1998dq}, one
can show that these satisfy the Yang-Baxter 
algebra of the form
\bea\label{YBeq1a}
 R_{jj'}
 \big(\sqrt{\lambda_1/\lambda_2}\,\big)\ 
 \big({\boldsymbol L}_{j}
 (\lambda_1)
 \otimes {\boldsymbol 1 }\big)\big({\boldsymbol 1}\otimes {\boldsymbol L}_{j'}
(\lambda_2)\big)&=&
 \big({\boldsymbol 1}\otimes {\boldsymbol L}_{j'}
 (\lambda_2)\big)\big({\boldsymbol L}_{j}
(\lambda_1)\otimes {\boldsymbol 1 }\,\big)\,
 R_{jj'}
 \big(\sqrt{\lambda_1/\lambda_2}\,\big) \nonumber \\[0.2cm]
 R_{jj'}
 \Big(\sqrt{\bar{\lambda}_2/\bar{\lambda}_1}\,\Big)\ 
 \big(\bar{\boldsymbol L}_{j}
 (\bar{\lambda}_1)
 \otimes {\boldsymbol 1 }\big)\big({\boldsymbol 1}\otimes \bar{\boldsymbol L}_{j'}
(\bar{\lambda}_2)\big)&=&
 \big({\boldsymbol 1}\otimes \bar{\boldsymbol L}_{j'}
 (\bar{\lambda}_2)\big)\big(\bar{\boldsymbol L}_{j}
(\bar{\lambda}_1)\otimes {\boldsymbol 1 }\,\big)\,
 R_{jj'}
 \Big(\sqrt{\bar{\lambda}_2/\bar{\lambda}_1}\,\Big) \nonumber \\[0.2cm]
 \big({\boldsymbol L}_{j}
 (\lambda)
 \otimes {\boldsymbol 1 }\big)\,
\big({\boldsymbol 1}\otimes \bar{\boldsymbol L}_{j'}
(\bar{\lambda})\big)&=&
\big({\boldsymbol 1}\otimes \bar{\boldsymbol L}_{j'}
(\bar{\lambda})\big)\,
\big({\boldsymbol L}_{j}
 (\lambda)
 \otimes {\boldsymbol 1 }\big)
\ .
\eea
Here $ R_{jj'}(\lambda)$ is the trigonometric solution to the Yang-Baxter equation
which acts in the tensor product $\pi_j\otimes\pi_{j'}$ and in particular
 \bea\label{YBeq2a}
 R_{\text{\textonehalf\textonehalf}}(\lambda)
  \,=\, 
 \begin{pmatrix}
 q^{-1}\lambda-q\lambda^{-1}&0&0&0\\
 0&\lambda-\lambda^{-1} &q^{-1}-q &0\\
 0&q^{-1}-q & \lambda-\lambda^{-1}& 0\\
 0&0&0&q^{-1}\lambda-q\lambda^{-1}
 \end{pmatrix}
  \ .
 \eea
Notice that 
in the first line of \eqref{YBeq1a} the $R$-matrix depends on the ratio $\lambda_1/\lambda_2$,
while in the second line it depends on $\bar{\lambda}_2/\bar{\lambda}_1$.
 This is because in the definition \eqref{Laltcase1a},
the path ordering for ${\boldsymbol L}(\lambda)$ is opposite to that of
$\bar{{\boldsymbol L}}(\bar{\lambda})$. 
An immediate consequence of the algebraic relations \eqref{YBeq1a}
is that the operators
\bea\label{asusuya}
{\boldsymbol{\tau}}(\lambda)={\rm Tr}\Big[\re^{\frac{\ri\pi}{\sqrt{n+2}}\,a_0\sigma^3}\,
{\boldsymbol L}_{\text{\textonehalf}}(\lambda)\Big]\,,
\qquad\qquad
\bar{{\boldsymbol{\tau}}}(\bar{\lambda})={\rm Tr}\Big[\,
\bar{\boldsymbol L}_{\text{\textonehalf}}(\bar{\lambda})\,\re^{-\frac{\ri\pi}{\sqrt{n+2}}\,{\bar a}_0\sigma^3}\,\Big]
\eea
obey the commutativity conditions
\be
\big[{\boldsymbol{\tau}}(\lambda),\,{\boldsymbol{\tau}}(\lambda')\big]=
\big[\bar{{\boldsymbol{\tau}}}(\bar{\lambda}),\,\bar{{\boldsymbol{\tau}}}(\bar{\lambda}')\big]=
\big[{\boldsymbol{\tau}}(\lambda),\,\bar{{\boldsymbol{\tau}}}(\bar{\lambda})\big]=0\ .
\ee
\smallskip

As in the homogeneous case,  ${\boldsymbol{\tau}}(\lambda)$
commutes with ${\mathlarger{\mathlarger{\mathlarger{\mathlarger {\bf \it a}}}}}_\pm(\lambda')$
and
satisfies the relation
  \bea\label{iaisaisa}
  \boldsymbol{\tau}(\lambda){\mathlarger{\mathlarger{\mathlarger {\bf \it a}}}}_+(\lambda)=
  \re^{+\frac{2\ri\pi}{\sqrt{n+2}}\, a_0}\,
{\mathlarger{\mathlarger{\mathlarger {\bf \it a}}}}_+(q^{+2}\lambda)  +
  \re^{- \frac{2\ri\pi}{\sqrt{n+2}}\, a_0}\,
{\mathlarger{\mathlarger{\mathlarger {\bf \it a}}}}_+(q^{-2}\lambda)\ .
  \eea
However, it deserves to be mentioned that now 
$ \boldsymbol{\tau}(\lambda)$ and 
${\mathlarger{\mathlarger{\mathlarger{\mathlarger {\bf \it a}}}}}_\pm(\lambda)$  
possess a power series expansion in $\lambda$ rather than $\lambda^2$.
The similar statements hold true for the barred counterparts $\bar{\bm{\tau}}(\bar{\lambda})$
and $\bar{{\mathlarger{\mathlarger{\mathlarger{\mathlarger {\bf \it a}}}}}}_\pm(\bar{\lambda})$.
The latter
is defined by the formulae analogous to \eqref{vert1ad}-\eqref{Lop2ad}.
\bigskip

It is instructive to consider explicitly the first few terms in the Taylor series for the vacuum eigenvalues
of $ \boldsymbol{\tau}(\lambda)$. From the definition \eqref{asusuya} it follows that
\bea\label{texpeq1a}
\tau^{(\rm vac)}(\lambda\,|\,p,s)&=&2\, \cos(\tfrac{2\pi p}{n+2})-2s\, Q_1\big( \tfrac{p}{n+2},  \tfrac{1}{n+2}\big)\,
\lambda\\
&+&
\Big(4 s^2\ { Q}_2\big( \tfrac{p}{n+2},  \tfrac{1}{n+2}\big) 
-2n\,{\tilde Q}_2\big( \tfrac{p}{n+2},  \tfrac{1}{n+2}\big)\Big)\ \lambda^2+O(\lambda^3)\ ,\nonumber
\eea
where $Q_{1,2}(h,g)$ are given in eq.\,\eqref{QQeig1a}, while
\bea\label{oiaso81}
 {\tilde Q}_2(h,g)&=&\int_0^{2\pi}\rd u_1\int_0^{u_1}\rd v_1\int_0^{v_1}\rd u_2\int_0^{u_2}\rd v_2\,
 \Big(2\sin\big(\tfrac{u_1-u_2}{2}\big)\Big)^{2g-2}  \Big(2\sin\big(\tfrac{v_1-v_2}{2}\big)\Big)^{2g}
 \nonumber\\[0.2cm]
 &\times&\Big(2\sin\big(\tfrac{u_1-v_1}{2}\big)\Big)^{-2g}  \Big(2\sin\big(\tfrac{u_1-v_2}{2}\big)\Big)^{-2g}
 \Big(2\sin\big(\tfrac{v_1-u_2}{2}\big)\Big)^{-2g}  \Big(2\sin\big(\tfrac{u_2-v_2}{2}\big)\Big)^{-2g}
 \nonumber\\[0.2cm]
 &\times&2\, \cos\big(2 h(\pi-u_1-u_2+v_1+v_2)\big)\ .
 \eea
Remarkably, the four-fold integral $ {\tilde Q}_2(h,g)$ may be computed analytically.
Indeed,  using eqs.\,\eqref{texpeq1a} and \eqref{logaseries1a} to expand
 both sides of \eqref{iaisaisa}  in $\lambda$,
and comparing the coefficient of $\lambda^2$ from both sides of that equation, one can
express $ {\tilde Q}_2(h,g)$ in terms of the vacuum eigenvalues of ${\bf J}_1^{(+)}$ and ${\bf J}_2^{(+)}$.
Then with  eq.\,\eqref{Jvaceig1a} at hand  one finds
 \bea\label{paospo1a}
 {\tilde  Q}_2(h,g)=\frac{g}{2(2g-1)}\,
\frac{\pi^2\ \Gamma(1-4 g)}
{\Gamma(1-2 g+2 h)\Gamma(1-2 g-2 h)} \ \frac{\Gamma^4(-g)}
{\Gamma^2(-2 g)}\ .
 \eea
The last formula shows that the integral in the r.h.s. of \eqref{oiaso81} converges only
in the left half plane $\Re e(g)<0$. Nevertheless, \eqref{paospo1a} provides an analytic continuation
of this multi-fold integral to the whole complex plane. Note that $Q_1(h,g)$ and $Q_2(h,g)$ \eqref{QQeig1a}
converge for any $\Re e(g)<\frac{1}{2}$.
\bigskip

The analysis of the vacuum eigenvalues  leads one to conclude that
the definition \eqref{asusuya}, understood as a series expansion
involving ordered integrals over the vertex operators,
can not be taken literally for any $n>0$.
This is an important difference to the homogeneous case, where the 
 expression
\eqref{asusuya231} makes sense in the domain $0<\beta^2<\frac{1}{2}$.
For the ${\cal Z}_2$ invariant model,
the formulae for $\bm{\tau}(\lambda)$ and $\bar{\bm{\tau}}(\bar{\lambda})$  
 \eqref{asusuya} as well as eq.\,\eqref{soso1aa} that defines
${\mathlarger{\mathlarger{\mathlarger{\mathlarger {\bf \it a}}}}}_\pm(\lambda)$
may only be understood via analytic continuation in complex $n$.
The latter is achieved   by re-writing the ordered integrals in terms of the contour integrals 
following the procedure explained in the work \cite{Bazhanov:1998dq}.

\bigskip

The scaling limit of the eigenvalue of the
transfer matrix corresponding to the RG trajectory $\bm{\Psi}_N$
may be obtained through a comparison of 
eqs.\,\eqref{TArel1a},\,\eqref{ffunc1a} where $\eta_J = \ri\,(-1)^{J+1}$
and the scaling counterpart \eqref{iaisaisa}. Keeping in mind the formula
\eqref{as56d1a}  describing the scaling limit of ${A}_+(\zeta)$ as well as 
 \eqref{iaosid9898}, one finds
\be\label{sysaysaddd}
{\rm s}\!\!\!\!\lim_{N\to \infty\atop b(N)\to s} G^{(N/2)}\big(-q^2 \mu^2 \,|\,\tfrac{2}{n+2}\big)
G^{(N/2)}\big( -q^{-2} \mu^2 \,|\, \tfrac{2}{n+2}\big)\,
T^{(N)}\big(\,  \big(N/(2N_0)\big)^{-\frac{n}{n+2}}\  \ri\mu\big)=(-1)^{\tt w}\,{ \tau}(\lambda)\, ,
\ee
where ${\tau}(\lambda)={\tau}(\lambda\,|\,\bm{w},p,s)$ stands for the eigenvalue of the operator $\bm{\tau}(\lambda)$
on the state $\bm{\psi}_{p,s}(\bm{w})$.
Recall that $q=\re^{\frac{\ri\pi}{n+2}}$ while   \eqref{mulambda1a} provides the relation between
 $\lambda$ and $\mu$. Contrary to the homogeneous case the sign factor $(-1)^{\tt w}$ does not show
up in the formula \eqref{tower1b} for the eigenvalues of the lattice translation operator $\mathbb{K}$
on $\bm{\Psi}_N$.
However, in the sector of low energy states, one can define the operator $\sqrt{\mathbb{K}}$,
which belongs to the commuting family of operators and whose eigenvalues on $\bm{\Psi}_N$
are given by\footnote{%
Notice that $\sqrt{\mathbb{K}}$ does not coincide with the one\,-\,site lattice translation operator ${\cal K}$,
whose matrix elements are
\be
\big({\cal K}\big)_{a_{N} a_{N-1}\ldots
  a_1}^{b_{N}b_{N-1}\ldots b_1}=\re^{\ri\pi{\tt k}\,a_1}
\,\delta_{a_N}^{b_{N-1}}\,\delta_{a_{N-1}}^{b_{N-2}}\,\ldots\,
\delta_{a_1}^{b_{N}}\ .\nonumber
\ee
 Despite that ${\cal K}^2=\mathbb{K}$, the one\,-\,site translation does not commute with the transfer matrix.
As is discussed  in sec.\,7 of ref.\cite{Bazhanov:2020new},
\be
{\cal K}^{-1}\,{\mathbb{T}}(\zeta)\,{\cal K}=\hat{{\cal D}}\ \mathbb{T}(\zeta)\,\hat{{\cal D}}=
{\mathbb{T}}(-\zeta)\ .\nonumber
\ee}
\be\label{eigsqrtK}
\sqrt{K}= (-1)^{\tt w}
\exp\bigg(\frac{2\pi\ri}{N}\,\bigg( \frac{{p}^2-{\bar p}^2}{n+2}+ {\tt L}-\bar{\tt L}\bigg)\bigg)\ .
\ee
Then  eq.\,\eqref{sysaysaddd}   may be rewritten in the operator form
\be\label{8s9saa}
{\rm s}\!\!\!\!\lim_{N\to \infty\atop b(N)\to s}G^{(N/2)}\big(-q^2 \mu^2 \,|\,\tfrac{2}{n+2}\big)\ 
G^{(N/2)}\big( -q^{-2} \mu^2 \,|\, \tfrac{2}{n+2}\big)\,
\mathbb{T}\big(\,  \big(N/(2N_0)\big)^{-\frac{n}{n+2}}\ \ri\mu\big)\,\sqrt{\mathbb{K}}=\bm{ \tau}(\lambda)\ .
\ee
The ``barred'' version of the above relation reads as
\be\label{8s9saabar}
{\rm s}\!\!\!\!\lim_{N\to \infty\atop b(N)\to s}G^{(N/2)}\big(-q^2\bar{ \mu}^{-2} \,|\,\tfrac{2}{n+2}\big)\ 
G^{(N/2)}\big( -q^{-2} {\bar \mu}^{-2} \,|\, \tfrac{2}{n+2}\big)\,
\mathbb{T}\big(\, \big(N/(2N_0)\big)^{+\frac{n}{n+2}}\ 
(\ri {\bar \mu})^{-1}\big)\,\sqrt{\mathbb{K}}=\bar{\bm{ \tau}}(\bar{\lambda})\,,
\ee
where ${\bar\mu}$ is given in terms of ${\bar\lambda}$ similar to \eqref{mulambda1a}:
\be\label{mulambda1ass}
{\bar \mu}=-\ri\ (n+2)^{-\frac{2}{n+2}}\ \Gamma^2\big(-\tfrac{1}{n+2}\big)\ {\bar \lambda}\ .
\ee
\medskip

There is an alternative way to define $\sqrt{\mathbb{K}}$.
To this end, consider the operators
\bea\label{oiasoid1a}
\mathbb{K}^{(\pm)}&=&\re^{\ri\pi{\tt k}}\,q^{-\frac{N}{2}+\mathbb{S}^z}
\ \mathbb{A}_+(\mp\ri q^{+1})\,\big[\mathbb{A}_+(\mp\ri q^{-1})\big]^{-1}\ .
\eea
As it follows from eqs.\,\eqref{Keigeq1a},\,\eqref{Beq1} the lattice translation operator $\mathbb{K}$ and the
quasi-shift $\mathbb{B}$ are expressed in terms of \eqref{oiasoid1a} as follows
\bea\label{pospo1a}
\mathbb{B}=\mathbb{K}^{(+)}\,\big(\mathbb{K}^{(-)}\big)^{-1}\ ,\ \ \ \ \ \ {\mathbb K}=
\mathbb{K}^{(+)}\,\mathbb{K}^{(-)}\ .
\eea
It was discussed  in sec.\,\ref{sec8} that
for the low energy states  one can unambiguously introduce the operator
$\frac{n}{4\pi}\log\mathbb{B}$. Its eigenvalues 
are equal to $b(N)$ that appears in eq.\,\eqref{tower1a} describing the low energy spectrum of the lattice Hamiltonian
and lie in the strip $|\Im m\big(b(N)\big)|<\frac{n}{4}$ (see \eqref{aisodio1231}).
This allows one to define the operator
$\sqrt{\mathbb{B}}$, acting on the low energy states,
with eigenvalues given by $\re^{\frac{2\pi}{n}b}$. 
Our numerical work confirms the relation
\be\label{idosi9a9a}
\mathbb{K}^{(\pm)}=\sqrt{\mathbb{K}}\,\big(\sqrt{\mathbb{B}}\,\big)^{\pm 1}\,.
\ee
The latter, instead of \eqref{eigsqrtK},
may be used to introduce the operator  $\sqrt{\mathbb{K}}$.

\bigskip

\section{Local integrals of motion and the chiral states $\bmT{\psi}_{p,s}(\bmT{w})$\label{sec15}}
In the scaling limit the low energy Bethe states take the form 
$\bar{\bm{\psi}}_{\bar{p},s}(\bar{\bm{w}})\otimes {\bm{\psi}}_{{p},s}(\bm{w})$.
The chiral states
may be interpreted as states in the Fock spaces, based on
the diagonalization problem of ${\mathlarger{\mathlarger{\mathlarger{\mathlarger {\bf \it a}}}}}_\pm(\lambda)$
and $\bar{{\mathlarger{\mathlarger{\mathlarger{\mathlarger {\bf \it a}}}}}}_\pm(\bar{\lambda})$.
The  latter, being defined in terms of a path-ordered exponential, 
 are difficult to work with for any practical calculations.
As in the homogeneous case, for the explicit construction of ${\bm{\psi}}_{{p},s}(\bm{w})\in{\cal F}_{{\bf P}}$
it turns out to be most convenient to diagonalize the operators which occur 
 in the large $\lambda$ asymptotic of
${\mathlarger{\mathlarger{\mathlarger{\mathlarger {\bf \it a}}}}}_\pm(\lambda)$ and/or $\bm{\tau}(\lambda)$.
Among these are the so-called local Integrals of Motion (IM).
 For $n>2$ they are the only operators
that appear in the large $\lambda$ expansion of $\bm{\tau}(\lambda)$.
It follows from the results of the work \cite{Fateev:2005kx} that as $\lambda\to\infty$
\be\label{9sd98f989sd1a}
\log\bm{\tau}(\lambda)\asymp 
\begin{cases}
-
2\pi \sum\limits_{m=-1}^\infty c_m\,{\bf I}_m\ \big(+(n+2) \lambda\big)^{-\frac{(n+2) m}{n}}\ \ \ \ \ \ \ &\Re e(\lambda)>0\\[0.4cm]
+
2\pi \sum\limits_{m=-1}^\infty c_m\,(-1)^{m}\ {\bf I}
_m\ \big(-(n+2) \lambda\big)^{-\frac{(n+2) m}{n}}\ \ \ \ \ \ \ &\Re e(\lambda)<0
\end{cases}\ \ \ \ \ \ \ (n>2)\ .
\ee
Here
${\bf I}_{-1}=1$, ${\bf I}_0=\int_{0}^{2\pi}\frac{{\rm d} u}{2\pi}\ \partial\vartheta=b_0$,
while the non-trivial local IM $\{{\bf I}_m\}_{m=1}^\infty$ have the form
\bea\label{Idef1b}
{\bf I}_{m}=\int_0^{2\pi}\frac{\rd u}{2\pi}\ T_{m+1}(u)
\eea
with $T_{m+1}(u)$ being a chiral local density, i.e.,  a differential polynomial in 
$\partial\varphi(u)$ and $\partial\vartheta(u)$,
 of Lorentz spin $m+1$.
Another way to formulate the last condition
is to assign a grade $1$ to $\partial\vartheta(u)$, $\partial\varphi(u)$ as well as the
derivative. Then $T_{m+1}$ is a homogeneous polynomial
in the chiral fields and their derivatives of grade $m+1$.
The first few densities read explicitly as \cite{Fateev:1995ht,Fateev:2005kx}
\bea\label{ioasido}
T_2&=&(\partial\vartheta)^2 +(\partial \varphi)^2\nonumber\\[0.2cm]
T_3&=&(\partial\vartheta)^3+\frac{3(n+2)}{3n+4}\ (\partial\varphi)^2\partial\vartheta+\frac{3\ri(n+1)\sqrt{n+2}}{3n+4}\ 
\partial^2\varphi\,\partial\vartheta\\[0.2cm]
T_4&=&(\partial\vartheta)^4-\frac{n^2-2}{5n+6}\,(\partial^2\vartheta)^2\ +
\frac{6\,(n+2)}{5n+6}\,(\partial\vartheta)^2\,(\partial\varphi)^2+\frac{6\ri\,(n+1)\sqrt{n+2}}{5n+6}\ 
(\partial\vartheta)^2\,\partial^2\varphi
\nonumber\\[0.2cm]
&-&\frac{(n+1)^2}{5n+6}\ (\partial^2\varphi)^2+\frac{n+2}{5n+6}\ (\partial\varphi)^4\ .
\nonumber
\eea
Notice that the local IM are defined up to an overall normalization. 
If we take 
$
T_{m+1}=(\partial\vartheta)^{m+1}+\ldots\ ,
$
 where the ``$\ldots$''
denote terms containing lower powers of $\partial\vartheta$,
then the numerical coefficients $c_m$ in \eqref{9sd98f989sd1a} are given by
\bea\label{oiasiod8989}
c_m=
\frac{2^m\,\Gamma(\frac{1}{2}+\frac{n+1}{n} m)}{\sqrt{\pi}\,(m+1)!\,\Gamma(1+\frac{m}{n})}\ 
\big(1+\tfrac{2}{n}\big)^{-m}\ \Big[ \Gamma\big(1-\tfrac{1}{n+2}\big)\Big]^{-\frac{2(n+2)m}{n}}\ 
\,n^{-\frac{m+1}{2}}\ .
\eea
\bigskip

The local IM act invariantly in the level subspace of the Fock space.
Restricted to ${\cal F}_{{\bf P}}^{({\tt L})}$ they are given by 
a  sum of a finite number of terms involving the Heisenberg generators 
\eqref{hasaast}. This makes the computation of the
matrix elements of ${\bf I}_m$ and, in turn, the diagonalization problem
\bea\label{898s981a}
{\bf I}_{m}\, \bm{\psi}_{p,s}(\bm{w})=I_m({\bm w},p,s)\, \bm{\psi}_{p,s}(\bm{w})
\eea
a straightforward task -- much simpler than the spectral problem of
${\mathlarger{\mathlarger{\mathlarger{\mathlarger {\bf \it a}}}}}_\pm(\lambda)$.
The Fock highest state is, of course, an eigenstate and the 
corresponding eigenvalues for the first few IM can be easily extracted
from the explicit formulae \eqref{ioasido}:
\bea\label{ciics98as}
I_1^{({\rm vac})}(p,s)&=&\frac{p^2}{n+2}+\frac{s^2}{n}-\frac{1}{12}  \nonumber\\[0.2cm]
I_2^{({\rm vac})}(p,s)&=& \frac{s}{\sqrt{n}}\ 
\Big(\,\frac{3p^2}{3n+4}+\frac{s^2}{n}-\frac{(2n+3)}{4(3n+4)}\,\Big)
 \\
I_3^{({\rm vac})}(p,s)&=&
\frac{p^4}{(5n+6)(n+2)}-\frac{p^2}{2(5n+6)}+
\frac{6p^2\,s^2}{n(5n+6)}+\frac{s^4}{n^2}-\frac{(3n+4)\,s^2}{2n(5n+6)} 
\nonumber \\[0.2cm]
&-&\frac{(n-6)\,(2n+3)}{240\,(5n+6)}\ .\nonumber
\eea
The vacuum and higher level 
eigenvalues may be alternatively obtained through a WKB analysis of the ODE \eqref{aisausau}.
It turns out that $I_m({\bm w},p,s)$
is a symmetric polynomial in $\bm{w}=\{w_j\}_{j=1}^{{\tt L}}$ of degree $m-1$.
For instance,
\bea\label{9d8s9889a}
I_1({\bm w},p,s)&=&I_1^{(\rm vac)}\big(\sqrt{p^2+(n+2)\, {\tt L} },s\, \big)\nonumber\\
I_2({\bm w},p,s)&=&I_2^{(\rm vac)}\big(\sqrt{p^2+(n+2)\, {\tt L} }, s\, \big)+\frac{3\ri \sqrt{n}}{3n+4}\ \sum_{j=1}^{\tt L}w_j\\
I_3({\bm w},p,s)&=&I_3^{(\rm vac)}\big(\sqrt{p^2+(n+2)\, {\tt L}},s\, \big)-
\frac{4}{(5n+6)(n+2)}\ \bigg(n\sum_{j=1}^{\tt L}w_j^2-
\ri s\,(n+4)\, \sum_{j=1}^{\tt L}w_j\bigg)\ .
\nonumber
\eea
\bigskip

It is expected that the joint spectrum of the local IM lifts all the degeneracies in ${\cal F}_{\bf P}^{({\tt L})}$
so that the state $\bm{\psi}_{p,s}(\bm{w})$ is uniquely
specified by the eigenvalues  $I_m({\bm w},p,s)$.
For  ${\tt L}\le 5$
we found it sufficient to use just the first three IM, along with formulae \eqref{9d8s9889a},
to obtain $\bm{\psi}_{p,s}(\bm{w})$ for some given set $\bm{w}$.
\bigskip

The local IM  also appear in the large $\lambda$ asymptotic expansion
for ${\mathlarger{\mathlarger{\mathlarger{\mathlarger {\bf \it a}}}}}_\pm(\lambda)$. 
However, unlike  eq.\,\eqref{9sd98f989sd1a} for $\bm{\tau}(\lambda)$,
the latter involves the so-called dual non-local IM  as well. 
The simplest of these are the
operators $\check{{\bf C}}^{(\pm)}$:
\be\label{aosido1201}
\check{{\bf C}}^{(\pm)}\,\bm{\psi}_{p,s}(\bm{w})=\check{{   C}}_{p,s}^{(\pm)}(\bm{w})\,\bm{\psi}_{p,s}(\bm{w})
\qquad \quad {\rm with} \qquad\quad\check{{   C}}_{p,s}^{(\pm)}(\bm{w})=
{ \mathfrak C}_{p,s}^{(\pm)}(\bm{w})/{\mathfrak C}_{p,s}^{(0,\pm)}\ .
\ee
Here ${ \mathfrak C}_{p,s}^{(\pm)}(\bm{w})$  are the coefficients that enter into the asymptotic formula
\eqref{Apeq1}, while
 ${\mathfrak C}_{p,s}^{(0,\pm)}={ \mathfrak C}_{p,s}^{(\pm)}(\bm{w})\big|_{{\tt L}=0}$ 
are given  in \eqref{iaosdi012891}. This way, as indicated by the ``check'' symbol, the
operators  are normalized so that their eigenvalue on  the Fock highest state is one. 
Earlier, we used the functions $\check{D}_{p,s}(\bm{w})$ and $\check{R}_{p,s}(\bm{w})$
(see, e.g., eqs.\,\eqref{oiaodi1a1a} and \eqref{oasid89129812}), which coincide with
 the eigenvalues of the reflection operators 
\be\label{DRope0823}
\check{{\bf D}}=\check{{\bf C}}^{(+)}\,\big(\check{{\bf C}}^{(-)}\big)^{-1}\,,\qquad\quad
\qquad
\check{{\bf R}}=\check{{\bf C}}^{(+)}\,\big(\check{{\bf C}}^{(-)}\big)^{-1}\ .
\ee
The construction of  $\check{\bf C}^{(\pm)}$,
$\check{{\bf D}}$ and $\check{{\bf R}}$ as operators acting in the Fock space, as well as 
a closed analytic expression for their eigenvalues in terms of the sets
$\bm{w}$  is given in sec.\,3 of ref.\,\cite{Kotousov:2019nvt}.
For the reader's convenience, we quote the formulae
for $\check{D}_{p,s}(\bm{w})$ and $\check{R}_{p,s}(\bm{w})$ in
Appendix \ref{app2}.
We  found the diagonalization problem
of the reflection operators useful for the construction of the
chiral states $\bm{\psi}_{p,s}(\bm{w})$.

\section{Extended conformal symmetry\label{sec16}}

\subsection{The $W_\infty$\,-\,algebra}

The graded linear space ${\cal V}_{p,s}=\bigoplus_{\tt L}{\cal V}_{p,s}^{({\tt L})}$ \eqref{FOck39933}
 for  real $s$  as well as the spaces 
${\cal V}_{p,s}$ with $s=\pm\ri\mathfrak{q}_a,\pm\ri\bar{\mathfrak{q}}_a$
and
${\cal V}_{\rho,\nu}$ with $(\rho,\nu)=(p_+\pm\ri\mathfrak{q}_a'),\,(p_-\pm\ri\bar{\mathfrak{q}}_a')$
\eqref{iaoisd8182}  
 are the building blocks of  the right chiral components of
the conformal towers in
${\cal H}_{S^z}^{({\rm cont})}$ and ${\cal H}_{S^z}^{({\rm disc},\pm)}$. 
The operators from the commuting family  generated by 
${\mathlarger{\mathlarger{\mathlarger{\mathlarger {\bf \it a}}}}}_\pm(\lambda)$,
including the local IM, act invariantly inside these spaces. However
the local densities $T_{m+1}(u)$, which occur in the definition of the local  IM \eqref{Idef1b} do not,
in general, act invariantly
in ${\cal V}_{p,s}$ for pure imaginary $s$ when
the graded space does not coincide with the Fock space.
Since the local densities are defined up to a total derivative one could try to
choose them in such a way so that the chiral spaces ${\cal V}_{p,s}$
are invariant w.r.t. their action both for real or pure imaginary $s$.
The most general form of the spin $2$ density is
$W_2=T_2+\alpha_1\,\partial^2\vartheta+\alpha_2\,\partial^2\varphi$,
where $T_2$ is given in \eqref{ioasido} and $\alpha_{1},\,\alpha_2$ are arbitrary constants. This local field would leave
${\cal V}_{p,s}$ invariant provided it commutes with the two screening charges \eqref{ioasd90023}, i.e.,
\be
W_2(u)\,{\tt Q}_{\sigma} = {\tt Q}_\sigma\, W_2(u)\qquad \qquad(\sigma=\pm)\ .
\ee
The commutativity condition fixes $W_2(u)$ to be
\be\label{w2iosdi}
W_2=(\partial\vartheta)^2+(\partial \varphi)^2+\frac{\ri}{\sqrt{n+2}}\ \partial^2\varphi\ .
\ee 
Then a simple calculation shows that  $W_2(u)$ satisfies the Operator Product Expansion (OPE) 
\be\label{oias90190321}
 W_2(u)\,W_2(0)=\frac{c}{2 u^4}-\frac{2}{u^2}\ W_2(0)-\frac{1}{u}\ \partial W_2(0)+O(1)
\ee
with
\be\label{iaosid321}
c=\frac{2\,(n-1)}{n+2}\ .
\ee
In turn the modes ${\widetilde W}_2(m)$, defined through the Fourier series
 \bea\label{Wmodes1aa}
 W_{2}(u)=-\frac{c}{24}+\sum_{m=-\infty}^\infty {\widetilde W}_2(m)\ \re^{-\ri m u}\,,
 \eea
form the Virasoro algebra with the central charge \eqref{iaosid321}.
Thus the chiral spaces ${\cal V}_{p,s}$ can be classified according to the irreps
of this conformal symmetry algebra. Note that the local IM ${\bf I}_1$
coincides with the zero mode ${\widetilde W}_2(0)$ up to an additive constant and
its eigenvalue is  related to the conformal dimension of a state as
\be\label{aiosd10920091}
I_1=\Delta-\frac{c}{24}\,.
\ee
\bigskip

The conformal algebra admits a natural extension. Clearly a local field defined
through the commutator  $\big[W_2(u),\,{\bf I}_2\big]$ 
acts invariantly in ${\cal V}_{p,s}$ for any values of real or pure imaginary $s$.
An explicit calculation shows that
\be\label{02ioasid90}
\partial W_3(u)=\frac{3n+4}{3\ri(n+2)}\ \big[W_2(u),\,{\bf I}_2\big]\ ,
\ee
for\footnote{%
The currents $W_2$ \eqref{w2iosdi} and $W_3$ \eqref{W3def1a},
along with the screening charge ${\tt Q}_\sigma$ \eqref{ioasd90023},
were originally obtained in the work \cite{Fateev:1987vh}.
See the footnote on page 649 therein.}
\be\label{W3def1a}
W_3=
\frac{6n+8}{3n+6}\, (\partial \vartheta)^3+2\,
 (\partial \varphi)^2\partial \vartheta+\ri\sqrt{n+2}\ \partial^2 \varphi\,\partial\vartheta
-\frac{\ri n}{\sqrt{n+2}}\ \partial\varphi\,\partial^2\vartheta+\frac{n}{6(n+2)}\  \partial^3\vartheta \ .
\ee
The choice of the overall factor in the definition of $W_3(u)$ is somewhat arbitrary 
and we take it to be $\frac{6n+8}{3n+6}$ for future convenience.
Computing the OPE of $W_2$ and $W_3$ yields  
\be
W_2(u)\,W_3(0)=-\frac{3}{u^2}\ W_3(0)-\frac{1}{u}\ \partial W_3(0)+O(1)\,,
\ee
which means that $W_3(u)$ is a primary chiral field of spin $3$.
Similar to \eqref{Wmodes1aa} it can be expanded in the Fourier series
 \bea\label{Wmodes2}
 W_{3}(u)=\sum_{m=-\infty}^\infty {\widetilde W}_3(m)\ \re^{-\ri m u}\ .
 \eea
Notice that
the zero mode ${\widetilde W}_3(0)$ coincides with the local IM ${\bf I}_2$ 
\eqref{Idef1b},\,\eqref{ioasido} 
up to an overall factor,
\be\label{iaosid198289}
{\bf I}_2=\frac{3n+6}{6n+8}\ \, {\widetilde W}_3(0)\,,
\ee
so that the operators ${\widetilde W}_2(0)$ and ${\widetilde W}_3(0)$
can be diagonalized simultaneously.
\bigskip

The linear space of local spin $3$ fields, invariantly acting in ${\cal V}_{p,s}$, is 
generated by $W_3$ and $\partial W_2$.
As for the spin $4$ fields acting in ${\cal V}_{p,s}$, they include
the derivatives
 $\partial^2 W_2$, $\partial W_3$ as well as
the composite field $W_2^2$, which is defined as the first regular term
in the OPE \eqref{oias90190321}. There is one more linearly independent
spin $4$ field $W_4(u)$, which we introduce through
the OPE:
 \bea\label{aiisaisa1a2a}
W_3(u)\,W_3(0)&=&-\frac{c(c+7)(2c-1) }{9(c-2)u^6}+\frac{ (c+7)(2c-1)}{3(c-2)u^4}\ 
\big(W_2(u)+W_2(0)\big)-\frac{1}{u^2}\ \Big( W_4(u)+W_4(0)\nonumber \\[0.2cm]
&+&W^2_2(u)+W^2_2(0)
+\frac{2c^2+22c-25}{30 (c-2)}\,\big (\partial^2W_2(u)+\partial^2W_2(0)\big)\Big)+O(1)\ .
\eea
The definition of $W_4(u)$ is not unique and it is fixed as in \eqref{aiisaisa1a2a}
for the following reason. A priori, it would be natural to have 
$W_4(u)$ be a spin $4$ primary field. However it turns out that this is impossible 
to achieve  for any linear combination of $W_4$, $\partial^2 W_2$, $\partial W_3$ and
 $W_2^2$. With $W_4$ defined through  \eqref{aiisaisa1a2a}, the OPE of $W_4$ and
$W_2$,
\be\label{8989aioiaso}
 W_2(u)\,W_4(0)=\frac{(c+10)(17c+2)}{15 (c-2)\, u^4}
 \ W_2(0)-\frac{4}{u^2}\ W_4(0)-\frac{1}{u}\ \partial W_4(0)+O(1)\ ,
\ee
 does not contain the singular terms $\propto u^{-6}$ and $u^{-3}$.
Since the densities for the local IM are defined up to a total derivative, ${\bf I}_3$ \eqref{Idef1b},\,\eqref{ioasido} 
must be expressible as an integral over a  linear combination of 
 $W_4(u)$ and  $W_2^2(u)$.
A straightforward calculation yields
\be\label{iaosid198289A}
{\bf I}_3=
\frac{n+2}{(2n+3)(5n+6)}\int_0^{2\pi}\frac{{\rm d}u}{2\pi}\ \Big((n+2)\,W_4+(2n+3)\,W_2^2\Big)\ .
\ee
\bigskip

Continuing the process one can describe 
the linear space of local spin $j=2,3,4,\ldots$ fields that act invariantly in 
${\cal V}_{p,s}$. It turns out that a basis would consist of 
composite fields built from the $W$ fields of lower spin and their derivatives
as well as one extra field $W_j$. 
The latter, of course, is not uniquely defined
and can be generated through
the OPE of the $W$ fields of lower spin, similar to how $W_4$ is generated in the OPE
\eqref{aiisaisa1a2a}.
For generic values of $n>0$ the total number of linearly independent spin $j$ fields is
$N(j)={ d}_{0}(j)-{ d}_{1}(j-1)=1,2,4,6,11,\ldots\ $, where
the integers $d_a({\tt L})$ 
are described by eq.\,\eqref{Zdef1b}.\footnote{The 
expression for $N(j)$ follows from the formula for 
the character of the $W_\infty$\,-\,algebra  \eqref{charadegen1v} 
specialized to the case $|{\rho}|=\half,\ {\nu}=0$.}
In turn the densities for the  local IM  \eqref{Idef1b} are
 expressible as a  linear combination of such fields.
Since $T_{m+1}$ is defined up to a total derivative,
it can  be written as a 
sum of
 $N(m+1)- N(m)$ terms.
The corresponding coefficients may be fixed through the commutativity condition 
$[{\bf I}_m,\,{\bf I}_2]=0$.
One of them would remain undetermined, which  manifests
 the freedom in the overall 
normalization of ${\bf I}_m$.

\bigskip
There exists a simple way of obtaining the ``independent'' set of
local fields $\{W_j(u)\}_{j=2}^\infty$, which is based on the following observation.
Consider the pair of chiral  non-local fields of Lorentz spin $1-\frac{1}{n}$,
\bea\label{isisaiasi}
\xi_\pm(u)=n^{-1}\ \Big(\sqrt{n}\,\partial\vartheta\pm \ri \sqrt{n+2}\
\partial\varphi\Big)\ \re^{\pm \frac{2\vartheta}{\sqrt{n}}}(u)\ .
\eea
It is simple to check that they commute with the screening charges \eqref{iaosid09120912}
\be
\xi_+(u)\,{\tt Q}_{\sigma}= {\tt Q}_{\sigma}\,\xi_+(u)\,,\,\qquad
\xi_-(u)\,{\tt Q}_{\sigma}= {\tt Q}_{\sigma}\,\xi_-(u)\qquad\qquad (\sigma=\pm)\ .
\ee
Hence  the local fields occurring in the OPE
\bea\label{iaosido89812}
&&\!\!\!\!\!\xi_+(u)\,\xi_-(0)=-n^{-1}\, u^{-2(1-\frac{1}{n})}\ \bigg[1-\frac{n+2}{2 n}\ \big( W_2(u)+W_2(0)\big)\, u^2-
\frac{n+2}{2n\sqrt{n}}\ \big( W_3(u)+W_3(0)\big)\, u^3\nonumber \\[0.2cm]
&&\!\!\!\!\!- \frac{(n+2)^2}{4n^2\,(2n+3)}\ \Big( W_4(u)+W_4(0)-
\frac{n\,(2n+3)}{5\,(n+2)}\ \big(\partial^2\,W_2(u)+\partial^2\,W_2(0)\big)\Big)\, u^4+\ldots\bigg]
\eea
would also commute with ${\tt Q}_\sigma$.
A straightforward calculation shows that the coefficients of $u^2$ and $u^3$ involve the fields
 $W_2$ from \eqref{w2iosdi} and $W_3$ in \eqref{W3def1a}. 
The spin $4$ field $W_4$ appearing in the coefficient of $u^4$ is the same as the one
defined via the OPE \eqref{8989aioiaso}. 
One can choose the fields $W_j$ with $j\ge 5$ in such a way that the remaining terms in the expansion in 
\eqref{iaosido89812}, denoted by the ellipsis, coincide with
$\sum_{j\ge 5}\big(W_j(u)+W_j(0)\big)\,u^j$.
Notice that \eqref{isisaiasi} is the well known bosonization formula \cite{Zam1986,Wakimoto,Gerasimov:1989mz,Jayaraman:1989tu,Griffin:1990fg}, which  extends the notion
of the Fateev-Zamolodchikov 
${\cal Z}_n$ parafermions \cite{Fateev:1985mm} to the case when $n$ is non-integer.

\bigskip

The infinite set of local chiral fields $\{W_j(u)\}_{j=2}^\infty$ form a closed
operator algebra, in the sense that the singular part of the OPE of
$W_j(u)W_{j'}(0)$ is expressible in terms of composite fields built
out of the $W$ fields and their derivatives.
This  algebra was discussed in the work \cite{Bakas:1991fs}, 
and we will refer to it as the $W_\infty$\,-\,algebra.
Repeating all the above for the left chirality one would arrive at
a barred copy of the algebra, $\overline{W}_\infty$, for the currents
$\{\overline{W}_j(\bar{u})\}_{j=2}^\infty$. Thus the
algebra of extended conformal symmetry underlying the critical
behaviour of the ${\cal Z}_2$ invariant inhomogeneous six-vertex model
is  $\overline{W}_\infty\otimes W_\infty$. 
\bigskip

Let's make some comments
regarding our terminology, which was borrowed from
CFT \cite{Zamolodchikov:1985wn}. 
In the description of a Lorentz invariant $1+1$D quantum field theory one employs
the space-time co-ordinates $x^\mu=(t,x)$.  
For a CFT in finite volume,
the space co-ordinate can be rescaled so that $x$
belongs to the segment of length $2\pi$.
Moreover it is always possible to choose
the unit measurement of time
such that the ``speed of light'' is one.
Then it is convenient to use the light cone co-ordinates
\be\label{hasyya}
u=t+x\,,\qquad \qquad \bar{u}=t-x\qquad \qquad (0\le  x\leq 2\pi)\ .
\ee
A theory with extended conformal symmetry possesses 
 chiral currents, which are local fields such that
$W_j(t,x)=W_j(u)$
and $\overline{W}_j(t,x)=\overline{W}_j(\bar{u})$
as a consequence of the equations of motion.
We use the convention that the (half-)integer $j$ coincides with the Lorentz spin 
in the case of $W_j(t,x)$ and minus the spin for $\overline{W}_j(t,x)$.
The theory with $\overline{W}_\infty\otimes W_\infty$ extended conformal symmetry
 contains, among the local fields, two infinite sets
currents
with $j=2,3,4,\ldots\ $, which are independent in the sense that no one current can be expressed
as a differential polynomial in the others.
The fields 
$W_2(u)$ and $\overline{W}_2(\bar{u})$ are naturally identified with
the holomorphic and anti-holomorphic components of the energy momentum tensor,
respectively.
Assuming  the boundary conditions  of the theory are such that 
the chiral currents are periodic, as is the case here,
\be
W_j(t,x)=W_j(t,x+2\pi)\,,\qquad\qquad \overline{W}_j(t,x)=\overline{W}_j(t,x+2\pi)\,,
\ee
they can be expanded in the Fourier series:
 \be\label{aoisd182918}
 W_{j}=-\frac{c}{24}\,\delta_{j,2}+\sum_{m=-\infty}^\infty {\widetilde W}_j(m)\ \re^{-\ri m u} \ ,\qquad
 \overline{W}_{j}=-\frac{c}{24}\,\delta_{j,2}+\sum_{m=-\infty}^\infty \widetilde{\overline{W}}_j(m)\ \re^{-\ri m \bar{u}}\ .
 \ee
As usual, the modes
$\widetilde{W}_2(m)$ and $\widetilde{\overline{W}}_2(m)$
generate two independent copies of the Virasoro algebra 
with central charge  $c$ and the CFT Hamiltonian is given by
\be
\hat{H}=\widetilde{W}_2(0)+\widetilde{\overline{W}}_2(0)-\frac{c}{12}\,\ .
\ee
The states in a $2$D CFT can be chosen to have a definite value of the pair of conformal dimensions
$(\bar{\Delta},\Delta)$.
 The corresponding CFT energy reads as
\be
E=\Delta+\bar{\Delta}-\frac{c}{12} \ ,
\ee
while the Lorentz spin coincides with the difference $\Delta-\bar{\Delta}$.
\bigskip

In the case of the ${\cal Z}_2$ invariant inhomogeneous  six-vertex model,
with the anisotropy parameter $q=\re^{\frac{\ri\pi}{n+2}}$ and $n>0$, 
the central charge is given by
$c=\frac{2(n-1)}{n+2}$ and lies in the interval
\be
-1<c<2\ .
\ee
The CFT energy appears in the large $N$ asymptotics of the eigenvalues of the lattice
Hamiltonian
while the Lorentz spin is related to the eigenvalue of the lattice translation operator.
Namely  \eqref{tower31} can be written as \cite{Cardy:1986ie}
\bea
{\cal E}&=&N e_{\infty}+\frac{4\pi v_{\rm F}}{N}\,
\Big(\,\Delta+\bar{\Delta}-\frac{c}{12}\,\Big)+o(N^{-1}) \nonumber\\[0.2cm]
K&=&\exp\bigg(\frac{4\pi\ri}{N}\,\big(\Delta-\bar{\Delta}\big)\bigg)\ .
\eea

\subsection{Highest weight irreps of the $W_\infty$\,-\,algebra\label{sec162}}

For a theory possessing extended conformal symmetry, the space
of states is naturally classified according to the highest weight irreps
of the symmetry algebra.
It is convenient to describe the latter in terms of the Verma module.
The Verma module of the $W_\infty$\,-\,algebra  contains the highest state
which is defined by 
the conditions
\be\label{aios3298}
\widetilde{W}_j(m)\,|\bm{\omega}\rangle=0 \qquad\qquad (\forall m>0)\,,\qquad\qquad
\widetilde{W}_j(0)\,|\bm{\omega}\rangle=\omega_j\,|\bm{\omega}\rangle\,,
\ee
where $\bm{\omega}=(\omega_2,\omega_3)$ is the highest weight.
The component $\omega_2$ is equal to the conformal dimension of the 
highest state and is
simply related to the eigenvalue of the local IM ${\bf I}_1$ \eqref{aiosd10920091}, 
while $\omega_3$ coincides up to an overall factor with the eigenvalue of  ${\bf I}_2$ \eqref{iaosid198289}.
It turns out that the 
 highest state $|\bm{\omega}\rangle$ is fully specified by the relations \eqref{aios3298}
 with $j=2,3$.
Moreover,
the Verma module is spanned by the states of the form
\be
\widetilde{W}_{2}(-l_1)\ldots\widetilde{W}_{2}(-l_m)\,
\widetilde{W}_{3}(-l_{1}')\ldots\widetilde{W}_{3}(-l_{m'}')|\bm{\omega}\rangle
\ee
with $1\le l_1\le l_2\le \ldots\le l_m$ and  $1\le l_{1}'\le l_{2}'\le \ldots\le l_{m'}'$,
which contain  the Fourier modes of the spin $2$ and spin $3$ currents only.
It  is a naturally graded linear space and the dimensions of its 
level subspace with  $\ell=\sum_i l_i+\sum_{i'} l'_{i'}$ is given by ${\rm par}_2(\ell)$.
Formulae \eqref{w2iosdi} and \eqref{W3def1a} introduce the structure of the $W_\infty$
Verma module in the Fock space ${\cal F}_{\bf P}$ with the highest weight
$\bm{\omega}=(\omega_2,\omega_3)$  related to  ${\bf P}=(\frac{\rho}{\sqrt{n+2}},\frac{\nu}{\sqrt{n}})$ as
\bea\label{Deltvarpi1a}
\omega_2&=&\frac{\rho^2-\frac{1}{4}}{n+2}+\frac{\nu^2}{n}  \\[0.2cm]
\omega_3&=&
\frac{2\nu}{\sqrt{n}}\,\Big(\,\frac{\rho^2}{n+2}+\frac{(3n+4)\,\nu^2}{3n\,(n+2)}-\frac{2n+3}{12\,(n+2)}\,\Big)\nonumber
\ .
\eea
In fact, it is convenient to use $\rho$ and $\nu$ to parameterize $\bm{\omega}$ 
without necessarily any reference to the Fock space. 
 Note that in this parameterization
the highest weight   depends only on $\rho^2$ so that $\rho$ should be identified with $-\rho$.
The  highest weight irrep of the $W_\infty$\,-\,algebra, with $\bm{\omega}$
parameterized by the pair $(\rho,\nu)$ as in \eqref{Deltvarpi1a},
will be denoted by  ${\cal W}_{\rho,\nu}\equiv{\cal W}_{-\rho,\nu}$.
\bigskip

For generic complex values of  $\rho$ and $\nu$ the Verma module is
an irrep of the $W_\infty$\,-\,algebra. Its character, 
\be\label{chdef1a}
{\rm ch}_{\rho,\nu}({\tt q})\equiv{\rm Tr}_{{\cal W}_{\rho,\nu}}\Big[\,{\tt q}^{\widetilde{W}_2(0)-\frac{c}{24}}\,\Big]\,,
\ee
with $c=\frac{2(n-1)}{n+2}$ is given by 
\be\label{iasodi1092091}
{\rm ch}_{\rho,\nu}({\tt q})=\frac{{\tt q}^{-\frac{1}{12}+\frac{\nu^2}{n}+\frac{\rho^2}{n+2}}}{({\tt q},{\tt q})_\infty^2}
\  \ \qquad\qquad \big(\,\rho,\,\nu\ \ {\rm generic}\,\big)\ .
\ee
When certain constraints are imposed on $\rho$ and $\nu$,
the Verma module contains null vectors
-- highest states occurring at non-zero levels. 
In this case the  highest weight irrep can be obtained
from the Verma module by factoring out 
all of the  invariant subspace(s) generated by the null vector(s).
As was demonstrated  in sec.\,\ref{iaso132218s}  
using  the Fock space realization of the Verma module,
 when
$\rho+\frac{1}{2}+\ri \nu=-a_+=0,\pm1,\pm2,\ldots$ 
there is a null vector $|\chi_+\rangle$ at the level $\big|a_++\frac{1}{2}\big|+\frac{1}{2}$.
Similarly if $-\rho+\frac{1}{2}+\ri \nu=-a_-=0,\pm1,\pm2,\ldots\ $, a null-vector $|\chi_-\rangle$
occurs at the level $\big|a_-+\frac{1}{2}\big|+\frac{1}{2}$.
Such Verma modules are usually referred to as degenerate.
It turns out that if either $ \rho+\tfrac{1}{2}+\ri\nu=-a\in\mathbb{Z}$ or
 $ \rho+\tfrac{1}{2}-\ri\nu=-a\in\mathbb{Z}$
and $2\rho\notin\mathbb{Z}$, the character \eqref{chdef1a} is given by \cite{Jayaraman:1989tu}
\be\label{charadegen1a}
{\rm ch}_{\rho,\nu}({\tt q})=\frac{{\tt q}^{-\frac{1}{12}+\frac{\nu^2}{n}+\frac{\rho^2}{n+2}}}{ ({\tt q},{\tt q})_\infty^{2}}
\ \
\sum_{m=0}^\infty (-1)^{m}\ {\tt q}^{m|a+\frac{1}{2}|+\frac{m^2}{2}}\qquad\qquad
\begin{array}{l}
\rho+\tfrac{1}{2}\pm\ri\nu\in\mathbb{Z}\  \\[0.2cm]
\rho\ \ {\rm generic}
\end{array}\,.
\ee
Note that when $2\rho,2\ri \nu\in\mathbb{Z}$, while $2(\rho+\ri \nu)$ is an odd integer then
the Verma module  contains both null-vectors 
$|\chi_\pm\rangle$.
In this case, 
assuming $n$ is irrational,%
\footnote{For integer $n=2,3,\ldots$ the corresponding formula for the character
was first obtained in ref.\cite{Gepner} (see also \cite{Jayaraman:1989tu}).
In addition note that eqs.\eqref{iasodi1092091}-\eqref{charadegen1vAA},
which 
assume that $c=2-\frac{6}{n+2}<2$,  can be applied 
to the case $c>2$ if one makes the formal substitutions 
$n\to-n-2$, $\rho\to \ri s$, $\nu\to\ri p$.
The central charge and highest weight of the  irrep would be parameterized as
in \eqref{Deltvarpi1aas}
and \eqref{Deltvarpi1aasa} below, see refs.\cite{Griffin:1990fg,Bakas:1991fs}}
\be\label{charadegen1v}
{\rm ch}_{\rho,\nu}({\tt q})=\frac{{\tt q}^{-\frac{1}{12}+\frac{\nu^2}{n}+\frac{\rho^2}{n+2}}}
{({\tt q},{\tt q})_\infty^{2}} \ 
\sum_{m=0}^\infty (-1)^{m}\, {\tt q}^{\frac{m^2}{2}}\big(\, {\tt q}^{m |\, |\rho|-|\nu|\, |}-{\tt q}^{(m+1)
(|\rho|+|\nu| +1 )-\frac{1}{2}}\,\big)\ ,
\ee
where $\Im m(\rho)=\Re e(\nu)=0$ such that
\bea\label{charadegen1vAA}
|\rho|\pm|{\mathfrak \nu}|\in\tfrac{1}{2}+{\mathbb Z}\   .
\eea

\bigskip

The chiral subspaces  ${\cal V}_{p,\sigma\ri{\mathfrak{q}}_a}$,
${\cal V}_{p,\sigma\ri\bar{\mathfrak{q}}_a}$,
${\cal V}_{p_+,\sigma\ri\mathfrak{q}_a'}$ and
${\cal V}_{p_-,\sigma\ri\bar{\mathfrak{q}}_a'}$
\eqref{iaoisd8182}
of the conformal towers in ${\cal H}_{S^z}^{({\rm disc},\pm)}$,
that were discussed in sec.\,\ref{iaso132218s}, are
highest weight irreps  of the $W_\infty$\,-\,algebra.
Namely ${\cal V}_{\rho,\nu}\cong {\cal W}_{\rho,\nu}$,
where for the four spaces $(\rho,\nu)$ should be replaced by $(p,\sigma\ri\mathfrak{q}_a),\,
(p,\sigma\ri\bar{\mathfrak{q}}_a),\,
(p_+,\sigma\ri{\mathfrak{q}}_a')$ and
$(p_-,\sigma\ri\bar{\mathfrak{q}}_a')$, respectively.
The admissible values of $\rho=p,p_\pm$  for the lattice model
has the form $2\rho=m_1+(n+2)({\tt k}+m_2)$, where $m_1,m_2$ are integers.
We will mainly
focus on the case when  the twist parameter ${\tt k}$ and/or  anisotropy 
parameter $n$ are generic and assume that $2\rho\notin\mathbb{Z}$. 
When $\nu$ is real and $2\rho=m_1+(n+2)({\tt k}+m_2)$ with 
$(n+2)\,{\tt k}\notin\mathbb{Z}$ the chiral subspace ${\cal V}_{\rho,\nu}={\cal F}_{\bf{P}}$
(see eq.\,\eqref{FOck39933})
is an irreducible representation of the $W_\infty$\,-\,algebra, i.e.,
${\cal V}_{\rho,\nu}\cong{\cal W}_{\rho,\nu}$.

\bigskip

In the case of generic $\nu$ but with 
$2\rho=m_1+(n+2)\,m_2$, i.e., ${\tt k}=0$,  the Verma module may
  become degenerate.
This is  related to the existence of the ``bosonic'' screening charge (see, e.g.,
\cite{Gerasimov:1989mz,Jayaraman:1989tu,Griffin:1990fg}):
 \bea\label{ioasd900231a}
\hat{{\tt Q}}=\int\limits_{u_0}^{u_0+2\pi}\rd u \  \partial\vartheta\,\re^{-\frac{2\ri \varphi}{\sqrt{n+2}}}(u)
\eea
(in the physical slang the formal operators  $\hat{{\tt Q}}_\sigma$ \eqref{ioasd90023} are referred to
as ``fermionic'' screening charges).
Similar to $\hat{{\tt Q}}_\sigma$,  the integrand  here has conformal dimensions $\Delta=1$
w.r.t. the chiral component of the energy momentum tensor $W_2(u)$ \eqref{w2iosdi}. Thus, being a $1$-form,
the screening charge density can be integrated so that the action of $\hat{{\tt Q}}$
is   formally defined  on any Fock space ${\cal F}_{\bf P}$. For
generic values of ${\bf P}$ 
 the integration contour in \eqref{ioasd900231a} is not closed, i.e., $\hat{{\tt Q}}$ depends on the arbitrarily chosen
initial  integration point $u_0$.
However, when restricted to the Fock space 
${\cal F}_{{\bf P}}$ with $P_1=\tfrac{1}{2}\,  (m(n+2)+r)$ and  arbitrary $P_2$,
one can show that the action of the
$r$-th power of  $\hat{{\tt Q}}$ is well defined and does not depend on the choice of $u_0$.
It is not difficult to see that for positive integers $m$ and $r$
the  state
\bea
|\chi\rangle=\lim_{{\Im  m(v)}\to+\infty}\re^{-\ri v\Delta_\chi }\ \hat{{\tt Q}}^{r}\ \re^{\ri (m\sqrt{n+2}+
\frac{r-1}{\sqrt{n+2}})\varphi}(v)\, |{\bf P}_0\rangle\,,\ \ \ \ 
{\rm  where}\ \ \ \ {\bf P}_0=\big(\tfrac{1}{2\sqrt{n+2}},P_2\big)
\eea
and $\Delta_\chi=\frac{(m(n+2)+r)^2-1}{4(n+2)}$ 
is non-trivial and belongs to
the level subspace
 ${\cal F}^{({\tt L})}_{{\bf P}}$ with ${\bf P}=
(\frac{m(n+2)-r}{2\sqrt{n+2}}, P_2)$ at level ${\tt L}=mr$.
Furthermore it  turns out to be a highest state of the $W_\infty$\,-\,algebra.
Once the invariant subspace generated by this null vector is factored out one obtains an
irrep whose character is given by
\be\label{charadegen1aaa}
{\rm ch}_{\rho,\nu}({\tt q})={\tt q}^{-\frac{1}{12}+\frac{\nu^2}{n}+\frac{\rho^2}{n+2}}
\  \frac{1-{\tt q}^{mr}}{({\tt q},{\tt q})_\infty^{2}}\ \ ,\  \ \  \ \qquad
\begin{array}{l}
\rho=\pm \tfrac{1}{2}\,  \big(m(n+2)-r\,\big)\,,\ \ \ m,r=1,2,\ldots\  \\[0.2cm]
\nu,\,n\ \ {\rm generic}
\end{array}
\ee

\bigskip

A final comment is in order regarding the intertwiner 
 $\hat{{\tt C}}_{\rm R}:\,
 {\cal F}_{(\pm P_1,P_2)}\mapsto {\cal F}_{(\mp P_1,P_2)}$,
which was used in the description of the irreps of the $W_\infty$\,-\,algebra in terms of the Fock spaces
for some cases with pure imaginary $\nu$. 
This operator was introduced through the
eigenbasis of 
${\mathlarger{\mathlarger{\mathlarger{\mathlarger {\bf \it a}}}}}_\pm(\lambda)$ \eqref{iaos98a9s8d}.
An alternative definition is based on the fact that, as it follows from \eqref{Deltvarpi1a}, the Fock spaces
${\cal F}_{(+ P_1,P_2)}$ and ${\cal F}_{(-P_1,P_2)}$ are equivalent
highest weight representations
of the $W_\infty$\,-\,algebra.
Then the intertwiner $\hat{{\tt C}}_{\rm R}$ can be unambiguously defined by the commutativity condition
with the $W$ currents
\be
\hat{{\tt C}}_{\rm R}\, W_j(u)=W_j(u)\,\hat{{\tt C}}_{\rm R}\ \qquad\qquad (j=2,3)
\ee
supplemented  by its action  on the highest weight:
$\hat{{\tt C}}_{\rm R}|(\pm P_1,P_2)\rangle= |(\mp P_1,P_2)\rangle$.

\section{The space of states in the scaling limit\label{sec17}}
We are now ready to
synthesize the analyses of the  previous sections and
describe the linear space of states occurring in the scaling limit of the
low energy sector of the ${\cal Z}_2$ invariant inhomogeneous
six-vertex model. Some of the formulae presented here
constitute the main results of our study of the lattice model and will be referred
back to in the later part of the paper.

\subsection{The sectors with $S^z=0,1,2,\ldots$ and $(n+2)\,{\tt k}\notin\mathbb{Z}$ \label{sec171}}
Recall our working definition
of a low energy state -- a state which can be assigned the
quantum numbers $S^z$, ${\tt w}$, ${\tt L}$ and $\bar{\tt L}$,
such that the energy and eigenvalue of the lattice translation operator follow 
the large $N$ asymptotics \eqref{oaisoi1093}-\eqref{uassaysa},
where $b=b(N)$ is defined by \eqref{poapso1a} along with
the condition  $|\Im m\big(b(N)\big)|<\frac{n}{4}$.
In the scaling limit the states with fixed value of $S^z$ were
organized into the three sectors ${\cal H}_{S^z}^{({\rm cont})}$,
${\cal H}_{S^z}^{({\rm disc},+)}$
and ${\cal H}_{S^z}^{({\rm disc},-)}$. 
Each of these is further split into the subsectors labeled by the
winding number ${\tt w}=0,\pm1,\pm2,\ldots\ $:
\be\label{iaosido1902099102}
{\cal H}^{({\rm cont})}_{S^z}=\bigoplus_{{\tt w}\in\mathbb{Z}}{\cal H}^{({\rm cont})}_{S^z\!,{\tt w}}\,,\qquad
\qquad
{\cal H}^{({\rm disc},\pm)}_{S^z}=\bigoplus_{{\tt w}\in\mathbb{Z}}{\cal H}^{({\rm disc},\pm)}_{S^z\!,{\tt w}}\ .
\ee
The subsector ${\cal H}^{({\rm cont})}_{S^z\!,{\tt w}}$ is described through a direct integral as
\be\label{iaosidoi192009}
{\cal H}^{({\rm cont})}_{S^z\!,{\tt w}}=
\int^{\oplus}_{\mathbb{R}}\!\rd s\
\bar{{\cal V}}_{\bar{p},s}\otimes {\cal V}_{p,s}\,,\qquad  
{\rm where}
\qquad 
\begin{array}{l} p=\frac{1}{2}\,S^z+\frac{1}{2}\,(n+2)\,({\tt k}+{\tt w})\\[0.2cm]
                                       \bar{p}=\frac{1}{2}\,S^z-\frac{1}{2}\,(n+2)\,({\tt k}+{\tt w})
\end{array}
\ee
and $\bar{{\cal V}}_{\bar{p},s}\otimes{\cal V}_{p,s}$ is isomorphic to a highest weight
irrep of the $\overline{W}_\infty\otimes W_\infty$\,-\,algebra.
Contrary to ${\cal H}^{({\rm cont})}_{S^z\!,{\tt w}}$,
the decomposition of the linear space ${\cal H}_{S^z\!,{\tt w}}^{({\rm disc},+)}$ 
involves a direct sum over the discrete set of pure imaginary admissible values of $s$.
It reads as
\be\label{8se94398ir}
 {\cal H}_{S^z\!,{\tt w}}^{({\rm disc},+)}=
\bigoplus_{\sigma=\pm1} \Big({\cal H}^{(1,+)}_{S^z\!,{\tt w},\sigma}\,\oplus\,
{\cal H}^{(2,+)}_{S^z\!,{\tt w},\sigma}\Big)
\ee
with
\be\label{iao989382931}
{\cal H}^{(1,+)}_{S^z\!,{\tt w},\sigma}=\bigoplus_{a\in\Sigma(p)}
\bar{{\cal V}}_{\bar{p},\sigma\ri\mathfrak{q}_a}\otimes{\cal V}_{p,\sigma\ri\mathfrak{q}_a}\,,
\qquad \qquad
{\cal H}^{(2,+)}_{S^z\!,{\tt w},\sigma}=
\bigoplus_{a\in{\Sigma(\bar{p})}}
\bar{{\cal V}}_{\bar{p},\sigma\ri\bar{\mathfrak{q}}_a}\otimes{\cal V}_{p,\sigma\ri\bar{\mathfrak{q}}_a}\ .
\ee
Here
\be
\mathfrak{q}_a=-p-\tfrac{1}{2}-a,\qquad\qquad
\bar{\mathfrak{q}}_a=-\bar{p}-\tfrac{1}{2}-a
\ee
and the summation is taken over the non-negative integer
$a$ restricted to the sets
\be\label{sigmaP1a}
\Sigma(p)=\Big\{a:\ a\in\mathbb{Z}_+,\ 
-p-\tfrac{n+2}{4}\le a<-\tfrac{1}{2}-p\Big\}
\ee
as well as
 $\Sigma(\bar{p})$, which is given by the same formula with $p$ substituted by $\bar{p}$.
Each of the  components $\bar{{\cal V}}_{\bar{p},s}\otimes{\cal V}_{p,s}$
from \eqref{iaosidoi192009},\,\eqref{iao989382931},
being a highest weight irrep of the $\overline{W}_\infty\otimes{W}_\infty$\,-\,algebra,
is a naturally graded linear space. 
The pair of non-negative quantum numbers $(\bar{{\tt L}},{\tt L})$   for a state
coincides with its level in the highest weight irrep.
\bigskip

The subsector ${\cal H}^{({\rm disc},-)}_{S^z\!,{\tt w}}$  is also decomposed into the irreps
of the  $\overline{W}_\infty\otimes{W}_\infty$\,-\,algebra. However 
an important difference   from the cases ${\cal H}^{({\rm cont})}_{S^z\!,{\tt w}}$
 and ${\cal H}^{({\rm disc},+)}_{S^z\!,{\tt w}}$ is that the pair $(\bar{\tt L},{\tt L})$
does not coincide with the level of the state in the highest weight irrep.
The 
linear structure of ${\cal H}^{({\rm disc},-)}_{S^z\!,{\tt w}}$ is more involved.
To describe it, in addition to $p$, $\bar{p}$, $\mathfrak{q}_a$ and 
$\bar{\mathfrak{q}}_a$, we use the notation
\be\label{sdkkjaqw}
\arraycolsep=0.8cm
\begin{array}{ll}
p_+=\tfrac{1}{2}\,S^z+\tfrac{1}{2}\,(n+2)({\tt k}+{\tt w}+1)\,, & 
\bar{p}_+=\tfrac{1}{2}\,S^z-\tfrac{1}{2}\,(n+2)({\tt k}+{\tt w}+1) \\[0.4cm]
p_-=\tfrac{1}{2}\,S^z+\tfrac{1}{2}\,(n+2)({\tt k}+{\tt w}-1)\,, & 
\bar{p}_-=\tfrac{1}{2}\,S^z-\tfrac{1}{2}\,(n+2)({\tt k}+{\tt w}-1) \\[0.4cm]
\mathfrak{q}_a'=-p-\tfrac{n+1}{2}-a\,, &  
\bar{\mathfrak{q}}_a'=-\bar{p}-\tfrac{n+1}{2}-a\ .
\end{array}
\ee
Then
\be\label{ioasiod1298aaab}
 {\cal H}_{S^z\!,{\tt w}}^{({\rm disc},-)}=\bigoplus_{\sigma=\pm1}\Big(\,
{\cal H}_{S^z\!,{\tt w},\sigma}^{(1,-)}\oplus{\cal H}_{S^z\!,{\tt w},\sigma}^{(2,-)}
\oplus{\cal H}_{S^z\!,{\tt w},\sigma}^{(3,-)}\oplus{\cal H}_{S^z\!,{\tt w},\sigma}^{(4,-)}\,\Big)
\ee
and the decomposition 
of each of the four spaces ${\cal H}^{(i,-)}_{S^z\!,{\tt w},\sigma}$  into irreps of the
 $\overline{W}_\infty\otimes{W}_\infty$\,-\,algebra reads explicitly as
\bea\label{iaosido89812aaa}
&&{\cal H}_{S^z\!,{\tt w},\sigma}^{(1,-)}=\bigoplus\limits_{a\in\Sigma_1(p)}
\bar{{\cal V}}_{\bar{p}_+,\sigma\ri\mathfrak{q}_a'}\otimes
{\cal V}_{p_+,\sigma\ri\mathfrak{q}_a'}\,,\qquad \qquad
{\cal H}_{S^z\!,{\tt w},\sigma}^{(2,-)}=
\bigoplus\limits_{a\in\Sigma_2(p)}
\bar{{\cal V}}_{\bar{p},\sigma\ri\mathfrak{q}_a}\otimes
{\cal V}_{p_+,\sigma\ri\mathfrak{q}_a'}
\nonumber
\\[-0.4cm]
&& \hspace{13.5cm} .\\
&&{\cal H}_{S^z\!,{\tt w},\sigma}^{(3,-)}=\bigoplus\limits_{a\in\Sigma_2(\bar{p})}
\bar{{\cal V}}_{\bar{p}_-,\sigma\ri\bar{\mathfrak{q}}_a'}\otimes
{\cal V}_{p,\sigma\ri\bar{\mathfrak{q}}_a}\,,\qquad \qquad\ \,
{\cal H}_{S^z\!,{\tt w},\sigma}^{(4,-)}=\bigoplus\limits_{a\in\Sigma_1(\bar{p})}
\bar{{\cal V}}_{\bar{p}_-,\sigma\ri\bar{\mathfrak{q}}_a'}\otimes
{\cal V}_{p_-,\sigma\ri\bar{\mathfrak{q}}_a'}\nonumber
\eea
Here the summation index $a$  takes negative integer values and runs over the sets
\bea\label{oaspdoaposd}
\Sigma_1(p)&=&\Big\{a:\ a\in\mathbb{Z}_-,\ 
-p-\tfrac{n+2}{4}\le a<-\tfrac{1}{2}-p\ { \&}\  a<-S^z\Big\}
\nonumber\\[-0.2cm]
\\[-0.1cm]
\Sigma_2(p)&=&\Big\{a:\ a\in\mathbb{Z}_-,\ 
-p-\tfrac{n+2}{4}\le a<-\tfrac{1}{2}-p\ \&\  a\ge -S^z\Big\}
\nonumber
\eea
and $\Sigma_1(\bar{p})$, $\Sigma_2(\bar{p})$ which are defined by the analogous formulae.
The levels w.r.t. the $\overline{W}_\infty\otimes W_\infty$\,-\,algebra
of the components in the r.h.s. of \eqref{iaosido89812aaa} do not coincide with $(\bar{{\tt L}},{\tt L})$. 
The relation between them
depends on the case being considered, and can be read off from eqs.\,\eqref{aiso102} and \eqref{aiso103}. 
For example, for the right chiral component ${\cal V}_{p_+,\sigma\ri\mathfrak{q}_a'}$ 
the level w.r.t. the $W_\infty$\,-\,algebra, denoted by ${\tt L}_+$, is expressed in terms of
${\tt L}$ as ${\tt L}_+={\tt L}-|a|$.
\bigskip

In the linear decompositions \eqref{iaosidoi192009},\,\eqref{iao989382931} and \eqref{iaosido89812aaa},
each of the chiral components ${\cal V}_{p,s}$,
${\cal V}_{p,\ri\mathfrak{q}_a},\ldots$ is
isomorphic to ${\cal W}_{\rho,\nu}$, the highest weight irrep of the $W_\infty$\,-\,algebra,
whose  highest weight  is given by \eqref{Deltvarpi1a}
with $(\rho,\nu)=(p,s),\,(p,\ri\mathfrak{q}_a),\ldots\ $, respectively.

\subsection{Global symmetries\label{sec172}}
The  ${\cal Z}_2$ invariant inhomogeneous six-vertex model
 possesses  global ${\cal CPT}$ and ${\cal Z}_2$ symmetry.
Since their generators ${\cal \hat{C}\hat{P}\hat{T}}$ and ${\cal \hat{D}}$
 commute with the lattice Hamiltonian, they 
preserve the low energy sector of the  model.
The action of the symmetry transformations
on the low energy states in the scaling limit can be deduced from 
eqs.\,\eqref{CPT1asdasda} and \eqref{DTApmcomm1}, which 
describe the commutation relations of 
${\cal \hat{C}\hat{P}\hat{T}}$ and ${\cal \hat{D}}$ with the lattice 
operators $\mathbb{A}_\pm(\zeta)$, $\mathbb{T}(\zeta)$.
Combining them with the scaling relations \eqref{scalingrel1a},\,\eqref{8s9saa} 
yields
\be\label{i89a89}
\arraycolsep=0.6cm
\begin{array}{cc}
{\cal \hat{C}\hat{P}\hat{T}}\,{\mathlarger{\mathlarger{\mathlarger{\mathlarger {\bf \it a}}}}}_\pm(\lambda)
\,{\cal \hat{C}\hat{P}\hat{T}}={\mathlarger{\mathlarger{\mathlarger{\mathlarger {\bf \it a}}}}}_\pm(\lambda^*)\ ,& 
{\cal \hat{D}}\, {\mathlarger{\mathlarger{\mathlarger{\mathlarger {\bf \it a}}}}}_\pm(\lambda)
\,{\cal \hat{D}}={\mathlarger{\mathlarger{\mathlarger{\mathlarger {\bf \it a}}}}}_\pm(-\lambda)\ 
\\[0.2cm]
{\cal \hat{C}\hat{P}\hat{T}}\,\bm{\tau}(\lambda)\,{\cal \hat{C}\hat{P}\hat{T}}=\bm{\tau}(\lambda^*)\
,&
{\cal \hat{D}}\,\bm{\tau}(\lambda)\,{\cal \hat{D}}=\bm{\tau}(-\lambda)
\end{array}\!\!\!\!\! .
\ee
The similar formulae also hold true for 
$\bar{{\mathlarger{\mathlarger{\mathlarger{\mathlarger {\bf \it a}}}}}}_\pm(\bar{\lambda})$ and
$\bar{\bm{\tau}}(\bar{\lambda})$. 
For our purposes it is sufficient to focus on the commutation relations
of the global symmetry generators with $\bm{\tau}(\lambda)$ and
$\bar{\bm{\tau}}(\bar{\lambda})$. Keeping in mind that the local
IM ${\bf I}_m$   $\big(\,\bar{{\bf I}}_m\big)$ occur in the large 
$\lambda$ $(\bar{\lambda})$ 
 asymptotic expansion 
for $\bm{\tau}(\lambda)$ $\big(\bar{\bm{\tau}}(\bar{\lambda})\big)$ as in  eq.\,\eqref{9sd98f989sd1a},
one concludes that
\be\label{CPTIM1}
\arraycolsep=0.6cm
\begin{array}{ll}
{\cal \hat{C}\hat{P}\hat{T}}\ {\bf I}_m\,{\cal \hat{C}\hat{P}\hat{T}}={\bf I}_m\,, &
{\cal \hat{D}}\ {\bf I}_m\,{\cal \hat{D}}=(-1)^{m+1}\,{\bf I}_m \\[0.3cm]
{\cal \hat{C}\hat{P}\hat{T}}\ \bar{{\bf I}}_m\,{\cal \hat{C}\hat{P}\hat{T}}=\bar{{\bf I}}_m\,, &
{\cal \hat{D}}\ \bar{{\bf I}}_m\,{\cal \hat{D}}=(-1)^{m+1}\,\bar{{\bf I}}_m 
\end{array}\!\!\!\! .
\ee
The  densities for the local IM can be expressed in terms of the $W$ currents,  so that
the above 
relations would follow from
\be\label{oapsd0092}\arraycolsep=0.6cm
\begin{array}{ll}
{\cal \hat{C}\hat{P}\hat{T}}\,{W}_j(u)\,
{\cal \hat{C}\hat{P}\hat{T}}={W}_j(-u^*)\,,&
\qquad
\hat{{\cal D}}\,{W}_j({u})\,\hat{{\cal D}}=(-1)^j\,{W}_j({u}) \\[0.4cm]
{\cal \hat{C}\hat{P}\hat{T}}\,\overline{W}_j(\bar{u})\,
{\cal \hat{C}\hat{P}\hat{T}}=\overline{W}_j(-\bar{u}^*)\,,&
\qquad
\hat{{\cal D}}\,\overline{W}_j(\bar{u})\,\hat{{\cal D}}=(-1)^j\,\overline{W}_j(\bar{u})
\end{array} \!\!\!\!\!.
\ee
These immediately imply that  the symmetry transformations act, in general,
as the intertwiners between the highest weight irreps 
appearing in the decompositions
 \eqref{iaosidoi192009},\,\eqref{iao989382931} and \eqref{iaosido89812aaa}.
Namely,
\be
{\cal \hat{C}\hat{P}\hat{T}}\ : \ 
{\cal V}_{\bar{\rho},\bar{\nu}}\otimes {\cal V}_{\rho,\nu}\mapsto  
{\cal V}_{\bar{\rho},\bar{\nu}^*}\otimes{\cal V}_{\rho,\nu^*}\,,\qquad\qquad
\hat{{\cal D}}\ : \ {\cal V}_{\bar{\rho},\bar{\nu}}\otimes {\cal V}_{\rho,\nu}\mapsto  
{\cal V}_{\bar{\rho},-\bar{\nu}}\otimes{\cal V}_{\rho,-\nu}\ ,
\ee
where $(\rho,\nu)=(p,s),\,(p,\sigma\ri\mathfrak{q}_a),\ldots\ $
and $(\bar{\rho},\bar{\nu})=(\bar{p},s),\,(\bar{p},\sigma\ri\mathfrak{q}_a),\ldots\ $.
Notice that 
 each of  the subsectors ${\cal H}_{S^z\!,{\tt w}}^{({\rm cont})}$
and  ${\cal H}_{S^z\!,{\tt w}}^{({\rm disc},\pm)}$
turn out to be invariant under the ${\cal CPT}$ and ${\cal D}$ 
transformations.
In order to specify the action of the global symmetries on the states
from the irrep ${\cal V}_{\bar{\rho},\bar{\nu}}\otimes {\cal V}_{\rho,\nu}$  one should return to the
lattice system.
The scaling limit of the low energy Bethe states 
yields the basis states 
\be\label{iaosdioa19209a}
\bm{\psi}_{\bar{\rho},\rho,\bar{\nu},\nu}(\bar{\bm{w}},\bm{w})\equiv
\bar{\bm{\psi}}_{\bar{\rho},\bar{\nu}}(\bar{\bm{w}})\otimes {\bm{\psi}}_{{\rho},\nu}(\bm{w})\in
 \bar{{\cal V}}_{\bar{\rho},\bar{\nu}}\otimes {{\cal V}}_{{\rho},{\nu}}\ .
\ee
Formulae \eqref{CPTBethe1} and
\eqref{8s887f87d87fd}, that describe the action of the 
${\cal {C}{P}{T}}$ and ${\cal{ D}}$
conjugations
on $\bm{\Psi}_N$, allow one to deduce how
the global symmetries act
on $\bm{\psi}_{\bar{\rho},\rho,\bar{\nu},\nu}$.
 In particular, for the $\overline{W}_\infty\otimes W_\infty$ primary  states 
\be
{\cal \hat{C}\hat{P}\hat{T}}\,\bm{\psi}_{\bar{\rho},\rho,\bar{\nu},\nu}^{({\rm vac})}=
+\,\bm{\psi}_{\bar{\rho},\rho,\bar{\nu}^*,\nu^*}^{({\rm vac})}\,,\qquad\qquad
{\cal \hat{D}}\,\bm{\psi}_{\bar{\rho},\rho,\bar{\nu},\nu}^{({\rm vac})}
=+\,\bm{\psi}_{\bar{\rho},\rho,-\bar{\nu},-\nu}^{({\rm vac})}\ .
\ee
The latter, combined with the commutation relations \eqref{oapsd0092},
unambiguously defines the symmetry transformations
for any state in ${\cal V}_{\bar{\rho},\bar{\nu}}\otimes {\cal V}_{\rho,\nu}$.
\bigskip

Formula \eqref{oapsd0092} involves the left and right $W$ currents
separately, so that the action of the ${\cal CPT}$ and ${\cal Z}_2$ symmetries 
may be naturally defined for each chiral component of the $\overline{W}_\infty\otimes W_\infty$
irrep. 
For instance, for the right chiral component:
\be
{\cal \hat{C}\hat{P}\hat{T}}\ : \ {\cal V}_{\rho,\nu}\mapsto  {\cal V}_{\rho,\nu^*}\,,\qquad\qquad
\hat{{\cal D}}\ : \ {\cal V}_{\rho,\nu}\mapsto  {\cal V}_{\rho,-\nu}\ .
\ee
For the chiral  primary state
$\bm{\psi}_{\rho,\nu}^{({\rm vac})}\in{\cal V}_{\rho,\nu}$, 
by choosing a proper 
normalization including the phase assignment, one can arrange that
\be
{\cal \hat{C}\hat{P}\hat{T}}\,\bm{\psi}_{\rho,\nu}^{({\rm vac})}=
\bm{\psi}_{\rho,\nu^*}^{({\rm vac})}\,,\qquad\qquad
{\cal \hat{D}}\,\bm{\psi}_{\rho,\nu}^{({\rm vac})}=\bm{\psi}_{\rho,-\nu}^{({\rm vac})}\ .
\ee
\bigskip

Recall that
the chiral state $\bm{\psi}_{\rho,\nu}(\bm{w})\in{\cal V}_{\rho,\nu}$ in eq.\,\eqref{iaosdioa19209a}
 is an eigenvector of 
${\mathlarger{\mathlarger{\mathlarger{\mathlarger {\bf \it a}}}}}_\pm(\lambda)$ with eigenvalue
$D_\pm(\mu\,|\,\bm{w},\rho,\nu)$ \eqref{iaosid9898}
and similarly for $\bar{\bm{\psi}}_{\bar{\rho},\bar{\nu}}(\bar{\bm{w}})$.
Again,  with a proper choice of the normalization, the action of the 
${\cal CPT}$ and ${\cal Z}_2$ conjugations on the eigenstates 
can be taken to be
\be\label{CPT*8787aaa}
{\cal \hat{C}\hat{P}\hat{T}}\,\bm{\psi}_{\rho,\nu}(\bm{w})=\bm{\psi}_{\rho,\nu^*}(-\bm{w}^*)\,,\qquad\qquad
\hat{{\cal D}}\,\bm{\psi}_{\rho,\nu}(\bm{w})=\bm{\psi}_{\rho,-\nu}(-\bm{w})\ .
\ee
The above is motivated through an examination of the algebraic system 
satisfied by the set $\bm{w}=\{w_a\}_{a=1}^{\tt L}$ \eqref{sksksk1}.
Given a solution,  the set 
$-\bm{w}^*\equiv\{-w_a^*\}_{a=1}^{\tt L}$ solves
the same equations with the parameter 
$s$ substituted by its complex conjugate,
while $-\bm{w}\equiv\{-w_a\}_{a=1}^{\tt L}$
is a solution of \eqref{sksksk1} with $s$ replaced by $-s$.
In turn the eigenvalues of ${\mathlarger{\mathlarger{\mathlarger{\mathlarger {\bf \it a}}}}}_\pm(\lambda)$
corresponding to $\bm{\psi}_{\rho,\nu}(\bm{w})$  obey 
\be\label{DCPTeq1a}
\!\!\!\!\!\big(D_\pm(\mu\,|\,\bm{w},\rho,\nu)\big)^*=D_\pm(-\mu^*\,|\,-\bm{w}^*,\rho,\nu^*)\,,\quad
D_\pm(\mu\,|\,\bm{w},\rho,\nu)=D_\pm(-\mu\,|\,-\bm{w},\rho,-\nu)
\ee
where we take into account the imaginary unit entering
into the $\lambda$-$\mu$ relation \eqref{mulambda1a}.
\bigskip

The lattice model also possesses  ${\cal CP}$ invariance which, in turn, becomes a symmetry that acts in  
the space of states occurring in the scaling limit.
The key relation for defining the ${\cal CP}$ conjugation is
\be\label{iaosio11212}
{\cal \hat{C}\hat{P}}\,{W}_j(u)={\overline{W}}_j(u)\,{\cal \hat{C}\hat{P}}\ .
\ee
It may be advocated for using the similar arguments that led to \eqref{oapsd0092}.
Namely, one should start  with the commutation relation of 
${\cal \hat{C}\hat{P}}$ with the lattice transfer matrix,
\be
{\cal \hat{C}\hat{P}}\,\mathbb{T}(\zeta)\,{\cal \hat{C}\hat{P}}=\zeta^{N}\,\mathbb{T}(\zeta^{-1})\ ,
\ee
which was already quoted in the Preliminaries.
This, in view of eqs.\,\eqref{8s9saa} and \eqref{8s9saabar},
in the scaling limit becomes
\be\label{asidoi232}
{\cal \hat{C}\hat{P}}\ \bm{\tau}(\lambda)\ {\cal \hat{C}\hat{P}}=
\bar{\bm{\tau}}(\bar{\lambda})\,.
\ee
The latter, combined with the large $\lambda$ asymptotic formula \eqref{9sd98f989sd1a} and the similar one for 
$\bar{\bm{\tau}}(\bar{\lambda})$, results in
$
{\cal \hat{C}\hat{P}}\ {\bf I}_m\ {\cal \hat{C}\hat{P}}=
\bar{{\bf I}}_m$,  which is clearly consistent with \eqref{iaosio11212}.
\bigskip

Contrary to the  other global symmetries, the ${\cal CP}$ conjugation does
not commute with the lattice total spin operator $\mathbb{S}^z$. 
As a result, it acts invariantly only in the subsectors 
${\cal H}_{S^z\!,{\tt w}}^{({\rm cont})}$ and ${\cal H}_{S^z\!,{\tt w}}^{({\rm disc},\pm)}$
with $S^z=0$.  In this case, the action of ${\cal CP}$  on the
$\overline{W}_\infty\otimes W_\infty$ irreps is described by
\be\label{isoaido1212}
{\cal CP}\,: \ \bar{{\cal V}}_{\bar{\rho},\bar{\nu}}\otimes{\cal V}_{\rho,\nu}\mapsto
\bar{{\cal V}}_{-\rho,\nu}\otimes{\cal V}_{-\bar{\rho},\bar{\nu}}\,,
\ee
where again $(\rho,{\nu})=(p,s),(p,\sigma\ri\mathfrak{q}_a),\ldots $ and
$(\bar{\rho},\bar{\nu})=(\bar{p},s),(\bar{p},\sigma\ri\mathfrak{q}_a),\ldots\ $. 
 Recall 
that the space ${\cal V}_{\rho,\nu}$, being considered as a highest weight
irrep of the $W_\infty$\,-\,algebra, is isomorphic to ${\cal V}_{-\rho,\nu}$
as the highest weight is not sensitive to a flip of the sign of $\rho$, see eq.\,\eqref{Deltvarpi1a}.
This makes \eqref{isoaido1212} consistent with the relations \eqref{iaosio11212}.
\bigskip

The components  $\bar{{\cal V}}_{\bar{\rho},\bar{\nu}}\otimes{\cal V}_{\rho,\nu}$ 
occurring in the decomposition of ${\cal H}_{0,{\tt w}}^{({\rm cont})}$ \eqref{iaosidoi192009}
and ${\cal H}_{0,{\tt w}}^{({\rm disc},+)}$ \eqref{8se94398ir},
are always such that $\rho+\bar{\rho}=0$ and $\nu=\bar{\nu}$ so that the ${\cal CP}$ conjugation
acts invariantly in each of them.
At first glance, this  property does not seem to hold true for the case of
${\cal H}_{0,{\tt w}}^{({\rm disc},-)}$.
The direct sum  \eqref{ioasiod1298aaab} for the 
subsector ${\cal H}_{0,{\tt w}}^{({\rm disc},-)}$  in general
contains eight terms. 
However, when $S^z=0$ the sets $\Sigma_2(p)$ and $\Sigma_2(\bar{p})$ are empty and
the linear spaces ${\cal H}_{S^z\!,{\tt w},\sigma}^{(2,-)}$ and ${\cal H}_{S^z\!,{\tt w},\sigma}^{(3,-)}$ \eqref{iaosido89812aaa} become trivial.
In addition $\bar{p}_\pm+p_\pm=0$
so that the components
$\bar{{\cal V}}_{\bar{p}_+,\sigma\ri{\mathfrak{q}}_a'}\otimes
{\cal V}_{p_+,\sigma\ri{\mathfrak{q}}_a'}$ and
$\bar{{\cal V}}_{\bar{p}_-,\sigma\ri\bar{\mathfrak{q}}_a'}\otimes
{\cal V}_{p_-,\sigma\ri\bar{\mathfrak{q}}_a'}$,
appearing in the decomposition of the remaining four spaces ${\cal H}_{S^z\!,{\tt w},\sigma}^{(1,-)}$
and ${\cal H}_{S^z\!,{\tt w},\sigma}^{(4,-)}$, 
respectively,
are preserved under the  ${\cal CP}$ conjugation.
\bigskip

Similar as for the other global symmetries discussed above,
the ${\cal CP}$ conjugation
in ${\cal H}_{0,{\tt w}}^{({\rm cont})}$ and ${\cal H}_{0,{\tt w}}^{({\rm disc},\pm)}$
may be determined by considering its action on the low energy Bethe states of the finite lattice system.
A numerical analysis suggests that 
for the $\overline{W}_\infty\otimes W_\infty$ primary states
\be\label{pasodpo12-01}
{\cal \hat{C}\hat{P}\,\bm{\psi}_{\bar{\rho},\rho,\bar{\nu},\nu}^{({\rm vac})}=
+\,\bm{\psi}_{-{\rho},-\bar{\rho},{\nu},\bar{\nu}}^{({\rm vac})}}\qquad\qquad (\rho+\bar{\rho}=0)\ .
\ee
Together with  the relation \eqref{iaosio11212}, this unambiguously defines
the action of the ${\cal CP}$ transformation for any state from
${\cal H}_{S^z\!,{\tt w}}^{({\rm cont})}$
and ${\cal H}_{S^z\!,{\tt w}}^{({\rm disc},\pm)}$ with $S^z=0$.
\bigskip

In our study of the scaling limit
we have focused on
the case with $S^z\ge0$.
Since  ${\cal \hat{C}\hat{P}}\,\mathbb{S}^z=-\mathbb{S}^z\,{\cal \hat{C}\hat{P}}$,
one can make use of ${\cal CP}$ invariance
to  describe the scaling limit of the low energy states  with $S^z<0$.
These would organize into the subsectors 
 ${\cal H}_{S^z\!,{\tt w}}^{({\rm cont})}$
and ${\cal H}_{S^z\!,{\tt w}}^{({\rm disc},\pm)}$,
which are the 
${\cal {C}{P}}$ image of the corresponding spaces having the opposite sign of $S^z$:
\be\label{ioasido129012}
{\cal H}_{S^z\!,{\tt w}}^{({\rm cont})}\equiv {\cal \hat{C}\hat{P}} \big({\cal H}_{-S^z\!,{\tt w}}^{({\rm cont})}\big)\,,
\qquad \qquad
{\cal H}_{S^z\!,{\tt w}}^{({\rm disc},\pm)}\equiv {\cal \hat{C}\hat{P}} \big({\cal H}_{-S^z\!,{\tt w}}^{({\rm disc,\pm})}\big) 
\qquad\qquad (S^z<0)\ .
\ee
Supplementing the $\overline{W}_\infty\otimes W_\infty$ decomposition of 
${\cal H}_{S^z\!,{\tt w}}^{({\rm cont})}$ and ${\cal H}_{S^z\!,{\tt w}}^{({\rm disc,\pm})}$
given in the previous subsection with eq.\,\eqref{iaosio11212} 
provides a classification of
 the states from \eqref{ioasido129012}  w.r.t. the irreps of the conformal symmetry algebra. 
Note that formula \eqref{pasodpo12-01} 
for $\rho+\bar{\rho}>0$ can be taken as the definition of 
$\bm{\psi}_{-{\rho},-\bar{\rho},{\nu},\bar{\nu}}^{({\rm vac})}$,
which are the primary  $\overline{W}_\infty\otimes W_\infty$  states in the irreps with $S^z<0$.
This way the full space of states occurring in the scaling limit of the 
${\cal Z}_2$ invariant inhomogeneous six-vertex model  
is split into the continuous and discrete components of the form
\be\label{iasodioi12233}
{\cal H}^{({\rm cont})}=\bigoplus_{S^z\!,{\tt w}\in\mathbb{Z}}{\cal H}_{S^z\!,{\tt w}}^{({\rm cont})}\,,\qquad\qquad
{\cal H}^{({\rm disc},\pm)}=\bigoplus_{S^z\!,{\tt w}\in\mathbb{Z}} {\cal H}_{S^z\!,{\tt w}}^{({\rm disc},\pm)}\ .
\ee

\subsection{Partition function  in the scaling limit \label{oasid90120912}}
The linear decomposition of the spaces ${\cal H}^{({\rm cont})}$
and ${\cal H}^{({\rm disc},\pm)}$ described above allows one to study
the scaling behaviour of the lattice partition function. 
For a lattice with $N$ horizontal sites, we define the partition function associated with the
Hamiltonian $\mathbb{H}$ \eqref{aioiisa}
and the shift operator $\mathbb{K}$, given by \eqref{Kformula1} with $r=2$, via the formula 
\be
Z^{({\rm lattice})}_N(M_1,M_2)={\rm Tr}_{{\mathscr V}_N}\Big[
\re^{-M_1\mathbb{H}}\ \mathbb{K}^{M_2}\Big]
\ee
with the trace being taken over the $2^N$ dimensional space ${\mathscr V}_N=\mathbb{C}^2_N\otimes
 \mathbb{C}^2_{N-1}\otimes\cdots\otimes\mathbb{C}^2_1$.
Keeping fixed the ratios
\be
\tau=\frac{2\ri}{N}\,\big(v_{\rm F}\, M_1-\ri M_2\big)\,,\qquad
\bar{\tau}=\frac{2\ri}{N}\,\big(v_{\rm F}\, M_1+\ri M_2\big)
\ee
the large $N$ behaviour of the lattice partition function
is described as
\be
Z^{({\rm lattice})}_N(M_1,M_2)\asymp
\re^{-M_1 Ne_\infty}\, { Z}^{({\rm scl})}\ .
\ee
Here $Z^{({\rm scl})}$ is given in terms of a trace 
over the full space of states occurring in the scaling limit
of the lattice model ${\cal H}={\cal H}^{({\rm cont})}\oplus {\cal H}^{({\rm disc},+)}\oplus
{\cal H}^{({\rm disc},-)}$. Namely, 
\be\label{iasoiao444s}
{ Z}^{({\rm scl})}={\rm Tr}_{{\cal H}}\Big[\,
\bar{\tt q}^{\widetilde{\overline{W}}_2(0)-\frac{c}{24}}\
{\tt q}^{\widetilde{W}_2(0)-\frac{c}{24}}
\Big]\qquad\quad {\rm with} \qquad\quad
 {\tt q}=\re^{2\pi\ri \tau}\,,\ \bar{{\tt q}}=\re^{2\pi\ri \bar{\tau}}\ .
\ee
\bigskip

The  trace in \eqref{iasoiao444s}  is naturally split into the contributions of 
the states from the continuous and discrete components:
\be
{ Z}^{({\rm scl})}={ Z}^{({\rm cont})}+{ Z}^{({\rm disc})}\ .
\ee
It is straightforward to calculate ${ Z}^{({\rm disc})}$ using the 
formulae \eqref{8se94398ir}-\eqref{oaspdoaposd}, as well as 
the explicit expression \eqref{charadegen1a} for
the character of the highest weight irrep of the $W_\infty$\,-\,algebra.
To write the result in a compact way
we borrow the notation $\chi^d_{(\mathfrak{j}, a-\mathfrak{j})}({\tt q})$
from ref.\cite{Ribault:2003ss}. Up to a simple factor, this function coincides with  $\chi_{a}({\tt q})$ defined in eq.\,\eqref{Zdef1b}
(see also footnote \ref{ft4}):
\be
\chi^d_{(\mathfrak{j}, a-\mathfrak{j})}({\tt q})\equiv
{\tt q}^{-\frac{1}{12}-\frac{(\mathfrak{j}+\frac{1}{2})^2}{n}+\frac{({\mathfrak j}-a)^2}{n+2}}\
\chi_{a}({\tt q}) \qquad (a\in\mathbb{Z})\ .
\ee
It is related to the character of the irrep as
\be
\chi^d_{(\mathfrak{j}, a-\mathfrak{j})}({\tt q})=
{\rm ch}_{a-\mathfrak{j},\ri(\mathfrak{j}+\frac{1}{2})}({\tt q})  \times \begin{cases}
1  & \quad{\rm for}\qquad  a\ge 0 \\[0.2cm]
{\tt q}^{-a}& \quad{\rm for}\qquad a<0
\end{cases}\, .
\ee
Also  introduce the notation ${\mathfrak  J}({\tt v},{\tt u})$
for the finite set of all  real numbers  belonging to the half-open segment
$[-\frac{n+1}{2},-\frac{1}{2})$ such that
\be\label{iaosid982981}
{\mathfrak  J}({\tt v},{\tt u})\equiv\Big\{{\mathfrak{j}}:\  {\mathfrak{j}}\in\big[-\tfrac{n+1}{2},-\tfrac{1}{2}\big)\ \&\ {\mathfrak{j}}-\tfrac{1}{2}\,{\tt v}-\tfrac{1}{2}\,(n+2)({\tt k}+{\tt u})
\in{\mathbb Z}\,\Big\}\ .
\ee
Then the calculation of the
 trace over the space ${\cal H}^{({\rm disc})}={\cal H}^{({\rm disc},+)}\oplus{\cal H}^{({\rm disc},-)}$ yields
\be\label{oaspdo121}
{ Z}^{({\rm disc})}=
2\sum_{{\tt v},{\tt u}\in{\mathbb Z}}\ 
\sum_{\mathfrak{j}\in {\mathfrak  J}({\tt v},{\tt u}) }\ 
\chi^d_{(\mathfrak{j},\bar{\mathfrak{p}})}(\bar{{\tt q}})\,\chi^d_{(\mathfrak{j},-\mathfrak{p})}({\tt q})\,,
\ee
where\footnote{%
In the  formula \eqref{oaspdo121}
for $Z^{({\rm disc})}$, the integers ${\tt v}$ and ${\tt u}$ are formal summation variables,
which can not be identified with the eigenvalue of $\mathbb{S}^z$ and the winding number ${\tt w}$.
In turn the notation ${\mathfrak  p}$ and $\bar{{\mathfrak  p}}$ in \eqref{aoisdo1902asas}
 should not be confused with $p$ and $\bar{p}$ from \eqref{oaisoi1093}. 
}
\be\label{aoisdo1902asas}
\bar{\mathfrak{p}}=\tfrac{1}{2}\,{\tt v}-\tfrac{1}{2}\,(n+2)\,({\tt k}+{\tt u})\,,\qquad\qquad
\mathfrak{p}=\tfrac{1}{2}\,{\tt v}+\tfrac{1}{2}\,(n+2)\,({\tt k}+{\tt u})\ .
\ee
The overall factor of $2$ in the formula for ${ Z}^{({\rm disc})}$
occurs due to the global ${\cal Z}_2$ invariance of the model.
\bigskip

The following comment is in order here.
For arbitrary values of ${\tt k}$, the inclusion of 
the endpoints into  the interval for $\mathfrak{j}$  in \eqref{iaosid982981} 
 has no effect on the set $\mathfrak{J}({\tt v},{\tt u})$.
However for ${\tt k}=0$ and with  $n$ generic, which is of special interest, $\mathfrak{j}$ may
coincide with $-\frac{n+1}{2}$ or $-\frac{1}{2}$. 
Taking the limit ${\tt k}\to 0$ of $Z^{({\rm disc})}$ one finds that in order for 
\eqref{oaspdo121} to correctly describe the contribution of the discrete spectrum
to the partition function $Z^{({\rm scl})}$ for the model with periodic boundary 
conditions, one  of the endpoints  in \eqref{iaosid982981} must be included.  
The choice of whether to include $\mathfrak{j}=-\frac{n+1}{2}$ or
$\mathfrak{j}=-\frac{1}{2}$ does not matter, since they  correspond to the contribution of
the same states to $Z^{({\rm disc})}$.

\bigskip

The contribution of the continuous spectrum to the partition function ${ Z}^{({\rm scl})}$
 is simply obtained by combining the
decomposition of ${\cal H}^{({\rm cont})}$ 
into the direct integral \eqref{iaosidoi192009}
 with the density of states
 \eqref{aisodio12311}.
For future reference, we write it in the form
\bea\label{iaosido12032}
Z^{({\rm cont})}&=&
\sqrt{\frac{n}{\Im m(\tau)}}\ \ 
\frac{\log\big({2^{\frac{2}{n}}\,N}/{N_0}\big)}{\pi\, (\bar{{\tt q}},{\bar{\tt q}})_\infty^{2}({\tt q},{\tt q})_\infty^{2}}\ 
\sum_{S^z\!,{\tt w}=-\infty}^\infty
\bar{{\tt q}}^{-\frac{1}{12}+\frac{\bar{p}^2}{n+2}}\ 
{\tt q}^{-\frac{1}{12}+\frac{p^2}{n+2}}\\[0.2cm]
&+&
\sum_{S^z\!,{\tt w}=-\infty}^\infty\int_{-\infty}^{+\infty}\rd s\
\sum_{{\tt L},\bar{\tt L}\ge 0}\tilde{\rho}_{\bar{p},p}^{(\bar{\tt L},{\tt L})}(s)\ 
\bar{{\tt q}}^{-\frac{1}{12}+\frac{s^2}{n}+\frac{\bar{p}^2}{n+2}+\bar{{\tt L}}}\ 
{\tt q}^{-\frac{1}{12}+\frac{s^2}{n}+\frac{p^2}{n+2}+{\tt L}}\ .\nonumber
\eea
Here we take into account that $\bar{\tau}=-\tau^*$ so that
\bea
{\bar {\tt q}}{\tt q}=\re^{-4\pi\Im m(\tau)}\ .
\eea
The summand in the second line of  \eqref{iaosido12032}
 is naturally interpreted
 as the regularized matrix elements
 of a certain density matrix and
the expansion coefficients  $\tilde{\rho}_{\bar{p},p}^{(\bar{\tt L},{\tt L})}$
 read explicitly  as
\bea\label{aasas312321A}
\tilde{\rho}_{\bar{p},p}^{(\bar{\tt L},{\tt L})}(s)
&=&\frac{1}{2\pi\ri}\ \partial_s
\log\bigg[\,
\big(\mathfrak{D}^{(\bar{\tt L})}_{\bar{p}}(s)\big)^{{\rm par}_2({{\tt L}})}\
\big(\mathfrak{D}^{({{\tt L}})}_{{p}}(s)\big)^{{\rm par}_2(\bar{{\tt L}})}\,\bigg]
\eea
with
\be\label{aasas312321B}
{\mathfrak D}^{({\tt L})}_p(s)=
\bigg(\frac{\Gamma(\frac{1}{2}+p-\ri s)}{\Gamma(\frac{1}{2}+p+\ri s)}
\bigg)^{{\tt par}_2({\tt L})}\  
\prod_{a=0}^{{\tt L}-1}
\Bigg[\frac{\big(\tfrac{1}{2}+a+p-\ri s\big)\,\big(\tfrac{1}{2}+a-p-\ri s\big)}
{\big(\tfrac{1}{2}+a+p+\ri s\big)\,\big(\tfrac{1}{2}+a-p+\ri s\big)}\Bigg]^{{\rm par}_2({\tt L})-d_{a}({\tt L})} \ .
\ee
The integers $d_{a}({\tt L})$, appearing in the exponent, are defined  in \eqref{Zdef1b}. Due to the property
$d_{a}({\tt L})={\rm par}_2({\tt L})$ for $a\ge {\tt L}$, the upper limit  in the product
in \eqref{aasas312321B} may be set to infinity.  This allows one to perform the sum over
 ${\tt L}$ and $\bar{\tt L}$  
in the second line of \eqref{iaosido12032} 
and bring it to the form, which is convenient for numerical calculations:
\bea
\sum_{{\tt L},\bar{\tt L}\ge 0}
\tilde{\rho}_{\bar{p},p}^{(\bar{\tt L},{\tt L})}(s)\ 
\bar{{\tt q}}^{\bar{{\tt L}}} {\tt q}^{{\tt L}}=-
\frac{r_{\bar p}(s,\bar{\tt q})+r_{ p}(s,{\tt q})}{\pi\, (\bar{{\tt q}},{\bar{\tt q}})_\infty^{2}({\tt q},{\tt q})_\infty^{2}}
\eea
with
\bea\label{iaosid1298}
r_{ p}(s,{\tt q})&=&\frac{1}{2}\ \sum_{\sigma=\pm}\psi\big(\tfrac{1}{2}+p+\ri\sigma s\big)\\
&+& \oint_{|z|<1}\frac{\rd z}{2\pi\ri}\ \frac{({\tt q},{\tt q})^2_\infty}{
(z,{\tt q})_\infty (z^{-1}\,{\tt q},{\tt q})_\infty}\ 
\ \frac{1}{2}\sum_{\sigma,\sigma'=\pm }\Phi(z,1,\tfrac{1}{2}+\sigma' p+\ri\sigma s)\  .\nonumber
\eea
Here $\psi(\alpha)=\partial_\alpha\log \Gamma(\alpha)$,  while  $\Phi(z,1,\alpha)$ stands for the  Lerch transcendent,
\bea
\Phi(z,s,\alpha)=\sum_{m=0}^\infty\frac{z^m}{(m+\alpha)^s}\ .
\eea
\bigskip

Similar to the contribution of the discrete spectrum to the partition function,
 the formula for $Z^{({\rm cont})}$ requires special attention 
for the case of periodic boundary conditions.
This is because at ${\tt k}=0$, 
$p$ ($\bar{p}$) can take half integer values for which
 the function $r_p(s,{\tt q})$ ($r_{\bar{p}}(s,{\tt q})$) \eqref{iaosid1298}
contains simple poles at $s=0$. 
These poles $\propto\frac{\sigma'}{s}$  formally cancel out after summation over the sign factor $\sigma'$.
However, the naive cancellation does not take into account the possibility of contact terms 
proportional to the Dirac delta function $\delta(s)$
which would give a finite contribution to the integral in \eqref{iaosido12032}. To resolve the ambiguity
one should start with $Z^{({\rm cont})}$ for  non-vanishing ${\tt k}$ 
and then  perform the limit
${\tt k}\to 0$ using the Sokhotski-Plemelj formula.

\bigskip

\subsection{The case of ${\tt k}= 0$  with generic $n>0$\label{sec173}}

In view of applications to local quantum field theory, of special interest
is when the spectrum of the Lorentz spin in ${\cal H}^{({\rm cont})}$ 
and ${\cal H}^{({\rm disc},\pm)}$ 
consists of (half-)integers.
Since the Lorentz spin of the states, characterized by the quantum numbers
$S^z$, ${\tt w}$, ${\tt L}$ and $\bar{\tt L}$, reads as
\be\label{oasdpo120903}
\Delta-\bar{\Delta}=S^z\,({\tt k}+{\tt w})+{\tt L}-\bar{\tt L}
\ee
this motivates  a detailed study of a few special cases.
Among them is the ${\cal CP}$ invariant
sector of the model,  where $S^z=0$.
If in addition one sets
${\tt k}=\pm\frac{1}{n+2}$, then the space of states contains 
a ${\cal CP}$ and 
${\cal Z}_2$ invariant $\overline{W}_\infty\otimes W_\infty$ primary state with conformal dimensions
$\Delta=\bar{\Delta}=0$. Thus  the sector $S^z=0$
with ${\tt k}=\pm\frac{1}{n+2}$ (and, perhaps, with $n$ a positive integer)
is interesting to study in the context of 
 the RSOS reductions of the inhomogeneous six-vertex model.\footnote{%
In fact, RSOS reductions of the inhomogeneous six-vertex model for various boundary conditions have been already considered in 
refs.\cite{Robertson:2019eam,Robertson:2020eri}.
}
However, this will not be considered here. Instead we'll focus 
on another situation when the Lorentz spin \eqref{oasdpo120903}
takes integer values, namely, when ${\tt k}=0$.
\bigskip

The case ${\tt k}=0$, i.e., periodic boundary conditions for 
$\sigma^a_m$ entering into the Hamiltonian $\mathbb{H}$
\eqref{aioiisa},\,\eqref{BC1a},
has a special feature. 
As discussed in ref.\cite{Bazhanov:2020new} for arbitrary ${\tt k}$
the  matrix
\be\label{oaspodp9012}
\hat{{\cal C}}=c_N
\prod_{J=1}^N\,(\eta_J)^{\frac{1}{2}\sigma^z_J}\,\sigma^x_J \qquad \qquad\qquad \big(\,\eta_J=(-1)^{J+1}\,\ri\,\big)\,,
\ee
where $c_N^2=1$,
satisfies the following commutation relations with 
$\mathbb{A}_\pm(\zeta)$ and the transfer matrix:
\bea\label{Ctrans}
{\hat {\cal C}}\,{\mathbb A}_\pm(\zeta\,|\,{\tt k})\, 
{\hat {\cal C}}={\mathbb A}_\mp(\zeta\,|-{\tt k})\,, \qquad
{\hat {\cal C}} \,{\mathbb T}(\zeta\,|\,{\tt k})\, {\hat {\cal C}}
={\mathbb T}(\zeta\,|-{\tt k})
\ .
\eea
In turn the  Hamiltonian $\mathbb{H}$
does not commute with  $\hat{\cal C}$ when the twist is non-trivial.
However, for ${\tt k}=0$ the system possesses
an additional global symmetry -- ${\cal C}$ invariance.
The space of states $\mathscr{V}_N$ \eqref{vec1}
can be split into two components distinguished by their ${\cal C}$ parity.
Numerical work shows that for  ${\tt k}=0$ and \emph{generic} values of the anisotropy parameter $n>0$
the transfer matrix resolves all the degeneracies in
the energy spectrum in each component. This implies that one can introduce a basis in
the finite dimensional space $\mathscr{V}_N$, which diagonalizes $\mathbb{T}(\zeta)$ and
$\hat{\cal C}$ simultaneously.
Though the latter  commutes with the transfer matrix
for ${\tt k}=0$, it
 anti-commutes with the total
spin operator $\mathbb{S}^z=\frac{1}{2}\sum_J\sigma_J^z$.
Hence each basis state would no longer have a definite value of 
$S^z$, except for the states with $S^z=0$.
Note that the matrices $\mathbb{A}_\pm(\zeta)$ restricted to this sector coincide,
so that
\be\label{aiosid19821}
A_+(\zeta)=A_-(\zeta)\qquad \qquad ({\tt k}=S^z=0)\ .
\ee
This follows from the first equation in \eqref{Ctrans}, specialized to ${\tt k}=0$, and
the fact that the transfer matrix, which commutes with $\mathbb{A}_\pm(\zeta)$, by itself lifts all the degeneracies 
in the  $S^z=0$ sector.
Also it turns out that in this sector the generator $\hat{\cal C}$, up
to a sign factor, coincides with $\mathbb{A}_+^{(\infty)}$ \eqref{8d8d91029a}.
Namely, one can show that
\be\label{aoisdo981892}
\hat{{\cal C}}\,\bm{\Psi}={\cal C}_{\bm{\Psi}}\,\bm{\Psi}\,,\qquad\qquad
{\cal C}_{\bm{\Psi}}=c_N\prod_{m=1}^{N/2}\zeta_m^{-1}\ \qquad\qquad \big(\,{\tt k}=S^z=0\,\big)\,.
\ee
\bigskip

Since $\hat{{\cal C}}$ anticommutes with $\mathbb{S}^z$, the  ${\cal C}$  even and odd components of
 $\mathscr{V}_N$  do not possess the ${\rm U}(1)$ symmetry.
Nevertheless, these sectors are still invariant w.r.t. 
the subgroup of ${\rm U}(1)$,
 whose generator corresponds to a $180^\circ$ rotation and 
may be chosen to be
\be\label{Ugen1a}
\hat{{\cal U}}=(-1)^{N/2}\ \re^{\ri\pi \mathbb{S}^z}\ ,\qquad\qquad
\hat{{\cal U}}^2=1\ .
\ee
The extra  factor $(-1)^{N/2}$ has been included 
so that the eigenvalues of $\hat{{\cal U}}$ 
coincide with  the sign factor $\sigma$ \eqref{sigmadef1a} entering 
into the asymptotic relation \eqref{quantC1}.
 Recall that ${\cal \hat{C}\hat{P}}$ commutes with the generator $\hat{\cal D}$
of the ${\cal Z}_2$ symmetry
for arbitrary values of the twist parameter ${\tt k}$.
However the matrix $\hat{{\cal C}}$
satisfies the
 commutation relations
\be
\hat{{\cal C}}\,\hat{{\cal D}}=\hat{{\cal U}}\,\hat{{\cal D}}\,\hat{{\cal C}}\ ,\qquad\qquad
\big[\,\hat{{\cal C}}\,,\,\hat{{\cal U}}\,\big]=\big[\,\hat{{\cal D}}\,,\,\hat{{\cal U}}\,\big]=0\ .
\ee

The following comment is in order here.
The definition of the ${\cal C}$ conjugation
 \eqref{oaspodp9012}
contains the sign factor $c_N=\pm1$, which may depend on the number of sites.
We found it convenient to set
\be\label{aosid9818921}
c_N=\begin{cases}(-1)^{N/4} & \quad N/2\,-\,{\rm even}\\[0.2cm]
1 & \quad N/2\,-\,{\rm odd}
\end{cases}\  .
\ee
For $N/2$ even  the ground state (the state with the lowest possible 
energy)
of the lattice Hamiltonian with periodic boundary conditions
is non-degenerate. With the choice  of the sign factor as in \eqref{aosid9818921},
 its ${\cal C}$ parity is equal to $+1$.
When $N/2$ is odd the ground state is a  ${\cal Z}_2$ doublet and the ${\cal C}$
parity of the two states is $+1$ and $-1$.
\bigskip

In taking the scaling limit, one can apply the same arguments
that lead to eqs.\,\eqref{oapsd0092} and \eqref{iaosio11212} for
the previously discussed global symmetries. This way one finds
\be\label{aisud9812}
{\cal \hat{C}}\ {W}_j(u)\,{\cal \hat{C}}={W}_j(u)\,,\qquad \qquad
{\cal \hat{C}}\ \overline{W}_j(\bar{u})\,{\cal \hat{C}}=\overline{W}_j(\bar{u})\qquad
\qquad
({\tt k}=0)\ .
\ee
Hence  $\hat{\cal C}$  maps a
 $\overline{W}_\infty\otimes W_\infty$ highest weight irrep to 
an equivalent representation, i.e., one that is characterized by the same
 highest weight. For the
components 
${\cal H}_{S^z\!,{\tt w}}^{({\rm cont})}$ and 
${\cal H}_{S^z\!,{\tt w}}^{({\rm disc},+)}$ occurring in the linear decomposition
\eqref{iasodioi12233}
 the ${\cal C}$ conjugation acts as
\be\label{aisd89129812}
{\cal \hat{C}}\ : \ \ \
\begin{array}{l}
{\cal H}^{({\rm cont})}_{+S^z,+{\tt w}}\mapsto {\cal H}^{({\rm cont})}_{-S^z,-{\tt w}} \\[0.4cm]
{\cal H}^{({\rm disc},+)}_{+S^z,+{\tt w}}\mapsto {\cal H}^{({\rm disc},+)}_{-S^z,-{\tt w}}
\end{array}\qquad \quad {\rm for} \qquad \quad  S^z>0\ .
\ee
The case of ${\cal H}_{S^z\!,{\tt w}}^{({\rm disc},-)}$ is more involved.
It turns out that the action of  $\hat{\cal C}$  is described by the relations 
\be\arraycolsep0.6cm
{\cal \hat{C}}\ : \
{\cal H}^{(i,-)}_{+S^z,+{\tt w},\sigma}\mapsto {\cal H}^{(5-i,-)}_{-S^z,-{\tt w},\sigma}
\qquad \quad {\rm for} \qquad \quad  S^z>0\ .
\ee
Here ${\cal H}_{S^z\!,{\tt w},\sigma}^{(i,-)}$  with $i=1,\ldots,4$ are given by \eqref{iaosido89812aaa}
for $S^z> 0$, while
$${\cal H}_{S^z\!,{\tt w},\sigma}^{(i,-)}\equiv{\cal CP}\big({\cal H}_{-S^z\!,{\tt w},\sigma}^{(i,-)}\big)\
\qquad {\rm for}\qquad S^z<0\ .$$ 

\bigskip

Special attention is required for the ${\cal CP}$ invariant sector where $S^z=0$. 
First we note that the scaling limit of the eigenvalues of $\mathbb{A}_+(\zeta)$ 
described by \eqref{as56d1a} involves the connection coefficients 
$D_+(\mu\,|\,\bm{w},+\frac{1}{2}(n+2)\,{\tt w},s)$ and 
$D_+(\bar{\mu}\,|\,\bar{\bm{w}},-\frac{1}{2}(n+2)\,{\tt w},s)$, which do not depend on the sign of ${\tt w}$.
This can be seen from \eqref{aiosid19821} and that
$D_+(\mu\,|\,\bm{w},+\frac{1}{2}(n+2)\,{\tt w},s)=D_-(\mu\,|\,\bm{w},-\frac{1}{2}(n+2)\,{\tt w},s)$.
Thus in our prescription,  the sign of the winding number ${\tt w}\ne 0$
for  an RG trajectory $\bm{\Psi}_N$ remains undetermined when
 ${\tt k}=S^z=0$.
Nevertheless we found that the pair of low energy states which become indistinguishable in
the scaling limit 
have different energies for finite $N$. This allows one to set, by definition, that the state with $-|{\tt w}|$
and $+|{\tt w}|$  has the  lower and higher value of $|{\cal E}|$,
respectively. Another way to resolve the ambiguity in the sign of the winding number is to
start with the Bethe state with ${\tt k}\ne0$  and consider the limit ${\tt k}\to 0$.
It follows from the formula for the energy \eqref{oaisoi1093},\,\eqref{tower1a}
that for small positive ${\tt k}$ and $S^z=0$ the state with ${\tt w}>0$ 
will be of higher energy than the corresponding state having the opposite sign of ${\tt w}$.
In the limit ${\tt k}\to 0^+$ the Bethe states with $+|{\tt w}|$ and $-|{\tt w}|$ would
become the states with higher and lower energy, respectively, and the two ways of
specifying the sign of the winding number turn out to be equivalent.
Having resolved the issue with the sign, each of the spaces 
${\cal H}^{({\rm cont})}_{0,{\tt w}}$, ${\cal H}^{({\rm disc},+)}_{0,{\tt w}}$
and ${\cal H}^{({\rm disc},-)}_{0,{\tt w}}$
occurring in the scaling limit become invariant w.r.t. the ${\cal C}$ conjugation.
\medskip

\begin{figure}
\centering
\begin{tikzpicture}
\node at (0,0) {\includegraphics[width=15cm]{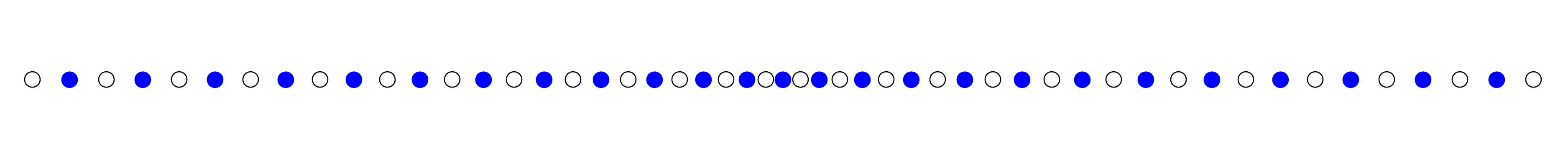}};
\draw (-7.5,0) -- (7.5,0);
\node at (0,-0.5) { $0$};
\draw (0,0.2) -- (0,-0.2);
\draw (7.5,0.2) -- (7.5,-0.2);
\node at (7.5,-0.5) { $10$};
\draw (-7.5,0.2) -- (-7.5,-0.2);
\node at (-7.7,-0.5) { $-10$};
\draw (3.75,0.2) -- (3.75,-0.2);
\node at (3.75,-0.5) { $5$};
\draw (-3.75,0.2) -- (-3.75,-0.2);
\node at (-3.95,-0.5) { $-5$};
\draw (0.75,0.12) -- (0.75,-0.12);
\draw (1.5,0.12) -- (1.5,-0.12);
\draw (2.25,0.12) -- (2.25,-0.12);
\draw (3.0,0.12) -- (3.0,-0.12);
\draw (4.5,0.12) -- (4.5,-0.12);
\draw (5.25,0.12) -- (5.25,-0.12);
\draw (6.0,0.12) -- (6.0,-0.12);
\draw (6.75,0.12) -- (6.75,-0.12);
\draw (-0.75,0.12) -- (-0.75,-0.12);
\draw (-1.5,0.12) -- (-1.5,-0.12);
\draw (-2.25,0.12) -- (-2.25,-0.12);
\draw (-3.0,0.12) -- (-3.0,-0.12);
\draw (-4.5,0.12) -- (-4.5,-0.12);
\draw (-5.25,0.12) -- (-5.25,-0.12);
\draw (-6.0,0.12) -- (-6.0,-0.12);
\draw (-6.75,0.12) -- (-6.75,-0.12);
\end{tikzpicture}
\caption{\small
For fixed $N=100$ and $S^z=0$ the value of real $b(N)$ 
 is plotted on the number line for the class of Bethe states described in sec.\,\ref{sec31}.
These states are such that all the corresponding Bethe roots are real,
and are distinguished by the difference between the number of positive roots $M_+$ and the number of
negative roots $M_-$, which for the $51$ states used to produce the figure varies from
$M_+-M_-=-50,-48,\ldots,-2,0,2,\ldots,50$. The parameter ${\tt k}$ was set to zero so that, since $S^z=0$,
the states have a definite ${\cal C}$ parity. The latter, computed from the Bethe roots using 
 formula \eqref{aoisdo981892}, is indicated by the solid fill for the ${\cal C}$ even states and
no fill for the ${\cal C}$ odd ones. The number of ${\cal C}$ even (odd) states with
$b(N)\in (s,s+\Delta s)\subset (-b_{{\rm max}},+b_{{\rm max}})$ 
is approximately
$\frac{1}{2}\,\rho_{0,0}^{(0,0)}(s)\,\Delta s$ with $\rho_{\bar{p},p}^{(0,0)}$ being the density of primary Bethe
states given in eq.\,\eqref{rho01a}. The anisotropy parameter was taken to be $n=2.93\,$.
\label{figEa}}
\end{figure}
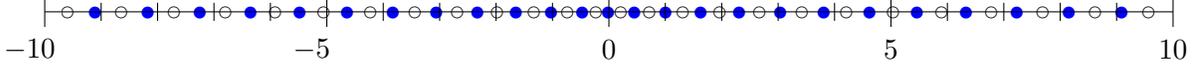

Recall that the space  ${\cal H}^{({\rm cont})}_{0,{\tt w}}$
is formed by the scaling limit of the low energy Bethe states 
$\bm{\Psi}_N$
with $S^z=0$, such that 
 $\lim_{N\to\infty}\Im m\big(b(N)\big)=0$.
It turns out that  for finite $N$ the difference between
the number of ${\cal C}$ even  and ${\cal C}$ odd 
 states  that become part of ${\cal H}_{0,{\tt w}}^{({\rm cont})}$
is an order one number as $N\to\infty$,
see fig.\,\ref{figEa}. 
For the low energy Bethe states with given ${\cal C}$ parity and fixed  $(\bar{\tt L},{\tt L})$,
the corresponding values of $\Re e\big(b(N)\big)$ become densely distributed
within the segment $(-\Lambda_N,\Lambda_N)$ with 
$\lim_{N\to\infty}\Lambda_N=\infty$, and the density of states
for $\Re e\big(b(N)\big)\in (s,s+\Delta s)$ turns out to be half the total density, 
$\rho_{\bar{p},p}^{(\bar{\tt L},{\tt L})}(s)$, from  eq.\,\eqref{aisodio12311}.

\medskip

In the case of ${\cal H}^{({\rm disc},+)}_{0,{\tt w}}=
\bigoplus\limits_{\sigma=\pm1} \Big({\cal H}^{(1,+)}_{0,{\tt w},\sigma}\,\oplus\,
{\cal H}^{(2,+)}_{0,{\tt w},\sigma}\Big)$ our numerical work shows
that 
the value of the ${\cal C}$ parity is the same for all the states in each component
${\cal H}^{(i,+)}_{0,{\tt w},\sigma}$ with ${\tt w}\ne 0$.
However, it depends on whether the scaling limit with $N\to\infty$ is taken such that 
$N/2$ is kept fixed to be even or odd:
\be
\hat{{\cal C}}\big({\cal H}_{0,{\tt w},\sigma}^{(i,+)}\big)=
-\,\sigma^{N/2}\ {\rm sgn}({\tt w})\,{\cal H}_{0,{\tt w},\sigma}^{(i,+)}
\qquad\qquad (i=1,2)\ .
\ee
Note that for ${\tt w}=0$ the space ${\cal H}_{0,0}^{({\rm disc},+)}=0$ as follows from 
eqs.\,\eqref{iao989382931} and \eqref{sigmaP1a}.
For ${\cal H}^{({\rm disc},-)}_{0,{\tt w}}$
the subspaces ${\cal H}^{(i,-)}_{0,{\tt w},\sigma}$ with $i=2,3$ are trivial
since the summation index $a$ runs over the empty 
sets $\Sigma_2(p)$ and $\Sigma_2(\bar{p})$  in \eqref{iaosido89812aaa}.
For  similar reasons 
${\cal H}^{(4,-)}_{0,{\tt w},\sigma}$ and 
${\cal H}^{(1,-)}_{0,{\tt w},\sigma}$ are also
trivial for 
 ${\tt w}>0$  and ${\tt w}<0$, respectively.
It is expected that
\bea\label{aosid912981}
&&\hat{\cal C}\big({\cal H}^{(1,-)}_{0,{\tt w},\sigma}\big)=
+ \,c^{(1)}_{{\tt w}}\,\sigma^{N/2}\ {\cal H}^{(1,-)}_{0,{\tt w},\sigma} \,,\qquad
{\cal H}^{(4,-)}_{0,{\tt w},\sigma}=0
\qquad\qquad  ({\tt w}>0)
\nonumber\\[-0.3cm]
\\[0.1cm]
&&
\hat{\cal C}\big({\cal H}^{(4,-)}_{0,{\tt w},\sigma}\big)=
 -\,c^{(4)}_{{\tt w}}\,\sigma^{N/2}\ {\cal H}^{(4,-)}_{0,{\tt w},\sigma}\,,\qquad
{\cal H}^{(1,-)}_{0,{\tt w},\sigma}=0
\qquad\qquad ({\tt w}<0)\ , \nonumber
\eea
where $c^{(1)}_{{\tt w}}$ and $c^{(4)}_{{\tt w}}$ are some signs
that could depend on ${\tt w}$. 
However the value of these sign factors is still unknown to us as their determination
involves the analysis of the Bethe states,
which are of rather high energy compared to the ground state.
The subspace ${\cal H}^{({\rm disc},-)}_{0,0}$ contains the two 
non-trivial components
${\cal H}^{(1,-)}_{0,0,\sigma}$ and ${\cal H}^{(4,-)}_{0,0,\sigma}$,
which are classified identically w.r.t.  the algebra of extended conformal symmetry.
In turn, there is an ambiguity in assigning a low energy Bethe state to either one of these components.
However this can be resolved by making use of ${\cal C}$ parity for finite $N$
 and then taking the  large $N$ limit, so that
by definition
\be
\hat{\cal C}\big({\cal H}^{(1,-)}_{0,0,\sigma}\big)=+\,\sigma^{N/2}\ {\cal H}^{(1,-)}_{0,0,\sigma}\,,\qquad
\qquad
\hat{\cal C}\big({\cal H}^{(4,-)}_{0,0,\sigma}\big)=-\,\sigma^{N/2}\ {\cal H}^{(4,-)}_{0,0,\sigma}\ .
\ee

\medskip

This way each of the spaces ${\cal H}^{({\rm cont})}$ and ${\cal H}^{({\rm disc},\pm)}$ 
is splitted into two sectors characterized by the value of the ${\cal C}$ parity. 
The decompositions of the even and odd components
into the highest weight irreps of the
$\overline{W}_\infty\otimes W_\infty$\,-\,algebra are identical. 
We'll restrict our further  discussion and only focus on the ${\cal C}$
even (or equivalently odd) sector of ${\cal H}^{({\rm cont})}$.
\bigskip

Let's turn to  formula \eqref{iaosidoi192009}, which describes the 
 decomposition of ${\cal H}_{S^z,{\tt w}}^{({\rm cont})}$
into the highest weight irreps for $(n+2)\,{\tt k}\notin \mathbb{Z}$.
Each of the chiral components in the integrand therein coincides with the
Verma module, which is an irreducible representation of the chiral $W_\infty$\,-\,algebra.
However, as was discussed in sec.\,\ref{sec162}, for ${\tt k}=0$ some of these Verma modules
become reducible. 
The 
degenerate Verma module, ${\cal V}er_{ \rho, s}$, splits into the two irreps
\bea\label{oasid89189212}
{\cal V}er_{ \rho, s}={\cal W}_{ \rho,s}\oplus {\cal W}_{ \rho+m(n+2),s}\ ,\ \ \  {\rm where}\ \ \  \rho=\tfrac{1}{2}\,\big(r-m\,(n+2)\,\big)\, ,\ \ \  m,r=1,2,\ldots
\eea
and $s$ is an arbitrary real number.
With this in mind, 
 it is straightforward to obtain  from eq.\,\eqref{iaosidoi192009}
the decomposition into the highest weight irreps
of the ${\cal C}$ even sector of ${\cal H}^{({\rm cont})}$
for ${\tt k}=0$:
\bea\label{ioasid1892981}
{\cal H}^{({\rm cont})}_{\rm even}=\tilde{\cal H}^{({\rm cont})}_{\rm even}
\oplus {\cal H}^{(\rm null)}\ .
\eea
Here
\be\label{iaosidoi1920091a}
\tilde{\cal H}^{({\rm cont})}_{\rm even}=
\bigoplus_{{\tt v}= 0}^{\infty}\Bigg[
\bigoplus_{{\tt w}=-\infty}^{\infty}\tilde{\cal H}^{({\rm cont})}_{{\tt v},{\tt w}}\Bigg]
\ee
with
\be\label{oaisd988921}
\tilde{\cal H}^{({\rm cont})}_{{\tt v},{\tt w}}=\int^{\oplus}_{\mathbb{R}}\!\rd s\
\overline{{\cal W}}_{\bar{\rho},s}\otimes {\cal W}_{\rho,s}
\qquad \quad{\rm and}
\qquad \quad
\begin{array}{l}
\rho=\tfrac{1}{2}\,{\tt v}+\tfrac{1}{2}\,(n+2)\,{\tt w}\\[0.2cm]
\bar{\rho}=\tfrac{1}{2}\,{\tt v}-\tfrac{1}{2}\,(n+2)\,{\tt w}
\end{array} \ ,
\ee
while the space  ${\cal H}^{(\rm null)}$ is a direct sum of two components,
\bea\label{aosid9182233}
{\cal H}^{(\rm null)}={\cal H}^{(\rm null)}_{+}\oplus{\cal H}^{(\rm null)}_{-} \,,
\eea
that  are decomposed identically into the irreps of the  algebra of extended conformal symmetry
\bea\label{aopsid9120asas}
{\cal H}^{(\rm null)}_{\pm}
=\bigoplus_{{\tt v},{\tt w}=1}^{+\infty}\int^{\oplus}_{\mathbb{R}}\!\rd s\
\overline{{\cal W}}_{{\rho},s}\otimes {\cal W}_{\rho,s}\ \ \ \ \ \ \ \ \ \ \ \  \big(\,\rho=\tfrac{1}{2}\,{\tt v}+\tfrac{1}{2}\,(n+2)\,{\tt w}\, \big)\ .
\eea
The superscript ``null'' emphasizes that the highest state 
in either one of the chiral irreps 
occurring in the decomposition of ${\cal H}^{(\rm null)}_{\pm}$  coincides with the null
vector in the original Verma module (see \eqref{oasid89189212}).
\bigskip

Similar to  ${\cal H}^{(\rm null)}_{\pm}$
 the subspaces $\tilde{\cal H}^{({\rm cont})}_{0,+{\tt w}}$   and $\tilde{\cal H}^{({\rm cont})}_{0,-{\tt w}}$  
also possess identical  
decompositions w.r.t. the $\overline{W}_\infty\otimes W_\infty$ algebra.
This way
${\cal H}^{({\rm cont})}_{\rm even}$ contains degeneracies, which are not present 
in the ${\cal C}$ even sector of $\mathscr{V}_N$
for any finite $N$. 
As a result, at least at the formal level, one can introduce
two extra ${\cal Z}_2$ symmetry transformations in
${\cal H}^{({\rm cont})}_{\rm even}$, which commute with the algebra of 
 extended conformal symmetry.
The first one, $\hat{{\cal X}}^{({\tt w})}$, acts as the identity operator 
on all the subspaces appearing
in the linear decompositions \eqref{ioasid1892981} and \eqref{iaosidoi1920091a}
except for $\tilde{\cal H}^{({\rm cont})}_{0,{\tt w}}$ with ${\tt w}\ne 0$.
In the latter case, it intertwines 
the subspaces with opposite signs of ${\tt w}$:
\bea\label{z2sybaif891}
&&\!\!\!\!\!\!\!\hat{{\cal X}}^{({\tt w})}\big(\tilde{\cal H}^{({\rm cont})}_{{\tt v},{\tt w}}\big)=
 \tilde{\cal H}^{({\rm cont})}_{{\tt v},{\tt w}}\qquad {\rm for}
 \qquad {\tt v}={\tt w}=0\ \ \&\ \ {\tt v}\ge 1,\,{\tt w}\in\mathbb{Z}\nonumber \\[0.2cm]
&&\!\!\!\!\!\!\!\hat{{\cal X}}^{({\tt w})}\big({\cal H}^{({\rm null})}\big)={\cal H}^{({\rm null})}\\[0.2cm]
&& \!\!\!\!\!\!\!\!\!\!\!\hat{{\cal X}}^{({\tt w})}:\ \ \tilde{\cal H}^{({\rm cont})}_{0,{\tt w}}\mapsto
\tilde{\cal H}^{({\rm cont})}_{0,-{\tt w}}\qquad\quad\ \  \ \quad ({\tt w}\ne 0)\ .\nonumber
\eea
The second ${\cal Z}_2$ transformation, $\hat{{\cal X}}^{({\rm null})}$,  acts between
the  ``$\pm$'' components of the space ${\cal H}^{({\rm null})}$ \eqref{aosid9182233},
\be\label{asodi809ioas}
\hat{{\cal X}}^{({\rm null})}\big(\tilde{\cal H}^{({\rm cont})}_{\rm even}\big)=
\tilde{\cal H}^{({\rm cont})}_{\rm even}\,,\qquad\qquad
\hat{{\cal X}}^{({\rm null})}:\ \ {\cal H}^{({\rm null})}_{\pm}\mapsto{\cal H}^{({\rm null})}_{\mp}\ .
\ee
\bigskip

For any values of the twist parameter ${\tt k}$
the lattice system possesses ${\cal CP}$ symmetry.
Thus when ${\tt k}=0$ not only ${\cal C}$, but also
the ${\cal P}$ conjugation becomes a global symmetry of
the model. The
generator $\hat{\cal P}\in{\rm End}(\mathscr{V}_N)$ can be chosen to be
\be\label{oaisodi899832}
(\hat{\cal P})^{b_Nb_{N-1}\ldots b_1}_{a_Na_{N-1}\ldots a_1}=
c_N
\ \delta_{a_N}^{b_1}\,\delta_{a_{N-1}}^{b_2}\ldots \delta_{a_1}^{b_N}\  \prod_{J=1}^N\eta_J^{a_J/2}
\qquad \qquad\qquad \big(\,\eta_J=\ri\,(-1)^{J-1}\,\big)\,,
\ee
where $a_J,b_J=\pm1$ and
 $c_N$ is the same sign factor as in \eqref{aosid9818921}. 
Though $\hat{\cal P}$
 commutes with the lattice Hamiltonian
subject to periodic boundary conditions, 
in view of the relations \eqref{CP2}
and \eqref{Ctrans},  it does not commute with the transfer matrix. 
Instead,
\be
{\hat  {\cal P}}
\,{{\mathbb T}}(\zeta)\,{\hat  {\cal P}}=
\zeta^{N}\ {{\mathbb T}}\big(\zeta^{-1}\big)\qquad \qquad ({\tt k}=0)\ .
\ee
Since $[\,\hat{\cal C},\,\hat{\cal P}\,]=0$ 
the ${\cal C}$ even and odd components of 
the finite dimensional space $\mathscr{V}_N$ are ${\cal P}$ invariant. 
However it turns out that there are some subtleties in taking
the scaling limit of the operator \eqref{oaisodi899832}.
Assuming that the limit exists,
 eqs.\,\eqref{iaosio11212} and \eqref{aisud9812}
would imply that
\be\label{aoisd91821}
{\cal \hat{P}}\ {W}_j(u)=\overline{W}_j(u)\,{\cal \hat{P}}
\qquad\qquad\qquad ({\tt k}=0)\,.
\ee
To determine the action of the parity conjugation in
${{\cal H}}^{({\rm cont})}_{\rm even}$,
all that remains is to find how it acts
on the $\overline{W}_\infty\otimes W_\infty$
primary states in the decompositions \eqref{oaisd988921},\,\eqref{aopsid9120asas}.
Without loss of generality, one can always set
\be
\hat{{\cal P}}\,\bm{\psi}_{\bar{\rho},\rho,\nu}^{({\rm vac})}=+\bm{\psi}_{\rho,\bar{\rho},\nu}^{({\rm vac})}
\qquad\qquad
{\rm for} \qquad\qquad  \rho\ne\pm\bar{\rho}\ .
\ee
Otherwise, when $|\bar{\rho}|=|\rho|$,
the primary state $\bm{\psi}_{\bar{\rho},\rho,\nu}^{({\rm vac})}$ 
is an eigenvector of $\hat{{\cal P}}$,\footnote{%
Recall that the low energy Bethe states, which become the
primary states $\bm{\psi}_{\bar{\rho},\rho,\nu}^{({\rm vac})}\in \tilde{{\cal H}}^{({\rm cont})}_{\rm even}$  with
${\tt v}=0$ and having opposite signs of the winding number
have different energies on the finite lattice. Thus, despite that such primary states 
correspond to
equivalent irreps of the $\overline{W}_\infty\otimes W_\infty$ algebra, each of them is an eigenvector of 
$\hat{\cal P}$. Similarly the primary states 
$\bm{\psi}_{{\rho},\rho,\nu}^{({\rm vac},\pm)}\in{\cal H}^{({\rm null})}_\pm$
are also eigenvectors of $\hat{{\cal P}}$.
}
\be\label{oaiasdsd9812}
\hat{{\cal P}}\,\bm{\psi}_{\bar{\rho},\rho,\nu}^{({\rm vac})}\ =\ 
\sigma_{\bm{\psi}}\ \bm{\psi}_{\rho,\bar{\rho},\nu}^{({\rm vac})}\qquad\qquad\qquad
(\rho=\pm\bar{\rho})\ .
\ee
Here the sign factor $\sigma_{\bm{\psi}}$ can not be eliminated by a change of the normalization
of the state and its determination requires  a numerical study of the lattice system.
As it follows from the result quoted in \eqref{pasodpo12-01},
$\sigma_{\bm{\psi}}=+1$ for ${\rho}=-\bar{\rho}=\tfrac{1}{2}\,(n+2)\,{\tt w}$.
 In sec.\,\ref{sec31} 
the primary Bethe states $\bm{\Psi}_N$
with vanishing winding number and $S^z\ge 0$ were discussed.
The ${\cal C}$ even combination, $\bm{\Psi}_N+\hat{{\cal C}}\,\bm{\Psi}_N$,
which has the same ${\cal P}$ parity as $\bm{\Psi}_N$ itself,
in the scaling limit
becomes the primary state in the sector $\tilde{\cal H}^{({\rm cont})}_{{\tt v},0}$.
We found that the parity of $\bm{\Psi}_N$ is given by
\be\label{paspdo01921}
\hat{\cal P}\,\bm{\Psi}_N=c_N\,(-1)^{\frac{1}{2}(N/2-S^z+{\tt m})}\,
\bm{\Psi}_N\ \qquad \qquad ({\tt w}=0)\,,
\ee
where ${\tt m }$ is  the integer that 
 coincides with the difference between the number of 
negative and positive Bethe roots.
This formula implies that as $N\gg 1$ the value of $b(N)$ for
the  ${\cal P}$ even and odd  low energy primary Bethe states
 is densely distributed within the segment 
$(-\Lambda_N,\Lambda_N)$  with equal densities.
A similar situation occurs for $\bm{\Psi}_N$, which become the primary states in
${\cal H}_\pm^{({\rm null})}$.
\bigskip

To summarize the space ${\cal H}_{\rm even}^{({\rm cont})}$ 
possesses  ${\cal P}$, ${\cal T}$ and ${\cal Z}_2$ 
global symmetries while ${\cal C}$, by definition,
acts trivially inside it.
Moreover, there is another ${\cal Z}_2$ symmetry
${\cal U}$, which comes from the invariance 
of the ${\cal C}$ even sector of $\mathscr{V}_N$
w.r.t. to the transformation \eqref{Ugen1a}.
Being restricted to any irrep occurring in the decomposition of ${\cal H}^{({\rm cont})}_{\rm even}$,
the ${\cal U}$ transformation 
acts as the identity modulo a sign factor:
\be
\hat{{\cal U}}\,\big(\,\overline{{\cal W}}_{\bar{\rho},s}\otimes {\cal W}_{\rho,s}\,\big)=
\pm (-1)^{\rho+\bar{\rho}}\  \overline{{\cal W}}_{\bar{\rho},s}\otimes {\cal W}_{\rho,s}\subset 
{\cal H}^{({\rm cont})}_{\rm even}\ .
\ee
Here ``$\pm$''  depends on whether,
 for the construction
of the RG trajectories $\bm{\Psi}_N$, $N/2$ is kept  to be an even or an odd integer.
Finally, there are the two formal ${\cal Z}_2$ symmetries $\cal{X}^{({\tt w})}$
and $\cal{X}^{({\rm null})}$ acting in ${\cal H}_{\rm even}^{({\rm cont})}$, which are broken 
in  the lattice system.

\section{Numerical work\label{sec18}}
Our analysis of the scaling limit
is based  on a definition
of a low energy state which
was referred to as a ``working'' one.
This was to emphasize that it contains
several non-trivial
assumptions  regarding
the spectrum of the  ${\cal Z}_2$ invariant
inhomogeneous six-vertex model. 
Among the strongest of them is that the pair of integers
 $(\bar{\tt L},{\tt L})$ in \eqref{tower31}
 may only take non-negative values.
This would be natural to assume once
that pair has been identified with
the levels of the state  in the highest weight irrep of
the extended conformal symmetry algebra.
However, as was pointed out in sec.\,\ref{sec171},
the space of states in the scaling limit contains
the sector ${\cal H}^{({\rm disc},-)}$, where this identification does not hold true.
The condition $\bar{\tt L},{\tt L}\ge 0$ was motivated 
through a numerical study of the low energy spectrum of the lattice Hamiltonian.
\bigskip

\begin{figure}
\centering
\scalebox{0.9}{
\begin{tikzpicture}
\node at (0,0) {\includegraphics[width=7.3cm]{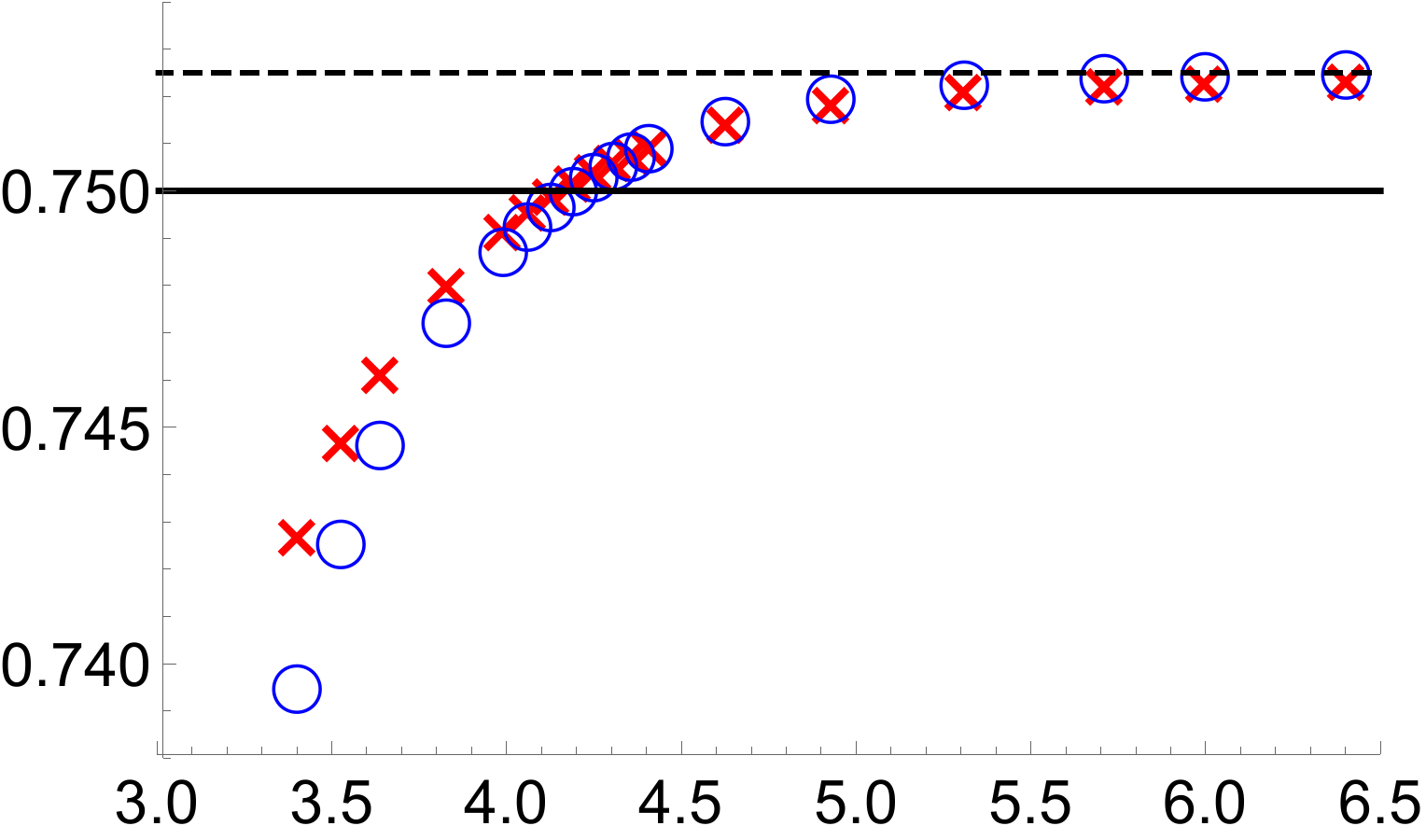}};
\node at (9.5,-0.05) {\includegraphics[width=7.3cm]{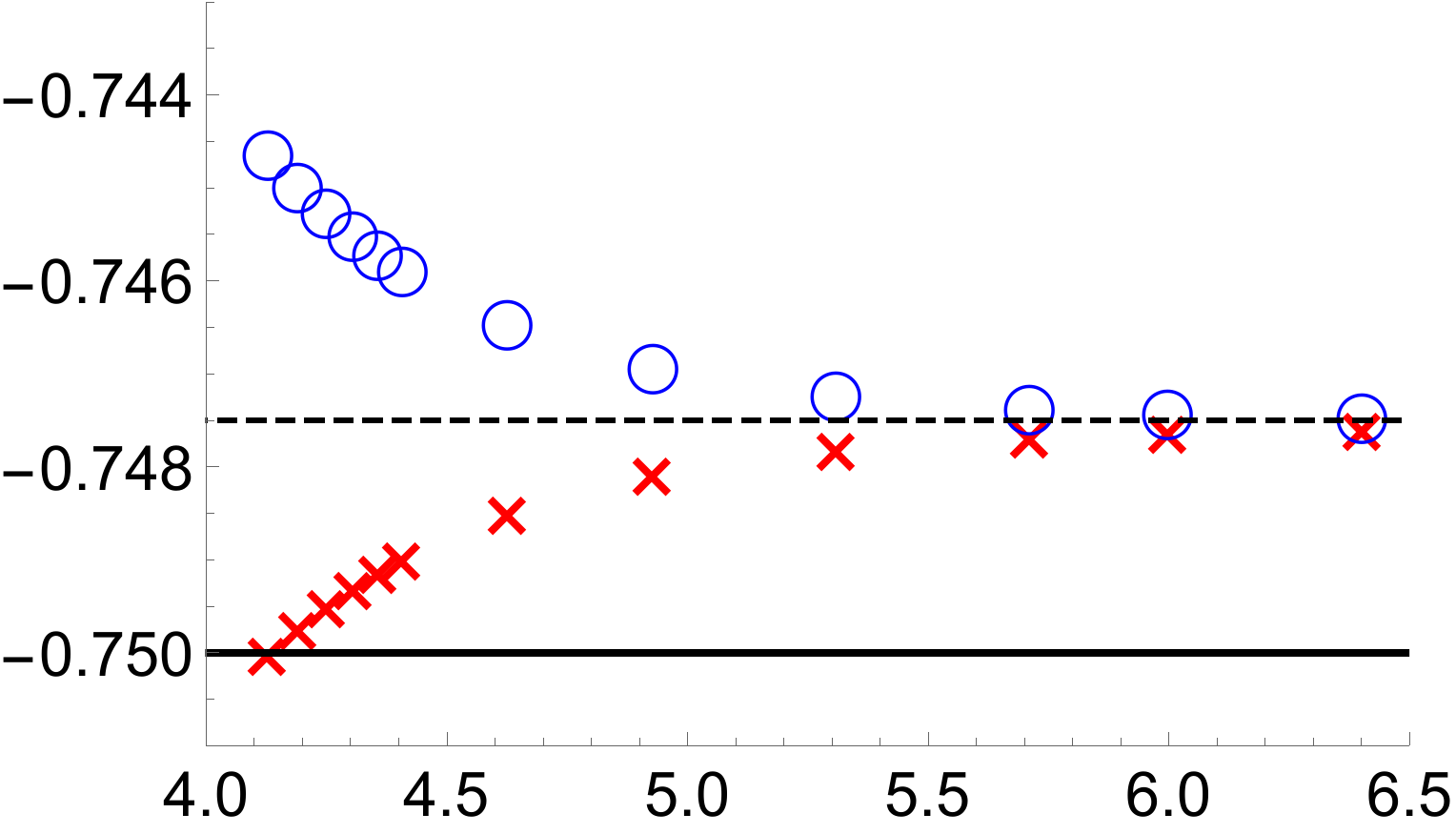}};
\node at (-3,2.4) {\small $-\ri b(N)$};
\node at (6.7,2.4) {\small $-\ri b(N)$};
\node at (4.4,-1.7) {\small $\log(N)$};
\node at (13.9,-1.75) {\small $\log(N)$};
\end{tikzpicture}
}
\caption{\small 
The red crosses depict the imaginary part of $b(N)=\frac{n}{4\pi}\log(B)$,
where $B$ is the eigenvalue of the quasi-shift operator \eqref{qshift},
 for an RG trajectory $\bm{\Psi}_N$ with the parameters set to be $n=3$, ${\tt k}=0.099$ and $S^z=0$.
For the left panel the logarithm was defined so that $b(N)$ is a continuous function
of $N$. As $N$ increases $b(N)$ leaves the strip $\big|\Im m\big(b(N)\big)\big|<\frac{n}{4}$,
whose boundary is marked  by the solid black line, and tends to the limiting value 
$\lim_{N\to\infty}b(N)=0.7525\ri$ corresponding to the dashed line.
It turns out that this way of specifying $b(N)$ is consistent with 
the asymptotic formula \eqref{tower1a} for the energy provided that
one takes ${\tt w}=-1$ and ${\tt L}=\bar{\tt L}=0$.
This is illustrated by the blue open circles, which correspond to $b(N)$
obtained from the energy ${\cal E}$ by 
 inverting eq.\,\eqref{tower1a}  with the correction terms ignored.
For the right panel, the branch of the  logarithm for $b(N)=\frac{n}{4\pi}\log(B)$  
(red crosses)
was taken such that
$\big|\Im m\big(b(N)\big)\big|<\frac{n}{4}$.
The blue circles depict $b(N)$ calculated  from the energy ${\cal E}$,
where in inverting eq.\,\eqref{tower1a} we set
 ${\tt w}=0$ and ${\tt L}=\bar{\tt L}=1$.
\label{fig12}}
\end{figure}
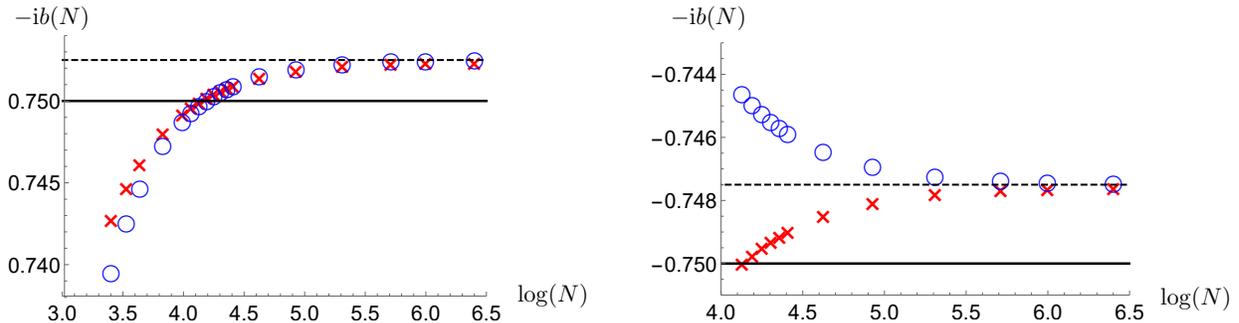

A related question concerns the constraint $\big|\Im m\big(b(N)\big)\big|<\frac{n}{4}$. 
Recall that the latter
was introduced so that $b(N)$, which is proportional to the logarithm of the eigenvalue of the quasi-shift
 operator \eqref{poapso1a}, would be defined  unambiguously
and  in a way that is consistent with the large $N$ asymptotic formula for the low
energy spectrum \eqref{tower1a}.
 However it turns out that one can choose the branch of the logarithm in 
\eqref{poapso1a} differently, such that
$b(N)$ is  continuous  in $N$
and \eqref{tower1a} is still valid.
These two ways of specifying the branch are equivalent 
in the case of the continuous spectrum where  $\lim_{N\to\infty}\Im m\big(b(N)\big)=0$.
However
there exist RG trajectories with
$\big|\Im m\big(b(N)\big)\big|<\frac{n}{4}$
for sufficiently small $N$, while if $b(N)$ is defined to be continuous 
in $N$ its limiting value as $N\to\infty$  lies outside of this strip.
An example is depicted in fig.\,\ref{fig12}, where
 $n=3$, ${\tt k}=0.099$, ${\tt w}=-1$, 
$S^z={\tt L}=\bar{\tt L}=0$, while
$\lim_{N\to\infty}\Im m\big(b(N)\big)= -\frac{3}{2}-{p}=0.7525>\frac{n}{4}$.
Remarkably, the trajectory still fits within our working definition of
the low energy state. This is due to the simple identity
\be\label{iaosid9012}
\frac{p^2+\bar{p}^2}{n+2}-\frac{2\,(p+\frac{1}{2}+a)^2}{n}+{\tt L}+\bar{\tt L}=
\frac{p_+^2+\bar{p}_+^2}{n+2}-\frac{2\,(p+\frac{1+n}{2}+a)^2}{n}+{\tt L}_++\bar{\tt L}_+
\ee
with 
\be
p_+=p+\tfrac{1}{2}\,(n+2)\,,\qquad \bar{p}_+=\bar{p}-\tfrac{1}{2}\,(n+2)\,,\qquad {\tt L}_+={\tt L}+a\,,\qquad
\bar{\tt L}_+=\bar{\tt L}+a+S^z\ .
\ee
Thus 
the RG trajectory from fig.\,\ref{fig12}
may be equivalently  assigned  $S^z={\tt w}=0$, ${\tt L}=\bar{\tt L}=1$ and
with the limiting value of $b(N)$
belonging to the strip,
$\lim_{N\to\infty}b(N)= \big(-\frac{3}{2}-p-\frac{n}{2}\big)\,\ri=-0.7475\,\ri$.
Notice that the levels ${\tt L}$ and $\bar{\tt L}$ have increased by one rather
than becoming negative.

\bigskip

That every low energy state of the lattice Hamiltonian fits within
the ``working'' definition is difficult to justify rigorously.
The same can also be said for the conjectures from sec.\,\ref{sec10} 
concerning the space of low energy states 
of the lattice model with $N\gg 1$. 
Nevertheless the assumptions turn out to be
 in full accordance with the results of a
 numerical investigation of the low energy spectrum of $\mathbb{H}$.
The latter is based on the following procedure.
First, for sufficiently small $N$ we performed the numerical 
diagonalization of the Hamiltonian, the lattice translation, 
the quasi-shift and other operators belonging 
to the commuting family
in a  sector with given $0\le S^z\ll N$.
We consider only those states below a certain cutoff
in the energy and lattice momentum.
To each of them we tried to assign the winding number ${\tt w}$
and a pair of non-negative integers ${\tt L}$, $\bar{\tt L}$ such that 
eq.\,\eqref{tower31} with $|\Im m\big(b(N)\big)|<\frac{n}{4}$
approximately holds true.
As will be discussed in a moment
we found that practically the most effective way to make such an identification was using the
subleading corrections to that formula.
Once the main characteristics of a low energy state are specified, 
we try to match the value of $b(N)$ with some solution of  \eqref{quantC1},
considered as an equation for $b(N)$ with the finite size correction terms ignored.\footnote{%
For given $p=\frac{1}{2}\big(S^z+(n+2)({\tt k}+{\tt w})\big)$, 
$\bar{p}=\frac{1}{2}\big(S^z-(n+2)({\tt k}+{\tt w})\big)$, ${\tt L}$ and $\bar{\tt L}$, the number of solution sets
$\bm{w}$, $\bar{\bm{w}}$ of \eqref{sksksk10} (with $s$ 
substituted by $b(N)$) 
is finite, 
so that there are a finite number of equations
\eqref{quantC1} with
$\delta=\delta(\bar{\bm{w}},\bm{w}\,|\,\bar{p},p,s)$
to be checked. In fact, an analysis of the subleading corrections to the energy
are  typically enough to determine the sets $\bar{\bm{w}}$, $\bm{w}$
for a given low energy state.}
Having a solution to \eqref{quantC1} for given $N$,
it can be continued  for increasing values of $N$
without numerical diagonalization of the lattice operators
and the numerical solutions of the Bethe ansatz  equations.
Thus the properties of the space of states in the scaling limit
may be determined from an analysis of eq.\,\eqref{quantC1}
alone. This was the way in which we arrived at the conjectures in
sec.\,\ref{sec10}.

\bigskip

Let's illustrate the above procedure on a concrete example.
Among others we numerically analyzed the  first 400 low energy eigenstates of the Hamiltonian
${\mathbb H}$ \eqref{aioiisa},\,\eqref{BC1a} 
with 
\be
N=22\,,  \qquad\quad q=\re^{\frac{\ri\pi}{5}}
\quad\quad (n=3)\,,\,\qquad\quad {\tt k}=-9/50
\ee
in the sector $S^z=1$.
The total number of states in this sector is over
$6\times10^5$ so a brute force numerical diagonalization  is not
possible. 
However the Hamiltonian is a sparse  matrix 
with a significant number of vanishing elements.
This allows one to find the first few hundred low energy eigenvectors and eigenvalues  
using the Krylov-Arnoldi method \cite{Krylov1,Arnoldi1}
within a  reasonable computer time. 
It turned out that among the first 400 eigenstates of ${\mathbb H}$,
 ordered
according to  the real part of the energy,
there were only four non-degenerate  
eigenstates, while  the
remaining ones formed the ${\mathcal
  Z}_2$ doublets. 
Having at hand the eigenvectors of the Hamiltonian,  the computation of the eigenvalues of
${\mathbb A}_+(\zeta)$ becomes a relatively easy
task. 
For the case of a doublet one needs to calculate two rows of the matrix ${\mathbb
  A}_+(\zeta)$, contract them with the eigenvectors and then
diagonalize the  resulting $2\times 2$ matrix.
Note that,
unlike the Hamiltonian, ${\mathbb
  A}_+(\zeta)$ is a dense matrix having no vanishing entries.
Thus even the calculation of all of its $4.1\times 10^{11}$ matrix elements
in the $S^z=1$ sector for $N=22$ would be simply  impossible.
 The same also applies to the transfer matrix ${\mathbb T}(\zeta)$.
For the lattice model with $N=22$ and $S^z=1$ the eigenvalues of ${\mathbb A}_+(\zeta)$ are
tenth degree polynomials in $\zeta$, whose zeroes solve
the Bethe ansatz equations \eqref{bae}. This allows one to calculate
the Bethe roots for all  the 400 eigenstates.
In turn, the eigenvalues of the lattice translation and quasi-shift operators 
are obtained from the Bethe roots using 
formulae \eqref{Keigeq1a} specialized to $r=2$, $\eta_J=(-1)^{J+1}\,\ri$
and \eqref{Beq1}, respectively.
Apart from these we also found it useful to consider the eigenvalues of the
operators $\mathbb{H}^{(\pm)}$. They belong to the commuting family,
are related to each other
through the ${\cal Z}_2$ transformation ${\cal D}$ and their sum is equal 
to the Hamiltonian:
\be\label{iasoid91829812}
{\mathbb H} ={\mathbb H} ^{(+)}+{\mathbb H} ^{(-)}\,,\qquad
\big[{\mathbb H}^{(\pm)},\,{\mathbb H} \big]=0
\,,\qquad
\mathbb{H}^{(\mp)}={\cal D}\ \mathbb{H}^{(\pm)}\, {\cal D}
\ .
\ee
The explicit formula for the matrices ${\mathbb H}^{(\pm)}$ 
 is quoted in sec.\,8.2 in the work \cite{Bazhanov:2020new}.
Their eigenvalues are expressed in terms of the Bethe roots as
\be
{\mathbb H}^{(\pm)}\,\boldsymbol{\Psi}\big(\{\zeta_m\}\big)
={\mathcal E}^{(\pm)}\,\boldsymbol{\Psi}\big(\{\zeta_m\}\big)\,,\qquad\qquad
{\mathcal E}^{(\pm)}=\pm\!\!\sum_{m=1}^{N/2-S^z}\,\frac{2\,(q-q^{-1})}{\zeta_m-\zeta^{-1}_m\,\mp
\ri\,(q+q^{-1})}\ \ .
\ee

\bigskip

For the considered set of 400 eigenstates the absolute value of the 
scaled energy
\be\label{aisodi19829}
\delta {\mathcal E}=\frac{N}{4\pi v_F}\ \big({\mathcal
  E}-N\,e_\infty\big)
\ee
varies between $0$ and $|\delta
{\mathcal E}_{\rm max}|\approx 3.3$. Therefore, in view of the large $N$
asymptotic formula \eqref{tower1a} our analysis could only apply to the case 
\be\label{energyset}
0\le{\tt L}+{\tt \bar{L}}\le 3\,, \qquad\qquad {\tt L},{\tt \bar{L}}\ge0\,.
\ee 
For the low energy states
the branch of the logarithm of the eigenvalues of the lattice translation operator \eqref{tower1b}
can be chosen such that 
\be\label{lorentz}
\frac{N}{4\pi \ri} \log (K)-S^z\,{\tt k}={\tt L}-{\tt \bar{L}}+S^z\,{\tt w}\, .
\ee
The r.h.s. is an integer, which for
 $N=22$  can take any values in the range 
$-5\le {\tt L}-{\tt \bar{L}}+S^z\,{\tt w}\le 5$. However
we imposed the momentum cut-off
\be\label{asoido1892}
\Big|\,\frac{N}{4\pi \ri}\, \log (K)-S^z\,{\tt k}\,\Big|\le 3
\ee
since otherwise this would require considerations of the states with
${\tt L},\bar{\tt L}=4,5$ which are excluded by the relation \eqref{energyset}.
Among the original 400 states, 
338 of them satisfy this condition.
\bigskip

We now come  up against the problem of 
assigning each of the 338 states 
the non-negative integers ${\tt L}$ and $\bar{\tt L}$
as well as
the winding number ${\tt w}$.
For this purpose we used the finite size correction formulae
to the eigenvalues of $\mathbb{H}^{({\pm})}$
presented in the work \cite{Bazhanov:2019xvy}. 
In the scaling limit the low energy Bethe states  $\bm{\Psi}_N$
 take the form
 $\bar{\bm{\psi}}_{\bar{p},s}(\bar{\bm{w}})\otimes\bm{\psi}_{{p},s}({\bm{w}})$.
As was already  mentioned
for ${\tt L},{\tt\bar{L}}\le5$ the chiral states
$\bm{\psi}_{{p},s}({\bm{w}})$ and $\bar{\bm{\psi}}_{\bar{p},s}(\bar{\bm{w}})$
 are completely 
determined by the eigenvalues of the local IM 
$I_m({\bm w},p,s)$ and ${I}_m(\bar{\bm w},\bar{p},s)$,
respectively,  with $m=1,2,3$  \eqref{9d8s9889a}.
For finite $N$
the subleading correction to the
scaled energy
$\delta {\mathcal E}=\frac{N}{4\pi v_F}\big({\mathcal
  E}-N e_\infty\big)$,
corresponding to $\bm{\Psi}_N$,
is described by the formula
\be
\label{asympeq1a}
\delta {\mathcal E}=
I_{1,N}+\,\bar{I}_{1,N}-
\frac{4 n^2}{N^2}\ \Big(\, 2\pi^2\,g_1 \, I_{1,N}\ 
\bar{I}_{1,N}+g_3 \,
 \big(\, { I}_{3,N}+
 {{ \bar{I}}}_{3,N}\, \big)\, \Big)
+O\big(N^{-4},N^{-2n}\big)\ .
\ee
Here $g_1$, $g_3$ stand for the numerical constants
\be
g_1=-\frac{\cot(\frac{\pi}{n})}{2\pi\,n^2} \,, \ \ \ \ \ \ \ \ \ \ \ 
g_3= \frac{\pi \Gamma(\frac{7}{2}+\frac{3}{n})\Gamma^3(1+\frac{1}{n})}
{18\, \Gamma(\frac{3}{n})\Gamma^3(\frac{3}{2}+\frac{1}{n})}\,,
\ee
while
\be
I_{m,N}=I_m(\bm{w},p,s)\big|_{s=b(N)}\,,\qquad \qquad
\bar{I}_{m,N}=I_m(\bar{\bm{w}},\bar{p},s)\big|_{s=b(N)}
\ee
and the sets $\bm{w}$, $\bar{\bm{w}}$  solve
the algebraic system \eqref{sksksk10} with $s$ replaced by the ``running coupling'' $b(N)$.
In \eqref{asympeq1a}
the notation $O\big(N^{-a},N^{-b}\big)$ stands for
$ o(N^{-c})$,
where $c={\rm min}(a, b)-\epsilon$ 
for all  $\epsilon>0$. It should be pointed out that 
the large $N$ asymptotic formula for $\delta{\cal E}$ is not literally
applicable when $n\le1$.
In this case  the description of the finite size corrections is more involved
and includes a contribution from the
so-called dual non-local IM (see ref.\cite{Bazhanov:2019xvy} for a further discussion).
As it follows from \eqref{iasoid91829812} the energy ${\cal E}$
coincides with the sum of the eigenvalues of $\mathbb{H}^{(+)}$ and
$\mathbb{H}^{(-)}$. The finite size corrections for the difference
${\cal E}^{(+)}-{\cal E}^{(-)}$ is expressed in terms of the
eigenvalues of the local IM ${\bf I}_2$ and $\bar{\bf I}_2$:
\bea\label{Edifeq1}
\frac{{\mathcal E}^{(+)}-{\mathcal E}^{(-)}}{4\pi v_{\rm F}} =-\frac{2 \ri
  n^{3/2}}{N^2}\ g_2\,\big(I_{2,N}-\bar{I}_{2,N}\big)+
o\big(N^{-2}\big)\,,
\eea
where
\be
 g_2=
\frac{\sqrt{\pi}\,\Gamma(\frac{5}{2}+\frac{2}{n})\Gamma^2(1+\frac{1}{n})}{3\,\Gamma(\frac{2}{n})
\Gamma^2(\frac{3}{2}+\frac{1}{n})}\  .\nonumber
\ee

\bigskip

\begin{table}
\label{tabstates}
\centering
\scalebox{0.9}{
\begin{tabular}{|c|c|c|c|}
\multicolumn{4}{c}{}\\[-0.45cm]
\multicolumn{4}{c}{${\tt w}=0$}\\[0.2cm]
\hline
& &  & \\[-0.45cm]
$({\tt L},\bar{\tt L})$ & 
${\mathcal H}^{(\rm cont)}_{N|S^z}$ & 
${\mathcal H}^{({\rm disc},+)}_{N|S^z}$ & 
${\mathcal H}^{({\rm disc},-)}_{N|S^z}$ \\[0.05cm]
\hline
 & & & \\[-0.45cm]
  (0,0) & 9& 0 & 0\\[0.05cm]
  & & & \\[-0.45cm]
 (1,0)  &  18& 0 & 2 \\[.05cm]
  & & & \\[-0.45cm]
 (0,1) &  20 & 0 & 0 \\[.05cm]
  & & & \\[-0.45cm]
 (1,1)&  32& 0 & 2\\[.05cm]
  & & & \\[-0.45cm]
  (2,0)  & 36& 0 & 4\\[.05cm]
  & & & \\[-0.45cm]
  (0,2)& 40& 0 & 0\\[.05cm]
  & & & \\[-0.45cm]
 (1,2)&  22& 0 & 8\\[.05cm]
  & & & \\[-0.45cm]
(2,1)& 16& 0 & 4 \\[.05cm]
  & & & \\[-0.45cm]
 (3,0)& 34& 0 & 8  \\[.05cm]
  & & & \\[-0.45cm]
 (0,3)&  46&  0 & 0 \\[.05cm]
\hline
\end{tabular}
\begin{tabular}{|c|c|c|c|}
\multicolumn{4}{c}{}\\[-0.1cm]
\multicolumn{4}{c}{${\tt w}=1$}\\[0.2cm]
\hline
& &  & \\[-0.45cm]
$({\tt L},\bar{\tt L})$ & 
${\mathcal H}^{(\rm cont)}_{N|S^z}$ & 
${\mathcal H}^{({\rm disc},+)}_{N|S^z}$ & 
${\mathcal H}^{({\rm disc},-)}_{N|S^z}$ \\[0.05cm]
\hline
 & & & \\[-0.45cm]
 (0,0) &  5& 2 & 0 \\[.05cm]
  & & & \\[-0.45cm]
  (1,0)  &  6& 4 & 0 \\[.05cm]
  & & & \\[-0.45cm]
   (0,1)&  8& 4 & 0 \\[.05cm]
  & & & \\[-0.45cm]
  (2,0)& 2& 6 & 0 \\[.05cm]
\hline
\multicolumn{4}{c}{}\\[3.2cm]
\end{tabular}
}
\caption{\small 
A classification of the 338 lowest energy states,
subject to the momentum cut-off \eqref{asoido1892},
 of the lattice Hamiltonian 
$\mathbb{H}$
 with $N=22$ in the sector $S^z=1$.
The states are assigned to ${\mathcal H}^{(\rm cont)}_{N|S^z}$
or ${\mathcal H}^{(\rm disc,\pm)}_{N|S^z}$ based on the predictions
of the asymptotic formula \eqref{quantC1}, see also figs.\,\ref{sdist1},\,\ref{fig130} and
those contained in Appendix \ref{AppC}.
In  the case ${\tt w}=1$, ${\tt L}=2$, $\bar{\tt L}=0$
the number of states that were delegated to ${\cal H}^{({\rm disc},+)}_{N|S^z}$
is six, which is less than what is predicted by eq.\,\eqref{99898dsjksj} (ten).
Note that  
for these states $|\delta{\cal E}|$ \eqref{aisodi19829}  is close to $|\delta{\cal E}_{{\rm max}}|\approx 3.3$.
In all other cases the number of states in ${\mathcal H}^{(\rm disc,\pm)}_{N|S^z}$
agrees  with \eqref{99898dsjksj}.
The parameters entering into the Hamiltonian were taken to be
$q=\re^{\frac{\ri\pi}{5}}$ ($n=3$) and
 ${\tt k}=-0.18$.
\label{tab001}} 
\end{table}

The procedure that we used for 
assigning the full set of RG invariants to 
the low energy Bethe states 
is the following. 
For a given state $\bm{\Psi}_N$ 
the eigenvalues 
${\mathcal E}^{(\pm)}$, $K$ and $B$ are
 calculated  from the Bethe roots   obtained 
via the diagonalization of ${\mathbb H}$ and
${\mathbb A}_+(\zeta)$ described above.
 Then   $b(N)=\frac{n}{4\pi}\log(B)$ is used to 
compute the  r.h.s. of \eqref{asympeq1a} for all possible
  pairs $({\tt L},\bar{\tt L})$  satisfying
  \eqref{energyset}, with ${\tt w}$ 
  determined through the relation \eqref{lorentz}. 
This involves  solving the algebraic system \eqref{sksksk10},
where
$p=\frac{1}{2}\big(S^z+(n+2)({\tt k}+{\tt w})\big)$, 
$\bar{p}=\frac{1}{2}\big(S^z-(n+2)({\tt k}+{\tt w})\big)$ and $s$ is swapped for $b(N)$. 
The obtained values of the r.h.s. of \eqref{asympeq1a} are then matched with
$\delta{\cal E}=\frac{N}{4\pi v_{\rm F}}\big({\mathcal
  E}^{(+)}+{\mathcal
  E}^{(-)}-N e_\infty\big)$.
 In almost all cases the procedure allows one to
  unambiguously
determine the integers 
 ${\tt L}$, $\bar{\tt L}$, ${\tt w}$ as well as the sets
$\bm{w}$ and $\bar{\bm{w}}$ associated  to the state $\bm{\Psi}_N$. It should be mentioned
that we encountered  about a half dozen cases, out of the 338, where we could not
unambiguously identify 
the states with the help of \eqref{asympeq1a} alone. In all these cases the issue was resolved by 
employing
the relation \eqref{Edifeq1} and the product rule \eqref{oasodi1121}.
The results of the above procedure are 
summarized in tab.\,\ref{tab001}. 
\bigskip

\begin{figure}
\centering
\scalebox{1}{
\begin{tikzpicture}
\node at (4,3) {$b$};
\draw  (4,3) circle [radius=0.3];
\node at (0,0) {\includegraphics[trim = {0.1cm 0  0 0}, clip,width=10.5cm]{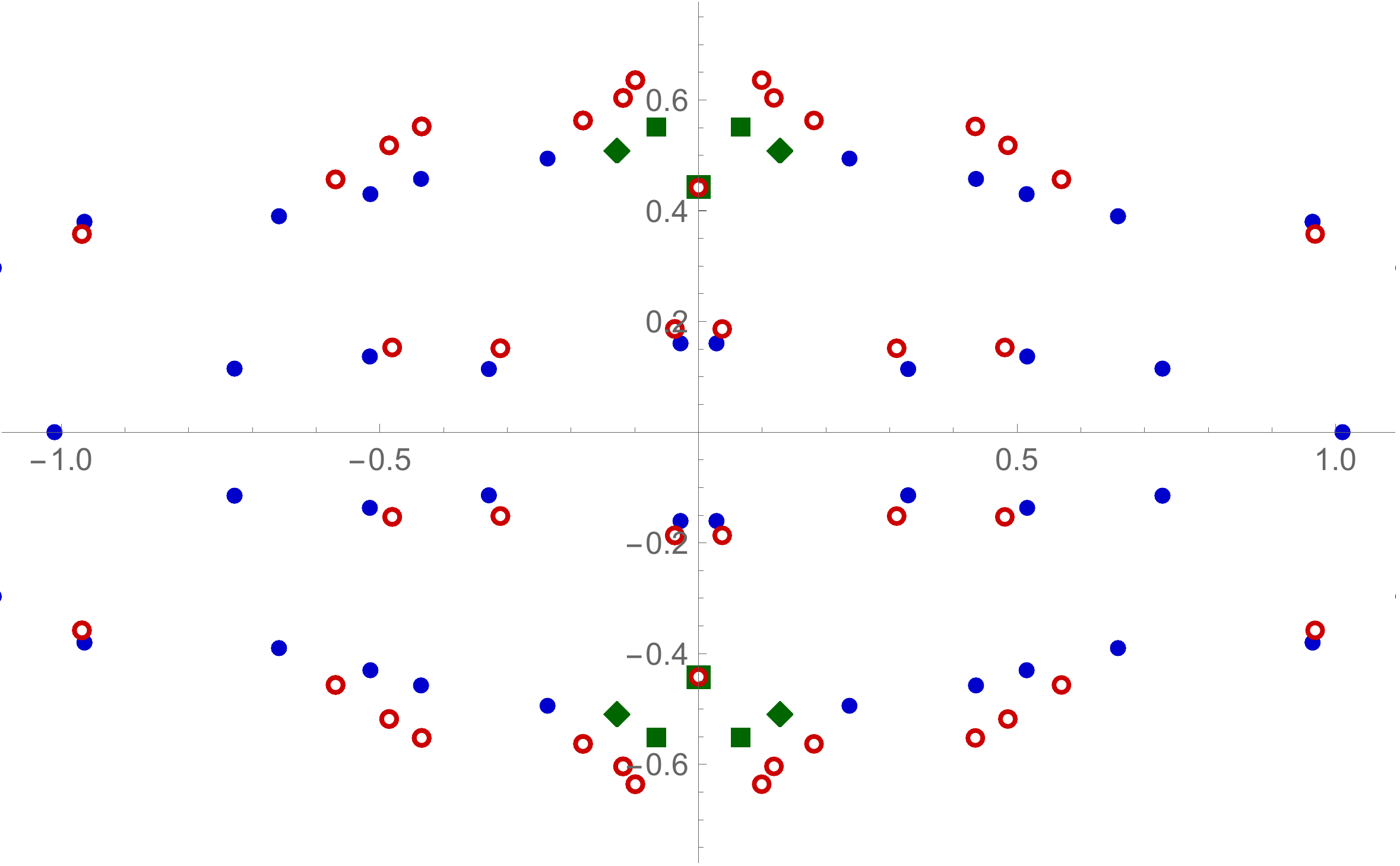}};
\end{tikzpicture}
}
\caption{\label{fig130}
\small
The open circles depict the distribution  of
$b(N)=\frac{n}{4\pi}\log(B)$ in the complex plane,
where $B$ is the eigenvalue of the quasi-shift operator, for the 
$42=34+8$ states with ${\tt L}=3$, $\bar{\tt L}=0$ and ${\tt w}=0$
as indicated in tab.\ref{tab001}. 
The filled circles, squares  and diamonds correspond to the solutions
$b_*$ of eq.\,\eqref{iasodi12190}. For the circles 
$\lim_{N\to \infty}\Im m\big(b_*(N)\big)=0$,
while for the squares, $b_*(N)\to\pm\frac{9\ri}{20}$
as $N\to\infty$. The filled diamonds form two pairs
which have the same value of $\Im m\big(b_*(N)\big)$
and opposite real part. At large but finite $N$
the  diamonds from the upper pair collide at the imaginary axis
at which point for one of  the  diamonds
$b_*(N)\to+\frac{9\ri}{20}$ while for the other
one $b_*(N)\to 0$.
The $N$ dependence of $b_*(N)$ for the lower pair of diamonds
is obtained from that of the upper pair via complex conjugation.
}
\end{figure}

The table also contains a classification of the states according to whether
they become part of the continuous or discrete spectrum in the 
scaling limit. This is achieved by matching  $b(N)$ with a certain
solution of  the equation
\bea\label{iasodi12190}
\bigg(\frac{N}{2N_0}\bigg)^{4\ri b}\ \exp\big(\,\tfrac{\ri}{2}\,\delta(\bar{\bm{w}}(b),\bm{w}(b)\,|\,\bar{p},p,b)\,\big)=
(-1)^{N/2-S^z}\ ,
\eea
which comes from dropping the correction terms in the asymptotic formula \eqref{quantC1}.
Here the phase shift depends on $b$ explicitly as well as implicitly through the sets 
$\bm{w}(b)$ and $\bar{\bm{w}}(b)$ which solve \eqref{sksksk10} 
with $s$ substituted by $b$. For fixed ${\tt L}$ and $\bar{\tt L}$ 
there are ${\rm par}_2({\tt L})\times {\rm par}_2(\bar{\tt L})$  pairs $(\bm{w}(b),\bar{\bm{w}}(b))$
so that  ${\rm par}_2({\tt L})\times {\rm par}_2(\bar{\tt L})$
 equations of the form \eqref{iasodi12190} need to be considered. Among the solutions of all of these equations,
one should choose the particular one, $b_*$, which is closest to the value of $b(N)$ corresponding to the
 Bethe
state $\bm{\Psi}_N$. In practice this is not too difficult a task.
The initial approximation for finding the solution $b_*$ may be taken to be
$b_{\rm in}=b(N)=\frac{n}{4\pi}\log(B)$.
The proper sets $\bm{w}_*(b)$, $\bar{\bm{w}}_*(b)$,
are the ones which at $b=b_{\rm in}$ coincide with the $\bm{w}$, $\bar{\bm{w}}$ 
that were assigned to the Bethe state $\bm{\Psi}_N$ 
through the examination of the finite size corrections.
Once $b_*$ is determined its $N$ dependence,
with $b_*=b_*(N)$ being a continuous function, is obtained
by means of varying $N$   in eq.\,\eqref{iasodi12190} with  the sign factor $(-1)^{N/2-S^z}$  kept fixed.
Then the state $\bm{\Psi}_N$ is delegated to 
${\mathcal H}^{(\rm cont)}_{N|S^z}$ or ${\mathcal H}^{(\rm disc)}_{N|S^z}$
depending on whether or not $\lim_{N\to\infty}\Im m\big(b_*(N)\big)$ vanishes.
Note that for any given Bethe state one can always verify through the explicit construction of the
corresponding RG trajectory,
by  solving the Bethe ansatz equations,  that 
$b(N)$ asymptotically approaches  $b_*(N)$ as $N\to\infty$ see, e.g., fig.\ref{fig999}.

\bigskip

For the 34 states in the sector ${\tt L}=\bar{\tt L}=1$ and ${\tt w}=0$
the correspondence between $b(N)$, obtained via 
the numerical diagonalization of the lattice operators, and
the solutions $b_*$ of eq.\,\eqref{iasodi12190} has 
already been illustrated in fig.\,\ref{sdist1}. 
 Of these states, 32  are predicted to
form part of the continuous spectrum in the scaling limit, i.e,
belong to ${\cal H}^{({\rm cont})}_{N|S^z}$, while the two
states which have been matched to $b_*$ depicted by 
the solid squares in the figure are part of ${\cal H}^{({\rm disc})}_{N|S^z}$.
We also confirmed this by explicitly constructing the RG trajectories corresponding
to these two states and verifying that $\lim_{N\to\infty}\Im m\big(b(N)\big)\ne 0$.
The results of a similar analysis for the states in the sector
with ${\tt L}=3$ and $\bar{{\tt L}}={\tt w}=0$   are presented in
 fig.\,\ref{fig130}. 
 Note that in this case, since the sum ${\tt L}+\bar{\tt L}$
reaches the upper bound in \eqref{energyset},
for many of the  solutions $b_*$
there is no corresponding lattice state
among the 400 lowest energy states of 
$\mathbb{H}$. 
An additional six figures that cover the remaining cases
listed in tab.\,\ref{tab001} are contained in Appendix \ref{AppC}.
\bigskip

Analogous computations were performed for 
the low energy states of the lattice Hamiltonian
for different sectors $S^z$ as well as various values
of the anisotropy parameter $q=\re^{\frac{\ri\pi}{n+2}}$ and the twist parameter  ${\tt k}$.

\section{Hermitian structure of the space of states in the scaling limit}
\subsection{Hermitian versus integrable structure\label{sec191}}
Up till now we have been focused on describing the linear
structure of the space of states occurring in the scaling limit
of the ${\cal Z}_2$ invariant inhomogeneous six-vertex model.
This was achieved through the decomposition 
of ${\cal H}$
into the highest weight irreps of the $\overline{W}_\infty\otimes W_\infty$\,-\,algebra,
accounting for the superselection rules imposed by the global symmetries along the way.
The space of states also possesses an integrable structure.
It  is inherited from  the finite dimensional (pseudo-)Hilbert space of the
lattice model, where there exists
a basis of Bethe states diagonalizing the matrices
$\mathbb{A}_\pm(\zeta)$. 
The latter, in the scaling limit, become the operators
${\mathlarger{\mathlarger{\mathlarger{\mathlarger {\bf \it a}}}}}_\pm(\lambda)$
and $\bar{{\mathlarger{\mathlarger{\mathlarger{\mathlarger {\bf \it a}}}}}}_\pm(\bar{\lambda})$,
while the scaling limit of the low energy Bethe states   yields the states
\be\label{iaosdioa19209}
\bm{\psi}_{\bar{\rho},\rho,\bar{\nu},\nu}(\bar{\bm{w}},\bm{w})\equiv
\bar{\bm{\psi}}_{\bar{\rho},\bar{\nu}}(\bar{\bm{w}})\otimes {\bm{\psi}}_{{\rho},\nu}(\bm{w})\in
 \bar{{\cal V}}_{\bar{\rho},\bar{\nu}}\otimes {{\cal V}}_{{\rho},{\nu}}\ .
\ee
In turn, these form a basis  
for ${\cal H}$
diagonalizing ${\mathlarger{\mathlarger{\mathlarger{\mathlarger {\bf \it a}}}}}_\pm(\lambda)$,
 $\bar{{\mathlarger{\mathlarger{\mathlarger{\mathlarger {\bf \it a}}}}}}_\pm(\bar{\lambda})$.
Here we discuss the Hermitian structures consistent with the integrable one.
\bigskip

We'll consider two types of sesquilinear forms in ${\cal H}$.
The first one, labeled by the subscript ``$+$'', is such that
the eigenstates
$\bm{\psi}_{\bar{\rho},\rho,\bar{\nu},\nu}(\bar{\bm{w}},\bm{w})$
satisfy the condition\footnote{%
Although the normalization for the eigenstates has yet to be fixed,
in writing formulae \eqref{iaosid912981a} and \eqref{iaosid912981b} 
we assume  that
${\cal \hat{C}\hat{P}\hat{T}}\,\bm{\psi}_{\bar{\rho},\rho,\bar{\nu},\nu}(\bar{\bm{w}},\bm{w})=
\bm{\psi}_{\bar{\rho},\rho,\bar{\nu}^*,\nu^*}(-\bar{\bm{w}}^*,-\bm{w}^*)
$
and
${\cal \hat{D}}\,\bm{\psi}_{\bar{\rho},\rho,\bar{\nu},\nu}(\bar{\bm{w}},\bm{w})=
\bm{\psi}_{\bar{\rho},\rho,-\bar{\nu},-\nu}(-\bar{\bm{w}},-\bm{w})$
(see eq.\,\eqref{CPT*8787aaa}).
}
\be\label{iaosid912981a}
\big(\bm{\psi}_{\bar{\rho}',\rho',\bar{\nu}',\nu'}(\bar{\bm{w}}',\bm{w}'),
\bm{\psi}_{\bar{\rho},\rho,\bar{\nu},\nu}(\bar{\bm{w}},\bm{w})\big)_+=0\qquad {\rm unless} \qquad
\bm{\psi}_{\bar{\rho}',\rho',\bar{\nu}',\nu'}={\cal \hat{C}\hat{P}\hat{T}}\,
\bm{\psi}_{\bar{\rho},\rho,\bar{\nu},\nu}\ .
\ee
Recall that the ${\cal CPT}$ conjugation acts as
\be
{\cal \hat{C}\hat{P}\hat{T}}: \ 
 \bar{{\cal V}}_{\bar{\rho},\bar{\nu}}\otimes {{\cal V}}_{{\rho},{\nu}}\mapsto
 \bar{{\cal V}}_{\bar{\rho},\bar{\nu}^*}\otimes {{\cal V}}_{{\rho},{\nu}^*}\ .
\ee
Hence for the continuous component of the space of states, ${\cal H}^{({\rm cont})}$, where
$\rho=p$, $\bar{\rho}=\bar{p}$ and $\nu=\bar{\nu}=s$ are all real 
 the states  $\bm{\psi}_{\bar{\rho},\rho,\bar{\nu},\nu}(\bar{\bm{w}},\bm{w})$
and
$\bm{\psi}_{\bar{\rho}',\rho',\bar{\nu}',\nu'}(\bar{\bm{w}}',\bm{w}')$  
in \eqref{iaosid912981a} belong
to the same irrep of the $\overline{W}_\infty\otimes W_\infty$\,-\,algebra. 
On the other hand for the irreps from
 ${\cal H}^{({\rm disc})}$,  where
$\nu^*=-\nu$ and $\bar{\nu}^*=-\bar{\nu}$,
the ${\cal CPT}$ conjugated space, ${\cal V}_{\bar{\rho},-\bar{\nu}}\otimes {\cal V}_{{\rho},-{\nu}}$,
does not coincide with the initial one ${\cal V}_{\bar{\rho},\bar{\nu}}\otimes {\cal V}_{{\rho},{\nu}}$.
 This makes it natural  to  introduce
another sesquilinear form, using the ${\cal Z}_2$ symmetry, such that
\be\label{iaosid912981b}
\big(\bm{\psi}_{\bar{\rho}',\rho',\bar{\nu}',\nu'}(\bar{\bm{w}}',\bm{w}'),
\bm{\psi}_{\bar{\rho},\rho,\bar{\nu},\nu}(\bar{\bm{w}},\bm{w})\big)_-=0\qquad {\rm unless} \qquad
\bm{\psi}_{\bar{\rho}',\rho',\bar{\nu}',\nu'}=\hat{{\cal D}}{\cal \hat{C}\hat{P}\hat{T}}\,
\bm{\psi}_{\bar{\rho},\rho,\bar{\nu},\nu}\ .
\ee
Since
\be
\hat{{\cal D}}{\cal \hat{C}\hat{P}\hat{T}}: \ 
 \bar{{\cal V}}_{\bar{\rho},\bar{\nu}}\otimes {{\cal V}}_{{\rho},{\nu}}\mapsto
 \bar{{\cal V}}_{\bar{\rho},-\bar{\nu}^*}\otimes {{\cal V}}_{{\rho},-{\nu}^*}
\ee
any irrep of the $\overline{W}_\infty\otimes W_\infty$\,-\,algebra  occurring in the decomposition of
 ${\cal H}^{({\rm disc})}$ would coincide with its conjugate.
\bigskip

 For each of the  sesquilinear forms consistent with the integrable structure there is an evident candidate, which
is defined through the conjugation conditions for the $W$ currents:
\begin{subequations}\label{haysays}
\be\label{haysaysa}
\big(\bm{\chi}_2,W_j(u)\bm{\chi}_1\big)_\pm=(\pm 1)^j\,\big(W_j(u^*)\bm{\chi}_2,\bm{\chi}_1\big)_\pm
\ee
\vspace{-1cm}

\be\label{haysaysb}
\ \ \big(\bm{\chi}_2,\overline{W}_j(\bar{u})\bm{\chi}_1\big)_\pm=
(\pm1)^j\,\big(\overline{W}_j(\bar{u}^*)\bm{\chi}_2,\bm{\chi}_1\big)_\pm\ ,
\ee
\end{subequations}
where $\bm{\chi}_1\in \bar{{\cal V}}_{\bar{\rho},\bar{\nu}}\otimes {{\cal V}}_{{\rho},{\nu}}$
 and $\bm{\chi}_2\in \bar{{\cal V}}_{\bar{\rho},\pm\bar{\nu}^*}\otimes {{\cal V}}_{{\rho},\pm{\nu}^*}$
are arbitrary states.
Indeed since the local IM are expressed as integrals over the local densities built from the $W$ currents, the
above conjugation conditions imply
\begin{subequations}\label{haysays1a}
\be\label{haysays1aa}
\big(\bm{\chi}_2,{\bf I}_m\bm{\chi}_1\big)_\pm=(\pm1)^{m+1}\,\big({\bf I}_m\bm{\chi}_2,\bm{\chi}_1\big)_\pm
\ee
\vspace{-0.9cm}

\be\label{haysays1ab}
\ \ \big(\bm{\chi}_2,\bar{{\bf I}}_m\bm{\chi}_1\big)_\pm=
(\pm1)^{m+1}\,\big(\bar{{\bf I}}_m\bm{\chi}_2,\bm{\chi}_1\big)_\pm\ .
\ee
\end{subequations}
For generic values of the twist and anisotropy parameters ${\tt k}$ and $n$ it is expected that the 
set of eigenvalues  $\{{\bf I}_m\}_{m=1}^\infty$ and $\{\bar{{\bf I}}_m\}_{m=1}^\infty$  
unambiguously specifies the states in ${\cal H}$.
Then \eqref{haysays1a} together with 
the commutation relations ${\cal \hat{C}\hat{P}\hat{T}}\,{\bf I}_m={\bf I}_m\,{\cal \hat{C}\hat{P}\hat{T}}$,
 ${\cal \hat{D}}\,{\bf I}_m=(-1)^{m+1}\,{\bf I}_m\,{\cal \hat{D}}$
and similarly for $\bar{{\bf I}}_m$  (see \eqref{CPTIM1})
leads to the orthogonality condition \eqref{iaosid912981a} or \eqref{iaosid912981b}. 
\bigskip

The relations  \eqref{haysays} do not define the sesquilinear forms unambiguously.
They should be supplemented by the value of the  forms on the $\overline{W}_\infty\otimes {W}_\infty$ 
primary states. 
Independently of this choice,  
the orthogonality condition  \eqref{iaosid912981a}
along with the commutation relations of
 ${\cal \hat{C}\hat{P}\hat{T}}$, $\hat{{\cal D}}$
with ${\mathlarger{\mathlarger{\mathlarger{\mathlarger {\bf \it a}}}}}_\pm(\lambda)$,
${\bm{\tau}}({\lambda})$  \eqref{i89a89} 
and  $\bar{{\mathlarger{\mathlarger{\mathlarger{\mathlarger {\bf \it a}}}}}}_\pm(\bar{\lambda})$,
 $\bar{\bm{\tau}}(\bar{\lambda})$
imply that for the ``$+$'' form
\begin{subequations}\label{Hermapm21a}
\be\label{Hermapm21aa}
\big(\bm{\chi}_2,{\mathlarger{\mathlarger{\mathlarger {\bf \it a}}}}_\pm(\lambda)\,\bm{\chi}_1\big)_+
=\big({\mathlarger{\mathlarger{\mathlarger {\bf \it a}}}}_\pm(\lambda^*)\,\bm{\chi}_2,\bm{\chi}_1\big)_+\,,\qquad
\big(\bm{\chi}_2,\bm{\tau}(\lambda)\,\bm{\chi}_1\big)_+
=\big(\bm{\tau}(\lambda^*)\,\bm{\chi}_2,\bm{\chi}_1\big)_+
\ee
\vspace{-1.2cm}

\be\label{Hermapm21ab}
\ \ \big(\bm{\chi}_2,\bar{{\mathlarger{\mathlarger{\mathlarger {\bf \it a}}}}}_\pm(\bar{\lambda})\,\bm{\chi}_1\big)_+
=\big(\bar{\mathlarger{\mathlarger{\mathlarger {\bf \it a}}}}_\pm(\bar{\lambda}^*)\,\bm{\chi}_2,\bm{\chi}_1\big)_+\,,\qquad
\big(\bm{\chi}_2,\bar{\bm{\tau}}(\bar{\lambda})\,\bm{\chi}_1\big)_+
=\big(\bar{\bm{\tau}}(\bar{\lambda}^*)\,\bm{\chi}_2,\bm{\chi}_1\big)_+\ .
\ee
\end{subequations}
Similarly for the ``$-$'' sesquilinear form one has
\begin{subequations}\label{Hermapm22a}
\be\label{Hermapm22aa}
\big(\bm{\chi}_2,{\mathlarger{\mathlarger{\mathlarger {\bf \it a}}}}_\pm(\lambda)\,\bm{\chi}_1\big)_-
=\big({\mathlarger{\mathlarger{\mathlarger {\bf \it a}}}}_\pm(-\lambda^*)\,\bm{\chi}_2,\bm{\chi}_1\big)_-\,,\qquad
\big(\bm{\chi}_2,\bm{\tau}(\lambda)\,\bm{\chi}_1\big)_-
=\big(\bm{\tau}(-\lambda^*)\,\bm{\chi}_2,\bm{\chi}_1\big)_-
\ee
\vspace{-1.2cm}

\be\label{Hermapm22ab}
\ \ \big(\bm{\chi}_2,\bar{{\mathlarger{\mathlarger{\mathlarger {\bf \it a}}}}}_\pm(\bar{\lambda})\,\bm{\chi}_1\big)_-
=\big(\bar{\mathlarger{\mathlarger{\mathlarger {\bf \it a}}}}_\pm(-\bar{\lambda}^*)\,\bm{\chi}_2,\bm{\chi}_1\big)_-\,,\qquad
\big(\bm{\chi}_2,\bar{\bm{\tau}}(\bar{\lambda})\,\bm{\chi}_1\big)_-
=\big(\bar{\bm{\tau}}(-\bar{\lambda}^*)\,\bm{\chi}_2,\bm{\chi}_1\big)_-\ .
\ee
\end{subequations}
To establish the above formulae, it is sufficient to check them in the eigenbasis \eqref{iaosdioa19209}.
\bigskip

The sesquilinear forms  allow one to introduce an inner product for
the continuous and discrete components of ${\cal H}$.
For the case of ${\cal H}^{({\rm cont})}$ we take it to be
\be\label{oaisod9182921}
\big\langle\bm{\chi}_1,\,\bm{\chi}_2\big\rangle_{{\rm cont}}=\big(\bm{\chi}_1,\,\bm{\chi}_2\big)_+\,,
\qquad\qquad \bm{\chi}_1,\,\bm{\chi}_2\in{\cal H}^{({\rm cont})}\ .
\ee
As it follows from \eqref{haysays} the Hermitian conjugation for the Fourier coefficients 
of the
$W$ currents \eqref{aoisd182918} corresponding to such an inner product
is
\bea\label{aoisdo8923}
\big[\,\widetilde{W}_j(m)\,\big]^\star=\widetilde{W}_j(-m)\ ,\qquad 
\big[\,\widetilde{\overline{W}}_j(m)\,\big]^\star=\widetilde{\overline{W}}_j(-m)\ .
\eea
Once the norms  of the highest states are specified, this condition
allows one to calculate the inner product for any given states 
from the same irrep of the $\overline{W}_\infty\otimes W_\infty$\,-\,algebra.
Two states which belong to irreps that are not isomorphic to
each other, are orthogonal.
In the case under consideration with the central charge
$-1<c<2$ the inner product is not positive definite so that ${\cal H}^{({\rm cont})}$
equipped with $\big\langle\cdot,\,\cdot\big\rangle_{{\rm cont}}$ becomes a pseudo-Hilbert space.
\bigskip

The structure of the pseudo-Hilbert space for ${\cal H}^{({\rm disc})}$
can be introduced
using the inner product
\be\label{iaosid98121}
\big\langle\bm{\chi}_1,\,\bm{\chi}_2\big\rangle_{{\rm disc}}=\big(\bm{\chi}_1,\,\bm{\chi}_2\big)_-\,,
\qquad\qquad \bm{\chi}_1,\,\bm{\chi}_2\in{\cal H}^{({\rm disc})}\ .
\ee
In this case the conjugation condition \eqref{aoisdo8923} is replaced by
\bea\label{aoisdo8923A}
\big[\,\widetilde{W}_j(m)\,\big]^{\text{\tiny\ding{105}}}=(-1)^j\ \widetilde{W}_j(-m)\ ,\qquad 
\big[\,\widetilde{\overline{W}}_j(m)\,\big]^{\text{\tiny\ding{105}}}=(-1)^j\ \widetilde{\overline{W}}_j(-m)\ .
\eea
Likewise 
the inner product $\big\langle\cdot,\,\cdot\big\rangle_{{\rm disc}}$
is not positive definite when $c<2$.

\subsection{Chiral sesquilinear forms}
In the above discussion of the Hermitian structure for 
${\cal H}$
it was sufficient to focus on an irrep 
 occurring in the  $\overline{W}_\infty\otimes W_\infty$ decomposition of this space.
Recall that the ${\cal CPT}$ and ${\cal D}$ transformations, 
required for defining the conjugated irrep,
can be introduced for each left and right chiral factor 
in $\bar{{\cal V}}_{\bar{\rho},\bar{\nu}}\otimes{\cal V}_{\rho,\nu}$ 
separately (see sec.\ref{sec172}).
This makes it possible to  restrict the
sesquilinear forms  to the chiral components.
For instance, for the ``$+$'' sesquilinear form
 on the right chiral spaces, relations \eqref{haysays1aa} and \eqref{Hermapm21aa}
would continue
to hold true with  $\bm{\chi}_1\in{\cal V}_{\rho,\nu}$,  
$\bm{\chi}_2\in{\cal V}_{\rho,\nu^*}={\cal \hat{C}\hat{P}\hat{T}}\,\big({\cal V}_{\rho,\nu}\big)$ 
and similarly for the left chiral ones. 
\medskip

The spaces ${\cal V}_{\rho,\nu}$ were originally realized in terms of the Fock spaces.
For real $\rho$ and $\nu$, this irrep of the $W_\infty$\,-\,algebra coincides as a linear space with ${\cal F}_{\bf P}$, where
${\bf P}=\big(\frac{\rho}{\sqrt{n+2}},\frac{\nu}{\sqrt{n}}\big)$. When $\nu$ is pure imaginary
and $\rho+\frac{1}{2}\pm\ri\nu\in\mathbb{Z}$,
the corresponding Fock space becomes reducible w.r.t. the $W_\infty$\,-\,algebra and
${\cal V}_{\rho,\nu}$ is obtained by factoring ${\cal F}_{\bf P}$ over the invariant subspace generated
by the null vector as in eqs.\,\eqref{ioasido1212} and \eqref{ioasido1212a}.
Despite this, the conjugation conditions \eqref{haysaysa},  combined with 
the bosonization formulae for the
$W$ currents \eqref{w2iosdi},\,\eqref{W3def1a}, allow one to lift
the sesquilinear forms to the complex bilinear maps
${\cal \hat{C}\hat{P}\hat{T}}\big({\cal F}_{{\bf P}}\big)\times {\cal F}_{{\bf P}}
\mapsto\mathbb{C}$ for the ``$+$'' form and
$\hat{\cal D}{\cal \hat{C}\hat{P}\hat{T}}\big({\cal F}_{{\bf P}}\big)\times {\cal F}_{{\bf P}}
\mapsto\mathbb{C}$  in the case of the ``$-$'' one,
for any ${\bf P}$. 
Note that the action of the ${\cal CPT}$ and ${\cal D}$ transformations in the Fock
space
can be defined through the relations
\be\label{osaoid19020912}
\arraycolsep=0.6cm
\begin{array}{ccc}
{\cal \hat{C}\hat{P}\hat{T}}: {\cal F}_{\bf P}\mapsto {\cal F}_{{\bf P}^*}\,,&
{\cal \hat{C}\hat{P}\hat{T}}\,a_m=a_m\,{\cal \hat{C}\hat{P}\hat{T}}\,,&
{\cal \hat{C}\hat{P}\hat{T}}\,b_m=b_m\,{\cal \hat{C}\hat{P}\hat{T}} \\[0.2cm]
 \ \ \ {\cal \hat{D}}: {\cal F}_{\bf P}\mapsto {\cal F}_{{\bf P}'}\,,&
{\cal \hat{D}}\,a_m=a_m\,{\cal \hat{D}}\,,&
\ \ \,{\cal \hat{D}}\,b_m=-b_m\,{\cal \hat{D}}
\end{array}
\ee
along with
\be
{\cal \hat{C}\hat{P}\hat{T}}\,|{\bf P}\rangle =|{\bf P}^*\rangle\,,\qquad 
{\cal{\hat D}}\,|{\bf P}\rangle =|{\bf P}'\rangle\,, \qquad \quad{\rm where} \qquad\quad
\begin{array}{l}
{\bf P}^*=\big(\frac{\rho^*}{\sqrt{n+2}},+\frac{\nu^*}{\sqrt{n}}\big) \\[0.2cm]
{\bf P}'\,=\big(\frac{\rho}{\sqrt{n+2}},-\frac{\nu}{\sqrt{n}}\big) 
\end{array}\, .
\ee
It is straightforward to check that 
$
{\cal \hat{C}\hat{P}\hat{T}}\,
{\mathlarger{\mathlarger{\mathlarger{\mathlarger {\bf \it a}}}}}_\pm(\lambda)=
{\mathlarger{\mathlarger{\mathlarger{\mathlarger {\bf \it a}}}}}_\pm(\lambda^*)\,{\cal \hat{C}\hat{P}\hat{T}}
$ and ${\cal\hat{D}}\,{\mathlarger{\mathlarger{\mathlarger{\mathlarger {\bf \it a}}}}}_\pm(\lambda)=
{\mathlarger{\mathlarger{\mathlarger{\mathlarger {\bf \it a}}}}}_\pm(-\lambda)\,{\cal\hat{D}}$
from the definition \eqref{soso1aa}-\eqref{Lop2ad} of 
${\mathlarger{\mathlarger{\mathlarger{\mathlarger {\bf \it a}}}}}_\pm(\lambda)$ as an operator acting in
 the Fock space as well as
${\cal \hat{C}\hat{P}\hat{T}}\, W_j(u)=W_j(-u^*)\,{\cal \hat{C}\hat{P}\hat{T}}$,
${\cal \hat{D}}\, W_j(u)=(-1)^j\,W_j(u)\,{\cal \hat{D}}$
using the bosonization formulae for $W_j(u)$.
\bigskip

Remarkably there exists another pair of chiral sesquilinear forms
for which relations \eqref{haysays1aa} and \eqref{Hermapm21aa}  remain true,
while the conjugation condition for the $W$ currents \eqref{haysaysa}
is no longer valid.
This may be motivated through the following observation.
A computation based on the explicit formulae \eqref{Idef1b} and \eqref{ioasido}
shows that the first three local IM can be written as \cite{Kotousov:2019nvt}
\bea\label{hsasysay}
{\bf I}_1&=&\int_0^{2\pi}\frac{\rd u}{2\pi}\ \Big((\partial\vartheta)^2+T\Big)\nonumber\\[0.2cm]
{\bf I}_2&=&\int_0^{2\pi}\frac{\rd u}{2\pi} \ \Big((\partial\vartheta)^3+\frac{3\, (n+2)}{3n+4}\ (\partial\vartheta)\, T\Big)\\[0.2cm]
{\bf I}_3&=&\int_0^{2\pi}\frac{\rd u}{2\pi}\ \Big( (\partial\vartheta)^4-\frac{n^2-2}{5n+6}\,(\partial^2\vartheta)^2+
\frac{6\,(n+2)}{5n+6}\,(\partial\vartheta)^2\ T+\frac{n+2}{5n+6}\ T^2\Big)\nonumber \ ,
\eea
where $T(u)$ stands for the chiral field
\bea
T(u)=(\partial\varphi)^2+\ri\,\frac{n+1}{\sqrt{n+2}}\,\partial^2\varphi\ .
\eea
It turns out to be possible to choose the densities for all the ${\bf I}_m$ 
to be a  local field built from $\partial\vartheta(u)$ and
$T(u)$. 
Since  the latter satisfy the commutation relations
\be
\arraycolsep=0.8cm
\begin{array}{cc}
{\cal \hat{C}\hat{P}\hat{T}}\,T(u)=T(-u^*)\,{\cal \hat{C}\hat{P}\hat{T}}\,, &
{\cal \hat{C}\hat{P}\hat{T}}\,\partial\vartheta(u)=+\partial\vartheta(-u^*)\,{\cal \hat{C}\hat{P}\hat{T}}\\[0.2cm]
{\cal \hat{D}\hat{C}\hat{P}\hat{T}}\,T(u)=T(-u^*)\,{\cal \hat{D}\hat{C}\hat{P}\hat{T}}\,, &
{\cal \hat{D}\hat{C}\hat{P}\hat{T}}\,\partial\vartheta(u)=-\partial\vartheta(-u^*)\,{\cal \hat{D}\hat{C}\hat{P}\hat{T}}
\end{array}
\ee
with the ${\cal CPT}$ and ${\cal D}$ transformations  defined  as in \eqref{osaoid19020912},
one can introduce the sesquilinear forms through the relations
\be\label{oioias887cAAV}
\big(\!\!\big(\bm{\chi}_2,T(u)\,\bm{\chi}_1\big)\!\!\big)_{\pm}=
\big(\!\!\big(T(u^*)\,\bm{\chi}_2,\bm{\chi}_1\big)\!\!\big)_{\pm}\ ,\qquad 
\big(\!\!\big(\bm{\chi}_2,\partial\vartheta(u)\,\bm{\chi}_1\big)\!\!\big)_{\pm}=
\pm\,\big(\!\!\big(\partial\vartheta(u^*)\,\bm{\chi}_2,\bm{\chi}_1\big)\!\!\big)_{\pm}\ .
\ee
Here $\bm{\chi}_1$,$\bm{\chi}_2$ are arbitrary states such that
$\bm{\chi}_{1}\in{\cal F}_{{\bf P}}$, while $\bm{\chi}_{2}\in{\cal F}_{{\bf P}^*}$
for the ``$+$'' case and $\bm{\chi}_{2}\in{\cal F}_{({\bf P}')^*}$ for  the ``$-$'' one
with $({\bf P}')^*\,=\big(\frac{\rho}{\sqrt{n+2}},-\frac{\nu^*}{\sqrt{n}}\big)$.
\bigskip

To see that \eqref{oioias887cAAV} 
indeed defines the sesquilinear forms on the Fock spaces
let's introduce a suitable basis for ${\cal F}_{\bf P}$.
The coefficients $\{L_m\}$ occurring in the expansion
of $T(u)$ in the Fourier series \eqref{Texpansion1} 
generate the Virasoro algebra 
with central charge $c=1-\frac{6(n+1)^2}{n+2}$. 
The number of states of the form
\be\label{Loio9basis}
L_{-m_1}\ldots L_{-m_j}\,b_{-m_{1}'}\ldots b_{-m_{j'}'}\,|{\bf P}\rangle  \,,\qquad 
1\le m_1\le m_2\le \ldots\le m_j\,,\ \  1\le m_{1}'\le m_{2}'\le \ldots\le m_{j'}'
\ee
with $\sum_{j} m_j+\sum_{j'}m_{j'}'=\ell$  is given by ${\rm par}_2(\ell)$,
so that they provide a basis in the level subspace of the Fock space ${\cal F}_{{\bf P}}^{(\ell)}$.
Then \eqref{oioias887cAAV},
together with the commutation relations for the Virasoro
and Heisenberg algebra generators   $L_m$ and $b_m$,  determine the sesquilinear 
form  in the basis \eqref{Loio9basis} up to an overall multiplicative constant. 
The latter is fixed by specifying the form on the Fock highest states
$\bm{\psi}_{\rho,+\nu}^{({\rm vac})}\equiv|{\bf P}\rangle$.
In view of what follows we'll take it to be
\be\label{normalization129a}
\big(\!\!\big(\bm{\psi}_{\rho,+\nu^*}^{({\rm vac})},\bm{\psi}_{\rho,\nu}^{({\rm vac})}\big)\!\!\big)_+=
\big(\!\!\big(\bm{\psi}_{\rho,-\nu^*}^{({\rm vac})},\bm{\psi}_{\rho,\nu}^{({\rm vac})}\big)\!\!\big)_-=
Z_+\big(\tfrac{\rho}{\sqrt{n+2}}\,\big|\, \sqrt{n+2}\,\big) \ ,
\ee
where the  function $Z_+(P\,|\,\beta)$ is given in eq.\eqref{asusausss}.
\bigskip

Let's consider the chiral sesquilinear forms, 
defined through eqs.\,\eqref{oioias887cAAV} and \eqref{normalization129a},
 in the eigenbasis
of the operator
${\mathlarger{\mathlarger{\mathlarger{\mathlarger {\bf \it a}}}}}_+(\lambda)\in{\rm End}\big({\cal F}_{\bf P}\big)$.
The forms 
are constructed in such a way that the local IM satisfy the relations
\be
\big(\!\!\big(\bm{\chi}_2,{\bf I}_m\bm{\chi}_1\big)\!\!\big)_\pm=
(\pm1)^{m+1}\,\big(\!\!\big({\bf I}_m\bm{\chi}_2,\bm{\chi}_1\big)\!\!\big)_\pm\ .
\ee
Then it follows that the eigenstates 
$\bm{\psi}_{\rho,\nu}(\bm{w})$ obey the orthogonality conditions
\be\label{oiaso8c8ca}
\begin{array}{c}
\big(\!\!\big(\bm{\psi}_{\rho,+\nu^*}(\bm{w}_2),\bm{\psi}_{\rho,\nu}(\bm{w}_1)\big)\!\!\big)_{+}=
U_{\rho,\nu}(\bm{w}_1)\ 
\delta_{\bm{w}_2^{\vphantom{2}},-\bm{w}_1^*} \\[0.4cm]
\big(\!\!\big(\bm{\psi}_{\rho,-\nu^*}(\bm{w}_2),\bm{\psi}_{\rho,\nu}(\bm{w}_1)\big)\!\!\big)_{-}=
U_{\rho,\nu}(\bm{w}_1)\ 
\delta_{\bm{w}_2^{\vphantom{2}},+\bm{w}_1^*}
\end{array}\,,
\ee
where we've taken into account formula \eqref{CPT*8787aaa}
describing  the action of the ${\cal CPT}$ and ${\cal D}$ transformation
on the eigenstates.
The function $U_{\rho,\nu}(\bm{w})$ depends on the normalization of 
$\bm{\psi}_{\rho,\nu}(\bm{w})$. The latter, up till now,
have been considered as eigenstates of 
${\mathlarger{\mathlarger{\mathlarger{\mathlarger {\bf \it a}}}}}_+(\lambda)$ without reference
to their overall normalization.
It turns out to be convenient to set this using the 
basis \eqref{Loio9basis}. Namely we'll take
\be\label{oio888s}
\bm{\psi}_{\rho,\nu}(\bm{w})
=\big((L_{-1})^{\ell}+\ldots\big)\,
\bm{\psi}_{\rho,\nu}^{({\rm vac})}\in {\cal F}_{\bf P}^{(\ell)}
\qquad\quad \big(\bm{w}=\{w_a\}_{a=1}^{\ell}\big)\ ,
\ee
where the dots stand for the terms, which contain lower powers of $L_{-1}$.

\bigskip

With the normalization of the states $\bm{\psi}_{\rho,\nu}(\bm{w})$
fixed, the functions $U_{\rho,\nu}(\bm{w})$ in \eqref{oiaso8c8ca} are 
defined unambiguously.
Here, as an illustration, we quote some explicit formulae for
the basis states in the  level subspace ${\cal F}_{\bf P}^{(1)}$.
Normalized as in \eqref{oio888s},
 they are given by
\be\label{oiasoi1992839a}
{\bm{\psi}}_{\rho,\nu}(w_\pm)=
\bigg(L_{-1}+\frac{2\ri\sqrt{n}}{n+2}\ w_\pm\ b_{-1}\bigg)\bm{\psi}_{\rho,\nu}^{({\rm vac})}
\ee
with $w_\pm$ being the two solutions of \eqref{sksksk1}, which for ${\tt L}=1$ becomes a quadratic equation,
\be\label{eqw1a}
w_\pm=-\frac{n+1}{2n}\,\Bigg(2\ri\,\nu\pm\sqrt{n(n+2)}\ \sqrt{1-\frac{4\rho^2}{(n+1)^2}-\frac{4\nu^2}{n(n+2)}}\ \Bigg)\ .
\ee
It is simple to check that the orthogonality conditions \eqref{oiaso8c8ca} are satisfied 
and find 
\be\label{asdio1i2121}
U_{\rho,\nu}({\bm w})\, = \, 
Z_+\big(\tfrac{\rho}{\sqrt{n+2}}\,\big|\, \sqrt{n+2}\,\big) \ \,
\frac{2n}{(n+2)^2}\ \times\ \begin{cases} w_+\,(w_{-}-w_{+})\,, &\qquad \bm{w}=\{w_+\} \\[0.2cm]
w_-\,(w_{+}-w_{-})\,,
&\qquad  \bm{w}=\{w_-\}
\end{cases}\ \ .
\ee

\bigskip

As was pointed out in the work \cite{Kotousov:2019nvt},
the chiral sesquilinear forms
$\big(\!\!\big(\cdot,\cdot\big)\!\!\big)_\pm$
and $\big(\cdot,\cdot\big)_\pm$
are related
through the reflection operator:
\be\label{ooa99s99s1a}
\big(\!\!\big(\bm{\chi}_2,\bm{\chi}_1\big)\!\!\big)_{\pm}=
f_{\rho,\nu}^{(\pm)}\,\times\, \big(\bm{\chi}_2,\check{{\bf R}}\,\bm{\chi}_1\big)_{\pm}\ .
\ee
Here $\bm{\chi}_1$, $\bm{\chi}_2$ are arbitrary states belonging to the
Fock space  and conjugated Fock space, respectively,
while 
the factor $f_{\rho,\nu}^{(\pm)}$ is the same for all the states in
${\cal F}_{{\bf P}}$.
The  reflection operator $\check{\bf R}$
was already discussed at the end of sec.\,\ref{sec15}. 
Its explicit construction
as an operator in the Fock space is given in  ref.\cite{Kotousov:2019nvt}.
\bigskip

Considering \eqref{ooa99s99s1a} in the eigenbasis of 
${\mathlarger{\mathlarger{\mathlarger{\mathlarger {\bf \it a}}}}}_\pm(\lambda)$, one obtains
\be\label{oiaso8c8}
\big(\bm{\psi}_{\rho,\pm\nu^*}(\bm{w}_2),\bm{\psi}_{\rho,\nu}(\bm{w}_1)\big)_{\pm}=
F_{\rho,\nu}^{(\pm)}(\bm{w}_1)\ 
\delta_{\bm{w}_2^{\vphantom{2}},\mp\bm{w}_1^*} 
\ee
with
\be\label{aisod1032w}
F_{\rho,\nu}^{(\pm)}(\bm{w})=f^{(\pm)}_{\rho,\nu}\,U_{\rho,\nu}(\bm{w})/\check{R}_{\rho,\nu}(\bm{w})\ .
\ee
Taking into account that $\check{\bf{R}}$ acts as the identity on the Fock highest states,
the above equations imply that
\bea\label{iasodio12}
\big(\bm{\psi}_{\rho,\pm\nu^*}^{({\rm vac})},\,\bm{\psi}_{\rho,\nu}^{({\rm vac})}\big)_{\pm}=
F^{(\pm,{\rm vac})}_{\rho,\nu}=
Z_+\big(\tfrac{\rho}{\sqrt{n+2}}\,\big|\, \sqrt{n+2}\,\big)\ f^{(\pm)}_{\rho,\nu}\ .
\eea
This way the functions $f^{(\pm)}_{\rho,\nu}$ are determined once
the value of the sesquilinear forms $\big(\cdot,\cdot\big)_\pm$
are fixed on the Fock vacua. We'll postpone 
making this choice till the next subsections.
\bigskip

In practice the calculation of the chiral sesquilinear forms
$\big(\!\!\big(\cdot,\cdot\big)\!\!\big)_\pm$ on two given states is significantly simpler than that of
$\big(\cdot,\cdot\big)_\pm$.
For instance expressing ${\bm{\psi}}_{\rho,\nu}(w_\pm)$  \eqref{oiasoi1992839a}
as states in the Verma module of the $W_\infty$\,-\,algebra
results in the  more cumbersome formulae
\bea\label{osiod8f0}
\bm{\psi}_{\rho,\nu}(w_\pm)=
\frac{ C_{2,\pm}\,\widetilde{W}_2(-1)\,+\,
4\sqrt{n}\  C_{3,\pm}\,\widetilde{W}_3(-1)}
{n\,(1+n+2\rho)\,(1-2\rho+2\ri \nu)\,(1-2\rho-2\ri \nu)}\ \ \bm{\psi}_{\rho,\nu}^{({\rm vac})}
\eea
with the coefficients
\bea
C_{2,\pm}&=&(1+n-2\rho)\big(n\,(1-2\rho)\,(1+n+2\rho)-4(3n+4)\,\nu^2\,\big)+4\ri n\,\nu\,(n+2-4\rho)\,w_\pm\nonumber 
 \\[0.2cm]
C_{3,\pm}&=& (n+2)(n+1-2\rho)\,\nu-\ri n\,(1-2\rho)\,w_{\pm}\ .
\eea
The functions $F_{\rho,\nu}^{(\pm)}(\bm{w})$ in \eqref{oiaso8c8}
may be computed directly from the definition \eqref{haysaysa},\,\eqref{iasodio12}
or obtained via \eqref{aisod1032w}:
\bea
F_{\rho,\nu}^{(\pm)}\big(\{ w\}\big)&=& 
F^{(\pm,{\rm vac})}_{\rho,\nu}\ 
\frac{2n}{(n+2)^2}\ 
\frac{(n+1-2 p-2 w)(n+1-2 p+2 w)}{(n+1+2 p-2 w)(n+1+2 p+2 w)}\nonumber \\[0.2cm]
&\times&\ \begin{cases} w_+\,(w_{-}-w_{+})\,, &\qquad w=w_+ \\[0.2cm]
w_-\,(w_{+}-w_{-})\,,
&\qquad  w=w_-
\end{cases}\ \ .
\eea

\subsection{Scaling limit of the Bethe states with real $s$   and $(n+2)\,{\tt k}\notin \mathbb{Z}$\label{193}}
The space of states of the ${\cal Z}_2$ invariant inhomogeneous six-vertex model, i.e., the
finite dimensional space 
${\mathscr V}_N=\mathbb{C}^2_N\otimes
 \mathbb{C}^2_{N-1}\otimes\cdots\otimes\mathbb{C}^2_1$,
admits a variety of Hermitian structures for which the Bethe states satisfy the
orthogonality condition 
$\big(\bm{\Psi}^{(2)}_N,\,\bm{\Psi}^{(1)}_N\big)=0$ unless 
$\bm{\Psi}^{(2)}_N\propto{\cal \hat{C}\hat{P}\hat{T}}\,\bm{\Psi}^{(1)}_N$. 
These are distinguished by the value of the  ``norms''
$({\cal \hat{C}\hat{P}\hat{T}}\bm{\Psi}_N,\bm{\Psi}_N)$.
For the description  of the Hermitian structures consistent with the integrable structure a
fundamental r$\hat{\rm{o}}$le belongs to the sesquilinear form
$(\cdot,\cdot)_{\star}$,
which was
mentioned in the Preliminaries (see eq.\,\eqref{is8s8s8dia}).
Here we present
the results of our numerical study of the  norm \eqref{FinalNorm} 
for the RG trajectories characterized by the real  RG invariant  $s$.
They enable one to establish a precise relation between the sesquilinear form
 $(\cdot,\cdot)_{\star}$
and those that are  induced in the space ${\cal H}^{({\rm cont})}$. 
This, in turn, completes our description of the scaling limit of the
 low energy Bethe states with real $s$.

\bigskip

We performed a numerical study of the  norm
$({\cal \hat{C}\hat{P}\hat{T}}\bm{\Psi}_N,\bm{\Psi}_N)_{\star}$
of the Bethe states \eqref{Bstate1}
for a wide range of RG trajectories
with $S^z=0,1,2,\ldots\ $, ${\tt w}=0,\pm1,\pm2,\ldots\ $, 
${\tt L},\bar{{\tt L}}=0,1,2,\ldots\ $.
It was found that the combination
\be\label{aoisd89213hsd}
G[{\boldsymbol \Psi}_N]\equiv
\big({\cal\hat{ C}\hat{P}\hat{T}}
\bm{\Psi}_N,\bm{\Psi}_N\big)_{\star}\
(N/2)^{-\frac{1}{3}+f(p)+f({\bar p})+4{\tt L}+4{\bar {\tt L}}}\ \re^{-\frac{1}{2}{\cal A}_2 N^2}
\,,  
\ee
where $p$ and $\bar{p}$ are given by  \eqref{oaisoi1093},
satisfies the asymptotic condition 
\bea\label{hasysayas}
G[{\boldsymbol \Psi}_N]= O\big(\log(N)\big) \qquad {\rm as} \qquad N\to\infty\qquad
{\rm for}\ {\rm real}\ s\, .
\eea
The constant ${\cal A}_2$ coincides with \eqref{aisiasjjs} upon the substitution 
$\beta^2\mapsto\frac{2}{n+2}$, i.e., 
\bea
{\cal A}_2=
\int_0^\infty\frac{\rd t}{t}\ \frac{\sinh(\frac{ 2t}{n})\sinh(t)}{2\sinh\big(  (1+\frac{2}{n})\, t
\big)\cosh^2(t)}\ ,
\eea
while
\bea
f(p)=\frac{4p^2}{n+2}+\frac{1}{6(n+2)}+\frac{n+2}{6}-\frac{1}{2}\ .
\eea
To provide a more precise description of the asymptotic behaviour \eqref{hasysayas}, introduce
\bea\label{aoiisiasa}
g_N\big(\bar{\bm{w}},\bm{w}\,|\,\bar{p},p,s)=
\frac{1}{2\pi}\ \bigg[\, 4\,\log\bigg(\frac{N}{2N_0}\bigg)-\ri\ \frac{\rd}{\rd s}\, 
\log\Big({ D}_{{\bar{p}},s}(\bar{{\boldsymbol  w}})\,{ D}_{p,s}({\boldsymbol  w})\Big)\, \bigg]
\eea
with ${ D}_{p,s}({\boldsymbol  w})$ being given by \eqref{oiaodi1a1a}. 
Then 
for the RG trajectory $\bm{\Psi}_N$, which in the scaling limit becomes the state 
$\bar{\bm{\psi}}_{\bar{p},s}(\bar{\bm{w}})\otimes\bm{\psi}_{{p},s}({\bm{w}})$,
our numerical study led us to the following asymptotic formula
\bea\label{isisaisa}
G[{\boldsymbol \Psi}_N]&\asymp& 
 \big( {C}^{(\rm alt)}_0\big)^2\  
\bigg(\frac{C}{\sqrt{2}}\bigg)^{\frac{8s^2}{n}+f(p)+f({\bar p})+4{\tt L}+4{\bar {\tt L}}}\ \
\Big(2^{-1-\frac{2}{n}}\sqrt{n+2}\, N_0 \Big)^{-\frac{8s^2}{n}}\nonumber\\[0.2cm]
&\times&\ U_{{\bar p},s}(\bar{{\boldsymbol w}})\ U_{p,s}({\boldsymbol w})\ 
\Big(g_N(\bar{\bm{w}},\bm{w}\,|\,\bar{p},p,s)+o(1)\Big)\ .
\eea
Here
$U_{p,s}({\boldsymbol w})$ is the same function as in \eqref{oiaso8c8ca}.
As was previously discussed, it is unambiguously determined through relations 
\eqref{oioias887cAAV},\,\eqref{normalization129a} specifying the chiral sesquilinear form 
$\big(\!\!\big(\cdot,\cdot\big)\!\!\big)_+$ and the 
normalization condition \eqref{oio888s} for the chiral states $\bm{\psi}_{p,s}(\bm{w})$.
Formula \eqref{isisaisa}
also involves the positive constants $C^{({\rm alt})}_0$ and $C$   depending only on $n$.
Their numerical values at different $n$ are presented  in Appendix \ref{app1}.
Note that the constant $C$ is the same as $C(\beta)$ from eq.\,\eqref{iasisaias}
provided that $\beta$ and $n$ are identified as
$\beta=\sqrt{\frac{2}{n+2}}\, $. 
\bigskip

It is possible to give a natural explanation of the asymptotic formula 
\eqref{isisaisa} if we make the following assumptions concerning 
the scaling limit of the Bethe states.
\begin{enumerate}[(i)]
\item
There exists the limit
\bea\label{scaling1a}
{\rm s}\!\!\!\!\lim_{N\to \infty\atop b(N)\to s}\ 
\Big({\cal  K}^{(\bar{{\tt L}})}_{N}(\bar{p},s)\,{\cal  K}^{({\tt L})}_{N}(p,s)\Big)^{-\frac{1}{2}}\ 
\bm{\Psi}_N\,=\,\bar{\bm{\psi}}_{\bar{p},s}(\bar{\bm{w}})\otimes\bm{\psi}_{{p},s}({\bm{w}})\equiv
\bm{\psi}_{\bar{p},p,s}(\bar{\bm{w}},\bm{w})\, ,
\eea
where
\bea\label{isisaisa2}
{\cal  K}^{({\tt L})}_{N}(p,s)&=&{C}^{(\rm alt)}_0\
\bigg(\frac{C}{\sqrt{2}}\bigg)^{\frac{4s^2}{n}+f(p)+4{\tt L}}\ 
\big(2^{-1-\frac{2}{n}}\sqrt{n+2}\, N_0 \big)^{-\frac{4s^2}{n}}\nonumber\\[0.2cm]
&\times& 
(N/2)^{\frac{1}{6}-f(p)-4{\tt L}}\ \re^{\frac{1}{4}{\cal A}_2 N^2} \ .
\eea
For given $S^z\ge 0$ the set of all possible states $\{\bm{\psi}_{\bar{p},p,s}(\bar{\bm{w}},\bm{w})\}$   form a basis 
in ${\cal H}^{({\rm cont})}_{S^z}$.
\item The space ${\cal H}^{({\rm cont})}_{S^z}$ is equipped with the inner product 
$\big\langle\!\!\big\langle\cdot,\cdot\big\rangle\!\!\big\rangle_{{\rm cont}}$,
which in the basis $\bm{\psi}_{\bar{p},p,s}(\bar{\bm{w}},\bm{w})$   is given by
\bea\label{otho99r}
\big\langle\!\!\big\langle\bm{\psi}_{\bar{p}',p',s'}(\bar{\bm{w}}',\bm{w}'),\,
\bm{\psi}_{\bar{p},p,s}(\bar{\bm{w}},\bm{w})\big\rangle\!\!\big\rangle_{\rm cont}&=&
\delta_{\bm{w}', -\bm{w}^*}\, \delta_{\bar{\bm{w}}', -\bar{\bm{w}}^*}\, 
\delta_{p',p}\,\delta_{\bar{p}',\bar{p}}\
\delta(s'-s)\nonumber\\[0.4cm]
&\times&U_{{\bar p},s}(\bar{{\boldsymbol w}})\ U_{p,s}({\boldsymbol w})\ .
\eea
\end{enumerate}
\bigskip

To describe how \eqref{isisaisa} arises from (i)-(ii),
let's for simplicity focus on the case of the primary Bethe states 
having ${\tt L}=\bar{\tt L}=0$. 
Since for fixed $N$ and $\bar{p}$, $p$ these states are distinguished
by  the integer ${\tt m}$ entering into \eqref{poaospo1qa1a},
we'll denote them as $\bm{\Psi}_N^{({\tt m})}$.
Taking  an arbitrary linear combination,
\be\label{Zidef1}
\bm{\Xi}_N=\sum_{\tt m} C_{{\tt m}}\,\bm{\Psi}_N^{({\tt m})}\,,
\ee
 consider the sesquilinear form $(\cdot,\cdot)_{\star}$ of $\bm{\Xi}_N$ and the state
$\bm{\Psi}_N^{({\tt m_0})}$ with given integer ${\tt m_0}$. The orthogonality condition
\eqref{ortho1} implies
\be\label{xixixi99d}
\big(\bm{\Xi}_N,\bm{\Psi}_N^{({\tt m_0})}\big)_{\star}
=C_{{\tt m_0}}\,
\big({\cal\hat{ C}\hat{P}\hat{T}}
\bm{\Psi}_N^{({\tt m_0})},\bm{\Psi}_N^{({\tt m_0})}\big)_{\star}\ .
\ee
At large $N$  it follows from  (i) that the  state $\bm{\Psi}_N^{({\tt m})}$
is approximated by
\be\label{0s90d0s}
\bm{\Psi}_N^{({\tt m})}\approx \Big({\cal  K}^{(0)}_{N}(\bar{p},s)\,{\cal  K}^{(0)}_{N}(p,s)\Big)^{\frac{1}{2}}\,
\bm{\psi}_{\bar{p},p,s}^{({\rm vac})}\ .
\ee
Moreover the sum in \eqref{Zidef1} may be replaced by the integral
\be\label{sumreplace9s}
\sum_{\tt m}\mapsto \int{\rm d}s\,\rho(s)\,,
\ee
where the density of states $\rho(s)$ comes from eq.\,\eqref{rho01a}. The latter
coincides with  $g_N$
\eqref{aoiisiasa} specialized to the primary Bethe states.
Evaluating the l.h.s. of \eqref{xixixi99d} using eqs.\,\eqref{0s90d0s} and
\eqref{sumreplace9s} as well as  \eqref{otho99r}
leads to the asymptotic formula \eqref{isisaisa}. 
\bigskip

The similar arguments may be applied for the Bethe states with 
${\tt L},\bar{\tt L}\ge 0$. However, it should be pointed out 
that $g_N$, in general, 
takes complex values. Therefore
$g_N\big(\bar{\bm{w}},\bm{w}\,|\,\bar{p},p,s)\,\Delta s$  can not be interpreted as the
number of low energy Bethe states $\bm{\Psi}_N$ with fixed 
$p$, $\bar{p}$, ${\tt L}$, $\bar{\tt L}$, $\bm{w}$ and $\bar{\bm{w}}$
having $\Re e\big(b(N)\big)$ belonging to the segment $(s,s+\Delta s)$. Nevertheless,
the sum of $g_N\big(\bar{\bm{w}},\bm{w}\,|\,\bar{p},p,s)$ over all the 
${\rm par}_2({\tt L})\times {\rm par}_2(\bar{{\tt L}})$ solutions
sets $\bm{w}$ and $\bar{\bm{w}}$  with fixed ${\tt L}$ and $\bar{\tt L}$
turns out to be a real positive function of $s$ that coincides with the density  \eqref{aisodio12311}:
\be
\rho_{\bar{p},p}^{({\tt L},\bar{\tt L})}(s)=\sum_{(\bar{\bm{w}},\bm{ w})\atop {\tt L},\bar{\tt L} - {\rm fixed}}\,
g_N(\bar{\bm{w}},\bm{w}\,|\,\bar{p},p,s)\ .
\ee 

\bigskip

The above analysis suggests
that  the sesquilinear form $(\cdot,\cdot)_{\star}$ 
in the lattice model induces the inner product
$\big\langle\!\!\big\langle\cdot,\cdot\big\rangle\!\!\big\rangle_{\rm cont}$ 
for the states in ${\cal H}^{({\rm cont})}_{S^z}$ with $S^z\ge 0$.
The latter is defined through \eqref{otho99r} 
in the eigenbasis diagonalizing the operators
 ${\mathlarger{\mathlarger{\mathlarger{\mathlarger {\bf \it a}}}}}_\pm(\lambda)$ and
 $\bar{{\mathlarger{\mathlarger{\mathlarger{\mathlarger {\bf \it a}}}}}}_\pm(\bar{\lambda})$.
A basis independent description is provided by the relations \eqref{oioias887cAAV}
for the ``$+$'' case and the similar ones involving $\overline{T}$, $\partial\bar{\vartheta}$,
along with  the value of the inner product for the
highest states:
\be\label{iaosid12212221}
\big\langle\!\!\big\langle
\bm{\psi}_{\bar{p}',p',s}^{({\rm vac})},\bm{\psi}_{\bar{p},p,s}^{({\rm vac})}
\big\rangle\!\!\big\rangle_{\rm cont}=
\delta_{\bar{p}',\bar{p}}\,\delta_{{p}',p}\,\delta(s'-s)\ 
Z_+\big(\tfrac{\bar{p}}{\sqrt{n+2}}\,\big|\, \sqrt{n+2}\,\big)\,Z_+\big(\tfrac{p}{\sqrt{n+2}}\,\big|\, \sqrt{n+2}\,\big)\ .
\ee
For the sectors ${\cal H}_{S^z}^{({\rm cont})}$ with $S^z<0$, the inner product
$\big\langle\!\!\big\langle\cdot,\cdot\big\rangle\!\!\big\rangle_{\rm cont}$
 is defined  using the ${\cal CP}$ invariance of the model. 
With the same line of arguments that led to \eqref{iaosio11212} one can show
that
$
{\cal \hat{C}\hat{P}}\,T(u)=\overline{T}(u)\,{\cal \hat{C}\hat{P}}$ and 
${\cal \hat{C}\hat{P}}\,\partial\vartheta(u)=\partial\bar{\vartheta}(u)\,{\cal \hat{C}\hat{P}}$.
Thus 
for the sectors with $S^z<0$, the defining relations \eqref{oioias887cAAV} for the case
``$+$'',
its barred counterpart and \eqref{iaosid12212221}
 remain valid. In turn, formula \eqref{otho99r} is applicable for any  
$S^z=0,\pm1,\pm2,\ldots\ $.

\bigskip
Eq.\,\eqref{scaling1a} describes a scaling limit of the
Bethe states that leads to a Hermitian structure in 
the linear space ${\cal H}^{({\rm cont})}$ that is consistent with the
integrable structure. However the inner product 
$\big\langle\!\!\big\langle\cdot,\cdot\big\rangle\!\!\big\rangle_{\rm cont}$
\eqref{otho99r}
 is not consistent with the natural conjugation \eqref{aoisdo8923} in the
$\overline{W}_\infty\otimes W_\infty$\,-\,algebra.
It  turns out to be possible to modify  the definition of the scaling limit
such that the inner product $\big\langle\cdot,\cdot\big\rangle_{\rm cont}$
\eqref{oaisod9182921} is induced in ${\cal H}^{({\rm cont})}$.
This can be done in the following way.
\bigskip

Let's change the normalization of the Bethe states prescribed by eq.\,\eqref{Bstate1}
and  introduce
\be\label{psiNprimedef1a}
\bm{\Psi}_N'\big(\{\zeta_m\}\big)=\alpha(\zeta_1,\ldots,\zeta_M)\,\bm{\Psi}_N\big(\{\zeta_m\}\big)
\ee
with
\be\label{iaosid81289812}
\alpha(\zeta_1,\ldots,\zeta_M)=\re^{\ri\pi{\tt k}}\, q^{-\frac{N}{2}+S^z}
{ A}^{(\infty)}_+\ \big[{ A}_+(+\ri q ){A}_+(-\ri q )\big]^{-1}\nonumber
\ee
and
$$A_+(\pm\ri q)=\prod_{m=1}^M\big(1\mp\ri q/\zeta_m\big)\,, \qquad\qquad
{ A}^{(\infty)}_+=\prod_{m=1}^M(-1/\zeta_m) \ .$$
Recall that for given $N$ the Bethe states 
as defined by \eqref{Bstate1} satisfy the condition
${\cal \hat{C}\hat{P}\hat{T}}\,\bm{\Psi}_N\big(\{\zeta_m\}\big)=\bm{\Psi}_N\big(\{\zeta_m^*\}\big)$. 
This is no longer true for $\bm{\Psi}_N'$. Instead,
in view of
formula \eqref{Keigeq1a} for the eigenvalues of the lattice translation operator,
\be\label{oaisd9819281}
{\cal\hat{ C}\hat{P}\hat{T}}\bm{\Psi}'_N\big(\{\zeta_m\}\big)=
{\mathbb K}^{-1}\,\bm{\Psi}_N'\big(\{\zeta_m^*\}\big)\,.
\ee
The proportionality factor $\alpha$  in \eqref{psiNprimedef1a} has been chosen 
so that as $N\to\infty$, 
\bea\label{oaisodias}
\alpha(\zeta_1,\ldots,\zeta_M)\,\big(\alpha(\zeta_1^*,\ldots,\zeta_M^*)\big)^*&\asymp&
\big(
{ R}_{\bar{p},s}(\bar{{\boldsymbol w}})\,
{ R}_{p,s}({\boldsymbol w})
\big)^{-1} \\[0.2cm]
&\times&
\bigg(\frac{N}{2N_0}\bigg)^{\frac{n(\bar{p}+p)}{n+2}}\  \bigg(\frac{n+2}{4n}\bigg)^{N}\ \big(1+o(1)\big)\ .
\nonumber
\eea
This  follows by considering the product of the ``$+$'' and ``$-$''
 cases in \eqref{aiosdi1209112} and taking into account
eq.\,\eqref{iaosido9898}.
Then combining \eqref{oaisodias} with \eqref{isisaisa} one obtains
\bea\label{isoi9890s}
\big({\mathbb K}\,{\cal\hat{ C}\hat{P}\hat{T}}
\bm{\Psi}_N',\bm{\Psi}_N'\big)_{\star}\,&\asymp&
\   U_{{\bar p},s}(\bar{{\boldsymbol w}})\ U_{p,s}({\boldsymbol w})\ 
\ \big(
{ R}_{\bar{p},s}(\bar{{\boldsymbol w}})\,
{ R}_{p,s}({\boldsymbol w})
\big)^{-1}\
\\[0.3cm]
&\times&  {\cal N}_N^{(  \bar{{\tt L}})}({\bar{p}},s)\ { \cal N}_N^{(  {\tt L})}({p,s})
 \ \Big(g_N\big(\bar{\bm{w}},\bm{w}\,|\,\bar{p},p,s\big)+o(1)\Big)\nonumber
\eea
with
\bea\label{isisaisa3}
{ \cal N}^{(  {\tt L})}_N({p,s})
&=&{C}^{(\rm alt)}_0\
\bigg(\frac{C}{\sqrt{2}}\bigg)^{\frac{4s^2}{n}+f(p)+4{\tt L}}\ 
\big(2^{-\frac{n+2}{n}}\sqrt{n+2}\, N_0 \big)^{-\frac{4s^2}{n}}\nonumber\\[0.2cm]
&\times& 
(N/2)^{\frac{1}{6}-f(p)+\frac{2n p}{n+2} -4{\tt L}}\ \re^{\frac{1}{4}{\cal A}_2 N^2} \ \bigg(\frac{n+2}{4n}\bigg)^{N/2}\ .
\eea
Similar arguments that lead us to \eqref{scaling1a}
suggest
 that there exists the limit
\bea\label{scaling1b}
{\rm s}\!\!\!\!\lim_{N\to \infty\atop b(N)\to s}\ 
\Big({ \cal N}^{(\bar{{\tt L}})}_{N}(\bar{p},s)\,{ \cal N}^{({\tt L})}_{N}(p,s)\Big)^{-\frac{1}{2}}\
\bm{\Psi}_N'\,=\,\bm{\psi}_{\bar{p},p,s}(\bar{\bm{w}},\bm{w})
\eea
and the sesquilinear form for the lattice model induces an inner product in
${\cal H}^{({\rm cont})}$  such that
\bea\label{otho99r2a}
\big\langle\bm{\psi}_{\bar{p}',p',s'}(\bar{\bm{w}}',\bm{w}'),\,
\bm{\psi}_{\bar{p},p,s}(\bar{\bm{w}},\bm{w})\big\rangle_{\rm cont}&=&
\delta_{\bar{\bm{w}}', -\bar{\bm{w}}^*}\,\delta_{\bm{w}', -\bm{w}^*}\, 
\delta_{\bar{p}',\bar{p}}\,\delta_{p',p}\
\delta(s'-s)\nonumber \\[0.2cm]
&\times&
F^{(+)}_{\bar{p},s}(\bar{\bm{w}})\ F^{(+)}_{p,s}({\bm{w}})\ ,
\eea
where
\be
F^{(+)}_{\bar{p},s}(\bar{\bm{w}})=U_{\bar{p},s}(\bar{\bm{w}})/{ R}_{\bar{p},s}(\bar{{\boldsymbol w}})\,,
\qquad \qquad
F^{(+)}_{p,s}({\bm{w}})=U_{{p},s}({\bm{w}})/{ R}_{{p},s}({{\boldsymbol w}})\ .
\ee
Notice that the states appearing in the scaling limit \eqref{scaling1b} satisfy
the ${\cal {C}{P}{T}}$ conjugation condition
${\cal \hat{C}\hat{P}\hat{T}}\,\bm{\psi}_{\bar{p},p,s}(\bar{\bm{w}},\bm{w})=
\bm{\psi}_{\bar{p},p,s}(-\bar{\bm{w}}^*,-\bm{w}^*)$.
Though the action of the ${\cal {C}{P}{T}}$
transformation on the Bethe states $\bm{\Psi}_N'$ involves a phase factor 
${ K}^{-1}={ K}^*$, as it follows from \eqref{tower1b}, 
for any low energy Bethe state 
${ K}$ tends to one as $N\to\infty$.
\bigskip

We now observe that, since
${ R}_{{p},s}({{\boldsymbol w}})={ R}_{p,s}^{(0)}\,\check{R}_{p,s}(\bm{w})$,
the function $F^{(+)}_{p,s}({\bm{w}})$ can be expressed  as in  \eqref{aisod1032w} 
with $f^{(+)}_{p,s}=1/{ R}_{p,s}^{(0)}$.
From the relation between the chiral sesquilinear forms
$\big(\cdot,\cdot\big)_+$ and $\big(\!\!\big(\cdot,\cdot\big)\!\!\big)_+$, see eq.\,\eqref{ooa99s99s1a},
one concludes
that \eqref{otho99r2a}
defines an inner product in the space ${\cal H}^{(\rm cont)}$, which is
consistent with the conjugation conditions \eqref{aoisdo8923}.
Furthermore, for the
 $\overline{W}_\infty\otimes W_\infty$ 
primary states, in view of the explicit formula  \eqref{opsod898s} for ${ R}_{p,s}^{(0)}$,
one finds
\bea\label{iaosido891221}
\big\langle\bm{\psi}_{\bar{p}',p',s'}^{({\rm vac})},\,
\bm{\psi}_{\bar{p},p,s}^{({\rm vac})}\big\rangle_{\rm cont}=
\delta_{p',p}\,\delta_{\bar{p}',\bar{p}}\
\delta(s'-s)\ \ 
\frac{\Gamma(1+\frac{2\bar{p}}{n+2})\,\Gamma(1+\frac{2p}{n+2})}
{\Gamma(1+2\bar{p})\,\Gamma(1+2p)} \ \
\big|Z_{{\bar p},s}\, Z_{p,s}\big|^2
\eea
where
\bea\label{isoiod888}
Z_{p,s}&=&
\frac{2^{p}}{\sqrt{2\pi}}\ (n+2)^{\frac{1}{4}-\frac{p\,(n-1)}{2\,(n+2)}}\ 
\Gamma(\tfrac{1}{2}+p+\ri s)\
Z\big(\tfrac{p}{\sqrt{n+2}}\,\big|\, \sqrt{n+2}\,\big)\ .
\eea
Recall that $Z(P\,|\,\beta)$ is defined  by \eqref{asss}. 
\bigskip

Thus, with a  proper taking of the scaling limit,
the conjugation conditions for the $W$ currents
\be\label{asodi1982981}
\big[W_j(u)\big]^\star=W_j(u^*)\,,\qquad\qquad\ \ 
 \big[\overline{W}_j(\bar{u})\big]^\star=\overline{W}_j(\bar{u}^*)
\ee
are induced by
\be\label{oaisod91289aaaaa}
\hat{{\mathsf O}}^\star=
\hat{{\mathsf X}}_\star^{-1}\ \hat{{\mathsf O}}^\dag\  \hat{{\mathsf X}}_\star^{}\ ,\qquad
\quad \hat{{\mathsf O}}\in{\rm End}\big(\mathscr{V}_N\big) \,,
\ee
where 
$
\hat{{\mathsf X}}_\star^{}=\hat{\mathsf{X}}\ \re^{\ri\pi ({\mathbb S}^z-N/2)}\, 
\mathbb{A}^{({\infty})}_+
$
and ``$\dag$'' stands for the standard matrix Hermitian conjugation (see formulae
\eqref{odoappo}-\eqref{is8s8s8dia} in the Preliminaries).
The  matrix $\mathbb{A}^{({\infty})}_+$ 
is diagonal in the basis of Bethe states and its eigenvalues
are given by
$\prod_{m=1}^{M} (-1/\zeta_m)$. For general values of the twist and
anisotropy  parameters ${\tt k}$ and $n$,
$\mathbb{A}^{({\infty})}_+$  is invertible.
However for certain values of the parameters, some
of the Bethe states $\bm{\Psi}_N\big(\{\zeta_m\}\big)$
may be such that one of the Bethe roots  become zero or infinity (in the last case
the corresponding eigenvalue $A_+(\zeta)$ is a polynomial of order $M-1$).
Then $\mathbb{A}^{({\infty})}_+$  is singular and special consideration is required to define
the $\star$\,-\,conjugation.
Also the matrix  $\hat{{\mathsf X}}$ \eqref{Xcase1}
in the case of the ${\cal Z}_2$ invariant model
may be expressed in terms of the generator of the ${\cal Z}_2$ symmetry \eqref{Dz2case},
the total spin operator $\mathbb{S}^z$ and also
$\Sigma^z=\sigma^z_{N}\,\sigma^z_{N-2}\ldots \sigma^z_{2}$ as
$
\hat{{\mathsf X}}=\Sigma^z\,\hat{{\cal D}}\,\re^{\frac{\ri\pi}{2}({\mathbb S}^z-N/2)}
$.
In turn $\hat{{\mathsf X}}_\star^{}$ entering into the conjugation condition \eqref{oaisod91289aaaaa}
can be written as
\be\label{aoisdo8912}
\hat{{\mathsf X}}_\star^{}=\hat{{\mathsf X}}_\star^{\dag}=
\Sigma^z\,\hat{{\cal D}}\,\re^{\frac{\ri\pi}{2}(N/2-{\mathbb S}^z)}
\,\mathbb{A}^{({\infty})}_+\ .
\ee
Finally note that w.r.t. the $\star$\,-\,conjugation the  Hamiltonian $\mathbb{H}$
 and lattice translation operator $\mathbb{K}$
 satisfy the conditions
\be\label{congu7a878sa}
\mathbb{H}^\star=\mathbb{H}\,,\qquad\qquad
 \mathbb{K}^\star=\mathbb{K}^{-1}\ .
\ee
To avoid confusion, let's reiterate that this conjugation does not correspond
to a positive definite inner product.

\subsection*{Comments on the case ${\tt k}=0$ with $s$ real}
Our considerations regarding the scaling limit of
the Bethe states 
explicitly assumed that $(n+2)\,{\tt k}$ is not an integer.
However, for the later parts of this work, the
case ${\tt k}=0$ and $s$ a real number is of special interest.
As was already pointed out in sec.\,\ref{sec173}, some of the Verma modules 
of the chiral $W_\infty$\,-\,algebra,
which appears
in the decomposition of ${\cal H}^{({\rm cont})}_{S^z,{\tt w}}$ \eqref{iaosidoi192009}, 
 become reducible at ${\tt k}=0$.
This way
the space ${\cal H}^{({\rm cont})}_{S^z,{\tt w}}$ splits into two sectors
${\cal H}^{({\rm cont})}_{S^z,{\tt w}}=\tilde{{\cal H}}^{({\rm cont})}_{S^z,{\tt w}}\oplus 
{\cal H}^{({\rm null})}_{S^z,{\tt w}}$ similar to \eqref{ioasid1892981}, where
 the $\overline{W}_\infty\otimes W_\infty$ decomposition of 
${\cal H}^{({\rm null})}_{S^z,{\tt w}}$
contains the irreps $\overline{{\cal W}}_{{\rho},s}\otimes {\cal W}_{\rho,s}$,
for which the highest state of either $\overline{{\cal W}}_{{\rho},s}$ 
or ${\cal W}_{\rho,s}$  is 
a null vector in the Verma module.
The two sectors $\tilde{{\cal H}}^{({\rm cont})}_{S^z,{\tt w}}$ and ${\cal H}^{({\rm null})}_{S^z,{\tt w}}$
are orthogonal w.r.t. the inner product $\langle\cdot,\cdot\rangle_{\rm cont}$. 
Our analysis above
is adapted most straightforwardly to those  low energy Bethe states, 
which become part of $\tilde{{\cal H}}^{({\rm cont})}_{S^z,{\tt w}}$
in the scaling limit,
and we'll only comment on this case.
\bigskip

Assuming that $n$ is generic, 
the relations \eqref{scaling1b}-\eqref{isoiod888}
are applicable 
for describing the scaling limit of the low energy Bethe 
states forming the sector $\tilde{{\cal H}}^{({\rm cont})}_{S^z,{\tt w}}$
for the case  $S^z= 1,2,3,\ldots$ as well as $S^z={\tt w}=0$. 
However for $S^z=0$ the product of the two $\Gamma$-functions 
in the numerator in eq.\,\eqref{iaosido891221}  becomes 
$\frac{\pi{\tt w}}{\sin(\pi{\tt w})}$
, i.e.,  is singular 
for non-zero integer ${\tt w}$.
This is related to the fact, discussed in sec.\,\ref{sec173},
that the two states
 $\bm{\psi}_{\bar{p},p,s}(\bar{\bm{w}},\bm{w})$ with
$p=-\bar{p}=\pm\frac{1}{2}\,(n+2)|{\tt w}|$
are indistinguishable.
Nevertheless, similar to the Bethe states for finite $N$,
 one can resolve the ambiguity by starting with
$\bm{\psi}_{\bar{p},p,s}(\bar{\bm{w}},\bm{w})$ with non-zero ${\tt k}>0$ and then setting
${\tt k}\to 0^+$. 
For taking this limit it is useful to
change the normalization of the states and define
$\widetilde{\bm{\psi}}_{s}^{({\tt w})}(\bar{\bm{w}},\bm{w})=\sqrt{{\tt k}}\ 
\bm{\psi}_{\bar{p},p,s}(\bar{\bm{w}},\bm{w})$
where $p=-\bar{p}=\frac{1}{2}(n+2)({\tt k}+{\tt w})$. Then
any inner product involving the states $\widetilde{\bm{\psi}}_{s}^{({\tt w})}(\bar{\bm{w}},\bm{w})$  
remains well defined as ${\tt k}\to 0^+$. In particular for the primary $\overline{W}_\infty\otimes W_\infty$ states
it follows from eq.\,\eqref{iaosido891221} that
\bea\label{aoisido1920}
\big\langle\widetilde{\bm{\psi}}_{s'}^{({\tt w}',{\rm vac})},\,
\widetilde{\bm{\psi}}_{s}^{({\tt w},{\rm vac})}\big\rangle_{\rm cont}&=&
\delta_{{\tt w}',{\tt w}}\
\delta(s'-s)\ (-1)^{{\tt w}}\  \, 
\frac{\sin\big(\pi\,(n+2)\,{\tt w}\big)}
{\pi\,(n+2)}\ \\[0.2cm] 
 &\times &\big|Z_{p,s}\, Z_{-p,s}\big|^2 \ \Big|_{p=\frac{1}{2}\,(n+2)\,{\tt w}}\qquad \qquad
(S^z={\tt k}=0,\,{\tt w}\ne 0)\ .\nonumber 
\eea
This way the
subspaces $\tilde{{\cal H}}^{({\rm cont})}_{0,{\tt w}}$ with ${\tt w}\ne 0$
become equipped with the inner product determined through \eqref{aoisido1920}
as well as the conjugation conditions \eqref{asodi1982981} for the $W$ currents.

\subsection{Scaling limit of the Bethe states with  pure imaginary  $s$ and $(n+2)\,{\tt k}\notin \mathbb{Z}$
\label{194}}
In sec.\,\ref{sec191} we introduced two sesquilinear forms.
For the form  $(\cdot,\cdot)_+$ the irrep conjugated to
${\cal V}_{\bar{\rho},\bar{\nu}}\otimes{\cal V}_{{\rho},{\nu}}$ coincides
with
${\cal V}_{\bar{\rho},\bar{\nu}^*}\otimes{\cal V}_{{\rho},{\nu}^*}$,
while for $(\cdot,\cdot)_-$ the conjugated irrep is
${\cal V}_{\bar{\rho}-,\bar{\nu}^*}\otimes{\cal V}_{{\rho},-{\nu}^*}$. 
Equipping the sector ${\cal H}^{({\rm cont})}$
with the ``plus'' form and   ${\cal H}^{({\rm disc})}$
with the ``minus'' one, each of the irreps occurring in their
decomposition w.r.t. the $\overline{W}_\infty\otimes W_\infty$\,-\,algebra
would be self-conjugated. The form $(\cdot,\cdot)_+$ is consistent with the 
formal anti-involution
\eqref{asodi1982981} in the 
$\overline{W}_\infty\otimes W_\infty$\,-\,algebra.
For the other form, the corresponding conjugation
reads as
\be\label{aisodi1212AA}
\big[W_j(u)\big]^{\text{\tiny\ding{105}}}=(-1)^j\,W_j(u^*)\,,\qquad\qquad\ \ 
 \big[\overline{W}_j(\bar{u})\big]^{\text{\tiny\ding{105}}}=(-1)^j\,\overline{W}_j(\bar{u}^*)\ .
\ee
The two anti-involutions 
are related through the ${\cal Z}_2$ transformation:
\be\label{aoisdoias9013}
\big[W_j(u)\big]^{\text{\tiny\ding{105}}}=\hat{\cal D}\,\big[W_j(u)\big]^\star\,\hat{\cal D}\,,\qquad\qquad
\big[\overline{W}_j(\bar{u})\big]^{\text{\tiny\ding{105}}}=\hat{\cal D}\,\big[\overline{W}_j(\bar{u})\big]^\star\,\hat{\cal D}\ .
\ee
In the previous subsection
it was pointed out that with a proper taking of the scaling limit
the $\star$\,-\,conjugation for the $W$ currents is induced from
the lattice one defined by eqs.\,\eqref{oaisod91289aaaaa} and \eqref{aoisdo8912}.
Therefore one might expect that the lattice version of \eqref{aisodi1212AA} 
is given by
\be\label{iasodi1212}
\hat{{\mathsf O}}^{\text{\tiny\ding{105}}}=\hat{{\cal D}}\,
\hat{{\mathsf O}}^{\star}\,
\hat{{\cal D}}
\ee
for an arbitrary operator $\hat{{\mathsf O}}$  acting in the 
finite dimensional space $\mathscr{V}_N$.
Since the lattice translation operator and  Hamiltonian both commute with
$\hat{{\cal D}}$, the relations \eqref{congu7a878sa} carry over to
\be
\mathbb{H}^{\text{\tiny\ding{105}}}=\mathbb{H}\ ,\qquad\qquad
 \mathbb{K}^{\text{\tiny\ding{105}}}=\mathbb{K}^{-1}\ .
\ee
Combining formulae \eqref{iasodi1212} with \eqref{oaisod91289aaaaa},\,\eqref{aoisdo8912}
and using that $\hat{\cal D}\,\mathbb{A}^{({\infty})}_+\,\hat{\cal D}=\re^{\ri\pi(N/2-\mathbb{S}^z)}\,
\mathbb{A}^{({\infty})}_+$
as well as $\hat{\cal D}^2=1$
one finds
\be\label{oaisod91289aaa}
\hat{{\mathsf O}}^{\text{\tiny\ding{105}}}
=\hat{{\mathsf X}}_{\text{\tiny\ding{105}}}^{-1}\,\hat{{\mathsf O}}^\dag\, 
\hat{{\mathsf X}}_{\text{\tiny\ding{105}}}^{}\ ,\qquad
\quad \hat{{\mathsf O}}\in{\rm End}\big(\mathscr{V}_N\big)
\ee
with
\be
\hat{{\mathsf X}}_{\text{\tiny\ding{105}}}^{}=\hat{{\mathsf X}}_{\text{\tiny\ding{105}}}^{\dag}=
\Sigma^z\,\re^{\frac{\ri\pi}{2}(\mathbb{S}^z-N/2)}\,
\mathbb{A}_+^{({\infty})}\ .
\ee
\bigskip

Let $(\cdot,\cdot)_{\text{\tiny\ding{105}}}$ be the sesquilinear form 
in the finite dimensional space $\mathscr{V}_N$ that is consistent with the conjugation 
\eqref{oaisod91289aaa} and such that its value on the pseudovacuum, 
$|\uparrow\rangle\otimes |\uparrow\rangle\otimes \ldots\otimes |\uparrow\rangle$, is one.
It is easy to see that the form in the basis of Bethe states is described via
the relations
\be\label{ortho1AA}
\big({\boldsymbol \Psi}^{(2)},{\boldsymbol \Psi}^{(1)} \big)_{\text{\tiny\ding{105}}}=0\ \ \ \ \ {\rm unless} \ \ \ \ \ 
\ {\boldsymbol \Psi}^{(2)}={ \hat {\cal D}}{ \hat {\cal C}}{\hat  {\cal P}}{ \hat {\cal T}}\,{\boldsymbol \Psi}^{(1)}\ ,
\ee
and
\be\label{aiosdiuasd0098}
\big({\cal \hat{D}}{\cal\hat{ C}}{\cal \hat{P}}{\cal \hat{T}}{\boldsymbol \Psi},
{\boldsymbol \Psi}\big)_{\text{\tiny\ding{105}}}=
\big({\cal\hat{ C}}{\cal \hat{P}}{\cal \hat{T}}{\boldsymbol \Psi},{\boldsymbol \Psi}\big)_{\star}\ .
\ee
The r.h.s. in the last equation is given by  \eqref{FinalNorm}.

\bigskip

The result of our numerical investigation of 
$\big({\cal\hat{ C}}{\cal \hat{P}}{\cal \hat{T}}{\boldsymbol \Psi}_N,{\boldsymbol \Psi}_N\big)_{\star}$
for the low energy Bethe states is 
summarized by the formulae  \eqref{aoisd89213hsd}-\eqref{isisaisa}. However
a literal attempt to apply them for the RG trajectories characterized by
pure imaginary $s$ meets an immediate problem. 
In view of the condition \eqref{pdpd000d}
the function $g_N$ \eqref{aoiisiasa}   develops a 
simple pole whenever $s$ belongs to the 
admissible set of pure imaginary values. Nevertheless we checked
that \eqref{isisaisa} remains valid for finite $N\gg 1$ provided
$s$ is substituted by the ``running coupling'' $b(N)$. 
Then combining the relation with formula  \eqref{quantC1}
that describes the large $N$ asymptotics of $b(N)$, one 
can obtain the large $N$ asymptotic behaviour of 
\eqref{aiosdiuasd0098}
for a   Bethe state that becomes part of the discrete
spectrum in the scaling limit. The result may be formulated in the following way.
\bigskip

 Let $\bm{\Psi}_N$ be the RG trajectory, which in the scaling limit
becomes the state $\bm{\psi}_{\bar{\rho},\rho,\bar{\nu},\nu}(\bar{\bm{w}},\bm{w})$ 
from the space ${\cal H}^{({\rm disc})}$.
To be precise, suppose 
that state belongs to the level subspace
${\cal V}_{\bar{\rho},\bar{\nu}}^{(\bar{\ell} )}\otimes{\cal V}_{{\rho},{\nu}}^{({\ell} )}$
of a highest weight irrep which occurs in the $\overline{W}_\infty\otimes W_\infty$  decomposition of 
${\cal H}^{({\rm disc},\pm)}$ described in sec.\,\ref{sec171}. The
levels of the state in the irrep, $\bar{\ell}$ and $\ell$, 
could be any one of ${\tt L}$, ${\tt L}_\pm$ and $\bar{{\tt L}}$, $\bar{{\tt L}}_\pm$,
respectively, depending on the situation at hand, see eq.\,\eqref{iaoisd8182}.
Then a straightforward calculation shows that as $N\to\infty$,
\bea\label{aisoid91023oiwasjknmd}
\big({\cal\hat{D}}{\cal\hat{ C}\hat{P}\hat{T}}
\bm{\Psi}_N,\bm{\Psi}_N\big)_{\text{\tiny\ding{105}}}&\asymp&
 {\cal  M}^{(\bar{\ell})}_{N}({\bar \rho},{\bar \nu}) \  {\cal  M}^{({\ell})}_{N}(\rho,\nu)
\ 
\big(1+o(1)\big)
\\[0.3cm]
&\times &
\ \sigma\,f_{\bar{\rho},\bar{\nu}}^{(-)}\ f_{\rho,\nu}^{(-)}\
 \big(\,\check{C}^{(\bar{A})}_{\bar{\rho},\bar{\nu}}(\bar{\bm{w}})\,\big)^2
\ \big(\,\check{C}^{(A)}_{\rho,\nu}({\bm{w}})\,\big)^2
\  \frac{{U}_{{\bar \rho},\bar{\nu}}(\bar{ \bm{w}})}
 {\check{R}_{{\bar \rho},\bar{\nu}}(\bar{\bm{w}})}\ 
 \frac{{U}_{\rho ,\nu}(\bm{w})}{\check{R}_{\rho ,\nu}(\bm{w})}\ ,\nonumber
\eea
where 
\be
A=\sgn(\ri\nu)\,,\qquad\qquad \bar{A}=\sgn(\ri\bar{\nu})\ .
\ee
All of the dependence on $N$ is contained 
in the sign factor $\sigma=(-1)^{N/2-S^z}$,
as well as the first line of this equation and
\bea\label{Maisd7812}
{\cal  M}^{(\ell)}_{N}(\rho,\nu)&=& {C}^{(\rm alt)}_0\ \bigg(\frac{N}{2}\bigg)^{\frac{1}{6}}\ 
\bigg(\frac{\sqrt{2}C}{N}\bigg)^{\frac{4\nu^2}{n}+f(\rho)+4\ell}\ 
\bigg(\frac{2^{\frac{2}{n}}N}{N_0}\bigg)^{\frac{2|\nu|}{n}(n-2|\nu|)}\
 \re^{\frac{1}{4}{\cal A}_2 N^2} \ .
\eea
The functions $\check{C}^{(\pm)}_{\rho,\nu}({\bm{w}})$ and 
$\check{R}_{\rho ,\nu}(\bm{w})$ are the eigenvalues of the operator
$\check{{\bf C}}^{(\pm)}$ \eqref{aosido1201} and the reflection operator $\check{\bf R}$ \eqref{DRope0823},
respectively.
Since
the latter are operators acting in the Fock spaces, ${\cal V}_{{\rho},{\nu}}$  
 should be understood as a subspace of the Fock space
according to eqs.\,\eqref{ioasido1212} and \eqref{ioasido1212a}.
Finally   $f_{\rho,\nu}^{(-)}$
 in \eqref{aisoid91023oiwasjknmd} 
does not depend on
the chiral state in the irrep ${\cal V}_{{\rho},{\nu}}$ 
 and reads explicitly as
\be\label{aisod09123}
f_{\rho,\nu}^{(-)}=
\frac{\Gamma\big(\tfrac{1}{2}+\rho-|{\nu}|\big)}{2\pi\,
(n+2)^{2\nu^2/n}}\ \times \ \begin{cases}
(-1)^{a}\,a!\,
 &\ \ \ \  {\rm if} \qquad \frac{1}{2}+{\rho}+|{\nu}|=-a=0,-1,-2,\ldots\\[0.3cm]
\dfrac{2\pi}{\Gamma(\frac{1}{2}+{\rho}+|{\nu}|)} &\ \ \ \   {\rm otherwise}
\end{cases}
\ee
The same holds true for the barred counterparts. 
\bigskip

The factor
$\big(\,\check{C}^{(\bar{A})}_{\bar{\rho},\bar{\nu}}(\bar{\bm{w}})\,\big)^2\ 
\big(\,\check{C}^{(A)}_{\rho,\nu}({\bm{w}})\,\big)^2$ 
prevents one from interpreting the second line of \eqref{aisoid91023oiwasjknmd} as the inner product 
$\langle\cdot,\cdot\rangle_{\rm disc}$, consistent with the conjugation conditions \eqref{aisodi1212AA}
in the $\overline{W}_\infty\otimes W_\infty$\,-\,algebra,  evaluated on
 the eigenstate 
$\bm{\psi}_{\bar{\rho},\rho,\bar{\nu},\nu}(\bar{\bm{w}},\bm{w})$ and its ${\cal {D}{C}{P}{T}}$ conjugate.
However, 
the eigenvalues of the operators $\check{\bf{C}}^{(\pm)}$ satisfy the relations 
\bea
\big(\check{C}^{(-)}_{{\rho},{\nu}}({\bm{w}})\big)^*=\check{C}^{(+)}_{{\rho},{\nu^*}}(-\bm{w}^*)\,,
\qquad\qquad
\check{C}^{(+)}_{{\rho},{\nu}}(\bm{w})=\check{C}^{(-)}_{{\rho},{-\nu}}(-\bm{w})\ ,
\eea
which follow from \eqref{DCPTeq1a} and \eqref{Apeq1}. 
Hence
$
\big(\check{C}^{(\pm)}_{{\rho},{\nu}}({\bm{w}})\big)^*=\check{C}^{(\pm)}_{{\rho},{\nu}}({\bm{w}}^*)
$ for pure imaginary $\nu$.
Introduce  the inner product in ${\cal H}^{({\rm disc})}$ using the basis 
$\bm{\psi}_{\bar{\rho},\rho,\bar{\nu},\nu}(\bar{\bm{w}},\bm{w})\equiv
\bar{\bm{\psi}}_{\bar{\rho},\bar{\nu}}(\bar{\bm{w}})\otimes {\bm{\psi}}_{{\rho},\nu}(\bm{w})$
with the chiral eigenstates  normalized as in \eqref{oio888s},
via the formula
\bea\label{otho99raaa}
\big\langle\bm{\psi}_{\bar{\rho}',\rho',\bar{\nu}',\nu'}(\bar{\bm{w}}',\bm{w}'),\,
\bm{\psi}_{\bar{\rho},\rho,\bar{\nu},\nu}(\bar{\bm{w}},\bm{w})\big\rangle_{\rm disc}&=&
\sigma\,f_{\bar{\rho},\bar{\nu}}^{(-)}\ f_{\rho,\nu}^{(-)}\ \
\frac{U_{{\bar \rho},\bar{\nu}}(\bar{{\boldsymbol w}})}
{\check{R}_{{\bar \rho},\bar{\nu}}(\bar{{\boldsymbol w}})}\ 
\frac{U_{\rho,\nu}({\boldsymbol w})}{\check{R}_{{ \rho},{\nu}}({{\boldsymbol w}})}
\\[0.3cm]
&\times&
\delta_{\bar{\bm{w}}', \bar{\bm{w}}^*}\, 
\delta_{\bm{w}', \bm{w}^*}\, 
\delta_{\bar{\rho}',\bar{\rho}}\,
\delta_{\rho',\rho}\, 
\delta_{\bar{\nu}',\bar{\nu}}\,
\delta_{\nu',\nu}\ .\nonumber
\eea 
 Then it is easy to see that the second line in 
\eqref{aisoid91023oiwasjknmd} coincides with 
$\big\langle
{\cal \hat{D}\hat{C}\hat{P}\hat{T}}\bm{\psi}_{\bar{\rho},\rho,\bar{\nu},\nu}',\,
\bm{\psi}_{\bar{\rho},\rho,\bar{\nu},\nu}'\big\rangle_{\rm disc}$
for the state
\be\label{aiosd8912}
\bm{\psi}_{\bar{\rho},\rho,\bar{\nu},\nu}'(\bar{\bm{w}},\bm{w})=
\big(\check{\bar{{\bf C}}}^{(\bar{A})}\otimes\check{{\bf C}}^{(A)}\big)
\ \bm{\psi}_{\bar{\rho},\rho,\bar{\nu},\nu}(\bar{\bm{w}},\bm{w})\,\qquad
\qquad \big(A=\sgn(\ri\nu)\,,\ \bar{A}=\sgn(\ri\bar{\nu})\big)\ .
\ee
This way we   conclude that there exists the limit
\bea\label{scaling1c}
{\rm s}\!\!\!\lim_{N\to \infty}\ 
\Big({\cal  M}^{(\bar{\ell})}_{N}(\bar{\rho},\bar{\nu})\,{\cal  M}^{(\ell)}_{N}(\rho,\nu)\Big)^{-\frac{1}{2}}\
\bm{\Psi}_N
&=&
\ \bm{\psi}_{\bar{\rho},\rho,\bar{\nu},\nu}'(\bar{\bm{w}},\bm{w})\ .
\eea
As was already discussed,
the inner product described in the eigenbasis by formula
\eqref{otho99raaa} 
may be equivalently introduced
 through the conjugation conditions \eqref{aisodi1212AA} supplemented by its value on
the primary $\overline{W}_\infty\otimes W_\infty$ states:
\bea\label{otho99raaa77}
\big\langle\bm{\psi}_{\bar{\rho}',\rho',\bar{\nu}',\nu'}^{({\rm vac})},\,
\bm{\psi}_{\bar{\rho},\rho,\bar{\nu},\nu}^{({\rm vac})}\big\rangle_{\rm disc}&=&
\sigma\, f_{\bar{\rho},\bar{\nu}}^{(-)}\ f_{\rho,\nu}^{(-)}\ 
Z_+\big(\tfrac{\bar{\rho}}{\sqrt{n+2}}\,\big|\, \sqrt{n+2}\,\big)\,Z_+\big(\tfrac{\rho}{\sqrt{n+2}}\,\big|\, \sqrt{n+2}\,\big)
\nonumber\\[0.4cm]
&\times&
\delta_{\bar{\rho}',\bar{\rho}}\,
\delta_{\rho',\rho}\, 
\delta_{\bar{\nu}',\bar{\nu}}\,
\delta_{\nu',\nu}\ .
\eea
Here $f_{\rho,\nu}^{(-)}$ is given in \eqref{aisod09123},
while $Z_+(P|\beta)$ was defined in \eqref{asusausss}.
As for the sign factor $\sigma=(-1)^{N/2-S^z}$ it depends on whether, in constructing
the RG trajectories, we keep $N/2-S^z$ to be an even or an odd number.

\bigskip

Let's highlight an important point to take away from our investigation. 
We found that the scaling limit should be defined differently for the
low energy states, which become part of the spaces 
${\cal H}^{({\rm cont})}$ and ${\cal H}^{({\rm disc})}$.
These sectors  
are naturally equipped by different inner products, which are induced
by different conjugation conditions for the operators in the 
${\cal Z}_2$ invariant inhomogeneous six-vertex model.
All this suggests that if a description of the critical behaviour 
of the lattice system within the framework of a local CFT exists,
the
states from  ${\cal H}^{({\rm cont})}$ and
${\cal H}^{({\rm disc})}$
 can not be interpreted simultaneously as  normalizable states 
within a single field theory.

\pagebreak

\part{Towards the QFT \label{sec4}}

\section{Integrable and Hermitian  structures  for $c\to 2^-$\label{sec5}}
The integrable structure which occurred
 in  our study
 of the scaling limit of the  inhomogeneous six\,-\,vertex model with  ${\cal Z}_2$ symmetry, has a deep relation
to the AKNS classical integrable hierarchy. 
The latter includes
some famous  classically integrable partial differential equations
such as the non-linear Schr\"{o}dinger and the Lund-Regge 
(complex sin(h)-Gordon I) equation.  
To explain this relation, one should consider the $n\to+\infty$ limit, which can be 
understood as  a classical limit with
 \bea\label{hassaysa}
\hbar=\frac{2\pi}{n}
\eea
playing the r$\hat{{\rm o}}$le of the Planck constant.
\bigskip

Let us rescale the field $\varphi$ \eqref{bosefiled1} and introduce 
$\phi(u)=\frac{1}{\sqrt n}\ \varphi(u)$ as well as the similarly defined field
\bea\label{oiso9090}
\theta(u)=\frac{1}{\sqrt n}\ \bigg(\vartheta_0+b_0\, u+\ri \sum_{m\not=0}\frac{b_{m}}{m}\ \re^{-\ri m u}\bigg)\ \ \ \ \ 
\eea
with $[\,\vartheta_0,b_m\,]=\tfrac{\ri}{2}\ \delta_{m,0}$.
A simple calculation shows that
\bea
[\,\phi(u_1),\phi(u_2)\,]=[\,\theta(u_1),\theta(u_2)\,]=-\ri\hbar \ \tfrac{1}{4 }\ \epsilon(u_1-u_2)\ ,\ \ \ \qquad\  \ [\,\phi(u_1),\theta(u_2)\,]=0\ ,
\eea
where $\epsilon(u)=2m+1$ for $2\pi m<u<2\pi(m+1)\ \ (m\in \mathbb{Z})$.
Applying  the correspondence principle,
$\ri \hbar^{-1}[\cdot, \cdot]\mapsto \{\cdot,\cdot\}$,
one  concludes that $\phi$ and $\theta$  become classical fields in the large $n$ limit
 subject to the Poisson Bracket (PB) relations
\bea\label{isaisai11112}
\{\phi(u_1),\phi(u_2)\}=\{\theta(u_1),\theta(u_2)\}=\tfrac{1}{4}\ \epsilon(u_1-u_2)\ ,\ \ \ \  \qquad\{\phi(u_1),\theta(u_2)\}=0\ .
\eea
For the $W_j$ currents, the bosonization formulae \eqref{w2iosdi},\,\eqref{W3def1a}
 imply that as $n\to \infty$ they
become classical fields built form $\partial\phi$ and $\partial\theta$:
\bea\label{90s9889dfsa}
W_j\to n^{j/2}\ W_j^{(cl)}\,,
\eea
where explicitly
 \bea\label{Weqaa1}
&&W^{(cl)}_2=(\partial \phi)^2+(\partial \theta)^2
\\[0.2cm]
&&W^{(cl)}_3=2\, (\partial \theta)^3+2\,
 (\partial \phi)^2\partial \theta+\ri\, \big( \partial^2 \phi\,\partial\theta
- \partial\phi\,\partial^2 \theta\big) \ .
 \nonumber
\eea
Recall that all the $W$ currents can be generated from the parafermion
fields. In turn, the fields $W_j^{(cl)}$ are conveniently
expressed in terms of the classical counterparts of \eqref{isisaiasi}:\footnote{%
We will use the same symbol for the quantum and classical fields $ \xi_\pm$,
similar as with $\phi$ and $\theta$.}
\bea\label{aisisa}
 \xi_\pm=  (\partial\theta\pm \ri \partial\phi)\,\re^{\pm 2\theta}\ \ \ \ \ \ \  \ (n\to \infty)\ .
\eea
In particular,
\bea\label{isisai112aaa}
W^{(cl)}_2&=&\xi_+\,\xi_-\ ,\ \ \ \ \ W^{(cl)}_3=\tfrac{1}{2}\ \big(\xi_-\,\partial \xi_+-
\xi_+\,\partial \xi_-\big)\\[0.2cm]
W^{(cl)}_4&=&\tfrac{2}{5}\ \big(\xi_+\,\partial^2\xi_-+
\xi_-\,\partial^2 \xi_+\big)-\tfrac{6}{5}\ \partial\xi_+\partial\xi_-\nonumber \ .
\eea
Using  eqs.\,\eqref{aisisa} and \eqref{isaisai11112},
it is straightforward to compute the PBs involving $\xi_+$, $\xi_-$ and show that 
\bea\label{jsasusa}
\big\{\xi_\pm (u_1),\xi_\pm (u_2)\big\}&=&\epsilon(u_1-u_2)\ \xi_\pm (u_1)\,\xi_\pm (u_2)\\[0.2cm]
\big\{\xi_\pm(u_1),\xi_\mp (u_2)\big\}&=&-\delta'(u_1-u_2)-\epsilon(u_1-u_2)\ \xi_\pm (u_1)\,\xi_\mp (u_2)\ .\nonumber
\eea
The above relations  combined with 
the formulae expressing $W_j^{(cl)}$ in terms of  $\xi_\pm$ such as
\eqref{isisai112aaa}
are sufficient for deriving the Poisson algebra for
 the classical $W$ currents.
 They provide a short cut to this algebra automatically
 satisfying the Jacobi and skew symmetry conditions, that would otherwise need to be
obtained from the $c=2-\frac{6}{n+2}\to 2^-$ limit of the OPEs such as \eqref{aiisaisa1a2a}.
 In particular, it is straightforward to show that
\bea\label{jasususa}
&&\big\{W^{(cl)}_2(u_1),W^{(cl)}_2(u_2)\big\}=-\big(W^{(cl)}_2(u_1)+W^{(cl)}_2(u_2)\big)\ \delta'(u_1-u_2)\nonumber\\[0.2cm]
&&\big\{W^{(cl)}_3(u_1),W^{(cl)}_2(u_2)\big\}= -3\ W^{(cl)}_3(u_1)\ \delta'(u_1-u_2)-\partial W^{(cl)}_3(u_1)\ \delta(u_1-u_2)\\[0.2cm]
&&\big\{W^{(cl)}_3(u_1),W^{(cl)}_3(u_2)\big\}=-
\tfrac{1}{4}\,\big(W^{(cl)}_2(u_1)+W^{(cl)}_2(u_2)\big)\ \delta'''(u_1-u_2)
-\delta'(u_1-u_2)\times  \nonumber\\[0.3cm]
&& \Big(W^{(cl)}_4(u_1)+W^{(cl)}_4(u_2)+
2\,W^{(cl)}_2(u_1)\, W^{(cl)}_2(u_2)-\tfrac{3}{20}\,\big(\,\partial^2\, W^{(cl)}_2(u_1)+\partial^2\,
W^{(cl)}_2(u_2)\,\big)\Big)\ .\nonumber
\eea
Taking into account that $u=t+x$ and
$W^{(cl)}_j(t,x)=W^{(cl)}_j(t+x)$, the latter may be understood as an infinite  system of equal-time PB  relations
for the classical $W$ currents.

\bigskip

One should keep in mind that
$\xi_\pm(u)$  are quasiperiodic fields contrary to the $W_j^{(cl)}(u)$, which are periodic:
\bea\label{90s9d0909f}
W_j^{(cl)}(u+2\pi)=W_j^{(cl)}(u)\,,\qquad \xi_\pm (u+2\pi)=B^{\pm 1}\ \xi_\pm (u)\  .
\eea
Using the ```bosonization'' formulae \eqref{aisisa} one finds
\bea\label{iaosd899382}
\big\{ B , \xi_\pm (u)\big\}=\pm 2B\,\xi_\pm (u)\  ,
\eea
while
\bea
\big\{B,W_j^{(cl)}(u)\big\}=0\ .
\eea
From the definition of the quantum field
$\theta$ \eqref{oiso9090} and eq.\,\eqref{aisisa}
it is easy to see that  
$B$ is the classical counterpart of $\re^{\frac{4\pi}{\sqrt{n}}\,b_0}$, whose 
eigenvalues coincide with $\re^{\frac{4\pi s}{n}}$ \eqref{oasidoioais311}.
The latter is equal to the eigenvalues of the  quasi-shift operator  in the scaling limit
(see eq.\,\eqref{poapso1a}). 
For this reason, with  some abuse of notation, we use the same symbol $B$ for the dynamical variable
defined through \eqref{90s9d0909f} as the one denoting the eigenvalues of ${\mathbb B}$.

\bigskip
Let's turn to the classical limit of the local IM \eqref{iaosid198289},\,\eqref{iaosid198289A}.
As it follows from \eqref{90s9889dfsa},
\bea\label{oios09090as}
{\bf I}_m\to  n^{(m+1)/2}\,  I^{(cl)}_m\,,
\eea
where the explicit formula for the first few $I^{(cl)}_m$, expressed 
in terms of $\xi_\pm$, may be obtained from \eqref{isisai112aaa}
\bea\label{oisodio909}
I^{(cl)}_1&=&\int_0^{2\pi}\frac{\rd u}{2\pi}\ \xi_+\xi_-\nonumber \\[0.2cm]
I^{(cl)}_2&=&\frac{1}{4}\ \int_0^{2\pi}\frac{\rd u}{2\pi}\ \big(\xi_-\,\partial \xi_+-
\xi_+\,\partial\xi_-\big) \\[0.2cm]
I^{(cl)}_3&=& \frac{1}{5}\int_0^{2\pi}\frac{\rd u}{2\pi}\ \big( \, (\xi_+\xi_-)^2
-\partial\xi_+\partial\xi_-
\big)\ . \nonumber
\eea
In general  $I^{(cl)}_m$ are given by an integral over a real local density built from
$\xi_\pm$ and their derivatives.
In this work, we always assumed that the  quantum IM
were normalized as
${\bf I}_{m}=n^{\frac{m+1}{2}}\ \int_0^{2\pi}
\frac{{\rm d} u}{2\pi}\, \big((\partial\theta)^{m+1}+\ldots\big)
$.
One can show that this translates to
\bea\label{oisodio909AAB}
I^{(cl)}_m&=&\frac{m\ \Gamma^2(\frac{m}{2})}{2\sqrt{\pi}\ \Gamma(\frac{1}{2}+m)}
\ \int_0^{2\pi}\frac{\rd u}{2\pi}\ \Big(\big(\xi_+\xi_-\big)^{\frac{m+1}{2}}+\ldots\,\Big)
 \qquad\qquad (m-{\rm odd}) \\[0.2cm]
I^{(cl)}_m&=&\frac{(m+1)\,\Gamma^2(\frac{1}{2}+\frac{m}{2})}{4\sqrt{\pi}\ \Gamma(\frac{1}{2}+m)}
\ \int_0^{2\pi}\frac{\rd u}{2\pi}\ \Big(
\big(\xi_+\xi_-\big)^{\frac{m}{2}-1}\ \big(\,\xi_-\partial \xi_+-
\xi_+\,\partial \xi_-\,\big)+\ldots\Big)  \qquad (m-{\rm even})\ .\nonumber
\eea
Here the ``$\ldots$'' stands for 
the monomials
which are of lower power in $\xi_\pm$ and their derivatives. 
Of course, all the $I^{(cl)}_m$ mutually Poisson commute with each other.
This set coincides with the commuting family of local IM for the 
AKNS integrable hierarchy.
\bigskip

The key ingredient in the theory of classically integrable
partial differential equations is the zero curvature
or Lax representation.
In the case under consideration
the auxiliary linear problem is given by
\be
\big(\partial-{\bm A}(u\,|\,\lambda_c)\big)\bm{\Phi}=0
\ee
with
\be\label{osaoi90000a}
{\bm A}(u\,|\,\lambda_c)=\xi_-\,{\tt e}_--\xi_+\,{\tt e}_++\lambda_c\,{\tt h}\, .
\ee
Here ${\tt e}_\pm$ and ${\tt h}$ are the generators of the $\mathfrak{sl}_2$ algebra, 
$[{\tt h},\,{\tt e}_\pm]=\pm 2{\tt e}_\pm$ and
$[{\tt e}_+,\,{\tt e}_-]={\tt h}$, while
$\lambda_c$ is the auxiliary spectral parameter.
We define the classical transfer matrix as the trace
\be\label{iaois898}
\tau^{(cl)}(\lambda_c)={\rm Tr}_{\text{\textonehalf}}\Big[\, B^{-\frac{{\tt h}}{2}}
\ \overset{\leftarrow}{{\cal P}}\exp\Big(\int_{0}^{2\pi}\,\rd u\,{\bm A}(u\,|\,\lambda_c)\Big)\Big]
\ee
taken over the fundamental representation, which is indicated by the subscript $\frac{1}{2}$.
The factor $B^{-\frac{{\tt h}}{2}}$ is inserted to take into account that
\be
\bm{A}(u+2\pi\,|\,\lambda_c)=B^{\frac{{\tt h}}{2}}\,\bm{A}(u\,|\,\lambda_c)\,B^{-\frac{{\tt h}}{2}}\ ,
\ee
which is a consequence of the quasiperiodicity condition \eqref{90s9d0909f}.
The classical transfer matrix \eqref{iaois898} appears in  the $n\to\infty$ limit
of  $\bm{\tau}(\lambda)$ \eqref{asusuya}.
The precise relation may be motivated via a comparison of 
their large $\lambda$ asymptotic expansions.
Representing $\tau^{(cl)}(\lambda_c)$ in the form
\bea\label{as0a0123}
\tau^{(cl)}(\lambda_c)=2\cos\big(\nu(\lambda_c)\big)\,,
\eea
one can show  (see, e.g., \cite{Fateev:2005kx,Faddeev:1987ph})
\bea
\nu(\lambda_c)\asymp -2\pi\ri \lambda_c+\tfrac{\ri}{2}\, \log(B)+2\pi\ri \sum_{m=1}^\infty 
\frac{2^m\,\Gamma(\frac{1}{2}+m)}{\sqrt{\pi}\, (m+1)!}\ I_m^{(cl)}\,\lambda^{-m}_c\ \ \
\ \ \ (\lambda_c\to\infty)\ ,
\eea
where $I_m^{({\rm cl})}$ are the classical local IM \eqref{oisodio909AAB}.
The similar expansion for the quantum transfer matrix \eqref{9sd98f989sd1a} involves the quantum IM.
In view of the  relation between ${\bf I}_m$ and $I_m^{({ cl})}$ \eqref{oios09090as},
this suggests
\be\label{90s9d8f9dsf}
\bm{\tau}(\lambda)\to \tau^{(cl)}(\lambda_c)\ \ \ \ \ \ {\rm as} \qquad n\to \infty\quad \ \ {\rm with}\ \ 
\quad \lambda_c=(n+2)\,\lambda\quad {\rm fixed}\ .
\ee
\bigskip

It is possible to justify the relation
 \eqref{90s9d8f9dsf} by explicitly  calculating the classical limit
of $\bm{\tau}(\lambda)$  order  by order in $\lambda$
following the lines of the work \cite{Bazhanov:2018xzh}.
This requires a study of
 ${\boldsymbol L}_{\text{\textonehalf}}(\lambda)\equiv \pi_{\text{\textonehalf}}\big({\boldsymbol L}(\lambda)\big)$,
entering into the definition of the transfer matrix \eqref{asusuya},
in the large $n$ limit.
 Eq.\eqref{Laltcase1a}
 gives ${\boldsymbol L}(\lambda)$ 
as a path ordered exponent, i.e.,
a series expansion in $\sqrt{\lambda}$  whose coefficients are ordered integrals over the
vertex operators. However, as was already mentioned, the ordered integrals diverge for 
any $n>0$ and hence \eqref{Laltcase1a} is not literally applicable for taking the classical limit.
 Instead, each coefficient of the formal series
 ${\boldsymbol L}(\lambda)$  should be understood
via analytic continuation in complex $n$, which may be achieved by re-writing the ordered 
integrals  in terms of the contour integrals. 
In ref.\cite{Bazhanov:2018xzh} it was explained how to take the 
 $n\to\infty$ limit of expressions involving the contour integrals.
This results in 
the classical version of ${\boldsymbol L}(\lambda)$
 as a series expansion in $\sqrt{\lambda}$ whose coefficients involve
 multifold integrals over the classical fields $\re^{\pm2\ri\phi}$ and 
$\partial\theta\re^{-2\ri\phi}$.
To compare the Taylor series for the classical limit of $\bm{\tau}(\lambda)$ obtained in this way
 with the r.h.s. of \eqref{90s9d8f9dsf},
one should apply a gauge transformation to the connection \eqref{osaoi90000a},
\be
\bm{A}\mapsto \bm{G}^{-1}\bm{A}\,\bm{G}-\bm{G}^{-1}\partial\bm{G}\,,
\ee
and rewrite $\tau^{(cl)}(\lambda_c)$ in a way that is
suitable for a small $\lambda_c$ expansion. 
Using the matrix
\be
\bm{G}(u)=\re^{{\theta(u)}\,{\tt h}}\,\re^{\frac{\pi}{4}({\tt e}_+-{\tt e}_-)}\,\re^{\ri\phi(u)\,{\tt h}}\ ,
\ee
a simple calculation shows that
\be\label{90a9s0d9as}
B^{-\frac{{\tt h}}{2}}\ 
 \overset{\leftarrow}{{\cal P}}\exp\bigg(\int_{0}^{2\pi}\,\rd u\,{\bm A}(u\,|\,\lambda_c)\bigg)=\bm{G}(0)
\ \lambda_c^{-\frac{{\tt h}}{4}}\ \big(\,  \re^{\ri \pi P {\tt h}}\ {\boldsymbol L}^{(cl)}(\lambda_c)\,\big)\
\lambda_c^{+\frac{{\tt h}}{4}}\ \bm{G}^{-1}(0)
\ee
with 
\be\label{Mlambdac1a}
{\boldsymbol L}^{(cl)}(\lambda_c)=
\lambda_c^{+\frac{{\tt h}}{4}}
\ \re^{\ri \pi P {\tt h}}\  \overset{\leftarrow}{{\cal P}}\exp\bigg(\int_0^{2\pi}\rd u\ 
\Big(\!-2\partial\theta\, \re^{-2\ri\phi}\, {\tt e}_++ \lambda_c\, 
 \big(\re^{+2\ri\phi} {\tt e}_- 
 +\re^{-2\ri\phi} {\tt e}_+\, \big)\Big)\bigg)
\ \lambda_c^{-\frac{{\tt h}}{4}}
\ee
and $P$ stands for the zero-mode momentum of the field $\phi$:
\bea
P=\int_0^{2\pi}\frac{\rd u}{2\pi}\ \partial\phi\ .
\eea
This way one obtains
\bea\label{asusuyddda}
{\tau}^{(cl)}(\lambda_c)={\rm Tr}_{\text{\textonehalf}}\Big[\re^{\ri\pi P {\tt h}} \ 
{\boldsymbol L}^{(cl)}(\lambda_c)\Big]\ .
\eea
Notice that
formally setting $n\to\infty$ into the path-ordered exponential \eqref{Laltcase1a}
would reproduce the r.h.s. of \eqref{Mlambdac1a} without 
the last term in the exponent, $\lambda_c\,\re^{-2\ri\phi}{\tt e}_+$.
However taking the classical limit of 
${\boldsymbol L}(\lambda)$ 
as outlined above, 
with the ordered integrals 
being analytically regularized, we have checked that the first few terms in the Taylor series expansion of
the classical limit of $\bm{\tau}(\lambda)$ reproduce the corresponding terms
for $\tau^{(cl)}(\lambda_c)$
extracted from eqs.\,\eqref{asusuyddda} and \eqref{Mlambdac1a}.
\bigskip

The monodromy matrix in the gauge \eqref{Mlambdac1a}
possesses a remarkable property.
Namely, it obeys the Sklyanin exchange relations \cite{Sklyanin:1979gh,Faddeev:1987ph}
\be\label{PBrel1}
\big\{{\boldsymbol L}^{(cl)}(\lambda_c) \begin{array}{ccc} \\[-0.4cm] \otimes \\[-0.35cm] , 
\end{array}{\boldsymbol L}^{(cl)}(\lambda_c')\,\big\}=
\Big[\,{\boldsymbol L}^{(cl)}(\lambda_c)\,{\otimes}\, {\boldsymbol L}^{(cl)}(\lambda_c'),
\,{\boldsymbol r}\big(\sqrt{\lambda_c/\lambda_c'}\ \big)\,\Big]
\ee
with the classical $R$-matrix
\be\label{CRmatrix111}
{\boldsymbol r}(\rho) = \frac{1}{\rho-\rho^{-1}}\,
\big(\,{\tt e}_+\otimes {\tt e}_-+{\tt e}_-\otimes {\tt e}_++
                                                 \tfrac{1}{4}(\rho+\rho^{-1})\,{\tt h}\otimes {\tt h}\,\big)\ .
\ee
 By expanding ${\boldsymbol L}^{(cl)}(\lambda_c)$
and ${\boldsymbol L}^{(cl)}(\lambda_c')$
as a series in $\sqrt{\lambda_c}$ and $\sqrt{\lambda_c'}$, respectively, 
and also $\bm{r}\big(\sqrt{\lambda_c/\lambda_c'}\big)$ 
say in the domain $|\lambda_c|<|\lambda_c'|$,
eq.\,\eqref{PBrel1} can be checked order by order in these two variables.
Note that \eqref{PBrel1} does not assume 
any choice of representation for the $\mathfrak{sl}_2$
generators. Specialized  to the 
finite dimensional representation $\pi_j\otimes\pi_{j'}$, 
it becomes the classical counterpart
of the Yang-Baxter algebra \eqref{YBeq1a} with
$\bm{L}^{(cl)}(\lambda_c)$  being the classical version of the operator 
\eqref{Laltcase1a}, i.e.,
\be
\bm{L}(\lambda)\to \bm{L}^{(cl)}(\lambda_c)
\ \ \ \ \ \ ( n\to \infty\quad {\rm with}
\quad \lambda_c=(n+2)\,\lambda\quad {\rm fixed})\ .
\ee

\bigskip

In the above discussion of the classical limit,
$\xi_\pm$  have been treated as 
unrelated complex fields. There are two natural reality constraints which can
be imposed on them that are consistent with
 the Poisson algebra  \eqref{jsasusa}. Namely, 
 \bea\label{hasysa1AA}
{\rm (I)}\ &:&\ \ \ \  \big(\xi_\pm(u)\big)^*=\xi_\pm (u^*)\ ,\ \ \ \  B^*=B\\[0.2cm]
{\rm (II)}\ &:&\ \ \ \  \big(\xi_\pm(u)\big)^*=\xi_\mp(u^*)\ ,\ \ \ \  B^*=B^{-1}\ .\nonumber
\eea
These imply the following reality conditions for the classical $W$-currents
 \bea\label{hasysads}
{\rm (I)}\ &:&\ \ \ \  \big(W_j^{(cl)}(u)\big)^*=W_j^{(cl)}(u^*)\\[0.2cm]
{\rm (II)}\ &:&\ \ \ \ \big(W_j^{(cl)}(u)\big)^*=(-1)^j\ W_j^{(cl)}(u^*) \nonumber
\eea
and for the classical transfer matrix
 \bea\label{hasysadssss}
{\rm (I)}\ &:&\ \ \ \  \big(\tau^{(cl)}(\lambda_c)\big)^*=\tau^{(cl)}(\lambda_c^*)\\[0.2cm]
{\rm (II)}\ &:&\ \ \ \ \big(\tau^{(cl)}(\lambda_c)\big)^*=\tau^{(cl)}(-\lambda_c^*) \ .\nonumber
\eea

\bigskip

Several comments are in order here.
The conjugation (I) in \eqref{hasysads}
corresponds to the classical limit of the conjugation condition 
$\big[W_j(u)\big]^\star=W_j(u^*)$, which occurred in our study of the
Hermitian structure for the space ${\cal H}^{({\rm cont})}$
(see eq.\,\eqref{asodi1982981}). Similarly, 
conjugation (II) is the classical version of
$\big[W_j(u)\big]^{\text{\tiny\ding{105}}}=(-1)^j\,W_j(u^*)$ \eqref{aisodi1212AA} for the spaces
${\cal H}^{({\rm disc},\pm)}$.
It should also be pointed out that  
for both reality conditions
${\tau}^{(cl)}(0)$ is real.
Furthermore, from \eqref{asusuyddda}
it follows that ${\tau}^{(cl)}(0)=2\, \cos(2\pi P)$.
In light of  our previous  discussion of the spectrum of 
the quantum transfer matrix $\bm{\tau}(\lambda)$ in the spaces
${\cal H}^{({\rm cont})}$ and ${\cal H}^{({\rm disc},\pm)}$  we'll take $P$ to be 
real and assume that
\bea
 -2<\tau^{(cl)}(0)< 2\ .
\eea
The last comment serves to make a link to integrable partial differential equations.
When 
 reality  condition (II) is imposed, i.e., $\xi_+=\xi,\ \xi_-=\xi^*$, the Hamiltonian flow 
generated by the classical local IM
 $I_2^{(cl)}=-\big(I_2^{(cl)}\big)^*$
w.r.t. the Poisson structure \eqref{jsasusa}
 coincides with
 the nonlinear Schr$\ddot{\rm o}$dinger equation in the attractive (focusing) regime \cite{Magri,Faddeev:1987ph}:
 \bea\label{poapsodp}
\ri \partial_\tau \xi=
\{\xi,  I_2^{(cl)}\}
=-\partial^2\xi-|\xi|^2\,\xi\ .
\eea
The repulsive regime,
where  the sign in front of the non-linear term is flipped,  is related to the $c\to 2^+$
limit of the $W_\infty$\,-\,algebra.

\bigskip

The classical limit of the $\overline{W}_\infty$\,-\,algebra
is described in the same way.
In particular, for 
$\bar{\boldsymbol\tau}({\bar\lambda})$ \eqref{asusuya}, whose action is
non-trivial on the left irrep of $\overline{W}_\infty\otimes W_\infty$, one has
\bea
\bar{\boldsymbol\tau}({\bar\lambda})\to \bar{\tau}^{(cl)}({\bar \lambda}_c)
\ \ \ \ \ \ \qquad ( n\to \infty\quad {\rm with}
\quad \bar{ \lambda}_c=(n+2)\,\bar{\lambda}\quad {\rm fixed})\ .
\eea
Here the classical transfer matrix   reads as
\bea\label{asusuydddaAA}
\bar{\tau}^{(cl)}({\bar \lambda}_c)={\rm Tr}_{\text{\textonehalf}}\Big[ 
\bar{{\boldsymbol L}}^{(cl)}(\bar{\lambda}_c)\ \re^{-\ri\pi \bar{P} {\tt h}}\Big]\qquad\quad
{\rm with}\qquad \quad
\bar{P}=\int_0^{2\pi}\frac{{\rm d} \bar{u}}{2\pi}\ \bar{\partial}\bar{\phi}\ \,,
\eea
 while $\bar{{\boldsymbol L}}^{(cl)}(\bar{\lambda}_c)$ stands for the path ordered
exponent that appears in the 
classical limit of 
$\bar{\bm{L}}(\bar{\lambda})$ \eqref{Laltcase1a}:
\be\label{Mlambdac1aAA}
\bar{{\boldsymbol L}}^{(cl)}(\bar{\lambda}_c)=
\bar{\lambda}_c^{+\frac{{\tt h}}{4}}
\ \overset{\rightarrow}{{\cal P}}\exp\bigg(\int_0^{2\pi}\rd \bar{u}\ 
\Big(\!-2\bar{\partial}\bar{\theta}\, \re^{-2\ri\bar{\phi}}\, {\tt e}_++ \bar{\lambda}_c\, 
 \big(\re^{+2\ri\bar{\phi}} {\tt e}_- 
 +\re^{-2\ri\bar{\phi}} {\tt e}_+\, \big)\Big)\bigg)\ \re^{-\ri \pi \bar{P} {\tt h}}\  
\ \bar{\lambda}_c^{-\frac{{\tt h}}{4}}
\ee
It follows from  the Yang-Baxter algebra
\eqref{YBeq1a}, that
$\bar{{\boldsymbol L}}^{(cl)}(\bar{\lambda}_c)$
 satisfies the Sklyanin exchange relations
\be\label{PBrel1q}
\big\{\bar{{\boldsymbol L}}^{(cl)}(\bar{\lambda}_c)\begin{array}{ccc} \\[-0.4cm] \otimes \\[-0.35cm] , 
\end{array}\bar{{\boldsymbol L}}^{(cl)}(\bar{\lambda}_c')\,\big\}=
-\Big[\,\bar{{\boldsymbol L}}^{(cl)}(\bar{\lambda}_c)\,{\otimes}\, \bar{{\boldsymbol L}}^{(cl)}(\bar{\lambda}_c'),
\,{\boldsymbol r}\Big(\sqrt{\bar{\lambda}_c/\bar{\lambda}_c'}\ \Big)\,\Big]
\ee
with ${\boldsymbol r}(\rho)$  the same as in \eqref{CRmatrix111}.
\bigskip

The classical transfer matrix may be expressed
 in terms of the quasiperiodic fields
\be\label{89sd89sd89sd}
\bar{\xi}_\pm=(\bar{\partial}\bar{\theta}\pm \ri \bar{\partial}\bar{\phi})\,\re^{\pm 2\bar{\theta}}\ : \ \ \qquad 
\bar{\xi}_\pm(u+2\pi)={\bar B}^{\pm 1}\ \bar{\xi}_\pm(u)\ .
\ee
This is achieved
through the relation similar to \eqref{90a9s0d9as}. Namely,
\be\label{90a9s0d9asA}
 \overset{\rightarrow}{{\cal P}}\exp\bigg(\int_{0}^{2\pi}\,\rd \bar{u}\,
\bar{{\bm A}}(\bar{u}\,|\,\bar{\lambda}_c)\bigg)\,\bar{B}^{+\frac{{\tt h}}{2}}=\bar{\bm{G}}(0)
\ \bar{\lambda}_c^{-\frac{{\tt h}}{4}}\ \big(\,\bar{{\boldsymbol L}}^{(cl)}(\bar{\lambda}_c)\ 
\re^{-\ri \pi \bar{P} {\tt h}}\,\big)\
\bar{\lambda}_c^{+\frac{{\tt h}}{4}}\ \bar{\bm{G}}^{-1}(0)
\ee
with $\bar{\bm{G}}(\bar{u})=\re^{\bar{\theta}{\tt h}}\,\re^{-\frac{\pi}{4}({\tt e}_+-{\tt e}_-)}\, 
\re^{\ri\bar{\phi}{\tt h}}$ and
\be\label{osaoi900003a333}
\bar{{\bm A}}({\bar u}\,|\,\lambda_c)=\bar{\xi}_-\,{\tt e}_--\bar{\xi}_+\,{\tt e}_+-{\bar \lambda}_c\,{\tt h}\,.
\ee
The above equations, together with \eqref{asusuydddaAA}, yield 
\be\label{iaois89dd8} 
\bar{\tau}^{(cl)}({\bar \lambda}_c)={\rm Tr}_{\text{\textonehalf}}\Big[\,
\, \overset{\rightarrow}{{\cal P}}\exp\Big(\int_{0}^{2\pi}\,\rd {\bar u}\,\bar{\bm A}({\bar u}\,|\, \bar{\lambda}_c)\Big)\, 
 \bar{B}^{+\frac{{\tt h}}{2}}
\Big]\ .
\ee
Finally, for the left chirality 
the reality conditions  are only notationally different from eqs.\,\eqref{hasysa1AA}-\eqref{hasysadssss}.

\section{Lorentzian black hole NLSM}
\subsection{The classical field theory}
Let's  consider anew the  Poisson algebra, whose first few PBs are given in
eq.\,\eqref{jasususa}
 with the $W^{(cl)}_j(u)$
being real classical fields.
As it follows from the  first two equations in \eqref{isisai112aaa}
one can introduce, at least locally, the real fields $\xi_\pm$ through
the relation
\bea\label{hasay}
\big(\xi_\pm\big)^2=W_2^{({ cl})}\ 
\exp\bigg(\pm 2 \int^u\rd u\ W_3^{({ cl})}\big/ W_2^{({ cl})}\,\bigg)\ .
\eea
Together  with  $ W_2^{({ cl})}$ and $ W_3^{({ cl})}$, 
the above formula involves a real  integration constant. It
can be interpreted  as a dynamical variable conjugated to
\bea
\log(B)=\oint\rd u\ W_3^{({ cl})}\big/ W_2^{({ cl})}\ .
\eea 
The latter belongs to the  center of the 
classical  $W_\infty$ algebra.
Note that $B$ must be  real and, furthermore, 
we take it to be positive.
Then  the Poisson structure for the classical $W$ currents is  lifted to
the  Poisson algebra \eqref{jsasusa} for  the real quasiperiodic fields $\xi_\pm$. The center of  \eqref{jsasusa} is generated by
the real constant
\bea\label{Tcentre1}
\tau^{(cl)}(0)={\rm Tr}_{\text{\textonehalf}}\Big[\,B^{-\frac{{\tt h}}{2}}\ \boldsymbol{\Omega}(2\pi)
\Big]\ :\ \  \ \  -2<  \tau^{(cl)}(0)=2\,\cos(2\pi P)< 2\ ,
\eea
where we use the notation
\bea\label{isisai4421}
\boldsymbol{\Omega}(u)=\overset{\leftarrow}{{\cal P}}
\exp\bigg(\int_{0}^{u}\,\rd u\, \big(\xi_-\,{\tt e}_--\xi_+\,{\tt e}_+\big)\bigg)\ .
\eea
All the above holds true for the left chirality. In particular, the fields $\bar{\xi}_\pm$
satisfy the Poisson algebra similar to \eqref{jsasusa},
\bea\label{jsasusss1sa}
\big\{\bar{\xi}_\pm (\bar{u}_1),\bar{\xi}_\pm (\bar{u}_2)\big\}&=&
\epsilon(\bar{u}_1-\bar{u}_2)\ \bar{\xi}_\pm (\bar{u}_1)\,\bar{\xi}_\pm (\bar{u}_2)\, ,
\nonumber\\[0.2cm]
\big\{\bar{\xi}_\pm(\bar{u}_1),\bar{\xi}_\mp (\bar{u}_2)\big\}&=&-\delta'(\bar{u}_1-\bar{u}_2)-\epsilon(\bar{u}_1-\bar{u}_2)\ 
\bar{\xi}_\pm (\bar{u}_1)\,\bar{\xi}_\mp (\bar{u}_2)
\eea
 and Poisson commute with $\xi_\pm$:
\be\label{jsasusss1saC}
\big\{\xi_\pm(u_1),\,\bar{\xi}_\pm(\bar{u}_2)\big\}=\{\xi_\pm(u_1),\,\bar{\xi}_\mp(\bar{u}_2)\}=0\ .
\ee
The center of the Poisson algebra for $\xi_\pm$, $\bar{\xi}_\pm$ is generated by 
$\tau^{(cl)}(0)$ together with
\bea\label{Tcentre2}
\bar{\tau}^{(cl)}(0)={\rm Tr}_{\text{\textonehalf}}\Big[\, 
\bar{\boldsymbol{\Omega}}(2\pi)
\ \bar{B}^{+\frac{{\tt h}}{2}}\,\Big]\ :\ \  \ \  -2<  \bar{\tau}^{(cl)}(0)=2\,\cos(2\pi {\bar P})< 2\ ,
\eea
where 
\be\label{isisai4421a}
\bar{\bm{\Omega}}(\bar{u})=\overset{\rightarrow}{{\cal P}}
\exp\bigg(\int_{0}^{\bar{u}}\,\rd \bar{u}\, \big(\bar{\xi}_-\,{\tt e}_--\bar{\xi}_+\,{\tt e}_+\big)\bigg)\ .
\ee
As with $B$ we assume that  ${\bar B}$ is positive and, moreover,
we'll impose the constraint
\bea\label{BBbar1a}
{\bar B}=B>0\ .
\eea

The above gives a sketch of the basic  
properties of the phase space (more precisely the algebra of functions on
the phase space) for a class of dynamical systems. Having in mind
our purpose of identifying
the CFT governing the critical behaviour of   the ${\cal Z}_2$ invariant inhomogeneous six-vertex model,
we take the  classical Hamiltonian as
\be\label{ioaisdoi1a}
H^{(cl)}=\int_0^{2\pi}\rd x\ \Big(W_2^{(cl)}(x)+\overline{W}_2^{(cl)}(x)\Big)\ .
\ee
An immediate question arises as to the possibility of a Lagrangian description of 
such a dynamical system. In connection with this, it is useful
to turn to the known classical Lagrangian
 field theory possessing the same type of Hamiltonian structure.
\bigskip

Consider the path ordered exponents
 $\boldsymbol{\Omega}(u)$ \eqref{isisai4421}
and $\bar{\bm{\Omega}}(\bar{u})$ \eqref{isisai4421a} with the $\mathfrak{sl}_2$
generators specialized to be in the fundamental representation
$\pi_{\text{\textonehalf}}({\tt e}_\pm)=\sigma^\pm$ and 
$\pi_{\text{\textonehalf}}({\tt h})= \sigma^3$.
Since $\xi_\pm$, $\bar{\xi}_\pm$ are real fields, 
the $2\times 2$ matrices $\bm{\Omega}_{\text{\textonehalf}}=\pi_{\text{\textonehalf}}\big(\bm{\Omega}\big)$,\,
$\bar{\bm{\Omega}}_{\text{\textonehalf}}=\pi_{\text{\textonehalf}}\big(\bar{\bm{\Omega}}\big)$
have real elements and their determinants are equal to one. 
With $\bm{g}_{\text{\textonehalf}}(0)\in {\rm SL}(2,\mathbb{R})$ being an arbitrary constant matrix, introduce
$\bm{g}_{\text{\textonehalf}}(t,x)\in \rm{SL}(2,\mathbb{R})$:
\be\label{9d0s9d0f}
\bm{g}_{\text{\textonehalf}}(t,x)=\bm{\Omega}_{\text{\textonehalf}}(t+x)\,\bm{g}_{\text{\textonehalf}}(0)\,
\bar{\bm{\Omega}}_{\text{\textonehalf}}(t-x)\ .
\ee
Writing it in the form
\be\label{iosoid1a}
\bm{g}_{\text{\textonehalf}}=\left(
\begin{array}{cc}
A& U \\[0.2cm]
-V & D
\end{array}\right)
\ee
one finds via a straightforward computation that the real functions
$U=U(t,x)$ and $V=V(t,x)$ satisfy the closed system of partial differential equations
\bea\label{jausussau}
(1-UV)\,\partial\bar{\partial}U=-V\,\partial U\bar{\partial}U\,,\qquad\qquad  \
(1-UV)\,\partial\bar{\partial}V=-U\,\partial V\bar{\partial}V\,,
\eea
where $\partial=\frac{1}{2}\, (\partial_t+\partial_x)$
and $\bar{\partial}=\frac{1}{2}\,(\partial_t-\partial_x)$.
The diagonal entries are determined  through the relations
 \bea\label{asoidoi1212}
\partial \log\Big(\frac{A}{D}\Big)=\frac{U\partial V-V\partial U}{1-UV}
\,,\qquad 
\bar{\partial} \log\Big(\frac{A}{D}\Big)=-\frac{U{\bar \partial} V-V\bar{\partial} U}{1-UV}\ ,\ \ \ 
AD=1-UV\ .
\eea
The equations of motion \eqref{jausussau} are the Euler-Lagrange equations corresponding to
the Lagrangian density
\bea\label{oiaodisa090}
{\cal L}=
\frac{1}{2}\, \frac{\partial_t U{\partial}_t V-\partial_x V{\partial}_x U}{1-UV}
\eea
and the fields satisfy the reality conditions
\bea\label{jasysa}
\big(U(t,x)\big)^*=U(t,x)\,,\qquad\qquad  \big(V(t,x)\big)^*=V(t,x)\ .
\eea
The latter is the Lagrangian density  for the 
Non-Linear Sigma Model (NLSM) whose target space coincides with  the so-called 
Lorentzian black hole.
In the work \cite{Witten:1991yr} this model was obtained by gauging 
a  non-compact one dimensional subgroup
of the classical ${\rm SL}(2,\mathbb{R})$ 
WZW model.
\bigskip

Let's  explain how the Lagrangian density \eqref{oiaodisa090} leads
to the Poisson structure \eqref{jsasusa},\,\eqref{jsasusss1sa} and \eqref{jsasusss1saC}.
As it follows from eq.\,\eqref{9d0s9d0f} the fields
$\xi_\pm$ and $\bar{\xi}_\pm$ are given by
\be\label{oaop9203}
\xi_-{\tt e}_--\xi_+{\tt e}_+=\partial{\bm{g}}\,\bm{g}^{-1}\,,\qquad
\bar{\xi}_-{\tt e}_--\bar{\xi}_+{\tt e}_+=\bm{g}^{-1}\,\bar{\partial}{\bm{g}}\,.
\ee
This allows one to express $\xi_\pm$, $\bar{\xi}_\pm$ in terms of $U$ and $V$:
\bea\label{isaodi9090a}
&&\xi_+=U\,\partial A-A\,\partial U\,,\qquad \,  \xi_-=V\,\partial D- D\,\partial V\\[0.2cm]
&&\bar{\xi}_+=U\,\bar{\partial}D-D\,\bar{\partial} U\,,\qquad  \bar{\xi}_-=
V\,\bar{\partial}A-A\,\bar{\partial} V\nonumber
\eea
and the equations of motion \eqref{jausussau} as well as \eqref{asoidoi1212} imply that
$\xi_\pm$, $\bar{\xi}_\pm$ are chiral fields:
\be
\bar{\partial}\xi_\pm=0\,,\qquad \partial\bar{\xi}_\pm=0\ .
\ee
Note that in \eqref{oaop9203}, we dropped the index $\frac{1}{2}$ denoting the fundamental representation
of  $\mathfrak{sl}_2$, since it remains valid as a relation in the Lie algebra  without reference to a particular representation.
The Lagrangian density \eqref{oiaodisa090} induces a canonical Poisson structure:
\be\label{iaosid9818921}
\big\{\Pi_U(t,x_1),\,U(t,x_2)\big\}=\delta(x_1-x_2)\,,\qquad \big\{\Pi_V(t,x_1),\,V(t,x_2)\big\}=\delta(x_1-x_2)\,,
\ee
where 
\be\label{iaosid9818921a}
\Pi_U=\frac{1}{2}\ \frac{\partial_t V}{1-UV}\,,\qquad \Pi_V=\frac{1}{2}\ \frac{\partial_t U}{1-UV}\ .
\ee
Combining this with \eqref{isaodi9090a} 
we indeed obtain the PB relations 
\eqref{jsasusa},\,\eqref{jsasusss1sa},\,\eqref{jsasusss1saC},
where $u=t+x$, $\bar{u}=t-x$ and $t$ is assumed to be fixed.
\bigskip

Having at hand the explicit formula \eqref{isaodi9090a} one can construct 
out of the  fundamental fields $U$ and $V$
the 
classical $W$ currents. 
Clearly  they are local chiral fields:
\be
W_j^{(cl)}(t,x)=W_j^{(cl)}(t+x)\,,\qquad 
\overline{W}_j^{(cl)}(t,x)=\overline{W}_j^{(cl)}(t-x)\ .
\ee
 Since in all our previous 
discussions the $W^{(cl)}_j(u)$ were assumed to be periodic, 
we supplement  \eqref{oiaodisa090} 
by the periodic boundary conditions 
\be\label{89a8s9d89}
U(t,x+2\pi)=U(t,x)\,,\qquad\qquad  V(t,x+2\pi)=V(t,x)
\ee
and take the classical action to be
\be\label{iasdioi120143}
S_{\rm \scriptscriptstyle LBH}=\frac{1}{2\hbar}\int\rd t \int_0^{2\pi}\rd x\ \
\frac{\partial_t U{\partial}_t V-\partial_x V{\partial}_x U}{1-UV}\qquad\quad
\qquad \big(\,\hbar=2\pi/n\to  0^+\,\big)\ .
\ee
The Hamiltonian is then given by  \eqref{ioaisdoi1a}.
That the fields $U$ and $V$  are real
is consistent with the reality condition
$\big(\xi_\pm(u)\big)^*=\xi_\pm (u^*)$, 
$\big(\bar{\xi}_\pm({\bar u})\big)^*=\bar{\xi}_\pm ({\bar u}^*)$
and, in turn,
\bea\label{iaosid9891}
 \big(W_j^{(cl)}(t+x)\big)^*=W_j^{(cl)}(t+x)\ ,\ \ \ \ \  
\big(\overline{ W}_j^{(cl)}(t-x)\big)^*=\overline{ W}_j^{(cl)}(t-x)\ .
\eea
This way the classical field theory 
defined by the action \eqref{iasdioi120143} where
$U$ and $V$ are real periodic fields  reproduces the Poisson structure,
Hamiltonian and reality conditions occurring in the scaling limit of the lattice model
in the sector  ${\cal H}^{({\rm cont})}$ and with the central charge 
$c=2-\frac{6}{n+2}\to 2^-$ as $n\to+\infty$.
\bigskip

Our qualitative discussion of the Poisson structure
suggests that the phase space for the Lorentzian black hole NLSM \eqref{iasdioi120143}
is made up of the symplectic leaves, $\Gamma_{\bar{P},P,B}$, labeled by the real numbers 
$P$, $\bar{P}$ and $B$.
On each leaf the symplectic form is non-degenerate.
The algebra of functions on the leaf, $\Gamma_{\bar{P},P,B}^\star$,
is generated by the currents $W^{(cl)}_j(u)$ and $\overline{W}^{(cl)}_j(\bar{u})$,
subject to the reality conditions \eqref{iaosid9891},
while  the Poisson algebra on $\Gamma_{\bar{P},P,B}^\star$ is fully
specified by the PBs  \eqref{jasususa} for the 
$W$ currents, the similar relations
for the left chiral currents as well as $\{W^{(cl)}_j(u),\,\overline{W}^{(cl)}_{j'}(\bar{u})\}=0$.
To get some  insight into the global structure of the phase space 
of the model, it is useful to consider
 basic solutions of the classical equations of motion.
These may be constructed using \eqref{9d0s9d0f} and \eqref{iosoid1a}.
First of all, in the ``bosonization'' formulae \eqref{aisisa},\,\eqref{89sd89sd89sd} we set
$\theta=\bar{\theta}=-\frac{\ri\pi}{4}$ and $\partial\phi=P,\ {\bar \partial}\bar{\phi}={\bar P}$.
Then 
$\xi_\pm=P$ and ${\bar \xi}_\pm={\bar P}$
become space-time independent real constants.
Eqs.\,\eqref{9d0s9d0f},\,\eqref{iosoid1a} with $\bm{g}_{\text{\textonehalf}}(0)=-\bf{1}$ yield
\bea\label{jsausay}
U(t,x)=V(t,x)=\sin\big(\,(P+{\bar P})\, t+(P-{\bar P})\, x\, \big)
\eea
and one can easily see that the equations of motion \eqref{jausussau} are indeed satisfied.
The periodic boundary condition \eqref{89a8s9d89} requires that the difference $P-{\bar P}$
be an integer.
It hints that the real numbers $P$ and $\bar{P}$ labeling the symplectic leaves
might not be arbitrary, but obey the condition:
\be\label{iasodiao3109}
P-\bar{P}=0,\pm 1,\pm 2,\ldots\ .
\ee
\bigskip

Other hints provided by the explicit solutions \eqref{jsausay} 
concern the  action of the global symmetries
 of the classical field theory on its phase space.
There are two evident space-time symmetry transformations, ${\cal T}$ and ${\cal P}$,
which are defined as
\bea\label{iaosid892}
&&{\cal T}\ : \ \ \  U(t,x)\mapsto \phantom{-}U(-t,x)\,,\qquad  V(t,x)\mapsto \phantom{-} V(-t,x)
\\[0.2cm]
&&{\cal P}\ : \ \  \ U(t,x)\mapsto -U(t,-x)\,,\qquad V(t,x)\mapsto -V(t,-x)\,.\nonumber
\eea
The extra sign in the definition of 
${\cal P}$  is a matter of convention
since the transformation
\bea\label{iaosid8981}
{\cal U}\ : \ \ \ U\mapsto -U\,,\qquad V\mapsto -V
\eea
also leaves the action \eqref{iasdioi120143} invariant.
The basic solutions \eqref{jsausay}
are unchanged under the ${\cal P T}$ transformation. More generally, 
we will assume that two solutions related via ${\cal P T}$ belong
to the same symplectic leaf, i.e.,
\be\label{oasiod192aa}
{\cal PT}\ : \ \ \ \Gamma_{\bar{P},P,B}\mapsto \Gamma_{\bar{P},P,B}\ .
\ee
Since \eqref{iaosid8981}
does not affect 
 the $W$ currents, it will likewise be assumed that
\be
{\cal U}\ : \ \  \ \Gamma_{\bar{P},P,B}\mapsto \Gamma_{\bar{P},P,B}\ .
\ee
The action of ${\cal P}$ and ${\cal T}$  on the fundamental  fields, as described by formula \eqref{iaosid892},  induces
the action of these transformations 
on $\Gamma_{\bar{P},P,B}$. 
We make the assumption that two solutions 
related through  ${\cal P}$ or ${\cal T}$
separately belong to different symplectic leaves.
A brief examination of \eqref{jsausay}
motivates that 
\bea\label{aisodi1902}
&&{\cal T}\ : \ \ \Gamma_{\bar{P},P,B}\mapsto \Gamma_{-{P},-\bar{P},B}\nonumber \\[0.2cm]
&&{\cal P}\ : \ \ \Gamma_{\bar{P},P,B}\mapsto \Gamma_{{P},\bar{P},B}\ .
\eea
An immediate consequence is that ${\cal PT}$ maps
 $\Gamma_{\bar{P},P,B}$ to  $\Gamma_{-\bar{P},-P,B}$.
Consistency with the condition \eqref{oasiod192aa} requires  
the following identification to be made
\be
\Gamma_{\bar{P},P,B}\equiv \Gamma_{-\bar{P},-P,B}\ .
\ee
With this important property one can always take
\be\label{iasoid18392}
P+\bar{P}\ge 0
\ee
without loss of generality.
Formula \eqref{aisodi1902} in addition  implies that if the phase space contains the leaf
$\Gamma_{\bar{P},{P},B}$ it must also contain $\Gamma_{{P},\bar{P},B}$.
\bigskip

Another global symmetry of the action is the ${\cal Z}_2$ transformation
which interchanges the fields $U$ and $V$:
\be
{{\cal D}}\,:\ \ \ U\mapsto V\ ,\qquad V\mapsto U\ .
\ee
In turn, $\xi_\pm\mapsto \xi_\mp$ and taking into account eq.\,\eqref{90s9d0909f}, 
its action  on the symplectic leaves is given by
\be
{{\cal D}}\,: \ \ \ \Gamma_{\bar{P},P,B}\mapsto \Gamma_{\bar{P},P,B^{-1}}\ .
\ee
\bigskip

Finally, there is one more evident symmetry. The classical action \eqref{iasdioi120143}
 remains unchanged under the transformation
\bea\label{hssattas}
 {\cal R}_a\,:\ \ \  U\mapsto a^{+1}\,U\,,\qquad V\mapsto a^{-1}\,V\qquad {\rm with} \qquad a> 0\,.
\eea
This  acts on the non-local fields as $\xi_\pm\mapsto a^{\pm 1} \xi_\pm$, 
${\bar  \xi}_\pm\mapsto a^{\pm 1} \bar{\xi}_\pm$ and has no effect on the $W$ currents.
The symmetry is a continuous one and,
in view of the PB relations \eqref{iaosd899382}
and the constraint $B=\bar{B}$ (see eq.\,\eqref{BBbar1a}),
the associated Noether charge may be
identified with $\log (B)$.

\bigskip

\begin{figure}
\centering
\scalebox{1.15}{
\includegraphics[width=7.5cm]{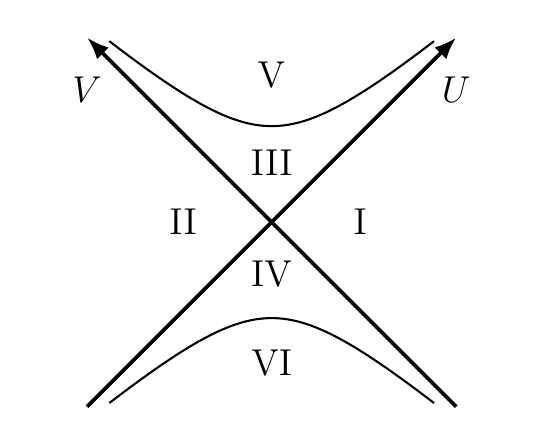}
}
\caption{\label{iaosi1212}\small Space-time diagram for the Lorentzian black hole \eqref{hsasaysaty}.
The cross 
defined by the equation $UV=0$ is a horizon, while the metric 
possesses a physical singularity on the hyperbola 
$UV=1$.
\label{fig20}}
 \end{figure}

Our intuition regarding the global properties of the 
field theory  phase space was in a large part motivated 
through an examination of the basic solutions \eqref{jsausay}.
These satisfy the inequality
\bea\label{iussusau}
0\leq UV<1\ .
\eea
In all likelihood,
for the phase space made up from the symplectic leaves $\Gamma_{\bar{P},P,B}$
with $P$, $\bar{P}$ and $B$ subject to the conditions
\eqref{iasodiao3109},\,\eqref{iasoid18392}  and $B>0$ this constraint
should be imposed on all  the classical field configurations.
In ref.\cite{Witten:1991yr} it was observed that  the Lorentzian target space metric corresponding to the action \eqref{iasdioi120143},
\bea\label{hsasaysaty}
(\rd\sigma)^2=\frac{{\rm d} U{\rm d} V}{1-UV}\ ,
\eea
exhibits the characteristic features of a black hole geometry.
In particular, as depicted in the space-time diagram in fig.\ref{fig20}, it possesses a horizon at $UV=0$ as well
as a curvature singularity at $UV=1$ just as the Schwarzschild black hole
in terms of Kruskal coordinates. An important property of the metric is that
there is no globally defined time coordinate. There is, however, a non-trivial
Killing vector
which is time-like only in regions I and II of
fig.\ref{fig20} and space-like in regions III and IV.
The restriction \eqref{iussusau} means that we are focusing on the Lorentzian 
NLSM
with the fields $U$ and $V$ taking values 
in the domain, which is the union of regions III and IV in fig.\ref{fig20}.

\subsection{Quantization\label{sec212}}
One can proceed with the study of the quantum NLSM
through the  quantization of the
algebra of functions on the symplectic leaves.
Identifying the Planck constant
as $\hbar=\frac{2\pi}{n}$,
 leads us to the quantum $\overline{W}_\infty\otimes W_\infty$\,-\,algebra with central charge
$c=2-\frac{6}{n+2}<2$.
The parameters $(\bar{P},P,B)$ labeling the symplectic leaves
are related to the highest weights of 
the  irreducible representation  of the 
quantum algebra.
The  highest weight of the $W_\infty$\,-\,algebra, 
$\bm{\omega}=(\omega_2,\omega_3)$, may be parameterized by $(\rho$, $\nu)$ as in 
\eqref{Deltvarpi1a} and similarly $\bar{\bm{\omega}}$ is swapped for
$(\bar{\rho},\bar{\nu})$.
Based on the previous discussion,
 the following identification of the parameters can be made
\bea\label{ususauas}
\rho=(n+2)\, P\ ,\ \  \ \ \ \ \ \ \bar{\rho}= (n+2)\,\bar{P}\ ,\qquad
\nu=\bar{\nu}=\frac{n}{4\pi}\log(B)\ .
\eea
Then formulae \eqref{iasodiao3109},\,\eqref{iasoid18392} and $B>0$ translate to 
the conditions
\be
\rho+\bar{\rho}\ge 0,\,\qquad \qquad \rho-\bar{\rho}=(n+2)\,{\tt w}\,,\qquad\qquad -\infty<\nu=\bar{\nu}<+\infty
\ee
with ${\tt w}=0,\pm1,\pm2,\ldots\ $.
Recall that the components $\bar{\omega}_2$ and $\omega_2$ of the highest weight 
labeling the  $\overline{W}_\infty\otimes W_\infty$ irrep coincide with
the conformal dimensions of the highest  state, so that
\bea
\bar{\Delta}_{\bar{\rho},\nu}=\frac{\bar{\rho}^2-\frac{1}{4}}{n+2}+\frac{\nu^2}{n}  \,,\qquad\qquad
{\Delta}_{\rho,\nu}=\frac{\rho^2-\frac{1}{4}}{n+2}+\frac{\nu^2}{n}\ .
\eea
 Assuming that $\rho+\bar{\rho}$ is a non-negative integer,
the Lorentz spin of these states, $\Delta_{\rho,\nu}-\bar{\Delta}_{\bar{\rho},\nu}$,
would only take integer values.
Then $\rho$ and $\bar{\rho}$ would have the form
\be\label{asudiui89129}
\rho=\tfrac{1}{2}\,{\tt v}+\tfrac{1}{2}\,(n+2)\,{\tt w}\,,\qquad \bar{\rho}=\tfrac{1}{2}\,{\tt v}-\tfrac{1}{2}\,(n+2)\,{\tt w}\,,
\ee
where ${\tt v}=0,1,2,\ldots\ $ and ${\tt w}\in\mathbb{Z}$. 
This way we come to expect that the space of states 
 of the Lorentzian black hole NLSM,
${\cal H}_{\scriptscriptstyle{\rm  LBH}}$,
is decomposed into irreps of the
$\overline{W}_\infty\otimes W_\infty$ algebra as 
\be\label{iaosidoi192009AA}
{\cal H}_{\scriptscriptstyle{\rm  LBH}}=
\bigoplus_{{\tt v}= 0}^{\infty}\Bigg[
\bigoplus_{{\tt w}=-\infty}^{\infty}\int^{\oplus}_{\mathbb{R}}\!\rd \nu\
\overline{{\cal W}}_{\bar{\rho},\nu}\otimes {\cal W}_{\rho,\nu}\Bigg]\ .
\ee
The latter is identical to the
 linear decomposition \eqref{iaosidoi1920091a} of the space
$\tilde{{\cal H}}_{\rm even}^{({\rm cont})}$,
which is a sector of ${\cal H}^{({\rm cont})}_{\rm even}$ \eqref{ioasid1892981}
-- the space of states occurring
 in the scaling limit of the ${\cal C}$ even sector
of the ${\cal Z}_2$ invariant inhomogeneous six-vertex model subject to periodic
boundary conditions  $({\tt k}=0)$.

\bigskip

The Hermitian structure in the space ${\cal H}_{\scriptscriptstyle{\rm  LBH}}$
should be consistent with the  Hermitian
conjugation \eqref{asodi1982981},
which as was already mentioned  is the quantum version of the 
classical reality condition \eqref{iaosid9891} (see the comments below \eqref{hasysadssss}).
This leads us to propose that, not only the linear structure of the spaces
${\cal H}_{\scriptscriptstyle{\rm  LBH}}$ and $\tilde{{\cal H}}^{({\rm cont})}_{\rm even}$
coincide,
 but also their Hermitian structures.
In particular  the inner product of the $\overline{W}_\infty\otimes W_\infty$ primary
states ${\Psi}_{{\tt v},{\tt w},\nu}$ of the irreps appearing in the decomposition \eqref{iaosidoi192009AA},
 up to real positive constants that depend on the overall normalization of ${\Psi}_{{\tt v},{\tt w},\nu}$,
would be equal to $\langle\cdot,\cdot\rangle_{\rm cont}$ computed on
the corresponding states from $\tilde{{\cal H}}^{({\rm cont})}_{\rm even}$.
It follows from eqs.\,\eqref{iaosido891221} and \eqref{aoisido1920}
that one can set
\be\label{iaosido891221aa}
\big\langle{\Psi}_{{\tt v}',{\tt w}',\nu'}\,,
{\Psi}_{{\tt v},{\tt w},\nu}\big\rangle=\delta_{{\tt v}',{\tt v}}\,\delta_{{\tt w}',{\tt w}}\, 
\delta(\nu'-\nu)\, N_{{\tt v},{\tt w}}
\ee
with
\be\label{aosidi39034sa}
N_{0,0}=1\,,\qquad N_{{\tt v},{\tt w}}=
\begin{cases}
(-1)^{\tt w}\ \,\dfrac{\sin(\pi(n+2){\tt w})}{\pi (n+2)}\ \ \ &
\quad {\tt v}=0,\,{\tt w}\ne 0\\[0.6cm]
\dfrac{\Gamma(1-{\tt w}+\frac{{\tt v}}{n+2})\,
\Gamma(1+{\tt w}+\frac{{\tt v}}{n+2})}{\Gamma(1+{\tt v}-(n+2)\,{\tt w})\,
\Gamma(1+{\tt v}+(n+2)\,{\tt w}) } 
\  \ \  &\quad {\tt v }\geq 1,\,{\tt w}\in\mathbb{Z}
\end{cases}
\ee
Then ${\cal H}_{\scriptscriptstyle{\rm  LBH}}$  is a pseudo-Hilbert space equipped with a non-positive
definite inner product.
This would reflect the fact that 
the target space for the  NLSM \eqref{iasdioi120143} has Lorentzian signature.

\bigskip

The identification of the pseudo-Hilbert spaces ${\cal H}_{\scriptscriptstyle{\rm  LBH}}$
and $\tilde{{\cal H}}^{({\rm cont})}_{\rm even}$ turns out to be 
consistent with all the global symmetries. 
First of all, as was discussed in sec.\,\ref{sec173}, the full linear space
${{\cal H}}^{({\rm cont})}_{\rm even}$ admits the two formal ${\cal Z}_2$
symmetries, ${\cal X}^{({\tt w})}$  \eqref{z2sybaif891} and
${\cal X}^{({\rm null})}$
\eqref{asodi809ioas}, having
no counterparts in the lattice system. 
The transformation  ${\cal X}^{({\rm null})}$ acts trivially in
$\tilde{{\cal H}}^{({\rm cont})}_{\rm even}$, or equivalently, 
 ${\cal H}_{\scriptscriptstyle{\rm  LBH}}$
 and thus should be ignored. 
The other global symmetry ${\cal X}^{({\tt w})}$
arises due to the degeneracy in the decomposition \eqref{iaosidoi192009AA} in which the irreps,
whose highest states are ${\Psi}_{0,+{\tt w},\nu}$ and ${\Psi}_{0,-{\tt w},\nu}$ with ${\tt w}\ne 0$ are
equivalent.
However,  formula \eqref{aosidi39034sa} shows
 that the ``norms'' of  these two primary states, $N_{0,\pm{\tt w}}$, differ in their sign, which can not be 
eliminated by a change of their normalization.
Hence we conclude that the ${\cal Z}_2$ transformation ${\cal X}^{({\tt w})}$,
in spite that it commutes with the generators of the $\overline{W}_\infty\otimes W_\infty$\,-\,algebra, is not actually
a  symmetry
of the pseudo-Hilbert space $\tilde{{\cal H}}^{({\rm cont})}_{\rm even}\cong{\cal H}_{\scriptscriptstyle{\rm  LBH}}$.
In connection to this, let's mention that 
in all likelihood 
the  states ${\Psi}_{0,\pm{\tt w},0}$ correspond to 
the solutions $U=V=\cos({\tt w}x)$  
and $U=V=\sin({\tt w}x)$
of the classical equations of motion
\eqref{jausussau},\,\eqref{89a8s9d89}. These are distinguished by
their properties w.r.t. the parity transformation $x\to-x$.
\bigskip

The classical NLSM possesses ${\cal P}$ invariance as well as the
time-reversal symmetry. The action of the ${\cal P}$ and ${\cal T}$
transformations on the fundamental fields $U$ and $V$ is described by
eq.\,\eqref{iaosid892}. It is natural to expect that 
these global symmetries are present in the quantum NLSM as well.
The corresponding generators would satisfy the following commutation
relations with the $W$ currents
\be\label{aisodi1212}
{\cal \hat{P}}_{\scriptscriptstyle{\rm  LBH}}\ {W}_j(u)=\overline{W}_j(u)\,{\cal \hat{P}}_{\scriptscriptstyle{\rm  LBH}}\,,
\qquad
\qquad
{\cal \hat{T}}_{\scriptscriptstyle{\rm  LBH}}\ {W}_j(u)=\overline{W}_j(u)\,{\cal \hat{T}}_{\scriptscriptstyle{\rm  LBH}}\ .
\ee
Their action on the primary states, without loss of generality, can be chosen to be
\bea\label{aisodi1212aaaa}
\hat{{\cal P}}_{\scriptscriptstyle{\rm  LBH}}\,{\Psi}_{{\tt v},{\tt w},\nu}=
\hat{{\cal T}}_{\scriptscriptstyle{\rm  LBH}}\,{\Psi}_{{\tt v},{\tt w},\nu}=\sigma_{{\tt v},{\tt w}}\ {\Psi}_{{\tt v},-{\tt w},\nu}
\eea
with the sign factor being
such that $\sigma_{{\tt v},{\tt w}}=+1$ for ${\tt v}={\tt w}=0$ and ${\tt v}\cdot {\tt w}\not=0$.
For the case when ${\tt v}=0$ and ${\tt w}\ne 0$ or ${\tt v}\ge 1$ and ${\tt w}=0$, the states 
${\Psi}_{{\tt v},{\tt w},\nu}$ are eigenstates of the parity generator and the corresponding sign factors
$\sigma_{0,{\tt w}}$ and $\sigma_{{\tt v},0}$  are as yet undetermined.
\bigskip

Finally it remains to discuss the 
${\cal Z}_2$ symmetry ${\cal D}$ as well as the 
symmetry generated by $\hat{\cal U}$ \eqref{Ugen1a}, which is a remnant of the 
broken ${\rm U}(1)$. In the previous subsection we described the manifestation
of these global symmetries in the classical field theory, where they were denoted by the same
symbols. 
Note that for the quantum NLSM, 
the generator $\hat{{\cal U}}$ acts on any state in the irrep
$\overline{{\cal W}}_{\bar{\rho},\nu}\otimes {\cal W}_{\rho,\nu}$ 
via multiplication by the sign factor
 $\pm(-1)^{\tt v}$. The extra sign ``$\pm$'' may be chosen at will as it
remains the same for all the states in ${\cal H}_{\scriptscriptstyle{\rm  LBH}}$.
The action of the global symmetries in ${\cal H}_{\scriptscriptstyle{\rm  LBH}}$
is defined through the commutation relations of their generators with the $W$ currents
supplemented by the action of the symmetries 
on the $\overline{W}_\infty\otimes W_\infty$ primary states.
For the ${\cal Z}_2$ symmetry transformation
${\cal D}$  the relevant formulae are \eqref{oapsd0092} and
\bea
\hat{{\cal D}}\,{\Psi}_{{\tt v},{\tt w},\nu}={\Psi}_{{\tt v},{\tt w},-\nu}\,,
\eea
while  $\hat{\cal U}$
commutes with the $W$ currents and
\be\label{sigmadef1aaa}
\hat{{\cal U}}\,{\Psi}_{{\tt v},{\tt w},\nu}=\sigma\, {\Psi}_{{\tt v},{\tt w},\nu}\quad\qquad {\rm with}\quad\qquad
\sigma=\pm\, (-1)^{\tt v}\ .
\ee
There is no apparent candidate for a non-trivial ${\cal C}$ conjugation for the classical action \eqref{iasdioi120143},
which is consistent with the fact that, by construction,
$\hat{\cal C}={\rm id}$ for the space  $\tilde{{\cal H}}^{({\rm cont})}_{\rm even}$.

\subsection{Minisuperspace approximation\label{sec213}}

For a better qualitative understanding of the quantum NLSM it is useful to consider the model within the
so-called minisuperspace approximation. This entails 
taking into account only those
field configurations that do not depend on the space co-ordinate $x$,
such as the classical solutions \eqref{jsausay}  with $P=\bar{P}$.
We still take $U$ and $V$ to satisfy the constraint $0\le UV<1$ corresponding  to the
union of regions  III and IV in fig.\,\ref{fig20}.
For a  preliminary analysis 
it is convenient to parameterize $U,\,V$ from this domain as
\bea\label{iaoisod1}
U=\re^{\Theta}\ \sin(\Phi)\ ,\ \ \ \ \ V=\re^{-\Theta}\ \sin(\Phi)\ ;\ \ \ \  \Phi\in(-\tfrac{\pi}{2},\tfrac{\pi}{2})\ ,
\ \ \ \Theta\in(-\infty,\infty)\ .
\eea
Then the minisuperspace version of the  classical action   \eqref{iasdioi120143}  reads as
\bea\label{aksisa}
{ S}^{({\rm ms})}_{\scriptscriptstyle{\rm LBH}}
=\frac{\pi}{\hbar}\int \rd t \ \big(\,\dot{\Phi}^2-\tan^2(\Phi)\ \dot{\Theta}^2\,\big)\ .
\eea
Since the generalized coordinate  $\Theta$ is cyclic, its conjugate  momentum 
$\Pi_\Theta=-\tan^2(\Phi)\, \dot{\Theta}$  is an integral of motion. 
The  effective  Lagrangian (the Routhian)  for  
the non-cyclic degree of freedom  is given by
\bea
L_{\rm eff}=\frac{1}{2}\ \big(\dot{\Phi}^2-V_{\rm eff}(\Phi)\,\big)\ , \ \ \ \  \     \qquad  
V_{\rm eff}(\Phi)=-\Pi_\Theta^2\ \cot^2(\Phi) \ .
\eea
 The latter  describes a 1D particle falling to the origin $\Phi=0$.
An elementary calculation shows  that for any value $\Pi_\Theta\not= 0$ the particle, starting  its motion at
 $t=0$,  reaches  the origin 
in a finite amount of time $t_{\rm fall}<+\infty$.   For $t>t_{\rm fall}$ the motion remains undetermined.
Thus  the action \eqref{aksisa} specifies the time evolution of the mechanical system   only within 
a finite time interval
(except for the trajectories with  $\Pi_\Theta=0$). To continue the classical trajectories  for $t>t_{\rm fall}$
the unbounded effective potential should be somehow regularized. There are of course numerous ways
of doing this.
A simple minded one is to  replace  $ V_{\rm eff}(\Phi)=-\Pi_\Theta^2\ \cot^2(\Phi)$
by a  smooth  potential $V^{(\rm reg)}_{\rm eff}(\Phi)$, which together with 
its 
derivative is bounded from below 
within the infinitesimal interval $\Phi\in (-\epsilon,\epsilon)$.
Outside this interval 
$V^{(\rm reg)}_{\rm eff}(\Phi)=V_{\rm eff}(\Phi)$.
To keep the original symmetry of the potential we  assume  that the regularized one is an
even function:
\bea\label{aisaisai}
V^{(\rm reg)}_{\rm eff}(\Phi)=V^{(\rm reg)}_{\rm eff}(-\Phi)\ .
\eea
Then the motion of $\Phi$
becomes   globally defined and periodic  for any values of $\Pi_\Theta\ne 0$.

\bigskip

With basic intuition from quantum mechanics,
we can  predict  the symmetry properties of the minisuperspace stationary wave functions.
First of all, that the regularized potential is an even function of $\Phi$   implies that 
the stationary states may be assigned a parity $\sigma=\pm 1$,
\bea\label{usassayt}
\hat{\cal U}\,\Psi^{(\sigma)}(U,V)\equiv\Psi^{(\sigma)}(-U,-V) =\sigma\, \Psi^{(\sigma)}(U,V)\ ,
\eea
where we now switch to the original target space coordinates $(U,V)$.
This relates the values of the wave function in the domains III 
and IV from fig.\,\ref{fig20}.
Next, $\Psi^{(\sigma)}$ can be chosen to be  an eigenfunction  of the operator
$\hat{\Pi}_\Theta=\frac{\hbar}{\ri}\,\partial_\Theta=\frac{\hbar}{\ri}\,(U\partial_U-V\partial_V)$.
Since $\hat{\Pi}_{\Theta}$ is the infinitesimal generator of the continuous symmetry \eqref{hssattas},
its eigenvalue is related to the conserved charge $\nu$ \eqref{ususauas}:
\bea
\hat{\Pi}_\Theta\,  \Psi^{(\sigma)}_\nu=2\hbar \nu\ \Psi^{(\sigma)}_\nu\ .
\eea
It follows that
\bea\label{hassysa}
\Psi^{(\sigma)}_\nu(U,V)= \bigg(\frac{U}{V}\bigg)^{\ri \nu}\ F^{(\sigma)}_\nu(UV)\ .
\eea
The minisuperspace approximation ignores  the presence of the oscillatory  modes
so that  the  wave functions $\Psi^{(\sigma)}_\nu$ are  expected to correspond
 to   $\overline{W}_\infty\otimes W_\infty$ primary  states,
characterized by $\rho=\bar{\rho}$ and $\nu$.
In turn the minisuperspace energy  becomes  
$\Delta_{\rho,\nu}+\Delta_{{\bar \rho},\nu}=2\Delta_{\rho,\nu}$ in the leading non-vanishing order of $\hbar=\frac{2\pi}{n}$
(the approximation is  reliable only in the limit  $n\to\infty$). Namely,
\bea\label{iassausa}
E^{({\rm ms})}=\tfrac{\hbar}{\pi}\ \big({\rho}^2+\nu^2-
\tfrac{1}{4}\big)\ .
\eea
At this point, $\rho$ can be thought  of as a real number  
parameterizing the minisuperspace
energy $E^{({\rm ms})}$  and the corresponding wavefunction $\Psi^{(\sigma)}_{\rho,\nu}$.
Since the highest weight  is an even function of $\rho$,
\bea
\Psi^{(\sigma)}_{-\rho,\nu}(U,V)=\Psi^{(\sigma)}_{\rho,\nu}(U,V)\,.
\eea
As we mentioned before, one can assume that
$\rho={\bar \rho}\geq 0$.

\bigskip

Though the 
highest weight of the $W_\infty$ irrep $\bm{\omega}=(\omega_2,\omega_3)$ 
is not sensitive to the sign of $\rho$, 
as follows from \eqref{Deltvarpi1a}
it does depend on the sign of  $\nu$: $\omega_3(\rho,-\nu)=-\omega_3(\rho,\nu)$.
 Thus  
 the primary states characterized by $(\rho,\nu)$ and $(\rho,-\nu)$
are distinguishable. They are related
through the ${\cal Z}_2$ transformation, so that
\bea\
{\hat {\cal D}}\, \Psi^{(\sigma)}_{\rho,\nu}(U,V)=\Psi^{(\sigma)}_{\rho,-\nu}(U,V)\ .
\eea
On the other hand, by definition, this symmetry  interchanges $U$ and $V$:
\bea
{\hat {\cal D}}\, \Psi^{(\sigma)}_{\rho,\nu}(U,V)\equiv\Psi^{(\sigma)}_{\rho,\nu}(V,U)\ .
\eea
Combining the above two relations with \eqref{hassysa} one  concludes that
\bea\label{isaisisaa}
\Psi^{(\sigma)}_{\rho,\nu}(U,V)= \bigg(\frac{U}{V}\bigg)^{\ri \nu}\ F^{(\sigma)}_{\rho,\nu}(UV)\ ,\ \ \ \ \ {\rm where}\ \ \ \ 
 F^{(\sigma)}_{\rho,\nu}(z)= F^{(\sigma)}_{-\rho,\nu}(z)= F^{(\sigma)}_{\rho,-\nu}(z)\ .
\eea

\bigskip

Having described the symmetry properties
of the stationary wave functions, we turn to deriving them explicitly.
In the work \cite{Dijkgraaf:1991ba}, a minisuperspace analysis was performed for the NLSM 
\eqref{iasdioi120143} with the fields $U$, $V$ belonging to region I  from fig.\,\ref{fig20} (or equivalently II).
Though this is not the domain of interest,  
we can still follow the same line of arguments of that paper.
In particular, up to a trivial factor,  the minisuperspace  Hamiltonian
coincides with the ``dilatonic'' Laplacian:
\bea\label{isaiasias2a}
\hat{H}^{({\rm ms})}=-\frac{\hbar }{4\pi}\ \triangle_D\ ,\ \ \ \ \ \triangle_D=\frac{1}{\re^{D}\sqrt{-G}}\ \partial_i
\big(\re^D\sqrt{-G}\,G^{ij}\partial_j\big)\ ,
\eea
where  the metric is the one in  \eqref{hsasaysaty} and the dilaton field is given by
\bea
 D= \log(1-UV)\ .
\eea
The stationary Schr${\rm{\ddot o}}$dinger equation $\hat{H}^{({\rm ms})}\,\Psi=E^{({\rm ms})}\, \Psi$
reads explicitly as
\bea\label{ajsaasu}
-\big(\, (1-UV)\ \partial_U\partial_V-\tfrac{1}{2}
\ ( U\partial_U+V\partial_V)\, \big)\, \Psi=\tfrac{\pi}{\hbar} \,E^{({\rm ms})}\ \Psi\ .
\eea
Using the general form   \eqref{isaisisaa}  for the stationary  wave functions  and
parameterizing the energy as in \eqref{iassausa},  it is straightforward to
show that $F(z)=z^{-\ri \nu}\, F^{(\sigma)}_{\rho,\nu}(z)$ obeys the 
 Gauss  hypergeometric equation
\bea\label{isaisai}
z\,(1-z)\,F''+\big(1+2\ri \nu-2\,(1+\ri \nu)\,z\,\big)\,
F'-\big( \tfrac{1}{2}+\ri \nu+{ \rho}\big)\big( \tfrac{1}{2}+\ri \nu-{ \rho}\big)\,F=0\, .
\eea
Keeping in mind our preliminary analysis,
the ODE  \eqref{isaisai}  is applicable only in the
domain $\epsilon^2<z<1$ with  a 
small  regularization parameter $\epsilon\ll 1$ (recall that $z=UV=\sin^2(\Phi)$).

\bigskip

The function
$F^{(\sigma)}_{\rho,\nu}(z)$  is a certain linear combination of 
$z^{\pm \ri \nu}\ {}_2F_{1}\big(\tfrac{1}{2}\pm \ri \nu+{ \rho},
\tfrac{1}{2}\pm\ri \nu-{ \rho}, 1\pm 2\ri \nu, z)$, which
can be specified as follows.  Applying the
elementary identity
\bea\label{hasysa}
 \re^{D}\sqrt{-G}\, \big(\,\Psi^*_1\,  \hat{H}^{({\rm ms})}\,\Psi_2-\Psi_2 \,\hat{H}^{({\rm ms})} \, \Psi^*_1\,\big)=
\frac{\hbar}{4\pi }\ \partial_i \Big(   \re^{D} \sqrt{-G}\, G^{ij} \big(\Psi_2 \partial_j\Psi_1^*-\Psi_1^* \partial_j\Psi_2\,\big)\Big)
\eea
to the pair of stationary wave functions $\Psi_1$, $\Psi_2$
corresponding to the energies $E^{({\rm ms})}_1$, $E^{({\rm ms})}_2$, 
and then integrating
the result  over the domain ${\rm B}_\epsilon: \ \epsilon^2<UV<1$, one obtains
\bea\label{hsasaysay}
\big(\,E^{\rm ( ms)}_2-E^{\rm ( ms)}_1\,\big)\ 
\int_{{\rm B}_\epsilon}\rd U\rd V\  \re^{D}\sqrt{-G}\  \Psi^*_1  \Psi_2=\frac{\hbar}{4\pi }\
\int_{\partial {\rm B}_\epsilon}\rd\ell \,  
\re^{D}\, \big(\Psi_2 \partial_n\Psi_1^*-\Psi_1^* \partial_n\Psi_2\big)\ .
\eea
Here the integral in the r.h.s. is taken over the boundary of ${\rm B}_\epsilon$, 
which is the union of $UV=\epsilon^2$ and
$UV=1$. Also, 
$\partial_n$ stands for the normal derivative to $\partial {\rm B}_\epsilon$. 
 As was discussed before, the wave functions possess a definite parity. Due to this
either the wave function or its normal derivative vanishes at $UV=0$. Hence as $\epsilon\to 0$
the horizon  $UV=0$
 does not contribute to the r.h.s. of  eq.\,\eqref{hsasaysay}. Further, since the dilaton factor $\re^D$ 
vanishes at
 the black hole singularity  $UV=1$
one could make 
 the whole boundary integral vanish
 by imposing that both the eigenfunctions and their normal derivatives remain finite at 
$UV=1$. In this case the wave functions corresponding to different energies would be orthogonal w.r.t. the
inner product
\bea\label{ususu}
\big\langle\Psi_1,\Psi_2\big\rangle= \int_{0< UV<1}\rd U\rd V\  \re^{D}\sqrt{-G}\  \Psi^*_1  \Psi_2\ .
\eea
This suggests to take $F^{(\sigma)}_{\rho,\nu}(z)$ in \eqref{isaisisaa}
as
\bea
F^{(\sigma)}_{\rho,\nu}(z)=z^{\ri \nu}\ 
{}_{2}F_{1}\big(\tfrac{1}{2}+\ri \nu+{ \rho},
\tfrac{1}{2}+\ri \nu-{\rho}, 1; 1-z)\ \ \ \ \ \ \ (\epsilon^2<z<1)
\eea
or, equivalently,
\bea
F^{(\sigma)}_{\rho,\nu}(z)&=&A_{\rho,+\nu}\  z^{+\ri \nu}\ {}_2F_{1}\big(\tfrac{1}{2}+ \ri \nu+{ \rho},
\tfrac{1}{2}+\ri \nu-{ \rho}, 1+ 2 \ri \nu; z)\\[0.2cm]
&+
&A_{\rho,-\nu}\  z^{ -\ri \nu}\ {}_2F_{1}\big(\tfrac{1}{2}- \ri \nu+{ \rho},
\tfrac{1}{2}-\ri \nu-{\rho}, 1- 2 \ri \nu; z)\ ,\nonumber
\eea
where
\bea
A_{\rho,\nu}=\frac{\Gamma(-2\ri \nu)}
{\Gamma(\frac{1}{2}-\ri \nu-{ \rho}) \Gamma(\frac{1}{2}-\ri \nu+{ \rho})}\ .
\eea
\bigskip

For $z\ll 1$ it is convenient to use the variable $y$ such that $z=\re^{y}$. Then
$F^{(\sigma)}_{\rho,\nu}$ asymptotically approaches to a superposition of two plane waves
 \bea
 F^{(\sigma)}_{\rho,\nu}\,\asymp\, A_{\rho,+\nu}\ 
\re^{+\ri \nu y}+A_{\rho,-\nu}\ \re^{-\ri \nu y}\ \ \ \ \ \ \ \ \  \big( 1\ll (-y)<2\,\log(1/\epsilon)\big)\ .
 \eea
The regularized   interaction discussed before in the domain $(-y)>2\,\log(1/\epsilon)$  results in
 a quantization condition for $\nu$ 
 \bea\label{ioasio1290}
 \epsilon^{-4\ri \nu}\ \re^{\frac{\ri }{2}\delta^{({\rm ms})}(\rho,\nu)}\,\asymp \,\sigma\ .
 \eea
 The phase shift $\delta^{({\rm ms})}$ here depends on the precise form of the regularized potential.
As $\epsilon\to 0$, the spectrum of $\nu$ becomes continuous and is
characterized by the  density of states
\bea\label{oas091errr}
\rho^{({\rm ms})}(\nu)=\tfrac{2}{\pi}\ \log(1/\epsilon)+\tfrac{1}{4\pi}\ \partial_\nu\,\delta^{({\rm ms})}(\rho,\nu)\ .
\eea
The corresponding minisuperspace wave functions would be orthogonal w.r.t. the inner product \eqref{ususu}:
\bea\label{asusausa}
\big\langle\Psi_{\rho',\nu'}^{(\sigma')},\Psi_{\rho,\nu}^{(\sigma)}\,\big\rangle\propto
\delta_{\rho',\rho}\ \delta_{\sigma',\sigma}\ \delta(\nu'-\nu)\ .
\eea
Here we use the Dirac $\delta$-function for $\nu$ since the latter
can be any real number. 
At the same time the Kronecker symbol indicates that $\rho$ belongs to  some  discrete set.
The quantization of $\rho$ seems rather natural once we note that  the term 
$\frac{\hbar}{\pi}\,(\rho^2-\frac{1}{4}) $ in 
the formula for the minisuperspace energy  \eqref{iassausa} can be interpreted as the contribution
of the non-cyclic degree of freedom $\Phi$, which executes  periodic motion in the regularized 
effective potential. This is consistent with our general discussion of the quantization of the 
Lorentzian black hole NLSM. Setting  ${\tt w}=0$ in formula \eqref{asudiui89129}
for the admissible values of $\rho$ and $\bar{\rho}$, one has
$
2\rho=2\bar{\rho}={\tt v}=0,1,2,\ldots\ $.
Also $\delta_{\sigma',\sigma}$ in \eqref{asusausa} can be ignored --
 the sign factor $\sigma$ is not an independent quantum number
and is related to the parity of the integer ${\tt v}$ (see \eqref{sigmadef1aaa}).

\bigskip

Our analysis  within
 the minisuperspace approximation
 is in  agreement  with the conjectured link
between the Lorentzian black
hole NLSM and the ${\cal Z}_2$
invariant inhomogeneous six-vertex model in the 
scaling limit. Moreover it elucidates the occurrence
of the ``quantization condition'' \eqref{quantC1},
which plays a central r$\hat{{\rm o}}$le
in the description of the critical behaviour of the lattice system.
Namely, the density of states is not an immanent property of the
CFT, but rather, a result of the  regularization of the theory.
The lattice model provides a particular 
integrable regularization,  which yields a density of states as in
eqs.\,\eqref{aisodio12311},\,\eqref{rho01a} with $\epsilon\propto N^{-1}$ being the regularization parameter.
Note that the spectrum of states with pure imaginary $s$ depends on the precise form of the phase shift $\delta$.
We interpret them as
non-normalizable virtual states appearing as a result of the regularization
and not belonging to the set of normalizable states from the pseudo-Hilbert space of the Lorentzian black hole NLSM.

\pagebreak

\section{Partition function for the Euclidean black hole NLSM}

In the work \cite{Ikhlef:2011ay} the 
authors put forward the pioneering conjecture that the
Euclidean black hole NLSM is
the  CFT governing the scaling limit of the ${\cal Z}_2$ invariant
inhomogeneous six-vertex model in the domain 
of the anisotropy parameter $\arg(q^2)\in(0,\pi)$. This is not quite in line with the results
of our study.
Here
we'd like to critically re-examine the arguments 
from ref.\cite{Ikhlef:2011ay}.

\bigskip
Let's recall the definition of the Euclidean black hole NLSM.
The corresponding target space metric has Euclidean signature and is given by
 \bea\label{hsasaysaaaty}
(\rd\sigma_{\rm\scriptscriptstyle EBH})^2=\frac{{\rm d} U{\rm d} U^*}{1+UU^*}\ .
\eea
It may be obtained from the metric $(\rd\sigma)^2$ \eqref{hsasaysaty}
in the following way.
For the co-ordinates $U$ and $V$ taking values  in region I from fig.\,\ref{iaosi1212},
one performs the ``Wick rotation'' in the target space, which makes them satisfy the
reality condition
 \bea\label{iaosid12123}
 V=-U^*\ .
 \eea
Then ignoring the overall negative sign, the metric
 \eqref{hsasaysaty} becomes  $(\rd\sigma_{\rm\scriptscriptstyle EBH})^2$.
The classical action for the NLSM is now given by
 \be\label{iasdioi12014sa3}
S_{\rm \scriptscriptstyle EBH}=\frac{1}{2\hbar}\int\rd t \int_0^{2\pi}\rd x\ \
\frac{\partial_t U{\partial}_t U^*-\partial_x U{\partial}_x U^*}{1+UU^*}\qquad\quad
\qquad \big(\,\hbar\to 0^+\,\big)\ .
\ee
In this case instead of imposing periodic boundary conditions,
it is useful to consider the more general quasiperiodic ones
\bea\label{iasod129102}
U(t,x+2\pi)=\re^{2\pi\ri {\tt k}}\ U(t,x)\ .
\eea
The model \eqref{iasdioi12014sa3} possesses ${\rm U}(1)$ symmetry and the Noether current is given by
\bea\label{hsysya}
I_\mu=\frac{1}{2\ri}\ \frac{U^*\partial_\mu U-U^*\partial_\mu U}{1+UU^*}\ .
\eea
\bigskip

The Euclidean black hole NLSM has been well studied\cite{Elitzur:1991cb,Mandal:1991tz,Witten:1991yr,Dijkgraaf:1991ba,ZAM,Maldacena:2000hw,
Maldacena:2000kv,Hanany:2002ev,Ribault:2003ss,Schomerus:2005aq}. In particular,
the classical field theory \eqref{iasdioi12014sa3} still
possesses an infinite set of chiral currents, which form the 
classical $\overline{W}_\infty\otimes W_\infty$ Poisson algebra. 
The  quantization of the latter leads to 
the   algebra of extended conformal symmetry with central charge $c>2$.
The Hilbert space  can be classified according to the highest weight irreps of the
$\overline{W}_\infty\otimes W_\infty$\,-\,algebra.
It is convenient to parameterize the central charge and the highest weight 
of the irreps $\bm{\omega}=(\omega_2,\omega_3)$  using $n$, $s$ and $p$ as
 \bea\label{Deltvarpi1aas}
 c=2+\frac{6}{n}\, >\, 2
 \eea
and
 \bea\label{Deltvarpi1aasa}
\omega_2&=&\frac{s^2+\frac{1}{4}}{n}+\frac{p^2}{n+2}  \\[0.2cm]
\omega_3&=&
\frac{2p}{\sqrt{n+2}}\,\Big(\,\frac{s^2}{n}+\frac{(2+3 n)\,p^2}{3n\,(n+2)}-\frac{2n+1}{12\, n}\,\Big)\nonumber\ .
\eea
To avoid confusion let us emphasize that in these relations $n>0$, $s$ and $p$ are formal parameters,
without the meaning that was assigned to them in the previous sections.
The Hilbert space of the NLSM
 contains both a continuous ${\cal H}_{{\rm\scriptscriptstyle EBH}}^{({\rm cont})}$ 
and a discrete component ${\cal H}_{{\rm\scriptscriptstyle EBH}}^{({\rm disc})}$.
Let's first focus on the continuous one.
Its linear
decomposition into the irreps of the $\overline{W}_\infty\otimes W_\infty$\,-\,algebra is given by
\cite{Dijkgraaf:1991ba,ZAM,Maldacena:2000hw,Maldacena:2000kv,Hanany:2002ev,Ribault:2003ss,Schomerus:2005aq}
 \bea\label{iaosasdidoi192009AA}
{\cal H}_{{\rm\scriptscriptstyle EBH}}^{({\rm cont})}=
\bigoplus_{{\tt v},{\tt w}=-\infty}^{+\infty}\ \int^{\oplus}_{s>0}\!\rd s\
\overline{{\cal W}}_{\bar{p},s}^{(c>2)}\otimes {\cal W}_{{ p},s}^{(c>2)}\,,\qquad
 {\rm where}
\qquad 
\begin{array}{l} p=\frac{1}{2}\, {\tt v}+\frac{1}{2}\,(n+2)\,({\tt k}+{\tt w})\\[0.2cm]
                                       \bar{p}=\frac{1}{2}\,{\tt v} -\frac{1}{2}\,(n+2)\,({\tt k}+{\tt w})
\end{array}
 \eea
Here ${\tt v}$ 
is the eigenvalue of the ${\rm U}(1)$ conserved charge 
$\hbar^{-1}\oint\rd x I_0$ associated with the Noether current \eqref{hsysya}.
It takes integer values 
 provided that the Planck constant is 
identified with $n$ as 
\be
\hbar=\frac{2\pi}{n+2}\ .
\ee
The integer ${\tt w}$
may be interpreted as 
a winding number related to the fact that the boundary condition \eqref{iasod129102}
is invariant w.r.t. the substitution ${\tt k}\mapsto {\tt k}+{\tt w}$
 with ${\tt w}\in\mathbb{Z}$. Let us note that 
the highest weight \eqref{Deltvarpi1aasa} is not sensitive to the sign of $s$.
Due to this the direct integral in  \eqref{iaosasdidoi192009AA}
is restricted to positive values of $s$.
 For the states at the level $\bar{\tt  L}$ and ${\tt L}$ 
in the irrep
$\overline{{\cal W}}_{\bar{p},s}^{(c>2)}\otimes {\cal W}_{{ p},s}^{(c>2)}$, the corresponding energy 
$E=\Delta+{\bar\Delta}-\frac{c}{12}$
in terms of the parameters $n$, $s$, $p$  and $\bar{p}$ reads  as
 \bea\label{iaosid1212}
 E=-\frac{1}{6}+\frac{2s^2}{n}+\frac{p^2+{\bar p}^2}{n+2}+{\tt L}+{\bar {\tt L}}\ .
 \eea

\bigskip

The study of the low energy spectrum of the
Hamiltonian $\mathbb{H}$ \eqref{aioiisa} with $\arg(q^2)\in(0,\pi)$
was initiated in the work \cite{Jacobsen:2005xz}. 
Within the Bethe ansatz approach,  the
leading $1/N$ correction to the energy 
was considered. 
Formula \eqref{tower1a} was obtained, where
$p$ and $\bar{p}$ are given in \eqref{oaisoi1093}, while $b(N)$
is the eigenvalue of the quasi-shift operator \eqref{poapso1a}.
Then it was  understood in \cite{Ikhlef:2008zz} that the scaling limit of the low energy states
could be organized so that 
$b(N)$ is replaced by the RG invariant $s$, which becomes a continuous parameter
in the scaling limit (see also the discussion in sec.\,\ref{sec31} from this paper). 
The observation that the universal correction term in \eqref{tower1a}
coincides with \eqref{iaosid1212}
was among the original arguments that the critical
behaviour of the lattice system is described by the Euclidean black hole NLSM.
With such an identification the ${\rm U}(1)$ symmetry of the action \eqref{iasdioi12014sa3} is 
interpreted as the counterpart of the lattice ${\rm U}(1)$ symmetry, so that the quantum number
${\tt v}$ coincides with $S^z$. Needless to say that ${\tt k}$ in the twisted
boundary conditions \eqref{iasod129102} corresponds to the
twist parameter ${\tt k}$ from \eqref{BC1a} \cite{Candu:2013fva}.
\bigskip

There are two immediate concerns to the above identification.
The first is regarding the ${\cal Z}_2$ symmetry of the lattice model.
In the scaling limit, the states 
related through the ${\cal Z}_2$ transformation are 
characterized  by the RG invariant $+s$ and $-s$  and
should be considered as distinct states in the Hilbert space of the CFT.
On the other hand,  the highest weight irrep
$\overline{{\cal W}}_{\bar{p},s}^{(c>2)}\otimes{\cal W}_{p,s}^{(c>2)}$ is identical to
$\overline{{\cal W}}_{\bar{p},-s}^{(c>2)}\otimes{\cal W}_{p,-s}^{(c>2)}$
 and in the Euclidean black hole NLSM the states with $\pm s$
must be identified. For this reason the domain of integration in \eqref{iaosasdidoi192009AA}
is $s>0$. The second concern is that the NLSM is a unitary field theory. Its
Hilbert is equipped with a positive definite inner product \cite{Dixon:1989cg} such that the Fourier modes
of the $W$ and $\overline{W}$ currents, generating the  $\overline{W}_\infty\otimes W_\infty$\,-\,algebra, 
satisfy the conjugation conditions
 \bea
\big[\widetilde{W}_j(m)\big]^\dagger=\widetilde{W}_j(-m)\,,\qquad \big[\,  \widetilde{\overline{W}}_j(m)\, \big]^{\dagger}=
 \widetilde{\overline{W}}_j(-m)\ .
\eea
Contrary to this, since the spectrum of the lattice Hamiltonian $\mathbb{H}$ is not real,
there does not exist a positive definite inner product for the lattice system w.r.t. which the matrix $\mathbb{H}$
would be Hermitian.
\bigskip

In principle, the objections may be addressed as follows. Instead of considering
the full Hilbert space ${\cal H}$ occurring in the scaling limit of the spin chain, one could
focus on its ${\cal Z}_2$ invariant sector and identify this with the 
space of states of  the Euclidean black hole NLSM.
 Also, it is a rather common
situation when the lattice (regularized) system is equipped with  a non-positive definite 
inner product, but unitary is restored in the scaling limit. 
\bigskip

In the consequent work \cite{Ikhlef:2011ay} an additional argument
was presented in support of the relation between the lattice model
and the quantum Euclidean black hole NLSM.
It  uses the results of refs.\,\cite{Maldacena:2000kv,Hanany:2002ev}.
In these papers  an explicit formula was presented for the  partition function 
of the black hole NLSM with periodic boundary conditions (${\tt k}=0$) and
 the Euclidean world-sheet compactified on the torus.
It was argued that the contribution of the continuous spectrum to the partition function
takes the from (see formula (4.17) from ref.\cite{Hanany:2002ev})
\be\label{pasod190291}
Z_{{\rm\scriptscriptstyle EBH}}^{({\tt k}=0)}
=\sum_{{\tt v},{\tt w}=-\infty}^\infty\int_{0}^{\infty}
\rd s\, \rho(s)\
 {\rm ch}_{\bar{p},s}(\bar{{\tt q}})\,{\rm ch}_{p,s}({\tt q})\ +\ \ldots\ ,
\ee 
where the contribution  of the discrete spectrum is denoted by the ellipsis.
The product ${\rm ch}_{\bar{p},s}(\bar{{\tt q}})\,{\rm ch}_{p,s}({\tt q})$ is 
 the character of the highest weight irrep 
$\overline{{\cal W}}_{\bar{p},s}^{(c>2)}\otimes {\cal W}_{{ p},s}^{(c>2)}$   appearing in the decomposition \eqref{iaosasdidoi192009AA}
and explicitly
\be
{\rm ch}_{p,s}({\tt q})=
{\tt q}^{-\frac{1}{12}+\frac{s^2}{n}+\frac{p^2}{n+2}}
\ ({\tt q},{\tt q})_\infty^{-2}\ .
\ee
The density of states reads as
\be\label{iasoid10928}
\rho(s)=
\frac{2}{\pi}\, \log (1/\epsilon)+\frac{1}{2\pi\ri}\ 
\partial_s\log\bigg[
\, \frac{\Gamma(\frac{1}{2}+p-{\ri s})\,\Gamma(\frac{1}{2}+{\bar p}-{\ri s})}
{\Gamma(\frac{1}{2}+p+{\ri s})\,\Gamma(\frac{1}{2}+{\bar p}+{\ri s})}\,\bigg]+o(1)\,,
\ee
where $\epsilon^{-1}\gg 1$ is a regularization parameter
(for an explanation see  the original work \cite{Maldacena:2000kv}).
\bigskip

Based on a 
numerical  study of the Bethe ansatz equations,
the  observation was made in ref.\cite{Ikhlef:2011ay}
that with a proper understanding of the 
scaling limit, the density  of 
primary Bethe states
is given by the function \eqref{iasoid10928}.
The r$\hat{\rm o}$le of the 
  regularization parameter is played by the
number of lattice sites, i.e.,  $\epsilon^{-1}\propto N$.
Our interest in the ${\cal Z}_2$ invariant inhomogeneous
six-vertex model  was inspired by this remarkable observation. 
 
\bigskip

The arguments for formula \eqref{pasod190291}
rely essentially on the minisuperspace approximation.
We didn't find compelling reasons in the papers \cite{Maldacena:2000kv,Hanany:2002ev}
as to why the density of states remains the same for the excited states
with ${\tt L},\bar{\tt L}>0$.
At best, one might expect that formula \eqref{pasod190291} should be replaced by
\bea\label{asodi918921}
Z_{{\rm\scriptscriptstyle EBH}}&=&
\frac{1}{2}\ \sqrt{\frac{n}{\Im m(\tau)}}\ \ 
\frac{\log(1/\epsilon)}{\pi (\bar{{\tt q}},{\bar{\tt q}})_\infty^{2}({\tt q},{\tt q})_\infty^{2}}\ 
\sum_{{\tt v},{\tt w}=-\infty}^\infty
\bar{{\tt q}}^{-\frac{1}{12}+\frac{\bar{p}^2}{n+2}}\ 
{\tt q}^{-\frac{1}{12}+\frac{p^2}{n+2}} \\[0.4cm]
&+&
\!\!\!\!\sum_{{\tt v},{\tt w}\in\mathbb{Z}}\ \int_{0}^{\infty}\rd s\
\sum_{{\tt L},\bar{\tt L}\ge 0}{\rho}^{({\bar {\tt L}},{\tt L})}_{{\rm\scriptscriptstyle EBH}}(s\,|\,\bar{p},p)\ 
\bar{{\tt q}}^{-\frac{1}{12}+\frac{s^2}{n}+\frac{\bar{p}^2}{n+2}+\bar{{\tt L}}}\ 
{\tt q}^{-\frac{1}{12}+\frac{s^2}{n}+\frac{p^2}{n+2}+{\tt L}}+Z_{{\rm\scriptscriptstyle EBH}}^{(\rm disc)} \ .\nonumber
\eea
\bigskip

The divergent term in \eqref{asodi918921} admits a simple interpretation 
that has to do with the geometry of the target space of the NLSM.
The manifold equipped with the metric  $(\rd\sigma_{\rm\scriptscriptstyle EBH})^2$  \eqref{hsasaysaaaty}
may be embedded into three dimensional Euclidean space and visualized as a half-infinite cigar.
The tip is located at $U=0$ while in the domain
 $|U|\gg1$, where the metric becomes flat, the 
target manifold resembles a half-infinite cylinder.
The first term in \eqref{asodi918921} is the partition function
of two free bosons. One of them, $\arg(U)$, takes values in the interval $[-\pi,\pi]$,
and satisfies the quasiperiodic (if ${\tt k}\ne 0$) boundary conditions.
The other Bose field, $\log|U|$, takes values in a segment  of length $\propto\log(1/\epsilon)\to\infty$
as $\epsilon\to 0$. 
\bigskip

While the divergent term is somewhat universal the density matrix, whose matrix elements 
essentially coincide with
${\rho}^{({\bar {\tt L}},{\tt L})}_{{\rm\scriptscriptstyle EBH}}(s\,|\,\bar{p},p)$
 in \eqref{asodi918921},
depends on the IR regularization of the target manifold.
In the works \cite{Maldacena:2000hw,Maldacena:2000kv,Hanany:2002ev} the Euclidean black hole NLSM occurs in the context
of bosonic string theory on ${\rm AdS}_3$.
This provides a particular ``integrable'' IR regularization
for which the NLSM partition function
in the case of periodic boundary conditions   reads explicitly as\footnote{%
Formula (3.9) in \cite{Hanany:2002ev}
for the partition function contains an additional factor of $2$.
This is related to the fact that the corresponding NLSM was obtained
by gauging the  U(1) symmetry, ${\bf g}\mapsto {\bf h}\,{\bf g}\,{\bf h}$ $({\bf h}=\re^{\frac{\ri\alpha}{2}\sigma^y})$,
of the ${\rm SL}(2,{\mathbb R})$ WZW model.
This  results in two  copies of the Euclidean black hole NLSM
(see also the discussion in sec.\,\ref{sec231}).}
\be\label{parfun2}
Z_{{\rm \scriptscriptstyle EBH}}^{({\tt k}=0)}=\frac{\sqrt{n(n+2)}}{\Im m(\tau)}\,\sum_{{\tt a},{\tt b}\in\mathbb{Z}}\
\int_{D_{\epsilon}}\rd^2 z\
\re ^{-\frac{\pi(n+2)}{\Im m(\tau)}|z+{\tt a}+{\tt b}\tau|^2
+\frac{2\pi}{\Im m(\tau)}(\Im m(z))^2}\,
\left|\frac{\eta(\tau)}{\vartheta_1(z |\tau)}\right|^2\,.
\ee
Here   $\vartheta_1$ and $\eta$ are the standard 
elliptic theta and Dedekind eta functions:
\bea
\vartheta_1(u|\tau)&=&2 {\tt q}^{\frac{1}{8}}\,\sin(\pi u)
\ (\re^{2\pi\ri u}\,{\tt q},{\tt q})_\infty \ (\re^{-2\pi\ri u}\,{\tt q},{\tt q})_\infty \ ({\tt q},{\tt q})_\infty
  \\[0.4cm]
\eta(\tau)&=&{\tt q}^{\frac{1}{24}}\,({\tt q},{\tt q})_{\infty} \ \qquad \qquad \qquad\qquad\qquad\qquad
\qquad \qquad (\,{\tt q}=\re^{2\pi\ri \tau}\,)\ .\nonumber
\eea
The integral in \eqref{parfun2} is taken over the parallelogram $D$
in the complex $z$ plane 
with vertices at $z=\pm \half \pm \half\,\tau$. However since the
integrand is singular at $z=0$, a small neighbourhood
around the origin, whose size is controlled by the
parameter $\epsilon$, should be excluded from the integration domain.
For instance if one chooses
\be\label{iaosid182981}
D_\epsilon=D/\{z:\ |z|<\tfrac{1}{2\pi}\, \re^{-\gamma_{{\rm E}}}\,\epsilon\,\}\ ,
\ee
where $\gamma_{\rm E}$ denotes the Euler constant,
then  for $|{\tt q}|\to 0$
\be
Z_{{\rm \scriptscriptstyle EBH}}^{({\tt k}=0)}=\frac{1}{2\pi}\, \sqrt{\frac{n}{\Im m(\tau)}}\  |{\tt q}|^{-\frac{1}{6}} \ 
\Big(\log(4\re^{\gamma_{\rm E}}/\epsilon)+o\big(|{\tt q}|^0\big)\Big)\ .
\ee
This is consistent with 
formulae
 \eqref{pasod190291} and \eqref{iasoid10928}.
We define 
the regularized partition function of  the Euclidean black hole NLSM as
\be\label{asodi1288912}
Z_{{\rm \scriptscriptstyle EBH},{\rm reg}}^{({\tt k}=0)}= \lim_{\epsilon\to 0}\ 
\big(\,Z_{{\rm \scriptscriptstyle EBH}}^{({\tt k}=0)} - Z^{({\rm sing})}_{\epsilon}
\,\big)
\ee
with
\be\label{ia89989888912kja}
Z^{({\rm sing})}_{\epsilon}=\sqrt{\frac{n}{\Im m(\tau)}}\ \ 
\frac{\log(4\re^{\gamma_{\rm E}}/\epsilon)+\frac{1}{2}\log\big(\Im m(\tau)\big)}
{2\pi\, (\bar{{\tt q}},{\bar{\tt q}})_\infty^{2}({\tt q},{\tt q})_\infty^{2}}\ 
\sum_{{\tt v},{\tt w}\in\mathbb{Z}}
\bar{{\tt q}}^{-\frac{1}{12}+\frac{\bar{p}^2}{n+2}}\ 
{\tt q}^{-\frac{1}{12}+\frac{p^2}{n+2}}
\ee
and recall that ${\tt q}=\re^{2\pi\ri\tau}$, $\bar{{\tt q}}=\re^{-2\pi\ri\tau^*}$.
Here an extra term $\propto \log\big(\Im m(\tau)\big)$ was included into the definition of 
$ Z^{({\rm sing})}_{\epsilon}$ in order to ensure that the regularized partition function,
$Z_{{\rm \scriptscriptstyle EBH},{\rm reg}}^{({\tt k}=0)}$,
is modular invariant.

\bigskip

Now that the partition function has been specified, the 
finite part in \eqref{asodi918921} is defined unambiguously. The explicit formula for
$Z_{{\rm\scriptscriptstyle EBH}}^{(\rm disc)}$, which accounts for the contribution
of the discrete spectrum in the black hole NLSM,
 was presented in ref.\cite{Ribault:2003ss}.
It appears to be identical with $\frac{1}{2}\,Z^{({\rm disc})}$ 
from \eqref{oaspdo121} specialized to ${\tt k}=0$.\footnote{%
Apart from an obvious typo, formulae (2.5) and (2.10) from ref.\cite{Ribault:2003ss}
do not quite  correctly take into account the contribution of the states to
$Z_{{\rm\scriptscriptstyle EBH}}^{(\rm disc)}$
with $\mathfrak{j}=-\frac{n+1}{2},\,-\frac{1}{2}$
corresponding to the boundary of the interval in the set $\mathfrak{J}({\tt v},{\tt u})$ \eqref{iaosid982981}.}
Then one may guess that
\bea\label{iaosidoaisd12121}
Z_{{\rm\scriptscriptstyle EBH}}=\tfrac{1}{2}\,{ Z}^{({\rm scl})}
\eea
provided  the
 regularization parameter $\epsilon$ is related to $N$ as
\be
\epsilon^{-1}=
\frac{2^{\frac{n+2}{n}}\,\Gamma\big(\frac{3}{2}+\frac{1}{n}\big)}
{\sqrt{\pi}\,\Gamma\big(1+\frac{1}{n}\big)}\  N
\ee
(here we use the explicit expression \eqref{N0altdef1} for the constant $N_0$).
The relation \eqref{iaosidoaisd12121}
was  confirmed numerically in the work \cite{Bazhanov:2020uju}.  For 
the reader's convenience, we quote some of the numerical data that
was presented in that paper.
Tab.\,\ref{tab01} compares 
$2Z_{{\rm \scriptscriptstyle EBH},{\rm reg}}^{({\tt k}=0)}$ 
with $Z_{\rm reg}^{({\rm cont})}+Z^{({\rm disc})}$, where the  NLSM partition function
is regularized as in \eqref{asodi1288912}, while
\bea\label{iaosdi198291}
Z_{\rm reg}^{({\rm cont})}&=&{\rm second\ line\ of\ eq.\,\eqref{iaosido12032}} \\[0.4cm]
&-&
\sqrt{\frac{n}{\Im m(\tau)}}\ \ 
\frac{\log(4\re^{\gamma_{\rm E}})+\frac{1}{2}\log\big(\Im m(\tau)\big)}
{\pi\, (\bar{{\tt q}},{\bar{\tt q}})_\infty^{2}({\tt q},{\tt q})_\infty^{2}}\ 
\sum_{S^z\!,{\tt w}\in\mathbb{Z}}
\bar{{\tt q}}^{-\frac{1}{12}+\frac{\bar{p}^2}{n+2}}\ 
{\tt q}^{-\frac{1}{12}+\frac{p^2}{n+2}}\ .\nonumber
\eea
\bigskip

In ref.\cite{Bazhanov:2020uju}  a numerical study of $Z^{({\rm scl})}$
 is also performed for ${\tt k}\ne 0$.
It is found that the relation similar to  \eqref{iaosidoaisd12121} holds, but with
a simple modification of the formula \eqref{parfun2}:
\be\label{aois8912hkf}
Z^{({\tt k})}=\frac{\sqrt{n(n+2)}}{\Im m(\tau)}\,\sum_{{\tt a},{\tt b}\in\mathbb{Z}}\
\int_{D_{\epsilon}}\rd^2 z\  
\re ^{-\frac{\pi(n+2)}{\Im m(\tau)}|z+{\tt a}+({\tt k}+{\tt b})\,\tau|^2
+\frac{2\pi}{\Im m(\tau)}(\Im m(z))^2}\,
\left|\frac{\eta(\tau)}{\vartheta_1(z |\tau)}\right|^2\,.
\ee
This is illustrated in tab.\,\ref{tab02}. Thus it is expected that with a proper regularization
of the Euclidean black hole NLSM, \eqref{iaosidoaisd12121} holds true for any values
of the twist parameter ${\tt k}$ and furthermore
\be
Z_{{\rm \scriptscriptstyle EBH}}=Z^{({\tt k})}=\tfrac{1}{2}\,Z^{({\rm scl})} \ .
\ee
A one\,-\,loop calculation, which is almost identical to that from \cite{Maldacena:2000kv},
supports that $Z_{{\rm \scriptscriptstyle EBH}}$ for twisted boundary conditions
indeed coincides with \eqref{aois8912hkf}.
\bigskip

The highly non-trivial formula  \eqref{iaosidoaisd12121} is in full  
agreement with the original observation of ref.\cite{Ikhlef:2011ay}.
However, let's emphasize that in order to state
that the Euclidean black hole NLSM governs the critical behaviour of the
${\cal Z}_2$ invariant inhomogeneous six-vertex model, this relation is insufficient.
Our numerical study of the finite size corrections to the CFT Hamiltonian, which
 are controlled by irrelevant perturbations (see eqs.\,\eqref{asympeq1a} and \eqref{Edifeq1}), 
shows that the extended conformal symmetry algebra is the 
$\overline{W}_\infty\otimes W_\infty$\,-\,algebra with $c<2$.
 Accepting the latter also naturally resolves  the issues with the 
${\cal Z}_2$ symmetry and unitarity mentioned above.

\bigskip

\begin{table}
\centering
\scalebox{0.91}{
\begin{tabular}{|c|c|c|c|c|}
\hline
 & & & & \\[-0.3cm]
$\tau$ &$Z^{\rm (cont)}_{\rm reg}$ &$Z^{\rm (disc)}$&
$Z^{\rm (cont)}_{\rm reg}+Z^{\rm (disc)}$&
$2Z_{{\rm \scriptscriptstyle EBH},{\rm reg}}^{({\tt k}=0)}$\\[.2cm]
\hline
\hline
 & & & & \\[-0.3cm]
$\tau=.9 \ri$ &$ -3.9509313$ &$ 0.0210525$ &$-3.9298787$&$-3.9298786$
\\[.2cm]
$-1/\tau$&  $-3.9358543 $& $0.0059766$ &$-3.9298776$ & $-3.9298787$ \\[.2cm]
$\tau+1$& $-3.9509313$ &$ 0.0210525$ & $-3.9298787$ &$-3.9298786$\\[.2cm]
\hline
\hline
 & & & & \\[-0.3cm]
$\tau=.2+.9\ri$&$-3.8983544 $&$0.0065418 $&$-3.8918125$&$-3.8918125$\\[.2cm]
$-1/\tau$&$ -3.8925978 $&$ 0.0007853 $&$
   -3.8918125 $&$-3.8918124$\\[.2cm]
$\tau+1$&$ -3.8983544 $&$ 0.0065418 $&$
   -3.8918125 $&$-3.8918124$\\[.2cm]
\hline
\hline
 & & & & \\[-0.3cm]
$\tau=.66\ri$&$-4.4682528$&$0.0943594$&$-4.3738934$&$-4.3738934$\\[.2cm]
$-1/\tau$&$-4.3744476$&$0.0005542$&$-4.3738934$&$-4.3738933 $\\[.2cm]
$\tau+1$&$-4.4682528$&$0.0943594  $&$-4.3738934  $&$-4.3738933  $\\[.2cm]
\hline
\hline
 & & & & \\[-0.3cm]
$\tau=.5\ri$&$-5.7668560$&$0.2960118$&$-5.4708441$&$-5.4708421  $\\[.2cm]
$-1/\tau  $&$-5.4708761$&$0.0000322$&$-5.4708439$&$-5.4708437  $\\[.2cm]
$\tau+1$&$-5.7668560$&$0.2960118$&$-5.4708441$&$ -5.4708421 $\\[.2cm]
\hline
\hline
 & & & & \\[-0.3cm]
$\tau=.33\ri$&$-12.070612$&$1.5569389$&$-10.513673$&$-10.5129976$\\[.2cm]
$-1/\tau$&$-10.513561$&$7.662\cdot 10^{-8}$&$-10.513561$&$-10.5135606$\\[.2cm]
$\tau+1$&$-12.070612$&$1.5569389$&$-10.513673$&$-10.5129975  $\\[.2cm]
\hline
\end{tabular}
}
\label{tab2}
\caption{\small A comparison of the numerical data for $n=3$ of twice the 
regularized partition function of the Euclidean black hole NLSM 
\eqref{asodi1288912} with  $Z_{\rm reg}^{({\rm cont})}+Z^{({\rm disc})}$
for the lattice model with
periodic boundary conditions (${\tt k}=0$).
Here $Z^{({\rm disc})}$ is given by eqs.\,\eqref{iaosid982981}-\eqref{aoisdo1902asas}, while
$Z_{\rm reg}^{({\rm cont})}$ is defined by  \eqref{iaosdi198291}.
The table also illustrates modular invariance of the
regularized partition function for ${\tt k }=0$.
Note that in order to achieve good accuracy for decreasing values of  $\Im m(\tau)$ 
one must take into account an increasing number of terms in the sum over $S^z$, ${\tt w}$
for $Z^{({\rm cont})}$  as well as ${\tt a}$, ${\tt b}$ in eq.\,\eqref{parfun2}.
This significantly increases the computer time.
\label{tab01}}
\end{table}
\begin{table}
\begin{center}
\scalebox{0.91}{
\begin{tabular}{|c|c|c|c|c|}
\hline
 & & & & \\[-0.4cm]
$\tau$ & $Z_{\rm reg}^{({\rm cont})}$ & $Z^{({\rm disc})}$ & 
$Z_{\rm reg}^{({\rm cont})}+Z^{({\rm disc})}$ 
& $2Z_{\rm reg}^{({\tt k})}$ \\[.1cm]
\hline
&&&& \\[-0.3cm]
$0.9\,\ri$&$-3.1430392$&$0.0233941$&$-3.1196452$&$-3.1196450 $\\[.2cm]
\hline
&&&& \\[-0.3cm]
$0.2+0.9\,\ri$&$-3.0646040$&$0.0099983$&$-3.0546057$&$-3.0546064 $\\[.2cm]
\hline
&&&& \\[-0.3cm]
$0.66\,\ri$&$-3.7836669$&$0.1033699$&$-3.6802970$&$-3.6802972 $\\[.2cm]
\hline
&&&& \\[-0.3cm]
$0.2 + 0.66\,\ri$ & $-3.5074556$ & $0.0418838$ & $-3.4655718$ & $ -3.4655717 $\\[.2cm]
\hline
&&&& \\[-0.3cm]
$0.50\,\ri$&$-5.1054421$&$0.3209649$&$-4.7844771$&$-4.7844724 $\\[.2cm]
\hline
&&&& \\[-0.3cm]
$0.33\,\ri$&$-11.2855973$&$1.6391928$&$-9.6464045$& $-9.646289 $\\[.2cm]
\hline
&&&& \\[-0.3cm]
$0.25\,\ri$&$-26.5761236$&$5.4010183$&$-21.1751053$&$ -21.171536 $\\[.2cm]
\hline
\end{tabular}
}
\caption{\small
The last column contains numerical data for $2Z_{\rm reg}^{({\tt k})}$,
where 
$Z_{\rm reg}^{({\tt k})}=\lim_{\epsilon\to 0}\big(Z^{({\tt k})}-Z^{({\rm sing})}_{\epsilon}\big)$.
Here  $Z^{({\tt k})}$ is given by \eqref{aois8912hkf}, while $Z^{({\rm sing})}_{\epsilon}$  is
defined in \eqref{ia89989888912kja} with $p$, $\bar{p}$
 taken to be as in \eqref{iaosasdidoi192009AA}.
This is compared to the numerical values for
$Z_{\rm reg}^{({\rm cont})}+Z^{({\rm disc})}$,
where $Z^{({\rm disc})}$ was computed using eqs.\,\eqref{iaosid982981}-\eqref{aoisdo1902asas} and
$Z_{\rm reg}^{({\rm cont})}$ via  \eqref{iaosdi198291}.
The parameters 
were set to be ${\tt k}=-0.1$ and $n=3$. 
\label{tab02}}
\end{center}
\end{table}

\pagebreak

 \section{Gauged ${\rm SL}(2,\mathbb{R})$ WZW model\label{sec23}}

One should keep in mind the different status of the 
Euclidean and Lorentzian black hole NLSMs. The former is
a well defined quantum theory, and there are many ways to check its
consistency, including at the level of the conformal bootstrap 
\cite{Dijkgraaf:1991ba,ZAM,Maldacena:2000hw,
Maldacena:2000kv,Hanany:2002ev,Ribault:2003ss,Schomerus:2005aq}.
Contrary to this the status of the quantum Lorentzian 
NLSM is rather tentative. Our conjecture
is an attempt at  assigning a meaning
to the NLSM, which goes beyond the scope of the classical field theory.
It also provides one with a handle on how to proceed with the quantization
of some closely related models.

\subsection{The classical field theory\label{sec231}}

As was already mentioned the Lorentzian black hole NLSM can be obtained
by gauging 
a  non-compact one dimensional subgroup
of the classical ${\rm SL}(2,\mathbb{R})$ 
WZW model \cite{Witten:1991yr,Gawedzki:1988nj}.  Following the work \cite{Witten:1991yr}
consider the classical action
\bea\label{ahasysysadd}
S\!&=&\!\frac{1}{\hbar}\int\!\rd t\!\int_0^{2\pi}\!\rd x\,
\Big[\,
\partial U{\bar\partial V}+{\bar\partial U}\partial V+
\partial X{\bar\partial Y}+{\bar\partial X}\partial Y+\log(X/Y)\, \big(\partial U{\bar\partial V}-{\bar\partial U}\partial V\big)\\[0.2cm]
&+&\!{\bar a}\,\big(Y\partial X-X\partial Y-U\partial V+V\partial U\big)+
{ a}\, \big(Y{\bar \partial} X-X{\bar \partial} Y+U{\bar \partial} V-V{\bar \partial} U\big)-2 a{\bar a } \, (1-UV)
\,\Big] 
 .\nonumber
\eea
Here the integrand in the  first line  is just the classical Lagrangian density  of the  usual WZW model \cite{Witten:1983ar},
${\cal L}_{\scriptscriptstyle{\rm WZW}}[{\bf g}]$,
expressed via the
matrix entries of the  fundamental WZW   field
\bea\label{gaastsat}
{\bf g}=\begin{pmatrix}
X&U\\
-V&Y
\end{pmatrix}\ .
\eea
Note that
the term involving $\log(X/Y)$ comes from the Wess-Zumino term and, up to a total derivative, can be rewritten in various
ways by employing the constraint  
\bea\label{usausau}
XY+UV=1\ .
\eea
The second line  in  \eqref{ahasysysadd} contains the  fields $a$ and ${\bar a}$,
which  are the chiral components of
the gauge potential  $a_\mu$.  The action
is invariant w.r.t. the infinitesimal gauge  transformation of the form
\bea\label{aisiasi}
\delta X=\delta\omega\, X\ ,\ \ \  \delta Y=-\delta\omega\, Y\ ,\ \ \ \delta U=\delta V=0\ ;\ \ \ \delta a_\mu=\partial_\mu(\delta \omega)\ .
\eea
This can be seen by rewriting
the  Lagrangian density  corresponding to the action  \eqref{ahasysysadd} as
\bea\label{aoisid129812}
{\cal L}=\frac{1}{2}\
\bigg[\
\frac{\partial_\mu U\partial^\mu V}{1-UV}
- (1-UV)\  f_\mu f^\mu+\epsilon^{\mu\nu}\partial_\mu C_\nu
\,\bigg]
\eea
with
\bea\label{oasosaoas}
f_\mu=a_\mu-  \tfrac{1}{2}\  \partial_\mu  \log(X/Y)-\epsilon_{\mu\nu}\, J^\nu\ ,\ \  \ \ \ \ \ \ 
C_\mu=\tfrac{1}{2}\ \log(X/Y)\, (U\partial_\mu V-V\partial_\mu U)
\eea
and
\bea\label{hsysya11111}
J_\mu=\frac{1}{2}\ \frac{U\partial_\mu V-V\partial_\mu U}{1-UV}
\eea
(here $J^0=J_0,\ J^1=-J_1$ while
the Levi-Civita symbol  $\epsilon_{\mu\nu}=-\epsilon^{\mu\nu}$  is defined to be  $\epsilon_{01}=-\epsilon_{10}=1$).
The extremum condition $\frac{\delta }{\delta a_\mu} S=0$ leads to the equation
\bea\label{isisai1a}
a_\mu= \tfrac{1}{2}\  \partial_\mu  \log(X/Y) +\epsilon_{\mu\nu}\, J^\nu\ .
\eea
The field strength corresponding to this vector potential is given by
\bea
\partial_\mu a_\nu-\partial_\nu a_\mu=\partial_\sigma J^\sigma\ \epsilon_{\mu\nu}\ .
\eea
It vanishes 
for any solution of the classical equations of motion, 
which includes  the continuity equation $\partial_\mu J^\mu=0$.
\medskip

 In the orthodox formulation of the  gauged  ${\rm SL}(2,\mathbb{R})$ WZW model,
 the matrix valued field ${\bf g}$
 is  assumed to be  periodic:
 \bea\label{oisaiasias}
{\bf g}(t,x+2\pi)={\bf g}(t,x)\ .
 \eea
 If  we take $U$ and $V$  from the domain 
 \bea
 0\leq UV<1\ ,
 \eea
 it is natural to fix the
 gauge   by setting $X=Y$ \cite{Witten:1991yr} which, in view of eq.\,\eqref{isisai1a},
results in
 $a_\mu=\epsilon_{\mu\nu}\, J^\nu$.
Then, after eliminating the auxiliary field $a_\mu$, the action
$S$ \eqref{ahasysysadd}  becomes   that of the Lorentzian  black hole NLSM \eqref{iasdioi120143}.
Note that, as was also pointed out in \cite{Witten:1991yr},
 if  we take the  ${\rm SL}(2,\mathbb{R})$ picture literally the 
full target space of the Lorentzian black hole NLSM would contain two copies
of the  regions III and IV in fig.\,\ref{iaosi1212} corresponding to the cases $X,Y>0$ and $X,Y<0$.
In what follows we'll consider the same field theory, but
with more general boundary conditions than \eqref{oisaiasias}. It is expected to be
 applicable  for the description of the critical behaviour of the ${\cal Z}_2$ invariant
inhomogeneous six-vertex model with twisted boundary conditions.
\medskip

The gauged  ${\rm SL}(2,\mathbb{R})$
WZW model possesses an alternative formulation \cite{Gawedzki:1988nj,Dijkgraaf:1991ba}.
Consider the  Lagrange density  which is just   the sum of that 
 of the WZW  model and  the massless Gaussian theory:
\bea
\label{kakaiaia}
{\tilde {\cal L}}={\cal L}_{\scriptscriptstyle{\rm WZW}}[{\bf G}]+2\, \partial \eta{\bar \partial}\eta\ .
\eea
The interaction  is introduced through the  constraints
\bea\label{aoasoasoas}
\Upsilon\equiv\half\ {\rm Tr}\big[\sigma^z\,\partial {\bf G}\, {\bf G}^{-1}\big]-\partial\eta=0\ ,\ \ \ \  \ 
\bar{\Upsilon}\equiv\half\ {\rm Tr}\big[\sigma^z\,{\bf G}^{-1}\, {\bar  \partial} {\bf G}\big]+{\bar \partial}\eta=0\ .
\eea 
If  the  infinitesimal gauge transformation of the WZW  field  and
 the massless Gaussian field  is defined as
\bea
\delta {\bf G}=\tfrac{1}{2}\ \big(\sigma^z{\bf G}+{\bf G}\sigma^z\big)\  \delta\omega\ ,\ \ \  \  
\ \ \ \partial_\mu\delta\eta=\epsilon_{\mu\nu}\partial^\nu\, (\delta \omega)\ ,
\eea
then
$\delta{\tilde {\cal L}}$ turns out to be a total derivative
provided the constraints \eqref{aoasoasoas} are imposed.
The  classical  field theory, thus defined,
is equivalent to the gauged WZW model governed by the action \eqref{ahasysysadd}.
In particular, for any  field configuration 
satisfying the equations of motion for  \eqref{kakaiaia},\,\eqref{aoasoasoas},
\bea\label{isisaiaa}
{\bf g}= \re^{\frac{1}{2} \omega\sigma^z}\, {\bf G}\, \re^{\frac{1}{2}\omega\sigma^z}\ ,\ \ \ \ \ \ \ \  \ \ 
a_\mu=-\epsilon_{\mu\nu}\partial^\nu \eta+\partial_\mu\omega
\eea
would be a solution of the Euler-Lagrange equations  associated with the action \eqref{ahasysysadd}.
Here $\omega$ is an arbitrary periodic function $\omega(t,x+2\pi)=\omega(t,x)$, which  
appears as a manifestation
of  the gauge
invariance of the model.
\medskip

To specify  the boundary conditions,
let us first recall some basic facts concerning the  phase space of the  WZW model 
(see, e.g., \cite{Witten:1983ar,Faddeev:1987ph,Fateev:1991aw}). 
The latter is conveniently  described in terms of  the left and right WZW currents,\footnote{Here and below we use the
notation ${\bf t}_A$ for the
$2\times 2$ real traceless matrices,
$$
{\bf t}_3=\begin{pmatrix}
1&0\\
0&-1
\end{pmatrix}
\ ,\ \ \ {\bf t}_+=\begin{pmatrix}
0&1\\
0&0
\end{pmatrix}\ ,\ \ \ \ 
{\bf t}_-=\begin{pmatrix}
0&0\\
1&0
\end{pmatrix}\ :\ \ \  \ \ \ \ \ \ \  [{\bf t}_A,{\bf t}_B]={f_{AB}}^C\,{\bf t}_C\ .
$$
Indices are raised and lowered via the Killing form defined as
$$\kappa_{AB}=\half\, {\rm Tr}[\,{\bf t}_A{\bf t}_B\,]\ ,  \ \ \  \ \ \ \ \ \kappa_{AC}\, \kappa^{CB}=\delta_A^B\ .$$}
\bea
\partial{\bf G}{\bf G}^{-1}=L^{A}{\bf  t}_A\ ,\ \ \ \ \ \ \  \ \ {\bf G}^{-1}{\bar \partial}{\bf G}={\bar R}^{A} {\bf t}_A\ ,
\eea
which satisfy the   closed set of  equal-time Poisson bracket relations:
\bea\label{isisailaoi}
&&\big\{L^A(t,x),L^B(t,y)\big\}=-\half\, \kappa^{AB}\,\delta'(x-y)-\half\, {f^{AB}}_{C}\, L^C(t,x)\,\delta(x-y)\nonumber\\[0.2cm]
&&\big\{{\bar R}^A(t,x),{\bar R}^B(t,y)\big\}=+\half\, \kappa^{AB}\,\delta'(x-y)+\half\, {f^{AB}}_{C}\, {\bar R}^C(t,x)\,\delta(x-y)\ \ \ \ \\[0.2cm]
&&\big\{L^A(t,x),R^B(t,y)\big\}=0\ .\nonumber
\eea
Assuming that the currents are periodic fields,
\bea\label{aiosaisia}
L^A(t,x+2\pi)=L^A(t,x)\ ,\ \ \ \ \  \ \ \ \ \  \ \ {\bar R}^A(t,x+2\pi)={\bar R}^A(t,x)\ ,
\eea
the center of the  Poisson algebra is generated by two elements
\be\label{aoaosao}
{\mathfrak C}={\rm Tr}\bigg[{\overleftarrow{\cal P}}\exp\bigg(+\int_{x_0}^{x_0+2\pi}\rd x\,L^{A}\,{\bf t}_A\bigg)\bigg]\ ,\ \ \ \ 
\quad
{\bar {\mathfrak C}}={\rm Tr}\bigg[{\overrightarrow{\cal P}}\exp\bigg(-\int_{-x_0}^{-x_0-2\pi}\rd x\,{\bar R}^{A}\,
{\bf t}_A\bigg)\bigg]\ .
\ee
We will  focus on the field configurations such that  the   values of the central elements
are  restricted by the 
inequalities
\bea\label{jasuasu}
-2<{\mathfrak C},\,{\mathfrak{\bar C}}<2
\eea
and use the parameterization
\bea\label{aiuswqd891212}
{\mathfrak C}=2\cos(2\pi P)\ ,\  \ \ \ \ \ \ \ {\mathfrak{\bar C}}=2\cos(2\pi {\bar  P})
\eea
with real $P$ and $\bar{P}$.
In this case   the  path ordered exponentials  inside the traces in \eqref{aoaosao} may be expressed as
\bea\label{ususausa}
{\overleftarrow{\cal P}}\exp\bigg(+\int_{x_0}^{x_0+2\pi}\rd x\,L^{A}\,{\bf t}_A\bigg)&=&{\boldsymbol \Gamma}\,\re^{+2\pi\ri P \sigma^y}\,
{\boldsymbol \Gamma}^{-1}\\
{\overrightarrow{\cal P}}\exp\bigg(-\int_{-x_0}^{-x_0-2\pi}\rd x\,{\bar R}^{A}\,{\bf t}_A\bigg)&=&
{\bar {\boldsymbol \Gamma}}\,\re^{-2\pi\ri {\bar P}\sigma^y }\, {\bar {\boldsymbol\Gamma}}^{-1}\ ,\nonumber
\eea
where the $2\times 2$ real non-degenerate matrices ${\boldsymbol \Gamma }$ and 
${\bar {\boldsymbol \Gamma}}$
depend on the initial integration point $x_0$.
If we require them to be  ${\rm SL}(2,{\mathbb R})$  matrices,
 then $\re^{+2\pi\ri P \sigma^y}$ and $\re^{-2\pi\ri {\bar P}\sigma^y }$  are uniquely defined.
 At the same time there is an ambiguity in 
${\boldsymbol \Gamma }$ and ${\bar {\boldsymbol \Gamma}}$ of the form
 ${\boldsymbol \Gamma }\mapsto \pm {\boldsymbol \Gamma }\, \re^{\ri \gamma \sigma^y}$ and 
${\bar {\boldsymbol \Gamma}}\mapsto
  \pm {\bar {\boldsymbol \Gamma}}\,\re^{\ri {\bar \gamma} \sigma^y}$ with arbitrary real $\gamma$ 
and ${\bar\gamma}$.
  This   can be fixed using the Iwasawa decomposition for
${\rm SL}(2,{\mathbb R})$ matrices, which
  allows one to specify that
  \bea\label{usaususa}
  {\boldsymbol \Gamma }=\begin{pmatrix}
  d&0\\
  0&d^{-1}
  \end{pmatrix}\ \begin{pmatrix}
  1& b\\
  0&1
  \end{pmatrix}\ ,\ \ \ \ \ \ \ \ \bar{{\boldsymbol \Gamma }}=\begin{pmatrix}
  {\bar d}&0\\
  0&{\bar d}^{-1}
  \end{pmatrix}\ \begin{pmatrix}
  1& {\bar b}\\
  0&1
  \end{pmatrix}\ \ \ \ \quad {\rm with}\ \ \  \quad d,{\bar d}>0\ .
  \eea

\bigskip

The  values  of the currents at $t=0$ are not enough to fully define the  time dependence
of the matrix valued field ${\bf G}(t,x)$.  Indeed the equations  of motion
in the WZW model
are given by
 \bea
 {\bar \partial} L^A=0\ ,\ \ \ \ \ \ \ \ \ \partial {\bar R}^A=0\ .
 \eea
 This implies that 
\bea\label{hsyusyas}
{\bf G}(t,x)={\boldsymbol \Omega}(t+x)\,{\bf G}(0,x_0)\,{\bar{\boldsymbol\Omega}}(t-x)\ ,
\eea
where
\bea\label{ausauasuas}
{\boldsymbol \Omega} (u)&=&{\overleftarrow{\cal P}}\exp\bigg(+\int_{x_0}^{u}\rd x\,L^{A}\,{\bf t}_A\bigg)\\[0.2cm]
{\bar {\boldsymbol \Omega}}({\bar u})&=&{\overrightarrow{\cal P}}\exp\bigg(-\int_{-x_0}^{
{\bar u}}\rd x\,{\bar R}^{A}\,{\bf t}_A\bigg)\ ,
\nonumber
\eea
while  ${\bf G}(0,x_0)$ is an arbitrary ${\rm SL}(2,{\mathbb R})$ matrix. Its   entries, 
together with  the  initial values of the currents, constitute  the full set of the  initial data.
We consider the field configurations at $t=0$ to be such that
\bea\label{saoosao}
{\bf G}(0,x_0)={\boldsymbol \Gamma}\ \re^{\ri \alpha\sigma^y }\ {\bar {\boldsymbol \Gamma}}^{-1}\ ,
\eea
where ${\boldsymbol \Gamma},\, {\bar {\boldsymbol \Gamma}}$  are
the same as in \eqref{ususausa},\,\eqref{usaususa} and $\alpha$ is some real number.
This is motivated through the following arguments.
Assuming $L^A$ are given,
the path ordered exponent  ${\boldsymbol \Omega} (u)$ \eqref{ausauasuas}
solves  the linear   differential equation
\bea\label{aososao}
\partial {\boldsymbol \Psi}=L^{A}\,{\bf t}_A\, {\boldsymbol \Psi}\ .
\eea
However ${\boldsymbol \Omega}(u)$, apart from the WZW currents,
also depends on
an arbitrarily chosen initial integration  point $x_0$  at which   it becomes   the identity  matrix.
At the same time 
$\bm{\Psi}_P={\boldsymbol \Omega} (u)\,{\boldsymbol \Gamma}$  
is  the Floquet solution of  the matrix ODE \eqref{aososao},
which is fixed unambiguously provided 
${\boldsymbol \Gamma}$ is taken to be of the form \eqref{usaususa}.
A change in the initial integration point $x_0$ to $x_0'$ would result in 
the transformation 
$\bm{\Psi}_P\mapsto \bm{\Psi}_P\,\re^{\ri \alpha_0\sigma^y }$,
where $\alpha_0=\alpha_0(x_0,x_0')$.
The solutions of the ODE with periodic coefficients
 possess the band structure. Thus the parameter $P$ labeling the 
Floquet solutions $\bm{\Psi}_P$ 
can be defined such that $P\in\mathbb{R}$ and  $2P\notin{\mathbb Z}$,
where the band number coincides with the greatest integer less than  $P+\frac{1}{2}$.
 The above carries over to  the Floquet solution 
 $\overline{\bm{\Psi}}_{\bar{P}}=
{\bar {\boldsymbol \Gamma}}^{-1}\,{\bar {\boldsymbol \Omega}} (\bar{u})$ 
of the barred counterpart of the  ODE \eqref{aososao}.
 This way the  construction of the WZW field ${\bf G}(t,x)$ given  by eqs.\,\eqref{hsyusyas},\eqref{saoosao} 
 involves  the  Floquet solutions  as well as an additional variable 
 $\alpha\sim \alpha+2\pi$. 
  Thus  the algebra of functions on the phase space of the WZW model, generated by the 
  currents $L^A$ and ${\bar R}^A$  subject to the periodic boundary conditions
  \eqref{aiosaisia}, should be extended by the inclusion of  the compact  generalized coordinate $\alpha$. The latter  can be 
 viewed  as  a dynamical variable  
canonically conjugated to the sum $2\pi (P+{\bar P})$.
As for their difference, having in mind the study of the lattice model,
we assume that $\re^{2\pi\ri (P-{\bar  P})}=\re^{2\pi\ri {\tt k}}$,
with  $\half<{\tt k}\leq \half$ being a  fixed parameter.  
Equivalently,
\bea\label{isiasaisi}
P-{\bar P}={\tt k}+{\tt w}\ \ \ \ \  \ \ \ ({\tt w}\in{\mathbb{Z}})
\eea
and the
integer ${\tt w}$ labels different disjoint components of the phase space.

\bigskip

The 
 boundary values of the WZW field  at $t=0$, defined by the formulae  \eqref{hsyusyas} and \eqref{saoosao},
satisfy  the relations
\bea\label{aiosd89211}
{\bf G}(0,x_0+2\pi)={\boldsymbol \Gamma}\, \re^{2\pi\ri{\tt  k} \sigma^y }\, {\boldsymbol \Gamma}^{-1}\ {\bf G}(0,x_0)
={\bf G}(0,x_0)\ {\bar {\boldsymbol \Gamma}}\ \re^{2\pi\ri {\tt  k}\sigma^y}\ {\bar {\boldsymbol \Gamma}}^{-1}\ .
\eea
This  implies 
\bea\label{isisai}
{\rm{Tr}}\Big[\,{\bf G}(t,x+2\pi)\,\big({\bf G}(t,x)\big)^{-1}\,\Big]=2\,\cos(2\pi{\tt k})\ ,
\eea
which  should be imposed 
along  with the periodicity condition  for   the currents \eqref{aiosaisia}. 
In fact there is an extra condition that needs be taken into account.
Substituting the matrix ${\boldsymbol \Gamma}$ \eqref{usaususa} into 
eq.\eqref{aiosd89211}, one finds 
\bea\label{iasisisawwA}
{\rm{Tr}}\Big[\,(-\ri\,\sigma^y)\ {\bf G}(0,x_0+2\pi)\,\big({\bf G}(0,x_0)\big)^{-1}\,\Big]=\sin(2\pi {\tt k})\ \big(d^2+d^{-2}+d^2b^2\big)\ .
\eea
 This results in the inequality
 \bea\label{iasisisaww}
 {\rm{Tr}}\Big[\,(-\ri\,\sigma^y)\ {\bf G}(t,x+2\pi)\,\big({\bf G}(t,x)\big)^{-1}\,\Big]\big/\sin(2\pi {\tt k})>0\ .
 \eea

\bigskip

The constraints  \eqref{aoasoasoas}  will only make sense when  both derivatives $\partial\eta$ and ${\bar \partial}\eta$ are periodic:
\bea\label{aiasiasi}
\partial\eta(t,x+2\pi)=\partial\eta(t,x)\ ,\ \ \ \ \ \ \ \ \ \ \ \ {\bar \partial}\eta(t,x+2\pi)={\bar \partial}\eta(t,x)\ .
\eea
In view of the relation  \eqref{isisaiaa},  the gauge field $a_\mu(x,t)$ in the original
formulation of the gauged WZW model is also periodic,
\bea
a_\mu(t,x+2\pi)=a_\mu(t,x)\ ,
\eea
as was  implicitly  assumed  in our initial  analysis of the model.
The boundary condition \eqref{isisai} as well as  the inequality  \eqref{iasisisaww} 
are invariant under  the gauge transformation and  therefore
  the field 
${\bf g}$ satisfies the similar relations
\begin{subequations}\label{isisaiuy}
\bea\label{isisaiuyA}
&&{\rm{Tr}}\Big[\,{\bf g}(t,x+2\pi)\,\big({\bf g}(t,x)\big)^{-1}\,\Big]=2\,\cos(2\pi{\tt k})\\[0.2cm]\label{isisaiuyB}
&&{\rm{Tr}}\Big[\,(-\ri\,\sigma^y)\,{\bf g}(t,x+2\pi)\,\big({\bf g}(t,x)\big)^{-1}\,\Big]\big/\sin(2\pi {\tt k})>0\ .
\eea
\end{subequations}

\bigskip

Let us make the following important point.
In the case of the gauged ${\rm SL}(2,\mathbb{R})$ WZW model 
with ${\tt k}=0$, the  condition \eqref{isisaiuy}
yields   ${\bf g}(t,x+2\pi)={\bf g}(t,x)$, i.e., periodicity of
 all the matrix  elements $X,Y,U,V$. In turn one can 
 use  the gauge fixing condition
$X=Y$. 
However for 
${\tt k}\not=0$,
 since  $X$ and $Y$ are no longer periodic fields,  
the same gauge fixing condition is not applicable.
This makes  the model with   ${\tt k}=0$
(which is equivalent to the Lorentzian black hole NLSM) a very special one that  
is not obtainable literally     through
a  naive  ${\tt k}\to 0$ limit.
\bigskip

The Poisson structure of
 the massless Gaussian model, whose Lagrange density is given by
the second term in the r.h.s. of \eqref{kakaiaia},
reads as
\be\label{aisaiasi}
\{\partial\eta(t,x),\partial\eta(t,y)\}=- \{{\bar \partial}\eta(t,x),{\bar \partial}\eta(t,y)\}=\half\ \delta'(x-y)\ ,\ \ \ \ 
\{\partial\eta(t,x),{\bar \partial}\eta(t,y)\}=0\ .
\ee
With  the boundary conditions \eqref{aiasiasi} imposed,
the center  of this Poisson algebra is generated by
\bea\label{asodioaisd109212}
P_\eta=\int_0^{2\pi}\frac{\rd x}{2\pi}\ { \partial}\eta\ ,\ \ \ \ \ \qquad\qquad
{\bar P}_\eta=\int_0^{2\pi}\frac{\rd x}{2\pi}\  {\bar \partial}\eta\ .
\eea
The general solution of the equation of motion $\partial{\bar\partial}\eta=0$
is
\bea\label{isaisaia}
\eta(t,x)=\tfrac{1}{2}\, \big(f(t+x)-{\bar f}(t-x)\big)
\eea
where, in view of the boundary conditions, the
 arbitrary   functions $f$ and ${\bar f}$ are quasiperiodic:
\bea\label{sasausau}
f(u+2\pi)=f(u)+P_\eta\ ,\ \ \ \ \ \qquad \qquad \bar{f}({\bar u}+2\pi)={\bar f}({\bar u})+{\bar P}_\eta\ .
\eea
\smallskip

The constraints \eqref{aoasoasoas} imposed on the WZW field ${\bf G}$ and
the Gaussian field, combined with \eqref{isaisaia}, yield the relations
\bea
L^3=-\half\, \partial f\ ,\ \ \ \ \ \ \ \ \ \ \ {\bar R}^3=-\half\, {\bar \partial} {\bar f}\ .
\eea
It is  easy to see now that the matrix ${\bf G}$, 
satisfying the equations of motion,
can be brought to the form
\bea
{\bf G}(t,x)=\re^{-\frac{1}{2} f(t+x)\sigma^z}\, \bm{g}_{\text{\textonehalf}}(t,x)\, 
\re^{-\frac{1}{2}{\bar f}(t-x)\sigma^z}\ \,,
\eea
where   $\bm{g}_{\text{\textonehalf}}$ is  defined as in  eq.\,\eqref{oaop9203} specialized to the fundamental
representation of $\mathfrak{sl}_2$. Namely,
\bea
\partial {\bm g}_{\text{\textonehalf}}\ {{\bm g}_{\text{\textonehalf}}}^{-1}=
\xi_-{\bf t}_--\xi_+{\bf  t}_+\ ,\ \ \ \ \ \ \ \ \ \ \ \ 
{{\bm g}_{\text{\textonehalf}}}^{-1}\,
{\bar\partial}{\bm g}_{\text{\textonehalf}}={\bar \xi}_-{\bf  t}_--{\bar \xi}_+{\bf t}_+
\eea
with
\bea\label{jsusausa}
&&\xi_-=\re^{-f}\, L^-\ ,\ \ \ \qquad\qquad\xi_+=-\re^{+f}\ L^+\nonumber\\[0.2cm]
&&{\bar \xi}_-=\re^{+{\bar f}}\, {\bar R}^-\ ,\ \ \ \qquad\qquad{\bar \xi}_+=-\re^{-{\bar f}}\ {\bar R}^+\ .
\eea
The latter  are real  chiral fields, 
${\bar \partial}\xi_\pm=\partial {\bar\xi}_\pm=0$,
 subject to   the quasiperiodic boundary conditions
\bea\label{aususalk}
 \xi_\pm (u+2\pi)=B^{\pm 1}\ \xi_\pm (u)\ ,\ \ \ \ \ \ 
\qquad
\qquad{\bar \xi}_\pm ({\bar u}+2\pi)={\bar B}^{\pm 1}\ {\bar  \xi}_\pm ({\bar u})\,,
\eea
where $B=\re^{2\pi P_\eta}$ and ${\bar B}=\re^{2\pi {\bar P}_\eta}$.
Making contact with our  analysis of the lattice model, we set 
$B={\bar B}$ or, equivalently, $P_\eta= {\bar P}_\eta$ 
(assuming that $P_\eta$ and ${\bar P}_\eta$ are real).
In this case, as it follows from 
eqs.\,\eqref{isaisaia} and \eqref{sasausau},
the field $\eta$ is periodic:
\bea\label{hsahsay}
\eta(t,x+2\pi)=\eta(t,x)\,.
\eea
Note that  the on-shell gauge potential $a_\mu$, entering into  the initial
formulation of the ${\rm SL}(2,\mathbb{R})$ gauged WZW  model,
satisfies  the condition
\bea
B={\bar B}=\exp\bigg(\oint\rd x^\mu\,a_\mu\bigg)\ .
\eea

\bigskip

Consider now  the classical $W$ currents defined through the relations
\be\label{isisai112assaa}
\arraycolsep=0.7cm
\begin{array}{lll}
W^{(cl)}_2=\xi_+\,\xi_-\ ,& 
W^{(cl)}_3=\tfrac{1}{2}\ \big(\xi_-\,\partial \xi_+-
\xi_+\,\partial \xi_-\big)\ ,& \ldots
\\[0.4cm]
\overline{W}^{(cl)}_2=\bar{\xi}_+\,\bar{\xi}_-\ ,&
\overline{W}^{(cl)}_3=\tfrac{1}{2}\ \big(
{\bar \xi}_-\,\partial {\bar \xi}_+-
{\bar \xi}_+\,\partial {\bar \xi}_-\big)\, ,&\ldots
\end{array}\, .
\ee
Using \eqref{jsusausa} they can be  rewritten in terms of the WZW currents along with $\partial\eta$ and $\bar{\partial}\eta$:
\be\label{aspod9102123a}
\arraycolsep=0.4cm
\begin{array}{ll}
W^{(cl)}_2=(\partial\eta)^2-\big((L^3)^2+L^+L^-\big) \,, &
W^{(cl)}_3=2\,\partial\eta\, L^+L^-+\half \, (L^+\partial L^--L^-\partial L^+)\ \ \ 
\\[0.4cm]
{\overline W}^{(cl)}_2=({\bar \partial}\eta)^2-\big(({\bar R}^3)^2+{\bar R}^+{\bar R}^-\big)\,,
&
{\overline W}^{(cl)}_3=2\,{\bar \partial\eta}\, {\bar R}^+{\bar R}^-+\half \, ({\bar R}^+\partial{\bar  R}^--
{\bar R}^-{\bar \partial} {\bar R}^+)
\end{array} \!\!.
\ee
These  formulae show   that   the $W$ currents are real, chiral and periodic fields.
Furthermore it is straightforward to check
using the PB relations \eqref{isisailaoi} and \eqref{aisaiasi} that
all the $W$ currents Poisson commute  (in a weak sense) with  the constraints $\bar{\Upsilon}$ and ${ \Upsilon}$ 
\eqref{aoasoasoas},
\be\label{hasyasys}
\begin{array}{l}
\big\{W^{(cl)}_j(t,x),\Upsilon(t,y)\big\}\big|_{\Upsilon=0}=\big\{W^{(cl)}_j(t,x),\bar{\Upsilon}(t,y)\big\}=0\\[0.5cm]
\big\{\overline{W}^{(cl)}_j(t,x),\bar{\Upsilon}(t,y)\big\}\big|_{{\bar \Upsilon}=0}=\big\{\overline{W}^{(cl)}_j(t,x),{\Upsilon}(t,y)\big\}=0
\end{array}\ \ \  .
\ee
Since the fields $W^{(cl)}_2$ and $\overline{W}^{(cl)}_2$ coincide with  the nonvanishing components of the
stress energy tensor,   the Hamiltonian Poisson    commutes  with $\bar{\Upsilon}$ and ${ \Upsilon}$.
Also it is easy to see that
\bea
\big\{\Upsilon(t,x),{ \Upsilon}(t,y)\big\}=
\big\{{\bar \Upsilon}(t,x),{\bar \Upsilon}(t,y)\big\}=\big\{\Upsilon(t,x),{\bar \Upsilon}(t,y)\big\}=0
\eea
and, hence, the
  constraints \eqref{aoasoasoas}
are of the first class. The $W$ currents are ``classical observables''
and 
additional straightforward calculations show that
they  form  the closed   Poisson algebra which occurs in the
 $n\to\infty$ limit of the algebra of  extended conformal symmetry of the lattice model.

\subsection{BRST quantization}
Once the gauged ${\rm SL}(2,\mathbb{R})$ WZW model
is formulated as a classical dynamical  system possessing constraints 
of the first class
one can consider  the problem of its quantization within 
the BRST approach. Here
we  briefly sketch the
algebraic procedure for the construction of the chiral component of the space of states.

\bigskip

The chiral component of the energy momentum tensor of the quantum theory 
is  split into three terms:
\bea\label{isisaiasaaq}
T_{\scriptscriptstyle{\rm total}}=T_{\scriptscriptstyle{\rm WZW}}+T_{\scriptscriptstyle{\rm Gauss}}+T_{\scriptscriptstyle{\rm ghost}}\ .
\eea
The first one  is \cite{Knizhnik:1984nr}
\bea\label{aspodi0912r43}
T_{\scriptscriptstyle{\rm WZW}}=-\frac{n^2}{n+C_{\tt V}}\  \kappa_{AB}\, L^AL^B\ .
\eea
It is built from the  left currents of the WZW model
\bea
L^{A}(u)=n^{-1}\ \sum_{m=-\infty}^\infty j_m^A\ \re^{-\ri  m u}\ \ \ \ \ \ \ \  \ \ \quad (A=3,\pm)
\eea
whose Fourier coefficients obey  the commutation relations
\bea
\big[\,j^A_m,j^B_r\,\big]=  -n
\ \kappa^{AB}\ \, \tfrac{m}{2}\, \delta_{m+r,0}- \tfrac{\ri }{2}\,  {f^{AB}}_{C}\ j^C_{m+r}\ .
\eea
Here the  level (central element)
  of the  Kac-Moody algebra has been  denoted by $n$.  It is  related to the Plank constant as $\hbar=\frac{2\pi}{n}$.
The constant  $C_{\tt V}$ entering into \eqref{aspodi0912r43}
stands for the so-called dual Coxeter number:
\bea
C_{\tt V}\,\kappa^{AB}=\tfrac{1}{4}\, {f^{AC}}_D\,{f^{BD}}_C
\eea
and in the case under consideration $C_{\tt V}=2$.
The second term in \eqref{isisaiasaaq} represents the contribution of the massless Gaussian field,
\bea
T_{\scriptscriptstyle{\rm Gauss}}=n\ (\partial\eta)^2
\eea
with
\bea
\partial\eta(u)=n^{-\frac{1}{2}}\, \sum_{m=-\infty}^\infty d_m\ \re^{-\ri m u}\ :\ \ \ \ \ \big[d_m, d_r\big]=\tfrac{m}{2}\, \delta_{m+r,0}\ .
\eea
Finally $T_{\scriptscriptstyle{\rm ghost}}$ is the chiral component of the energy momentum tensor for the $bc$\,-\,system:
\bea
T_{\scriptscriptstyle{\rm ghost}}=\ri\, b\partial c\ .
\eea
The ghost   fields have 
conformal dimensions  $(\Delta_b,\Delta_c)=(1,0)$ and, as with the chiral fields  $L^A$ and $\partial\eta$,
can also be   expanded in the Fourier series
\bea
b(u)= \sum_{m=-\infty}^\infty b_m\ \re^{-\ri m u}\ ,\ \ \ \ \ \ \ \ \ c(u)= \sum_{m=-\infty}^\infty c_m\ \re^{-\ri m u}\,,
\eea
where
\bea
\big\{b_m, c_r\big\}=\delta_{m+r,0}\ ,\ \ \ \ \ \qquad \big\{b_m, b_r\big\}=\big\{c_m, c_r\big\}=0\ .\nonumber
\eea
The Virasoro  central charge of the $bc$\,-\,system is equal to $(-2)$, so that the total central charge associated with
the energy momentum tensor \eqref{isisaiasaaq} is given by
\bea
c_{\scriptscriptstyle{\rm total}}=c_{\scriptscriptstyle{\rm WZW}}+c_{\scriptscriptstyle{\rm Gauss}}
+c_{\scriptscriptstyle{\rm ghost}}=\frac{3 n}{n+2}+1-2=
2-\frac{6}{n+2}\ .
\eea
\bigskip

The highest weight  representation for the combined chiral algebra generated  by the
Fourier  coefficients $ j_m^A,\,d_m,b_m,c_m$ is constructed  in the usual manner. 
First of all  one requires  that the  highest state
is  annihilated by   all the positive   frequency modes with $m>0$.
Since  the zero modes of the WZW currents 
satisfy    the  commutation relations
 \bea
  \big[\,j^A_0,j^B_0\,\big]=-\tfrac{\ri }{2}\, {f^{AB}}_{C}\ j^C_0\ ,
 \eea
 the highest states  form a representation of
 the ${\mathfrak{sl}}_2$ algebra. It makes sense to require that 
it is an irreducible one,  characterized by the value of the Casimir operator
\bea
{\hat C}_{\tt G}=-\kappa_{AB}\,j^A_0j^B_0\  ,
\eea
which in the  ${\mathfrak{sl}}_2$ case is usually denoted as $\ell (\ell+1)$. 
In order to make a link with our previous  notations 
 we will employ  the parameter ${  p}=\ell-\half$.
Together with this quantum number 
the highest  states  can be  labeled by the eigenvalues of the zero modes $j^{3}_0$ and $d_0$:
\be
{\hat C}_{\tt G}\, |{ p},\mu,s\rangle=
\big({ p}^2-\tfrac{1}{4}\big)\, \, |{ p},\mu,s\rangle\, ,\ \ \ \ \quad  
j^{3}_0\,|{ p},\mu ,s\rangle= \mu\ |{ p},\mu,s\rangle\, ,\  \ \ \quad
d_0\,|{ p},\mu,s\rangle=\tfrac{s}{\sqrt{n}}\, |{ p},\mu,s\rangle\ .
\ee
The highest states form a  representation not only of the  ${\mathfrak{sl}}_2$ algebra
but also the simple fermionic one
\bea
\{b_0,c_0\}=1\ ,\ \ \ b_0^2=c_0^2=0\ .
\eea
Thus we  supplement   the set
of conditions defining  them with
\bea\label{aisiaias}
c_0\, |{ p},\mu,s\rangle_+=0\,,\qquad\qquad
|{ p},\mu,s\rangle_-\equiv b_0\,|{ p},\mu,s\rangle_+\ .
\eea
The   highest weight representation  is built by the action of 
the negative frequency modes
  $ j_m^A,\,d_m,b_m,c_m$ with $m<0$ 
on the  highest state  multiplet.
The corresponding linear space  will be denoted by
 ${\cal A}_{{ p},s}$. The latter possesses a  grading induced by 
the Virasoro algebra generator $L_0^{\scriptscriptstyle{\rm (total)}}$.
 For given  ${\tt L}=0,1,2,\ldots\,$, the level subspace 
${\cal A}_{{ p},s}^{({\tt L})}$  is  finite dimensional and all its states
 have  the same conformal dimension $\Delta_{{ p},s}+{\tt L}$ with 
\bea
\Delta_{{ p},s}=\frac{{ p}^2-\tfrac{1}{4}}{n+2}+\frac{s^2}{n}\ .
\eea
Note that the  conformal dimensions of the primary states
 do not depend on the quantum number $\mu$.

\bigskip

The parameter
  ${ p}$ and its barred counterpart $\bar{ p}$ are related to  the central
elements \eqref{aoaosao}-\eqref{aiuswqd891212}  of the  Poisson algebra of the WZW currents. In particular,
the sum ${ p}+\bar{ p}$ can be identified with the eigenvalues of the
operator $-\ri\, \frac{\partial}{\partial\alpha}$
with $\alpha$ being the dynamical variable from \eqref{saoosao}. 
Then the compactness condition $\alpha\sim \alpha+2\pi$ 
yields the quantization rule 
 ${  p}+\bar{  p}\in {\mathbb Z}$. This, in  view of  the  classical relation \eqref{isiasaisi}, leads to
\bea\label{rcaji8912A}
p=\half\, \big({\tt u}+(n+2)\, ({\tt k}+{\tt w})\, \big)\,,
\qquad \qquad
\bar{p}=\half\, \big({\tt u}-(n+2)\, ({\tt k}+{\tt w})\, \big)
\ \ \ \ \  \ \ \quad \big({\tt u}, {\tt w}\in{\mathbb Z}\big)\ .
\eea
At the same time $s$ may take any real value,
\bea\label{rcaji8912B}
s\in{\mathbb R}\ .
\eea
\bigskip

The central r$\hat{\rm o}$le in the BRST approach belongs to
the BRST charge and the ghost number operator. These obey the relations
\bea
\hat{Q}_{\scriptscriptstyle{\rm BRST}}^2=0\ ,\ \ \ \ \qquad
\big[\hat{q}_{\scriptscriptstyle{\rm ghost}},\hat{Q}_{\scriptscriptstyle{\rm BRST}}\big]=\hat{Q}_{\scriptscriptstyle{\rm BRST}}\ .
\eea
In the
case at hand they read explicitly as
\bea\label{asussauas}
&&\!\!\!\!\!\!\!\!\!\!\!\!\!\!\!\!\!\!\!\!\!\!
\hat{Q}_{\scriptscriptstyle{\rm BRST}}=\frac{1}{\hbar}\int_0^{2\pi}\rd u\ \big(L^3-\partial\eta\big)\,c(u)=\big(j^3_0-\sqrt{n}\, d_0\big)\, c_0+\sum_{m\not=0} \big(j^3_m-\sqrt{n}\, d_m\big)\, c_{-m}\\[0.2cm]
&&\!\!\!\!\!\!\!\!\!\!\!\!\!\!\!\!\!\!\!
\hat{q}_{\scriptscriptstyle{\rm ghost}}=\int_0^{2\pi}\frac{\rd u}{2\pi}\ b c(u)=\ b_0c_0+\sum_{m=1}^\infty
\big(\,b_{-m}c_m-c_{-m}b_m\, \big) \ .\nonumber
\eea
It is easy to see that both  operators commute with the zero mode of the current $L^3(u)$:
\bea
\big[ j^3_0,\hat{Q}_{\scriptscriptstyle{\rm BRST}}\big]=\big[ j^3_0,{\hat q}_{\scriptscriptstyle{\rm ghost}}\big]=0\ .
\eea
 The 
 dimensions of the  level subspaces $\widetilde{\cal A}_{p,\mu,s}^{({\tt L})}$
  depend
 essentially   on  whether or not   $\mu-s$ vanishes.
This difference 
is
the coefficient in front of the ghost zero mode  $c_0$ in \eqref{asussauas}
 when the  action of  the BRST charge
is restricted 
to  the eigenspace ${\cal A}_{p,\mu,s}$.
Consider the highest states $|p,\mu,s\rangle_\pm$. If $\mu\ne s$,
then the state $|p,\mu,s\rangle_+$ is annihilated by the BRST charge.
On the other hand  $\hat{Q}_{\scriptscriptstyle{\rm BRST}}|p,\mu,s\rangle_-\ne 0$
and is proportional to $|p,\mu,s\rangle_+$. This implies that
the level subspace $\widetilde{\cal A}_{p,\mu,s}^{(0)}$ with $\mu\ne s$  is trivial.
In the case when $\mu=s$ both highest states are annihilated by the BRST charge.
However only $|p,s,s\rangle_+$ has zero ghost number so that
$\dim\big(\widetilde{{\cal A}}_{p,s,s}^{(0)}\big)=1$. 
Recall that $|p,s,s\rangle_+$ is a state from 
a ${\mathfrak{sl}}_2$ irrep characterized by  $p$.
The eigenvalues of $j_0^3$ for the other highest states from the multiplet
are given by  $\mu=s+\ri\,m$, where $m$ is a nonzero integer, and hence
the difference $\mu-s$ for these states would be nonvanishing.
Proceeding further, it is straightforward to check at least for small values of ${\tt L}=0,1,2,\ldots\,$, that
all the spaces $\widetilde{\cal A}_{p,\mu,s}^{({\tt L})}$ are trivial for $\mu\ne s$, while
the dimensions of $\widetilde{\cal A}_{p,s,s}^{({\tt L})}$ with generic $p$ is equal to the number of bipartitions of ${\tt L}$.
  \bigskip

  Perhaps the  easiest way to explore the linear structure of the factor space $ \widetilde{\cal A}_{{ p},\mu,s}$
  is to bosonize the $\widehat{{\mathfrak {sl}}}(2,{\mathbb R})$ current algebra 
\cite{Zam1986,Wakimoto,Gerasimov:1989mz}.
  This allows one to isolate the physical states  in  ${\cal A}_{{ p},s}$
  and to show that
  $\dim\big(\widetilde{\cal A}_{{ p},s,s}^{({\tt L})}\big)$ 
  coincides with the corresponding dimensions of the
  level subspaces of  the highest weight representation of the $W_\infty$\,-\,algebra.
  
\bigskip

All the above  leads us to the conjecture that
the  space 
\be
{\cal H}^{({\rm cont})}=\bigoplus_{{\tt u},{\tt w}\in\mathbb{Z}}
\int^{\oplus}_{\mathbb{R}}\!\rd s\
\overline{{\cal W}}_{\bar{p},s}\otimes {\cal W}_{p,s}\qquad  
{\rm with}
\qquad 
\begin{array}{l} p=\frac{1}{2}\,{\tt u}+\frac{1}{2}\,(n+2)\,({\tt k}+{\tt w})\\[0.2cm]
                                       \bar{p}=\frac{1}{2}\,{\tt u}-\frac{1}{2}\,(n+2)\,({\tt k}+{\tt w})
\end{array}\,,
\ee
that appears in the scaling
limit of the lattice model with twisted boundary conditions,
is the pseudo-Hilbert space which arises in the quantization of the classical  field theory 
\eqref{kakaiaia},\eqref{aoasoasoas}
subject to 
the boundary conditions \eqref{aiosaisia},\,\eqref{isisai},\,\eqref{iasisisaww} and \eqref{hsahsay}.

\bigskip

Our study  of the Hermitian structures in the lattice model
led us to conclude that the spaces ${\cal H}^{({\rm cont})}$ and
${\cal H}^{({\rm disc})}$ should be understood
as a result of two differently defined scaling limits.
In all likelihood the states from ${\cal H}^{({\rm disc})}$
and ${\cal H}^{({\rm cont})}$
can not be interpreted simultaneously as  normalizable states 
within a single CFT.
Perhaps the simplest idea for the  field theory, whose quantization 
results in the pseudo-Hilbert space ${\cal H}^{({\rm disc})}$, is the model described 
by the same Lagrangian density and constraints \eqref{kakaiaia},\,\eqref{aoasoasoas}
as well as the boundary conditions \eqref{aiosaisia} for the WZW currents and \eqref{aiasiasi}
for   $\partial_\mu\eta$.
However  the  fields now are subject to different reality conditions. The
classical $W$ currents should satisfy
\bigskip
 \bea\label{xiasissssai}
\big(W^{(cl)}_j\big)^*=(-1)^j\ W^{(cl)}_j\ ,\ \ \ \ \ \ \ \big(\overline{W}^{(cl)}_j\big)^*=(-1)^j\ \overline{W}^{(cl)}_j\ .
\eea
In view of eq.\,\eqref{aspod9102123a} this would follow from the reality conditions
\be\label{pasodpo11212}
\big(L^3\big)^*=-L^3\,,\qquad
\big(L^{\pm}\big)^*=L^{\mp}\,,\qquad
 \big(R^3\big)^*=-R^3\,,\qquad
\big(R^{\pm}\big)^*=R^{\mp}
\ee
imposed on the classical  WZW currents and
\be
\big(\partial\eta\big)^*=-\partial\eta\,,\qquad \qquad
\big(\bar{\partial}\eta\big)^*=-\bar{\partial}\eta
\ee
for the Gaussian field.
Furthermore  $\ri\eta$ is expected to be a real and compactified field,
\be
\ri\eta\sim\ri\eta+2\pi\ .
\ee
 The latter  implies that  the zero mode momenta 
$P_\eta$ and $\bar{P}_\eta$ \eqref{asodioaisd109212} are no longer equal, but instead
\be
\ri\,(P_\eta-\bar{P}_\eta)\in\mathbb{Z} \ .
\ee
Notice that $B=\re^{2\pi P_\eta}$ and $\bar{B}=\re^{2\pi\bar{P}_\eta}$
appearing in the boundary conditions \eqref{aususalk} still coincide. 
Such reality and boundary conditions for the 
currents correspond to the ${\rm SU}(2)$ WZW model  
gauged over the compact subgroup. However, they are not
enough to fully specify the field theory. In the ${\rm SL}(2,\mathbb{R})$  case
there were the additional constraints \eqref{isisai} and \eqref{iasisisaww},
whose motivation relied on the fact that the WZW field ${\bf G}\in{\rm SL}(2,\mathbb{R})$.
At the moment, it is not clear to us what extra conditions need to imposed for
the ${\rm SU}(2)$ case.

\section{Lund-Regge model}

The Yang-Baxter integrability of the ${\cal Z}_2$ invariant inhomogeneous six-vertex model
made possible a detailed numerical
study of its critical behaviour.
On the basis of this we formulated 
our central conjecture in sec.\,\ref{sec212} regarding the space of states 
occurring in the scaling limit
and made the identification of the sector 
$\tilde{\cal H}^{({\rm cont})}_{\rm even}$
with the pseudo-Hilbert space of the Lorentzian black hole NLSM.
However the CFT itself does not assume any integrable structures
and its space of states was described in terms of the
representations  of the algebra of extended conformal symmetry  without reference to integrability.
Nevertheless we believe that, finishing the paper, it would be
 instructive to discuss the integrable structure in the NLSM, inherited from the lattice system,
within the context of the theory of partial differential equations solvable by the inverse scattering method.
\bigskip

\subsection{Sklyanin exchange relations for the Lund-Regge model\label{sec232}}

Consider the Lagrangian density 
\bea\label{oiaisa090}
{\cal L}_{\varepsilon}=
\frac{1}{2}\, \frac{\partial_t U{\partial}_t V-\partial_x V{\partial}_x U}{1-UV}-\frac{\varepsilon^2}{2}\ UV\ ,
\eea
which is a perturbation of that for the Lorentzian black hole NLSM \eqref{oiaodisa090} controlled by the
 parameter $\varepsilon$. 
We take the fields $U$ and $V$ to satisfy the periodic boundary conditions as in \eqref{89a8s9d89}.
The perturbation does not break the invariance
w.r.t. the transformation
 \eqref{hssattas} so that $J_\mu$ from \eqref{hsysya11111} is still a Noether current.
 The continuity equation $\partial_\mu J^\mu=0$ allows one to introduce the dual field 
 \bea\label{isisaijj}
 \tilde{\Theta}({\bf x})=  -\int_{C_{\bf x}}{\rm d} x^\mu \,\epsilon_{\mu\nu} J^\nu\ .
\eea
Here $C_{\bf x}$  denotes 
an open integration contour which starts at an arbitrary chosen  initial point and ends up at ${\bf x}=(t,x)$.
Then in the co-ordinate frame $(\Phi,\Theta)$ \eqref{iaoisod1}
the $\rm{SL}(2,\mathbb{R})$ matrix  \eqref{iosoid1a},
whose matrix entries satisfy the conditions \eqref{asoidoi1212},
has the form
\bea\label{aisisai2333}
\bm{g}_{\text{\textonehalf}}=
\begin{pmatrix}
\pm \cos(\Phi)\, \re^{+{\tilde \Theta}}&\phantom{\pm}\sin(\Phi)\,\re^{+\Theta}\\
-\sin(\Phi)\,\re^{-\Theta}&\pm \cos(\Phi)\, \re^{-{\tilde \Theta}}
\end{pmatrix}\ .
\eea
Since $U$ and $V$ take values in 
 regions III and IV from fig.\,\ref{iaosi1212}, the 
 coordinate $\Phi$ was restricted to the interval $(-\frac{\pi}{2},\frac{\pi}{2})$.
In this    domain $\cos(\Phi)$
is positive so that the different signs  ``$\pm$''
in  \eqref{aisisai2333} must be treated separately.
However, one can take into account both cases at once 
if the domain of $\Phi$ is extended to the segment $[-\pi,\pi]$. 
Then the regions III and IV in fig.\,\ref{iaosi1212} would be double covered.
In fact, for the purposes of this subsection one can assume that the field $\Phi$ takes 
all possible real values, corresponding to  a universal cover of $\rm{SL}(2,\mathbb{R})$.
\bigskip

 One may still use eq.\,\eqref{oaop9203} to introduce the non-local fields $\xi_\pm$.
The formal Lie group element,
which appears in that relation, 
coincides with $\bm{g}_{\text{\textonehalf}}$
when specialized to the fundamental representation.
It can  be written in the form of the Euler decomposition as
 \be\label{iosoid1aA}
\bm{g}=\re^{\frac{1}{2}(\tilde{\Theta}+\Theta)\, {\tt h}}\ \re^{\Phi \,( {\tt e}_+-{\tt e}_-)}
\, \re^{\frac{1}{2}(\tilde{\Theta}-\Theta) \,{\tt h}}\ .
\ee
The fields $\xi_\pm$ are no longer chiral.
Instead,
the Euler-Lagrange equations corresponding to 
${\cal L}_{\varepsilon}$  imply
\bea\label{aisisisa}
{\bar \partial}\xi_\pm=\frac{\varepsilon^2}{8}\ \sin(2\Phi)\ \re^{\pm (\Theta+\tilde{\Theta}) }\ ,\ \ \ \ \ \qquad
{ \partial}{\bar \xi}_\pm=
\frac{\varepsilon^2}{8}\ \sin(2\Phi)\ \re^{\pm (\Theta-\tilde{\Theta}) }\ .
\eea
The perturbation  does not change the canonical momenta \eqref{iaosid9818921a} 
and the Poisson bracket relations \eqref{iaosid9818921} continue to hold true.
Hence   $\xi_\pm$ and  ${\bar \xi}_\pm$ still satisfy the PBs
 \eqref{jsasusa},\,\eqref{jsasusss1sa} and \eqref{jsasusss1saC} provided they are
understood as equal-time relations.

\bigskip

Let $\bm{{\cal A}}_\mu$ be 
a space-time 1-form  
which  takes values in the Lie algebra ${\mathfrak {sl}}_2$ and
whose light-cone components, $\bm{{\cal A}}=\frac{1}{2}\, (\bm{{\cal A}}_{0}+\bm{{\cal A}}_{1})$ and   
$\,\,\,\bar{\!\!\!\bm{{\cal A}}}=\frac{1}{2}\,(\bm{{\cal A}}_{0}-\bm{{\cal A}}_{1})$, are
defined as
\bea\label{Afield1a}
\partial-\bm{{\cal A}}&=&
\ \ \ \ \partial-\xi_-\,{\tt e}_-+\xi_+\,{\tt e}_+-\lambda\,{\tt h}\nonumber\\[0.2cm]
{\bar \partial}-\,\,\,\bar{\!\!\!\bm{{\cal A}}}&=&\bm{g}\,
\big(\,{\bar \partial}+\bar{\xi}_-\,{\tt e}_--\bar{\xi}_+\,{\tt e}_+-{\bar \lambda}\,{\tt h}\, \big)\, \bm{g}^{-1}\ .
\eea
It is straightforward to check using \eqref{aisisisa}
that if the equations of motion
are satisfied, then the connection $\partial_\mu-\bm{{\cal A}}_\mu$ is flat, i.e.,
\bea\label{isodiao12}
[\,\partial-\bm{{\cal A}}\,,{\bar \partial}-\,\,\,\bar{\!\!\!\bm{{\cal A}}}\,]=0
\eea
provided the auxiliary spectral parameters $\lambda$ and ${\bar \lambda}$ are  related as
\bea\label{cont1ada}
\lambda\,{\bar \lambda}=-\frac{\varepsilon^2}{16}\ .
\eea
The connection is not single valued on the space-time cylinder even when
periodic boundary conditions are imposed on the fields $U$ and $V$.
Indeed as it follows from the definition of the dual field $\tilde{\Theta}$,
\be
\exp\big(\tilde{\Theta}(t,x+2\pi)-\tilde{\Theta}(t,x)\big)=\exp\bigg(\int_x^{x+2\pi} {\rm d}x'\,J_0\bigg)=B\,,
\ee
and therefore
\bea\label{iususausaaa}
\bm{g}(t,x+2\pi)=B^{\frac{{\tt h}}{2}}\, \bm{g}(t,x)\, B^{\frac{{\tt h}}{2}}\ .
\eea
In turn, $\bm{{\cal A}}_\mu$  obeys the
quasiperiodicity condition
\bea\label{oasid989212}
\bm{{\cal A}}_{\mu}(t,x+2\pi)=B^{\frac{{\tt h}}{2}}\ \bm{{\cal A}}_{\mu}(t,x)\ B^{-\frac{{\tt h}}{2}} \ .
\eea
Choosing some representation of $\mathfrak{sl}_2$, one can define the
classical transfer matrix:
\bea\label{iaosdi1982}
T_j={\rm Tr}_j[{\boldsymbol M}]\ ,\ \ \ \ \ \ \ {\boldsymbol M}=B^{-\frac{{\tt h}}{2}}\ 
\overset{\leftarrow}{{\cal P}}\exp\bigg(\int_{0}^{2\pi}\rd  x\, \bm{{\cal A}}_{1}\bigg)\ .
\eea
The zero curvature relation \eqref{isodiao12} implies that $T_j$ does not depend on the time slice
at which  the path ordered integration is taken, i.e., it is an Integral of Motion.
In fact, $T_j$  is a one  parameter  family of  IM, since the connection $\bm{{\cal A}}_\mu$ depends
on $\lambda$, $\bar{\lambda}$ satisfying the constraint \eqref{cont1ada}.
The latter may be resolved by means of a single complex parameter $\beta$:
\bea\label{ioasidio1928912}
\lambda=\tfrac{\ri}{4}\, \varepsilon\, \re^{2\beta}\ ,\ \ \ \
\qquad {\bar \lambda}=\tfrac{\ri}{4}\, \varepsilon\, \re^{-2\beta}\ .
\eea
\begin{figure}
\centering
\scalebox{0.7}{
\begin{tikzpicture}
\draw [line width = 1mm,red] (0,0) -> (4,4);
\draw [line width = 1.5mm,->,red] (2,2) -> (2.2,2.2);
\draw [line width = 1mm,blue] (-4,4) -> (0,0);
\draw [line width = 1.5mm,->,blue] (-2,2) -> (-1.8,1.8);
\draw [ line width = 1mm] (-4,4) -> (4,4);
\draw [ line width = 1.5mm,->] (0,4) -> (0.2,4);
\draw [line width = 0.5mm, ->] (-5,-1.5) -- (-5,5);
\draw [line width = 0.5mm,->] (-6,-1)-- (6,-1);
\draw [dashed] (-5,4) -- (5,4);
\node at (-5.5,4) {\Large $t_0$};
\node [right] at (-5.2,5.5) {\Large $t$};
\node [right] at (6.1,-1) {\Large $x$};
\node at (-4.3,4.4) {\large ${\bf a}$};
\node at (4.3,4.4) {\large ${\bf b}$};
\node at (0,-0.4) {\large ${\bf c}$};
\end{tikzpicture}
}
\caption{\small The integration along the time slice $t\,=\,t_0$ (the segment $\overset{\rightarrow}{{\bf ab}}$) for
the path ordered exponent $\bm{M}$ in eq.\,\eqref{iaosdi1982}
 can be replaced by an integration 
along the characteristics: $u\,=\,t_0$ with $t_0<\bar{u}<t_0-2\pi$ ($\overset{\rightarrow}{{\bf ac}}$) 
and $\bar{u}\,=\,t_0-2\pi$ with 
$t_0<u<t_0+2\pi$ ($\overset{\rightarrow}{{\bf cb}}$).\label{fig14}}
\end{figure}
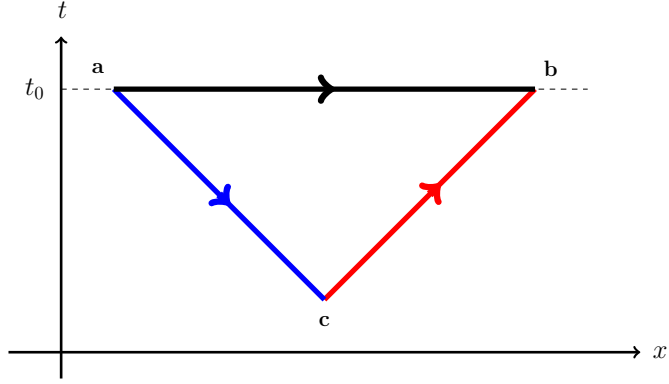
\smallskip

An immediate question arises as to the mutual Poisson commutativity of $T_j(\beta)$
for different values of the spectral parameter. 
This may be addressed by directly following the  line of arguments
 which were developed in the context of the sine-Gordon model
in the work \cite{Bazhanov:1996aq}.
Since the connection is flat, the path ordered exponent in \eqref{iaosdi1982} 
is unchanged under deformations of the integration contour that keep the endpoints fixed.
It is useful to swap the integration over the time slice to the one over the light-cone
pieces as depicted in fig.\,\ref{fig14}.
The contribution of the integration 
along each of the characteristics  $\bar{u}\equiv t-x=\overline{{\rm const}}$ and $u\equiv t+x={\rm const}$ 
is taken into account by 
 the two path ordered exponents $\bm{L}_\varepsilon^{(cl)}$
and $\bar{\bm{L}}_\varepsilon^{(cl)}$, respectively:
\bea\label{aoisdo1902}
\bm{L}_\varepsilon^{(cl)}(\lambda)&=&
{\lambda}^{-\frac{{\tt h}}{4}}\ 
\re^{\frac{\ri}{2}\Phi_{\scriptscriptstyle{\bf cb}}{\tt h}}\ 
\bm{G}_{{\bf b}}\
\overset{\leftarrow}{{\cal P}}
\exp\bigg(\int_{t_0}^{t_0+2\pi}\rd u\, \big(\xi_-{\tt e}_--\xi_+{\tt e}_++\lambda\,{\tt h}\big)\bigg)\bigg|_{\bar{u}}\ 
\bm{G}_{{\bf c}}^{-1}\ 
{\lambda}^{+\frac{{\tt h}}{4}}  \nonumber \\[0.0cm]
\\[0.1cm]
\bar{\bm{L}}_\varepsilon^{(cl)}(\bar{\lambda})&=&
\bar{\lambda}^{+\frac{{\tt h}}{4}}\
\bar{\bm{G}}_{{\bf c}}\,
\overset{\rightarrow}{{\cal P}}
\exp\bigg(\int_{t_0-2\pi}^{t_0}\rd \bar{u}\, 
\big(\bar{\xi}_-{\tt e}_--\bar{\xi}_+{\tt e}_+-\bar{\lambda}\,{\tt h}\big)\bigg)\bigg|_{u}\ 
\bar{\bm{G}}_{{\bf a}}^{-1}\ 
 \re^{\frac{\ri}{2}\Phi_{\scriptscriptstyle{\bf ca}}{\tt h}}\ 
\bar{\lambda}^{-\frac{{\tt h}}{4}}\nonumber\ .
\eea
Here $\Phi_{{\rm \bf{x}{\bf{y}}}}\equiv\Phi_{\bf{x}}-\Phi_{\bf{y}}$
and we use the shortcut notation $\Phi_{\bf{x}}$ to denote the value of the field $\Phi(\bf{x})$
at either one of the space-time points $\bf{x}=\bf{a},\bf{b},\bf{c}$ 
indicated in fig.\,\ref{fig14}.
Also $\bm{G}_{ {\bf x}}$ and $\bar{\bm{G}}_{{\bf x}}$ 
read explicitly as
\be
\bm{G}_{{\bf x}}=\re^{-\frac{\ri\pi}{4}\,({\tt e}_++{\tt e}_-)}\,
\re^{-\frac{1}{2}(\tilde{\Theta}_{\scriptscriptstyle{\bf x}}+\Theta_{\scriptscriptstyle{\bf x}}){\tt h}}\,,\qquad 
\bar{\bm{G}}_{{\bf x}}=
\re^{-\frac{\ri\pi}{4}\,({\tt e}_++{\tt e}_-)}\,
\re^{\frac{1}{2}(\tilde{\Theta}_{\scriptscriptstyle{\bf x}}-\Theta_{\scriptscriptstyle{\bf x}}){\tt h}}\qquad
\qquad
(\bf{x}=\bf{a},\bf{b},\bf{c})\ .
\ee
The above formulae  should be compared with
eqs.\,\eqref{90a9s0d9as} and \eqref{90a9s0d9asA} involving the path ordered exponents $\bm{L}^{(cl)}$ and 
$\bar{\bm{L}}^{(cl)}$.
In fact, the definition \eqref{aoisdo1902} has been arranged in such a way so that 
$\bm{L}_\varepsilon^{(cl)}$  and $\bar{\bm{L}}^{(cl)}_{\varepsilon}$ satisfy the same Poisson algebra as in
\eqref{PBrel1} and \eqref{PBrel1q}.
Namely,
 one can show 
that 
\bea\label{UIsd9812}
\big\{{{\boldsymbol L}}_\varepsilon^{(cl)}(\lambda) \begin{array}{ccc} \\[-0.4cm] \otimes \\[-0.35cm] , 
\end{array}{{\boldsymbol L}}_\varepsilon^{(cl)}(\lambda') \,\big\}&=&
+\Big[\,{{\boldsymbol L}}_\varepsilon^{(cl)}(\lambda) \,{\otimes}\, {{\boldsymbol L}}_\varepsilon^{(cl)}(\lambda') ,
\,{{\boldsymbol r}}\big(\sqrt{\lambda/\lambda'}\ \big)\,\Big] \nonumber\\[0.3cm]
\big\{\bar{{\boldsymbol L}}_\varepsilon^{(cl)}(\bar{\lambda}) \begin{array}{ccc} \\[-0.4cm] \otimes \\[-0.35cm] , 
\end{array}\bar{{\boldsymbol L}}_\varepsilon^{(cl)}(\bar{\lambda}') \,\big\}&=&
-\Big[\,\bar{{\boldsymbol L}}_\varepsilon^{(cl)}(\bar{\lambda}) \,{\otimes}\, 
\bar{{\boldsymbol L}}_\varepsilon^{(cl)}(\bar{\lambda}') ,
\,{{\boldsymbol r}}\big(\sqrt{\bar{\lambda}/\bar{\lambda}'}\ \big)\,\Big]  \\[0.3cm]
\big\{{{\boldsymbol L}}_\varepsilon^{(cl)}(\lambda) \begin{array}{ccc} \\[-0.4cm] \otimes \\[-0.35cm] , 
\end{array}\bar{{\boldsymbol L}}_\varepsilon^{(cl)}(\bar{\lambda}) \,\big\}&=&0\nonumber
\eea
with $\bm{r}$ being the  classical $R$-matrix \eqref{CRmatrix111}.
\bigskip

A straightforward calculation yields that the monodromy matrix \eqref{iaosdi1982} 
may be brought to the form
\bea
{{\boldsymbol M}}=\bm{C}\,{\tilde{\boldsymbol M}}\,{\boldsymbol C}^{-1}
\eea
with
\bea\label{oasiod09021}
{\tilde {\boldsymbol M}}=
\big(\tfrac{\ri\varepsilon}{4}\big)^{\frac{{\tt h}}{2}}\ \,
\bm{L}_{\varepsilon}^{(cl)}\ \,
\big(\tfrac{\ri\varepsilon}{4}\big)^{-\frac{{\tt h}}{2}}\ \ 
\re^{\ri\Phi_{\bf c}{\tt h}}\ \,
\bar{ \bm{L}}_{\varepsilon}^{(cl)}\ \,
\re^{-\ri\Phi_{\bf c}{\tt h}}\ .
\eea
The matrix $\bm{C}$  has no effect on the 
value of the trace in \eqref{iaosdi1982}
and is given by 
\be
\bm{C}=
\re^{\frac{1}{2}(\tilde{\Theta}_{\bf a}+\Theta_{\bf a}){\tt h}}\
\re^{\frac{\ri\pi}{4}({\tt e}_++{\tt e}_-)}\
\re^{\frac{\ri}{2}\Phi_{\bf ac}{\tt h}}\ 
\bar{\lambda}^{-\frac{\tt h}{4}}
\ .
\ee
Generally speaking  $\Phi_{\bf c}$, i.e., the value of the field $\Phi$ at the space-time point ${\bf c}$, as well as the
difference $\Phi_{\bf ab}=\Phi_{\bf a}-\Phi_{\bf b}$ are dynamical variables.
However, in order for the transfer matrix to be an IM, apart from the zero curvature relation
\eqref{isodiao12}, the connection must satisfy 
the quasiperiodic boundary condition \eqref{oasid989212} which, itself,
requires $\Phi$ to be a periodic field.
Due to this we need  to impose $\Phi_{\bf ab}=0$.
The latter 
is a first class constraint and the corresponding
gauge fixing condition can be achieved by assigning 
$\Phi_{\bf c}$ a certain value, say zero.
Then in view of \eqref{oasiod09021},\,\eqref{UIsd9812}
the matrix entries of $\tilde{\bm{M}}$,
considered as a function of the spectral parameter $\beta$ \eqref{ioasidio1928912},
turn out to satisfy the Sklyanin exchange relations 
\be\label{PBrel1ss}
\big\{\tilde{\boldsymbol M}(\beta) \begin{array}{ccc} \\[-0.4cm] \otimes \\[-0.35cm] , 
\end{array}\tilde{\boldsymbol M}(\beta')\,\big\}=
\Big[\,\tilde{\boldsymbol M}(\beta)\,{\otimes}\, \tilde{\boldsymbol M}(\beta'),
\,{{\boldsymbol r}}\big(\re^{\beta-\beta'}\ \big)\,\Big]\,.
\ee
Consequently,
\be
\big\{T_{j'}(\beta'),\,T_{j}(\beta)\big\}=0\ .
\ee
\bigskip

We have the following comments to make at this point.
For the derivation of the Sklyanin exchange relations
the model was considered, which is a deformation of the Lorentzian black 
hole NLSM. In this case the fundamental fields $U$ and $V$ are real. In fact, 
the reality conditions do not come into play and
the same arguments may be applied when these fields are complex conjugate to each other,
$V=U^*$ 
(the fields $\Theta$ and $\tilde{\Theta}$ are then pure imaginary so that
 $\bm{g}_{\text{\textonehalf}}$ from \eqref{aisisai2333} is a ${\rm SU}(2)$ matrix).
With this reality condition imposed the model is usually 
referred to as the complex sine-Gordon I,
or Lund-Regge model. The zero curvature relation, which
is also insensitive to the reality conditions, was  
proposed in the works \cite{Pohlmeyer:1975nb,Lund:1976ze,Getmanov:1979} 
for a connection in a gauge different to \eqref{Afield1a}.
\bigskip

The Lund-Regge model attracted a great deal of attention in the context of the so-called
non-ultralocality problem. 
In the ultralocal case the $x$-component of the flat connection satisfies the
equal-time Poisson bracket relations of the form
\be\label{PB0}
\big\{{\boldsymbol {A}}_1(x|\lambda_1) \begin{array}{ccc} \\[-0.4cm] \otimes \\[-0.35cm] , \end{array}  
{\boldsymbol { A}}_1 (y|\lambda_2)\big\}
= \big[
{\boldsymbol { A}}_1(x|\lambda_1)\otimes {\boldsymbol  1}+
 {\boldsymbol 1}\otimes {\boldsymbol {A}}_1(y|\lambda_2), {\boldsymbol r}(\lambda_1/\lambda_2)\big]\, \delta(x-y)\ .
\ee
Then the Sklyanin exchange relations for the monodromy matrix are easily derived
and the mutual Poisson commutativity condition for the transfer matrix for different values
of the spectral parameter follows. 
The ultralocality condition, of course, depends on the gauge of $\bm{A}_\mu$. 
Considerable effort was made to find an ``ultralocal'' flat connection 
for the Lund-Regge model. However, as with many other
interesting field theories admitting the
zero curvature relation, the attempts were met with failure.
This motivated the development of the so-called
``canonical $r-s$ matrix'' approach  for integrable two dimensional
models of non-ultralocal type \cite{Maillet:1985ek}. 
\bigskip

Inspired by the observation that the
 quantum monodromy matrix is somehow better
behaved than its classical limit, 
 in the work \cite{Bazhanov:2018xzh}  it was proposed to tackle the non-ultralocality
problem by starting with the quantum counterpart of 
the Sklyanin exchange relations --
the Yang-Baxter algebra. It was demonstrated on a specific example
that by taking the classical limit of the quantum algebra one could recover
the Sklyanin exchange relations for the classical monodromy matrix without 
reference to an ultralocal gauge.
Here  another illustration of this approach is given for the case of the
Lund-Regge model. Indeed, the starting point was the 
Yang-Baxter algebra \eqref{YBeq1a}
which in the classical limit becomes the Sklyanin exchange relations for 
$\bm{L}^{({cl})}$ and $\bar{\bm{L}}^{({cl})}$. Then the derivation
outlined above, which led to \eqref{PBrel1ss}, is 
 straightforward and almost
identical to that of the work \cite{Bazhanov:1996aq} for the sine-Gordon model.

\subsection{UV limit of the quantum complex sinh-Gordon I model}
Not much is known about the
 QFT corresponding to the classical Lagrangian density \eqref{oiaisa090} 
with real fields $U$ and $V$ as well as the quantum complex sine-Gordon I model, where
$U=V^*$. However the variant of the Lund-Regge model, which is sometimes referred to
as the  complex sinh-Gordon I model, is a well understood QFT now \cite{Fateev:1995ht}.
It is  an integrable deformation 
of the Euclidean black hole NLSM \eqref{iasdioi12014sa3}, whose
classical action is given by
 \be\label{iasdioi12014sa3A}
S_{\rm \scriptscriptstyle cshG}=\frac{n+2}{4\pi}\int\rd t \int_0^{2\pi}\rd x\ \ \bigg[\,
\frac{\partial_t U{\partial}_t U^*-\partial_x U{\partial}_x U^*}{1+UU^*}-\varepsilon^2\,U U^*\,\bigg]\qquad\qquad
\big(n\to +\infty\big)
\ee
and the complex field $U$ is subject to the quasiperiodic boundary conditions \eqref{iasod129102}.
\bigskip

The IR behaviour of the QFT is described in terms of 
the factorizable scattering theory.
The particle content  of the complex sinh-Gordon I model consists of
 a doublet of the same mass $m$
having opposite ${\rm U}(1)$ charges.
The two particle $S$-matrix is  diagonal (reflectionless).  Notice that in the
action \eqref{iasdioi12014sa3A}, $t$ and $x$ are assumed to be dimensionless world-sheet co-ordinates,
and $x$ has been brought to the standard segment $x\in[0,2\pi]$. If one were to restore
 the dimensions, $x$ would belong to the interval $[0,R]$ with $R$ having units of length. Then 
the parameter $\varepsilon$ is identified as
\be\label{aisodi1289asd}
\varepsilon=\frac{mR}{2\pi}\ .
\ee
\bigskip

To study the UV limit  it is convenient to make use of the dual description of the model.
The latter is based on the remarkable proposal originally put forward by Al.~B. Zamolodchikov \cite{ZAM},
that the so-called sine-Liouville
theory
provides a dual description of the quantum Euclidean black hole NLSM (see also \cite{Hikida:2008pe}).
 The duality relation is easily
 extended to the deformed model \eqref{iasdioi12014sa3A} and the dual action
reads as \cite{Fateev:1995ht}
\bea\label{aoosssaasok1a}
 {S}_{\rm \scriptscriptstyle cshG}^{({\rm dual})}=\int\rd t\int_0^{R}\rd x\ \Big(\tfrac{1}{4\pi}\big[\,(\partial_\mu\varphi)^2+
 (\partial_\mu\vartheta)^2\,\big]-2g\ \re^{-\sqrt{n}\,\varphi}\ \cos\big(\sqrt{n+2}\,\vartheta\big)
 -g'\,  \re^{\frac{2\varphi}{\sqrt{n}}}\,\Big)
 \eea
(the fields $\varphi$ and $\vartheta$ should not be confused with the chiral Bose fields
 used in the main body of this work).
Despite that the action \eqref{aoosssaasok1a} formally depends on three parameters, 
one of these may be eliminated by a constant
shift of the field $\varphi$. This way the relevant coupling constant in the theory is
the combination
$g^{\frac{2}{n}}{g'}$.
It turns out that  the particle mass $m$ in  \eqref{aisodi1289asd}
is related to the parameters in \eqref{aoosssaasok1a} as 
\be\label{Cajsdoi1212}
(Cm)^{\frac{2(n+2)}{n}}=\Big(\frac{\pi }{n}\Big)^{\frac{2}{n}}\ \frac{\pi\,\Gamma(\frac{1}{n})}{\Gamma(1-\frac{1}{n})}\ 
g^{\frac{2}{n}}\, g'\,,\qquad {\rm where}\ \ \ \ 
C=(2 n)^{-\frac{2}{n+2}}\ 
 \frac{\Gamma(\frac{1}{n+2})\Gamma(\frac{3}{2}-\frac{1}{n+2})}{\sqrt{\pi}}
\ . 
\ee

The existence of the dual description \eqref{aoosssaasok1a} allows one to adapt the arguments of
the work \cite{Zamolodchikov:1995aa}, which were applied for the quantum sinh-Gordon model.
Namely, one starts with the Euclidean black hole NLSM,
corresponding to the situation when the correlation length
is infinite, i.e., $m=0$. 
As was already mentioned, the space of states
of this CFT is classified according  to the
highest weight irreps 
of the $\overline{W}_\infty\otimes W_\infty$\,-\,algebra with the central charge $c>2$.
In particular, the decomposition of the  continuous component of the Hilbert space 
is given in \eqref{iaosasdidoi192009AA}.
Each state in the irrep is characterized by the values of 
 $s$, $p$, $\bar{p}$, parameterizing the 
highest weight \eqref{Deltvarpi1aasa}
 as well as their
levels  ${\tt L}$, $\bar{\tt L}$.
To resolve the degeneracy in the level subspaces, one 
can use the eigenbasis of the mutually commuting set of local IM.
Then for fixed $s,\,p,\,\bar{p}$, ${\tt L}$ and $\bar{\tt L}$ the states would be
specified by two sets $\bm{w}=\{w\}_{a=1}^{{\tt L}}$ and $\bar{\bm{w}}=\{\bar{w}\}_{a=1}^{\bar{\tt L}}$
solving a certain algebraic system. The latter coincides with  \eqref{sksksk10},
where $n$ is substituted by $-n-2$, while 
the pair $(s,p)$ is swapped for $(\ri p,\ri s)$  in \eqref{sksksk1} and $(s,\bar{p})\mapsto(\ri \bar{p},\ri s)$ 
in \eqref{sksksk1bar}. 
Following ref.\cite{Zamolodchikov:1995aa}, when the
correlation length $m^{-1}$ is much larger than $R$ but finite, the 
UV behaviour of the energy of 
a state from the eigenbasis is described by the formula
 \bea\label{iaosid12121aa}
 E_{\rm cshG}(R)=\frac{2\pi}{R}\,
\Big(-\frac{1}{6}+\frac{2s^2}{n}+\frac{p^2+{\bar p}^2}{n+2}+{\tt L}+{\bar {\tt L}}
+O\big((\log \varepsilon)^{-\infty}\big)\Big)\ .
 \eea
Here $s=s(R)$ and  the $R$ dependence, up to power law corrections, is  
determined through the quantization condition
\bea\label{quantC1ss1223}
(C\varepsilon)^{-\frac{4\ri s}{n}\ (n+2)}\ \re^{\frac{\ri}{2}\delta_{{\rm cshG}}}=1+
O\big((\log \varepsilon)^{-\infty}\big)
\eea
with  $\delta_{{\rm cshG}}=\delta_{{\rm cshG}}(\bar{\bm{w}},\bm{w}\,|\,\bar{p},p,s)$.
There are strong similarities between \eqref{iaosid12121aa},\,\eqref{quantC1ss1223} and the formulae
\eqref{tower1a},\,\eqref{quantC1} describing the scaling behaviour of the energy  for the 
 Bethe states in
 the  ${\cal Z}_2$ invariant inhomogeneous six-vertex model.
However in the lattice system the phase shift is
related to the eigenvalue of the reflection operator 
$\check{\bf{D}}$ considered in the 
parametric domain corresponding to $c<2$.
Contrary to this,  $\delta_{{\rm cshG}}$ is
expressed in terms of another reflection operator 
in the domain $c>2$:
\be\label{iaosid1828912}
\re^{\frac{\ri}{2}\delta_{{\rm schG}}}=
\frac{\Gamma(\frac{1}{2}+{ p}-\ri s)\Gamma(\frac{1}{2}+{ \bar p}-\ri s)\Gamma^2(1+2\ri s)\Gamma^2(1+
\frac{2\ri s}{n})}
{\Gamma(\frac{1}{2}+{ p}+\ri s)\Gamma(\frac{1}{2}+{\bar p}
+\ri s)\Gamma^2(1-2\ri s)\Gamma^2(1-\frac{2\ri s}{n})}\
\check{R}_{\bar{p},s}^{(c>2)}
(\bar{\bm{w}}) \ 
\check{R}_{p,s}^{(c>2)}({\bm{w}}) \ ,
\ee
where it is assumed that the constant $C$ in \eqref{quantC1ss1223} 
is the same as in \eqref{Cajsdoi1212}.
Recall,  the check notation means  the reflection operators have been normalized so that
their eigenvalue for the highest state is one. An explicit formula  for 
$\check{R}_{{p},s}^{(c>2)}({\bm{w}})$  was obtained in ref.\cite{Kotousov:2019nvt}.\footnote{%
The eigenvalues $\check{R}_{p,s}^{(c>2)}({\bm{w}})$ coincide with those of the
 reflection operator $\check{\mathbb{R}}^{({\rm AKNS})}$ 
defined by (3.36) in ref.\cite{Kotousov:2019nvt} provided the parameters are identified as
 $P_1=\frac{s}{\sqrt{n}}$, $P_2=\frac{p}{\sqrt{n+2}}$ and $\sqrt{k}=\sqrt{n}$.
Note that  $\check{ R}_{p,s}({\boldsymbol  w})$, 
 which is given explicitly in Appendix \ref{app2},
corresponds to the case $c<2$ and coincides  with the eigenvalues of
$\check{\mathbb{R}}^{({\rm AKNS})}$ with the
different identification of the parameters: $P_1=\frac{p}{\sqrt{n+2}}$, $P_2=\frac{s}{\sqrt{n}}$,
$\sqrt{k}=-\ri\sqrt{n+2}$.
}
In the case 
$p=\bar{p}={\tt L}=\bar{\tt L}=0$, the quantization condition \eqref{quantC1ss1223} 
reduces to the one from \cite{Zamolodchikov:1995aa}, for the ground state of the sinh-Gordon model
with the sinh-Gordon coupling constant $b=\sqrt{\frac{2}{n}}$.
This is related to the fact that  the system of coupled thermodynamic Bethe ansatz equations
describing the ground state energy of the complex sinh-Gordon I model with periodic boundary conditions
(${\tt k}=0$), boils down to a single integral equation which is identical to the one
describing the vacuum energy of the sinh-Gordon model.
\bigskip

Formulae \eqref{iaosid12121aa} and \eqref{quantC1ss1223} 
afford an interpretation that sheds some light on an important previously made point. 
They define a particular integrable IR regularization
for the target space manifold of the Euclidean black hole NLSM, which
is different to the one discussed in refs.\cite{Maldacena:2000kv,Hanany:2002ev}.
To illustrate, consider the form of the dual action \eqref{aoosssaasok1a}.
Setting $g'=0$ therein, one obtains the 
dual action for the NLSM.
In this case, in the domain of the configuration space with $\varphi\to +\infty$
the dual Lagrange density becomes that  of two non-interacting Bose fields.
This corresponds to the asymptotically flat domain for the target space manifold.
The addition of the Liouville wall potential $\propto g'\,  \re^{\frac{2\varphi}{\sqrt{n}}}$ into \eqref{aoosssaasok1a}
 works as an IR regularization for the NLSM target space. 
It effectively restricts the value of the non-compact 
field $\varphi$  to the finite interval $\propto \log(1/\varepsilon)$,
which results in the quantization of its zero-mode momentum according to \eqref{iaosid1828912}.
Taking the limit $\varepsilon\to 0$ the continuous spectrum is restored but with a certain density of states,
 similar to as in the lattice model. The latter comes up, for instance, in the computation
of the partition function for the Euclidean black hole NLSM, see eq.\,\eqref{asodi918921}.
The explicit formula for the density of states can be obtained using the identity
\bea
\prod_{\bm{w}\atop{\tt L}-{\rm fixed}}\check{R}_{p,s}^{(c>2)}(\bm{w}) &=&
\prod_{1\le j,m\atop jm \le {\tt L}}
\bigg[\frac{nj+m-2\ri s}{nj+m+2\ri s}\bigg]^{2\,{\rm par}_2({\tt L}-mj)} 
  \\[0.2cm]
&\times&
\prod_{a=0}^{{\tt L}-1}
\Bigg[\frac{\big(\tfrac{1}{2}+a+p-\ri s\big)\,\big(\tfrac{1}{2}+a-p-\ri s\big)}
{\big(\tfrac{1}{2}+a+p+\ri s\big)\,\big(\tfrac{1}{2}+a-p+\ri s\big)}\Bigg]^{{\rm par}_2({\tt L})-d_{a}({\tt L})}\ , 
 \nonumber
\eea
where the integers $d_{a}({\tt L})$ are defined via \eqref{Zdef1b}.
The above  relation is similar to \eqref{iaosidoi43232}, which was used in the derivation of the density of states
\eqref{aisodio12311}.
Thus apart from the  regularization of the Euclidean black hole considered in \cite{Maldacena:2000kv,Hanany:2002ev},
there is another integrable IR regularization of the target manifold, which yields a different density of states
for the continuous spectrum.
This illustrates the statement made at the end of sec.\,\ref{sec213}, that the density of states is not
an intrinsic property of the CFT but depends on the (IR) regularization of the model.
\bigskip

\section{Summary}
The work contains a detailed study of the scaling limit of 
the critical ${\cal Z}_2$ invariant inhomogeneous six-vertex model
in the parametric domain where $\arg(q^2)\in(0,\pi)$
and subject to twisted boundary conditions.
The Yang-Baxter integrability implies
the set of Bethe ansatz equations characterizing the
Bethe states. These form a basis in the $2^N$ dimensional space of states of 
the model defined on the lattice with $N$ columns. On the one hand, our analysis was based on a 
numerical study of the Bethe ansatz equations at large $N$.
On the other, we used the powerful analytical technique
of the ODE/IQFT correspondence. The combination
of numerical and analytical methods allows one to 
investigate in detail the scaling behaviour of not
only the vacuum state in each sector with fixed value of $S^z$, but
also  the excited states as well. 
Below is a summary of the main outcomes of our work.

\begin{itemize}
\item The linear space of states  ${\cal H}$ occurring in the scaling limit of the low
energy states of the lattice system is classified w.r.t. 
the highest weight irreps of the $\overline{W}_\infty\otimes W_\infty$\,-\,algebra
with central charge $-1<c<2$. The 
space splits into two components ${\cal H}^{({\rm cont})}$ and
${\cal H}^{({\rm disc})}$ depending on whether the
spectrum of the highest weights is continuous or discrete.
The linear decomposition of both these components
 into the irreps is provided in sections \ref{sec171} and \ref{sec172}
for generic values of the twist parameter ${\tt k}$.

\item 
The scaling limit of the lattice model yields a certain density of states for the sector ${\cal H}^{({\rm cont})}$.
With this at hand, an expression was obtained in sec.\,\ref{oasid90120912}
for the scaling limit of the
partition function $Z^{({\rm scl})}$ in the form of a series expansion in the modular nome(s),
which is applicable for numerical study.

\item
We confirmed the remarkable proposal of the work \cite{Ikhlef:2011ay}, that one half of the
 partition function
$Z^{({\rm scl})}$ for the case of periodic boundary conditions $({\tt k}=0)$  coincides with
the partition function $Z_{\rm \scriptscriptstyle EBH}$ of the Euclidean black hole Non-Linear Sigma Model,
which was
obtained in refs.\cite{Maldacena:2000kv,Hanany:2002ev}
assuming a certain IR regularization of the target manifold.
For generic ${\tt k}$ we  checked that $\frac{1}{2}\,Z^{({\rm scl})}$  likewise coincides with
the partition function of the properly regularized Euclidean black hole
NLSM with twisted boundary conditions.

\item The finite dimensional space of states of the lattice possesses
a variety of Hermitian structures with the inner product being 
such that the Bethe states obey a certain
orthogonality condition.
In sections \ref{193} and \ref{194} the Hermitian structures  were identified, which 
in the scaling limit induces  the inner products in ${\cal H}^{({\rm cont})}$ 
and ${\cal H}^{({\rm disc})}$ that are consistent with the natural conjugation
conditions in the $\overline{W}_\infty\otimes W_\infty$\,-\,algebra.
Since the central charge $c<2$, the inner products are  not positive definite  ones,
so that ${\cal H}^{({\rm cont})}$ and  ${\cal H}^{({\rm disc})}$ possess
the structure of the pseudo-Hilbert space.
We were led to conclude that the
states from  ${\cal H}^{({\rm cont})}$ and
${\cal H}^{({\rm disc})}$
can not be interpreted simultaneously as  normalizable states 
within a single CFT. We believe that these two spaces
should be considered
as being the result of different scaling limits.

\item 
The algebra of extended conformal symmetry of 
the Euclidean black hole NLSM is the
 $\overline{W}_\infty\otimes W_\infty$\,-\,algebra
but with the central charge $c>2$. Moreover, 
the Hilbert space is equipped with a positive definite inner
product and the QFT is unitary.
For these reasons we reject the proposal that the
Euclidean black hole NLSM  underlies the critical behaviour of
the ${\cal Z}_2$ invariant inhomogeneous six-vertex model,
despite the spectacular coincidence of the partition functions
$Z_{\rm \scriptscriptstyle EBH}$ and $\frac{1}{2}\,Z^{({\rm scl})}$.

\item We revised the original conjecture of \cite{Ikhlef:2011ay} in the following way.
In the case of periodic boundary conditions $({\tt k}=0)$ the 
lattice model possesses an additional global symmetry, that of ${\cal C}$ 
conjugation which, in turn,  is inherited by the space ${\cal H}$.
The ${\cal C}$-even sector of ${\cal H}^{({\rm cont})}$
contains the subspace $\tilde{{\cal H}}^{({\rm cont})}_{\rm even}$ and we propose that it
coincides with the pseudo-Hilbert space of the Lorentzian black hole NLSM
for the space-like domain of the target manifold.
Since the  status of the Lorentzian black hole NLSM is tentative,
our conjecture is essentially an attempt to assign a meaning to the
QFT that goes beyond the classical limit and minisuperspace approximation. 
In sec.\,\ref{sec23} some proposals are made concerning
the field theory interpretation for the case of twisted boundary conditions 
with ${\tt k}\ne 0$.
The local CFT, if it exists, whose pseudo-Hilbert space
coincides with  ${\cal H}^{({\rm cont})}$, in the classical limit with $c\to 2^-$
is described by the gauged ${\rm SL}(2,\mathbb{R})$ WZW model 
subject to certain boundary conditions imposed on the fields.

\item
Among the numerous offshoots of our study, of special mention is
the solution of the long standing non-ultralocality problem for the
complex sine-Gordon I (Lund-Regge) model. Adapting the ideas developed
in ref.\cite{Bazhanov:2018xzh}, we traced the appearance of the classical Sklyanin exchange relations
in this model (see sec.\,\ref{sec232} for details).

\end{itemize}

\section*{Acknowledgments}
The authors thank R.~J.~Baxter, H.~Saleur, V.~Schomerus  and A.~B.~Zamolodchikov for stimulating discussions.
They also gratefully acknowledge support from the Simons Center for Geometry and Physics,
Stony Brook University during the workshop “Exactly Solvable Models of Quantum Field
Theory and Statistical Mechanics” (September 4 -- November 30, 2018), where this work was
initiated. 

\medskip
\noindent
VB acknowledges the support of the Australian Research Council grant DP190103144.

\medskip
\noindent
The research of GK is funded by the Deutsche Forschungsgemeinschaft (DFG, German Research
Foundation) under Germany's Excellence Strategy -- EXC 2121 ``Quantum Universe'' -- 390833306.

\medskip
\noindent
The work of SL is supported by the 
Rutgers New High
Energy Theory Center.

\appendix

\pagebreak

\section{Appendix\label{app1}} 
Formula \eqref{iasisaias} involved in the
 description of the norms of 
the Bethe states in the homogeneous six-vertex model
contains the constants $C$ and $C_0$.
These depend only on
the anisotropy parameter $\beta^2$ entering into the Hamiltonian  $\mathbb{H}_{XXZ}$ 
\eqref{asiisaias}. Though the analytical form of this dependence is not yet known,
it is straightforward to obtain $C$ and $C_0$ numerically for any
$0<\beta^2<1$. Here we present the  corresponding numerical data, which was already quoted in 
sec.11.2 from ref.\cite{Kotousov:2019ygw}.
We also provide similar data for ${ C}^{(\rm alt)}_0$
appearing in eqs.\,\eqref{isisaisa},\,\eqref{isisaisa2},\,\eqref{isisaisa3}
and \eqref{Maisd7812},
that enter into the description of the scaling limit of the
Bethe states in the ${\cal Z}_2$ invariant inhomogeneous six-vertex model.
\bigskip

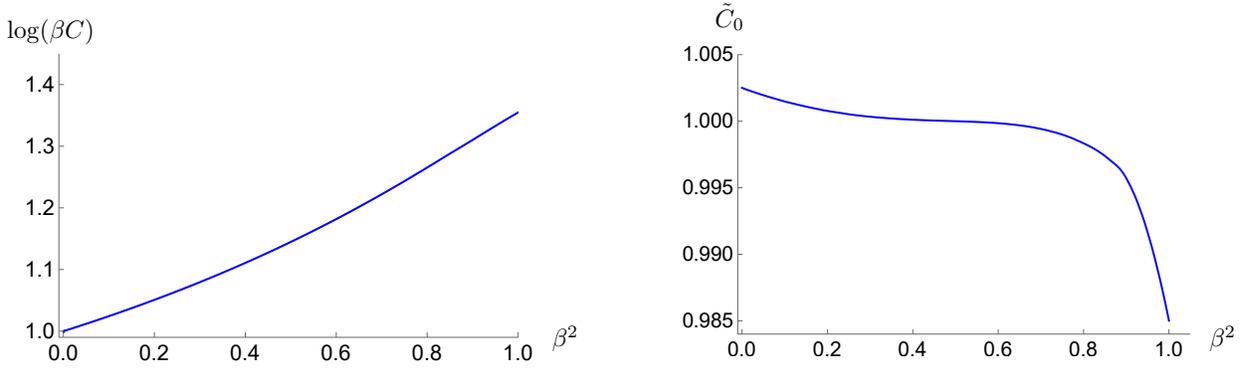
\begin{figure}
\centering
\scalebox{0.95}{
\begin{tikzpicture}
\node at (-3.2,2.5) {\small$\log(\beta C)$};
\node at (4,-1.8) {\small $\beta^2$};
\node at (0,0) {\includegraphics[width=0.42\textwidth]{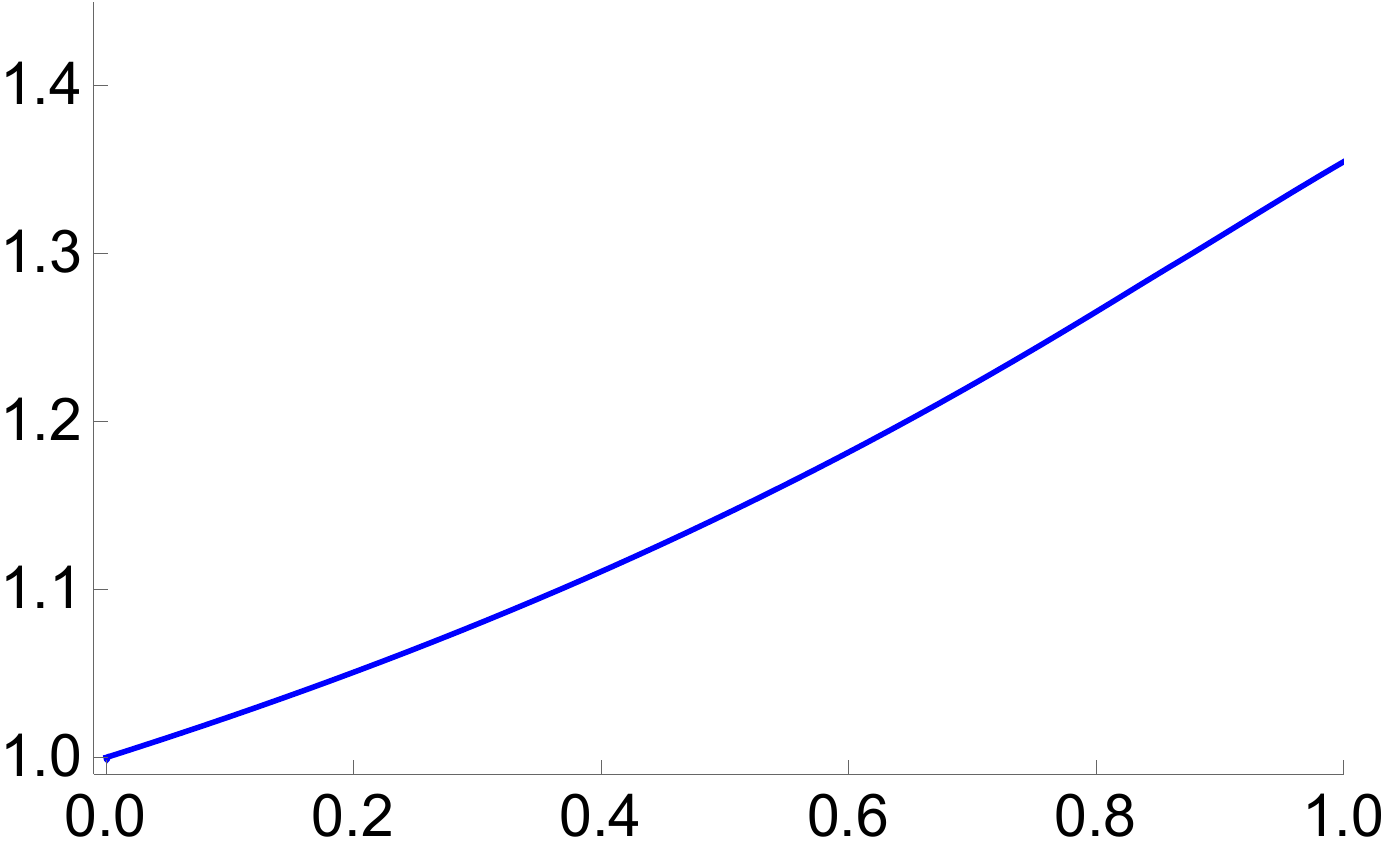}};
\node at (9.2,0.1)  {\includegraphics[width=0.42\textwidth]{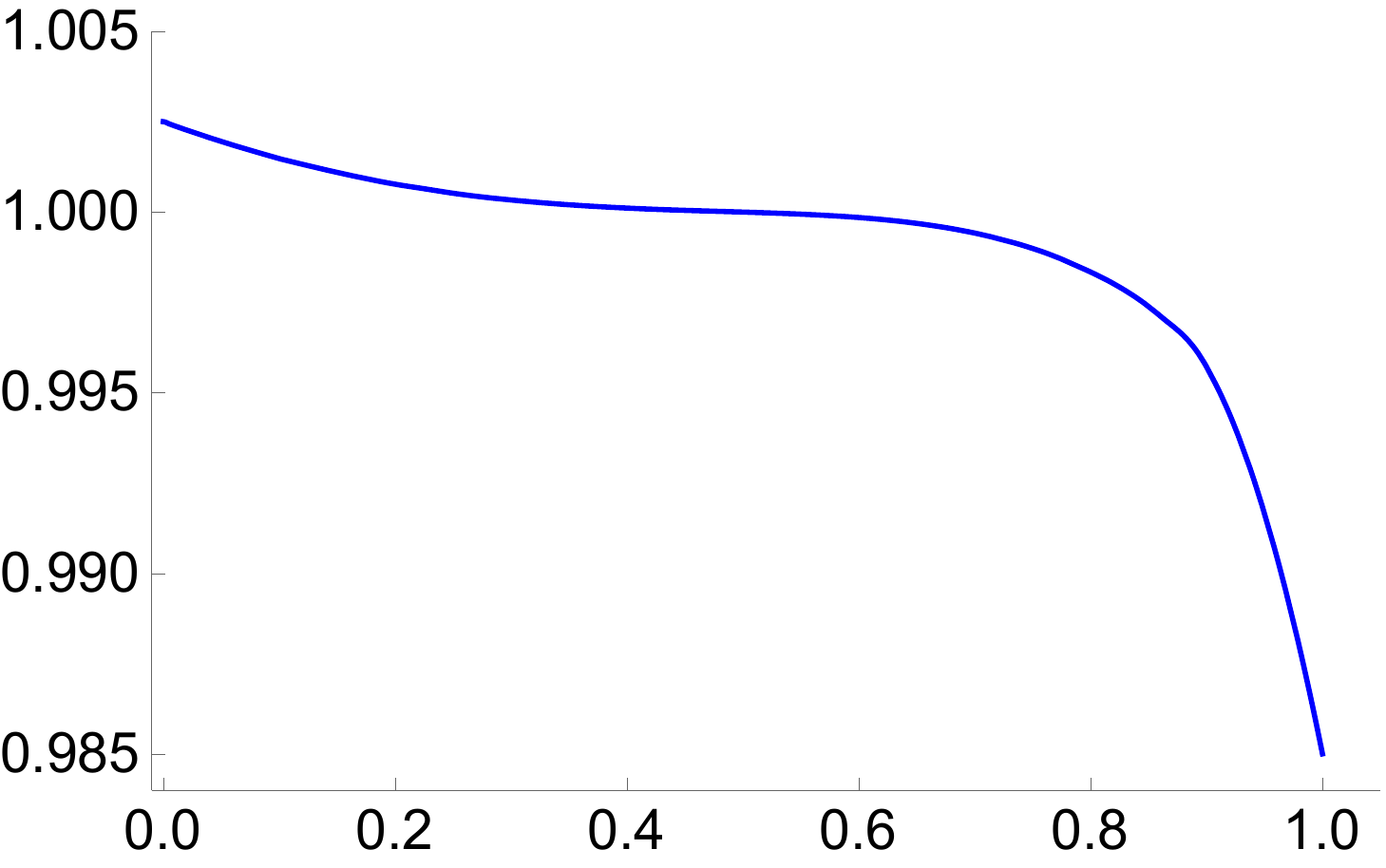}};
\node at (6.3,2.7) {\small${\tilde C}_0$};
\node at (13.2,-1.8) {\small $\beta^2$};
\end{tikzpicture}
}
\caption{\small\label{fig6} 
Numerical data for
$\log(\beta C)$ and  ${\tilde C}_0$ 
(for  $\tilde{C}_0$ see definition \eqref{haassta})
was interpolated and the result is 
plotted in the left and right panels, respectively, as a function of the parameter $\beta^2$.  
Some of the values from which
the interpolation   was obtained are listed in tab.\ref{tabC1a}.}
\end{figure}

Using the fact that for $\beta^2=\frac{1}{2}$
 the homogeneous six-vertex model can be reformulated as a 
non-interacting system of 1D lattice fermions, it is possible to show
that
\be
\beta C\big|_{\beta^2=\frac{1}{2}}=\pi\qquad {\rm and}\qquad
 C_0\big|_{\beta^2=\frac{1}{2}}=(\pi\re )^{-\frac{1}{12}}\ A_{\rm G}=1.07254
\ee
with  $A_{\rm G}$ being the Glaisher constant.
From the numerical data, one expects
\be\label{iasu8798}
\lim_{\beta\to 0}\beta C=\re \,,\qquad \lim_{\beta\to 1}\beta C=\Big(\frac{\pi}{2}\Big)^3\,,
\ee
whereas 
\be\label{isaoi9}
C_0\big|_{\beta=1}=C_0^2|_{\beta^2=\frac{1}{2}}=1.15034\ .
\ee
A plot of $\log(\beta C)$ as a function of $\beta^2$ is given in fig.\ref{fig6} 
and some numerical values are provided in tab.\ref{tabC1a}. 
As for $C_0$, following ref.\cite{Kotousov:2019ygw}, it is convenient to re-write it in the form
\bea\label{haassta}
C_0=\beta^{-\frac{1}{3}}\  \re^{-\frac{1}{6}\,(\beta^{-1}-\beta)^2}\  
\bigg(\frac{\re^{\gamma_{\rm E}+1}}{4\pi}\bigg)^{\frac{1}{12}}\ 
\big(\re^{-\frac{1}{6}(\gamma_{\rm E}+1)}\ A^2_{\rm G}\big)^{\beta^2}\ {\tilde C}_0\ ,
\eea
where for the free fermion case $\tilde{C}_0|_{\beta^2=\frac{1}{2}}=1$.
More generally $|\tilde{C}_0-1|<0.003$ within the domain $0\le\beta^2\le0.85$.
Fig.\ref{fig6} includes
a plot of $\tilde{C}_0$, while some of its numerical values are quoted in tab.\ref{tabC1a}.
\bigskip

\begin{figure}
\centering
\scalebox{1.0}{
\begin{tikzpicture}
\node at (-3.4,2.7) {\small $X$};
\node at (5,-2.1) {\small $\beta^2=\frac{2}{n+2}$};
\node at (0,0) {\includegraphics[width=0.45\textwidth]{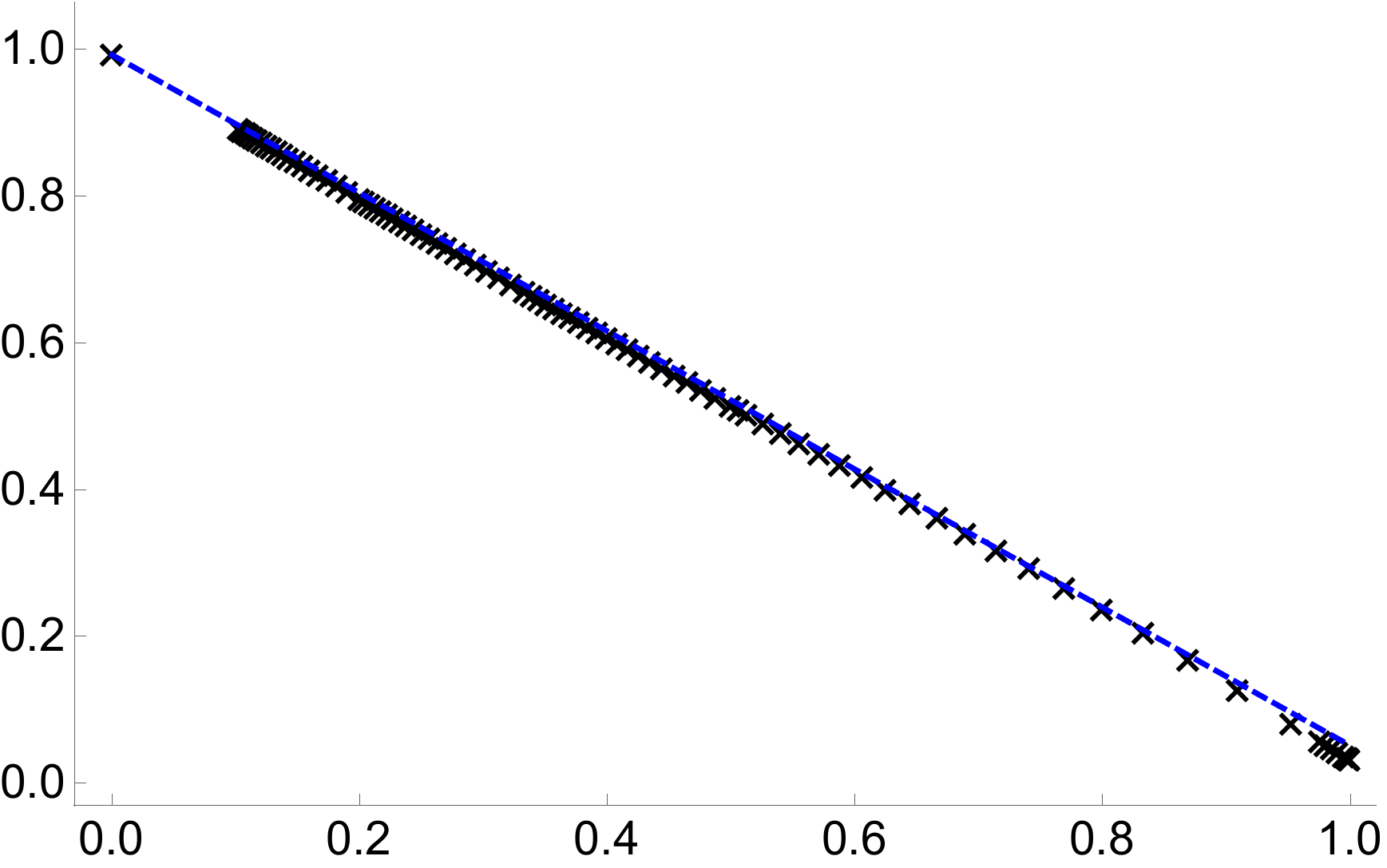}};
\end{tikzpicture}
}
\caption{\label{fig8}\small
A plot of $X$ from \eqref{iusuuas}
as a  function of  $\beta^2=\frac{2}{n+2}$.
The crosses correspond to the numerical data, 
a portion of which can be found in tab.\ref{tab8},
while the dashed line represents the linear
 fit 
 $ 0.991951 - 0.94084 \,\beta^2$. Notice that
 $X\big|_{\beta^2=1}=0.0391$. }
\end{figure}

The values of $C$ entering  into eqs.\,\eqref{isisaisa},\,\eqref{isisaisa2},\,\eqref{isisaisa3}
and \eqref{Maisd7812}
for the description of the norms of the
Bethe states in the ${\cal Z}_2$ invariant inhomogeneous six-vertex model,
are the same as those given above provided 
one sets $\beta^2=\frac{2}{n+2}$.
These formulae in addition contain the 
$n$ dependent constant $C_0^{({\rm alt})}$.
It turns out that $X$, defined through
\bea\label{iusuuas}
\big({ C}^{(\rm alt)}_0\big)^2=2X\,
(\beta C)^{-1+\frac{1}{2} \beta^2}\, C_0^4\ \ \ \ \ \ \ \ \ \Big(\,\beta=\sqrt{\tfrac{2}{n+2}}\ \Big)
\eea
is well approximated by a linear function of $\beta^2$, as shown in fig.\ref{fig8}.
Some numerics for $X$ is contained in tab.\ref{tab8}.

\begin{table}
\centering
\begin{tabular}{|c|c|c|}
\hline
 & & \\[-0.4cm]
$\beta^2$ & $\log(\beta C)$ & $\tilde{C}_0$\\
\hline
 & & \\[-0.4cm]
0.0 & $1.00000$  & $1.00250$ \\
\hline
 & & \\[-0.4cm]
0.1 & $1.02434$  & $1.00148$ \\
\hline
 & & \\[-0.4cm]
0.2 & $1.05074$  & $1.00077$ \\
\hline
 & & \\[-0.4cm]
0.3 & $1.07945$  &$ 1.00033$ \\
\hline
 & & \\[-0.4cm]
0.4 & $1.11070$  & $1.00011$ \\
\hline
 & & \\[-0.4cm]
0.5 & 1.14473 & 1.00000\\
\hline
 & & \\[-0.4cm]
0.6 &$1.18178$  & $0.99985$ \\
\hline
 & & \\[-0.4cm]
0.7 &$1.22201$  & $0.99942$ \\
\hline
 & & \\[-0.4cm]
0.8 & $1.26541$  & $0.99834$ \\
\hline
 & & \\[-0.4cm]
0.9 & $1.3102\overline{3}$ & $0.9957\overline{2}$  \\
\hline
 & & \\[-0.4cm]
1.0 & $1.35475$  & $0.98501$ \\
\hline
\end{tabular}
\caption{\small
Numerical data for
$\log(\beta C)$ and $\tilde{C}_0$ 
(the latter is defined  via \eqref{haassta}). 
The expected accuracy is indicated by the number 
of digits that are presented. Note that for $\beta^2=0.5$, corresponding
to the free fermion case, $\tilde{C}_0=1$ and $\log(\beta C)=\log(\pi)$. 
The values of  
 $\tilde{C}_0$ and $\log(\beta C)$ at
 $\beta^2=0$ given in the table are the result of interpolation, while for $\beta^2=1$
they follow from 
 \eqref{iasu8798} and
\eqref{isaoi9}.
\label{tabC1a}}
\end{table}

\begin{table}
\begin{center}
\begin{tabular}{|c|c||c|c|}
\hline
 & & & \\[-0.4cm]
$n$ & $X $ & $n$ & $X$ \\[0.1cm]
\hline
 & & & \\[-0.4cm]
$ 2.0 $ & $0.521203985$ & $4.0$ & 0.676335763 \\
\hline
 & & & \\[-0.4cm]
$ 2.5 $ & $0.572499215$ & $5.0$ &0.721418356 \\
\hline
 & & & \\[-0.4cm]
$ 3.0 $ & $0.613818210$ & $6.0$ & 0.755474688 \\
\hline
 & & & \\[-0.4cm]
$ 3.5 $ & $0.647834794$ & $7.0$ & 0.782110959 \\
\hline
\end{tabular}
\caption{\small Some numerical values for the constant $X$, 
defined through \eqref{iusuuas}.
\label{tab8}}
\end{center}
\end{table}

\pagebreak
\section{Appendix\label{app2}}
In this appendix we explain the assumptions and fill some gaps in the 
derivation that lead  to conjectures (A) and (B) from sec.\,\ref{sec102}.
As before we will always take 
${\tt L}$ and $\bar{\tt L}$ to be some fixed non-negative integers;
$p$ and $\bar{p}$ to be real numbers such that $p+\bar{p}=S^z=0,1,2,\ldots\ $;
and $s$ to belong to the strip $0<\Im m(s)\le\frac{n}{4}$. 
\bigskip

The 
function ${ D}_{p,s}({\boldsymbol  w})$
entering into \eqref{pdpd000d}
reads as
\be\label{oiaodi1a1a1b}
{ D}_{p,s}({\boldsymbol  w})=
2^{\frac{2\ri(n+2)s}{n}}\ \frac{\Gamma(\frac{1}{2}+p-\ri s)}{\Gamma(\frac{1}{2}+p+\ri s)}\  \ 
\check{ D}_{p,s}({\boldsymbol  w})\,,
\ee
where the explicit analytical expression for
$ \check{D}_{p,s}({\bm w})$ in
 terms of $p$, $s$ and the solution set 
$\bm{w}=\{w_a\}_{a=1}^{{\tt L}}$ 
was obtained in  ref.\cite{Kotousov:2019nvt}.
Formula (3.11) from that work gives
\bea\label{Dformula1}
 \check{D}_{p,s}({\bm w})&=&
(-1)^{\tt L}\ 
\prod_{a=1}^{\tt L}\frac{p+a-\frac{1}{2}-\ri s}{p+a-\frac{1}{2}+\ri s}\ \ \ 
\frac{\det\big(w_a^{b-1}\, V^{(+)}_a(b)\big)}{\det\big(w_a^{b-1}\,V^{(-)}_a(b)\big)}\,,
 \eea
where
\bea\label{aoiaiaskjjsss}
V^{(\pm)}_a(D)
&=&  (D-1)^2-
 \bigg( 2p+2+n\mp 2w_a+
\sum_{b\not= a}^{\tt L}\frac{4 w_a}{w_a-w_b}\bigg)\ (D-1)\nonumber\\
&+&\tfrac{1}{2}\, n^2+\big(p+\tfrac{3}{2}\big)\ n\mp (n+1+2p\pm 2\ri s)\ w_a+2 p+1
\\[0.2cm]
&+&\bigg(\sum_{b\not=a}^{\tt L}\frac{2w_a}{w_a-w_b}\,\bigg)^2
+\, \big(\,
4p+2\mp 4w_a+n \,\big)\,
\sum_{b\not=a}^{\tt L}\frac{w_a}{w_a-w_b}\nonumber\ .
\eea 
Note that 
there exists a similar expression for $R_{p,s}(\bm{w})$ --
the coefficient which appears in the asymptotic formula \eqref{eq3331}
for a product over the Bethe roots.
Namely $R_{p,s}(\bm{w})=R^{(0)}_{p,s}\,\check{R}_{p,s}(\bm{w})$
with $R^{(0)}_{p,s}$ from \eqref{opsod898s} and $\check{R}_{p,s}(\bm{w})$,
instead of a ratio of two determinants as in \eqref{Dformula1}, is given by
\be\label{aosid981212}
\check{R}_{p,s}(\bm{w})= \frac{(-1)^{\tt L}}{\prod_{a=1}^{\tt L}w_a^2}\ 
\frac{\det\big(w_a^{b-1}\,  V^{(+)}_a(b) \big)\,
\det\big(w_a^{b-1}\,V^{(-)}_a(b) \big)}
{\prod_{b>a}(w_b-w_a)^2 \prod_{a=1}^{\tt L}(2p+2a-1+2\ri s\big)
\big(2p+2a-1- 2\ri s\big)}\ \ .
\ee

\bigskip

It was pointed out in ref.\cite{Kotousov:2019nvt} that the product of $\check{ D}_{p,s}({\boldsymbol  w})$
over all the ${\rm par}_2({\tt L})$ solutions sets $\bm{w}$ of \eqref{sksksk1} with fixed ${\tt L}$ 
admits a simple form. 
According to formula (5.23) from that work one has
\bea\label{iaoio898}
\prod_{\bm{w}\atop{\tt L}-{\rm fixed}}\check{D}_{p,s}(\bm{w})
=\prod_{m=1}^{\tt L}\!
 \prod_{1\leq j,k\atop jk\leq m}\!
\bigg[\frac{(2p-2\ri s+2k-j)\,(2p+2\ri s-2k+j) }
{(2p-2\ri s-2k+j)\,(2p+2\ri s+2k-j)}\bigg]^{{\tt par}_1(m-kj)\,{\tt par}_1({\tt L}-m)}\ .
\eea
This expression can be rewritten  in a way which 
is more convenient for  an analysis of the condition \eqref{pdpd000d}.
First one should split the  r.h.s. of \eqref{iaoio898}
into two terms, where the index $j$ runs over even numbers
$j=2\ell$ and odd  numbers $j=2\ell-1$, respectively:
\bea
&&\prod_{\bm{w}\atop{\tt L}-{\rm fixed}}\check{D}_{p,s}(\bm{w})
=\prod_{m=1}^{\tt L}\
 \prod_{1\leq \ell,k\atop 2\ell k\leq m}\!
\bigg[\frac{(2p-2\ri s+2k-2\ell)\,(2p+2\ri s-2k+2\ell) }
{(2p-2\ri s-2k+2\ell)\,(2p+2\ri s+2k-2\ell)}\bigg]^{{\tt par}_1(m-2\ell k)\,{\tt par}_1({\tt L}-m)}\nonumber\\[0.2cm]
&&\times\prod_{m=1}^{\tt L}\!
 \prod_{1\leq \ell, k\atop (2\ell-1)k\leq m}\!
\bigg[\frac{(2p-2\ri s+2k-2\ell+1)\,(2p+2\ri s-2k+2\ell-1) }
{(2p-2\ri s-2k+2\ell-1)\,(2p+2\ri s+2k-2\ell+1)}\bigg]^{{\tt par}_1(m-(2\ell-1)k)\,{\tt par}_1({\tt L}-m)}
\ .\nonumber
\eea
The first line in the r.h.s. of the above equation is one since the numerator 
 coincides with the denominator when the dummy variables are swapped $k\leftrightarrow\ell$.
 As for the second line,
it contains poles and zeroes at $s=\pm\ri\,(p+\frac{1}{2}+a)$ with integer $a=-{\tt L},-{\tt L}+1,\ldots,{\tt L}$.
To compute their multiplicity consider,
for instance,
 the zero at $s=\ri\,(p+\frac{1}{2}+a)$ with $a\ge 0$.
The relevant terms are the  first factor in the numerator   with
$k=a+\ell$ and  the first factor of the denominator, where $\ell=a+k+1$. 
Counting the number of times they occur in the product 
leads to the following expression for the multiplicity of the zero
\bea
\sum_{m=1}^{{\tt L}}{\tt par}_1({\tt L}-m)\,
\sum_{j\ge 1}{\tt par}_1\big(m-(2j-1)(j+a)\big)-{\tt par}_1\big(m-j\,(2j+2a+1)\big) \nonumber \\[0.2cm]
=\sum_{m=1}^{{\tt L}}{\tt par}_1({\tt L}-m)\,
\sum_{j\ge 1}(-1)^{j+1}\ {\tt par}_1\big(m-j\,(a+\tfrac{j+1}{2})\big)\ .\nonumber
\eea 
This way, and using the identity  ${\rm par}_2({\tt L})=\sum_{m=0}^{\tt L}{\rm par}_1({\tt L}-m)\,{\rm par}_1(m)$,
one arrives at
\be\label{B5eq1}
\prod_{\bm{w}\atop{\tt L}-{\rm fixed}}\check{D}_{p,s}(\bm{w})=
\prod_{a=0}^{{\tt L}-1}
\Bigg[\frac{\big(\tfrac{1}{2}+a+p-\ri s\big)\,\big(\tfrac{1}{2}+a-p-\ri s\big)}
{\big(\tfrac{1}{2}+a+p+\ri s\big)\,\big(\tfrac{1}{2}+a-p+\ri s\big)}\Bigg]^{{\tt par}_2({\tt L})-d_{a}({\tt L})}
\ee
with
\be\label{iaopopopo}
d_a({\tt L})=\sum_{m=0}^{{\tt L}}\,{\tt par}_1({\tt L}-m)\,\sum_{j\ge 0}(-1)^{j}\ {\tt par}_1\big(m-j(a+\tfrac{j+1}{2})\big)\ .
\ee
A direct computation yields that 
the generating function for the integers \eqref{iaopopopo} 
is $\chi_{a}({\tt q})$ from  \eqref{Zdef1b}.
\bigskip

Conjecture (A) concerns the possible positions
 of the singularities  of 
${ D}_{{\bar{p}},s}(\bar{{\boldsymbol  w}})\,{ D}_{p,s}({\boldsymbol  w})$
as a function of $s$, where $\bm{w}$ and $\bar{\bm{w}}$
solve the algebraic system \eqref{sksksk10}  that contains $s$ as a parameter.
It is instructive to consider first the simplest case that is not the vacuum with ${\tt L}=1$ and $\bar{\tt L}=0$.
When ${\tt L}$ is set to one
 eq.\,\eqref{sksksk1} becomes a quadratic for $w\equiv w_1$, whose two solutions are:
\be\label{eqw1}
w_\pm=-\frac{n+1}{2n}\,\Bigg(2\ri\,s\pm\sqrt{n(n+2)}\ \sqrt{1-\frac{4p^2}{(n+1)^2}-\frac{4s^2}{n(n+2)}}\ \Bigg)\ .
\ee
In turn
\be\label{oapsod901920}
{ D}_{{\bar{p}},s}\,(\bar{\bm{w}}){ D}_{p,s}(\bm{w})\big|_{\bar{\tt L}=0,{\tt L}=1}
=2^{\frac{4\ri(n+2)s}{n}}\ 
\frac{\Gamma(\frac{1}{2}+p-\ri s)\,\Gamma(\frac{1}{2}+\bar{p}-\ri s)}
       {\Gamma(\frac{1}{2}+p+\ri s)\,\Gamma(\frac{1}{2}+\bar{p}+\ri s)}\
\ \check{ D}_{p,s}(w)
\ee
with
\be\label{opaop0302}
\check{ D}_{p,s}(w)=
\frac{(1+2p-2\ri s)(1-2p-2\ri s)}{(1+2p+2\ri s)(1-2p+2\ri s)}\ \frac{2n w-(n+2)(n-2\ri s)}{2n w+(n+2)(n+2\ri s)}\qquad
\qquad
(w=w_\pm)
\ee
(see also eq.\,(48) in \cite{Bazhanov:2019xvy}).
One can check that the singularities of \eqref{opaop0302} are simple poles
located at  $s=\ri\,(p+\frac{1}{2})$ and $s=-\ri\,(p-\frac{1}{2})$, while
 its zeroes  occur at $s=-\ri\,(p+\frac{1}{2})$ and $s=\ri\,(p-\frac{1}{2})$
also with multiplicity one.
Conjecture (A) is based on the assumption that 
for  given  ${\tt L}=0,1,2,\ldots$ and any set $\bm{w}$
all the singularities of
$\check{ D}_{p,s}({\boldsymbol  w})$ as a function of $s$ are poles 
and furthermore, no pole of $\check{ D}_{p,s}({\boldsymbol  w})$
coincides with a zero of another $\check{ D}_{p,s}({\boldsymbol  w}')$.
Then eqs.\,\eqref{oiaodi1a1a1b} and \eqref{B5eq1} imply that the possible values  of $s$ at which
the product ${ D}_{{\bar{p}},s}(\bar{{\boldsymbol  w}})\,{ D}_{p,s}({\boldsymbol  w})$ 
is singular is given by
$s=+\ri\,( p+\frac{1}{2}+a),+\ri\,(\bar{p}+\frac{1}{2}+a)$ with $a\ge 0$ 
(corresponding to the poles of the $\Gamma$\,-\,function) and 
$s=-\ri\,( p+\frac{1}{2}+a),-\ri\,(\bar{p}+\frac{1}{2}+a)$,
 where $a$ is any integer.
 Since the parameters $p$ and $\bar{p}$
are related as  $p+\bar{p}=S^z=0,1,2,\ldots\ $,
one has 
$\ri\,(p+\frac{1}{2}+a)=\ri\,(-\bar{p}-\frac{1}{2}-a')$ and
$\ri\,(\bar{p}+\frac{1}{2}+a)=\ri\,(-p-\frac{1}{2}-a')$ with
$a'=-a-S^z-1$. Thus it is sufficient to focus on the cases 
$s=\ri\mathfrak{q}_a\equiv\ri\,(-p-\frac{1}{2}-a)$ and $s=\ri\bar{\mathfrak{q}}_a\equiv\ri\,(-\bar{p}-\frac{1}{2}-a)$.
The additional requirement $0<\mathfrak{q}_a,\bar{\mathfrak{q}}_a\le\frac{n}{4}$ 
yields the inequality \eqref{oiasio8998}
 on the values of the integer $a$.
\bigskip

Making a further assumption that 
all the poles of $\check{ D}_{p,s}({\boldsymbol  w})$
are simple,
the number of solution sets $\bm{w}$ 
such that 
 $\check{ D}_{p,s}({\boldsymbol  w})$ is singular 
at $s=\ri\mathfrak{q}_a$ or $\ri\bar{\mathfrak{q}}_a$ may be determined
by counting the multiplicity of that pole occurring in the product
$\prod_{\bm{w}}\check{ D}_{p,s}({\boldsymbol  w})$ \eqref{B5eq1}.
Notice that, due to the relation
$\check{ D}_{p,s}({\boldsymbol  w})=\big(\check{ D}_{p,-s}(-{\boldsymbol  w})\big)^{-1}$,
it would also follow that 
the zeroes of $\check{ D}_{p,s}({\boldsymbol  w})$ are simple as well. 
For the case ${\tt L}=1$ the assumption can be checked
using the explicit formula \eqref{opaop0302}, while for
 ${\tt L}=2,3$ it has been numerically verified.
Once accepted, 
conjecture (B) in sec.\,\ref{sec102}, regarding
the number of solutions ${\cal N}_a^{(\bar{{\tt L}},{\tt L})}$ and $\bar{{\cal N}}_a^{(\bar{{\tt L}},{\tt L})}$ 
of the joint system \eqref{sksksk10}  for which
${ D}_{{\bar{p}},s}(\bar{{\boldsymbol  w}})\,{ D}_{p,s}({\boldsymbol  w})$
is singular at $s=\ri\mathfrak{q}_a$ and $s=\ri\bar{\mathfrak{q}}_a$, respectively, follows.
\bigskip

As an illustration, let's consider 
the case $s=\ri\mathfrak{q}_a=-\ri\,(p+\frac{1}{2}+a)$, where  $a\in\mathbb{Z}$.
The multiplicity of the corresponding pole in the product $\prod_{\bm{w}}{ D}_{p,s}({\boldsymbol  w})$ follows
from \eqref{oiaodi1a1a1b} and \eqref{B5eq1}.
It is given by $d_a({\tt L})$  provided that the definition 
\eqref{iaopopopo}  is extended to the case of negative $a$ as
\be\label{9s8d98sd9}
d_{-a-1}({\tt L})={\tt par}_2({\tt L})-d_{a}({\tt L})
\ee
(see footnote \ref{ft4}).
It is important to keep in mind that 
${ D}_{{\bar{p}},s}(\bar{{\boldsymbol  w}})$ 
could possess
a simple zero at  $s=\ri\mathfrak{q}_a$ that would cancel
the  pole of ${ D}_{p,s}({\boldsymbol  w})$
and render the product
${ D}_{{\bar{p}},s}(\bar{{\boldsymbol  w}})\,{ D}_{p,s}({\boldsymbol  w})$
finite.
The barred version of \eqref{oiaodi1a1a1b},\,\eqref{B5eq1} with  $p$ and  ${\tt L}$  replaced by
$\bar{p}= S^z-p$ and $\bar{\tt L}$, respectively, yields that the number of solution sets $\bar{\bm{w}}$
for which ${ D}_{{\bar{p}},s}(\bar{{\boldsymbol  w}})$ vanishes at $s=\ri\mathfrak{q}_a$
is $d_{-1-a-S^z}(\bar{\tt L})={\tt par}_2(\bar{{\tt L}})-d_{a+S^z}(\bar{\tt L})$. 
This way one obtains ${\cal N}_a^{(\bar{{\tt L}},{\tt L})}=d_{a+S^z}(\bar{\tt L})\,d_a({\tt L})$.
For the case $s=\ri\bar{\mathfrak{q}}_a$
one should simply interchange  ${\tt L}\leftrightarrow\bar{{\tt L}}$ 
in the r.h.s. of this relation, see eq.\,\eqref{99898dsjksj}.
\newpage

\section{Supplementary figures\label{AppC}}
The figures below, together with figs.\ref{sdist1} and \ref{fig130}
from the main body of the text,
present numerical 
data for the low energy states 
of the Hamiltonian $\mathbb{H}$ \eqref{aioiisa},\,\eqref{BC1a} with 
$N=22$,  $q=\re^{\frac{\ri\pi}{5}}$  ($n=3$), ${\tt k}=-0.18$ and in the sector $S^z=1$.
This was used to perform the
classification of the low energy states quoted in tab.\ref{tab001}.
The states are grouped according to their value of the winding number ${\tt w}$
as well as  the levels ${\tt L}$ and $\bar{{\tt L}}$, which were assigned to them using the procedure
described in sec.\ref{sec18}.
In all the figures the open circles
depict the distribution  of
$b(N)=\frac{n}{4\pi}\log(B)$ in the complex plane,
where $B$ is the eigenvalue of the quasi-shift operator $\mathbb{B}$ \eqref{qshift} that was obtained in the course
of the numerical diagonalization of the Hamiltonian $\mathbb{H}$.
The filled circles and squares correspond to the solutions 
$b_*$ of eq.\,\eqref{iasodi12190}, where
 for the circles 
$\lim_{N\to \infty}\Im m\big(b_*(N)\big)=0$,
while for the squares $\lim_{N\to \infty}\Im m\big(b_*(N)\big)\ne0$ and
$\lim_{N\to \infty}\Re e\big(b_*(N)\big)=0$.
\bigskip

\vskip1cm

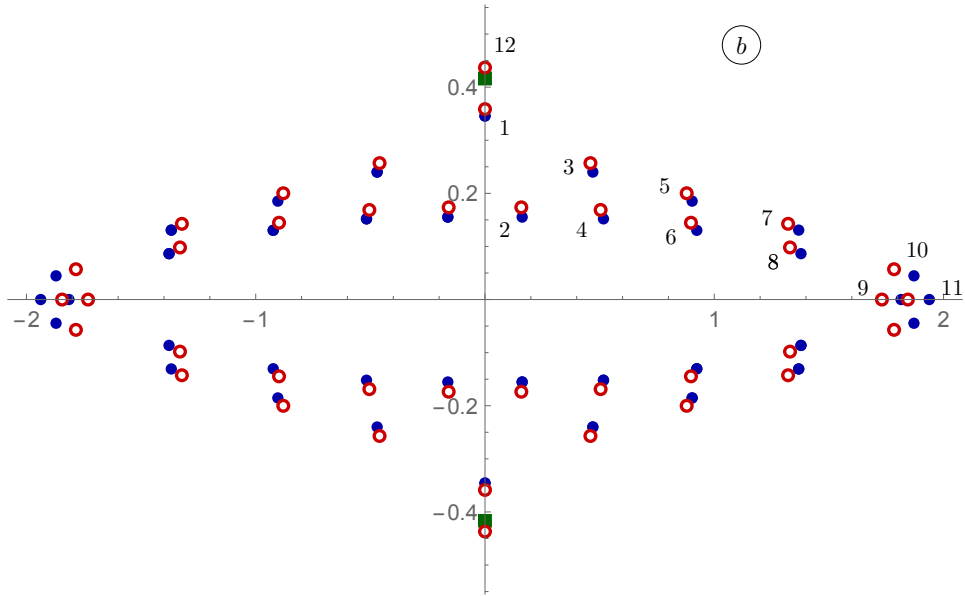
\begin{figure}[h!]
\centering
\scalebox{0.85}{
\begin{tikzpicture}
\node at (4,4) {$b$};
\draw  (4,4) circle [radius=0.3];
\node at (0,0) {\includegraphics[width=15cm]{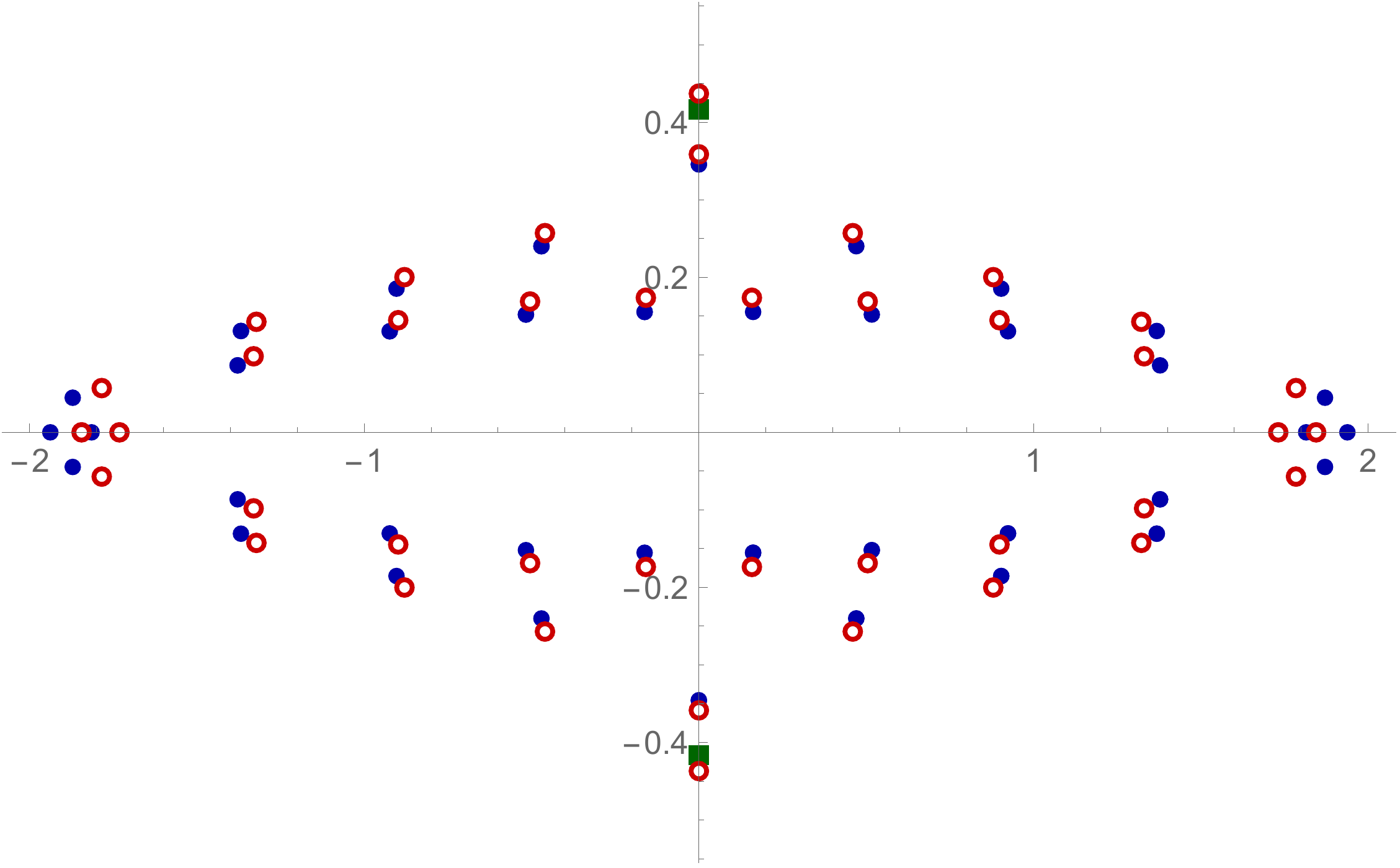}};
\node at (7.3,0.2) {\small $11$};
\node at (6.75,0.8) {\small $10$};
\node at (5.9,0.2) {\small $9$};
\node at (4.5,0.6) {\small $8$};
\node at (4.4,1.3) {\small $7$};
\node at (4.5,0.6) {\small $8$};
\node at (2.9,1.0) {\small $6$};
\node at (2.8,1.8) {\small $5$};
\node at (1.5,1.1) {\small $4$};
\node at (1.3,2.1) {\small $3$};
\node at (0.3,1.1) {\small $2$};
\node at (0.3,2.7) {\small $1$};
\node at (0.3,4) {\small $12$};
\end{tikzpicture}
}
\caption{
\small
The value of $b(N)$ for the  $40=38+2$ states (open circles) having ${\tt w}=0$ and  $({\tt L},\bar{\tt L})=(1,0)$,\,$(0,1)$.
The $12$ states 
with $\Re e\big(b(N)\big)\ge 0$ and $\Im m\big(b(N)\big)\ge 0$  are numbered
consistently with fig.\,\ref{fig999}, where the $N$ dependence of $b(N)$ 
for the corresponding RG trajectories is plotted.
\label{fig20AA}}
\end{figure}
\begin{figure}[h!]
\centering
\scalebox{0.85}{
\begin{tikzpicture}
\node at (4,4) {$b$};
\draw  (4,4) circle [radius=0.3];
\node at (0,0) {\includegraphics[width=15cm]{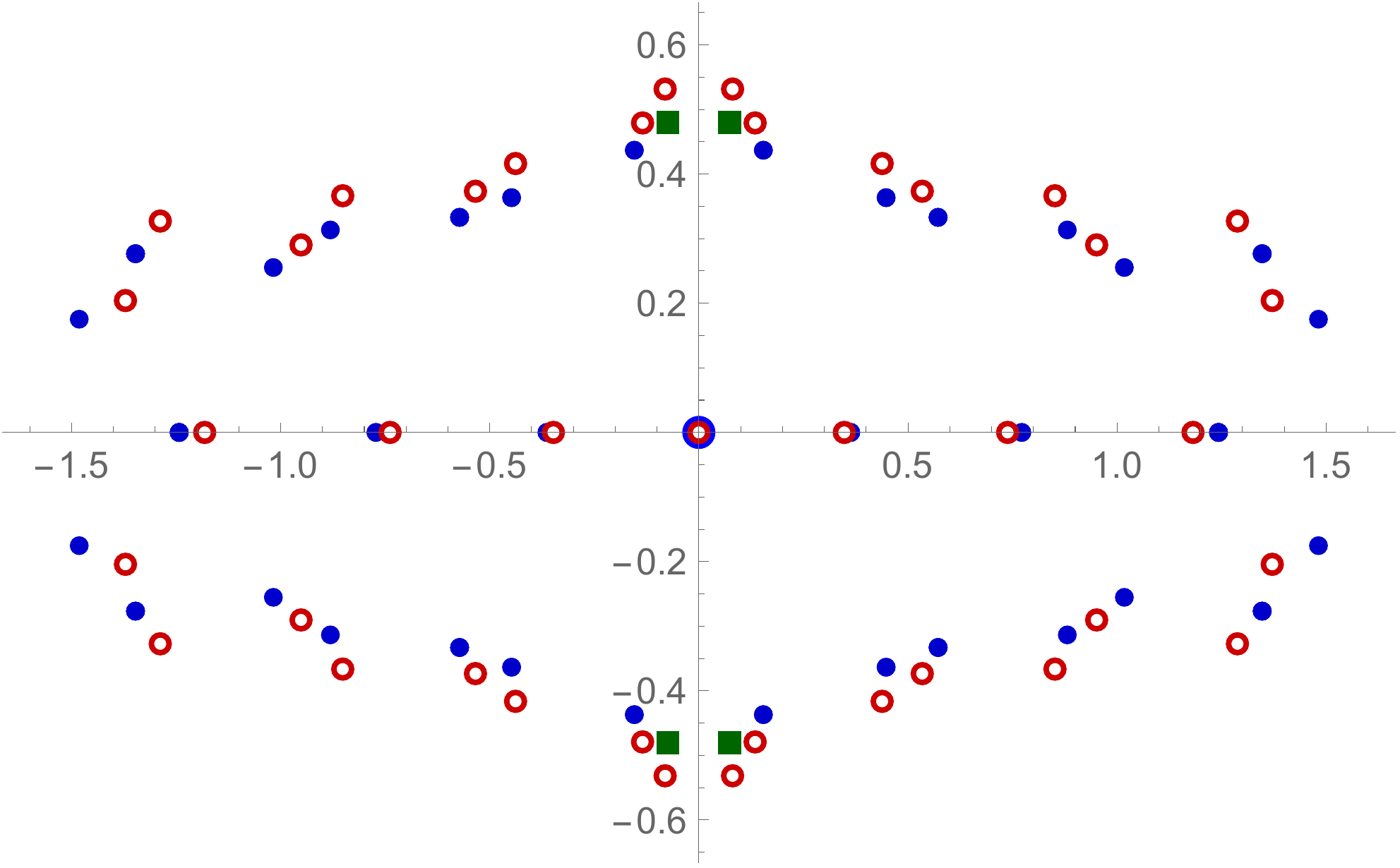}};
\end{tikzpicture}
}
\caption{
\small
The value of $b(N)$ for the  $40=36+4$ states 
(open circles) having ${\tt w}=0$ and  $({\tt L},\bar{\tt L})=(2,0)$.
}
\end{figure}
\begin{figure}[h!]
\centering
\scalebox{0.85}{
\begin{tikzpicture}
\node at (4,4) {$b$};
\draw  (4,4) circle [radius=0.3];
\node at (0,0) {\includegraphics[width=15cm]{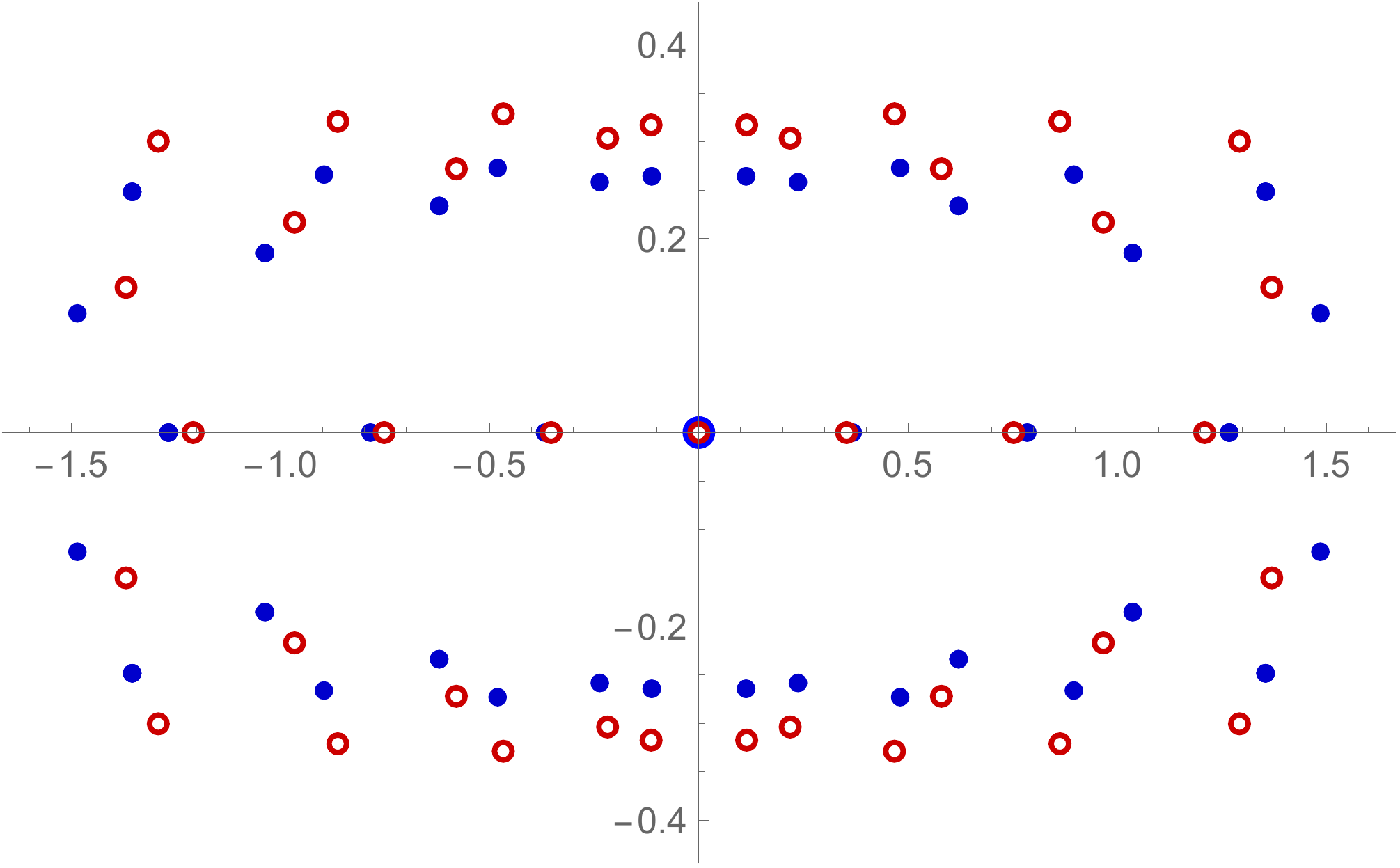}};
\end{tikzpicture}
}
\caption{
\small
The value of $b(N)$ for the  $40=40+0$ states 
(open circles) having ${\tt w}=0$ and  $({\tt L},\bar{\tt L})=(0,2)$.
}
\end{figure}
\begin{figure}[h!]
\centering
\scalebox{0.85}{
\begin{tikzpicture}
\node at (4,4) {$b$};
\draw  (4,4) circle [radius=0.3];
\node at (0,0) {\includegraphics[width=15cm]{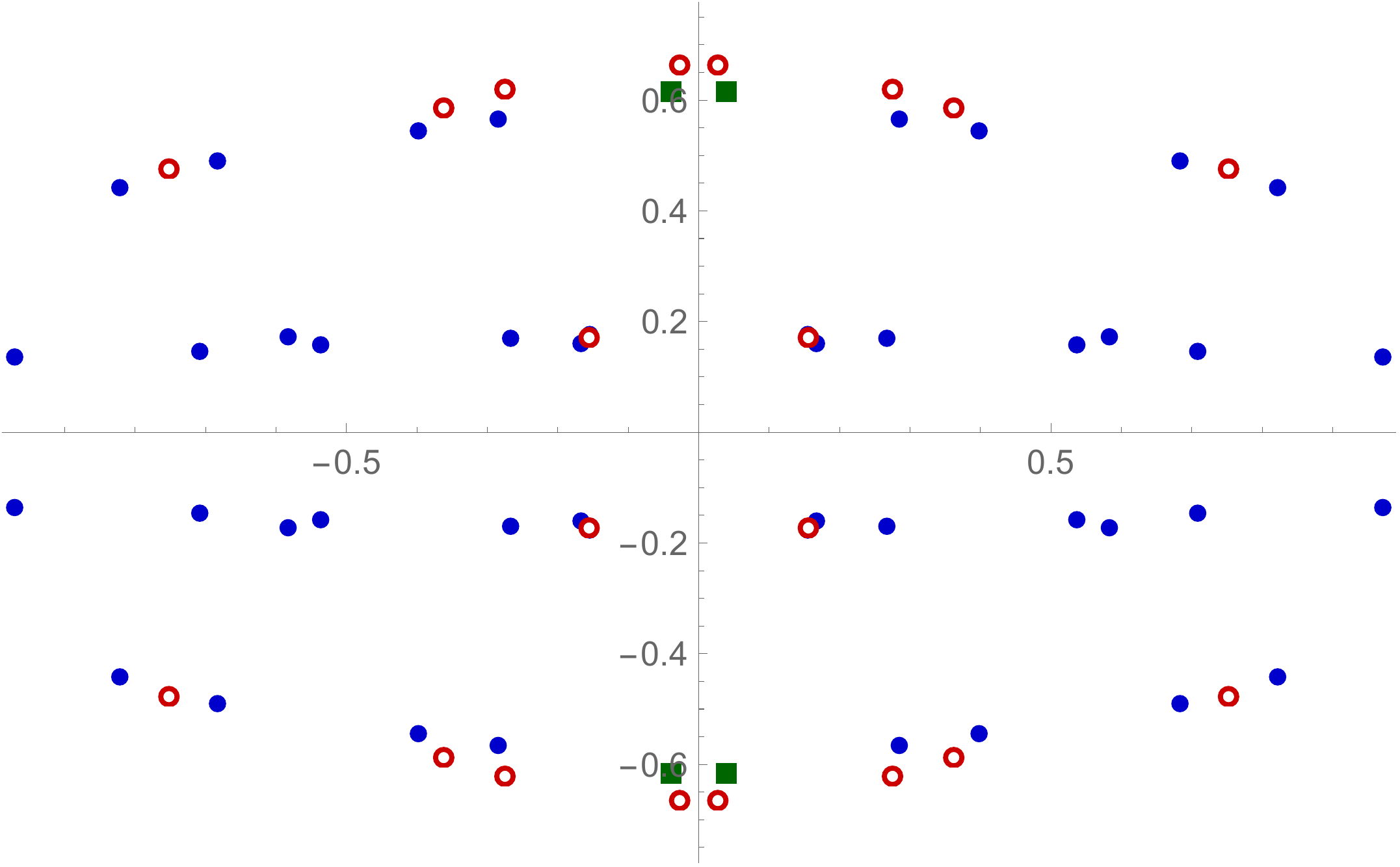}};
\end{tikzpicture}
}
\caption{
\small
The value of $b(N)$ for the  $20=16+4$ states (open circles) having ${\tt w}=0$ and  $({\tt L},\bar{\tt L})=(2,1)$.
}
\end{figure}
\begin{figure}[h!]
\centering
\scalebox{0.85}{
\begin{tikzpicture}
\node at (4,4) {$b$};
\draw  (4,4) circle [radius=0.3];
\node at (0,0) {\includegraphics[width=15cm]{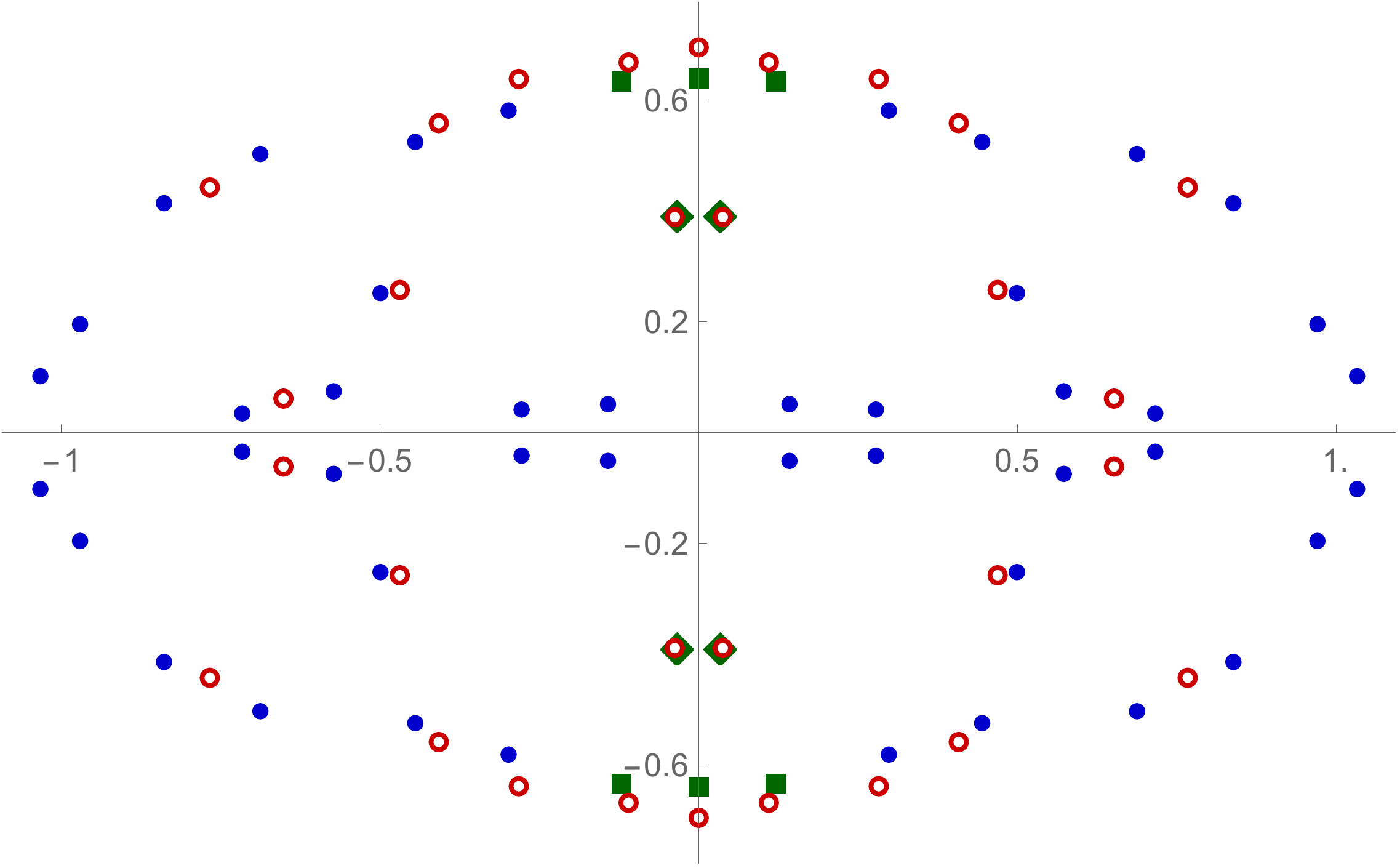}};
\end{tikzpicture}
}
\caption{
\small
The value of $b(N)$ for the  $30=22+8$ states (open circles) having ${\tt w}=0$ and  $({\tt L},\bar{\tt L})=(1,2)$.
As with the filled circles and squares, the filled diamonds correspond to solutions $b_*(N)$ of 
eq.\,\eqref{iasodi12190}.
These  form two pairs
which have the same value of $\Im m\big(b_*(N)\big)$
and opposite real part. At large but finite $N$
the  diamonds from the upper pair collide at the imaginary axis
at which point for one of  the  diamonds
$b_*(N)\to+\frac{9\ri}{20}$ while for the other
one $b_*(N)\to 0$.
The $N$ dependence of $b_*(N)$ for the lower pair of diamonds
is obtained from that of the upper pair via complex conjugation.
}
\end{figure}

\begin{figure}
\centering
\scalebox{0.9}{
\begin{tikzpicture}
\node at (4,4) {$b$};
\draw  (4,4) circle [radius=0.3];
\node at (0,0) {\includegraphics[width=15cm]{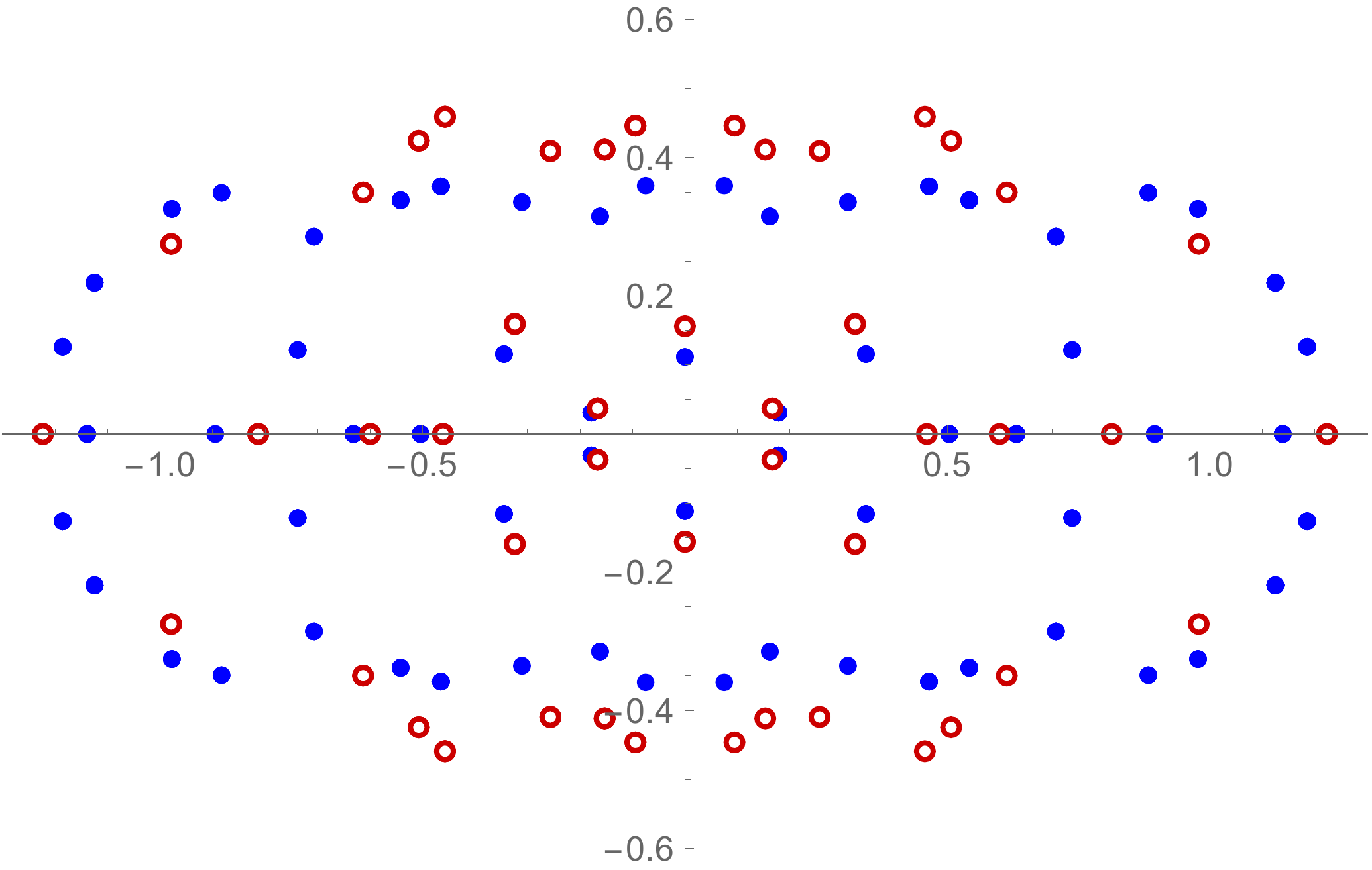}};
\end{tikzpicture}
}
\caption{
\small
The value of $b(N)$ for the  $46=46+0$ states (open circles) having ${\tt w}=0$ and  $({\tt L},\bar{\tt L})=(0,3)$.
}
\end{figure}

\vskip 2cm
\begin{figure}[h!]
\centering
\scalebox{0.9}{
\begin{tikzpicture}
\node at (4,4) {$b$};
\draw  (4,4) circle [radius=0.3];
\node at (0,0) {\includegraphics[width=15cm]{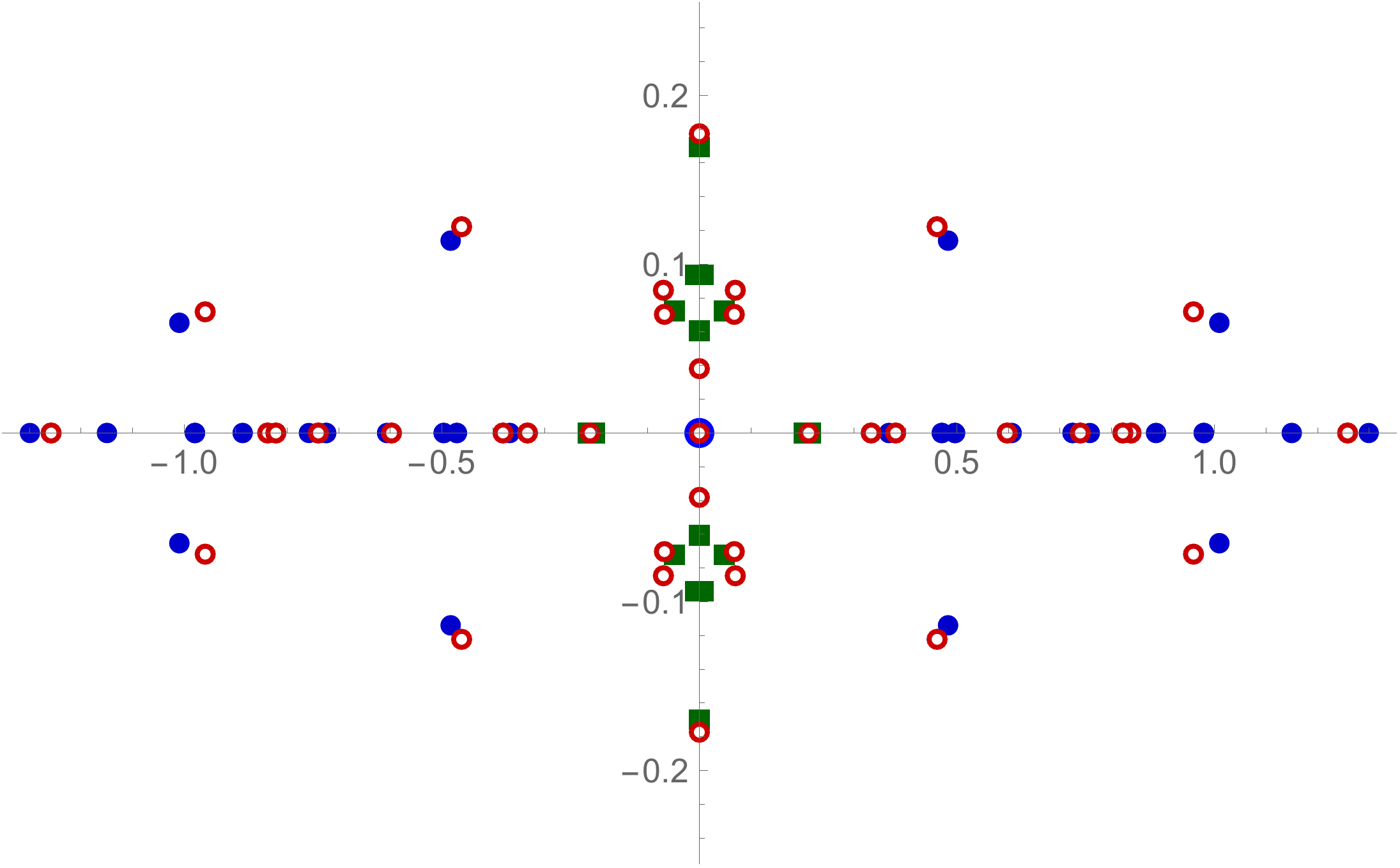}};
\end{tikzpicture}
}
\caption{
\small
The value of $b(N)$ for  the  $37=21+16$ states (open circles) with ${\tt w}=1$
and any values of ${\tt L}$ and $\bar{{\tt L}}$.
}
\end{figure}


\clearpage
\pagebreak

\end{document}